\documentclass[3p,times]{elsarticle}

\newcommand{\arxiv}[1]{}

\arxiv{\usepackage{ecrc}}
\usepackage{ecrc}

\usepackage[greek,english]{babel}

\arxiv{\volume{00}}

{\firstpage{1}}

\arxiv{\runauth{Allahverdyan, Balian and Nieuwenhuizen}}

\arxiv{\jnltitlelogo{Physics Reports}}  
\jnltitlelogo{Physics Reports}
\volume{00}
\runauth{Allahverdyan, Balian and Nieuwenhuizen}

\usepackage{amssymb}
 \usepackage{amsthm}

 \usepackage{lineno}

 \usepackage{epsfig}
 
 \usepackage{tipx}

\usepackage[figuresright]{rotating}

\newcommand{\ZeText}[1]{}
\renewcommand{\ZeText}{}
 
\newcommand{\myskipfigText}{}

\renewcommand{\ni}{\noindent}
  \newcommand{\myskip}[1]{}
  \newcommand{\mytext}[1]{}
  
 \newcommand{\scriptD}{{\hat{\cal D}}}
 \newcommand{\scriptR}{\hat{\cal R}}
\newcommand{\scriptE}{{\cal E}}
\newcommand{\scriptA}{{\cal A}}

\newcommand{\mF}{m_{\rm F}}

  \newcommand{\tildeK}{\tilde K}
  
 \newcommand{\up}{\uparrow}
  \newcommand{\down}{\downarrow}
    \newcommand{\tildeGamma}{\tilde\Gamma}
  \newcommand{\half}{\frac{1}{2}}
  
  \newcommand{\BEQ }{\begin{eqnarray}}
  \newcommand{\EEQ }{\end{eqnarray}}
  \newcommand{\BEA}{\begin{eqnarray}}
  \newcommand{\EEA}{\end{eqnarray}}
  \newcommand{\comment}[1]{}
  \newcommand{\nn}{\nonumber}
    \renewcommand{\d}{{\rm d}}
  \newcommand{\p}{\partial}
\newcommand{\reg}{{\rm reg}}
\newcommand{\split}{{\rm split}}
\newcommand{\sub}{{\rm sub}}

\newcommand{\tf}{t_{\rm f}}

  \newcommand{\truncated}{truncated}
  \newcommand{\trunc}{{\rm trunc}}

  \renewcommand{\thesection}{\arabic{section}}
   \renewcommand{\theequation}{\thesection.\arabic{equation}}
     \renewcommand{\thesection}{\arabic{section}}
 \renewcommand{\thesubsection}{\thesection\arabic{subsection}}
    \renewcommand{\theequation}{\thesection\arabic{equation}}

\pdfoutput=1

\begin{document}
    
  \begin{frontmatter}
  
 \arxiv{\dochead{}}

\title{Understanding quantum measurement from the solution of dynamical models}
 
  \author[LeMans,YerPhi]{Armen E. Allahverdyan}
 
 \address[LeMans] {Laboratoire de Physique Statistique et Syst\`emes Complexes, ISMANS, 44 Av. Bartholdi, 72000 Le Mans, France}
  
 \address[YerPhi] {On leave of:  Yerevan Physics Institute, Alikhanian Brothers Street 2, Yerevan 375036, Armenia}
  
 \author[IPhT]{Roger Balian}
 
 \address[IPhT] {Institut de Physique Th\'eorique, CEA Saclay, 91191 Gif-sur-Yvette  cedex, France}

 
\author[NYU,UvA] {Theo M. Nieuwenhuizen}
 
 \address[NYU]{
 Center for Cosmology and Particle Physics, New York University,  4 Washington Place, New York, NY 10003, USA}

  \address[UvA]{
  On leave of:  Institute for Theoretical Physics, University of Amsterdam, Science Park 904, Postbus 94485, 1090 GL Amsterdam, The Netherlands}
  

\begin{abstract}
\ZeText{
The quantum measurement problem, to wit,  understanding why a unique outcome is obtained in each individual experiment, is currently tackled 
by solving models. After an  introduction we review the many dynamical models proposed over the years for elucidating quantum  
measurements. The approaches range from standard quantum theory, relying for instance on quantum statistical mechanics or on decoherence,
 to quantum-classical methods, to consistent histories and to modifications of the theory. Next, a flexible and rather realistic 
 quantum model is  introduced, describing the measurement of the $z$-component of a spin through interaction with a magnetic memory
 simulated by a  Curie--Weiss magnet, including $N \gg1$ spins weakly coupled to a phonon bath. Initially prepared in a metastable paramagnetic state, 
 it may transit to its up or down ferromagnetic state, triggered by its coupling with the tested spin, so that its magnetization acts as a pointer. 
 A detailed solution of the dynamical equations is worked out, exhibiting several time scales. Conditions on the parameters of the model are found, 
 which ensure that the process satisfies all the features of ideal measurements. Various imperfections of the measurement are discussed, as well as 
 attempts of incompatible measurements. The first steps consist in the solution of the Hamiltonian dynamics for the spin-apparatus density matrix 
 $\scriptD(t)$. Its off-diagonal blocks  in a basis selected by the spin-pointer coupling, rapidly decay owing to the many degrees of 
 freedom of the pointer. Recurrences are ruled out either by some randomness of that coupling, or by the interaction with the bath.
 On a longer time scale, the trend towards equilibrium of the magnet produces a final state $\scriptD(t_{\rm f})$ that involves correlations between
  the system and the indications of the pointer, thus ensuring registration. Although $\scriptD(t_{\rm f})$ has the form expected for ideal measurements, 
  it only describes a large set of runs. Individual runs are approached 
   by analyzing the final states associated with all possible subensembles of runs, within a specified version of the statistical interpretation.
  There the difficulty lies in a quantum ambiguity: There exist many incompatible decompositions of the density matrix $\scriptD(t_{\rm f})$ 
 into a sum of sub-matrices, so that one cannot infer from its sole determination the states that would describe small subsets of runs. 
 This difficulty is overcome by  dynamics  due to suitable interactions within the apparatus, which produce a special combination 
 of relaxation and decoherence associated with the broken invariance of the pointer. Any subset of runs thus reaches over a brief delay a stable state 
 which satisfies the same hierarchic property as in classical probability theory; the reduction of the state for each individual run follows. 
  Standard quantum statistical mechanics alone appears sufficient to explain the occurrence of a unique answer in each run and the emergence 
  of classicality in a measurement process. Finally, pedagogical exercises are proposed and lessons for future works on models are suggested, 
  while  the statistical interpretation is promoted for teaching.
 }
  \end{abstract}

 
\end{frontmatter}

\vspace{-3mm}

\hfill{ Dedicated to our teachers and inspirers Nico G. van Kampen and Albert Messiah}

\vspace{3mm}

\hfill{\it Chi va piano va sano; 
chi va sano va lontano\footnote{Who goes slowly goes safely; who goes safely goes far}}

\hfill{ Italian saying}

\ZeText{
 \tableofcontents
}



 \setcounter{section}{0}

 \renewcommand{\thesection}{\arabic{section}}
 \section{ General features of quantum measurements}
 \setcounter{equation}{0} \setcounter{figure}{0}
 \renewcommand{\thesection}{\arabic{section}.}

\label{section.1}

\hfill{{\it For this thing}  {\rm is}  {\it too heavy for thee,}

\hfill{\it  thou art not able to perform it thyself  alone}}

 \hfill{ Exodus 18.18}
\vspace{3mm}

\ZeText{

In spite of a century of progress and success, quantum mechanics still gives rise to passionate discussions about its interpretation. 
Understanding quantum measurements is an important issue in this respect, since measurements are a privileged means to grasp the microscopic physical quantities. 
Two major steps in this direction were already taken in the early days. In 1926, Born gave the expression of the 
probabilities\footnote{Born wrote: {``Will man dieses Resultat korpuskular umdeuten, so ist nur eine Interpretation m\"oglich: $\Phi_{n,m}(\alpha,\beta,\gamma)$ 
bestimmt die  Wahrscheinlichkeit$^1$) daf\"ur, da\ss \  das aus der $z$-Richtung  kommende Elektron in die durch $\alpha,\beta,\gamma$  bestimmte Richtung
[$\cdots$] geworfen wird'', with the footnote: ``$^1$) Anmerkung bei der Korrektur: Genauere \"Uberlegung zeigt, da\ss  \ die Wahrscheinlichkeit dem Quadrat der 
Gr\"o\ss e $\Phi_{nm}$  proportional ist". In translation from Wheeler and Zurek~\cite{wh}: 
``Only one interpretation is possible: $\Phi_{n,m}$ gives the probability$^1$) for the electron \ldots''}, and the footnote:
`` $^1$) Addition in proof: More careful consideration shows that the probability is proportional to the square of the quantity $\Phi_{n,m}$.''}
 of the various possible outcomes of an ideal quantum measurement  \cite{Born}, thereby providing a probabilistic foundation for the wave mechanics.
In 1927 Heisenberg conceived the first models of quantum measurements \cite{heisenberg,heisenberg_book}
that were five years later extended and formalized by von Neumann \cite{vNeumann}. The problem was thus formulated as a mathematical 
contradiction:  the Schr\"odinger equation and the projection postulate of von Neumann are incompatible.
Since then, many theorists have worked out models of quantum measurements,  with the aim of understanding not merely the dynamics of such processes, 
but in particular solving the so-called  measurement problem. This problem is raised by a conceptual contrast between quantum theory, which is 
{\it irreducibly probabilistic}, and our macroscopic experience,  in which an {\it individual process results in a well defined outcome}.  
If a measurement is treated as a quantum physical process, in which the tested system interacts with an apparatus, the superposition 
principle seems to preclude the occurrence of a unique outcome, whereas each single run of a quantum measurement should yield a unique result.
The challenge has remained to fully explain how this property emerges, ideally without introducing new ingredients, that is, from the mere laws of quantum mechanics alone.  
Many authors have tackled this deep problem of measurements with the help of models so as to get insight on the interpretation of quantum mechanics.
For historical overviews of the respective steps in the development of the theory and its interpretation, see the books by 
Jammer \cite{Jammer1989,Jammer1974} and by Mehra and Rechenberg \cite{Mehra}.
The tasks we undertake in this paper are first to review these works, then to solve in full detail a specific family of dynamical models and to finally draw conclusions 
from their solutions.

}

\subsection{Measurements and interpretation of quantum mechanics}
\label{section.1.1.1}

 \hfill{ \it Quis custodiet ipsos custodes? \footnote{Who will watch the watchers themselves?}}

\vspace{3mm}

\ZeText{

 Few textbooks of quantum mechanics dwell upon questions of interpretation or upon quantum measurements, in spite of their importance in the comprehension 
 of the theory.  Generations of students have therefore stumbled over the problem of measurement,  before leaving it aside when they pursued
  research work. Most physicists have never returned to it, considering that it is not worth spending time on a problem which \textquotedblleft probably
  cannot be solved\textquotedblright\ and which has in practice little implication on physical predictions. 
  Such a fatalistic attitude has emerged after the efforts of the brightest physicists, including Einstein, Bohr,
  de~Broglie, von~Neumann and Wigner,  failed to lead to a universally accepted
  solution or even viewpoint; see for reviews \cite{vNeumann,wh,BallentineRMP,blokhintsev1,blokhintsev2,hepp,Wigner,vKampen}. 
However, the measurement problem has never been forgotten, owing to its intimate connection with the foundations of quantum mechanics, 
which it may help to formulate more sharply, and owing to its philosophical implications.

In this review we shall focus on the simplest measurements, ideal projective measurements~\cite{vNeumann},
and shall consider non-idealities and unsuccessful processes only occasionally and in section 8.
While standard courses deal mainly with this type of measurement,
it is interesting to mention that the first experiment based on a nearly ideal measurement 
was carried out only recently~\cite{Mooij2007}. An optical analog of a von Neumann measurement has been proposed too~\cite{D'Ariano}.

Experimentalists meet the theoretical discussions about quantum measurements with 
a feeling of speaking different languages.
While theorists ponder about the initial pure state of the apparatus, the collapse of its wave packet 
and the question ``when and in which basis does this collapse occur" and ``how does this collapse agree with 
the Schr\"odinger equation",  experimentalists  deal with different issues, such as choosing
an appropriate apparatus for the desired experiment or stabilizing it before the measurement starts.
If an experimentalist were asked to describe one cubic nanometer of his apparatus in theoretical
terms, he would surely start with a quantum mechanical approach. But this raises the question
whether it is possible to describe the whole apparatus, and also its dynamics, i. e., the 
dynamics and outcome of the measurement, by quantum mechanics itself. 
It is this question that we shall answer positively in the present work, thus closing the gap
between what experimentalists intuitively feel and the formulation of the theory of ideal quantum measurements. 
To do so, we shall consider models that encompass the points relevant to experimentalists.

As said above, for  theorists there has remained another unsolved paradox, 
even deeper than previous ones, the so-called  {\it quantum measurement problem}:
How can quantum mechanics,  with its superposition principle, be compatible with the fact
that {\it each individual run of a quantum measurement yields a well-defined outcome?}
This uniqueness is at variance with the description of the measurement process by means of a
pure state, the evolution of which is governed by the Schr\"odinger equation.
Many workers believe that the quantum measurement problem cannot be answered within
quantum mechanics.  Some of them then hope that a hypothetical ``sub-quantum theory",
more basic than standard quantum mechanics, might predict what happens in individual systems
~\cite{Bassi_Ghirardi,Bohm,BohmHolland,BeyondTheQuantum}. 
Our purpose is, however, to prove that the probabilistic framework of quantum mechanics is sufficient, in spite of conceptual difficulties, 
 to explain that the outcome of a single measurement is unique although unpredictable  within this probabilistic framework (section 11).
 We thus wish to show that quantum theory not only predicts the probabilities for the various possible outcomes of a set of measurements
  -- as a minimalist attitude would state -- but also accounts for the uniqueness of the result of each run. 
 
  A measurement is the only means through which information may be gained about
  a physical system ${\rm S}$ \cite{vNeumann,wh,BallentineRMP,blokhintsev1,blokhintsev2,hepp,Wigner,vKampen,daneri,PeresBook}. 
  Both in classical and in quantum physics, it is a dynamical process which couples this system ${\rm S}$ to another
  system, the apparatus ${\rm A}$. Some correlations are thereby generated between the initial (and possibly final) state of ${\rm S}$ and the final
  state of ${\rm A}$. Observation of ${\rm A}$, in particular the value indicated by its pointer, 
  then allows us to gain by inference some quantitative information about ${\rm S}$. A measurement thus  involves, 
  in one way or another, the observers\footnote{We shall make the case that observation itself does not influence the outcome of 
  the quantum measurement}. It also has statistical features, through unavoidable uncertainties and, 
  more  deeply, through the irreducibly probabilistic nature of our description of quantum systems.

  Throughout decades many thoughts were therefore devoted to quantum
  measurements in relation to the interpretation of quantum theory. Both
  Einstein \cite{einstein_dialectica} and de Broglie~\cite{deBroglie} spent much
  time on such questions after their first discovery; the issue of quantum measurements was formulated by 
  Heisenberg \cite{heisenberg,heisenberg_book} and put in a mathematically precise form by von Neumann \cite{vNeumann};
  the foundations of quantum
  mechanics were reconsidered in this light by people like
  Bohm \cite{Bohm,BohmHolland} or Everett~\cite{Everett,ManyWorlds} in the fifties; hidden
  variables were discussed by Bell in the sixties \cite{Bellhiddenvar}; 
  the use of a statistical interpretation to analyze quantum measurements was then
  advocated by Park \cite{X4}, Blokhintsev \cite{blokhintsev1,blokhintsev2} and Ballentine~\cite{BallentineRMP} 
  (subtleties of the statistical interpretation are underlined by Home and Whitaker \cite{HomeWhitaker}); 
  the most relevant papers were collected by Wheeler and Zurek in 1983~\cite{wh}. 
  Earlier reviews on this problem were given by London and Bauer \cite{london} and Wigner \cite{Wigner}.
  We can presently witness a renewed interest for measurement theory; among
  many recent contributions we may mention the book of de~Muynck~\cite{deMuynck}
  and the review articles by Schlosshauer~\cite{Schlosshauer} and Zurek \cite{zurek}. 
  Extensive references are given in the pedagogical article ~\cite{Laloe} and book \cite{LaloeBook} by Lalo\"e
  which review paradoxes and interpretations of quantum mechanics. 
  Indeed, these questions have escaped the realm of speculation
  owing to progresses in experimental physics which allow to tackle the
  foundations of quantum mechanics from different angles. Not only Bell's 
  inequalities~\cite{Bellhiddenvar,Laloe,bellac} but also the Greenberger--Horne--Zeilinger (GHZ) logical
  paradox \cite{GHZ} have been tested experimentally~\cite{Pan2000}. Moreover,
  rather than considering cases where quantum interference terms (the infamous \textquotedblleft Schr\"{o}dinger cat problem'' 
  \cite{wh,Wigner,Schroedinger}) vanish owing to decoherence processes \cite{Guilini}, experimentalists have become
  able to control these very interferences \cite{polzik}, which are essential to describe the
  physics of quantum superpositions of macroscopic states and to explore the new possibilities offered by quantum information~ \cite{PeresBook,galindo}.
  Examples include left and right going currents in superconducting circuits \cite{Mooij2007,nakamura,nakamura1,esteve}, 
  macroscopic atomic ensembles \cite{polzik} and entangled mechanical oscillators~\cite{wineland}.

}


\subsubsection{Classical versus quantum measurements: von Neumann-Wigner theory}

\label{section.1.1.2}

\hfill{\it When the cat and the mouse agree, }

\hfill{\it the grocer is ruined} 

\hfill{Iranian proverb} 

\vspace{3mm}

\ZeText{

  The difficulties arise from two major differences between
  quantum and classical measurements, stressed in most textbooks 
  \cite{vNeumann,heisenberg_book,BuschLahtiMittelstaedt,BallentineBook}.

  (\textit{i}) In classical physics it was legitimate to imagine that quantities such as the position and the momentum of a structureless particle like an
  electron might in principle be measured with increasingly large precision;  this allowed us to regard both of them as well-defined physical quantities.
  (We return in section 10 to the meaning of рphysical quantitiesс and of рstatesс within the statistical interpretation of quantum mechanics.)
  This is no longer true in quantum mechanics, where we cannot forget about the random nature of physical quantities. Statistical fluctuations are
  unavoidable, as exemplified by Heisenberg's inequality \cite{heisenberg,heisenberg_book}: we cannot even think
  of values that would be taken simultaneously by non-commuting quantities whether or not we measure them. In general both the theory and the measurements
  provide us only with \textit{probabilities}. 
  
  Consider a measurement of an observable $\hat{s}$ of the system ${\rm S}$ of interest\footnote{The eigenvalues of $\hat s$ are assumed here to be 
  non-degenerate. The general case will be considered in \S~1.2.3}, having eigenvectors $|s_{i}\rangle$ and eigenvalues $s_{i}$. It is an experiment in
  which ${\rm S}$ interacts with an apparatus ${\rm A}$ that has the  following property \cite{vNeumann,Wigner,london,BuschLahtiMittelstaedt}. 
  A physical quantity $\hat{A}$ pertaining to the apparatus ${\rm A}$ may take at the end of the process one value among a set $A_{i}$ which are
  in one-to-one correspondence with $s_{i}$. If initially ${\rm S}$ lies in the state $\left\vert s_{i}\right\rangle $, the final value $A_{i}$ will be produced 
  with certainty, and a repeated experiment will always yield the observed result  ${A}_{i}$, informing us that ${\rm S}$ was in $\left\vert s_{i}\right\rangle $ 
  However, within this scope, S should generally lie initially in a state represented by a wave function which is a linear combination,
 
  \begin{equation}
  \left\vert \psi\right\rangle
  =\sum_{i}\psi_{i}\left\vert s_{i}\right\rangle,
   \label{Bal0}
  \end{equation}

  \ni of the eigenvectors $\left\vert s_{i}\right\rangle $. \textit{Born's rule} then states that the probability of observing in a given experiment the result
  $A_{i}$ equals $\left\vert \psi_{i}\right\vert ^{2}$ \cite{Born}.  A prerequisite to the explanation of this rule is the solution of the measurement problem, 
  as it implicitly involves the uniqueness of the outcome of the apparatus in each single experiment. An axiomatic derivation of 
  Born's rule is given in \cite{gleason}; see
  \cite{Schlosshauer,zurek} for a modern perspective on the rule. Quantum mechanics does not allow us to predict which will be the outcome $A_i$ of an
  individual measurement, but provides us with the full statistics of repeated measurements of $\hat{s}$ performed on elements of
  an ensemble described by the state $\left\vert \psi\right\rangle $. The frequency of occurrence of each $A_{i}$ in repeated experiments informs us
  about the moduli $\left\vert \psi_{i}\right\vert ^{2}$, but not about the phases of these coefficients. In contrast to a classical state,\ a quantum
  state $|\psi\rangle$, even pure, always refers to an ensemble, and cannot be determined by means of a unique measurement performed on a single 
  system~\cite{DArianoYuen}.  
  It cannot even be fully determined by repeated measurements of the single observable $\hat{s}$, since
  only the values of the amplitudes $\left\vert \psi_{i}\right\vert $ can thus be estimated.

  (\textit{ii}) A second qualitative difference between classical and quantum
  physics lies in the \textit{perturbation of the system} ${\rm S}$ brought
  in by a measurement. Classically one may imagine that this perturbation could
  be made weaker and weaker, so that ${\rm S}$ is practically left in its
  initial state while ${\rm A}$ registers one of its properties. However, a
  quantum measurement is carried on with an apparatus ${\rm A}$ much larger
  than the tested object ${\rm S}$; an extreme example is provided by the huge
  detectors used in particle physics. Such a process may go so far as to destroy
  ${\rm S}$, as for a photon detected in a photomultiplier. 
  It is natural to wonder whether the perturbation of ${\rm S}$ has a lower bound. Much work
  has therefore been devoted to the \textit{ideal measurements}, 
  those which preserve at least the statistics of the observable $\hat{s}$ in the final state of S, 
    also referred to as non-demolition experiments or as measurements of the first kind~\cite{deMuynck}. 
  Such ideal measurements are often described by assuming that the apparatus A starts in a pure state\footnote{\label{pureA} Here we follow 
  a current line of thinking in the literature called von Neumann-Wigner theory of ideal measurements.
   In subsection 1.2 we argue that it is not realistic to assume that  A may start in a pure state and end up in a pure state}.     
  Then by  writing that, if ${\rm S}$ lies initially in the state $\left\vert s_{i}  \right\rangle $ and ${\rm A}$ in the state $\left\vert 0\right\rangle $,
  the measurement leaves ${\rm S}$ unchanged: the compound system  ${\rm S}+{\rm A}$ evolves from $\left\vert s_{i}\right\rangle \left\vert
  0\right\rangle $ to $\left\vert s_{i}\right\rangle \left\vert A_{i}  \right\rangle $, where $\left\vert A_{i}\right\rangle $ is an eigenvector of
  $\hat{A}$ associated with $A_{i}$. If however, as was first discussed by  von~Neumann, the initial state of ${\rm S}$ has the general form
  (\ref{Bal0}), ${\rm S}+{\rm A}$ may reach any possible final state   $\left\vert s_{i}\right\rangle \left\vert A_{i}\right\rangle $ depending on
  the result $A_{i}$ observed. In this occurrence the system ${\rm S}$ is left in $\left\vert s_{i}\right\rangle $ and ${\rm A}$ in $\left\vert
  A_{i}\right\rangle $, and according to Born's rule, this occurs with the probability $\left\vert \psi_{i}\right\vert ^{2}$.  
  As explained in \S~1.1.2, for this it is necessary (but not sufficient) to require that the
  final density operator describing S + A for the whole set of runs of the measurement has the diagonal form$^{\ref{pureA}}$
  
  \begin{equation}
  \mytext{\textcurrency Bal2\textcurrency \qquad}
  \sum_{i}\left\vert s_{i}  \right\rangle \left\vert A_{i}\right\rangle \left\vert \psi_{i}\right\vert  ^{2}\left\langle A_{i}\right\vert \left\langle s_{i}\right\vert {,}
  \label{Bal2}
  \end{equation}
  rather than the full form (1.3) below. Thus, not only is the state of the apparatus modified in a way controlled 
  by the object, as it should in any classical or quantal measurement that provides us
  with information on ${\rm S}$, but the marginal state of the quantum system
  is also necessarily modified  (it becomes $  \sum_{i}\left\vert s_{i}  \right\rangle \left\vert \psi_{i}\right\vert  ^{2}\left\langle s_{i}\right\vert$), 
  even by such an ideal measurement (except in the
  trivial case where (\ref{Bal0}) contains a single term, as it happens when one repeats the measurement).

 }


 \subsubsection{Truncation versus reduction}
 \label{section.1.1.3}
 
 \hfill{\it Ashes to ashes, }
 
 \hfill{\it dust to dust}
 
\hfill{Genesis 3:19}
 
 \vspace{3mm}
 
 \ZeText{

The rules of quantum measurements that we have recalled display a well known contradiction between the principles of quantum mechanics. On the one hand, 
 if the measurement process leads the initial pure state $|s_i\rangle|0\rangle$ into $|s_i\rangle |A_i\rangle$,  the linearity of the wave functions of the compound
  system ${\rm S}+{\rm A}$ and the unitarity of  the evolution of the wave functions of S + A governed by the Schr\"odinger equation
  imply that the final density operator of ${\rm S}+{\rm A}$  issued from (\ref{Bal0}) should be$^{\ref{pureA}}$

  \begin{equation}
  \mytext{\textcurrency Bal1\textcurrency \qquad}
  \sum_{ij}\left\vert s_{i}\right\rangle \left\vert A_{i}\right\rangle \psi_{i}\psi_{j}^{\ast}
  \langle A_{j}\vert \langle s_{j}\vert.
  \label{Bal1}
  \end{equation}
  
  \ni On the other hand, according to Born's rule \cite{Born} and von~Neumann's analysis \cite{vNeumann}, each run of an ideal measurement should 
  lead from the initial pure state $\left\vert \psi\right\rangle \left\vert 0\right\rangle$ to one or another of the pure states 
  $\left\vert s_i\right\rangle \left\vert A_i\right\rangle$ with the probability $|\psi_i|^2$; the final density operator accounting for a large statistical ensemble 
  $\scriptE$ of runs should be the mixture (1.2) rather than the superposition (1.3).
   In the orthodox Copenhagen interpretation, two separate postulates of evolution are 
  introduced, one for the hamiltonian motion governed by the Schr\"odinger equation, the other for measurements which lead the system from $|\psi\rangle$ 
  to one or the other of the states $|s_i\rangle$, depending on the value $A_i$ observed. This lack of
  consistency is unsatisfactory and other explanations have been searched for (\S ~\ref{section.1.3.1} and section 2).
  
  It should be noted that the loss of the off-diagonal elements takes place in a well-defined basis, the one in which both the tested observable $\hat s$ of S 
  and the pointer variable $\hat A$ of A are diagonal (such a basis always exists since the joint Hilbert space of S + A is the tensor product of the spaces of S and A). 
   In usual decoherence processes, it is the interaction between the system and some external bath which selects the basis in which off-diagonal 
  elements are truncated~\cite{Schlosshauer,zurek}. We have therefore to elucidate this {\it preferred basis paradox}, and to explain why the truncation 
  which replaces (1.3) by (1.2) occurs in the specific basis selected by the measuring apparatus.

  The occurrence in (\ref{Bal1}) of the off-diagonal $i\neq j$ terms is by itself an essential feature of an interaction process between two systems in
  quantum mechanics. There exist numerous experiments in which a pair of systems is left after interaction in a state of the form (\ref{Bal1}), not only at the
  microscopic scale, but even for macroscopic objects, involving for instance quantum superpositions of superconducting currents. 
  Such experiments allow us to observe purely quantum coherences represented by off-diagonal terms   $i\neq j$.
  
   However, such off-diagonal  \textquotedblleft Schr\"{o}dinger cat\textquotedblright\ terms, which contradict both Born's rule \cite{Born} and 
   von~Neumann's reduction \cite{vNeumann},  must disappear at the end of a measurement process. 
   Their absence is usually termed as the ``reduction'' or the ``collapse'' of the wave packet, or of the state. Unfortunately, depending on the authors, 
   these words may have different meanings; we need to define precisely our vocabulary. Consider first {\it a large set $\scriptE$ of runs} of a measurement performed 
   on identical systems S initially prepared in the state $|\psi\rangle$,  and interacting with A initially in the state $|0\rangle$.  The density operator of S + A should 
   evolve from  $|\psi\rangle|0\rangle\langle0|\langle\psi|$  to (1.2). We will term as ``{\it truncation}'' the elimination during the process 
   of the off-diagonal blocks $i\neq j$ of the density operator describing the {\it joint system} S + A {\it for the whole set $\scriptE$ of runs}. If instead of the full set $\scriptE$
    we focus on a {\it single run}, the process should end up in one among the states $\left\vert s_i\right\rangle \left\vert A_i\right\rangle$. 
    We will designate as ``{\it reduction}'' the transformation of the initial state of S + A into such a final ``{\it reduced state}'', {\it for a single run} of the measurement. 
    
	One of the paradoxes of the measurement theory lies in the existence of several possible final states issued from the same initial state. 
    Reduction thus seems to imply a {\it bifurcation} in the dynamics, whereas the Schr\"odinger equation entails a one-to-one correspondence between 
    the initial and final states of the isolated system S + A.
    
We stress that both above definitions refer to S + A. Some authors apply the words reduction or collapse to the {\it sole tested system} S. 
To avoid confusion,  we will call ``{\it weak reduction}'' the transformation of the initial state $|\psi\rangle\langle\psi|$ of S into the pure state 
    $|\psi_i\rangle\langle\psi_i|$  for a single run, 
      and ``{\it weak truncation}'' its transformation into the mixed state   $\sum_i |\psi_i\rangle\,|\psi_i|^2\, \langle\psi_i|$
  for a large ensemble $\scriptE$ of runs. In fact, the latter marginal density operator of S can be obtained by tracing out A, not only from the joint truncated state (1.2) 
  of S+A, but also merely from the non-truncated state (1.3),  so that the question seems to have been eluded. However, such a viewpoint,
  in which the apparatus is disregarded, cannot provide an answer to the measurement problem. The very aim of a measurement is to create correlations between 
  S and A and to read the indications of A so as to derive indirectly information about S; but the elimination of the apparatus suppresses both the correlations 
  between S and A and the information gained by reading A. 
       
   Physically, a set of repeated experiments involving interaction of S and A can be regarded as
   a measurement only if we observe on A in each run some well defined result $A_i$,
   with probability $|\psi_i|^2$. For an ideal measurement we should be able to predict that S is then left in the corresponding state $|s_i\rangle$.
    Explaining these features requires that the considered dynamical model produces in each run one of the reduced states
    $\left\vert s_i\right\rangle \left\vert A_i\right\rangle$. The quantum measurement problem thus amounts to the proof, not only of truncation, but also of reduction. 
    This will be achieved in section 11 for a model of quantum statistical mechanics.
  As stressed by Bohr and Wigner, the reduction, interpreted as expressing the ``uniqueness of physical reality'', is at variance with the superposition 
  principle which produces the final state (\ref{Bal1}). The challenge is to solve this contradiction, answering Wigner's wish: 
  ``{\it The simplest way that one may try to reduce the two kinds of changes of the state vector to a single kind is to describe the whole 
  process of measurement as an event in time, governed by the quantum mechanical 
  equations of motion}''. Our purpose is to show that this is feasible, contrary to 
  Wigner's own negative conclusion \cite{Wigner}.
  
  }

  \clearpage
  
\subsubsection{Registration and selection of outcomes }
   \label{section.1.1.4}
   
   \hfill{\it Non-discrimination is a cross-cutting principle}

\hfill{United Nations human rights, 1996}
   
   \vspace{3mm}
 
 \ZeText{
 
 When after a run of an ideal measurement, S is left in $|s_i\rangle$, a second measurement performed on the same system leaves this state unchanged 
 and yields the same indication $A_i$ of the pointer. Reduction, even weak, thus implies {\it repeatability}. Conversely, repeatability implies weak truncation, 
 that is, the loss in the marginal density of S of the elements $i\neq j$ during the first one of the successive measurement processes~\cite{BalianAJPh1989}.
 
 Apart from having been truncated, the final density operator (1.2) of S + A for the whole set $\scriptE$ of runs displays an essential feature, the complete correlation 
 between the indication $A_i$ of the pointer and the state $|s_i\rangle$ of S. We will term as ``{\it registration}'' the establishment of these correlations. 
 If they are produced, we can ascertain that, if the pointer takes a well defined value $A_i$ in some run, its observation will imply that $\hat s$ takes with certainty 
 the corresponding eigenvalue $s_i$ at the end of this run. Sorting the runs according to the outcome $A_i$ allows us to split the ensemble $\scriptE$ into 
 subensemble $\scriptE_i$, each one labelled by $i$ and described by the state $|s_i\rangle|A_i\rangle$$^{\ref{pureA}}$. 
 Selection of the subensemble $\scriptE_i$ by filtering the values $A_i$ therefore allows us to set S into this subensemble  ${\cal E}_i$ described by the 
 density operator $|s_i\rangle |A_i\rangle\angle  A_i|\langle  s_i |$. 
  It is then possible to sort the runs according to the indication $A_i$ of the pointer. Selecting thus the sub ensemble 
 ${\cal E}_i$ by filtering $A_i$ allows us to set S into the given state $\left\vert
  s_{i}\right\rangle $ with a view to future experiments on S. An ideal
  measurement followed by filtering can therefore be used as a
  \textit{preparation} of the state of  ${\rm S}$~\cite{balianveneroni}.
We will make the argument more precise in \S~\ref{fin10.2.2} and \S~\ref{fin11.3.2}.

  Note that some authors call \textquotedblleft measurement\textquotedblright\ a single run of the experiment, or a repeated experiment
   in which the occurrence of {\it some given eigenvalue of}
  $\hat{s}$ is detected, and in which only the corresponding events are selected for the outcoming system ${\rm S}$. Here we use the term \textquotedblleft
{\it   measurement}\textquotedblright\ to designate a repeated experiment performed on a {\it large ensemble of identically prepared systems}
  which informs us about {\it all possible values} $s_{i}$ of the observable $\hat{s}$ of ${\rm S}$, and the term \textquotedblleft ideal measurement\textquotedblright\ if the 
  process perturbs S as little as allowed by quantum mechanics, in the sense that it does not affect the statistics of the observables that commute with $\hat{s}$. 
  We do not regard the sorting as part of the measurement, but as a subsequent operation, and prefer to reserve the word
  \textquotedblleft {\it preparation through measurement}\textquotedblright\ to such processes including a selection.

}

\subsection{The need for quantum statistical mechanics}
\label{section.1.2}

 \hfill {{\it Om een paardendief te vangen heb je een paardendief nodig}\footnote{To catch a horse thief, you need a horse thief}}

\noindent\noindent
\hfill {{\it Un coupable en cache un autre}\footnote{ One culprit hides another}}

\hfill {Dutch and French proverbs}

\vspace{3mm}
\ZeText{

 We wish for consistency to use quantum mechanics for treating the dynamics of the interaction process between the apparatus and the tested system. 
 However, the apparatus must be a macroscopic object in order to allow the outcome to be read off from the final position of its pointer.  The natural framework 
 to reconcile these requirements is non-equilibrium quantum statistical mechanics, and not quantum mechanics of pure states as presented above. 
 It will appear that not only the registration process can be addressed in this way, but also the truncation and the reduction.
 
}

\subsubsection{Irreversibility of measurement processes}
\label{section.1.2.1}

 \hfill{\it The first time ever I saw your face} 
 
 \hfill{\it I thought the sun rose in your eyes } 
 
 \hfill{Written by  Ewan MacColl, sung by Roberta Flack}




\vspace{3mm}

\ZeText{

  Among the features that we wish to explain, the \textit{truncation} compels us  to describe states by means of density operators. The sole use of pure states
  (quantum states describable by a wave function or a ket), is prohibited by the form of (\ref{Bal2}), which is in general a statistical mixture. Even if we start from
  a pure state $\left\vert \psi\right\rangle \left\vert 0\right\rangle $, we must end up with the truncated mixed state (\ref{Bal2}) through an
  \textit{irreversible} process. This irreversibility is also exhibited by the fact the same final state (\ref{Bal2}) is reached if one starts from different
  initial states of the form (\ref{Bal0}) deduced from one another through changes of the phases of the coefficients $\psi_{i}$. Such a feature is associated with
  the disappearance of the specifically quantum correlations between ${\rm S}$ and ${\rm A}$ described by the off-diagonal terms of (\ref{Bal1}).

  Actually, there is a second cause of irreversibility in any effective measurement process. The apparatus ${\rm A}$ should
  \textit{register} the result $A_{i}$ in a robust and permanent way, so that it can be read off by any observer. Such a registration, which is often overlooked 
  in the literature on measurements, is needed for practical reasons especially since we wish to explore microscopic objects. 
  Moreover, its very existence allows us to disregard the observers in quantum measurements. Once
  the measurement has been registered, the result becomes {\it objective} and can be read off at any time by any observer.
 It can also be processed automatically, without being read off. Registration requires an \textit{amplification} within the
  apparatus of a signal produced by interaction with the microscopic system ${\rm S}$. For instance, in a bubble chamber, the
  apparatus in its initial state involves a liquid, overheated in a metastable phase. In spite of the weakness of the interaction
  between the particle to be detected and this liquid, amplification and registration of its track can be achieved owing to local
  transition towards the stable gaseous phase. This stage of the measurement process thus requires an irreversible phenomenon. It
  is governed by the kinetics of bubble formation under the influence of the particle and implies a dumping of free energy.
  Similar remarks hold for photographic plates, photomultipliers and other types of detectors.

  Since the amplification and the registration of the measurement results require the apparatus ${\rm A}$ to be a large object so as to behave
  irreversibly, we must use quantum statistical mechanics to describe ${\rm A}$. In particular, the above assumption that A lay initially in a pure state $|0\rangle$ 
was unrealistic -- nevertheless this assumption is frequent in theoretical works on measurements, see e.g. ~\cite{Everett,Schlosshauer,zurek}. 
Indeed, preparing an object in a pure state requires controlling a complete set of
commuting observables, performing their measurement and selecting the outcome (\S~\ref{section.1.1.4}). 
While such operations are feasible for a few variables, they cannot be carried out for a macroscopic
apparatus nor even for a mesoscopic apparatus involving, say, 1000 particles. 
What the experimentalist does in a quantum measurement is quite the opposite \cite{blokhintsev1,blokhintsev2,heisenberg_book,deMuynck}: 
rather than purifying the initial state of A, he lets it stabilize macroscopically by controlling a few 
collective parameters such as the temperature of the apparatus. The adequate theoretical representation of the
initial state of A, which is a {\it mixed state}, is therefore a {\it density operator} denoted as ${\hat {\cal R}(0)}$.
Using pure states in thought experiments or models would require averaging so as to reproduce the actual situation
(\S~\ref{fin10.2.3} and \S~\ref{fin12.1.4}). Moreover the initial state of A should be {\it metastable}, which requires a sudden change of, e.g., temperature.

  Likewise the final possible stable marginal states of A are not pure.  As we know from quantum statistical physics, 
  each of them, characterized by the value of the pointer variable $A_{i}$ that will be observed, should again be described by means 
  of a density operator $\hat{{\cal R}}_{i}$, and not by means of pure states $|A_i\rangle$ as in (\ref{Bal1}).
  Indeed, the number of state vectors associated with a sharp value of the \textit{macroscopic} pointer variable $A_{i}$ is huge for any actual
  measurement: As always for large systems, we must allow for small fluctuations, negligible in relative value,\ around the mean value
  $A_{i}={\rm {tr}}_{{\rm A}}\hat{A}\hat{{\cal R}}_{i} $. The fact that the possible final states $\hat{{\cal R}}_{i}$ are exclusive is expressed 
  by ${\rm {tr}}_{{\rm A}}\hat{{\cal R}}_{i}\hat{{\cal R}}_{j}\simeq 0$ for $j\neq i$,  which implies 
  
  \begin{equation}
  \mytext{\textcurrency RineqRj\textcurrency \qquad}
    \hat{{\cal R}}_{i}
  \hat{{\cal R}}_{j}{ \to 0} \quad 
  \textrm{for $N\to\infty$ when $i\neq j$.}
  \label{RineqRj}
  \end{equation}
In words, these macroscopic pointer states are practically orthogonal.

}

  \subsubsection{The paradox of irreversibility}
\label{section.1.2.2}

 \hfill{\it La vida es sue\~no\footnote{Life is a dream }} 

\hfill{Calder\'on de la Barca}

\vspace{3mm}

\ZeText{

  If we disregard the system ${\rm S}$, the irreversible process leading  ${\rm A}$ from the initial state $\hat{{\cal R}}\left(  0\right)  $ to
  one among the final states $\hat{{\cal R}}_{i}$ is reminiscent of relaxation processes in statistical physics, and the measurement problem
  raises the same type of puzzle as the paradox of irreversibility. In all problems of statistical mechanics, the evolution is governed at the
  microscopic level by equations that are invariant under time-reversal: Hamilton or Liouville equations in classical physics, Schr\"{o}dinger, or
  Liouville--von~Neumann equations in quantum physics. Such equations are reversible and conserve the von~Neumann entropy, which measures our missing
  information. Nevertheless we observe at our scale an irreversibility, measured by an increase of macroscopic entropy. The explanation of this paradox, see,
  e.g.,~\cite{mayer_mayer,krylov,Callen,BalianPoincare03,BalianUtrecht,AllahverdyanNieuwenhuizenPRE2006}, 
  relies on the {\it large number} of microscopic degrees of freedom of thermodynamic systems, on {\it statistical considerations} and on plausible assumptions
   about the {\it dynamics} and about the {\it initial state} of the system.
   
   Let us illustrate these ideas by recalling the historic example of a classical gas, for which the elucidation of the paradox was initiated by 
   Boltzmann \cite{mayer_mayer,krylov,Callen}. The microscopic state of a set of $N$ structureless particles enclosed in a vessel is represented at each time by a 
   point $\xi(t)$ in the $6N$-dimensional phase space, the trajectory of which is generated by Hamilton's equations, the energy $E$ being conserved. 
   We have to understand why, starting at the time $t=0$ from  a more or less arbitrary initial state with energy $E$, we always observe that the gas reaches at 
   the final time $t_{\rm f}$ a state which macroscopically has the equilibrium properties associated with $N$ and $E$, to wit, homogeneity and 
   Maxwellian distribution of momenta -- whereas a converse transformation is never seen in spite of the reversibility of the dynamics. 
   As we are not interested in a single individual process but in generic features, we can resort to statistical considerations. 
   We therefore consider an initial macroscopic state ${\cal S}_{\rm init}$ characterized by given values of the (non uniform) densities of particles, of energy, 
   and of momentum in ordinary space. 
    Microscopically, ${\cal S}_{\rm init}$ can be realized by any point $\xi_{\rm init}$ lying in some volume $\Omega_{\rm init}$ of phase space. 
    On the other hand, consider the volume ${ \Omega}_E$ in phase space characterized by the total energy $E$. 
    A crucial fact is that the immense majority of points $\xi$ with energy $E$ have macroscopically the equilibrium properties 
    (homogeneity and Maxwellian distribution): the volume ${ \Omega}_{\rm eq}$ of phase space associated with equilibrium occupies 
    nearly the whole volume $ {\Omega}_{\rm eq} /{\Omega}_E\simeq 1$.
     Moreover, the volume $\Omega_E$ is enormously larger than $\Omega_{\rm init}$. 
   We understand these properties by noting that the phase space volumes characterized by some macroscopic property are proportional to the 
   exponential of the thermodynamic entropy. In particular, the ratio $\Omega_{\rm eq}/\Omega_{\rm init}$ is the exponential of the increase of entropy 
   from ${\cal S}_{\rm init}$ to ${\cal S}_{\rm eq}$, which is large as $N$. We note then that Hamiltonian dynamics implies Liouville's theorem. 
   The bunch of trajectories issued from the points $\xi(0)$ in $\Omega_{\rm init}$ therefore reach at the time $t_{\rm f}$ a final volume 
   $\Omega_{\rm f} = \Omega_{\rm init}$
    that occupies only a tiny part of $\Omega _E$, but which otherwise is expected to have nothing special owing to the complexity of 
    the dynamics of collisions. Thus most end points $\xi(t_{\rm f})$ of these trajectories are expected to be typical points of $\Omega _E$, 
    that is, to lie in the equilibrium region $\Omega_{\rm eq}$. Conversely, starting from an arbitrary point of $\Omega _E$ or of $\Omega_{\rm eq}$, 
    the probability of reaching a point that differs macroscopically from equilibrium is extremely small, since such points occupy a negligible 
    volume in phase space compared to the equilibrium points. The inconceivably large value of Poincar\'e's {\it recurrence time} is also related 
    to this geometry of phase space associated with the macroscopic size of the system.
    
     The above argument has been made rigorous \cite{mayer_mayer,krylov,Callen} by merging the dynamics and the statistics, that is, by studying the 
     evolution of the density in phase space, the probability distribution which encompasses the bunch of trajectories evoked above. Indeed, it is 
     easier to control theoretically the Liouville equation than to study the individual Hamiltonian trajectories and their statistics for random initial 
     conditions. The initial state of the gas is now described by a non-equilibrium density in the $6N$-dimensional phase space. Our full information 
     about this initial state, or the full order contained in it, is conserved by the microscopic evolution owing to the Liouville theorem. However, the
      successive collisions produce correlations between larger and larger numbers of particles. Thus, while after some time the gas reaches at the 
      macroscopic scale the features of thermodynamic equilibrium, the initial order gets hidden into microscopic variables, namely many-particle 
correlations, that are inaccessible. Because the number of degrees of freedom is large --  and it is actually gigantic for any macroscopic object --
       this order cannot be retrieved (except in some exceptional controlled dynamical phenomena such as spin echoes 
       \cite{spin_echo,spin_echo1,spin_echo2a,spin_echo2b,spin_echo3,spin_echo4}).
       In any real situation, it is therefore impossible to recover, for instance, a non-uniform density from the very complicated correlations created during 
       the relaxation process. For all practical purposes, we can safely keep track, even theoretically, only of the correlations between a number of particles small
        compared to the total number of particles of the system: the exact final density in phase space cannot then be distinguished from a 
        thermodynamic equilibrium distribution. It is this dropping of information about undetectable and ineffective degrees of freedom, impossible
         to describe even with the largest computers, which raises the macroscopic entropy \cite{mayer_mayer,krylov,Callen,BalianPoincare03,BalianUtrecht}. 
 Such approximations can be justified mathematically through limiting processes where $N \to \infty$.

Altogether, irreversibility can be derived rigorously for the Boltzmann gas under assumptions of smoothness and approximate 
 factorization of the single particle density. The change of scale modifies qualitatively the properties of the dynamics, for all accessible 
  times and for all accessible physical variables. The {\it  emergence of an irreversible relaxation} from the reversible microscopic dynamics 
  is a statistical phenomenon which becomes nearly deterministic owing to the large number of particles. We shall encounter similar features
 in quantum measurement processes.

}


\vspace{3cm}

  \subsubsection{Quantum measurements in the language of statistical physics}
  \label{section.1.2.3}

 \hfill{\it  Now the whole earth was of one language and of one speech\footnote{
{Metaphorically, the discovery of quantum theory and the lack of agreement about its interpretation may be phrased in Genesis 11~\cite{Bible}:} \\
1.  Now the whole earth was of one language and of one speech. 
2. And it came to pass, as they journeyed from the east, that they found a plane in the land of Shinar; and they dwelt there.
 3. And they said one to another, Go to, let us make brick, and burn them throughly. And they had brick for stone, and slime had they for mortar.
 4. And they said, Go to, let us build a city, and a tower whose top {\it may reach} unto heaven; and let us make us a name, lest we be scattered
 abroad upon the face of the whole earth.
 5. And the Lord came down to see the city and the tower, which the children of men builded. 
 6. And the Lord said, Behold, the people {\it is} one, and they have all one language; and this they begin to do: and now nothing will be restrained 
 from them, which they have imagined to do.
 7. Go to, let us go down, and there confound their language, that they may not understand one another's speech.
 8. So the Lord scattered them abroad from thence upon the face of all the earth: and they left off to build the city.
 9. Therefore is the name of it called Babel; because the Lord did there confound the language of all the earth: and from thence did the Lord
 scatter them abroad upon the face of all the earth} }

\hfill{Genesis 11:1}

\vspace{3mm}

\ZeText{

 The theoretical description of a measurement process should be inspired by the above ideas.
  Actually, a measurement process looks like a relaxation process,
  but with several complications. On the one hand, the final stable state of
  ${\rm A}$ is not unique, and the dynamical process can have  {\it several possible outcomes} for $A$.
   In photodetection (the eye, a photomultiplier), one just detects whether an avalanche has or not been created 
   by the arrival of a photon. In a magnetic dot, one detects the  direction  of the magnetization.
  The apparatus is therefore comparable to a material
  which, in statistical physics, has a broken invariance and can relax towards
  one equilibrium phase or another, starting from a single metastable phase. On
  the other hand, the evolution of ${\rm A}$ towards one among the final
  states $\hat{{\cal R}}_{i}$ characterized by the variable $A_{i}$ should be
  triggered by interaction with ${\rm S}$, in a way depending on the initial
  microscopic state of ${\rm S}$ and, for an ideal measurement,
   the final microscopic state of S should be {\it correlated} to the outcome   $A_{i}$.  
  Thus, contrary to theories of standard relaxation processes in statistical
  physics, the theory of a measurement process will require a simultaneous
  control of {\it  microscopic and macroscopic} variables. In the coupled evolution of
  ${\rm A}$ and ${\rm S}$ which involves truncation and registration,
  coarse graining will be adequate for ${\rm A}$, becoming exact in the limit
  of a large ${\rm A}$, but not for ${\rm S}$. Moreover the final state of
  ${\rm S}+{\rm A}$ keeps \textit{memory} of the initial state of
  ${\rm S}$, at least partly. The very essence of a measurement lies in this
  feature, whereas memory effects are rarely considered in standard relaxation processes.

  Denoting by $\hat{r}\left(0\right)$ and $\hat {\cal R}(0)$ the density operators of the system
  ${\rm S}$ and the apparatus A, respectively, before the measurement, the initial state of ${\rm S}
  +{\rm A}$ is characterized in the language of quantum statistical
  mechanics\ by the density operator
  \begin{equation}
  \mytext{\textcurrency D(0)=\textcurrency \qquad}
  {\hat{{\cal D}}}\left(
  0\right)  =\hat{r}\left(  0\right)  \otimes\hat{{\cal R}}\left(  0\right)
  { .} \label{D(0)=}
  \end{equation}
  
  \ni In the Schr\"odinger picture, where the wave functions evolve according to the Schr\"odinger equation 
  while observables are time-independent, the density operator $\hat{\cal D}(t)=\exp(-i\hat Ht/\hbar)\hat{\cal D}(0)\exp(i\hat Ht/\hbar)$ 
  of the compound system S + A evolves according to the Liouville-von Neumann equation of motion 
  
  \renewcommand{\d}{{\rm d}}
  
  \begin{equation}  \label{LB001}
  i\hbar\frac{\d\hat{\cal D}}{\d t}=[\hat H,\hat{\cal D}]
 \equiv \hat H\hat{\cal D}-\hat{\cal D}\hat H,
    \end{equation} 
  
  \ni where $\hat H$ is the Hamiltonian of S + A including the interaction between S and A.
  By solving (1.6) with the initial condition (1.5), we find the expectation value  $\langle\hat A(t)\rangle$ 
  of any observable $\hat A$ of S + A at the time $t$ as ${\rm tr} [ \hat{\cal D}(t)\hat A]$  (see subsection 10.1 and Appendix G).

We first wish to show that, for an ideal measurement, the final density operator of S + A which represents the outcome af a large set 
$\scriptE$ of runs at the time $t_{\rm f}$ has the form

  \begin{equation}
  \mytext{\textcurrency Dtf=\textcurrency \qquad}
  {\hat{{\cal D}}}\left(
  {t_{{\rm f}}}\right)  =\sum_{i}\left(  \hat{\Pi}_{i}\hat{r}\left(
  0\right)  \hat{\Pi}_{i}\right)  \otimes\hat{{\cal R}}_{i}
  =\sum_ip_i\hat r_i\otimes\hat{\cal R}_i{ ,}
  \label{Dtf=}
  \end{equation}
  where $\hat{\Pi}_{i}$  denotes the projection operator  (satisfying $\hat\Pi_i\hat\Pi_j=\delta_{ij}\hat\Pi_i$)
on the eigenspace  $s_{i}$ of $\hat{s}$ in the Hilbert space of ${\rm S}$, with $\hat{s}=\sum_{i}s_{i}\hat{\Pi}_{i}$
and $\sum_i\hat\Pi_i=\hat{\rm I}$.   (If the eigenvalue $s_i$ is non-degenerate, $\hat{\Pi}_{i} $ is simply equal 
to $|s_i\rangle\langle s_i|$.)  We have denoted by
 
 \BEQ 
 \hat r_i=\frac{1}{p_i}\hat\Pi_i\hat r(0)\hat\Pi_i 
 \EEQ
 the corresponding normalized projected state (which reduces to $|s_i\rangle\langle s_i|$ if  $s_i$ is non-degenerate), and by
 
 \BEQ
 \label{p_i=}
 p_i\equiv {\rm tr}_{\rm S}\hat r(0)\hat\Pi_i
 \EEQ
 the normalizing factor (which reduces to $r_{ii}(0)$ if $s_i$ is non-degenerate). 
 The expression  (\ref{Dtf=}) generalizes (1.2) to arbitrary density operators; we will use the same vocabulary as in \S~1.1.2 to designate its various features. 
 This generalization was first conceived by L\"uders \cite{Lueders}. The lack in (\ref{Dtf=}) 
 of off-diagonal blocks $i\neq j$ in a basis where $\hat s$ is diagonal expresses {\it truncation}. The correlations between the states $\hat r_i$ for S and the states 
 $\scriptR_i$ for A, displayed in its diagonal blocks, express {\it registration}; they are encoded in $\langle\hat \Pi_i (\hat A-A_i)^2\rangle=0$
 for each $i$, a consequence of  (\ref{Dtf=}), which means that in an ideal measurement $\hat s$ takes the value $s_i$ when $\hat A$ takes the value $A_i$.
  
We further wish to show that {\it reduction} takes place, i.e., that the pointer takes for each run {\it a well-defined value} $A_i$ and 
that the set $\scriptE$ of runs can unambiguously be split into subsets $\scriptE_i$ including a proportion 
$p_i$ of runs, in such a way that {\it for each subset} $\scriptE_i$, characterized by the outcome $A_i$, the final state of S + A is $\scriptD_i =\hat r_i\otimes \scriptR_i$. 
This property obviously requires that  (\ref{Dtf=})  is satisfied, since by putting back together the subensembles $\scriptE_i$ we recover for $\scriptE$ 
the state $\sum_i p_i \scriptD_i$ of S + A. Nevertheless, due to a quantum queerness (\S~\ref{fin10.2.3}), we cannot conversely infer from the latter state the existence 
of physical subensembles $\scriptE_i$ described by the reduced states $\scriptD_i$. In fact, the very {\it selection} of some specific outcome labelled by the index
$i$ requires the reading of the indication $A_i$ of the pointer (\S~1.1.3), but it is not granted from  (\ref{Dtf=})  that each run provides such a well-defined indication. 
This problem will be exemplified by the Curie--Weiss model and solved in section 11. We will rely on a property of {\it arbitrary subsets of runs} of the measurement, 
their {\it hierarchic structure}. Namely, {\it any subset} must be described at the final time by a density operator of the form $\sum_i q_i \scriptD_i$ with arbitrary weights $q_i$. 
This property, which is implied by reduction, cannot be deduced from the sole knowledge of the density operator (1.7) that describes the final state of S + A 
{\it for the full set $\scriptE$ of runs}.

Tracing out the apparatus from  (\ref{Dtf=})  provides the marginal state for the tested system S after measurement, which is represented for the whole set of runs 
by the density operator

   \begin{eqnarray}   
   \label{rfin=}
   \hat{r}\left(  t_{\rm f}\right)\equiv
  {\rm  tr}_{\rm A} \hat{\cal D}(t_{\rm f}) =
   \sum_ip_i\hat r_i =\sum_i\hat\Pi_i\hat r(0)\hat\Pi_i  =\sum_ip_i  | s_i \rangle \langle s_i |
   =\sum_i  r_{ii}(0)  | s_i \rangle \langle s_i |.
     \end{eqnarray} 
   The last two expressions in (\ref{rfin=}) hold when the eigenvalues $s_i$ of $\hat s$ are non-degenerate. 
   Symmetrically, the final marginal state of the apparatus 

  \begin{equation}
  {\hat {\cal R}}\left(t_{{\rm f}}\right)  ={\rm tr}_{{\rm S}}\, \hat {\cal D}(t_{{\rm f}})  =\sum_{i}p_{i}{\hat {\cal R}}_{i} 
  \label{Rtf}
\end{equation}
is consistent with the occurrence with a probability $p_i$ of its indication $A_i$. The expression (1.10) is the result of {\it weak truncation}, while the selection of the runs 
characterized by the outcome $A_i$ produces for S the {\it weak reduction} into the state $\hat r_i$. The latter process constitutes a {\it preparation} of S. As already noted in 
\S~1.1.2, the fact that simply tracing out A may lead to a (weakly) truncated or a reduced state for S solves in no way the physics of the measurement process, 
a well known weakness of some models~\cite{blokhintsev1,blokhintsev2,deMuynck,ABNqm2003,requardt}.
 
}

\subsubsection{Entropy changes in a measurement}
\label{section.1.2.4}

\hfill{\it Discussions about entropy have produced quite some heat}

\hfill{Anonymous}
\vspace{3mm}

\ZeText{

When von Neumann set up in 1932 the formalism of quantum statistical mechanics \cite{vNeumann}, he introduced density operators $\scriptD$ as quantum analogues 
of probability distributions, and he associated with any of them a number, its entropy $S[\hat{\cal D}]=-{\rm tr} \, \hat{\cal D}\ln \hat{\cal D}$. In case $\scriptD$ describes a 
system in thermodynamic equilibrium, $S[\hat{\cal D}]$ is identified with the entropy of thermodynamics\footnote{With this definition, $S$ is dimensionless. In thermodynamic units, 
$S$ is obtained by multiplying its present expression by Boltzmann's constant $1.38 \cdot 10^{-23}$ JK$^{-1}$. Likewise, if we wish to express Shannon's entropy in bits, 
its expression should be divided by $\ln 2$}.  Inspired by these ideas, Shannon founded in 1948 the theory of communication, which relies on a quantitative estimate
 of the amount of information carried by a message \cite{ShannonWeaver1949}.
 Among the various possible messages that are expected to be emitted, each one $i$ has some probability $p_i$ to occur; by receiving the specific message $i$ 
 we gain an amount  $- \ln p_i$ of information. Shannon's entropy $S[p] = - \sum_i p_i \ln p_i$ characterizes the average amount of information which is missing when the 
 message has not yet been acknowledged. Returning to quantum mechanics, a new interpretation of von Neumann's entropy is thus obtained 
 \cite{Jaynes,lindblad_book,BalianBook}.
 When a system (or rather a statistical ensemble of systems prepared under similar conditions, in which the considered system is embedded) is described by some density 
 operator  $\hat{\cal D}$, the associated von Neumann entropy can be regarded as an extension of the Shannon entropy: it characterizes a lack of information due to the 
 probabilistic description of the system. It has thus a partly subjective nature, since it measures our uncertainty. 
One can also identify it with disorder ~\cite{BalianUtrecht,lindblad_book,BalianBook,BrillouinBook,BalianAJPh1999}.
 As measurement processes are means for gathering information, quantitative estimates of the amounts of information involved are provided by the changes of the 
 von Neumann entropies of the systems S, A and S + A. We gather below the various results found in the literature and their interpretation.               

The equation of motion of S + A is deterministic and reversible, and some manipulations justified by the large size of A are necessary, as in any relaxation
problem, to understand how the state of S + A may end as (\ref{Dtf=}). Strictly speaking, the Liouville-von Neumann evolution (\ref{LB001}) 
conserves the von Neumann entropy $-{\rm tr} \,\hat{\cal D}\ln\hat{\cal D}$ 
associated with the whole set of degrees of freedom of S + A; in principle no information is lost. 
However, in statistical physics, irreversibility means that information (identified with order) is transferred towards inaccessible 
degrees of freedom, in the form of many-particle correlations, without  possibility of return in a reasonable delay. A measure of this loss of information
is provided by the ``relevant entropy"  ~\cite{BalianUtrecht,lindblad_book,BalianBook,BrillouinBook,BalianAJPh1999},
which is the von Neumann entropy of the state that results from the elimination
of the information about such inaccessible correlations. Here the truncated state
$\hat{\cal D}(t_{\rm f})$ should have the latter status: As regards all accessible
degrees of freedom, $\hat{\cal D}(t_{\rm f})$ should be equivalent to the state
issued from $\hat{\cal D}(0)$ through the equation of motion (\ref{LB001}), but we
got rid in $\hat{\cal D}(t_{\rm f})$ of the irrelevant correlations involving a very
large number of elements of the macroscopic apparatus A; 
such correlations are irremediably lost.

We can therefore {\it measure the irreversibility} of the measurement process leading from $\hat{\cal D}(0)$ to $\hat{\cal D}(t_{\rm f})$ by the following 
entropy balance. The von~Neumann entropy of the initial state (\ref{D(0)=}) is split into contributions from ${\rm S}$ and ${\rm A}$, respectively, as
  
  \begin{equation}
  \mytext{\textcurrency entropie0\textcurrency \qquad}
  S\left[  {\hat{{\cal D}}
  }\left(  0\right)  \right]  =-{\rm {\rm tr}}\,{\hat{{\cal D}}
  }\left(  0\right)  \,\ln{\hat{{\cal D}}}\left(  0\right)  =S_{{\rm S}
  }\left[  \hat{r}\left(  0\right)  \right]  +S_{{\rm A}}\left[
  \hat{{\cal R}}\left(  0\right)  \right] { ,} \label{entropie0}
  \end{equation}
  whereas that of the final state (\ref{Dtf=}) is
  
  \begin{equation}
  \mytext{\textcurrency entropiefin\textcurrency \qquad}
  S\left[  {\hat {{\cal D}}}\left(  {t_{{\rm f}}}\right)  \right]  =
  S_{{\rm S}}\left[ \hat{r}\left( t_{\rm f}\right)  \right]  +\sum_i p_iS_A\left[\hat{\cal R}_i\right],
  \label{entropiefin}
  \end{equation}  
  where $ \hat{r}\left(  t_{\rm f}\right)$ is the marginal state (\ref{rfin=}) of S at the final time\footnote{The latter expression is found by using the orthogonality  
  $\hat{{\cal R}}_{i}\hat{{\cal R}}_{j}=0$ for $i\neq j$,  so that $-\hat{\cal D}(t_{\rm f})\ln \hat{\cal D}(t_{\rm f})$ is equal to the sum of its separate blocks, 
$\sum_ip_i\hat r_i\otimes\hat{\cal R}_i(- \ln p_i-\ln \hat r_i-\ln \hat{\cal R}_i)$, 
and hence the entropy of $\hat{\cal D}(t_{\rm f})$ is a sum of contributions arising from each $i$. 
The trace over A of the first two terms leads to $\sum_ip_i\hat r_i(- \ln p_i-\ln \hat r_i)$, the trace over S of 
which may be identified with the entropy $S_{\rm S}[  \hat {r}(t_{\rm f})] $ of (\ref{rfin=}); 
the trace of the last term leads to the last sum in (1.13)}.  This equality entails separate contributions from S and A.    
The increase of entropy from (\ref{entropie0}) to (\ref{entropiefin}) clearly arises from the two above-mentioned reasons, truncation and registration. 
  On the one hand, when the
  density operator $\hat{r}\left(  0\right)  $ involves off-diagonal blocks
  $\hat{\Pi}_{i}\hat{r}\left(  0\right)  \hat{\Pi}_{j}$ ($i\neq j$), their
  truncation raises the entropy. On the other hand, a robust registration
  requires that the possible final states $\hat{{\cal R}}_{i}$ of
  ${\rm A}$ are more stable than the initial state $\hat{{\cal R}}\left(
  0\right)  $, so that their entropy is larger. The latter effect dominates
  because the apparatus is large, typically $S_{\rm A}$ will be macroscopic and $S_{\rm S}$ microscopic.
  
  We can see that the state $\scriptD(t_{\rm f})$ expected to be reached at the end of the process is the one which maximizes von Neumann's entropy under the 
  constraints imposed by the conservation laws (\S~10.2.2). The conserved quantities are the energy $\langle\hat H\rangle$ (where 
  $\hat H=\hat H_{\rm S} + \hat H_{\rm A} -\hat s\,\hat A$ includes the coupling of the tested quantity $\hat s$ with the pointer observable $\hat A$) and the expectation 
  values of all the observables $\hat O_k$ of S that commute with $\hat s$ (we assume not only $[\hat H_{\rm S}, \hat s]=0$ but also $[\hat H_{\rm S},\hat O_k]=0$, 
  see \cite{Wigner,Yanase}. This maximization of entropy yields a density matrix proportional to $\exp(-\beta \hat H - \sum_k \lambda_k \hat O_k)$, 
  which has the form of a sum of diagonal blocks $i$, each of which factorizes as $p_i\hat r_i\otimes \scriptR_i$. The first factor $p_i \hat r_i$ associated with S, 
  obtained by adjusting the Lagrangian 
  multipliers $\lambda_k$, is identified with (1.8), due to the conservation of the diagonal blocks of the marginal density matrix of S. The second factor $\scriptR_i$ 
  associated with A is then proportional to $\exp[- \beta (\hat H_{\rm A} -s_i\hat A)]$, a density operator which for a macroscopic apparatus A describes one of its 
  equilibrium states characterized by the value $A_i$ of the pointer. Thus, the study of the evolution of S + A for a large statistical ensemble of runs (sections 4 to 7 for the 
  Curie--Weiss model) should amount to {\it justify dynamically the maximum entropy criterion} of equilibrium statistical mechanics. A further dynamical study is, 
  however, required in quantum mechanics to justify the assignment of one among the terms $\hat r_i \otimes\scriptR_i$ of (1.7) to 
  the outcome of an individual run (section 11 for the Curie--Weiss model).

  An apparatus is a device which allows us to {\it gain some information} on the state of ${\rm S}$ by reading the outcomes $A_{i}$. The {\it price we have to pay} for
  being thus able to determine the probabilities (\ref{p_i=}) is a complete {\it loss of information} about the off-diagonal elements $\hat{\Pi}_{i}\hat
  {r}\left(  0\right)  \hat{\Pi}_{j}$ ($i\neq j$) of the initial state of ${\rm S}$\footnote{In the language of section 1.1: 
  Loss of information about the phases of the $\psi_i$},  and a rise in the thermodynamic entropy of the apparatus.
  More generally, in other types of quantum measurements, some information about a system may be gained 
  only at the expense of erasing other information about this system~\cite{karen} (see subsection 2.5).  

The quantitative estimation of the gains and losses of information in the measurement process is provided by an entropic analysis,  reviewed in 
~\cite{PeresBook,lindblad_book,balian_1988}. Applications of entropy for quantifying the uncertainties in quantum measurements are  also discussed in 
~\cite{partovi_uncertainty}. We recall here the properties of the entropy of the marginal state of S and their interpretation in terms of information. 
We have just noted that  $S_{{\rm S}}\left[  \hat{r}\left(  t_{\rm f}\right)  \right]-S_{{\rm S}}\left[  \hat{r}\left(0\right)\right]$,
which is non-negative, measures the increase of entropy of S due to weak truncation. This means that,
in case we know $\hat r(0)$, the interaction with A ({\it without reading the pointer}) 
lets us loose the amount of information
$S_{{\rm S}}\left[  \hat{r}\left(  t_{\rm f}\right)  \right]-S_{{\rm S}}\left[  \hat{r}\left(0\right)\right]$
about all observables that do not commute with $\hat s$ ~\cite{lindblad_book,balian_1988}.
In fact, this loss is the largest possible among the set of states that preserve the whole information
about the observables commuting with $\hat s$. 
Any state of S that provides, for all observables commuting with $\hat s$, the same expectation values as $\hat r(t_{\rm f})$ 
is less disordered than $\hat r(t_{\rm f})$, and has an entropy lower than $S_{\rm S}[\hat r(t_{\rm f})]$. In other words, among all the 
processes that leave the statistics of the observables commuting with $\hat s$ unchanged, the ideal measurement of $\hat s$ 
is the one which {\it destroys the largest amount of information} (about the other observables of S).

{\it Reading the pointer value } $A_i$, which occurs with probability $p_i$, allows us to ascertain (for the considered ideal measurement) that S is 
in the state $\hat r_i$ after the measurement. By acknowledging the outcomes of a large sequence of runs of the measurement,
 we  {\it gain} therefore on average {\it some amount of information} given on the one hand by the Shannon entropy $-\sum_i p_i\ln p_i$,
 and equal on the other hand to the difference between the entropies of the final state and of its separate components,

\BEA
\label{Ssrtf}
S_{\rm S}\left[\hat r(t_{\rm f})\right]-\sum_ip_iS_{\rm S}\left[\hat r_i\right]=-\sum_i p_i\ln p_i\ge 0.
\EEA
The equality expresses {\it additivity} of information, or of uncertainty, at the end of the process, when we have not yet read the outcomes $A_i$: 
Our uncertainty $S_{\rm S}[\hat r(t_{\rm f})]$, when we know directly that $\hat r(t_{\rm f})$, the density operator of the final state,
encompasses all possible marginal final states $\hat r_i$, each with its probability $p_i$, is given by the left-hand side of (\ref{Ssrtf}).
It is the same as if we proceed in two steps. As we have not yet read $A_i$, we have a total uncertainty $S_{\rm S}\left[\hat r(t_{\rm f})\right]$ 
because we miss the corresponding amount of Shannon information $-\sum_i p_i \ln p_i$
 about the outcomes; and we miss also, with the probability $p_i$ for each possible occurrence of $A_i$, some information on S equal to 
$S_{\rm S}[\hat r_i]$,  the entropy of the state $\hat r_i$.
As it stands, the equality (1.14) also expresses the {\it equivalence between negentropy and information}~\cite{BrillouinBook,Brillouin,vedral}: 
{\it sorting} the ensemble of systems S according to the outcome $i$ lowers the entropy by a quantity equal on average to the left-hand side of (1.14), 
while {\it reading} the indication $A_i$ of the pointer provides, in Shannon's sense, an additional amount of information $-\ln p_i$, 
on average equal  to the right-hand side.

Two inequalities are satisfied in the whole process, including the sorting of results:

\BEA
-\sum_i p_i\ln p_i\ge 
S_{\rm S}\left[\hat r(0)\right]-\sum_ip_iS_{\rm S}\left[\hat r_i\right]\ge 0.
\EEA
The first inequality expresses that the {\it additivity} of the information gained on the {\it final} state $\hat r(t_{\rm f})$ of S by acknowledging the probabilities $p_i$, 
as expressed by  (\ref{Ssrtf}),  is {\it spoiled in quantum mechanics} when one considers the whole process, due to the quantum perturbation of the initial state of S 
which eliminates its off diagonal sectors. The second inequality, derived in  \cite{lindblad_1975}, expresses that measurements yield a {\it positive balance of information}
 about S  in spite of the losses resulting from the perturbation of S. Indeed, this inequality means that, on average over many runs of the measurement process, and after sorting 
of the outcomes, the entropy of S has decreased, i. e., more information on S is available at the time $t_{\rm f}$ than at the initial time. 
The equality holds only if all possible final states $\hat r_i$ of S have the same entropy.

Note finally that, if we wish to perform repeated quantum measurements in a closed cycle, we must reset the apparatus in its original metastable state. 
As for a thermal machine, this requires lowering the entropy and costs some supply of energy.

}

\subsection{Towards a solution of the measurement problem?}
\label{section.1.3}


\hfill{\includegraphics[width=4.5cm]{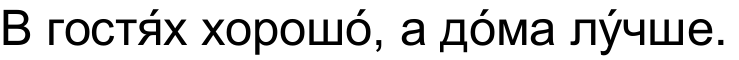} \hspace{-2mm}
 \footnote{Visiting is good, but home is better}}

\hfill{Russian proverb}

\vspace{3mm}

\ZeText{

The quantum measurement problem arises from the acknowledgement 
 that individual measurements provide well-defined outcomes. 
 Standard quantum mechanics yields only probabilistic results and thus seems
 unable to explain such a behavior. We have advocated above the use of quantum
 statistical physics, which seems even less adapted to draw conclusions about
 individual systems. Most of the present work will be devoted to show how a 
 statistical approach may nevertheless solve the measurement problem as will be discussed in section 11. 
 We begin with a brief survey of the more current approaches.

}

\subsubsection{Various approaches}
\label{section.1.3.1}

\hfill{{}\footnote{We need pluralism, there cannot be two opinions on that}}

\vspace{-0.8cm}
\myskipfigText{
\begin{figure}[h!h!h!]
\label{ArmProv2}
\hfill{\includegraphics[width=6cm]{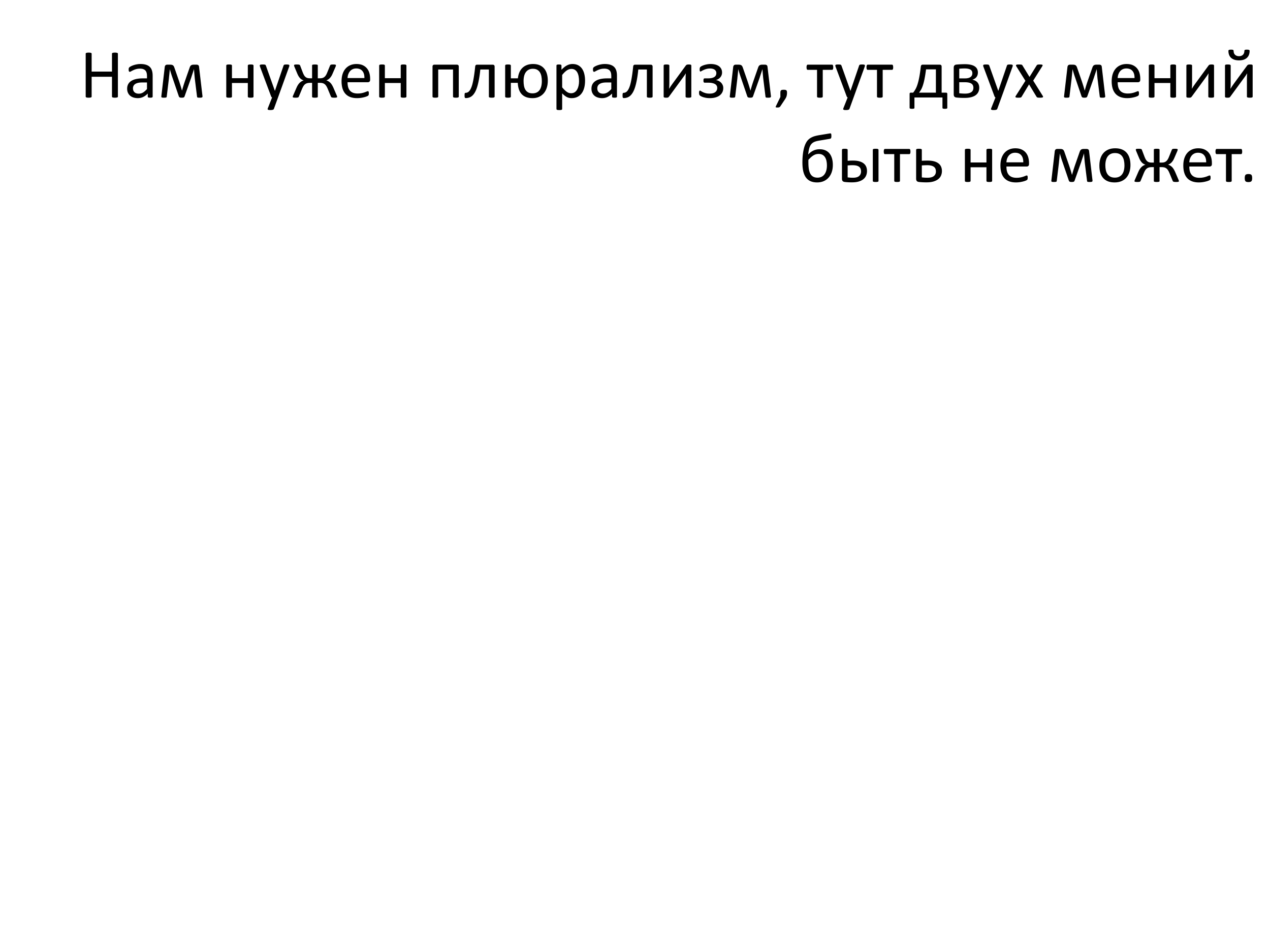}\hspace{2.6mm}}
\end{figure}}

\vspace{-4.cm}

\hfill{Mikhail Gorbachev}

\vspace{3mm}

 \ZeText{

In the early days of quantum mechanics, the apparatus was supposed to
behave classically, escaping the realm of quantum theory \cite{Bohr1928A,Bohr1928B,landau}. A similar idea
survives in theoretical or experimental works exploring the possible
existence of a border between small or large, or between simple and
complex objects, which would separate the domains of validity of quantum
and classical physics (Heisenberg's cut \cite{heisenberg_book}). 

Another current viewpoint has attributed the reduction\footnote{In order to distinguish two concepts often used in the literature, we use the word ``reduction'' as meaning 
 the transformation of the initial state of S + A into the final reduced state associated with one or another single run of the measurement, as specified in \S~1.1.2, although the 
 same word is often used in the literature to designate what we call ``truncation'' (decay of off-diagonal elements of the density matrix)} 
 in a measurement to the ``act of observing the result''.  
Again, the observer himself, who is exterior to the system, is not described in the framework of quantum mechanics. In Rovelli's relational interpretation \cite{Rovelli} 
a quantum mechanical description of some object is regarded as a codification of its properties which is ``observer-dependent'', that is, relative 
to a particular apparatus. Then, while a first ``observer'' A who gathers information about S regards reduction as real, a second observer testing 
S + A can consider that reduction has not taken place. In the many-worlds interpretation, reduction is even
denied, and regarded as a delusion due to the limitations of the human mind~\cite{Everett,ManyWorlds}. 
From another angle, people who wish to apply quantum theory to the whole universe, even have a non-trivial task in defining what is observation.
A more rational attitude is taken within the consistent histories approach,  in which  one is careful with defining when  and where the events
happen, but in which one holds that the measurements simply reveal the pre-existing values of events (this approach is discussed below in section 2.9).
For interpretations of quantum mechanics, see Bohm's textbook~\cite{BohmBookQM}
and for  interpretations based on entanglement and information, see Peres ~\cite{PeresBook} and Jaeger~\cite{Jaeger2009Book}.

The reduction may be regarded as a bifurcation in the evolution of the considered system, which may end up in different possible states 
$|s_i\rangle$ although it has been prepared in the single initial state $|\psi\rangle$. In the de Broglie--Bohm interpretation involving both waves 
and classical-like trajectories, the wave function $|\psi(t)\rangle$ appears both as arising from the density of trajectories and as guiding their dynamics. 
The randomness of quantum mechanics then arises merely from a randomness in the initial points of the set of trajectories. During a measurement process, 
the single initial bundle of trajectories, associated with $|\psi\rangle$, is split into separate bundles, each of which is associated with a wave function $|s_i\rangle$. 
While this interpretation accounts for the bifurcation and for  the uniqueness of the outcome of each run of a measurement process,
 it is not widely accepted \cite{Bohm,BohmHolland,deBroglie,LaloeBook,omnes}.

A more recent line of thought, going ``beyond the quantum'' ~\cite{BeyondTheQuantum}
relies on modification of the Schr\"odinger mechanics by additional non-linear and 
stochastic terms; see Refs.~\cite{Bassi_Ghirardi,pearle,grigorenko} for review.
Such generalizations are based in the belief, emphasized in the standard 
Copenhagen interpretation of quantum mechanics, that the Schr\"odinger equation
is unable to describe the joint evolution of a system S and an apparatus A, so that a 
special separate postulate is needed to account for the rules of quantum 
measurements, in particular reduction. Indeed, a hamiltonian evolution seems to
preclude the emergence of a single result in each single realization of a measurement
\cite{vNeumann,Wigner,london}.

We will focus below on the most conservative approach where S + A is treated as an
isolated quantum object governed by a Hamiltonian, and yet where reduction can be 
understood. The measurement is not considered on formal and general grounds as in 
many conceptual works aimed mainly at the interpretation of quantum mechanics,
but it is fully analyzed as a dynamical process. Unfortunately the theory of specific
experimental measurement processes based on hamiltonian dynamics is made difficult
by the complexity of a real measuring apparatus. One can gain full insight only by solving models that mimic actual measurements. 
The formal issue is first to show how S + A, which starts from the state (\ref{D(0)=}) and evolves along  (\ref{LB001}), may reach a final state of the truncated and 
correlated form (\ref{Dtf=}), then to explain how dynamics may provide for each run of the experiment one among the reduced states $\scriptD_i$.

The realization of such a program should meet the major challenge raised long ago by Bell \cite{Bell}:
   \textit{\textquotedblleft So long as the wave packet reduction is an essential component, and so long as we do not know
  exactly when and how it takes over from the Schr\"{o}dinger equation, we do not have an exact and unambiguous formulation of our most fundamental 
  physical theory\textquotedblright.} Indeed, a full understanding of quantum mechanics
  requires knowledge of the time scales involved in measurements. 
  
  Knowing how the truncation, then the reduction proceed in time, how long they take, is a prerequisite for
  clearing up the meaning of this phenomenon. On the other hand, the registration is part of the measurement; it is important to exhibit the
  time scale on which it takes place, to determine whether it interferes with the reduction or not, and to know when and how the correlations between
  ${\rm S}$ and ${\rm A}$ are established. These are the tasks we undertake in the body of this
work on a specific but flexible model. We resume in sections 9 and 11 how the solution of this model answers such questions.

}

\subsubsection{Glossary: Definition of the basic terms used throughout}

\hfill{\it Every word has three definitions }

\hfill{\it and three interpretations}

\hfill{Costa Rican proverb}

\vspace{3mm}


 \ZeText{

Authors do not always assign the same meaning to some current words. 
In order to avoid misunderstandings, we gather here the definitions that we are using throughout.

\begin{itemize}

\item{\it Observable}: an operator that represents a physical quantity of a system  (\S~10.1.1).

\item {\it Statistical ensemble}: a real or virtual set of systems prepared under identical conditions (\S~10.1.3).

\item{\it Subsensemble}: part of an ensemble, itself regarded as a statistical ensemble.

\item {\it Quantum state}: a mathematical object from which all the probabilistic properties of a statistical ensemble -- or subensemble -- of systems can be obtained.
(Strictly speaking, the state of an individual system refers to a thought ensemble in which it is embedded, since this state has a probabilistic nature.)
States are generally represented by a density operator (or, in a given basis,  a density matrix) which encompasses the expectation values of all the observables. 
Pure states are characterized by an absence of statistical fluctuations for some complete set of commuting observables (\S~10.1.4).

\item {\it Measurement}: a dynamical process which involves an apparatus A coupled to a tested system S and which provides information about one 
observable  $\hat s$ of S. The time-dependent state of the compound system S + A describes a statistical ensemble of runs, not individual runs. 
With this definition, the reading of the outcomes and the selection of the results are not encompassed in the ``measurement'', 
nor in the ``truncation" and the ``registration".

\item{\it Individual run of a measurement}:
a single interaction process between tested system and apparatus (prepared in a metastable state), followed by the reading of the outcome. 
 
\item {\it Ideal measurement}: a measurement which does not perturb the observables of S that commute with $\hat s$. 

\item {\it Pointer}; {\it pointer variable}:  a part of the apparatus which undergoes a change that can be read off or registered. 
 In general the pointer should be macroscopic and the pointer variable should be collective.

\item {\it Truncation; disappearance of Schr\"odinger cat states}: the disappearance, at the end of the measurement process, of the off-diagonal blocks of the
 density matrix of S + A describing the whole set of runs, in a basis where $\hat s$ is diagonal\footnote{\label{fn11} We will refrain from using popular terms 
 such as ``collapse of the wave function'' or ``reduction of the wave packet''}$^{,}$\footnote{\label{fn12} 
We use the terms ``weak truncation'' and ``weak reduction'' for the same operations as truncation and reduction, but performed on the marginal density matrix 
of the tested system S, and not on the density matrix of the compound system S+A} (sections 5 and 6).

\item{\it Dephasing}:
the decay of a sum of many oscillatory terms with different frequencies, arising from their mutual progressive interference in absence of a relevant coupling to an 
environment\footnote{An example is the relaxation due to an inhomogeneous magnetic field in NMR}.

\item {\it Decoherence}:  in general, a decay of the off-diagonal blocks of a density matrix under the effect of a random environment, such as a thermal bath.

\item {\it Registration}: the creation during a measurement of correlations between S and the macroscopic pointer of A.  Information is thus transferred 
to the apparatus, but becomes available only if uniqueness of the indication of the pointer is ensured for individual runs (section 7).

\item{\it Reduction}: for an individual run of the measurement, assignment of a state to S+A at the end of the process$^{\ref{fn11},\ref{fn12}}$.
Reduction is the objectification step, which reveals properties of a tested individual object. It requires truncation, registration, uniqueness of the indication 
of the pointer and selection of this outcome (\S\S~11.3.1 and 11.3.2). 

\item {\it Selection}: the sorting of the runs of an ideal measurement according to the indication of the pointer. 
The original ensemble that underwent the process is thus split into subensembles characterized by a well-defined value of $\hat s$. 
Measurement followed by selection may constitute a {\it preparation} (\S~10.2.2 and \S~\ref{finfin11.3.2}).

\item{ \it Hierarchic structure of subensembles}: a property required to solve the quantum measurement problem. Namely, the final state associated 
with any subset of runs of the measurement should have the same form as for the whole set but with different weights (\S~\ref{fin11.2.1}). 

\item{\it  Subensemble relaxation}: a dynamical process within the apparatus which leads the state of S + A to equilibrium, 
for an arbitrary subensemble of runs (\S\S~\ref{fin11.2.3} and \ref{fin11.2.4-5}).

\end{itemize}

}

\subsubsection{Outline}
\label{section.1.3.2}

\hfill{\it  Doorknob: Read the directions and directly}

\hfill{\it  you will be directed in the right direction}

\hfill{``Alice in Wonderland'', Walt Disney film}

\vspace{3mm}

 \ZeText{
 
    We review in section 2 the works that tackled the program sketched above, and discuss to which extent they satisfy 
    the various features that we stressed in the introduction. For instance, do they explain reduction by relying on a full 
    dynamical solution of the Liouville--von Neumann  equation for the considered model, or do they only invoke environment-induced decoherence? 
    Do they solve the preferred basis paradox? Do they account for a robust registration? Do they produce the time scales involved in the process? 
    
    In section 3 we present the Curie--Weiss model, which encompasses many properties of the previous models and 
    on which we will focus afterwards. It is sufficiently simple to be completely solvable, sufficiently elaborate to account
     for all characteristics of ideal quantum measurements, and sufficiently realistic to resemble actual experiments: 
     The apparatus simulates a magnetic dot, a standard registering device. 

The detailed solution of the equations of motion that describe a large set $\scriptE$ of runs for this model is worked out
     in sections 4 to 7, some calculations being given in appendices. 
    After analyzing the equations of motion of S + A (section 4), we exhibit several time scales. The truncation rapidly takes place 
    (section 5). It is then made irremediable owing to two alternative mechanisms (section 6). Amplification and registration 
    require much longer delays since they involve a macroscopic change of the apparatus and energy exchange with the bath   (section 7).
    
    Solving several variants of the Curie--Weiss model allows us to explore various dynamical processes which can be
     interpreted either as imperfect measurements or as failures (section 8). In particular, we study what happens when 
     the pointer has few degrees of freedom or when one tries to simultaneously measure non-commuting observables. 
     The calculations are less simple than for the original model, but are included in the text for completeness.
 
The results of sections 4 to 8 are resumed and analyzed in section 9, which also presents some simplified derivations suited for tutorial purposes.
However, truncation and registration, explained in sections 5 to 7 for the Curie--Weiss model, are only prerequisites for elucidating the quantum 
measurement problem,  which itself is needed to explain reduction. 

Before we tackle this remaining task, we need to make more precise the conceptual framework on which we rely, since reduction is tightly related
with the interpretation of quantum mechanics. The statistical interpretation (also called ensemble interpretation), 
in a form presented in section 10, appears as the most natural and consistent one in this respect. 
 
 We are then in position to work out the occurrence of reduction within the framework of the statistical interpretation by analyzing arbitrary subsets of runs.
 This is achieved in section 11 for a modified Curie--Weiss model, in which very weak but still sufficiently elaborate interactions within the 
 apparatus are implemented.  The uniqueness of the result of a single measurement, as well as the occurrence of classical probabilities, 
 are thus seen to emerge only from the dynamics of the measurement process.
 
Lessons for future work are drawn in section 12, and some open problems are suggested in section 13.
    
    
\vspace{3mm} 

The reader interested only in the results may skip the technical sections 4 to 8, and focus upon the first pages of section 9, which can be regarded 
as a self-contained reading guide for them, and upon section 11. The conceptual outcomes are gathered in sections 10 and 12.
     
 }

\renewcommand{\thesection}{\arabic{section}}
\section{The approach based on models}
\setcounter{equation}{0}\setcounter{figure}{0}
\renewcommand{\thesection}{\arabic{section}.}

\label{section.2}

\hfill{{\it Point n'est besoin d'esp\'erer pour entreprendre,}

\hfill{ \it ni de r\'eussir pour pers\'ev\'erer}\footnote{It is not necessary to hope for undertaking, neither to succeed for persevering}}

\hfill{Charles le T\'em\'eraire and William of Orange}

\vspace{3mm}


 \ZeText{

We have briefly surveyed in \S ~\ref{section.1.3.1} many theoretical ideas intended to elucidate the problem of quantum measurements.  
In \S~12.4.2 and \S~12.4.3 we mention a few other ideas about this problem. However, we
feel that it is more appropriate to think along the lines of an experimentalist who performs measurements in his laboratory.  For this
reason, it is instructive to formulate and solve models within this scope. We review in this section various models in which S + A is treated as a
compound system which evolves during the measurement process according to the standard rules of quantum mechanics.  The existing models are
roughly divided into related classes. Several models serve to elucidate open problems. Besides specific models, we shall discuss
several more general approaches to quantum measurements (e.g., the decoherence and consistent histories approaches). 

\clearpage

}
\subsection{Heisenberg--von Neumann setup }


\hfill{ {\it Quod licet Iovi, non licet bovi}
\footnote{What is allowed for Jupiter, is not allowed for the rind}}

\hfill{Roman proverb}

\vspace{3mm}

 \ZeText{

A general set-up of quantum measurement was proposed and analysed by Heisenberg \cite{heisenberg,heisenberg_book}. His ideas were formalized
by von Neumann who proposed the very first mathematically rigorous model of quantum measurement \cite{vNeumann}. An early review on this subject
is by London and Bauer \cite{london}, in the sixties it was carefully reviewed by Wigner \cite{Wigner}; see \cite{busch} for a modern review. 

Von Neumann formulated the measurement process as a coupling between two quantum systems with a specific interaction Hamiltonian that involves
the (tensor) product between the measured observable of the tested system and the pointer variable, an observable of the apparatus.  This
interaction conserves the measured observable  and ensures a correlation between the tested quantity and the pointer observable. In
one way or another the von Neumann interaction Hamiltonian is applied in all subsequent models of ideal quantum measurements. However, von
Neumannn's model does not account for the differences between the microscopic [system] and macroscopic [apparatus] scales. As a main
consequence, it does not have a mechanism to ensure the specific classical correlations (in the final state of the system + apparatus)
necessary for the proper interpretation of a quantum measurement. Another drawback of this approach is its requirement for the initial
state of the measuring system (the apparatus) to be a pure state (so it is described by a single wave function). Moreover, this should be a
specific pure state, where fluctuations of the pointer variable are small.  Both of these features are unrealistic. In addition, and most importantly, the von Neumann 
model does not account for the features of truncation and reduction; it only shows weak reduction (see terminology in \S~1.1.2 and \S~1.3.2). 
This fact led von Neumann (and later on Wigner~\cite{Wigner}) to postulate -- on top of the usual Schr\"odinger evolution -- a specific dynamic process
that is supposed to achieve the reduction \cite{vNeumann}. 

With all these specific features it is not surprising that the von Neumann model has only one characteristic time driven by the
interaction Hamiltonian. Over this time the apparatus variable gets correlated with the initial state of the measured system.

Jauch considers the main problem of the original von Neumann model, i.e. that in the final state it does not ensure specific classical correlations
between the apparatus and the system \cite{Jauch}. A solution of this problem is attempted within the lines suggested 
(using his words) during ``the heroic period of quantum mechanics'', 
that is, looking for classical features of the apparatus. To this end, Jauch introduces the concept of equivalence
between two states (as represented by density matrices):  two states are equivalent with respect to a set of
observables, if these observables cannot distinguish one of these states from another \cite{Jauch}. Next, he
shows that for the von Neumann model there is a natural set of commuting (hence classical) observables, so that with
respect to this set the final state of the model is not distinguishable from the one having the needed classical
correlations. At the same time Jauch accepts that some other observable of the system and the
apparatus can distinguish these states. Next, he makes an attempt to {\it define} the measurement event
via his concept of classical equivalence. In our opinion this attempt is interesting, but not successful.

}
\subsection{Quantum--classical models: an open issue?}

\hfill{\it Gooi geen oude schoenen weg }

\hfill{\it voor je nieuwe hebt\footnote{Don't throw away old shoes before you have new ones}}

\hfill{Dutch proverb}

\vspace{3mm}

  \ZeText{

Following suggestions of Bohr that the proper quantum measurement
should imply a classical apparatus \cite{Bohr1928A,Bohr1928B,landau}, there were several
attempts to work out interaction between a quantum and an explicitly classical system
\cite{picard,delos,stenholm,Diosi2000A,Diosi2000B,boucher,boucher1,boucher2,boucher3,boucher4,boucher5,PeresTerno,salcedo1,salcedo2,hall_reginatto}.  
(Neither Bohr \cite{Bohr1928A,Bohr1928B}, nor Landau and Lifshitz \cite{landau} who  present Bohr's opinion in quite detail, consider the proper interaction processes.)
This subject is referred to as hybrid (quantum--classical) dynamics.  Besides
the measurement theory it is supposed to apply in quantum chemistry \cite{picard,delos,boucher3} (where the full modeling of quantum degrees of
freedom is difficult) and in quantum gravity \cite{gravity}, where the proper quantum dynamics of the gravitational field is not known.  There
are several versions of the hybrid dynamics. The situation, where the classical degree of freedom is of a mean-field type is especially
well-known \cite{picard,delos}.  In that case the hybrid dynamics can be derived variationally from a simple combination of quantum and classical
Lagrangian. More refined versions of the hybrid dynamics attempt to describe interactions between the classical degree(s) of freedom and quantum fluctuations. 
Such theories are supposed to be closed and self-consistent, and (if they really exist) they would somehow get the same fundamental status as
their limiting cases, i.e., as quantum and classical mechanics. The numerous attempts to formulate such fundamental quantum--classical
theories have encountered severe difficulties \cite{Diosi2000A,Diosi2000B,boucher,boucher1,boucher2,boucher3,boucher4,boucher5,PeresTerno}.
There are no--go theorems showing in which specific sense such theories cannot exist \cite{salcedo1,salcedo2}. 

As far as the quantum measurement issues are concerned, the hybrid dynamical models have not received the attention they deserve. This
is surprising, because Bohr's insistence on the classicality of the apparatus is widely known and frequently repeated. The existing works are
summarized as follows.  Diosi and co-authors stated that their scheme for the hybrid dynamics is useful for quantum measurements \cite{Diosi2000A,Diosi2000B},
albeit that they did not come with a more or less explicit analysis. Later on Terno has shown that the problem of a quantum measurement
cannot be solved via a certain class of hybrid dynamic systems \cite{terno}.  His arguments rely on the fact that the majority of hybrid system have pathological 
features in one way or another. Terno also reviews some earlier attempts, in particular by him in collaboration with Peres~\cite{PeresTerno},
 to describe quantum measurements via hybrid dynamics; see the book of Peres \cite{PeresBook} for preliminary ideas within this approach.
However, recently Hall and Reginato \cite{hall_reginatto,hall+} suggest a scheme for the hybrid dynamics that seems to be free of pathological features.
This scheme is based on coupled quantum and classical ensembles. A related set-up of hybrid dynamics is proposed by Elze and coworkers
based on a path-integral formulation ~\cite{Elze1}, see also ~\cite{Elze3}. If Hall and Reginato's claim is true that such schemes can circumvent no-go theorems  
~\cite{hall_reginatto,hall+},  it should be interesting to look again at the features of quantum measurements from the perspective of an explicitly classical apparatus:
Bohr's program can still be opened! A modern view on the Copenhagen interpretation developed by (among
others) Bohr is presented by Grangier in Refs. \cite{X2,X3}.

Everitt, Munro and Spiller discuss a measurement model which, while fully quantum mechanical, makes use of analogy with classical features of the apparatus 
~\cite{Everitt}.  The model consists of a two-level system (the measured system), the apparatus, which is a one-dimensional quartic oscillator under external driving,
and an environment whose influence on the system + apparatus is described within the Lindblad master-equation approach and its quantum state diffusion 
unravelling~\cite{Percival}. The main point of this work is that the apparatus can display the chaotic features of a damped forced non-linear oscillator 
(and is thus not related to Hamiltonian chaos). Everitt, Munro and Spiller make use of this point for the following reason: The feature of chaos allows one 
to distinguish quantum from classical regimes for the apparatus (this is not fundamental - simply a convenience for demonstrating a quantum to classical transition).
The model reproduces certain features expected from individual measurement outcomes, but this happens at the cost of {\it unravelling} the master equation,
a relatively arbitrary procedure of going from density matrices to random wavefunctions. 
The authors of Ref. ~\cite{Everitt} are aware of this arbitrariness and attempt to minimize it. It should be noted that, as one would expect, in the classical limit 
the choice of how to unravel seems to have no effect ons the emergence of a classical dynamic (see, for example,~\cite{Everitt2}). 
This implies that the results of ~\cite{Everitt} may well be independent of the unravelling -- but this has yet to be demonstrated.


In Ref. \cite{A1} Blanchard and Jadczyk discuss a quantum-classical model for measurements. They present it as a minimal phenomenological
model for describing quantum measurements within the concept of an explicitly classical apparatus. In contrast to other quantum-classical
models, Blanchard and Jadczyk consider a dissipative interaction between the quantum and classical subsystems. This interaction is
modeled by a completely positive map. These maps are frequently applied for describing an open-system quantum dynamics, where the
target system couples with an external environment; see e.g. Refs. \cite{Weiss, Gardiner,Thirring}.  
(However, this is certainly not the only possibility for an open-system quantum dynamics; see in this context Ref. \cite{A4}.) 
Blanchard and Jadczyk find a simple form of the completely map that suffices for accounting (phenomenologically) for
certain features of quantum measurements, such the response of the pointer classical states to the initial state of the quantum system,
as well as the proper final state of the quantum system.

This approach is generalized in \cite{A2}, where Blanchard and Jadczyk account for the emergence of events during the quantum measurements.
This is done by introducing an additional phenomenological step thereby the quantum- classical dynamics for the quantum density matrix
and classical probability distribution is regarded as the result of averaging over the states of some underlying stochastic process (a
procedure akin to unraveling the open-system quantum master equation).
The stochastic process -- which gives rise to what Blanchard and Jadczyk call event-enhanced quantum theory -- is formulated in the
tensor product of the classical subsystem's event space and the quantum subsystem's Hilbert space.

In our opinion this approach to quantum measurements has an extensively phenomenological character, a fact well-admitted by
Blanchard and Jadczyk. On the other hand, its central idea that the emergence of measurement events should be related to specific features
of the measuring apparatus is certainly valuable and will be developed in the present work.

In closing this subsection we note that the relation between quantum and classical has yet another, {\it geometrical} twist, because the pure-state quantum dynamics 
(described by the Schr\"odinger equation) can be exactly mapped to a classical Hamiltonian dynamics evolving in a suitable
classical symplectic space \cite{Kibble, Heslot,Gibbons}. Quantum aspects (such as uncertainties and the Planck's constant) are then reflected 
via a Riemannian metrics in this space \cite{Heslot,Gibbons}; see also \cite{Ashtekar} for a recent review. This is a geometrical counterpart to the usual algebraic 
description of quantum mechanics, and is considered to be a potentially rich source for various generalizations of quantum mechanics \cite{Ashtekar,Kryukov}. A formulation
of the quantum measurement problem in this language was attempted in \cite{Kryukov}. We note that so
far this approach is basically restricted to pure states (see, however, \cite{Gibbons} in this context).

Further references on crucial aspects of the quantum-to-classical transition are~\cite{BallentineYangZibin1994,Habib2002,wiebe_ballentine}.

}

\subsubsection{Measurements in underlying classical theory}

  \hfill{\it Non quia difficilia sunt non audemus,}

\hfill{{\it sed quia non audemus, difficilia sunt}\footnote{It is not because things are difficult that we do not dare, but because we do not dare, things are difficult}}

\hfill{Seneca}


\vspace{3mm}

\ZeText{

The major part of this section is devoted to measurement models, where the measuring apparatus is modeled as a classical system. There is another line of research,
where quantum mechanics as such is viewed as as an approximation of a stochastic classical theory; see, e.g.
\cite{CettodelaPenaBook,cetto,TheoSED1,TheoSED2}, and \cite{Wetterich08,Khrennikov09,Khrennikov10,KhrennikovNilsson11,Khrennikov11} . 
The ultimate promise of such approaches is to go beyond the predictions of quantum mechanics; see, e.g. \cite{Khrennikov10}. Their basic problem is
to reconcile essential differences between the probability structures in quantum mechanics and classical mechanics.
There are numerous attempts of such effective classical descriptions, but many of them do not pay much attention
to those differences, focusing instead on deriving classically certain aspects of quantum theory (stochastic electrodynamics is a vivid example of such an attitude).

Recent works by Khrennikov and coauthors attempt to explain how an underlying classical theory can reproduce the probability rules of quantum mechanics without
conflicting with Bell theorems, contextuality etc. \cite{Khrennikov09,KhrennikovNilsson11,Khrennikov11}\footnote{For an (over)simplified discussion 
of the Bell theorem and related matters, see \cite{MerminPhysicsToday}}.
This is done by postulating specific scenarios for uncertainties produced during a measurement, by means of imprecise apparatuses, of the underlying classical objects
(random fields). In this sense the works by Khrennikov and coauthors \cite{KhrennikovNilsson11,Khrennikov11} belong to the realm of quantum measurements
and will be reviewed now. 

The starting point of the approach is based on the following observation \cite{Khrennikov09,Khrennikov10,KhrennikovNilsson11,Khrennikov11}. 
Let a classical random vector ($x_1,\cdots, x_n$) be given with zero average $\bar x_k = 0$ for $k = 1,\cdots, n$. Let  ($x_1,\cdots, x_n$)  be observed through
the mean value of a scalar function $f(x_1,\cdots, x_n)$. We assume that $f(0,\cdots,0) = 0$ and that fluctuations of $x_k$
around its average are small. Hence,  $\overline{f(x_1,\cdots, x_n)}=\sum_{i,j=1}^n \half \overline{x_ix_j}\p_{x_i}\p_{x_j}f(0,\cdots,0)$.
If the symmetric and positive matrix $\rho$ with elements $\rho_{ij}=\half \overline{x_ix_j}$ is regarded as a density matrix,
and the symmetric matrix $A$ with elements $A_{ij}=\p_{x_i}\p_{x_j}f(0,\cdots,0)$  as an observable,
one can write $\overline{f(x_1,\cdots, x_n)}={\rm tr}\,\rho A$, which has the form of Born's formula for calculating the average of  $A$
in the state $\rho$. By this principle all the quantum observables can be represented as averages over classical random fields.
Taking complex valued classical random fields one can make both $\rho$ and $A$ hermitean instead of just symmetric.
As it presently stands, this approach is purely phenomenological and is simply aimed at replacing quantum
observables by classical averages in a mathematically exact manner. No interpretation of the physical meaning of ($x_1,\cdots, x_n$)  is given\footnote{They may show up,
though, as the resonant modes in a dynamical path integral description of Stochastic Electrodynamics}.  In a way
this representation of quantum averages via classical random fields goes back to the wave-modus of accounting
for quantum effects. This is why it is important to see how experiments that demonstrate the existence of photon
as a corpuscle (particle) fit into this picture. Khrennikov and coauthors show that also experiments detecting
the corpuscular nature of light can be accommodated in this classical picture provided that one accounts for
the threshold of the detectors \cite{KhrennikovNilsson11,Khrennikov11}. Here the existence
of photon is a consequence of specific modifications introduced by threshold detectors when measuring classical
random fields. Khrennikov and co-authors stress that this picture is hypothetical as long as one has not verified
experimentally whether the threshold dependence of real experiments does indeed have this specific form \cite{KhrennikovNilsson11}. 
In their opinion this question is non-trivial and still awaits for its experimental resolution.

This resolution should also point out whether the idea of accounting for specific features of quantum probability (such as Bell's inequality) via classical models is tenable
\cite{Wetterich08, Khrennikov09, KhrennikovNilsson11,Khrennikov11}. It is currently realized that the violation of
Bell's inequalities \cite{Bellhiddenvar,HomeWhitaker,deMuynck,Laloe}   
should be attributed to the non-commutative nature of the distribution $\scriptD$ rather than to non-locality; quantum mechanics does 
not involve ordinary probabilities nor ordinary correlations. The violation of the classical inequality, observed experimentally 
\cite{Aspect1,Aspect2,Y1,Y2,Y3,Y4}  arises when one puts together outcomes of measurements performed in different experimental contexts, and this may itself be 
a problem~\cite{Accardi1981,Accardi,Khrennikov,TheoBell1,TheoBell2}.
The discussion of \S~8.3.4 shows how quantum and ordinary correlations may be reconciled in the context of a thought experiment where one attempts to measure simultaneously, 
with a unique setting, all spin components.

}

\subsection{Explicitly infinite apparatus: Coleman--Hepp and related models}

\hfill{\it Before you milk a cow,}

\hfill{\it tie it up}

 \hfill{South African proverb}


\vspace{0.25cm}

 \ZeText{

Several authors argued that once the quantum measurement apparatus is supposed to be a macroscopic system, the most natural framework  for 
describing measurements is to assume that it is explicitly infinite; see the review by Bub \cite{bub}. $C^*$-algebras is the
standard tool for dealing with this situation \cite{EmchBook}. Its main peculiarity is that there are (many) inequivalent unitary  representations of the
algebra of observables, i.e., certain superpositions between wavefunctions cannot be physical states (in contrast to
finite-dimensional Hilbert spaces) \cite{bub}. This is supposed to be helpful in constructing measurement models.  Hepp proposed first such models \cite{hepp}. 
He starts his investigation by stating some among the goals of quantum measurement models. In particular, he
stresses that an important feature of the problem is in getting classical correlations between the measured observable and the pointer variable of
the apparatus, and that quantum mechanics is a theory that describes probabilities of certain events.  Hepp then argues that the quantum
measurement problem can be solved, i.e., the required classical correlations can be established dynamically, if one restricts oneself to
macroscopic observables. He then moves to concrete models, which are solved in the $C^*$-algebraic framework. The infinite system approach is also 
employed in the quantum measurement model proposed by Whitten-Wolfe and Emch \cite{wolfe}. 
A $C^\ast$-algebraic framework was recently employed by Landsman for deriving
mathematically the classical limit of quantum mechanics and from it the Born rule \cite{Landsman}.

However, working with an infinite measuring apparatus hides the physical meaning of the approach, because some important
dynamic scales of the quantum measurement do depend on the number of degrees of freedom of the apparatus~\cite{ABNqm2003}.  
In particular, the truncation time may tend to zero in the limit of an infinite apparatus and cannot then be evaluated. 
Thus, making the apparatus explicitly infinite (instead of taking it large, but finite) misses an important piece of physics, and does not allow to
understand which features of the quantum measurement will survive for a apparatus having a mesoscopic scale. 

Hepp also studies several exactly solvable models, which demonstrate
various aspects of his proposal. One of them|proposed to Hepp by
Coleman and nowadays called the Coleman--Hepp model|
describes an ultra-relativistic particle interacting with a
linear chain of spins. Hepp analyzes this model in the infinite
apparatus situation; this has several drawbacks, e.g., the
overall measurement time is obviously infinite. The physical representation of
the Coleman--Hepp Hamiltonian is improved by Nakazato and Pascazio
\cite{pascazio}. They show that the basic conclusions on the
Coleman--Hepp Hamiltonian approach can survive in a more
realistic model, where the self-energy of the spin chain is taken into
account.  Nakazato and Pascazio also discuss subtleties involved in
taking the thermodynamic limit for the model \cite{pascazio}.  The
Coleman--Hepp model with a large but finite number of the apparatus
particles is studied by Sewell
\cite{sewell_2005,sewell_2007,sewell}. He improves on previous
treatments by carefully calculating the dependence of the characteristic
times of the model on this number, and discusses possible imperfections
of the measurement model arising from a finite number of particles. 

Using the example of the Coleman--Hepp model, Bell demonstrates
explicitly \cite{Bell} that the specific features of the quantum
measurement hold only for a certain class of observables, including
macroscopic observables \cite{requardt,sewell_2005,sewell_2007,sewell}.
It is then possible to construct an observable for which those features
do not hold \cite{Bell}. We recall that the same holds in the
irreversibility problem: it is always possible to construct an
observable of a macroscopic system (having a large, but finite number of
particles) that will not show the signs of irreversible dynamics, i.e., it will not be subject to relaxation.
  Bell takes this aspect as an
essential drawback and states that the quantum measurement was not and
cannot be solved within a statistical mechanics approach \cite{Bell}.
Our attitude in the present paper is different. We believe that although
concrete models of quantum measurements may have various drawbacks, the
resolution of the measurement problem is definitely to be sought along
the routes of quantum statistical mechanics. The fact that certain
restrictions on the set of observables are needed, simply indicates
that, similar to irreversibility, a quantum measurement is an emergent
phenomenon of a large system -- the tested system combined with the
apparatus -- over some characteristic time.  

}

\subsection{Quantum statistical models}

\hfill{\it If I have a thousand ideas and only one turns out to be good,}

\hfill{\it  I am satisfied}

\hfill{ Alfred Bernhard Nobel}

\vspace{3mm}

 \ZeText{

Here we describe several models based on quantum statistical mechanics.
In contrast to the previous chapter, these models do not invoke anything
beyond the standard quantum mechanics of finite though large systems. 

Green proposed a realistic model of quantum measurement~\cite{green}. He emphasizes the necessity of describing the
apparatus via a mixed, quasi-equilibrium state and stresses that the
initial state of the apparatus should be macroscopic and metastable. The
model studied in \cite{green} includes a spin-$\frac{1}{2}$ particle
interacting with two thermal baths at different temperatures. The
two-temperature situation serves to simulate metastability. The tested
particle switches interaction between the baths.  By registering the
amount of heat flow through the baths (a macroscopic pointer variable),
one can draw certain conclusions about the initial state of the spin.
Off-diagonal terms of the spin density matrix are suppressed via a
mechanism akin to inhomogeneous broadening. However, an explicit
analysis of the dynamic regime and its characteristic times is absent.

Cini studies a simple model for the quantum measurement process which
illustrates some of the aspects related to the macroscopic character of the
apparatus \cite{cini}. The model is exactly solvable and can be boiled
down to a spin-$\frac{1}{2}$ particle (tested spin) interacting with
a spin-$L$ particle (apparatus). The interaction Hamiltonian is
$\propto \sigma_z L_z$, where $\sigma_z$ and $L_z$ are, respectively,
the third components of the spin-$\frac{1}{2}$ and spin $L$. Cini shows
that in the limit $L\gg 1$ and for a sufficiently long interaction time, the
off-diagonal terms introduced by an (arbitrary) initial state of the tested spin
give negligible contributions to the observed quantities, i.e., to the
variables of the tested spin and the collective variables of the apparatus.
The characteristic times of this process are analyzed, as well
as the situation with a large but finite value of $L$.

In Refs.~\cite{blokhintsev1,blokhintsev2} Blokhintsev studies, within the statistical
interpretation of quantum mechanics, several interesting measurement
models with a metastable initial state of the apparatus: an incoming
test particle interacting with an apparatus-particle in a metastable
potential well, a test neutron triggering a nuclear chain reaction, et cetera. 
Though the considered models are physically appealing, the
involved measurement apparatuses are frequently not really macroscopic.
Neither does Blokhintsev pay proper attention to the correlations between
the system and the apparatus in the final state. 

Requardt studies a quantum measurement model, in which due to collisional
interaction with the tested system, the pointer variable of a
macroscopic measuring apparatus undergoes a coherent motion, in which
the momentum correlates with the values of the measured observable
(coordinate) \cite{requardt_1984}. It is stressed that for the approach
to have a proper physical meaning, the apparatus should have a large but finite number of degrees of freedom. 
However, no detailed account of characteristic measurement times is given. Requardt also assumes that
the initial state of the measurement apparatus is described by a wave
function, which is merely consistent with the macroscopic information
initially available on this apparatus. He focuses
on those aspects of the model which will likely survive in a
more general theory of quantum measurements; see in this context his
later work \cite{requardt} that is reviewed below.

An interesting statistical mechanical model of quantum measurement was
proposed and studied in Ref.~\cite{gaveau_schulman} by Gaveau and
Schulman. The role of apparatus is played by a one-dimensional Ising spin
model. Two basic energy parameters of the model are an external field
and the spin-spin coupling (exchange coupling). An external field is
tuned in such a way that a spontaneous flipping of one spin is energetically not
beneficial, while the characteristic time of flipping two spins
simultaneously is very large. This requirement of metastability puts an
upper limit on the number of spins in the apparatus. The tested spin
$\frac{1}{2}$ interacts only with one spin of the apparatus; this is
definitely an advantage of this model. The spin-apparatus interaction
creates a domino effect bringing the apparatus to a unique
ferrromagnetic state.  This happens for the tested spin pointing up. For
the tested spin pointing down nothing happens, since in this state the
tested spin does not interact with the apparatus.  Characteristic times
of the measurement are not studied in detail, though Gaveau and Schulman
calculate the overall relaxation time and the decay time of the
metastable state. It is unclear whether this model is supposed to work
for an arbitrary initial state of the tested spin. 

Ref.~\cite{merlin} by Merlin studies a quantum mechanical model for
distinguishing two different types of bosonic particles. The model is
inspired by Glaser's chamber device, and has the realistic feature that
the bosonic particle to be tested interacts only with one particle of
the apparatus (which by itself is made out of bosons).  The initial
state of the apparatus is described by a pure state 
and it is formally metastable (formally, because this is not a thermodynamic
metastability). The relaxation process is not accounted for explicitly;
its consequences are simply postulated. No analysis of characteristic
relaxation times is presented. Merlin analyses the relation of
measurement processes with the phenomenon of spontaneous symmetry breaking.

}
\subsubsection{Spontaneous symmetry breaking}


\hfill{\it Les miroirs feraient bien de r\'efl\'echir un peu plus}

\hfill{ \it avant de renvoyer les images\footnote{Mirrors 
should reflect some more before sending back the images}}

\hfill{Jean Cocteau, Le sang d'un po\`ete}

\vspace{3mm}

 \ZeText{

The role of spontaneous symmetry breaking as an essential ingredient of
the quantum measurement process is underlined in papers by Grady \cite{grady},
Fioroni and Immirzi \cite{fio} and Pankovic and Predojevic \cite{yugo}.
They stress that superpositions of vacuum states are not allowed in
quantum field theory, since these superpositions do not satisfy the
cluster property. All three approaches stay mainly at a qualitative level,
though Fioroni and Immirzi go somewhat further in relating ideas on
quantum measurement process to specific first-order phase transition
scenarios. An earlier discussion on symmetry breaking, quantum
measurements and geometrical concepts of quantum field theory is given
by Ne'eman \cite{neeman}. 

Ref.~\cite{hungary} by Zimanyi and Vladar also emphasizes the relevance
of phase transitions and symmetry breaking for quantum measurements.
They explicitly adopt the statistical interpretation of quantum
mechanics. General statements are illustrated via the Caldeira-Leggett
model \cite{Ullersma,ms,caldeira,NplusA}: a two-level system coupled to a bath of harmonic oscillators.
This model undergoes a second-order phase transition with relatively
weak decay of off-diagonal terms in the thermodynamic limit, provided
that the coupling of the two-level system to the bath is sufficiently
strong. The authors speculate about extending their results to
first-order phase transitions. A dynamical consideration is basically
absent and the physical meaning of the pointer variable is not clear.

Thus the concept of spontaneous symmetry breaking is frequently discussed in
the context of quantum measurement models (although it is not anymore
strictly spontaneous, but driven by the interaction with the system
of which the observable is to be measured). It is also an essential feature of
the approach discussed in the present paper.  It should however be noted
that so far only one scenario of symmetry breaking has been considered
in the context of quantum measurements (the one that can be called the
classical scenario), where the higher temperature extremum of the free
energy becomes unstable (or at least metastable) and the system moves to
another, more stable state (with lower free energy). Another scenario is
known for certain quantum systems (e.g., quantum antiferromagnets) with
a low-temperature spontaneously symmetry broken state; see, e.g.,
\cite{vanWezel}.  Here the non-symmetric state is not an eigenstate of the
Hamiltonian, and (in general) does not have less energy than the
unstable ground state. The consequences of this (quantum) scenario for
quantum measurements are so far not explored. However, recently van
Wezel, van den Brink and Zaanen study specific decoherence mechanisms
that are induced by this scenario of symmetry breaking.  \cite{vanWezel}. 

}
\subsubsection{System-pointer-bath models}

\hfill{\it Je moet met de juiste wapens ten strijde trekken\footnote{You must go into battle with 
the right weapons}}

\hfill{Dutch proverb}

\vspace{3mm}

 \ZeText{

Refs.~\cite{haake1} by Haake and Walls and ~\cite{haake2} by Haake and
Zukowski study a measurement of a discrete-spectrum variable coupled to
a single-particle apparatus (the meter). The latter is a harmonic
oscillator, and it interacts with a thermal bath, which is modeled via
harmonic oscillators. The interaction between the tested system and the
meter is impulsive (it lasts a short time) and involves the tensor
product of the measured observable and the momentum of the meter. There
are two characteristic times here: on the shorter time, the impulsive
interaction correlates the states of the object and of the meter, while on
the longer time scale the state of the meter becomes classical under the
influence of the thermal bath, and the probability distribution of the
meter coordinate is prepared via mixing well-localized probability
distributions centered at the eigenvalues of the measured quantity, with
the weights satisfying the Born rule \cite{Born}.  (This sequence of
processes roughly corresponds to the ideas of decoherence theory; see
below for more detail.) At an even longer time scale the meter will
completely thermalize and forget about its interaction with the tested
system. The authors of \cite{haake1,haake2} also consider a situation
where the meter becomes unstable under the influence of the thermal
bath, since it now feels an inverted parabolic potential. Then the
selection of the concrete branch of instability can be driven by the
interaction with the object. Since the initial state of such an unstable
oscillator is not properly metastable, one has to select a special
regime where the spontaneous instability decay can be neglected. 

The quantum measurement model studied in \cite{venugopalan} by
Venugopalan is in many aspects similar to models investigated in
\cite{haake1,haake2}. The author stresses relations of the studied model
to ideas from the decoherence theory. 

Ref.~\cite{ABNqm2001} by the present authors investigates a model of
quantum measurement where the macroscopic measurement apparatus is
modeled as an ideal Bose gas, in which the amplitude of the condensate
is taken as the pointer variable.  The model is essentially based on the
properties of irreversibility and of ergodicity breaking, which are
inherent in the model apparatus. The measurement process takes place in
two steps: First, the truncation of the state of the tested system takes
place, this process is governed by the apparatus-system interaction.
During the second step classical correlations are established between
the apparatus and the tested system over the much longer time scale of
equilibration of the apparatus. While the model allows to understand
some basic features of the quantum measurement as a driven
phase-transition, its dynamical treatment contains definite drawbacks.
First, the Markov approximation for the apparatus-bath interaction,
though correct for large times, is incorrectly employed for very short
times, which greatly overestimates the truncation time. Another drawback
is that the model is based on the phase transition in an ideal Bose gas.
This transition is known to have certain pathological features (as
compared to a more realistic phase-transition in a weakly interacting
Bose gas).  Though the authors believe that this fact would be repairable and does not influence
the qualitative outcomes of the model, it is certainly desirable to have
better models, where the phase transition scenario would be generic and
robust.  Such models will be considered in later chapters of this work.

In Ref.~\cite{SpehnerHaake1,SpehnerHaake2} Spehner and Haake present a measurement model
that in several aspects improves upon previous models.  The model includes 
the tested system, an oscillator (generally anharmonic),  which plays the role of apparatus,
and a thermal bath coupled to the oscillator. The
time scales of the model are set in such a way that the correlations between the
measured observable of the system and the  pointer variable of the apparatus (here
the momentum of the anharmonic oscillator) and the decay
of the off-diagonal terms of the tested system density matrix are
established simultaneously. This implies realistically that no
macroscopic superpositions are generated. In addition, the initial state
of the apparatus and its bath is not assumed to be factorized, which
makes it possible to study strong (and also anharmonic) apparatus-bath
couplings. 

Ref.~\cite{privman} by Mozyrsky and Privman studies a quantum
measurement model, which consists of three parts: the tested system,
the apparatus and a thermal bath that directly couples to the system (and
not to the apparatus). The initial state of the apparatus is not metastable, it
is chosen to be an equilibrium state. The dynamics of the measured
observable of the system is neglected in the course of measurement. The
authors of \cite{privman} show that after some decoherence time their
model is able to reproduce specific correlations that are expected for
a proper quantum measurement. 

Omn\`es recently studied a model for a quantum measurement \cite{omnes_model}.  The pointer variable of the apparatus is supposed
to be its (collective) coordinate. The introduction of the measurement process is accompanied by a discussion on self-organization. For
solving this Omn\`es partially involves the mean-field method, because the many-body apparatus density matrix is substituted by the tensor
product of the partial density matrices. The dynamics of the model involves both decoherence and reduction. These two
different processes are analysed together and sometimes in rather common terms, which can obscure important physical differences between them.
In the second part Omn\`es studies fluctuations of the observation probabilities for various measurement results. These fluctuations are said to arise due to
a coupling with an external environment modeled as a phonon bath.

Van Kampen stresses the importance of considering a macroscopic and
metastable measuring apparatus and proposes a model that is supposed to
illustrate the main aspects of the measurement process \cite{vKampen}.
The model consists of a single atom interacting with a multi-mode
electromagnetic field, which is playing the role of apparatus. The
emitted photon that is generated correlates with the value of the measured
observable.  The apparatus can be macroscopic (since the vacuum has many
modes), but its (thermodynamically) metastable character is questionable.
The measurement of a photon by one of the remote detectors is not solved in detail, and its main dynamical
consequences are not analyzed.  Nevertheless, van Kampen offers a
qualitative analysis of this model, which appears to support the common
intuition on quantum measurements. The resulting insights are
summarized in his ``ten theorems'' on quantum measurements.

}
\subsubsection{Towards model-independent approaches}

\hfill{\it  Qui se soucie de chaque petite plume}

\hfill{\it  ne devrait pas faire le lit\footnote{Who cares about every little feather should not make the bed}}

\hfill{Swiss proverb}

\vspace{3mm}

 \ZeText{

Sewell \cite{sewell_2005,sewell_2007,sewell} and independently Requardt
\cite{requardt} attempt to put the results obtained from several models
into a single model-independent approach, which presumably may
pave a way towards a general theory of quantum measurements. The basic
starting point of the approach is that the measuring apparatus, being a
many-body quantum system, does have a set of macroscopic, mutually
commuting observables $\{A_1,\ldots,A_{\cal M}\}$ with ${\cal M}$ a
macroscopic integer. The commutation is approximate for a large, but
finite number of reservoir particles, but it becomes exact in the
thermodynamic limit for the apparatus. Each $A_k$ is typically a
normalized sum over a large number of apparatus particles.  The set
$\{A_1,\ldots,A_{\cal M}\}$ is now partioned into macroscopic cells;
each such cell refers to some subspace in the Hilbert space formed by a common eigenvector. 
The cells are distinguished from each other by
certain combinations of the eigenvalues of $\{A_1,\ldots,A_{\cal M}\}$.
The purpose of partitioning into cells is to correlate each eigenvalue of
the microscopic observable to be measured with the corresponding cell.
In the simplest situation the latter set reduces to just one observable
$A$, while two cells refer to the subspace formed by the eigenvectors of
$A$  associated with positive or negative eigenvalues. Further derivations,
which so far are carried out on the levels of models only
\cite{requardt,sewell_2005,sewell_2007,sewell}, amount to showing that a
specific coupling between the system and the apparatus can produce their
joint final state, which from the viewpoint of observables $A_k\otimes S$ -- where $S$ is any observable of the microscopic measured system --
does have several features required for a good (or even ideal) quantum measurement. 

}
\subsubsection{Ergodic theory approach}

\hfill{\it Wenn i wieder, wieder komm~\footnote{When I come, come again}}

\noindent\noindent
\hfill{From the German folk song ``Mu\ss \, i denn''}

\vspace{3mm}

 \ZeText{

Daneri, Loinger and Prosperi approach quantum measurements via the
quantum ergodic theory \cite{daneri}. Such an approach is anticipated
in the late forties by the works of Jordan \cite{jordan} and Ludwig
\cite{ludwig}. Daneri, Loinger and Prosperi model the measuring
apparatus as a macroscopic system, which in addition to energy has
another conserved quantity, which serves the role of the pointer
variable. Under the influence of the system to be measured this
conservation is broken, and there is a possibility to correlate
different values of the measured observable with the pointer values.
Daneri, Loinger and Prosperi invoke the basic assumption of ergodic
theory and treat the overall density matrix via time-averaging
\cite{daneri}. The time-averaged density matrix satisfies the necessary
requirements for an ideal measurement. However, the use of the
time-averaging does not allow to understand the dynamics of the quantum
measurement process, because no information about the actual dynamical
time scales is retained in the time-averaged density matrix. Also,
although the initial state of the measuring apparatus does have some
properties of metastability, it is not really metastable in the
thermodynamic sense. 

The publication of the paper by Daneri, Loinger and Prosperi in early
sixties induced a hot debate on the measurement problem; see
\cite{tausk} for a historical outline. We shall not attempt to review
this debate here, but only mention one aspect of it: Tausk (see
\cite{tausk} for a description of his unpublished work) and later on
Jauch, Wigner and Yanase \cite{jauch} criticize the approach by Daneri,
Loinger and Prosperi via the argument of an interaction free
measurement.  This type of measurements is first discussed by Renninger
\cite{renninger}.  The argument goes as follows: sometimes one can
gather information about the measured system even without any
macroscopic process generated in the measuring apparatus. This can
happen, for instance, in the double-slit experiment when the apparatus
measuring the coordinate of the particle is placed only at one slit.
Then the non-detection by this apparatus will -- ideally --  indicate that
the particle passed through the other slit.  The argument thus intends
to demonstrate that quantum measurements need not be related to
macroscopic (or irreversible) processes. This argument however does not
present any special difficulty within the statistical interpretation of
quantum mechanics, where both the wavefunction and the density matrix
refer to an ensemble of identically prepared system. Although it is true
that not every single realization of the apparatus-particle interaction
has to be related to a macroscopic process, the probabilities of getting
various measurement results do rely on macroscopic processes in the
measuring apparatus. 

}
\subsection{No-go theorems and small measuring apparatuses}

\hfill{{\it Non ho l'et\`a,  per amarti}\footnote{I do not have the age to love you}}

 \hfill{Lyrics by Mario Penzeri, sung by Gigliola Cinquetti}

\vspace{0.3cm}
 \ZeText{

The quantum measurement process is regarded as a fundamental problem,
also because over the years several no--go theorems were established
showing that the proper conditions for quantum measurement cannot be
satisfied if they are demanded as exact features of the  final state of the apparatus
\cite{Wigner,abner_1,abner_2,abner_3}. The first such theorem was
established by Wigner \cite{Wigner}. Then several extensions of this
theorem were elaborated by Fine \cite{arthur_fine} and Shimony with co-authors
\cite{abner_1,abner_2,abner_3}. The presentation by Fine is especially
clear, as it starts from the minimal conditions required from a quantum
measurement \cite{arthur_fine}. After stating the no-go theorem, Fine
proceeds to discuss in which sense one should look for approximate
schemes that satisfy the measurement conditions, a general program  motivating
also the present study. The results of Refs.~\cite{abner_1,abner_2,abner_3} show that even when allowing
certain imperfections in the apparatus functioning, the quantum
measurement problem remains unsolvable in the sense that the existence
of specific classical correlations in the final state of the system + apparatus cannnot be ensured; see also in this context the recent review by
Bassi and Ghirardi \cite{Bassi_Ghirardi}. In our viewpoint, the no-go
theorems do not preclude approximate satisfaction of the quantum measurement requirements  -- 
owing to a macroscopic size of the apparatus.

Turning this point over, one may ask which features of proper quantum
measurements (as displayed by successful models of this phenomenon)
would survive for an apparatus that is not macroscopically large. There
are several different ways to pose this question, e.g., below we shall
study the measuring apparatus (that already performs well in the
macroscopic limit) for a large but finite number of particles.  Another approach was recently worked out
by Allahverdyan and Hovhannisyan \cite{karen}. They assume that the measuring apparatus is a
finite system, and study system-apparatus interaction setups that lead
to transferring certain matrix elements of the unknown density matrix
$\lambda$ of the system into those of the final state ${\widetilde r}$
of the apparatus. Such a transfer process represents one essential aspect
of the quantum measurement with a macroscopic apparatus. No further
limitations on the interaction are introduced, because the purpose is to
understand the implications of the transfer on the final state of the
system.  It is shown that the transfer process eliminates from the final state of the system the memory about
the transferred  matrix elements (or certain other ones) \cite{karen}.  
In particular, if one diagonal matrix element is transferred, ${\widetilde r}_{aa}=\lambda_{aa}$, the memory
on all non-diagonal elements $\lambda_{a\not=b}$ or $\lambda_{b\neq a}$ related to this
diagonal element is completely eliminated from the final density operator of the system
 (the memory on other non-diagonal elements $\lambda_{cd}$, where $c\not=a$ and $d\not =a$ may be preserved).
Thus, the general aspect of state disturbance in quantum
measurements is the loss of memory about off-diagonal elements,
rather than diagonalization (which means the vanishing of the
off-diagonal elements). 

}
\subsection{An open problem: A model for a non-statistical interpretation of the measurement process.}

\hfill{\it We can't go on forever, with suspicious minds}

\hfill{Written by Mark James, sung by Elvis Presley}

\vspace{0.3cm}

 \ZeText{

The statistical interpretation together with supporting models does
provide a consistent view on measurements within the standard quantum
mechanics. However, it should be important to understand whether there
are other consistent approaches {\it from within} the standard quantum
formalism that can provide an alternative view on quantum measurements.
Indeed, it cannot be excluded that the real quantum measurement is a
wide notion, which combines instances of different interpretations. In the present review we will not cover 
approaches that introduce additional ingredients to the standard quantum theory, and will only mention them in subsection 2.8.

We focus only on one alternative to the statistical interpretation,
which is essentially close to the Copenhagen interpretation
\cite{Bohr1928A,Bohr1928B,landau,krips} and is based on effectively non-linear
Schr\"odinger equation. We should however stress that so far the
approach did not yet provide a fully consistent and unifying picture of
quantum measurements even for one model. 

Recently Brox, Olaussen and Nguyen approached quantum measurements via a
non-linear Schr\"odinger equation \cite{brox}. The authors explicitly
adhere to a version of the Copenhagen interpretation, where the wave
function (the pure quantum state) refers to a single system. 
They present a model which is able to account for single measurement events. 
The model consists of a spin-$\frac{1}{2}$ (the system to be measured), a ferromagnet (the
measuring apparatus), and the apparatus environment. The overall system
is described by a pure wavefunction. The ferromagnetic apparatus is
prepared in an (unbiased) initial state with zero magnetization. The two
ground states of the ferromagnet have a lower energy and, respectively,
positive and negative magnetization. Moving towards one of these states
under influence of the tested system is supposed to amplify the weak
signal coming from this tested system.  (The latter features will also
play an important role in the models to be considered in detail later
on.) The environment is modeled as a spin-glass: environmental spins
interact with random (positive or negative) coupling constants. So far
all these factors are more or less standard, and -- as stressed by the
authors -- these factors alone cannot account for a solution of the
measurement problem within an interpretation that ascribes the
wavefunction to a single system. The new point introduced by Brox,
Olaussen and Nguyen is that the effective interaction between the apparatus and
the measured system is non-linear in the wavefunction: it contains an
analogue of a self-induced magnetic field \cite{brox}. In contrast to
the existing approaches, where non-linearity in the Schr\"odinger
equation are introduced axiomatically, Brox, Olaussen and Nguyen state
that their non-linearity can in fact emerge from the Hartree-Fock
approach: it is known that in certain situations (the Vlasov limit) the
many-body Schr\"odinger equation can be reduced to a non-linear  equation for the
single-particle wave function \cite{petr}. Examples of this are the
Gross-Pitaevskii equation for Bose condensates \cite{petr} or the non-linear
equation arising during quantum feedback control \cite{armen_mahler}.
However, the statement by Brox, Olaussen and Nguyen on the emergent
non-linearity is not really proven, which is an essential drawback.
 Leaving this problem aside, these authors show numerically that the
specific nonlinearity in the system-apparatus interaction may lead to a
definite, albeit random, measurement result. The statistics of this
randomness approximately satisfies the Born rule \cite{Born}, which
emerges due to the macroscopic size of the apparatus.  The cause of this
randomness is the classical randomness related to the choice of the
spin-glass interaction constants in the environment \cite{brox}, i.e.,
for different such choices (each one still ensuring the proper
relaxation of the apparatus) one gets different single-measurement
results. Thus in this approach the cause of the randomness in
measurement results is not the irreducible quantum randomness, but
rather the usual classical randomness, which is practically unavoidable
in the preparation of a macroscopic environment.  Brox, Olaussen and
Nguyen argue that the nonlinearity in the system-bath interaction --
which is crucial for obtaining all the above effects -- need not be
large, since the amplification may be ensured by a large size of the
ferromagnet \cite{brox}. Their actual numerical calculations are however
carried out only for moderate-size spin systems. 

\clearpage

}

\subsection{Decoherence theory}

\hfill{\it  Pure coherence is delirium,}

\hfill{\it  it is abstract delirium}

\hfill{Baruch Spinoza}

\vspace{3mm}

 \ZeText{

Presently it is often believed that decoherence theory solves the quantum measurement problem. So let us introduce this concept. 
Decoherence refers to a process, where due to coupling with an external environment, off-diagonal elements of the system density matrix decay in
time; see \cite{Schlosshauer,zurek,Guilini,Blanchard,walls,walls_book,Braun} for reviews. The basis where this decay
happens is selected by the structure of the system-environment coupling. In this way the system acquires some classical features. 

Decoherence is well known since the late 40's \cite{slichter}. One celebrated example is spin relaxation in NMR experiments.
The decay of the transverse polarization, perpendicular to the permanently applied field, is in general characterized by the relaxation time ${\cal T}_2$; 
it can be viewed as a decoherence of the spin system, since it exhibits the decay of the off-diagonal contributions to the spin density matrix in the representation 
where the applied Hamiltonian is diagonal ~\cite{Abragam1961,Abragam1982}.
Another standard example is related to the Pauli equation for an open quantum system weakly coupled to an external thermal bath
\cite{klim}. This equation can be visualized as a classical stochastic process during which the system transits from one energy level to another.

More recently decoherence has attracted attention as a mechanism of quantum-to-classical transition, and was applied to the quantum measurement
problem \cite{Schlosshauer,zurek,Guilini,Blanchard,walls,walls_book,Braun}. The standard pattern of such an application relies on an initial impulsive 
interaction of the von Neumann type which correlates (entangles) the measuring apparatus with the system to be measured.
Generally, this step is rather unrealistic, since it realizes macroscopic superpositions, which were never seen in any realistic  measurement or  any measurement model. 
Next, one assumes a specific environment for the apparatus, with the environment-apparatus interaction Hamiltonian directly related to
the variable to be measured. Moreover, within the decoherence approach it is stressed -- e.g., by Zurek in \cite{zurek} and by Milburn and Walls in \cite{walls_book} --
that the observable to-be-measured is determined during the process generated by the apparatus-environment interaction. The latter is
supposed to diagonalize the density matrix of the system plus the apparatus in a suitable basis.  This second step is again unrealistic, since it assumes
that the variable to be measured, which is normally under control of the experimentalist, must somehow correlate with the structure of the
system's environment, which --  by its very definition --  is out of direct control. To put it in metaphoric terms, decoherence theory asserts that {\it the surrounding air
measures a person's size}. But without explicit pointer variable that can be read off, this is not what one normally understands under {\it measuring a person's size};
we consider {\it measurement without a readable pointer variable} merely  as a linguistic redefinition of the concept, that obscures the real issue.
These criticisms of the decoherence theory approach agree with the recent analysis by Requardt \cite{requardt}.

One even notes that, as far as the problem of quantum-to-classical transition is concerned, the decoherence cannot be regarded as the only 
-- or even as the basic -- mechanism of this transition. As convincingly argued by Wiebe and Ballentine \cite{wiebe_ballentine} and Ballentine
\cite{ballentine}, realistic macroscopic Hamiltonian systems can -- and sometimes even should --  achieve the classical limit without invoking any
decoherence effect. This concerns both chaotic and regular Hamiltonian systems, although the concrete scenarios of approaching the classical
limit differ for the two cases. 

In spite of these caveats that prevent decoherence theory from providing {\it the} solution, it has been valuable in shaping the ideas on quantum
measurement models, In particular, this concerns a recent attempt by Omn\`es to develop a general theory of decoherence via ideas and methods
of non-equilibrium statistical mechanics \cite{omnes} (see also \cite{omnes_model} that we reviewed above).  Among the issues addressed
in \cite{omnes} is the generality of the system-environment structure that leads to decoherence, the physical meaning of separating the system
from the environment, and the relation of the decoherence theory to the hydrodynamic description. 

}

\subsubsection{``Envariance'' and Born's rule}

\hfill{\it Try to see it my way,}

\hfill{\it  Only time will tell if I am right or I am wrong}

\hfill{The Beatles, We can't work it out}

\vspace{3mm}

 \ZeText{

In recent papers \cite{zurek_prl_2003,zurek_pra_2005} Zurek attempts to derive Born's rule without a direct appeal to
measurement theory, but solely from features of transformations termed environment-assisted invariance (or envariance) plus a set of additional
assumptions. These assumptions are partially spelled out in \cite{zurek_prl_2003,zurek_pra_2005} and/or pointed out by other authors
\cite{schlosshauer_fine}. If successful, such a derivation will be of clear importance, since it will bypass the need for a theory of quantum
measurements. We would like now to review the premises of this derivation. 

Following the basic tenet of the decoherence theory, Zurek considers  an entangled pure state of the system S and its environment E 
\cite{zurek_prl_2003,zurek_pra_2005}:

\BEA
\label{en_1}
|\psi_{\rm SE}\rangle = \sum_{k=1}^n \alpha_k|s_k\rangle\otimes |\varepsilon_k\rangle,~~~ \sum_{k=1}^n |\alpha_k|^2=1.
\EEA 
This state is written is the so called Schmidt form with two orthonormal set of vectors $\{|s_k\rangle\}_{k=1,n}$ and
$\{|\varepsilon_k\rangle\}_{k=1,n}$ living in the Hilbert spaces of the system and environment, respectively. 

It is assumed that the pure state (\ref{en_1}) was attained under the effect of an interaction between S and E which was switched off before
our consideration. Any pure state living in the joint Hilbert of S + E can be represented as in (\ref{en_1}). 

Zurek now asks ``what can one know about the state of S given the joint state (\ref{en_1}) of S+E''? He states at the very beginning that he
refuses to trace out the environment, because this will make his attempted derivation of Born's rule circular
\cite{zurek_prl_2003,zurek_pra_2005}. This means that the wave function (\ref{en_1}) stands for Zurek as something that should describe
relations between observables and their probabilities. This description (Born's rule) is to be discovered, this is why one does not want to {\it
assume beforehand} its linearity over $|\psi_{\rm SE}\rangle\langle\psi_{\rm SE}|$. 

The core of Zurek's arguments is the following  particular case of (\ref{en_1}) for $n=2$
\cite{zurek_prl_2003,zurek_pra_2005}
\BEA
\label{en_2}
|\psi_{\rm SE}\rangle =\frac{1}{\sqrt{2}} \sum_{k=1}^2 |s_k\rangle\otimes |\varepsilon_k\rangle.
\EEA 
Using certain invariance features of $|\psi_{\rm SE}\rangle$ in (\ref{en_2}) | environment assisted invariance, or envariance | Zurek 
now attempts to derive that S is either in the state $|s_1\rangle$ or in the state $|s_2\rangle$ with probabilities
$\frac{1}{2}$ \cite{zurek_prl_2003,zurek_pra_2005}.  We stress here that $\alpha_1=\alpha_2=\frac{1}{\sqrt{2}}$ is really essential for the
derivation. A straightforward generalization of (\ref{en_2}) is employed by Zurek in his attempted derivation of Born's rule for rational
probabilities (on analogy to the classical definition of probability as a ratio of two integers), which is then extended to arbitrary
probabilities via a continuity argument. 

However, we do not need to go into details of this derivation to understand why it fails. 

First one notes that due to $\alpha_1=\alpha_2$ the representation (\ref{en_1}),  (\ref{en_2}) is not unique: any pair of orthonormal
vectors $\{|s_k\rangle\}_{k=1,2}$ can appear there. (This question about the derivation by Zurek was raised in \cite{schlosshauer_fine}.)
Hence it is not meaningful to say that S is with some probability in a definite state. 

The actual freedom in choosing the basis for S is even larger, because 
(\ref{en_2}) can be respresented as 
\BEA
\label{en_3}
|\psi_{\rm SE}\rangle = \sum_{k=1}^2
\kappa_k |\widetilde{s}_k\rangle\otimes |\varepsilon_k\rangle,
\EEA
where $|\widetilde{s}_1\rangle$ and $|\widetilde{s}_2\rangle$ are
normalized, but not orthogonal to each other. It is impossible to rule
out such non-orthogonal decompositions by relying on various invariance
features of (\ref{en_2}) (some arguments of Zurek seem to attempt this),
because these features are necessarily representation-independent. 

Admittedly, such non-orthogonal decompositions could be ruled out by {\it postulating beforehand} that S should be in a definite state|thus only
orthogonal $\{|s_k\rangle\}_{k=1,2}$ are accepted|and looking for the probability of these states.  But even under this orthogonality
condition the choice of $\{|s_k\rangle\}_{k=1,2}$ in (\ref{en_2}) is not unique due to degeneracy $\alpha_1=\alpha_2$. 

One may attempt to reformulate this statement by demanding that E is not an environment, but rather a measuring apparatus with a fixed basis
$\{|\varepsilon_k\rangle\}_{k=1,2}$. This then makes possible to fix $\{|s_k\rangle\}_{k=1,2}$. Such a reformulation,
natural in the context of measurement theory, does not seem to be acceptable for the following reason. 

If one now asserts that S is in the state $|s_1\rangle$ (or $|s_2\rangle$) with probability $\frac{1}{2}$ ($\frac{1}{2}$), then due to the symmetry between S and E, 
it is possible to assert that E is in the state $|\varepsilon_1\rangle$ (or $|\varepsilon_2\rangle$) with probability $\frac{1}{2}$ ($\frac{1}{2}$). 
This will then amount to stating that both S and E are in definite states with definite probabilities, which is not acceptable for a pure state $|\psi_{\rm SE}\rangle$, 
because there is no way to prepare the state (\ref{en_2})  of S + E by mixing (definite states with definite probabilities). For this  it would be necessary 
that the state of S + E be mixed, e.g. $\frac{1}{2}\sum_{k=1}^2 |s_k\rangle\otimes|\varepsilon_k\rangle \langle s_k|\otimes\langle \varepsilon_k|$. 
However, such mixed  states do not appear in the present theory that is based on pure states with the prohibition of taking partial traces. 

We conclude that the proposed derivation for Born's rule cannot work, because one cannot even state the probability of what is going to
be described by Born's rule. Even if one grants in the form of postulates various assumptions needed for the derivation -- i.e., one
postulates that S is indeed in a definite but unknown state according to a fixed basis $\{|s_k\rangle\}_{k=1,2}$ that is chosen {\it
somehow} -- even then the proposed derivation of Born's rule need not work, since it is not clear that the specific form (\ref{en_1}) of the
wave function S+E (without measuring apparatus?) is ever satisfied within realistic models of quantum measurements. 



} 

\subsection{Seeking the solution outside quantum mechanics}


\hfill{\it No, no, you're not thinking;}

\hfill{\it   you're just being logical}

 \hfill{Niels Bohr}

\vspace{3mm}
 \ZeText{

Though this review will restrict itself to approaches to quantum measurements within the standard quantum mechanics, we briefly list for
completeness a number of attempts to seek the solution for the quantum measurement problem beyond it. The de Broglie--Bohm approach
\cite{Bohm,BohmHolland,deBroglie} is currently one of the most popular alternatives to the standard quantum mechanics. It introduces an
additional set of variables (coordinates of the physical particles) and represents the Schr\"odinger equation as an equation of motion for those
particles, {\it in addition} to the motion of the wavefunction, which keeps the physical meaning of a separate entity (guiding field). Hence
in this picture there are two fundamental and separate entities: particles and fields. Recently Smolin attempted to construct a version
of the de Broglie--Bohm approach, where the wavefunction is substituted by certain phase-variables, which, together with coordinates, are supposed
to be features of particles \cite{smolin}. In this context see also a related contribution by Schmelzer, where the fundamental character of the wavefunction 
is likewise negated \cite{schmelzer}. The approach by Smolin is coined in terms of a real ensemble, which -- in contrast to ensembles of non-interacting
objects invoked for validation of any probabilistic theory -- does contain highly-nonlocal (distance independent) interactions between its constituents. 
It is presently unclear to which extent this substitution of the wavefunction by phase-variables will increase the eligibility of the de Broglie--Bohm approach, 
while Smolin does not discuss the issues of measurement that are known to be non-trivial within the approach \cite{Bohm,BohmHolland,BohmCushing}.

Another popular alternative is the spontaneous localization approach by Ghirardi, Rimini and Weber \cite{GRW}. This approach is based on a non-linear and stochastic
generalization of the Schr\"odinger equation such that the collapse of the wavefunction happens spontaneously (i.e., without any measurement) with
a certain rate governed by classical white noise. Bassi and Ghirardi recently reviewed this and related approaches in full detail
\cite{Bassi_Ghirardi}; other useful sources  are the book by Adler  \cite{adler} and the review paper by Pearle \cite{pearle_review}. 
 Spontaneous localization models in the energy basis are especially interesting, since they conserve the average energy of the quantum system; 
this subject is reviewed by Brody and Hughston \cite{BrodyHughston2006}. Non-linear modifications of the Schr\"odinger
equation have by now a long history \cite{pearle,grigorenko,bohm_bub,bb,gisin_1981,gisin_1989,weinberg}. All of them in one way or another
combine non-linearities with classical randomness. The first such model was introduced by Bohm and Bub \cite{bohm_bub} starting from certain
hidden-variables assumption. The approaches that followed were either oriented towards resolving the quantum measurement problem
\cite{pearle,gisin_1981,gisin_1989} or trying to obtain fundamentally nonlinear generalizations of the Schr\"odinger equation and quantum mechanics \cite{weinberg}.   
Several approaches of the former type were unified within a formalism proposed by Grigorenko \cite{grigorenko}. 
Recently Svetlichny presented a resource letter on fundamental (i.e., not emerging from the usual, linear theory) non-linearities in quantum
mechanics, where he also discusses their possible origin \cite{sveto}.  Some of those approaches based on nonlinear
generalizations of the Schr\"odinger equation were confronted to experiments, see e.g.  Refs.~\cite{experiments_nlsch,experiments_nlsch1}, 
but so far with negative result.

 A very different approach was taken by De Raedt and Michielsen,  who simulate the measurement process by specifying 
a set of simple rules that mimic the various components of the measurement setup, such as beam splitters, polarizers and detectors. 
They perform numerical simulations using algorithms that the mimic the underlying events, and are able to reproduce the
statistical distributions given by quantum mechanics ~\cite{DeRaedt1,DeRaedt2}.

 \newpage

}

\subsection{A short review on consistent histories}






\hfill{\it I shall make sure that EU action develops consistently over time}

\hfill{Herman Van Rompuy}

\vspace{3mm}

 \ZeText{

The consistent histories approach negates the fundamental need of measurements for understanding quantum measurements (quantum mechanics
without measurements). It was proposed by Griffiths \cite{griffiths_jsp} based on earlier ideas of Aharonov, Bergmann and Lebowitz
\cite{aharonov_bergmann_lebowitz}.  The approach is reviewed, e.g., by Griffiths \cite{griffiths_book}, Gell-Mann ~\cite{gellmann}, Hohenberg \cite{hohenberg}, 
and Omn\`es \cite{omnes_book}.  It aims to develop an interpretation of quantum mechanics valid for a closed system of any
size and any number of particles, the evolution of which is governed by the Liouville--von Neumann (or Heisenberg) equation.
Within this approach the notion of an event -- together with its probability -- is defined from the outset and ``measurements'',
which do not involve any interaction between the system and some apparatus,  simply reveal the pre-existing values of physical quantities. 
In particular, it is not necessary to invoke either the micro-macro separation or statistical assumptions on the initial states needed to derive the irreversibility 
aspect of quantum measurements. All of these may still be needed to describe concrete measurements, but the fundamental need for understanding 
quantum measurements from the viewpoint of statistical mechanics would be gone, if the consistent histories approach is viable or, at least, will turn out to 
be really viable in the end. 

However, as it stands presently the approach produces results at variance with predictions of the measurement-based quantum mechanics
\cite{kent_prl} (then it is not important which specific interpretation one prescribes).  Hence, within its present status, the consistent histories approach
cannot be a substitute for the statistical mechanics-based theory of quantum measurements. Some authors bypass problems of the consistent
histories approach and state that it is useful as a paradox-free reformulation of the standard mechanics; see e.g. the recent review by
Hohenberg \cite{hohenberg} and the book by Griffiths \cite{griffiths_book}. In fact the opposite is true: as we explain below, the
consistent histories approach adds paradoxes that do not exist within the statistical interpretation of quantum mechanics.

}
\subsubsection{Deeper into consistent histories}

\hfill{}\footnote{ Don't look for small bones in the wolf's den}


\vspace{-1.cm}
\myskipfigText{
\begin{figure}[h!h!h!]
\label{ArmProv2}
\hfill{\includegraphics[width=6cm]{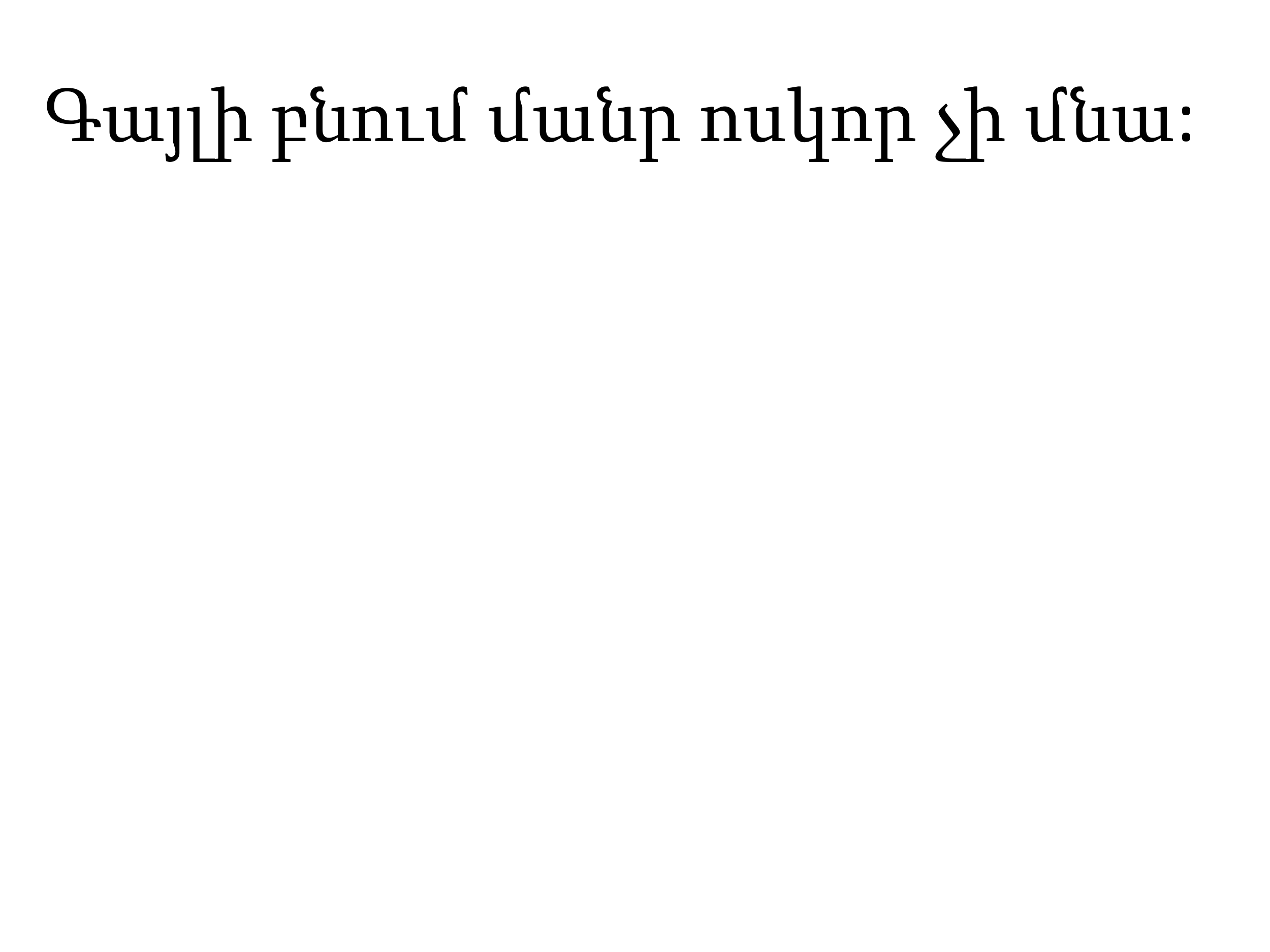}\hspace{2.3mm}}
\end{figure}}

\vspace{-4.2cm}


\hfill{Armenian proverb}

\vspace{3mm}

  \ZeText{

The easiest method of introducing the consistent histories approach is
to highlight as soon as possible its differences with respect to the standard
measurement-based approach. Let us start with the quantum mechanics
formula for the probability of two consecutive measurements ${\cal M}(t_1)$  and ${\cal M}(t_2)$ 
carried out at times $t_1$ and $t_2$ ($t_2>t_1$):
\BEA
p_{t_1,t_2}\left[i,j|{\cal M}(t_1), {\cal M}(t_2)\right]={\rm tr}\left[\Pi _j(t_2)\Pi _i(t_1)\rho \Pi _i(t_1)\Pi _j(t_2)  \right],
\label{cohi1}
\EEA
where $\rho$ is the initial state of the system, $\Pi _i(t_1)$ with $\sum_i \Pi _i(t_1)=1$
and $\Pi _j(t_2)$ with sum $\sum_i \Pi _i(t_2)=1$ are the projectors for
the physical quantities (represented by Hermitean operators) $A(t_1)$ and $B(t_2)$ measured
at the times $t_1$  and $t_2$, respectively. For simplicity we assume the Heisenberg
representation, and do not write in (\ref{cohi1}) explicit indices for
$A$ and $B$. What is however {\it necessary} to do is to indicate that
the joint probability in (\ref{cohi1}) is explicitly conditional on the
two measurements ${\cal M}(t_1)$ and ${\cal M}(t_2)$. As expected, the
meaning of (\ref{cohi1}) is that the measurement at time $t_1$ (with
probabilities given by Born's rule) is accompanied by selection of
the subensemble referring to the result $i$. The members of this
subensemble are then measured at the time $t_2$.  Generalization to
$n$-time measurements is obvious. 

What now the consistent histories approach does is to skip the
context-dependence in (\ref{cohi1}) and regard the resulting
probabilities $p[i,j]$ as a description of events taking place
spontaneously, i.e. {\it without any measurement and without any selection of outcome}. The cost to pay is
that the initial state $\rho$ and the projectors $\Pi _i(t_1)$ and $\Pi _j(t_2)$ have
to satisfy a special {\it consistency} condition that eliminates the dependence on the specific measurements ${\cal M}_{1,2}$ 
(without this condition the events are not defined):
\BEA
{\rm tr}\left[\Pi _j(t_2)\Pi _i(t_1)\rho \Pi _{i'}(t_1)\Pi _{j'}(t_2) \right ]=\delta_{ii'}\delta_{jj'}p_{t_1,t_2}[i,j],
\label{cohi2}
\EEA
where $\delta_{ii'}$ is the Kronecker delta. As a consequence of (\ref{cohi2}), one can 
sum out the first (i. e., the earlier) random variable and and using the completeness relation $\sum_i \Pi_i(t_1)=1$ get the probability 
for the second event alone: 

\BEA
\label{cohi3}
p_{t_2}[j] =\sum_i p_{t_1,t_2}[i,j]={\rm tr}\left[\Pi _j(t_2)\rho \Pi _j(t_2)  \right].
\EEA
To show this, one inserts $\sum_i \hat\Pi_i(t_1)=1$ left and right of $\rho$ and employs (2.5) with $j'=j$. Then Eq. (2.4) confirms that the calculated quantity 
is indeed the desired marginal probability. Note that without condition (\ref{cohi2}), i.e. just staying within the
standard approach (\ref{cohi1}), Eq. (\ref{cohi3}) would not hold,
e.g. generally $\sum_i p_{t_1,t_2}[i,j|{\cal M}(t_1), {\cal M}(t_2)]$
still depends on ${\cal M}(t_1)$ and is not equal to $p_{t_2}[j|{\cal
M}(t_2)]$ (probability of the outcome $j$ in the second measurement provided that no first
measurement was done). This is natural, since quantum measurements generally do
perturb the state of the measured system.  Hence (\ref{cohi2}) selects
only those situations, where this perturbation is seemingly absent. 

Any time-ordered sequence of events defines a history.
A set of histories satisfying (\ref{cohi2}) is called a {\it consistent histories} set. 
Due to (\ref{cohi2}), the overall probability of the consistent histories sums to one.

In effect (\ref{cohi2}) forbids superpositions; hence, it is called
decoherence condition \cite{griffiths_book,gellmann,hohenberg}. One
notes that (\ref{cohi2}) is sufficient, but not necessary for deriving
(\ref{cohi3}).  Hence, certain weaker conditions instead of
(\ref{cohi2}) were also studied \cite{griffiths_jsp}, but generally they
do not satisfy the straightforward statistical independence conditions
(independently evolving systems have independent probabilities)
\cite{diosi_consistent_history}. 

It was however noted that the consistent histories approach can produce
predictions at variance with the measurement based quantum mechanics
\cite{kent_prl}.  The simplest example of such a situation is given in
\cite{griffiths_hartle}. Consider a quantum system with zero Hamiltonian
in the pure initial state
\BEA
\rho=
|\phi\rangle\langle \phi|, ~~ |\phi\rangle=\frac{1}{\sqrt{3}}[|a_1\rangle+|a_2\rangle+|a_3\rangle ],
\EEA
where the vectors $\{|a_k\rangle\}_{k=1}^3$ are orthonormal: $\langle a_k|a_{k'}\rangle=\delta_{kk'}$.
Define a two-event history with projectors
\BEA
\label{mong1}
\{\Pi _1(t_1) = |a_1\rangle\langle a_1|, \Pi _2(t_1) =1-|a_1\rangle\langle a_1|\}  ~~{\rm and}~~ \{\Pi _1(t_2) = |\psi\rangle\langle \psi|, 
\Pi _2(t_2)=1-|\psi\rangle\langle \psi|  \}, ~~ t_2>t_1,
\EEA
where
\BEA
|\psi\rangle=\frac{1}{\sqrt{3}}[|a_1\rangle+|a_2\rangle-|a_3\rangle ].
\EEA
This history is consistent, since conditions (\ref{cohi2}) hold due to 
$\langle\phi|\psi\rangle=\langle\phi|a_1\rangle  \langle a_1|\psi\rangle=\frac{1}{3}$. 
One now calculates 
\BEA
p_{t_1,t_2}[a_1,\psi]= {\rm tr}\left[\Pi _1(t_2)\Pi _1(t_1)\rho \Pi _1(t_1)\Pi _1(t_2)  \right]=
\langle \psi|a_1\rangle \langle a_1|\phi\rangle \langle \phi|a_1\rangle \langle a_1|\psi\rangle=\frac{1}{9},
\label{oo1}
\EEA
\BEA
\label{oo2}
p_{t_2}[\psi]= {\rm tr}\left[\Pi _1(t_2)\rho \Pi _1(t_2)  \right]=
|\langle \psi|\phi\rangle|^2 =\frac{1}{9}.
\EEA
Given two probabilities (\ref{oo1}) and (\ref{oo2}) one can calculate 
the following conditional probability:
\BEA
\label{mo1}
p_{t_1|t_2}[a_1|\psi]=\frac{p_{t_1,t_2}[a_1,\psi]}{p_{t_2}[\psi]}=1.
\EEA

Yet another two-event consistent history is defined with projectors
\BEA
\label{mong2}
\{\widetilde{\Pi} _1(t_1) = |a_2\rangle\langle a_2|, \widetilde{\Pi} _2(t_1) =1-|a_2\rangle\langle a_2|\}  
~~{\rm and}~~ \{\Pi _1(t_2) = |\psi\rangle\langle \psi|, 
\Pi _2(t_2)=1-|\psi\rangle\langle \psi|  \}, ~~ t_2>t_1.
\EEA
Comparing (\ref{mong2}) with (\ref{mong1}) we note that the first measurement at $t_1$ is different, i.e. it refers to measuring
a different physical observable. Analogously to (\ref{mo1}) we calculate for the second consistent history
\BEA
\label{mo2}
p_{t_1|t_2}[a_2|\psi]=1.
\EEA
The consistent histories (\ref{mong1}) and (\ref{mong2}) share one event,
$\psi$, at the later time. On the basis of this event (\ref{mo1}) retrodicts with
probability one (i.e., with certainty) that $a_1$ happened. Likewise,
(\ref{mo2}) retrodicts with certainty that $a_2$ happened. But the
events $a_1$ and $a_2$ are mutually incompatible, since their projectors
are orthogonal, $\langle a_1|a_2\rangle =0$: if one happened, the other
one could not happen. 

Note that such an inconsistency is excluded within the measurement-based
approach. There the analogues of (\ref{mo1}) and (\ref{mo2}) refer to
different contexts [different measurements]: they read, respectively,
$p[a_1|\psi, {\cal M}(t_1), {\cal M}(t_2)]=1$ and $p[a_2|\psi,
\widetilde{{\cal M}}(t_1), {\cal M}(t_2)]=1$. It is not surprising that
different contexts, ${\cal M}(t_1)\not = \widetilde{{\cal M}}(t_1)$,
force conditional probabilities to retrodict incompatible events.
Naturally, if within the standard approach one makes the same
measurements the incompatible events cannot happen, e.g.  $p[a_1|\psi,
{\cal M}(t_1), {\cal M}(t_2)]\times p[a_2|\psi, {\cal M}(t_1), {\cal
M}(t_2)]=0$, because the second probability is zero. 

Following Kent \cite{kent_prl} we interpret this effect as a
disagreement between the predictions (or more precisely: the retrodictions)
of the consistent history approach versus those of the measurement-based
quantum mechanics. In response to Kent, Griffiths and Hartle suggested
that for avoiding above paradoxes, predictions and retrodictions of the
approach are to be restricted to a single consistent history
\cite{griffiths_hartle,grif_1998}. Conceptually, this seems to betray
the very point of the approach, because in effect it brings back the
necessity of fixing the context within which a given consistent history
takes place. And what fixes this context, once measurements are absent?
 
Another possible opinion is that condition (\ref{cohi2}) is not strong
enough to prevent a disagreement with the measurement based approach,
and one should look for a better condition for defining events
\cite{kent,debate_1}. To our knowledge, such a condition is so far not
found.  Bassi and Ghirardi \cite{BassiGhirardi1} pointed out another logical
problem with the consistent histories approach.  This produced another
debate on the logical consistency of the approach \cite{debate_2,BassiGhirardi2}, 
which we will not discuss here. 

We hold the opinion that in spite of being certainly thought-provoking and interesting, the consistent histories approach, as it presently stands, 
cannot be a substitute for the theory of quantum measurements: Both conceptually and practically we still need to understand what is going on 
in realistic measurements, with their imperfections, and what are the perturbations induced 
on the system by its interaction with a measuring apparatus.

}

\subsection{What we learned from these models}

\hfill{\footnote{Fisherman: ``What's the news from the sea?'' Fish: ``I have a lot to say,
   but my mouth is full of water''}}

\vspace{-7.5mm}

\myskipfigText{
\begin{figure}[h!h!h!]
\label{ArmProv}
\hfill{\includegraphics[width=6cm]{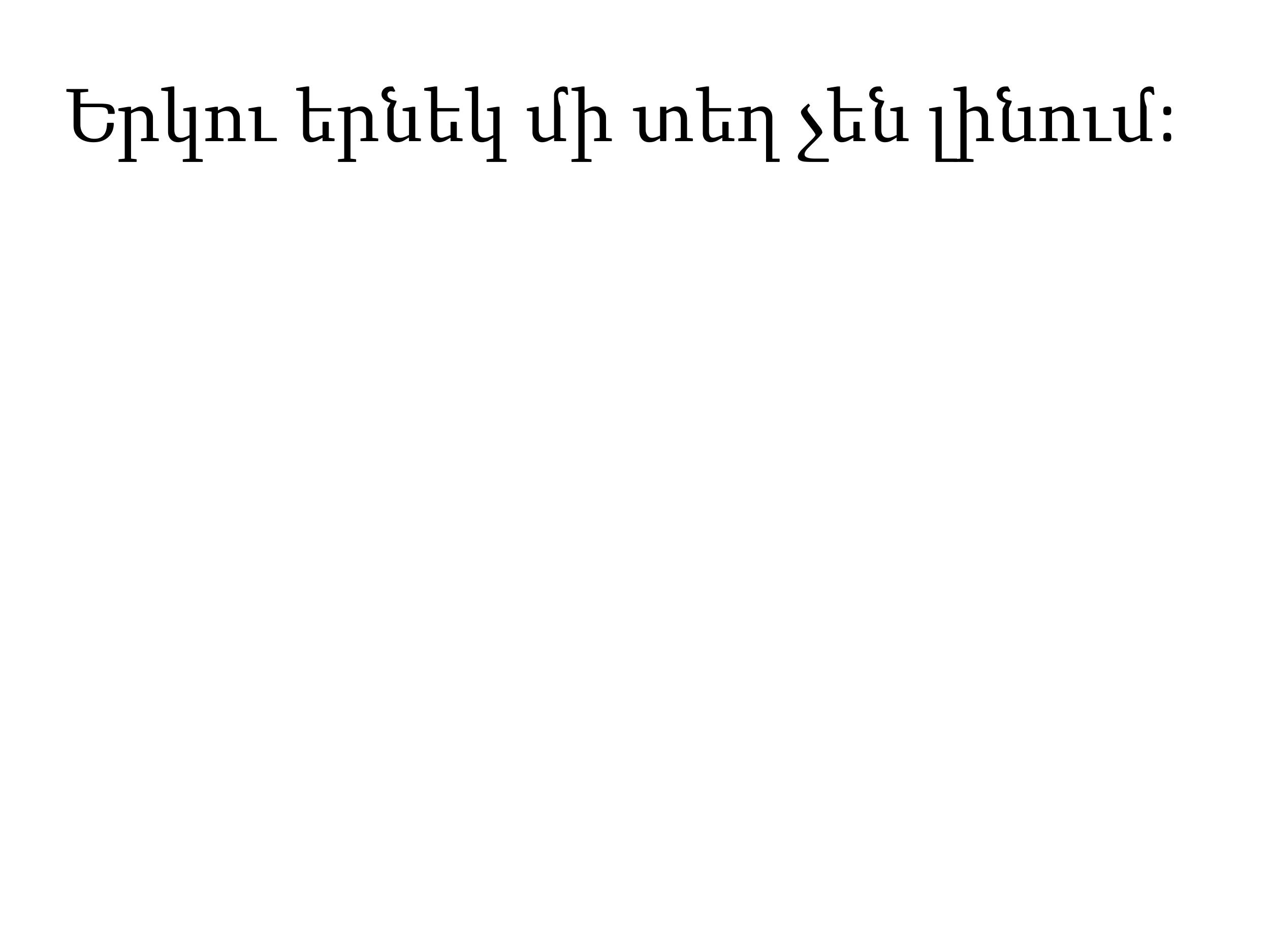}\hspace{3mm}}
\end{figure}
}

\vspace{-42.65mm}

\hfill{Armenian proverb}
\vspace{3mm}

 \ZeText{

Each one of the above models enlightens one or another among the many aspects of quantum measurements. However, none of them reproduces the
whole set of desired features: truncation and reduction of S + A, Born's rule, uniqueness of the outcome of a single process, 
complete scenario of the joint evolution of S + A, with an evaluation of its characteristic times, metastability of the initial state of A,
amplification within A of the signal, unbiased and robust registration by A in the final state, accurate establishment between S and the
pointer variable of A of the correlations that characterize an ideal measurement, influence of the parameters of the model on possible
imperfections of the measurement.  In particular, permanent registration requires the pointer to be macroscopic.  In the following we study in detail a
model, introduced in Refs.~\cite{ABNqm2003,ABNconf1,ABNconf2,ABNconf3,ABNconf4,ABNconf5,ABNconf6},
which encompasses these various requirements, and we extend the model in section 11.

\clearpage

}

\renewcommand{\thesection}{\arabic{section}}
\section{A Curie--Weiss model for quantum measurements}
\setcounter{equation}{0}\setcounter{figure}{0}
\renewcommand{\thesection}{\arabic{section}.}
\label{section.3}


\hfill{\it La vie humble aux travaux ennuyeux et faciles}

\hfill{\it Est une oeuvre de choix qui veut beaucoup d'amour\footnote{Humble life devoted to boring and easy tasks / Is 
a select achievement which demands much love}}

\hfill{Paul Verlaine, Sagesse} 

\vspace{3mm}

In this section we describe the model for a quantum measurement that was introduced by us in Ref.~\cite{ABNqm2003}.


\ZeText{
}

\subsection{General features}
\label{section.3.1}

\hfill{\it Perseverance can reduce an iron rod to a sewing needle}

\hfill{Chinese proverb}

\vspace{3mm}

\ZeText{

We take for S, the system to be measured, the simplest quantum system, namely a spin $\half$ 
(or any two-level system). The observable $\hat s$ to be measured is its
third Pauli matrix $\hat s_z$ = diag($1,-1$), with eigenvalues $s_i$ equal to $\pm1$. 
The statistics of this observable should not change significantly during the
measurement process \cite{vNeumann,Wigner,Yanase}. 
Hence $\hat s_z$ should be {\it conservative}, i.e., should commute with the Hamiltonian of S + A, at least nearly.

We have stressed at the end of \S ~\ref{section.1.2.1} that the apparatus A should lie initially in a metastable state \cite{ll_stat,lavis}, and finally 
in either one of several possible stable states (see section 2 for other models of this type).  This suggests to take for A,
as in several models described in section~\ref{section.2}, a quantum system that may undergo a phase transition with {\it broken invariance}. 
The initial state $\hat{{\cal R}}\left( 0\right) $ of ${\rm A}$\ is the metastable phase with unbroken invariance. The states $\hat{{\cal R}}_{i}$ 
represent the stable phases with broken invariance, in each of which registration can be permanent. The symmetry between the outcomes
prevents any bias. 

Here we need two such stable states, in one--to--one correspondence with the two
eigenvalues $s_i$ of $\hat s_z$. The simplest system which satisfies these properties
is an Ising model \cite{lavis}. Conciliating mathematical tractability and realistic features, we thus take as 
apparatus ${\rm A}={\rm M}+{\rm B}$, a model that
simulates a \textit{magnetic dot}: The magnetic degrees of freedom ${\rm M}$
consist of $N\gg1$ spins with Pauli operators $\hat{\sigma}_{a}^{\left(
n\right)  }$  ($n=1,2,\cdots,N$; $a=x$, $y$, $z$), which read for each $n$

\BEQ \label{sigmaxyz0}
\hat\sigma_x=\left( \begin{array}{ccc@{\ }r} 0 &1\\ 1&0 \end{array} \right),\qquad 
\hat\sigma_y=\left( \begin{array}{ccc@{\ }r} 0 &-i\\ i&0 \end{array} \right),\qquad 
\hat\sigma_z=\left( \begin{array}{ccc@{\ }r} 1 &0\\ 0&-1 \end{array} \right),\qquad 
\hat\sigma_0=\left( \begin{array}{ccc@{\ }r} 1 &0\\ 0&1 \end{array} \right),\qquad 
\EEQ
where $\hat\sigma_0$ is the corresponding identity matrix; 
 {\boldmath{$\hat \sigma$}} $=(\hat\sigma_x,\hat\sigma_y,\hat\sigma_z)$ denotes the vector spin operator.
The non-magnetic degrees of freedom such as phonons behave as a thermal bath ${\rm B}$ (Fig. 3.1). 
Anisotropic interactions between these spins can generate Ising ferromagnetism below the Curie temperature $T_{{\rm c}}$. As pointer variable $\hat{A}$\ we take the
order parameter, which is the magnetization in the $z$-direction (within normalization), as represented by the quantum observable\footnote{More
explicitly, the definition should involve the $\sigma_0^{(n')}$ for $n'\neq n$. E.g., for $N=3$ one has
$\hat m=\frac{1}{3}(\hat\sigma_z^{(1)}\hat\sigma_0^{(2)}\hat\sigma_0^{(3)}
+\hat\sigma_0^{(1)}\hat\sigma_z^{(2)}\hat\sigma_0^{(3)}+\hat\sigma_0^{(1)}\hat\sigma_0^{(2)}\hat\sigma_z^{(3)})$}

\begin{equation}
\mytext{\textcurrency hatm=\textcurrency \qquad}
\hat{m}=\frac{1}{N}\sum
_{n=1}^{N}\hat{\sigma}_{z}^{\left(  n\right)  }{ .} \label{hatm=}
\end{equation}

\ni We let $N$ remain finite, which will allow us to keep control of the equations 
of motion. It should however be sufficiently large so as to ensure the required
properties of phase transitions: The relaxation from $\hat{\cal R}(0)$ to either
one of the two states $\hat{\cal R}_i$, at the temperature $T$ (below $T_{\rm c}$)
imposed by the bath B, should be irreversible, the fluctuations of the order
parameter $\hat m$ in each equilibrium state $\hat{\cal R}_i$ should be weak 
(as $1/\sqrt{N}$), and the transition between these two states $\hat{\cal R}_i$
should be nearly forbidden.

The initial state ${\hat {\cal R}}\left(  0\right)  $\ of ${\rm A}$\ is the metastable paramagnetic state.
We expect the final state (\ref{Dtf=}) of S + A to involve for ${\rm A}$ the two stable ferromagnetic states
${\hat {\cal R}}_{i}$, $i=\,\uparrow$ or $\downarrow$, that we denote as $\hat {\cal R}_{\Uparrow}$ or $\hat {\cal R}_{\Downarrow}$, 
respectively\footnote{ Here and in the following, single arrows $\uparrow,\downarrow$ will denote the spin S,
while double arrows $\Uparrow,\Downarrow$ denote the magnet M}.
The equilibrium temperature $T$\ will be imposed to ${\rm M}$\ by the phonon bath \cite{petr,Weiss} through
weak coupling between the magnetic and non-magnetic degrees of freedom. Within
small fluctuations, the order parameter (\ref{hatm=}) vanishes
in ${\hat {\cal R}}\left(  0\right)  $\ and takes two opposite values in
the states ${\hat {\cal R}}_{\Uparrow}$ and ${\hat {\cal R}}_{\Downarrow}
$, $A_{i}\equiv\left\langle \hat{m}\right\rangle _{i}$ equal to
$+m_{{\rm F}}$ for $i=\uparrow$ and to $-m_{{\rm F}}$ for $i=\downarrow$\footnote{Note that the values $A_{i}=\pm m_{{\rm F}}$, which we wish to come out
associated with the eigenvalues $s_{i}=\pm1$, are determined from equilibrium
statistical mechanics; they are not the eigenvalues of $\hat{A}\equiv\hat{m}$,
which range from $-1$ to $+1$ with spacing $2/N$, but thermodynamic expectation values around which small fluctuations of order $1/\sqrt{N}$ occur.
For low $T$  they would be close to $\pm 1$}.
As in real magnetic registration devices \cite{nagaev}, information will be stored by ${\rm A}$\ in the form
of the sign of the magnetization.


}

\subsection{The Hamiltonian}
\label{section.3.2}

\hfill{\it I ask not for a lighter burden,}

\hfill{\it  but for broader shoulders}

\hfill{Jewish proverb}

\vspace{3mm}

\ZeText{

The full Hamiltonian can be decomposed into terms associated with the system,
with the apparatus and with their coupling:
\begin{equation}
\mytext{\textcurrency ham\textcurrency \qquad}
\hat{H}=\hat{H}_{{\rm S}}
+\hat{H}_{{\rm SA}}+\hat{H}_{{\rm A}}{ .} \label{ham}
\end{equation}

}
\subsubsection{The system}

\hfill{\it  A system that works is worth gold}

\hfill{Icelandic Proverb}

\vspace{3mm}

\ZeText{

Textbooks treat measurements as instantaneous, which is an idealization. If they are at least very fast, the tested system will hardly undergo dynamics by its own,
so the tested quantity $\hat s$ is practically constant.
As indicated above,  for an ideal measurement the observable $\hat{s}$ should commute with $\hat{H}$
\cite{Wigner, ABNqm2001, Yanase}. The simplest self-Hamiltonian that ensures this property (no evolution of S without coupling to A), is a constant one,
which is equivalent to the trivial one (since one may always add a constant to the energy)\footnote{As S is a spin $\half$, the only $\hat H_{\rm S}$ that commutes with the full 
Hamiltonian has the form $-b_z \hat s_z$, and the introduction of the magnetic field $b_z$ brings in only trivial changes (in sec 5)},

\begin{equation}
\mytext{\textcurrency HS=0\textcurrency \qquad}
\hat{H}_{{\rm S}}=0.
\label{HS=0}
\end{equation}
This commutation is required for ideal measurements, during the process of which the statistics of the tested observable should not be affected. 
More generally, in order to describe an imperfect measurement where $\hat{s}$
may move noticeably during the measurement (subsection 8.2), we shall introduce there
a magnetic field acting on the tested spin.

The coupling between the tested system and the apparatus,
\begin{equation}
\mytext{\textcurrency hamham\textcurrency \qquad}
\hat{H}_{{\rm SA}}=-g\hat
{s}_{z}\sum_{n=1}^{N}\hat{\sigma}_{z}^{\left(  n\right)  }=-Ng\hat{s}_{z}
\hat{m}{ ,} \label{HSA}
\end{equation}
has the usual form of a spin-spin coupling in the $z$-direction \cite{lavis}, and the constant $g>0$ characterizes its strength. As wished, it commutes with
$\hat{s}_{z}$. We have assumed that the range of the interaction between the spin ${\rm S}$ and the $N$ spins of ${\rm M}$ is large compared to the
size of the magnetic dot, so that we can disregard the space-dependence of the coupling. The occurrence of the factor $N$ in (\ref{HSA}) should not worry
us, since we will not take the thermodynamic limit $N\rightarrow\infty$. Although sufficiently large to ensure the existence of a clear phase
transition, $N$ is finite. We shall resume in  \S ~\ref{section.9.1.3} the conditions that $N$ should satisfy. In a realistic setting, the interaction
between ${\rm S}$ and ${\rm M}$ would first be turned on, then turned off continuously, while the tested spin approaches the dot then moves away. For
simplicity we assume $\hat{H}_{{\rm SA}}$ to be turned on suddenly at the initial time $t=0$, and it will be turned off at a final time $t_{\rm f}$, 
as we discuss below\footnote{ Contrary to the switching on, this switching off need not be performed suddenly since $m_{\rm F}$ is close to $m_{\Uparrow}$}.

}
\subsubsection{The magnet}

\ZeText{

\myskipfigText{
 \begin{figure}\label{figLoic1}
\centerline{ \includegraphics[width=8cm]{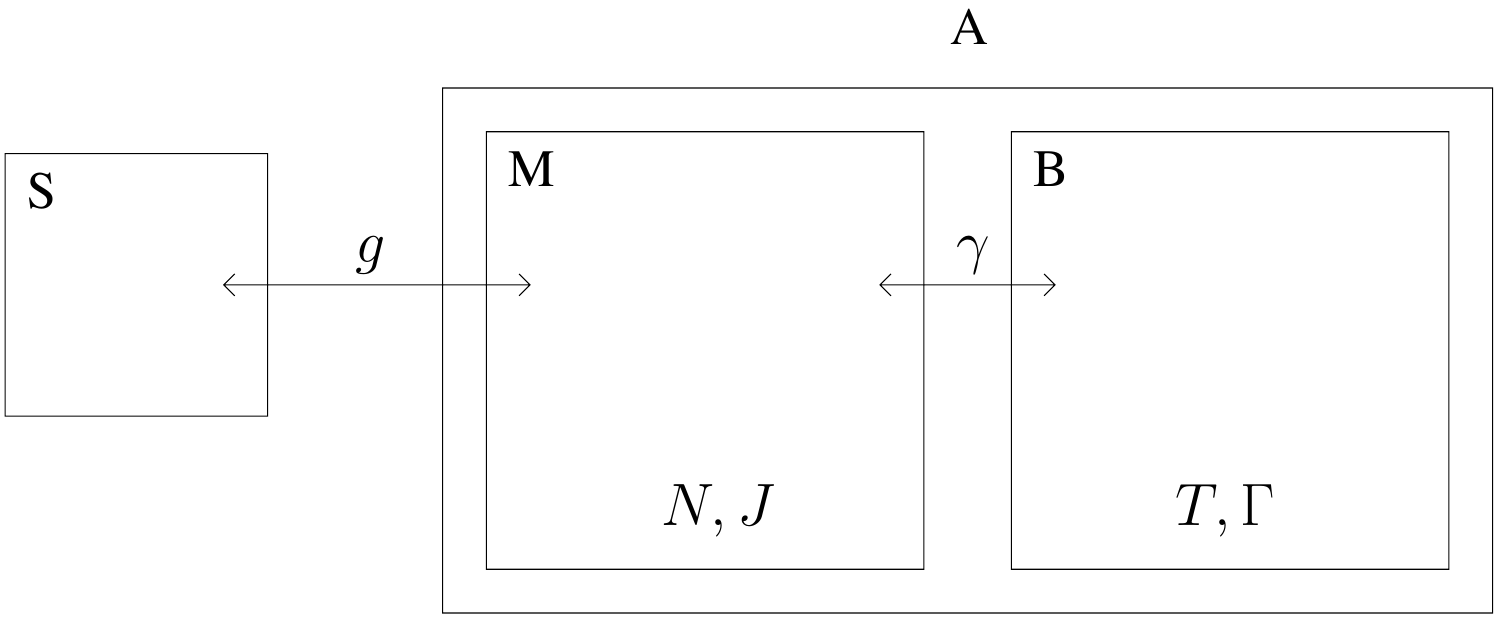}}
\caption{The Curie-Weiss measurement model and its parameters. The system S is a spin-$\half$ $\hat{\bf s}$.
The apparatus A includes a magnet M and a bath B. The magnet, which acts as a pointer, consists
of $N$ spins-$\half$ coupled to one another through an Ising interaction $J$. The phonon bath B is characterized
by its temperature $T$ and a Debye cutoff $\Gamma$. It interacts with M through a spin-boson coupling 
$\gamma$. The process is triggered by the interaction $g$ between the measured observable $\hat s_z$
and the pointer variable, the magnetization per spin, $\hat m$, of the pointer. 
}
\end{figure}
}

The apparatus ${\rm A}$ consists, as indicated above, of a magnet ${\rm M}$ and a phonon bath ${\rm B}$ 
(Fig. 3.1), and its Hamiltonian can be decomposed into

\begin{equation}
\hat{H}_{{\rm A}}=\hat
{H}_{{\rm M}}+\hat{H}_{{\rm B}}+\hat{H}_{{\rm M}{\rm B}}{ .}
\label{HA}
\end{equation}
The magnetic part is chosen as \cite{nagaev}%

\begin{equation}
\hat{H}_{{\rm M}}=-N\sum_{q=2,4}J_q
\frac{\hat{m}^{q}}{q}=-NJ_2\frac{\hat m^2}{2}-NJ_4\frac{\hat m^4}{4}
{ ,} \label{HM=}
\end{equation}
where the magnetization operator $\hat{m}$ was defined by (\ref{hatm=}). It couples all $q$-plets of spins {\boldmath{$\hat \sigma$}}$^{(n)}$
symmetrically, with the same coupling constant $J_qN^{1-q}$ for each $q$-plet.  (The factor $N^{1-q}$ is introduced only for convenience.) The
space-independence of this coupling is fairly realistic for a small magnetic dot, as in (\ref{HSA}). The interaction is fully anisotropic, involving
only the $z$-components. The exponents $q$ are even in order to ensure the up-down symmetry of the apparatus. 
The $q=2$ term is a standard spin-spin interaction.
The term $q=4$ describes so-called super-exchange interactions as realized for metamagnets \cite{nagaev}. 
We shall only consider ferromagnetic interactions ($J_{2}>0$ or $J_4>0$ or both).

We will see in \S~3.3.4 that the Hamiltonian (\ref{HA}) produces a paramagnetic equilibrium state at high temperature and
two ferromagnetic states at low temperature, with a transition of second order for $J_2 > 3J_4$, of first order for $3J_4 > J_2$. 
The former case is exemplified by the Curie--Weiss Ising model for an
anisotropic magnetic dot \cite{lavis}, with pairwise interactions in $\hat{\sigma}
_{z}^{\left(  n\right)  }\hat{\sigma}_{z}^{\left(  p\right)  }$, recovered here for $J_4=0$,

\begin{equation}
\mytext{\textcurrency HM2=\textcurrency \qquad}
\hat{H}_{{\rm M}}=-\frac{J_2N}{2}\hat{m}^{2}=
-\frac{J_2}{2N}\sum_{i,j=1}^N\hat\sigma_z^{(i)}\hat\sigma_z^{(j)}, \qquad (q=2).
\label{HM2=} \nn
\end{equation}
Likewise, the first-order case is exemplified by letting $J_2=0$, keeping in (\ref{HA}) only the quartic ``super-exchange'' term:

\begin{equation}
\mytext{\textcurrency HM4=\textcurrency \qquad}
\hat{H}_{{\rm M}}=-\frac{J_4N}{4}\hat{m}^{4}=-\frac{J_4}{4N^3}\sum_{i,j,k,l=1}^N
\hat\sigma_z^{(i)}\hat\sigma_z^{(j)}\hat\sigma_z^{(k)}\hat\sigma_z^{(l)}, \qquad (q=4).
\label{HM4=} \nn
\end{equation}

The more physical case (\ref{HA}) of mixtures of $q=2$ and $q=4$ terms will not differ qualitatively from either one of the two pure cases 
$q=2$ or $q=4$. It will therefore be sufficient for our purpose, in section 7, to illustrate the two situations $J_2 > 3J_4$ and $J_2<3J_4$ 
by working out the Hamiltonians (\ref{HM2=}) and (\ref{HM4=}), respectively.     
We may summarize these two cases by $H_{\rm M}=-(NJ/q)\hat m^q$ with $q=2$ or $4$, respectively.

Using ${\rm A}$ as a measurement apparatus requires the lifetime of the initial state to be larger than the
overall measurement time. An advantage of a first-order transition is the local stability of the paramagnetic state, even below the transition 
temperature, which ensures a much larger lifetime as in the case of a second order transition. 
We shall see, however (\S~7.3.2), that the required condition can be satisfied even for $q=2$ alone (i.e., for $J_4=0$).

\clearpage

}

\subsubsection{The phonon bath}

\hfill{\it It is not only one person who bathes in the witch's water}

\hfill{Ghanaian Proverb}

\vspace{3mm}

\ZeText{

The interaction between the magnet and the bath, which drives the apparatus to equilibrium, is taken
as a standard spin-boson Hamiltonian \cite{petr,Weiss,Gardiner}
\begin{equation}
\mytext{\textcurrency ham4\textcurrency \qquad}
\hat{H}_{{\rm M}{\rm B}
}=\sqrt{\gamma}\sum_{n=1}^{N}\left(  \hat{\sigma}_{x}^{\left(  n\right)  }
\hat{B}_{x}^{\left(  n\right)  }+\hat{\sigma}_{y}^{\left(  n\right)  }\hat
{B}_{y}^{\left(  n\right)  }+\hat{\sigma}_{z}^{\left(  n\right)  }\hat{B}
_{z}^{\left(  n\right)  }\right)  \equiv\sqrt{\gamma}\sum_{n=1}^{N}
\sum_{a=x,y,z}\hat{\sigma}_{a}^{\left(  n\right)  }\hat{B}_{a}^{\left(
n\right)  }{ ,} \label{ham4}
\end{equation}
which couples each component $a=x$, $y$, $z$ of each spin $\mathbf{\hat{\sigma}}^{\left(  n\right)}$ with some hermitean linear combination
$\hat{B}_{a}^{\left(  n\right)  }$ of phonon operators. The dimensionless
constant $\gamma\ll1$ characterizes the strength of the thermal coupling
between ${\rm M}$ and ${\rm B}$, which is weak.

For simplicity, we require that the bath acts independently for each spin degree of freedom $n$, $a$. 
(The so-called independent baths approximation.)  This can be achieved ($i$) by
introducing Debye phonon modes labelled by the pair of indices $k$, $l$, 
with eigenfrequencies $\omega_{k}$ depending
only on $k$, so that the bath Hamiltonian is
\begin{equation}
\mytext{\textcurrency Hbath\textcurrency \qquad}
\hat{H}_{{\rm B}}=\sum
_{k,l}\hbar\omega_{k}\hat{b}_{k,l}^{\dagger}\hat{b}_{k,l}{ ,}
\label{Hbath}
\end{equation}
and ($ii$) by assuming that the coefficients $C$ in
\begin{equation}
\mytext{\textcurrency B=\textcurrency \qquad}
\hat{B}_{a}^{\left(  n\right)
}=\sum_{k,l}\left[  C\left(  n,a;k,l\right)  \hat{b}_{k,l}+C^{\ast}\left(
n,a;k,l\right)  \hat{b}_{k,l}^{\dagger}\right]  \label{B=}
\end{equation}
are such that
\begin{equation}
\mytext{\textcurrency CC\textcurrency \qquad}
\sum_{l}C\left(  n,a;k,l\right)
C^{\ast}\left(  m,b;k,l\right)  =\delta_{n,m}\delta_{a,b}\,c\left(  \omega
_{k}\right)  { .} \label{CC}
\end{equation}
This requires the number of values of the index $l$ to be at least
equal to $3N$. For instance, we may associate with each component
$a$ of each spin $\mathbf{\hat{\sigma}}^{\left(  n\right)  }$ a
different set of phonon modes, labelled by $k$, $n$, $a$,
identifying $l$ as ($n$, $a$), and thus define
$\hat{H}_{{\rm B}}$ and $\hat{B}_{a}^{\left(  n\right)  }$ as
\begin{eqnarray}
\mytext{\textcurrency Hbaths\textcurrency \qquad}
\hat{H}_{{\rm B}}  &
=&\sum_{n=1}^{N}\sum_{a=x,y,z}\sum_{k}\hbar\omega_{k}\hat{b}_{k,a}
^{\dagger\left(  n\right)  }\hat{b}_{k,a}^{\left(  n\right)  }{
,}\label{Hbaths}\\
 \hat{B}_{a}^{\left(n\right)  }&=&\sum_{k}\sqrt{c\left(  \omega_{k}\right)  }\left(  \hat{b}
_{k,a}^{\left(  n\right)  }+\hat{b}_{k,a}^{\dagger\left(  n\right)  }\right)
{ .} \label{B=s}
\end{eqnarray}
We shall see in \S ~\ref{section.3.3.2} that the various choices of the phonon set, of the spectrum (\ref{Hbath}) and of the operators (\ref{B=}) coupled to the spins are
equivalent, in the sense that the joint dynamics of ${\rm S}+{\rm M}$ will depend only on the spectrum $\omega_{k}$ and on the coefficients $c\left(\omega_{k}\right)$.

The spin-boson coupling (\ref{ham4}) between M and B will be sufficient for our purpose up to section 9. This interaction, of the Glauber type, does not commute with 
$\hat H_{\rm M}$, a property needed for registration, since M has to release energy when relaxing from its initial metastable paramagnetic state to one of its final 
stable ferromagnetic states at the temperature $T$. However, the complete solution of the measurement problem presented in section 11 will require more
 complicated interactions. We will therefore introduce in \S~\ref{fin11.2.3} a small but random coupling between the spins of M, and in \S~11.2.5 a more realistic 
 small coupling  between M and B, of the Suzuki type, which produces flip-flops of the spins of M without changing the value of the energy that M 
 would have with only the terms (\ref{HM2=}) and/or (\ref{HM4=}).
 
 \clearpage

}

\subsection{Structure of the states}

\label{section.3.3}

\hfill{\it If you do not enter the tiger's cave, }

\hfill{\it you will not catch its cub}

\hfill{Japanese proverb}

\vspace{3mm}

\subsubsection{Notations}

\label{section.3.3.1}
\ZeText{

The full state ${\hat{{\cal D}}}$ of the system evolves according to the
Liouville--von~Neumann equation (\ref{LB001}), which we have to solve. It will
be convenient to define through partial traces, at any instant $t$, the
following marginal density operators: \\ 
$\hat{r}$ for the tested system ${\rm S}$, $\hat{{\cal R}}$ for the apparatus 
${\rm A}$, $\hat{R}_{\rm M}$ for the magnet ${\rm M}$, $\hat{R}_{{\rm B}}$ for the
bath, and $\hat{D}$ for ${\rm S}+{\rm M}$ after elimination of the bath 
(as depicted schematically in Fig. 3.2), according to

\begin{equation}
\hat{r}={\rm {\rm tr}}_{{\rm A}}{\hat{{\cal D}}}{,\qquad}
\hat{{\cal R}} ={\rm {\rm tr}}_{{\rm S}}{\hat{{\cal D}}}{,\qquad}
\hat{R}_{\rm M}={\rm tr}_{\rm B}\hat{{\cal R}} ={\rm tr}_{{\rm S},{\rm B}}{\hat{{\cal D}}}{,\qquad}\hat{R}_{{\rm B}
}={\rm {\rm tr}}_{{\rm S},\,{\rm M}}{\hat{{\cal D}}
}{\rm ,\qquad}\hat{D}={\rm {\rm tr}}_{{\rm B}}{\hat
{{\cal D}}}{\rm  .} \label{reducedDM=}
\end{equation}
The expectation value of any observable $\hat A$ pertaining, for instance, to the subsystem S + M of S + A 
(including products of spin operators $\hat s_a$ and $\hat\sigma^{(n)}_a$ ) can equivalently be evaluated as 
$\langle\hat A\rangle={\rm tr}_{S + A} \hat{\cal D}\hat A$ or as $\langle\hat A\rangle={\rm tr}_{S + M} \hat{D}\hat A$.

\myskipfigText{
 \begin{figure}
 \label{figLoic2}
\centerline{ \includegraphics[width=4cm]{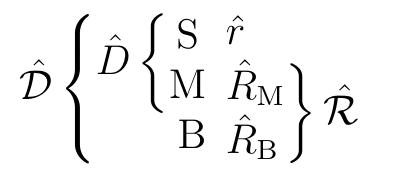}
\hspace{2cm} 
 \includegraphics[width=5.5cm]{FigOpMag1}}
\caption{
 Notations for the density operators of the system S + A and the subsystems M and B of A.
The full density matrix $\hat {\cal D}$ is parametrized by its submatrices $\hat{\cal R}_{ij}$ (with $i,j=\pm1$ or $\uparrow,\downarrow$),
the density matrix $\hat D$ of S + M by its submatrices $\hat R_{ij}$.  The marginal density operator of S is denoted as $\hat r$
and the one of A as $\hat {\cal R}$. The marginal density operator of M itself is denoted as $\hat R_{\rm M}$ and the one of B as $\hat R_{\rm B}$.
}
\end{figure}
}

As indicated in subsection~\ref{section.1.2}, the apparatus \textrm{A} is a large system,
treated by methods of statistical mechanics, while we need to follow in detail
the microscopic degrees of freedom of the system \textrm{S} and their
correlations with \textrm{A}. To this aim, we shall analyze the full state
${\hat{{\cal D}}}$ of the system into several sectors, characterized by the
eigenvalues of $\hat{s}_{z}$. Namely, in the two-dimensional eigenbasis of
$\hat{s}_{z}$ for ${\rm S}$, $\left\vert {\uparrow}\right\rangle $,
$\left\vert {\downarrow}\right\rangle $, with eigenvalues $s_{i}=+1$ for
$i=\uparrow$ and $s_{i}=-1$ for $i=\downarrow$, ${\hat{{\cal D}}}$ can be
decomposed into the four blocks

\begin{equation}
\hat{\cal D}=
\left( \begin{array}{ccc@{\ }r}
{\hat{\cal R}}_{\uparrow\uparrow}& {\hat{\cal R}}_{\uparrow\downarrow}\\
{\hat{\cal R}}_{\downarrow\uparrow}& {\hat{\cal R}}_{\downarrow\downarrow}
\end{array} \right)
{ ,} \label{CDelemts}
\end{equation}
where each ${\hat {\cal R}}_{ij}$ is an operator in the space of the apparatus.
We shall also use the partial traces (see again Fig. 3.2)

\begin{equation}
\mytext{\textcurrency Rij\textcurrency \qquad}
\hat{R}_{ij}={\rm {\rm tr}}_{{\rm B}}{\hat {\cal R}}_{ij},{\qquad
\qquad}\hat{D}={\rm {\rm tr}}_{{\rm B}}{\hat {\cal D}}=
\left( \begin{array}{ccc@{\ }r}
\hat{R}_{\uparrow\uparrow} & \hat{R}_{\uparrow\downarrow}\\
\hat{R}_{\downarrow\uparrow} & \hat{R}_{\downarrow\downarrow}
\end{array} \right)
\label{Rij}
\end{equation}
over the bath; each $\hat{R}_{ij}$ is an operator in the $2^{N}$-dimensional
space of the magnet. Indeed, we are not interested in the evolution of the
bath variables, and we shall eliminate ${\rm B}$ by relying on the weakness
of its coupling (\ref{ham4}) with ${\rm M}$. The operators $\hat{R}_{ij}$
code our full statistical information about ${\rm S}$ and ${\rm M}$. 
We shall use the notation $\hat R_{ij}$ whenever we refer to S + M and $\hat R_{\rm M}$ when referring 
to M alone. Tracing also over M, we are, according to (\ref{reducedDM=}), left with

\begin{equation}
\hat r
=\left( \begin{array}{ccc@{\ }r}
{{r}}_{\uparrow\uparrow}& {{ r}}_{\uparrow\downarrow}\\
{{r}}_{\downarrow\uparrow}& {{r}}_{\downarrow\downarrow}
\end{array} \right)= 
r_{\up\up}\,|\hspace{-1mm}\uparrow\rangle\langle\uparrow\hspace{-1mm}| +
r_{\up\down}\,|\hspace{-1mm}\uparrow\rangle\langle\downarrow\hspace{-1mm}| +
r_{\down\up}\,|\hspace{-1mm}\downarrow\rangle\langle\uparrow\hspace{-1mm}| +
r_{\down\down}\,|\hspace{-1mm}\downarrow\rangle\langle\downarrow\hspace{-1mm}| .
 \label{relements}
\end{equation}
 The magnet ${\rm M}$ is thus described by $\hat{R}_{\rm M}=\hat
{R}_{\uparrow\uparrow}+\hat{R}_{\downarrow\downarrow}$, the system
${\rm S}$ alone by the matrix elements $r_{ij}={\rm tr}
_{{\rm M}}\hat{R}_{ij}$ of $\hat{r}$. The correlations of $\hat{s}_{z}$,
$\hat{s}_{x} $ or $\hat{s}_{y}$ with any function of the observables
$\hat{\sigma}_{a}^{\left(  n\right)  }$ ($a=x,y,z$ , $n=1$ , \ldots$N$) are
represented by $\hat{R}_{\uparrow\uparrow}-\hat{R}_{\downarrow\downarrow}$,
$\hat{R}_{\uparrow\downarrow}+\hat{R}_{\downarrow\uparrow}$, $i\hat
{R}_{\uparrow\downarrow}-i\hat{R}_{\downarrow\uparrow}$, respectively. The
operators $\hat{R}_{\uparrow\uparrow}$ and $\hat{R}_{\downarrow\downarrow}$
are hermitean positive, but not normalized, whereas $\hat{R}_{\downarrow
\uparrow}=\hat{R}_{\uparrow\downarrow}^{\dagger}$.
Notice that we now have from (\ref{reducedDM=}) --  (\ref{Rij})

\begin{equation}
\mytext{\textcurrency reducedDM2=\textcurrency \qquad}
r_{ij} ={\rm {\rm tr}}_{{\rm A}}{\hat{{\cal R}}_{ij}} ={\rm {\rm tr}}_{{\rm M}}{\hat{{R}}_{ij}}   {,\qquad}
\hat{{\cal R}} = {\hat {\cal R}}_{\uparrow\uparrow} +{\hat {\cal R}}_{\downarrow\downarrow},\qquad
\hat{{R}}_{{\rm M}}={\hat{R}}_{\uparrow\uparrow} +{\hat{R}}_{\downarrow\downarrow}
{,\qquad}\hat{R}_{{\rm B}}= {\rm {\rm tr}}_{{\rm M}}({\hat {\cal R}}_{\uparrow\uparrow} +{\hat {\cal R}}_{\downarrow\downarrow}){\rm  .} 
\label{reducedDM2=}
\end{equation}

All these elements are functions of the time $t$ which elapses from the beginning of the measurement at  $t=0$ when $\hat H_{\rm SA}$ is switched on to 
the final value  $t_{\rm f}$ that we will evaluate in section 7. To introduce further notation, we mention that the combined system S + A = S + M + B should end up in

\begin{equation}
\hat{\cal D}(t_{\rm f})=
\left( \begin{array}{ccc@{\ }r}
p_{\uparrow}{\hat{\cal R}}_{\Uparrow}& 0 \\
0  & p_{\downarrow}{\hat{\cal R}}_{\Downarrow}
\end{array} \right)  
=p_{\uparrow}\,|\hspace{-1mm}\uparrow\rangle\langle\uparrow\hspace{-1mm}|\otimes {\hat{\cal R}}_{\Uparrow} 
+ p_\downarrow\, |\hspace{-1mm}\downarrow\rangle\langle\downarrow\hspace{-1mm}|\otimes{\hat{\cal R}}_{\Downarrow}
= \sum_i p_i \,\scriptD_i ,
\label{CDelemtsfin}
\end{equation}
where ${\hat{\cal R}}_{\Uparrow}$ (${\hat{\cal R}}_{\Downarrow}$) is density matrix of the thermodynamically 
stable state of the magnet and bath, after the measurement, in which the magnetization is up, taking the value  $m_{\Uparrow}(g)$ 
 (down, taking the value $m_{\Downarrow}(g)$); these events should occur with probabilities $p_{\uparrow}$ and $p_{\downarrow}$, 
respectively\footnote{Notice that in the final state we denote properties of the tested system by $\uparrow,\downarrow$
and of the apparatus by $\Uparrow,\Downarrow$. In sums like (\ref{Dtf=}) we will also use $i=\uparrow,\downarrow$, or sometimes $i=\pm1$}.
The Born rule then predicts that $p_{\uparrow}={\rm tr}_{\rm S}\hat r(0)\Pi_{\uparrow}=r_{\uparrow\uparrow}(0)$ and 
$p_{\downarrow}=r_{\downarrow\downarrow}(0)$.

Since no off-diagonal terms occur in (\ref{CDelemtsfin}), a point that we wish to explain, 
and since we expect B to remain nearly in its initial equilibrium state, 
we may trace out the bath, as is standard in classical and quantum thermal physics, without losing significant information. 
It will therefore be sufficient for our purpose to show that the final state is

\begin{equation}
\hat{D}(t_{\rm f})=
\left( \begin{array}{ccc@{\ }r}
p_{\uparrow}{\hat{ R}}_{{\rm M}\Uparrow}& 0 \\
0  & p_{\downarrow}{\hat{ R}}_{{\rm M}\Downarrow}
\end{array} \right)  
=p_{\uparrow}\, |\hspace{-1mm}\uparrow\rangle\langle\uparrow\hspace{-1mm}|\otimes {\hat{R}}_{{\rm M}\Uparrow}+
 p_\downarrow\, |\hspace{-1mm}\downarrow\rangle\langle\downarrow\hspace{-1mm}|\otimes {\hat{ R}}_{{\rm M}\Downarrow},
\label{Delemtsfin}
\end{equation}
now referring to the magnet M and tested spin S alone.

Returning to Eq. (\ref{reducedDM2=}), we note that from any density operator $\hat{R}_{}$ of the magnet we 
can derive the {\it probabilities $P_{\rm M}^{\rm dis}\left(  m\right)$ for $\hat{m}$ to take the eigenvalues}
$m$, where ``dis'' denotes their discreteness. These $N+1$ eigenvalues,

\begin{equation}
\mytext{\textcurrency eig\textcurrency \qquad}
m=-1{\rm ,\qquad}-1+\frac{2}
{N}{\rm ,\qquad}\ldots{\rm,\qquad}1-\frac{2}{N}{\rm ,\qquad}1{\rm ,}
\label{eig}
\end{equation}
have equal spacings $\delta m=2/N$ and multiplicities

\BEA
\mytext{\textcurrency deg\textcurrency \qquad}
\label{deg}
G\left(  m\right) =\frac{N!}{\left[  \frac{1}{2}N\left(  1+m\right)  \right]  !\left[  \frac{1}{2}N\left(  1-m\right)  \right]  !}
 =\hspace{-1mm} \sqrt{\frac{2}{\pi N\left(  1-m^{2}\right)  }}
\exp\left[N\left(-\frac{1+m}{2} \ln \frac{1+m}{2} -\frac{1-m}{2} \ln \frac{1-m}{2}\right)+{\cal O}\left(\frac{1}{N}\right)\right] . \,\,
\EEA
Denoting by $\delta_{\hat{m},m}$ the projection operator on the subspace $m$
of $\hat{m}$, the dimension of which is $G\left(  m\right)  $, we have

\BEQ
P_{\rm M}^{\rm dis}\left(  m,t\right)  ={\rm tr}_{\rm M}\hat{R}_{\rm M}(t)\delta_{\hat{m},m}.
\label{R->P}
\EEQ 
In the limit $N\gg1$,  where $m$ becomes basically a continuous variable, we shall later work with the functions $P_{\rm M}(m,t)$

\begin{equation}  
P_{\rm M}(m,t)=\frac{N}{2}P_{\rm M}^{\rm dis}(m,t),\qquad \int_{-1}^1 \d m\,P_{\rm M}(m,t)=\sum_m P_{\rm M}^{\rm dis}(m,t)=1,
\label{Pdis2cont}
\end{equation}
that have a finite and smooth limit for $N\to\infty$, and use similar relations between the functions $P_{ij}$ and $P^{\rm dis}_{ij}$,
and $C_a$ and $C_a^{\rm dis}$, introduced next.

In what follows, the density operators $\hat{R}_{\rm M}$ will most often depend only on the observables $\hat{\sigma}_{z}^{\left(  n\right)  }$ and be
symmetric functions of these observables. Hence, $\hat{R}_{\rm M}$ will reduce to a mere function of the operator $\hat{m}$ defined by (\ref{hatm=}).
In such a circumstance, eq. (\ref{R->P}) can be inverted: the knowledge of $P_{\rm M}\left(  m\right)  $ is then sufficient to determine the density operator $\hat
{R}_{\rm M}$, through a simple replacement of the scalar $m$ by the operator $\hat{m}$ in

\begin{equation}
\mytext{\textcurrency P->R\textcurrency \qquad}
\hat{R}_{\rm M}(t)=\frac{1}{G\left( \hat m\right)}P_{\rm M}^{\rm dis}\left(  \hat{m},t \right)  {\rm  .} 
\label{P->R}
\end{equation}
The expectation value of any product of operators $\hat{\sigma}_{a}^{\left(
n\right)  }$ of the magnet can then be expressed in terms of $P_{\rm M}^{\rm dis}\left(
m\right)  $. For instance, the two-spin correlations ($n\neq p$) are related
to the second moment of $P_{\rm M}^{\rm dis}\left(  m\right)  $ by

\begin{equation}
{\rm tr}_{\rm S,A}
{\hat {\cal D}}\hat{\sigma}_{a}^{\left(  n\right)  }\hat{\sigma}_{b}^{\left(  p\right)  }={\rm tr}_{\rm M}\hat{R}_{\rm M}\hat{\sigma}_{a}^{\left(  n\right)  }\hat{\sigma}_{b}^{\left(  p\right)
}=\frac{\delta_{a,z}\delta_{b,z}}{N-1}\left[  N\sum_{m}P_{\rm M}^{\rm dis}\left(  m\right)
m^{2}-1\right] {\rm  .} \label{sigmanp}%
\end{equation}

Likewise, when the operators $\hat{R}_{ij}$ in (\ref{Rij}) depend only on
$\hat{m}$, we can parameterize them at each time, according to
\begin{equation}
\hat{R}_{ij}(t)=\frac{1}{G\left(
\hat m\right)  }P_{ij}^{\rm dis}\left(  \hat{m},t\right)  {\rm  ,} \label{RijP}
\end{equation}
by functions $P_{ij}^{\rm dis}\left(  m\right)$ defined on the set (\ref{eig}) of
values of $m$, with $[P_{ij}^{\rm dis}\left(  m\right)]^{\ast}=P_{ji}^{\rm dis}\left(  m\right)$. 
(For the moment we refrain from denoting the explicit $t$ dependence.)
All statistical properties of ${\rm S}+{\rm M}$\ at the considered
time can then be expressed in terms of these functions $P_{ij}^{\rm dis}\left(
m\right)  $. Indeed the combinations

\begin{equation}
\mytext{\textcurrency P->C\textcurrency \qquad}
C_{x}^{\rm dis}\left(  m\right)  =P^{\rm dis}_{\uparrow\downarrow}(m)+
P^{\rm dis}_{\downarrow\uparrow}(m),{\qquad\qquad}
C^{\rm dis}_{y}=iP^{\rm dis}_{\uparrow\downarrow}-iP^{\rm dis}_{\downarrow\uparrow}
{,\qquad\qquad}
C^{\rm dis}_{z}=P^{\rm dis}_{\uparrow\uparrow}-P^{\rm dis}_{\downarrow\downarrow} 
\label{P->C}%
\end{equation}
encompass all the correlations between $\hat{s}_{x}$, $\hat{s}_{y}$ or
$\hat{s}_{z}$ and any number of spins of the apparatus. In particular, the
expectation values of the components of $\mathbf{\hat{s}}$ are given by
\begin{equation}
\mytext{\textcurrency s=C\textcurrency \qquad}
{\rm tr}{\hat {\cal D}}\hat{s}_{a}
=\sum_{m}C^{\rm dis}_{a}\left(  m\right) =\int_{-1}^1\d m\,C_{a}\left(  m\right)  {\rm  ,} \label{s=C}
\end{equation}
 with the continuous functions $C_a(m)=\half N C_a^{\rm dis}(m)$ as in (\ref{Pdis2cont}),
while the correlations between $\mathbf{\hat{s}}$ and one spin of ${\rm M}$\ are

\begin{equation}
\mytext{\textcurrency ssigma\textcurrency \qquad}
{\rm tr}{\hat {\cal D}}\hat{s}_{a}\hat{\sigma}_{b}^{\left(  n\right)  }
=\delta_{b,z}\sum_{m}C^{\rm dis}_{a}\left(  m\right)  m
=\delta_{b,z}\int_{-1}^1\d m\,C_{a}\left(  m\right)  m.  \label{ssigma}
\end{equation}
Correlations of $\mathbf{\hat{s}}$ with many spins of ${\rm M}$\ involve
higher moments of $C^{\rm dis}_{a}\left(  m\right)  $ as in eq. (\ref{sigmanp}). We can
interpret $P^{\rm dis}_{\uparrow\uparrow}\left(  m\right)  $ as the joint probability to
find ${\rm S}$\ in $\left\vert \uparrow\right\rangle $ and $\hat{m}$ equal
to $m$, so that $P^{\rm dis}_{\uparrow\uparrow}\left(  m\right)  +P^{\rm dis}_{\downarrow
\downarrow}\left(  m\right)  =P_{\rm M}^{\rm dis}\left(  m\right)  $ reduces to the probability
$P_{\rm M}^{\rm dis}\left(  m\right)  $ which characterizes through (\ref{P->R}) the marginal
state of ${\rm M}$.

}
\subsubsection{Equilibrium state of the bath}

\hfill{\it Motion is an illusion}

\hfill{Zeno of Elea}

\vspace{3mm}

\label{section.3.3.2}
\ZeText{

At the initial time, the bath is set into equilibrium at the temperature\footnote{We use units where Boltzmann's constant is equal to one \cite{ll_stat}; 
otherwise, $T$ and $\beta=1/T$ should be replaced throughout by  $k_BT$ and $1/k_B T$, respectively} $T=1/\beta$.
The density operator of the bath, 
\begin{equation}
\mytext{\textcurrency RB0=\textcurrency \qquad}
\hat{R}_{{\rm B}}\left(0\right)  =\frac{1}{Z_{{\rm B}}}e^{-\beta\hat{H}_{{\rm B}}}{\rm  ,}
\label{RB0=}%
\end{equation}
when $\hat{H}_{{\rm B}}$ is given by (\ref{Hbath}), describes the set of
phonons at equilibrium in independent modes.

As shown in section~\ref{section.4.1} the bath will be involved in our problem only through
its \textit{autocorrelation function} in the equilibrium state (\ref{RB0=}),
defined  in the Heisenberg picture (see \S~\ref{fin10.1.2}) by

\BEA 
\mytext{\textcurrency Kt-s}  \mytext{\textcurrency \qquad}
&{\rm tr}_{{\rm B}}\left[  \hat{R}_{{\rm B}}\left(  0\right)
\hat{B}_{a}^{\left(  n\right)  }\left(  t\right)  \hat{B}_{b}^{\left(
p\right)  }\left(  t^{\prime}\right)  \right] =\delta_{n,p}\delta
_{a,b}\,K\left(  t-t^{\prime}\right)  {\rm  ,}\label{Kt-s}\\
\mytext{\textcurrency Bt\textcurrency \qquad}
&\hat{B}_{a}^{\left(  n\right)
}\left(  t\right)    \equiv\hat{U}_{{\rm B}}^{\dagger}\left(  t\right)
\hat{B}_{a}^{\left(  n\right)  }\hat{U}_{{\rm B}}\left(  t\right)  {\rm 
,}\label{Bt}\\
\mytext{\textcurrency UB\textcurrency \qquad}
&\hat{U}_{{\rm B}}\left( t \right) = e^{-i\hat{H}_{{\rm B}}t/\hbar}{\rm  ,} \label{UB}%
\EEA
in terms of the evolution operator $\hat U_{\rm B}(t)$ of ${\rm B}$ alone. The bath operators
(\ref{B=}) have been defined in such a way that the equilibrium expectation
value of $B_{a}^{\left(  n\right)  }\left(  t\right)  $ vanishes for all $a=x,y,z$~\cite{petr,Weiss,Gardiner}. Moreover, the
condition (\ref{CC}) ensures that the equilibrium correlations between
different operators $\hat{B}_{a}^{\left(  n\right)  }\left(  t\right)  $ and
$\hat{B}_{b}^{\left(  p\right)  }\left(  t'\right)  $ vanish,  unless $a=b$ and $n=p$, and that the
autocorrelations for $n=p$, $a=b$ are all the same, thus defining a unique
function $K\left(  t\right)  $ in (\ref{Kt-s}).  We introduce the Fourier transform and its inverse,

\begin{equation}
\mytext{\textcurrency TF\textcurrency \qquad}
\tilde{K}\left(  \omega\right)
=\int_{-\infty}^{+\infty}{\rm d}t\ e^{-i\omega t}K\left(  t\right),\qquad
K(t)=\frac{1}{2\pi}\int_{-\infty}^{+\infty}{\rm d}\omega \,e^{i\omega t}\tilde K\left(  \omega\right)
\label{TF}
\end{equation}
and choose for $\tilde K(\omega) $ the simplest expression having the required properties, namely the quasi-Ohmic form
\cite{ms,caldeira,petr,Weiss,Gardiner}

\begin{equation}
\mytext{\textcurrency K\symbol{94}tilde\textcurrency \qquad}
\tilde{K}\left(
\omega\right)  =\frac{\hbar^{2}}{4}\frac{\omega e^{-\left\vert \omega
\right\vert /\Gamma}}{e^{\beta\hbar\omega}-1}{\rm  .} \label{Ktilde}
\end{equation}
The temperature dependence accounts for the quantum bosonic nature of the
phonons \cite{petr,Weiss,Gardiner}. The Debye cutoff $\Gamma$ characterizes the largest frequencies of
the bath, and is assumed to be larger than all other frequencies entering our problem.
 The normalization is fixed so as to let the constant $\gamma$
entering (\ref{ham4}) be dimensionless.

The form (\ref{Ktilde}) of $\tilde{K}\left(  \omega\right)  $ describes the
spectral function of the Nyquist-noise correlator, which is the quantum
generalization of the classical white noise. It can be obtained directly
through general reasonings based on the detailed balance and the approach to
equilibrium \cite{petr,Weiss}. We can also derive it from the expressions
(\ref{Hbath}) for $\hat{H}_{{\rm B}}$, (\ref{B=}) and (\ref{Bt}) for
$\hat{B}_{a}^{\left(  n\right)  }\left(  t\right)  $, and (\ref{RB0=}) for
$\hat{R}_{{\rm B}}\left(  0\right)  $, which under general conditions
provide a universal model for the bath \cite{petr,Weiss,Gardiner}. Indeed, by inserting
these expressions into the left-hand side of (\ref{Kt-s}), we recover the
diagonal form of the right-hand side owing to (\ref{CC}), which relates $c(\omega)$ to the bath Hamiltonian $\hat H_{\rm B}$, with the autocorrelation function 
$K\left(t\right)  $ given by

\BEA 
\mytext{\textcurrency Kt\textcurrency \qquad}
K\left(  t\right)   &  =&\sum
_{k}c\left(  \omega_{k}\right)  \left(  \frac{e^{i\omega_{k}t}}{e^{\beta
\hbar\omega_{k}}-1}+\frac{e^{-i\omega_{k}t}}{1-e^{-\beta\hbar\omega_{k}}
}\right) \nonumber\\
&  =&\int_{0}^{\infty}{\rm d}\omega\,\rho\left(  \omega\right)  c\left(
\omega\right)  \left(  \frac{e^{i\omega t}}{e^{\beta\hbar\omega}-1}
+\frac{e^{-i\omega t}}{1-e^{-\beta\hbar\omega}}\right)  {\rm  .} \label{Kt}
\EEA 
We have expressed above $K\left(  t\right)  $ in terms of the density of modes

\begin{equation}
\mytext{\textcurrency ro\textcurrency \qquad}
\rho\left(  \omega\right)
=\sum_{k}\delta\left(  \omega-\omega_{k}\right)  {\rm  ,} 
\label{ro}
\end{equation}
and this provides

\begin{equation}
\mytext{\textcurrency K\symbol{94}tilde2\textcurrency \qquad}
\tilde{K}\left(
\omega\right)  =2\pi\rho\left(  \left\vert \omega\right\vert \right)  c\left(
\left\vert \omega\right\vert \right)  \frac{{\rm sgn}\,\omega}
{e^{\beta\hbar\omega}-1}{\rm  .} \label{Ktilde2}
\end{equation}
In agreement with Kubo's relation, we also find for the dissipative response

\begin{equation}
\mytext{\textcurrency FD\textcurrency \qquad}
\int_{-\infty}^{+\infty}%
{\rm d}te^{-i\omega t}\, {\rm tr}_{{\rm B}}\left\{  \hat
{R}_{{\rm B}}\left(  0\right)  \left[  \hat{B}_{a}^{\left(  n\right)
}\left(  t\right)  ,\hat{B}_{b}^{\left(  p\right)  }\left(  0\right)  \right]
\right\}  =-2\pi\rho\left(  \left\vert \omega\right\vert \right)  c\left(
\left\vert \omega\right\vert \right)  {\rm sgn}\,\omega{\rm  .}
\label{FD}
\end{equation}
In the limit of a spectrum $\omega_{k}$ of the phonon modes sufficiently dense so that the relevant values of $t/\hbar$ and $\beta$ are small compared 
to the inverse level spacing of the phonon modes, we can regard $\rho\left(\omega\right)  c\left(  \omega\right)  $ as a continuous function.
In the quasi-Ohmic regime \cite{ms,caldeira,NplusA,petr,Weiss,Gardiner}, the dissipative response at low
frequencies is proportional to $\omega$, as obvious for a friction-dominated harmonic oscillator.  We thus take (for $\omega>0$)

\begin{equation}
\rho\left(  \omega\right)  c\left(  \omega\right)  =\frac{\hbar^{2}}{8\pi
}\omega e^{-\omega/\Gamma}{\rm  ,}
\end{equation}
where $\omega$ is called the Ohmic factor, and where we include a Debye cutoff $\Gamma$ on the phonon spectrum and a
proper normalization. Then (\ref{Ktilde2}) reduces to the assumed expression (\ref{Ktilde}).

}

\subsubsection{Initial state}

\hfill{\it In the beginning was the Word}

\hfill{Genesis 1.1}

\vspace{3mm}

\label{section.3.3.3}
\ZeText{

In the initial state ${\hat {\cal D}}\left(  0\right)  =\hat{r}\left(
0\right)  \otimes{\hat {\cal R}}\left(  0\right)  $ where ${\rm S}$ and
${\rm A}$ are statistically independent, the $2\times2$ density matrix
$\hat{r}\left(  0\right)  $ of \textrm{S} is arbitrary;  it has the form (\ref{relements}) with elements
$r_{\uparrow\uparrow}\left(  0\right)  $, $r_{\uparrow\downarrow}\left(
0\right)  $, $r_{\downarrow\uparrow}\left(  0\right)  $ and $r_{\downarrow
\downarrow}\left(  0\right)  $ satisfying

\begin{equation}
\mytext{\textcurrency Rij0\textcurrency \qquad}
\hat r(0)=
\left( \begin{array}{ccc@{\ }r}
{{r}}_{\uparrow\uparrow}(0)& {{r}}_{\uparrow\downarrow}(0)\\
{{r}}_{\downarrow\uparrow}(0)& {{r}}_{\downarrow\downarrow}(0)
\end{array} \right), \qquad
r_{\uparrow\uparrow}\left(
0\right)  +r_{\downarrow\downarrow}\left(  0\right)  =1{\rm ,\qquad
}r_{\uparrow\downarrow}\left(  0\right)  =r_{\downarrow\uparrow}^{\ast}\left(
0\right)  {\rm ,\qquad}r_{\uparrow\uparrow}\left(  0\right)  r_{\downarrow
\downarrow}\left(  0\right)  \geq r_{\uparrow\downarrow}\left(  0\right)
r_{\downarrow\uparrow}\left(  0\right)  {\rm  .} \label{rij0}
\end{equation}

According to the discussion of the section~\ref{section.3.1}, the initial density operator
$\hat{{\cal R}}\left(  0\right)  $ of the apparatus describes the magnetic
dot in a metastable paramagnetic state. As justified below, we take for it the
factorized form
\begin{equation}
\mytext{\textcurrency RA0=\textcurrency \qquad}
{\hat {\cal R}}\left(
0\right)  =\hat{R}_{\rm M}\left(  0\right)  \otimes\hat{R}_{{\rm B}
}\left(  0\right)  {\rm  ,} \label{RA0=}
\end{equation}
where the bath is in the equilibrium state (\ref{RB0=}), at the temperature
$T=1/\beta$ lower than the transition temperature of ${\rm M}$, while the
magnet with Hamiltonian (\ref{HA}) is in a paramagnetic equilibrium state at a temperature $T_{0}
=1/\beta_{0}$ higher than its transition temperature:
\begin{equation}
\mytext{\textcurrency HF0\textcurrency \qquad}
\hat{R}_{\rm M}\left(0\right)  =\frac{1}{Z_{{\rm M}}}e^{-\beta_{0}\hat{H}_{{\rm M}}}{\rm  .}
\label{HF0}
\end{equation}

How can the apparatus be actually initialized in the non-equilibrium state
(\ref{RA0=}) at the time $t=0$? This \textit{initialization} takes place
during the time interval $-\tau_{{\rm init}}<t<0$. The apparatus is first
set at earlier times into equilibrium at the temperature $T_{0}$. Due to the
smallness of $\gamma$, its density operator is then factorized and
proportional to $\exp[{-\beta_{0}(  \hat{H}_{{\rm M}}+\hat{H}_{{\rm B}})  }]$. At the time $-\tau_{{\rm init}}$ the phonon bath
is suddenly cooled down to $T$. We shall evaluate in \S~\ref{section.7.3.2} the
relaxation time of ${\rm M}$ towards its equilibrium ferromagnetic states
under the effect of ${\rm B}$ at the temperature $T$. Due to the weakness
of the coupling $\gamma$, this time turns out to be much longer than the
duration of the experiment. We can safely assume $\tau_{{\rm init}}$ to be
much shorter than this relaxation time so that ${\rm M}$ remains unaffected
by the cooling. On the other hand, the quasi continuous nature of the spectrum
of ${\rm B}$ can allow the phonon-phonon interactions (which we have
disregarded when writing (\ref{Hbath})) to establish the equilibrium of
${\rm B}$ at the temperature $T$ within a time shorter than $\tau
_{{\rm init}}$. It is thus realistic to imagine an initial state of the
form (\ref{RA0=}).

An alternative method of initialization consists in applying to the magnetic dot a strong radiofrequency field, which acts on M but not on B. 
The bath can thus be thermalized at the required temperature, lower than the transition temperature of M, while the populations of spins of M 
oriented in either direction are equalized. The magnet is then in a paramagnetic state, as if it were thermalized at an {\it infinite} temperature 
$T_0$ in spite of the presence of a cold bath. In that case we have the initial state (see Eq. (\ref{sigmaxyz0}))

\BEQ\label{purePM}
 \hat R_{\rm M}(0)=\frac{1}{2^N}\prod_{n=1}^N\hat\sigma_0^{(n)}.
\EEQ

The initial density operator (\ref{HF0}) of ${\rm M}$ being simply a
function of the operator $\hat{m}$, we can characterize it as in (\ref{R->P})
by the probabilities $P_{\rm M}^{\rm dis}\left(  m,0\right)  $ for $\hat{m}$ to take the values
(\ref{eig}). Those probabilities are the normalized product of the
degeneracy (\ref{deg}) and the Boltzmann factor,

\begin{equation}  \label{bzf}
P_{\rm M}^{\rm dis}(m,0)=\frac{1}{Z_0} G(m)\exp\left[\frac{N}{T_0}\left({\frac{J_2}{2}m^2+\frac{J_4}{4}m^4}\right)\right],
\qquad Z_0=\sum_m  G(m)\exp\left[\frac{N}{T_0}\left({\frac{J_2}{2}m^2+\frac{J_4}{4}m^4}\right)\right].
\end{equation} 
For sufficiently large $N$, the distribution $P_{\rm M}\left( m,0\right) =\half NP_{\rm M}^{\rm dis}(m,0) $ is peaked around $m=0$, with the Gaussian shape

\begin{equation}
\mytext{\textcurrency Pm0\textcurrency \qquad}
P_{\rm M}\left(  m,0\right)  \simeq
\frac{1}{\sqrt{2\pi}\,\Delta m}e^{-m^{2}/2\Delta m^{2}}=\sqrt{\frac{N}{2\pi\delta_0^2}}e^{-Nm^2/2\delta_0^2}.
\label{Pm0}
\end{equation}
This peak, which has a narrow width of the form
\begin{equation}
\mytext{\textcurrency DELTAm\textcurrency \qquad}
\Delta m=\sqrt{\left\langle
m^{2}\right\rangle }=\frac{\delta_{0}}{\sqrt{N}}{\rm  ,} \label{DELTAm}
\end{equation}
involves a large number, of order $\sqrt{N}$, of eigenvalues (\ref{eig}), so
that the spectrum can  be treated as a continuum (except in sections 5.3 and 6).
For  the Hamiltonian (\ref{HM4=}) with $q=4$, only the multiplicity  (\ref{deg}) contributes to $\Delta m$, so that the paramagnetic initial state (\ref{HF0})
is characterized  at any initial temperature $T_{0}$ by the distribution
$P_{\rm M}\left(  m,0\right)  $ equal to%
\begin{equation}
\mytext{\textcurrency P\_0\textcurrency \qquad}
P_{\rm M}\left(  m,0\right)
=P_{{\rm M}0}\left(  m\right) 
=\frac{1}{2^N}G(m)
 \equiv\sqrt{\frac{N}{2\pi }}e^{-Nm^{2}/2}{\rm  .}
\label{Pus0}%
\end{equation}
For  the general Hamiltonian (\ref{HM=}),  Eq. (\ref{bzf}) yields that the width is larger, due to correlations between spins, and given
by%
\begin{equation}
\delta_{0}=\sqrt{\frac{T_{0}}{T_{0}-J_2}},\qquad \Delta m=\sqrt{\frac{T_{0}}{N(T_{0}-J_2)}}.
\label{delta0=}%
\end{equation}
In the pure $q=2$ case with Hamiltonian (\ref{HM2=}), and in general in case $J_2>0$,  the temperature $T_{0}$ of initialization 
should be sufficiently higher than the Curie temperature so that $\delta_{0}^{2}\ll N$, which ensures the narrowness of the peak.
For an initialization caused by a radiofrequency, the initial distribution is again (\ref{Pus0}).

}
\subsubsection{Ferromagnetic equilibrium states of the magnet}

\hfill{\it  Je suis seul ce soir avec mes r\^eves}
 
\hfill{\it Je suis seul ce soir sans ton amour\footnote{ I am alone tonight with my dreams / I am alone tonight without your love}}

\hfill{Lyrics by Rose No\"el and Jean Casanova, music by Paul Durand, sung by Andr\'e Claveau}

\vspace{3mm}

\label{section.3.3.4}
\ZeText{

We expect the final state (\ref{Dtf=}) of ${\rm S}+{\rm A}$ after
measurement to involve the two ferromagnetic equilibrium states $\hat{\cal R}_{i}$, $i=\,\Uparrow$ or $\Downarrow$. 
As above these states $\hat {{\cal R}}_{i}$ of the apparatus factorize, in the weak coupling limit
($\gamma\ll1$), into the product of (\ref{RB0=}) for the bath and a  ferromagnetic equilibrium state $\hat{R}_{{\rm M}i}$ 
for the magnet ${{\rm M}}$. It is tempting to tackle broken invariance by means of the
mean-field approximation, which becomes exact at equilibrium for infinite $N$
owing to the long range of the interactions \cite{lavis,nagaev}. However, we are interested in a
finite, though large, value of $N$, and the probability distribution
$P_{{\rm M}i}\left(  m\right)  $ associated with $\hat{R}_{{\rm M}i}$ has a
significant width around the mean-field value for $m$. Moreover, we shall see
in subsection~\ref{section.7.3} that, in spite of the constancy of the interaction between
all spins, the {\it time-dependent mean-field approximation may fail} even for large
$N$. We will study there the dynamics of the whole distribution $P_{\rm M}\left(
m,t\right)  $ including the final regime where it is expected to tend to
$P_{{\rm M}\Uparrow}\left(  m\right)  $ or $P_{{\rm M}\Downarrow}\left(  m\right)  $,
and will determine in particular the lifetime of the metastable state (\ref{RA0=}). We
focus here on equilibrium only. For later convenience we include an external
field $h$ acting on the spins of the apparatus, so as to arrive from (\ref{HM=}) at\footnote{In section 7 we shall identify $h$ with $+g$ in the sector
 $\hat R_{\up\up}$ of $\hat D$, or with $-g$ in its sector $\hat R_{\down\down}$, 
where $g$ is the coupling between S and A, while a true field in the $y$-direction will be introduced in section 8.2 and denoted by $b$, see Eq. (\ref{HS})}

\begin{equation}
\mytext{\textcurrency HM=\textcurrency \qquad}
\hat{H}_{{\rm M}}=-Nh\hat m-NJ_2\frac{\hat m^2}{2}-NJ_4\frac{\hat m^4}{4}{.} \label{HMh=}%
\end{equation}

As in (\ref{P->R}) we characterize the canonical equilibrium density operator of the magnet
$\hat{R}_{\rm M} =(1/Z_{\rm M})\exp[-\beta\hat H_{\rm M}]$, 
 which depends only on the operator $\hat{m}$, by the probability distribution

\begin{equation}
\mytext{\textcurrency Peq\textcurrency \qquad}
P_{\rm M}\left(  m\right)  =
\frac{\sqrt{N}}{Z_{{\rm M}}\sqrt{8\pi }}e^{-\beta F\left(  m\right)  }{\rm  ,}
\label{Peq}%
\end{equation}
 where $m$ takes the discrete values $m_i$ given by (\ref{eig}); the exponent of  (\ref{Peq}) introduces the {\it free energy function}
                                    
\begin{equation}
\mytext{\textcurrency F=\textcurrency \qquad}
F\left(  m\right)  =-NJ_2\frac{m^2}{2}-NJ_4\frac{m^4}{4}-Nhm
+NT\left(\frac{1+m}{2} \ln \frac{1+m}{2} +\frac{1-m}{2} \ln \frac{1-m}{2}\right)
+\frac{T}{2}\ln\frac{1-m^{2}}{4}+{\cal O}\left(\frac{1}{N}\right),
 \label{F=}
\end{equation}
which arises from the Hamiltonian (\ref{HMh=}) and from the multiplicity $G(m)$ given by (\ref{deg}). 
The distribution (\ref{Peq}) displays narrow peaks at the minima of $F\left(  m\right)  $, and
the \textit{equilibrium free energy} $-T\ln Z_{{\rm M}}$ is equal for large $N$ to the absolute minimum of (\ref{F=}). The function $F\left(  m\right)  $
reaches its extrema at values of $m$ given by the self-consistent equation

\begin{equation}
\mytext{\textcurrency MF\textcurrency \qquad}
m\left(  1-\frac{1}{N}\right)
=\tanh\left[  \beta\left(  h+J_2m+J_4m^{3}\right)  \right]  {\rm  ,} \label{MF}
\end{equation}
which as expected reduces to the mean-field result for large $N$. In the vicinity of a minimum of $F\left(  m\right)  $ at $m=m_{i}$, the probability
$P_{\rm M}\left(  m\right)  $ presents around each $m_{i}$ a nearly Gaussian peak, given within normalization by%

\begin{eqnarray}
\mytext{\textcurrency Pim\textcurrency \qquad}
P_{{\rm M}i}\left(  m\right)   \propto
\exp\left\{ -\frac{N}{2} \left[  \frac{1}{1-m_{i}^{2}}-\beta J_2- 3\beta J_4m_i^2\right]  (  m-m_i)^2
 -\frac{N}{3}\left[  \frac{m_{i}}{(  1-m_{i}^{2})  ^{2}}-3\beta J_4m_{i}\right]  \left( m-m_{i}\right)  ^{3}\right\}.
  \label{Pim}%
\end{eqnarray}
This peak is located at a distance of order $1/N$ from the mean-field value,
it has a width of order $1/\sqrt{N}$ and a weak asymmetry. The possible values
of $m$ are dense within the peak, with equal spacing $\delta m=2/N$. With each
such peak $P_{{\rm M}i}\left(  m\right)  $ is associated through (\ref{Pdis2cont}), (\ref{P->R}) a
density operator $\hat{R}_{i}$ of the magnet ${\rm M}$ which may describe a locally stable equilibrium. 
Depending on the values of $J_2$ and $J_4$  and on the temperature, there may exist one, two or three such locally stable states. 
We note the corresponding average magnetizations $m_i$, for arbitrary $h$,  as $m_{\rm P}$ for a paramagnetic state and as 
$m_\Uparrow$ and $m_\Downarrow$  for the ferromagnetic states, with $m_\Uparrow>0$, $m_\Downarrow<0$.
We also note as $\pm m_{\rm F}$ the ferromagnetic magnetizations for $h=0$. 
When $h$ tends to $0$ (as happens at the end of the measurement where we set $g\to0$),  $m_{\rm P}$ tends to $0$, 
$m_\Uparrow$ to $+m_{\rm F}$ and $m_\Downarrow$ to $-m_{\rm F}$, namely

\BEQ \label{mUpDown}
m_\Uparrow(h>0)>0,\qquad  m_\Downarrow(h>0)<0,\qquad 
m_\Uparrow(-h)=-m_\Downarrow(h),\qquad  
m_{\rm F}=m_\Uparrow(h\hspace{-0.5mm}\to\hspace{-0.5mm}+0)=-m_\Downarrow(h\hspace{-0.5mm}\to\hspace{-0.5mm}+0). 
\EEQ

For $h=0$, the system ${\rm M}$ is invariant under change of sign of $m$ \cite{lavis}.
This invariance is spontaneously broken below some temperature \cite{lavis}. In the case
$q=2$ of the Ising interaction (\ref{HM2=}), there is above the Curie temperature $T_{\rm c}=J_2$ a
single paramagnetic peak $P_{{\rm M}0}\left(  m\right)  $ at $m_{\rm P}=0$, given by
(\ref{Pm0}), (\ref{delta0=}), and for $T<J_2$ two symmetric ferromagnetic peaks
(\ref{Pim}), $i=\,\Uparrow$ or $\Downarrow$, at the points $m_{\Uparrow}=m_{{\rm F}}$
and $m_{\Downarrow}=-m_{{\rm F}}$, given by $m_{{\rm F}}=\tanh\beta J_2m_{{\rm F}}$. 
These peaks are well separated provided%

\begin{equation}
\frac{N}{2}\left(  \frac{1}{1-m_{{\rm F}}^{2}}-\beta J_2\right)
m_{{\rm F}}^{2}\gg1{\rm  ,}%
\end{equation}
in which case they characterize two different equilibrium ferromagnetic states. This
condition is satisfied for large $N$ and $\beta J_2-1$ finite; near $\beta J_2=1$,
where $m_{{\rm F}}^{2}\sim3\left(  \beta J_2-1\right)  $, the two states
$\hat{R}_{\Uparrow}$ and $\hat{R}_{\Downarrow}$ still have
no overlap as soon as the temperature differs significantly from the critical
temperature, as%
\begin{equation}
\frac{J_2-T}{T}\gg\frac{1}{\sqrt{3N}}{\rm  .}%
\end{equation}
This property is needed to ensure a faithful registration by ${\rm M}$ of the measurement. 
Little is changed for the Hamiltonian (\ref{HM=}) with  $J_4>0$ but still $J_2 > 3J_4$. 
 
 \myskipfigText{
\begin{figure}[h!]
\label{figABN3}
\centerline{\includegraphics[width=8cm]{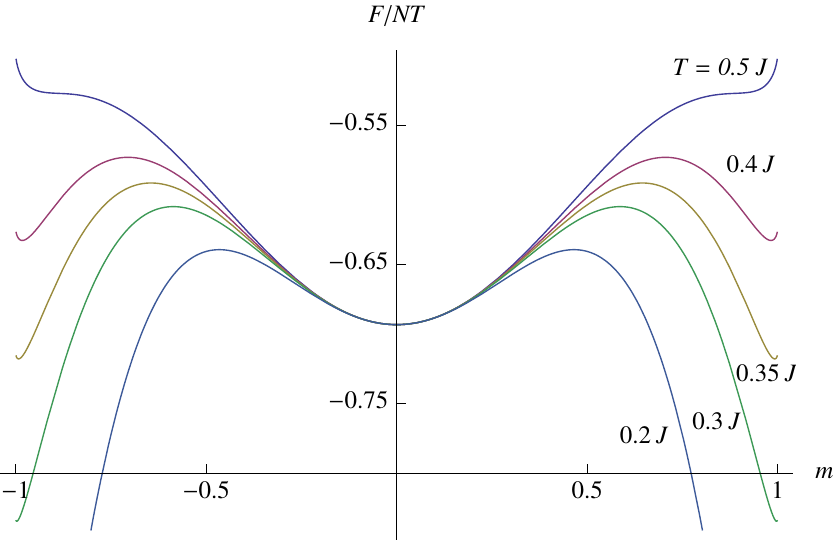}}
\caption{The free energy $F$ in units of $NT$  for a pure quartic interaction (eq. (\ref{HM4=}), evaluated from Eq.  (\ref{F=}) with $h=0$, as function of the 
magnetization $m$ at various temperatures. There is always a local paramagnetic minimum at $m=0$. A first-order transition occurs at $T_{\rm c}=0.363J_4$, 
below which the ferromagnetic states associated with the minima at $\pm \,m_{\rm F}$ near $\pm\,1$
become the most stable. 
 }
\end{figure}
}

Still for $h=0$, but in the case $3J_4 > J_2$ of a first-order transition, $F\left(  m\right)  $ has a minimum at $m=0$ if $T>J_2$  and hence (\ref{Peq}) 
has there a peak as (\ref{Pus0}) at $m=0$ whatever the temperature, see Fig. 3.3.
For the pure quartic interaction of Eq. (\ref{HM4=}), the two additional ferromagnetic peaks   
 $P_{{\rm M}\Uparrow}\left(  m\right)  $ and $P_{{\rm M}\Downarrow}\left(  m\right)  $
appear around $m_{\Uparrow}=m_{{\rm F}}=0.889$ and $m_{\Downarrow
}=-m_{{\rm F}}$ when the temperature $T$ goes below $0.496J_4$. As $T$
decreases, $m_{{\rm F}}$ given by $m_{{\rm F}}=\tanh\beta J_4m_{{\rm F}%
}^{3}$ increases and the value of the minimum $F\left(  m_{{\rm F}}\right)
$ decreases; the weight (\ref{Peq}) is transferred from $P_{{\rm M}0}\left(m\right)$ to 
$P_{{\rm M}\Uparrow}\left(  m\right)  $ and $P_{{\rm M}\Downarrow}\left(m\right)  $. 
A first-order transition occurs when $F\left(  m_{{\rm F} }\right)  =F\left(  0\right)  $, for $T_{\rm c}=0.363J_4$ and $m_{{\rm F}}=0.9906$,
from the paramagnetic to the two ferromagnetic states, although the
paramagnetic state remains locally stable. The spontaneous magnetization
$m_{{\rm F}}$ is always very close to $1$, behaving as $1-m_{{\rm F}}\sim2\exp({-2J_4/T})$.

 For the general Hamiltonian (\ref{HM=}), it is a simple exercise to study the cross-over between first and second-order transitions, 
 which takes place for $m_i \ll1$. To this aim, the free energy (\ref{F=}) is expanded as
 
 \BEQ
            \frac{F(m)-F(0)}{N}\approx  (T-J_2)\frac{m^2}{2} + (T-3J_4) \frac{m^4 }{ 12} +T \frac{m^6}{  30},    
            \EEQ
and its shape and minima are studied as function of $J_2$, $J_4$ and $T$. This approximation holds for $|T - J_2|\ll J_2$,  
$|3J_4 - J_2|\ll J_2$. For $J_2 > 3J_4$, the second-order transition takes place at $T_{\rm c} = J_2$ whatever $J_4$. 
For $3J_4 > J_2$, the first-order transition temperature $T_{\rm c}$ is given by $T_{\rm c} - J_2 \sim 5 (3J_4 - J_2)^2 / 48 J_2$, and the 
equilibrium magnetization jumps from $0$ to $\pm  m_{\rm F}$, with $m_{\rm F}^2  \sim 5 (3J_4 - J_2) / 4 J_2$. 
The paramagnetic state is locally stable down to $T > J_2$, the ferromagnetic states up to $T - J_2 < (4 /3) (T_{\rm c} - J_2 )$. 
When $3J_4 > J_2$, a metastability with a long lifetime of the paramagnetic state is thus ensured if the bath temperature satisfies $T_{\rm c} > T > J_2$.

Strictly speaking, the canonical equilibrium state of ${\rm M}$ below the
transition temperature, characterized by (\ref{Peq}), has for $h=0$
and finite $N$ the form

\BEQ \label{RMeq}
\hat{R}_{{\rm Meq}}=\frac{1}{2}(  \hat{R}_{{\rm M }\Uparrow}+\hat{R}_{{\rm M}\Downarrow}\,).
\EEQ
However this state is not necessarily the one reached at the end of a relaxation process governed by the bath ${\rm B}$, when a field $h$, even weak, is present: this field acts
as a source which breaks the invariance. The determination of the state $\hat{R}_{\rm M}\left(  t_{{\rm f}}\right)  $ reached at the end of a
relaxation process involving the thermal bath ${\rm B}$ and a weak field $h$ {\it requires a dynamical study} which will be worked out in subsection~\ref{section.7.3}. 
This is related to the ergodicity breaking: if a weak field is applied, then switched off, the full canonical state (\ref{RMeq})  is still recovered, but only after an unrealistically 
long time (for $N\gg 1$).  For finite times the equilibrium state of the magnet is to be found by restricting the full canonical state (\ref{RMeq})  to its component
having a magnetization with the definite sign determined by the weak external field. This is the essence of the spontaneous symmetry
breaking. However, for our situation this well-known recipe should be supported by dynamical considerations; see in this respect section 11.

In our model of measurement, the situation is similar, though slightly more
complicated. The system-apparatus coupling (\ref{HSA}) plays the r\^{o}le
of an operator-valued source, with eigenvalues behaving as a field $h=g$ or
$h=-g$. We shall determine in section~\ref{section.7} towards which state ${\rm M}$ is
driven under the conjugate action of the bath ${\rm B}$ and of the system
${\rm S}$, depending on the parameters of the model.

\myskipfigText{
\begin{figure}[h!]
\label{figABN4}
\centerline{\includegraphics[width=8cm]{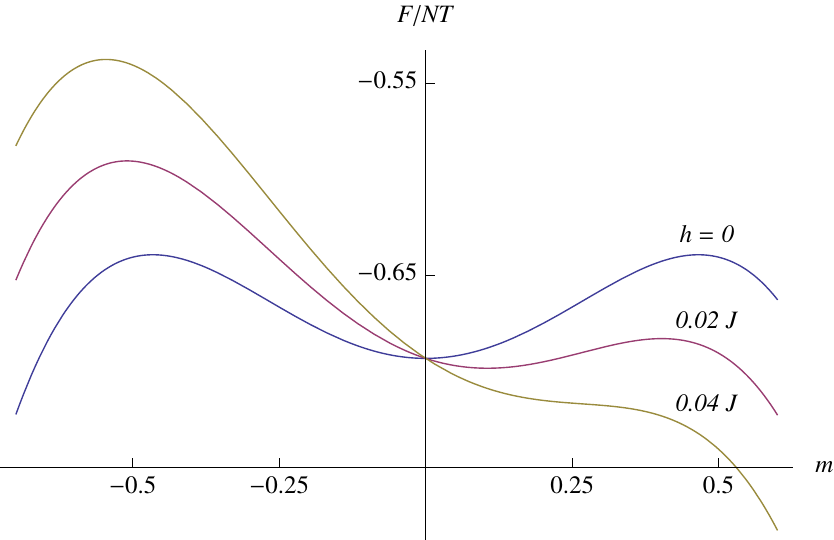}}
\caption{The effect of a positive field $h$ on $F(m)$ for $q=4$ at temperature $T=0.2J_4$.
As $h$ increases the paramagnetic minimum $m_{\rm P}$ shifts towards positive $m$. At the critical field $h_{\rm c}=0.0357J_4$ 
this local minimum disappears, and the curve has an inflexion point with vanishing slope at $m=m_{\rm c}=0.268$.  
For larger fields, like in the displayed case $g=0.04J_4$, the locally stable paramagnetic state disappears, and there remain only 
the two ferromagnetic states, the most stable one with positive magnetization $m_\Uparrow\simeq1$ and the metastable one
with negative magnetization $m_\Downarrow\simeq-1$.}
\end{figure}
}

As a preliminary step, let us examine here the effect on the free energy (\ref{F=}) of a small positive field $h$. Consider first the minima of
$F\left(  m\right)  $ \cite{ll_stat,lavis}. The two ferromagnetic minima $m_{\Uparrow}$ and $m_{\Downarrow}$ given by (\ref{MF}) are slightly shifted away from
$m_{{\rm F}}$ and $-m_{{\rm F}}$, and $F\left(  m_{\Uparrow}\right) -F\left(  m_{{\rm F}}\right)  $ behaves as $-Nhm_{{\rm F}}$. Hence, as soon as 
$\exp\{{-\beta\left[  F\left(  m_{\Uparrow}\right)  -F\left(m_{\Downarrow}\right)  \right]  }\}\sim \exp({2\beta Nhm_{{\rm F}}})\gg1$, 
only the single peak $P_{{\rm M}\Uparrow}\left(  m\right)  $ around $m_{\Uparrow}\simeq
m_{{\rm F}}$ contributes to (\ref{Peq}), so that the canonical equilibrium
state of ${\rm M}$ has the form $\hat{R}_{{\rm Meq}}=\hat{R}_{{\rm M}\Uparrow}$. The shape of $F\left(  m\right)  $ will
also be relevant for the dynamics. For a second order transition, although $F\left(m\right)  $ has when $h=0$ a maximum at $m=0$, its stationarity
allows the state $\hat{R}_{\rm M}\left( m,0\right)\propto P_{\rm M}(\hat m,0)$ given by (\ref{Pm0})
to have a long lifetime for $N\gg1$. The introduction of $h$ produces a negative slope $-Nh$ at $m=0$, which suggests that the dynamics will let
$\left\langle m\right\rangle $ increase. For a first order transition, the situation is different (Fig. 3.4). If $h$ is sufficiently small, $F\left(
m\right)  $ retains its paramagnetic minimum, the position of which is shifted as $m_{\rm P}\sim h/T$; the paramagnetic
state $\hat{R}_{\rm M}\left(  0\right)  $ remains locally stable. It may decay towards a stable ferromagnetic state only
through mechanisms of thermal activation or quantum tunneling, processes with very large characteristic times, of exponential
order in $N$. However, there is a threshold $h_{{\rm c}}$ above which this paramagnetic minimum of $F\left(  m\right)  $, which
then lies at $m=m_{{\rm c}}$, disappears. The value of $h_{{\rm c}}$ is found by eliminating $m=m_{{\rm c}}$
between the equations ${\rm d}^{2}F/{\rm d}m^{2}=0$ and ${\rm d} F/{\rm d}m=0$. In the pure $q=4$ case ($J_2=0$) on which we focus as an illustration 
for first order transitions, we find  $2m_{{\rm c}}^{2}=1-\sqrt{1-4T/3J_4}$, $h_{{\rm c}}=\frac{1}{2}T\ln[(1+m_{\rm c})/(1-m_{\rm c})]-J_4m_{{\rm c}}^{3}$. 
At the transition temperature $T_{\rm c}=0.363J_4$,  we have $m_{{\rm c}}=0.375$ and $h_{{\rm c}}=0.0904J_4$;  for $T=0.2J_4$, we obtain
$m_{{\rm c}}=0.268$ and $h_{{\rm c}}=0.036J_4$;  for $T\ll J_4$, $m_{{\rm c}}$\ behaves as $\sqrt{T/3J_4}$ and $h_{{\rm c}}$\ as
$\sqrt{4T^{3}/27J_4}$. Provided $h>h_{{\rm c}}$, $F\left(  m\right)  $ has now a negative slope in the whole interval $0<m<m_{{\rm F}}$. 

We can thus expect, in our measurement problem, that the registration will take place in a reasonable delay, either for a first order transition if the coupling
 $g$ is larger than $h_{{\rm c}}$, or for a second order transition. In the latter case, it will be necessary to
check, however, that the lifetime of the initial state is larger than the duration of the measurement. This will be done in \S~\ref{section.7.3.2}.

}
\renewcommand{\thesection}{\arabic{section}}
\section{Equations of motion}
\setcounter{equation}{0}\setcounter{figure}{0}\renewcommand{\thesection}{\arabic{section}.}
\label{section.4}

\hfill{ \foreignlanguage{greek}{T\`a p\'anta <re\"\i}\footnote{Everything flows}}

\hfill{Quoted from Heraklitos  by Plato and Simplicius}

\vspace{3mm}
\ZeText{

In this technical section, we rewrite the dynamical equations for our model in
a form which will help us, in the continuation, to discuss the physical
features of the solution. We will make no other approximation than the weak
spin-phonon coupling, $\gamma\ll1$, and will derive the equations up to first
order in $\gamma$. In subsection~\ref{section.4.4}, we take advantage of the large size of
the apparatus, $N\gg1$, to reduce the equations of motion into a pair of
partial differential equations.

\clearpage

}

\subsection{A conserved quantity, the measured component of the spin, and the Born rule}

\hfill{\it All the world's Great Journeys begin with the first step}

\hfill{\it A 1000 miles journey starts with a single step}

\hfill{Tibetan and Aboriginal Australian  proverbs}

\vspace{3mm}

\ZeText{

Since $\hat H_{\rm S}=0$ and since $\hat s_x$ and $\hat s_y$ do not occur in the coupling  (\ref{HSA})
between S and A, we can already conclude that $\hat s_z$ is conserved during the ideal
measurement, viz. $i\hbar \d\hat s_z/\d t=[\hat s_z,\hat H]=0$. This implies that the diagonal elements of the
density matrix of the spin are conserved, viz. $r_{\up\up}(t_{\rm f})=r_{\up\up}(t)=r_{\up\up}(0)$ and $r_{\down\down}(t_{\rm f})=r_{\down\down}(0)$.
The result is consistent with {\it Born's rule}: we expect the probabilities for the possible {\it outcomes} of an ideal measurement to be given by the diagonal elements 
of the {\it initial}  density matrix of S. But $r_{\up\down}$ and $r_{\down\up}$, on the other hand,  are not conserved (viz.  $[\hat s_{a},\hat H]\neq0$ for $a=x,y$),
 and they will evolve and ultimately vanish\footnote{This has the popular name ``decay of Schr\"odinger cat terms'',  or ``death of Schr\"odinger cats''}.


}

\subsection{Eliminating the bath variables}
\label{section.4.1}

\hfill{\it N\~ao chame o jacar\'e de boca-grande se voc\^e}

\hfill{\it  ainda n\~ao chegon na outra margem 
\footnote{Don't call the alligator a big-mouth till you have crossed the river}}

\hfill{Brazilian proverb}

\vspace{3mm}

\ZeText{

A complete description of the measurement process requires at least  the solution, in
the Hilbert space of ${\rm S}+{\rm A}$, of the Liouville--von~Neumann
equation of motion \cite{ll_stat}
\begin{equation}
\mytext{\textcurrency vN\textcurrency \qquad}
i\hbar\frac{{\rm d}
{\hat {\cal D}}}{{\rm d}t}=\left[  \hat{H},{\hat {\cal D}}\right]
{\rm  ,} \label{vN}
\end{equation}
with the initial condition
\begin{equation}
\mytext{\textcurrency ic\textcurrency \qquad}
{\hat {\cal D}}\left(  0\right)
=\hat{r}\left(  0\right)  \otimes\hat{R}_{\rm M}\left(  0\right)
\otimes\hat{R}_{{\rm B}}\left(  0\right)  =\hat{D}\left(  0\right)
\otimes\hat{R}_{{\rm B}}\left(  0\right)  {\rm  .} \label{ic}
\end{equation}
We are not interested, however, in the bath variables, and the knowledge of
$\hat{D}\left(  t\right)  ={\rm tr}_{{\rm B}}{\hat {\cal D}}\left(  t\right)  $ is sufficient for our purpose. As usual in
non-equilibrium statistical mechanics \cite{petr,Weiss,Gardiner,vKampenbook}, we rely on the weakness of the
coupling $\hat{H}_{{\rm MB}}$ between the magnet and the bath,
so as to treat perturbatively the dissipative effect of the bath.

Let us therefore split the Hamiltonian (\ref{ham}) into $\hat{H}=\hat{H}
_{0}+\hat{H}_{{\rm MB}}+\hat{H}_{{\rm B}}$ with $\hat{H}_{0}=\hat
{H}_{{\rm S}}+\hat{H}_{{\rm SA}}+\hat{H}_{{\rm M}}$. Regarding the
coupling $\hat{H}_{{\rm MB}}$ as a perturbation, we introduce the
unperturbed evolution operators, namely (\ref{UB}) for the bath, and
\begin{equation}
\mytext{\textcurrency U0\textcurrency \qquad}
\hat{U}_{0}\left(  t\right)=e^{-i\hat{H}_{0}t/\hbar},\qquad 
\hat H_0=-gN\hat s_z\hat{m}-N\sum_{q=2,4}\frac{J_q}{q}\hat{m}^{q},
 \label{U0}
\end{equation}
for ${\rm S}+{\rm M}$. 
We can then expand the full evolution operator in powers of the coupling $\sqrt{\gamma}$,
in the interaction picture, and take the trace over B of eq. (\ref{vN}) so as to generate finally an
equation of motion for the density operator $\hat D(t)$ of S + M. This calculation is worked out
in Appendix A.

The result involves the {\it autocorrelation function $K(t)$ of the bath}, defined by (\ref{RB0=}) -- (\ref{UB}) and expressed in our model 
by (\ref{TF}), (\ref{Ktilde}). It also involves the operators $\hat{\sigma}_{a}^{\left(n\right)  }\left(  u\right)$ in the space of S + M,
defined in terms of the memory time $u=t-t'$ by

\begin{equation}
\mytext{\textcurrency sigmau\textcurrency \qquad}
\hat{\sigma}_{a}^{\left(n\right)  }\left(  u\right)  \equiv\hat{U}_{0}\left(  t\right)  \hat{U}
_{0}^{\dagger}\left(  t^{\prime}\right)  \hat{\sigma}_{a}^{\left(  n\right)
}\hat{U}_{0}\left(  t^{\prime}\right)  \hat{U}_{0}^{\dagger}\left(  t\right)
=\hat{U}_{0}\left(  u\right)  \hat{\sigma}_{a}^{\left(  n\right)  }\hat{U}
_{0}^{\dagger}\left(  u\right)  \label{sigmau}.
\end{equation}
 It holds that $\hat\sigma^{(n)}_a(0)=\hat\sigma^{(n)}_a$.
Altogether we obtain a differential equation for $\hat{D}\left(  t\right)  $, the kernel of which involves
times earlier than $t$ through $K\left(  u\right)  $ and
$\hat{\sigma}_{a}^{\left(  n\right)  }\left( u\right)  $ \cite{petr,Weiss,Gardiner}:

\begin{equation}
\mytext{\textcurrency dD\textcurrency \qquad}
\frac{{\rm d}\hat{D}}
{{\rm d}t}-\frac{1}{i\hbar}\left[  \hat{H}_{0},\hat{D}\right]
=\frac{\gamma}{\hbar^{2}}\int_{0}^{t}{\rm d}u\sum_{n,a}\left\{  K\left(
u\right)  \left[  \hat{\sigma}_{a}^{\left(  n\right)  }\left(  u\right)
\hat{D},\hat{\sigma}_{a}^{\left(  n\right)  }\right]  +K\left(  -u\right)
\left[  \hat{\sigma}_{a}^{\left(  n\right)  },\hat{D}\hat{\sigma}_{a}^{\left(
n\right)  }\left(  u\right)  \right]  \right\}  +{\cal O}\left(  \gamma
^{2}\right)  {\rm  .} \label{dD}\label{dD1}
\end{equation}
As anticipated in \S~\ref{section.3.3.2}, the phonon bath occurs in this equation, which
governs the dynamics of ${\rm S}+{\rm M}$, only through the function
$K\left(  t\right)  $, the memory time being the time-range $\hbar/2\pi T$ of
$K\left(  t\right)  $ \cite{petr,Weiss,Gardiner}.

}
\subsection{Decoupled equations of motion}
\label{section.4.2}

\hfill{\it Married couples tell each other a thousand things,}

\hfill{\it  without speech}

\hfill{Chinese proverb}

\vspace{3mm}

\ZeText{

In our model, the Hamiltonian commutes with the measured observable $\hat
{s}_{z}$, hence with the projection operators $\hat{\Pi}_{i}$ onto the states
$\left\vert \uparrow\right\rangle $ and $\left\vert \downarrow\right\rangle $
of ${\rm S}$. The equations for the operators $\hat{\Pi}_{i}\hat{D}\hat
{\Pi}_{j}$ are therefore decoupled. We can replace the equation (\ref{dD}) for
$\hat{D}$ in the Hilbert space of ${\rm S}+{\rm M}$ by a set of four
equations for the operators $\hat{R}_{ij}$ defined by (\ref{Rij}) in the Hilbert space of ${\rm M}$.
We shall later see (section 8.2) that this simplification underlies the ideality of the measurement process.

The Hamiltonian $\hat{H}_{0}$ in the space ${\rm S}+{\rm M}$
gives rise to two Hamiltonians $\hat{H}_{\uparrow}$ and
$\hat{H}_{\downarrow}$ in the space ${\rm M}$,\ which according
to (\ref{HSA}) and (\ref{HM=}) are simply two functions of the observable $\hat{m}$, given by%
\begin{equation}
\mytext{\textcurrency Hi\textcurrency \qquad}
\hat{H}_{i}=H_{i}\left(  \hat
{m}\right)  =-gNs_{i}\hat{m}-N\sum_{q=2,4}\frac{J_q}{q}\hat{m}^{q}{\rm  ,} \qquad (i=\up,\down)
 \label{Hi}
\end{equation}
with $s_{i}=+1$ (or $-1$) for $i=\uparrow$ (or $\downarrow$).
These Hamiltonians $\hat{H}_{i}$, which describe interacting spins
$\mathbf{\hat {\sigma}}^{\left(  n\right)  }$ in an external field
$gs_{i}$, occur in
(\ref{dD}) both directly and through the operators
\begin{equation}
\mytext{\textcurrency sigmaui\textcurrency \qquad}
\hat{\sigma}_{a}^{\left(
n\right)  }\left(  u,i\right)  =e^{-i\hat{H}_{i}u/\hbar}\hat{\sigma}
_{a}^{\left(  n\right)  }e^{i\hat{H}_{i}u/\hbar}{\rm  ,} \label{sigmaui}
\end{equation}
obtained by projection of (\ref{sigmau}), use of (\ref{U0}) and reduction to the Hilbert space of ${\rm M}$.

The equation (\ref{dD}) for $\hat D(t)$ which governs the joint dynamics of ${\rm S}+{\rm M}$ 
thus reduces to the four differential equations in the Hilbert space of ${\rm M}$
(we recall that $i,j=\uparrow, \downarrow$ or $\pm1$):

\begin{equation}
\mytext{\textcurrency dRij\textcurrency \qquad}
\frac{{\rm d}\hat{R}_{ij}(t)}{{\rm d}t}-\frac{ \hat{H}_{i}\hat{R}_{ij}(t)-\hat{R}_{ij}(t)\hat{H}_{j}}{i\hbar} 
=\frac{\gamma}{\hbar^{2}}\int_{0}^{t}{\rm d}
u\sum_{n,a}\left\{  K\left(  u\right)  \left[  \hat{\sigma}_{a}^{\left(n\right)  }\left(  u,i\right)  \hat{R}_{ij}(t),\hat{\sigma}_{a}^{\left(n\right)  }\right]  +K\left(  -u\right)  
\left[  \hat{\sigma}_{a}^{\left(n\right)  },\hat{R}_{ij}(t)\hat{\sigma}_{a}^{\left(  n\right)  }\left(u,j\right)  \right]  \right\}. 
\label{dRij}
\end{equation}

}
\subsection{Reduction to scalar equations}
\label{section.4.3}

\subsubsection{Representing the pointer by a scalar variable}

\hfill{\it  Even a small star shines in the darkness}

\hfill{Finnish proverb}

\vspace{3mm}

\label{section.4.3.1}
\ZeText{

For each operator $\hat{R}_{ij}$, the initial conditions are given according
to (\ref{rij0}) and (\ref{RA0=}) by
\begin{equation}
\mytext{\textcurrency Rij0\textcurrency \qquad}
\hat{R}_{ij}\left(  0\right)
=r_{ij}\left(  0\right)  \hat{R}_{\rm M}\left(  0\right)  {\rm  ,}
\label{Rij0}
\end{equation}
and $\hat{R}_{\rm M}\left(  0\right)  $ expressed by the Gibbs state  (\ref{HF0}) is simply a {\it function of the operator} $\hat{m}$. We show in Appendix B that 
this property {\it is preserved for} $\hat{R}_{ij}\left(  t\right)  $ by the evolution (\ref{dRij}), owing to the form (\ref{Hi}) of $\hat{H}_{i}$ and in spite of
the occurrence of the separate operators $\hat{\sigma}_{a}^{\left(  n\right)  }$ in the right-hand side.

We can therefore parametrize, as anticipated at the end of \S~\ref{section.3.3.1}, at each $t$, 
the operators $\hat R_{ij}$ in the form $\hat R_{ij}= P_{ij}^{\rm dis}(\hat m)/G(\hat m)$.
Their equations of motion (\ref{dRij}) are then diagonal in the eigenspace of $\hat m$, and are therefore equivalent to scalar equations which govern the functions 
$P_{ij}(m)=(N/2)P^{\rm dis}_{ij}(m)$ of the variable $m$ taking the discrete values (\ref{eig}).

}
\subsubsection{Equations of motion for $P_{ ij}\left(  m,t\right)  $}
\label{section.4.3.2}

\ZeText{

The equations resulting from this parametrization are derived in Appendix B. The integrals over $u$ entering (\ref{dRij}) yield the functions 

\begin{eqnarray}
\mytext{K>}\qquad 
  \tilde{K}_{t>}\left(  \omega\right)&  =&\int_{0}
^{t}{\rm d}ue^{-i\omega u}K\left(  u\right)  =\frac{1}{2\pi i}\int
_{-\infty}^{+\infty}{\rm d}\omega^{\prime}\tilde{K}\left(  \omega^{\prime}\right)
\frac{e^{i\left(  \omega^{\prime}-\omega\right)  t}-1}{\omega^{\prime}-\omega}
 {\rm  ,}
\label{K>}
\EEA 
and

\BEA
\mytext{K<}\qquad
{\rm   \tilde{K}_{t<}}\left(  \omega\right)  &=&\int_{0}
^{t}{\rm d}ue^{i\omega u}K\left(  -u\right)  =\int_{-t}
^{0}{\rm d}ue^{-i\omega u}K\left(  u\right) =\left[  \tilde{K}_{t>}\left(
\omega\right)  \right]  ^{\ast}
 =\frac{1}{2\pi i}\int
_{-\infty}^{+\infty}{\rm d}\omega^{\prime}\tilde{K}\left(  \omega^{\prime}\right)
\frac{1-e^{i\left(  \omega-\omega'\right)  t}}{\omega^{\prime}-\omega}
  {\rm  ,}\quad \label{K<}
\end{eqnarray}
where $\omega$ takes, depending on the considered term, the values
$\Omega^+_\uparrow$, $\Omega^-_\uparrow$,  $\Omega^+_\downarrow$, $\Omega^-_\downarrow$, 
given by

\BEQ \label{Omidef}
\hbar\Omega^\pm_i(m)=H_i(m\pm\delta m)-H_i(m),\qquad (i=\up,\down),
\EEQ
in terms of the Hamiltonians (\ref{Hi}) and of the level spacing $\delta m=2/N$. They satisfy the relations

\BEQ 
\label{OmpmPMdm}
\Omega_i^\pm(m\mp\delta m)=-\Omega_i^\mp(m).
\EEQ
The quantities (\ref{Omidef}) are interpreted as excitation energies of the magnet M arising from the flip of one of its spins 
 in the presence of the tested spin S (with value $s_i$); the sign $+$ ($-$) refers to a down-up (up-down) spin flip. 
 Their explicit values are: 
 
\BEA
\label{ompmi=}
\hbar\Omega^\pm_i(m)=\mp2gs_i+2J_2(\mp m-\frac{1}{N})
+2J_4(\mp m^3-\frac{3m^2}{N}\mp\frac{4m}{N^2}-\frac{2}{N^3}),
\EEA
with $s_\uparrow =1$, $s_\downarrow =-1$. 

The operators $\hat\sigma^{(n)}_x$ and  $\hat\sigma^{(n)}_y$  which enter (\ref{dRij}) are shown in Appendix B to produce a flip of the spin 
{\boldmath{$\hat\sigma$}}$^{(n)}$, that is, a shift of the operator $\hat m$ into $\hat m\pm\delta m$. 
 We introduce the notations

\begin{eqnarray}
\Delta_{\pm}f\left(  m\right) =f\left(  m_{\pm}\right)  -f\left(  m\right),\qquad 
m_\pm=m\pm\delta m,\qquad \delta m=\frac{2}{N}  {\rm  .} \label{delta+-}
\end{eqnarray}

The resulting dynamical equations for $P_{ij}(m,t)$ take different forms for the diagonal and for
the off-diagonal components. On the one hand, the first \textit{diagonal
block} of $\hat{D}$ is parameterized by the \textit{joint probabilities}
$P_{\uparrow\uparrow}\left(  m,t\right)  $ to find ${\rm S}$ in $\left\vert
\uparrow\right\rangle $ and $\hat{m}$ equal to $m$ at the time $t $. These
probabilities evolve according to

\begin{equation}
\mytext{\textcurrency dPdiag\textcurrency \qquad}
\frac{{\rm d}P_{\uparrow
\uparrow}\left(  m,t\right)  }{{\rm d}t}=\frac{\gamma N}{\hbar^{2}}\left\{
\Delta_{+}\left[  \left(  1+m\right)  \tilde{K}_{t}\left(  \Omega_{\uparrow
}^{-}(m)\right)  P_{\uparrow\uparrow}\left(  m,t\right)  \right]  +\Delta
_{-}\left[  \left(  1-m\right)  \tilde{K}_{t}\left(  \Omega_{\uparrow}
^{+}(m)\right)  P_{\uparrow\uparrow}\left(  m,t\right)  \right]  \right\}  {\rm 
,} \label{dPdiag}
\end{equation}
with initial condition $P_{\uparrow\uparrow}\left(  m,0\right)
=r_{\uparrow\uparrow}\left(  0\right)  P_{\rm M}\left(  m,0\right)  $ given by
(\ref{Pm0}); likewise for $P_{\downarrow\downarrow}\left(  m\right)$,  which involves the frequencies $\Omega^\mp_\down(m)$.
 The factor $\tilde{K}_{t}\left(  \omega\right)$  is expressed by the combination of two terms,

\begin{equation}
\mytext{\textcurrency Ktomega\textcurrency \qquad}
\tilde{K}_{t}\left(\omega\right)\equiv\tilde{K}_{t>}\left(\omega\right)  +\tilde{K}_{t<}\left(  \omega\right) 
   =\int_{-t}^{+t}{\rm d}ue^{-i\omega u}K\left(  u\right)
=\int_{-\infty}^\infty \frac{{\rm d}\omega^{\prime}}{\pi}\frac{\sin\left(  \omega^{\prime
}-\omega\right)  t}{\omega^{\prime}-\omega}\tilde{K}\left(  \omega^{\prime
}\right)  {\rm  .} \label{Ktomega}
\end{equation}
It is real and tends to $\tilde{K}\left(  \omega\right)$, given in Eq. (\ref{Ktilde}), 
at times $t$ larger than the range $\hbar/2\pi T$ of $K\left(  t\right)  $ \cite{petr,Weiss,Gardiner}.
This may be anticipated from the relation  $\sin[(\omega'-\omega)t]/\pi(\omega'-\omega)\to \delta(\omega'-\omega)$ for $t\to\infty$
and it may be demonstrated with help of the contour integration techniques of Appendix D, which we leave as a student exercise, see \S~\ref{fin9.6.1}.

On the other hand, the sets $P_{\uparrow\downarrow}\left(  m,t\right)  $ and
$P_{\downarrow\uparrow}=P_{\uparrow\downarrow}^{\star}$ which parameterize the
\textit{off-diagonal blocks} of $\hat{D}$, and which are related through
(\ref{P->C}) to the correlations between $\hat{s}_{x}$ or $\hat{s}_{y}$ and
any number of spins of ${\rm M}$, evolve according to
\begin{eqnarray}
 \mytext{\textcurrency dPoff\textcurrency \qquad}
\frac{{\rm d}}{{\rm d}t}
P_{\uparrow\downarrow}\left(  m,t\right) -\frac{2iNgm}{\hbar}
P_{\uparrow\downarrow}\left(  m,t\right) 
=\frac{\gamma N}{\hbar^{2}}\left\{\Delta_{+}\left[  \left(  1+m\right) \tildeK_-(m,t)
P_{\uparrow\downarrow}\left(m,t\right)  \right]
+\Delta_{-}\left[  \left(  1-m\right)  \tildeK_+(m,t)
 P_{\uparrow\downarrow}\left(m,t\right) \right] \right\}  {\rm  ,} \label{dPoff}
\end{eqnarray}
with initial condition $P_{\uparrow\downarrow}\left(  m,0\right)
=r_{\uparrow\downarrow}\left(  0\right)  P_{\rm M}\left(  m,0\right)  $.
 Here $\tilde{K}_{t>}$ and $\tilde{K}_{t<}$ enter the combination
 
\BEQ \label{Kpm}
\tildeK_\pm(m,t) \equiv \tilde{K}_{t>}\left[  \Omega_{\uparrow}^{\pm}(m)\right]  +\tilde{K}_{t<}\left[\Omega_{\downarrow}^{\pm}(m)\right]. 
 \EEQ

}
\subsubsection{Interpretation as quantum balance equations}
\label{section.4.3.3}

\hfill{\it Je moet je evenwicht bewaren\footnote{You have to keep your balance}}

\hfill{Dutch expression}

\vspace{3mm}
\ZeText{

Our basic equations (\ref{dPdiag}) and (\ref{dPoff}) fully describe the dynamics of the measurement. The diagonal equation (\ref{dPdiag}) can be
interpreted as a balance equation \cite{petr,Weiss,Gardiner}. Its first term represents elementary
processes in which {\it one among the spins,} say $\mathbf{\sigma}^{\left(n\right)}$,  flips
 from $\sigma_{z}^{\left(  n\right)  }=+1$ to $\sigma_{z}^{\left(n\right)  }=-1$. For the value $m$ of the magnetization, a value taken with
probability $P_{\uparrow\uparrow}\left(  m,t\right)  $ at the time $t$, there are $\frac{1}{2}N\left(  1+m\right)  $ spins pointing upwards, and the
probability for {one of these spins to flip} down between the times $t$ and
$t+{\rm d}t$ under the effect of the phonon bath can be read off from  (\ref{dPdiag}) to be equal to
$2\gamma\hbar^{-2}\tilde{K}_{t}\left(  \Omega_{\uparrow}^{-}\right)  {\rm d}t$.
 This process produces a decrease of $P_{\uparrow\uparrow}\left(  m\right)  $ and it is accounted for by the negative contribution 
(which arises from the second part of $\Delta _+$ and is proportional to $- P_{\up\up}(m,t)$)
to the first term in the right-hand side of (\ref{dPdiag}). The coefficient $\tilde{K}_{t}\left(  \omega\right)  $ depends on the temperature $T$ of the
bath ${\rm B}$, on the duration $t$ of its interaction with ${\rm M}$, and on the energy $\hbar\omega$ that it has transferred to ${\rm M}$; this
energy is evaluated for $P_{\uparrow\uparrow}$ (or $P_{\downarrow\downarrow}$)
as if the spins of ${\rm M}$ were submitted to an external field $+g$ (or $-g$). The first term in (\ref{dPdiag}) also contains a positive contribution
arising from the same process, for which the magnetization decreases from $m+\delta m$ to $m$, thus raising
$P_{\uparrow\uparrow}\left(  m\right)$ by a term proportional to $P_{\uparrow\uparrow}\left(  m+\delta m\right)  $. Likewise,
the second term in the right-hand side of (\ref{dPdiag}) describes the negative and positive changes of $P_{\uparrow\uparrow}\left(  m\right)  $
arising from the flip of a single spin from $\sigma_{z}^{\left(  n\right)}=-1$ to $\sigma_{z}^{\left(  n\right)  }=+1$. Quantum mechanics occurs in
(\ref{dPdiag}) through the expression (\ref{Ktilde}) of $\tilde{K}\left(\omega\right)$;  the flipping probabilities do not depend on the factor $\hbar$,   
owing to the factor $\hbar^2$ that enters $\tilde K(\omega)$ and the fact that we have chosen the dimensionless coupling constant $\gamma$,
but their quantum nature is still expressed by the Bose-Einstein occupation number.

The equation (\ref{dPoff}) for $P_{\uparrow\downarrow}$ has additional quantum features. Dealing with an off-diagonal block,
it involves simultaneously the two Hamiltonians $\hat{H}_{\uparrow}$ and $\hat{H}_{\downarrow}$ of Eq. (\ref{Hi}) in the Hilbert
space of ${\rm M}$, through the expression (\ref{Omidef}) of $\Omega^\pm_{\uparrow,\downarrow}$. 
The quantities $P_{\uparrow\downarrow}$ and $P_{\downarrow\uparrow}$ are complex and cannot be interpreted as probabilities, although we recognize
in the right-hand side the same type of balance as in Eq. (\ref{dPdiag}). In fact, while  $\sum_{m}P^{\rm dis}_{\uparrow\uparrow}\left(  m\right)  
=1-\sum_{m}P^{\rm dis}_{\downarrow\downarrow}\left(m\right)$, or in the $N\gg1$ limit $\int\d m\,P_{\uparrow\uparrow}\left(  m\right)  
=1-\int \d{m}\,P_{\downarrow\downarrow}\left(m\right)$,  remains constant in time because the sum over $m$ of the right-hand side of (\ref{dPdiag}) vanishes, 
the term in the left-hand side of (\ref{dPoff}), which arises from $H_{i}-H_{j}$, prevents $\sum_{m}P^{\rm dis}_{\uparrow\downarrow}\left( m\right)$ 
from being constant; It  will, actually, lead to the disappearance of these ``Schr\"odinger cat'' terms.

Comparison of the right-hand sides of (\ref{dPdiag}) and (\ref{dPoff}) shows
moreover that the bath acts in different ways on the diagonal and off-diagonal 
blocks of the density operator $\hat{D}$ of ${\rm S}+{\rm M}$.

}

\subsection{Large $N$ expansion}
\label{section.4.4}

\ZeText{

Except in subsection 8.1 we shall deal with a magnetic dot sufficiently large so that $N\gg1$. The set of values (\ref{eig}) on which the
distributions $P_{ij}\left(  m,t\right)  $ are defined then become dense on the interval $-1\leq m\leq+1$. At the initial time, $P_{ij}\left(  m,0\right)
$, proportional to (\ref{Pm0}), extends over a range of order $1/\sqrt{N}$ while the spacing of the discrete values of $m$ is $\delta m=2/N$. The initial
distributions $P_{ij}$ are thus smooth on the scale $\delta m$, and $P_{\up\up}$ and $P_{\downarrow\downarrow}$ will
remain smooth at later times. It is therefore legitimate to
\textit{interpolate} the set of values of the diagonal quantities $P_{ii}\left(  m,t\right)$ defined
at the discrete points (\ref{eig}) into a continuous function of $m$. If we assume the two resulting functions $P_{ii}$ to be several times differentiable
with respect to $m$, the discrete equation (\ref{dPdiag}) satisfied by the original distributions will give rise to continuous equations, 
which we shall derive below, involving an asymptotic expansion in powers of $1/N$. Within exponentially small corrections, the characteristic functions
associated with $P_{ii}(m,t)$  then reduce to integrals:

\begin{equation}
\mytext{\textcurrency sum=int\textcurrency \qquad}
\Psi_{ii}\left(
\lambda,t\right)  \equiv\sum_{m}P^{\rm dis}_{ii}\left(  m,t \right)  e^{i \lambda  m}=
\int{\rm d}mP_{ii}\left(  m,t \right)  e^{i \lambda  m}{\rm  ,}
\label{sum=int}
\end{equation}
provided $\lambda\ll N$. The moments of $P_{ii}\left(  m\right)  $ of order
less than $N$ can also be evaluated as integrals.

However, the left-hand side of Eq. (\ref{dPoff})  generates for finite times rapid variations of $P_{\uparrow\downarrow}(m,t)$ and 
$P_{\downarrow\uparrow}(m,t)$ as functions of $m$, and it will be necessary in sections 5 and 6 to account for the discrete nature of $m$. 
When writing below the equations of motion for these quantities in the large $N$ limit, we will take care of this difficulty.

The differences $\Delta_{\pm}$ defined by (\ref{delta+-}) satisfy
 
 \BEQ    \label{4.a}
  \Delta_\pm [f(m)g(m)]=\left[\Delta_\pm f(m)\right] g(m) + f(m) \left[\Delta_\pm g(m)\right] +
 \left[\Delta_\pm f(m)\right]\left[\Delta_\pm g(m)\right],     
 \EEQ
and  give rise to
derivatives with respect to $m$ according to
\begin{equation}
\mytext{\textcurrency deltaappr\textcurrency \qquad}
\Delta_{\pm}f\left(
m\right)  \approx\pm\frac{2}{N}\frac{\partial f\left(  m\right)  }{\partial
m}+\frac{2}{N^{2}}\frac{\partial^{2}f\left(  m\right)  }{\partial m^{2}}
\pm\frac{4}{3N^{3}}\frac{\partial^{3}f\left(  m\right)  }{\partial m^{3}
}{\rm  .} \label{deltaappr}
\end{equation}
We can also expand the  excitation energies $\hbar\Omega_{i}^{\pm}$, defined by 
(\ref{Omidef})  and (\ref{Hi}), for large $N$  as 

\begin{equation}
\mytext{\textcurrency gaappr\textcurrency \qquad}
\Omega_{i}^{\pm}\left(m\right)  \approx
\mp2\omega_{i}-\frac{2}{N}\frac{{\rm d}\omega_{i}}{{\rm d}m}
=\left(1\pm\frac{1}{N} \frac{{\rm d}}{{\rm d}m}\right)(\mp2\omega_{i})
{\rm  ,} \label{gaappr}
\end{equation}
where we introduced the quantity

\BEQ  \label{omi=}
 \hbar \omega_i = -\frac{1}{N}\, \frac{\d H_i}{\d m} = gs_i +J_2m+J_4m^{3} ,           \qquad (s_i=\pm1),
\EEQ
interpreted as the effective energy of a single spin of M coupled to the other spins of M and to the tested spin S.

The above expansions will allow us to transform, for large $N$, the equations of motion for $P_{ij}$ into partial differential equations. 
In case $\p P_{ij} / \p m$ is finite for large $N$, we can simply replace in (\ref{dPdiag}) and (\ref{dPoff}) 
$N\Delta_\pm$ by $\pm 2\p/\p m$ and $\Omega_i^\pm$ by $\mp 2\omega_i$. However, such a situation is exceptional; we shall encounter it only in \S~7.3.2. 
In general $P_{ij}$ will behave for large $N$ as $A\exp NB$. This property, exhibited at $t=0$ in \S\S~3.3.3 and 3.3.4, is preserved by the dynamics. 
As $\p P_{ij}/\p t$ involves  leading contributions of orders $N$ and $1$, we need to include in the right-hand sides of
 (\ref{dPdiag}) and (\ref{dPoff})  contributions of the same two orders.  Let us therefore introduce the functions

\BEQ \label{4.b}
 X_{ij}(m,t) \equiv \frac{1}{N} \frac{\p \ln P_{ij}}{\p m}= \frac{1}{N P_{ij}} \frac{\p P_{ij}}{\p m},     
                \EEQ
which contain parts of order $1$ and $1/N$, and their derivatives

\BEQ\label{4.c}
 X'_{ij} \equiv\frac{1}{N} \frac{\p^2 \ln P_{ij}}{\p m^2}= \frac{\p X_{ij}}{\p m}, 
 \EEQ
which can be truncated at finite order in $N$.  The discrete increments of $P_{ij}$ are thus expanded as

\BEA\label{4.d}
\Delta_\pm P_{ij} &=&  P_{ij} \left[\exp (\Delta_\pm \ln P_{ij} )- 1\right] \approx
P_{ij} \left[\exp\left(\pm 2X_{ij} +\frac{2}{N}X'_{ij}\right)- 1+ {\cal O}\left(\frac{1}{N^2}\right)\right]
\nn\\ &\approx& P_{ij} \left[\exp\left(\pm 2X_{ij}\right)- 1 + \frac{2X'_{ij} }{N} \exp\left(\pm 2X_{ij}\right)+ {\cal O}\left(\frac{1}{N^2}\right)  \right].  
\EEA

We express (\ref{dPdiag})  by using the full relation (\ref{4.a}), with $f=P_{\up\up}$ and 
 $g=(1\pm  m)\tilde K_{t}(\Omega_\up^\mp)$, by evaluating $\Delta_\pm f$ from (\ref{4.d}),
 and by inserting  (\ref{OmpmPMdm}) into $\Delta_\pm g$. This yields
 
\BEA\label{4.e}
&&\frac{\p P_{\uparrow\uparrow}}{\p t}\approx\frac{2 \gamma}{\hbar^2} P_{\uparrow\uparrow} \left \{  \frac{}{}N \sinh X_{\uparrow\uparrow}
 \left[(1+m) \tilde K_t(\Omega_\uparrow^- ) e^{X_{\uparrow\uparrow} }- (1-m) \tilde K_t(\Omega_\uparrow^+) e^{-X_{\uparrow\uparrow}}\right] 
\right. \nn\\ &&  
\left.  + e^{X_{\uparrow\uparrow}}\frac{\partial}{\partial m} \left[(1+m) \tilde K_t(2\omega_\uparrow) e^{X_{\uparrow\uparrow}}\right]
 - e^{-X_{\uparrow\uparrow} }    \frac{\p}{\p m} \left[(1-m) \tilde K_t(-2\omega_\uparrow)e^{-X_{\uparrow\uparrow}}\right]  + {\cal O} \left(\frac{1}{N}\right)\right\}  .     
\EEA
The first term on the right-hand side determines the evolution of the exponent of $P_{\uparrow\uparrow}$, which contains parts of order $N$, 
but contains also contributions of order  $1$ arising from the terms of order $1/N$ of (\ref{gaappr}) and of $X_{\up\up}$. 
The remaining terms determine the evolution of the amplitude of $P_{\up\up}$. 
The bath term of the equation (\ref{dPoff}) for $P_{\uparrow\downarrow} (m,t)$ 
(and for $P_{\downarrow\uparrow}=P_{\uparrow\downarrow}^\ast$) has a similar form, again obtained from 
all the terms in (\ref{4.a}) and (\ref{4.d}),  namely, using the notation (\ref{Kpm}):

\BEA\label{4.f}\label{dPoff2}
&& \frac{\p P_{\uparrow\downarrow}}{\p t}- \frac{2iNgm}{\hbar} P_{\uparrow\downarrow}\approx
\frac{2 \gamma}{\hbar^2} P_{\uparrow\downarrow} \left \{ \frac{}{}N \sinh X_{\uparrow\downarrow}\left [(1+m) 
\tildeK_-(m,t)e^{X_{\uparrow\downarrow}}  - (1-m) 
\tildeK_+(m,t)
e^{-X_{\uparrow\downarrow}}\right]  \nn \,\,
 \right. \\&&
 \left.+   e^{X_{\uparrow\downarrow}}\frac{\partial}{\partial m} \left[(1+m) \tildeK_-(m,t)  e^{X_{\uparrow\downarrow}} \right] - e^{-X_{\uparrow\downarrow} } \frac{\p}{\p m} \left[(1-m) 
 \tildeK_+(m,t) e^{-X_{\uparrow\downarrow}}\right]  + {\cal O} \left(\frac{1}{N}\right)\right\}.   
\EEA          
      
 A further simplification occurs for large $N$ in the diagonal sector. Then $P_{\uparrow\uparrow}$, which is real, takes significant values only in the 
 vicinity of the maximum of $\ln P_{\uparrow\uparrow}$.
  This maximum is reached at a point $m=\mu(t)$, and $P_{\uparrow\uparrow}$ 
 is concentrated in a range for $|m - \mu(t)|$ of order $1/\sqrt{N}$\,\footnote{Numerically we find for $N=1000$ extended distributions, 
 see Figs. 7.5 and 7.6, since the typical peak width $1/\sqrt{N}$ is still sizable}. 
 In this range, $X_{\uparrow\uparrow}$ is proportional to $\mu(t) - m$, and it is therefore  of order $1/\sqrt{N}$\footnote{This property does not hold 
 for $P_{\uparrow\downarrow}$, since $X_{\uparrow\downarrow}$ contains a term $2igt/\hbar$ arising from the left hand side of (\ref{4.f})}.
 We can therefore expand (\ref{4.e}) in powers of
 $X_{\uparrow\uparrow}$, noting also that $X'_{\uparrow\uparrow}$ is finite, and collect the $X_{\uparrow\uparrow}$, $X^2_{\uparrow\uparrow}$,  
  $X'_{\uparrow\uparrow}$ and $X_{\uparrow\uparrow}X'_{\uparrow\uparrow}$ terms. 
  Thus, if we disregard the exponentially small tails of the distribution 
$P_{\uparrow\uparrow}$, which do not contribute to physical quantities, we find at the considered order, using (\ref{4.b}) and (\ref{4.c}),

\begin{equation}
\mytext{\textcurrency dPdiag2\textcurrency \qquad}
\frac{\partial P_{\uparrow
\uparrow}}{\partial t}\approx\frac{\partial}{\partial m}\left[  -v\left(  m,t\right)  P_{\uparrow\uparrow}\right]  +\frac{1}{N}
\frac{\partial^2}{\partial m^2}\left[  w\left(  m,t\right) P_{\uparrow\uparrow}\right]  {\rm  ,}
\label{dPdiag2}
\end{equation}
where 

\begin{eqnarray}
\mytext{\textcurrency vupup\textcurrency \qquad}
v\left(m,t\right)   &  =& \frac{2\gamma}{\hbar^{2}}\left[  \left(  1-m
\right)  \tilde{K}_{t}\left(  -2\omega_{\uparrow}\right)  -\left(1+m
\right)  \tilde{K}_{t}\left(  2\omega_{\uparrow}\right)
\right]  +{\cal O}\left(  \frac{1}{N^{}}\right)  {\rm  ,}\label{vupup}\\
\mytext{\textcurrency wupup\textcurrency \qquad}
w\left(m,t\right)   &  =& \frac{2\gamma}{\hbar^{2}}\left[  \left(  1-m\right)  \tilde
{K}_{t}\left(  -2\omega_{\uparrow}\right)  +\left(  1+m\right)  \tilde{K}
_{t}\left(  2\omega_{\uparrow}\right)  \right]  +{\cal O}\left(  \frac
{1}{N}\right)  {\rm  .} \label{wupup}
\end{eqnarray}   
The next contribution to the right hand side of (\ref{dPdiag2}) would be $- 2v X_{\uparrow\uparrow} X'_{\uparrow\uparrow} P_{\uparrow\uparrow}$,
 of order $1/\sqrt{N}$. We have replaced in $v$ and $w$ the frequencies $\Omega_\uparrow^\pm$ by $\mp2\omega_\uparrow$,  which has the sole 
 effect of shifting the position and width of the distribution $P_{\uparrow\uparrow}$ by a quantity of order $1/N$. 
As shown by the original equation (\ref{4.e}), the two terms of (\ref{dPdiag2}) have the same order of magnitude (in spite of the presence of the 
factor $1/N$ in the second one) when $P_{\uparrow\uparrow}$ has an exponential form in $N$. Only the first one contributes if 
$P_{\uparrow\uparrow}$ becomes smooth (\S~7.3.2).
The equation for $P_{\downarrow\downarrow}$\ is obtained from (\ref{dPdiag2}) by changing $g$\ into $-g$.

In the regime where the registration will take place (\S~\ref{section.7.1.1}), we shall be allowed to replace
$\tilde K_t (\pm 2\omega_i)$ by $\tilde K(\pm2\omega_i)$, which according to (\ref{Ktilde}) is equal to   

\BEQ \label{KOmipm=}
\tilde K(\pm2\omega_i)=
\frac{\hbar^2\omega_i}{4}\left[\coth(\beta \hbar\omega_i)\mp1\right]\, 
\exp\left(-\frac{2|\omega_i|}{\Gamma}\right),\qquad (i=\uparrow,\,\downarrow).   
\EEQ
Eqs. (\ref{vupup}) and (\ref{wupup}) will thereby be simplified.

\vspace{3mm}

The final equations (\ref{4.f}) and  (\ref{dPdiag2}), with the initial conditions $P_{ij}(m,0)=r_{ij} P_{\rm M}(m,0)$ expressed by (\ref{Pm0}), 
describe the evolution of S + M during the measurement process. We will work them out in sections 5 to 7. 
The various quantities entering them were defined by (\ref{4.b}) and (\ref{4.c}) for $X_{ij}$ and $X'_{ij}$, by (\ref{vupup}) and (\ref{wupup}) for 
$v$ and $w$, by (\ref{Ktilde}), (\ref{K>}), (\ref{K<}),  (\ref{Ktomega}) and (\ref{Kpm}) for $\tilde K_{t>}$, $\tilde K_{t<}$,  $\tilde K_t$ and $\tilde K_\pm$, respectively,
and by (\ref{omi=}) for $\omega _i$.

The dynamics of $P_{\uparrow\downarrow}$ has a {\it purely quantum} nature. The left-hand side of (\ref{4.f}) governs the evolution of the normalization
$\int \d m P_{\uparrow\downarrow}(m,t)$, equal to the off-diagonal element $r_{\uparrow\downarrow}(t)$ of the marginal state $\hat  r(t)$ of S. 
The bath gives rise on the right-hand side to a non-linear partial differential  structure, which arises from the discrete nature of the spectrum of $\hat m$.

 The final equation of motion (\ref{dPdiag2}) for $P_{\uparrow\uparrow}$ has the form of a {\it Fokker--Planck equation} \cite{vKampenbook,risken}, 
 which describes a stochastic 
 motion of the variable $m$. Its coefficient $v$, which depends on $m$ and $t$, can be interpreted as a {\it drift velocity}, while its coefficient $w$ characterizes 
 a diffusion process. This analogy with a classical diffusion process, should not, however, hide the {\it quantum origin of the diffusion term}, which is as sizeable 
 for large $N$ as the drift term. While the drift term comes out by bluntly taking the continuous limit of (\ref{dPdiag}), the diffusion term originates, as shown by the above derivation, 
 from the conjugate effect of two features: (i) the smallness of the fluctuations of $m$, and (ii) the discreteness of the spectrum of the pointer observable $\hat m$. 
 Although the pointer is macroscopic, its quantum nature is essential, not only in the off-diagonal sector, but also in the diagonal sector which accounts 
 for the registration of the result.

}
\renewcommand{\thesection}{\arabic{section}}
\section{Very short times: truncation} 
\setcounter{equation}{0}\setcounter{figure}{0}\renewcommand{\thesection}{\arabic{section}.}
\label{section.5}

\hfill{{\it Alea iacta est}\footnote{The die is cast}}

\hfill{ Julius Caesar}

\hspace{3mm}
\ZeText{

Since the coupling $\gamma$ of the magnet ${\rm M}$\ with the bath ${\rm B}$\ is weak, some time is required before ${\rm B}$\ acts
significantly on ${\rm M}$. In the present section, we therefore study the behavior of ${\rm S}+{\rm M}$ at times sufficiently short so that we can
neglect the right-hand sides of (\ref{dPdiag}) and (\ref{dPoff}). 
We shall see that the state $\scriptD(t)$ of S + A is then {\it truncated}, that is, its {\it off-diagonal blocks $\scriptR_{\up\down}$ and 
$\scriptR_{\down\up}$ rapidly decay}, while the diagonal blocks are still unaffected.

}
\subsection{The truncation mechanism} 
\label{section.5.1}

\subsubsection{The truncation time}
\label{section.5.1.1}

\hfill{\it An elephant does not get tired carrying his trunk}

\hfill{Burundian proverb}

\vspace{3mm}

\ZeText{

When their right-hand sides are dropped, the equations (\ref{dPdiag}) and
(\ref{dPoff}) with the appropriate boundary conditions are readily solved as
\begin{eqnarray}
\mytext{\textcurrency solPdiag\textcurrency \qquad}
P_{\uparrow\uparrow}\left(
m,t\right)   &  =&r_{\uparrow\uparrow}\left(  0\right)  P_{\rm M}\left(  m,0\right)
{\rm ,\qquad}P_{\downarrow\downarrow}\left(  m,t\right)  =r_{\downarrow
\downarrow}\left(  0\right)  P_{\rm M}\left(  m,0\right)  {\rm  ,}\label{solPdiag}\\
\mytext{\textcurrency solPoff\textcurrency \qquad}
P_{\uparrow\downarrow}\left(
m,t\right)   &  =&\left[  P_{\downarrow\uparrow}\left(  m,t\right)  \right]
^{\ast}=r_{\uparrow\downarrow}\left(  0\right)  P_{\rm M}\left(  m,0\right)
e^{2iNgmt/\hbar}{\rm  .} \label{solPoff}
\end{eqnarray}
From the viewpoint of the tested spin ${\rm S}$, these
equations describe a Larmor precession around the $z$-axis \cite{spin_echo}, under
the action of an effective magnetic field $Ngm$ which depends on
the state of ${\rm M}$. From the viewpoint of the magnet
${\rm M}$, we shall see in \S~\ref{section.5.1.3} that the phase occurring
in (\ref{solPoff}) generates time-dependent correlations between
${\rm M}$ and the transverse components of $\mathbf{s}$.

The expectation values $\left\langle \hat{s}_{a}\left(  t\right)
\right\rangle $ of the components of $\mathbf{s}$ are found from (\ref{P->C})
by summing (\ref{solPdiag}) and (\ref{solPoff}) over $m$. These equations are
valid for arbitrary $N$ and arbitrary time $t$ as long as the bath is
inactive. If $N$ is sufficiently large and $t$ sufficiently small so that the
summand is a smooth function on the scale $\delta m=2/N$, that is, if $N\gg1$
and $t\ll\hbar/g$, we can use (\ref{sum=int}) to replace the summation over
$m$ by an integration. These conditions will be fulfilled in subsections 5.1
and 5.2; we shall relax the second one in subsection~\ref{section.5.3} where we study the
effects of the discreteness of $m$. Using the expression (\ref{Pm0}),
(\ref{DELTAm}) of $P_{\rm M}\left(  m,0\right)  $, we find by integrating (\ref{solPoff}) over $m$:
\begin{equation}
\mytext{\textcurrency rofft\textcurrency \qquad}
r_{\uparrow\downarrow}\left(
t\right)  =r_{\uparrow\downarrow}\left(  0\right)  e^{-\left(  t/\tau
_{{\rm trunc}}\right)  ^{2}}{\rm  ,} \label{rofft}
\end{equation}
or equivalently
\begin{eqnarray}
\mytext{\textcurrency stransv\textcurrency \qquad}
\left\langle \hat{s}
_{a}\left(  t\right)  \right\rangle  &  =&\left\langle \hat{s}_{a}\left(
0\right)  \right\rangle e^{-\left(  t/\tau_{{\rm trunc}}\right)  ^{2}
}{, \qquad\qquad}(a=x{, }y){ ,}\label{stransv}\\
\mytext{\textcurrency slong\textcurrency \qquad}
\left\langle \hat{s}_{z}\left(
t\right)  \right\rangle  &  =&\left\langle \hat{s}_{z}\left(  0\right)
\right\rangle { ,} \label{slong}
\end{eqnarray}
where we introduced the truncation time 

\begin{equation}
\mytext{\textcurrency taured\textcurrency \qquad}
\tau_{{\rm trunc}}\equiv
\frac{\hbar}{\sqrt{2} \ Ng\Delta m}=\frac{\hbar}{\sqrt{2N}\ \delta_{0}g}{ .}
\label{taured}
\end{equation}
Although $P_{\uparrow\downarrow}(m,t)$ is merely an oscillating function of $t$ for each value of $m$, the summation over $m$ has 
given rise to extinction. This property arises from the dephasing that exists between the oscillations for different values of $m$. 

In the case $T_0=\infty$ of a fully disordered initial state, we may solve directly (\ref{dRij}) (without right-hand side) from the initial condition (\ref{Rij0}). 
We obtain, for arbitrary $N$, $\hat R_{\up\down}(t) = r_{\up\down} (0) 2^{-N}  \exp (2iNg \hat m t /\hbar)$,
whence by using the definition  (\ref{hatm=}) of $\hat m$ and taking the trace over M,  we find the exact result\footnote{An equivalent way to derive this result is to 
employ (\ref{RijP}) for making the identification 
$P_{\up\down}^{\rm dis}(m,t) = G(m)\times r_{\up\down} (0) 2^{-N} \exp (2iNg m t /\hbar)$, and to sum over the values (\ref{eig})  of $m$}

\BEQ \label{rupdown(t)}
r_{\up\down}(t)=r_{\up\down}(0)\left(\cos\frac{2gt}{\hbar}\right)^N,
\EEQ
which reduces to (\ref{rofft}) for times of order $\tau_{\rm trunc}$.

Thus, over a time scale of order $\tau_{{\rm trunc}}$, the transverse components of the spin ${\rm S}$ decay and vanish while the $z$-component
is unaltered: the off-diagonal elements $r_{\uparrow\downarrow}=r_{\downarrow\uparrow}^{\ast}$ of the marginal density matrix of ${\rm S}$ disappear
during the very first stage of the measurement process. It was to be expected that the apparatus, which is a large object, has a rapid and strong effect on
the much smaller system ${\rm S}$. In the present model, this rapidity arises from the {\it large number $N$ of spins of the magnet}, which shows up
through the factor $1/\sqrt{N}$ in the expression (\ref{taured}) of $\tau_{{\rm trunc}}$.

As we shall see in \S~\ref{section.5.1.3}, the off-diagonal block ${\hat {\cal R}}_{\uparrow\downarrow}={\hat {\cal R}}_{\downarrow\uparrow}^{\dag}$ of the
full density matrix ${\hat {\cal D}}$ of ${\rm S}+{\rm A}$ is proportional to $\hat{r}_{\uparrow\downarrow}\left(  t\right)  $ and its
elements also decrease as $\exp[{-(  t/\tau_{{\rm trunc}})  ^{2}}]$,
at  least those elements which determine correlations involving a number of spins of M small compared to $N$.
In the vocabulary of \S~1.3.2, {\it truncation} therefore takes place for the overall system ${\rm S}+{\rm A}$ over the brief initial time lapse
 $\tau_{{\rm trunc}}$, while Eq. (5.3) describes {\it weak truncation} for S.
 
The quantum nature of the truncation process manifests itself
through the occurrence of two different Hamiltonians
$\hat{H}_{\uparrow}$ and $\hat {H}_{\downarrow}$ in the Hilbert
space of ${\rm M}$. Both of them occur in the dynamical
equation (\ref{dPoff}) for $P_{\uparrow\downarrow}$, whereas only
$\hat{H}_{\uparrow}$ occurs in (\ref{dPdiag}) for $P_{\uparrow\uparrow}$  through $\Omega^\pm_\up$, 
and likewise only $\hat{H}_{\downarrow}$ for $P_{\downarrow\downarrow}$, through $\Omega^\pm_\down$.

The truncation time $\tau_{{\rm trunc}}$ is inversely proportional to the
coupling $g$ between $\hat{s}_{z}$ and each spin $\hat{\sigma}_{z}^{\left(
n\right)  }$ of the magnet. It does not depend directly on the couplings $J_q$ ($q=2,4$)
between the spins $\hat{\sigma}_{z}^{\left(  n\right)  }$. Indeed, the
dynamical equations (\ref{dPdiag}), (\ref{dPoff}) without bath-magnet coupling involve only
$H_{\uparrow}\left(  m\right)  -H_{\downarrow}\left(  m\right)  $, so that the
interactions $\hat{H}_{{\rm M}}$ which are responsible for ferromagnetism
cancel out therein. These interactions occur only through the right-hand side
 which describes the effect of the bath. They also appear
indirectly in $\tau_{{\rm trunc}}$ through the factor $\delta_{0}$ of $\Delta
m$ given by (\ref{delta0=}), in the case $q=2$ of an Ising magnet ${\rm M}$. 
When $J_2\neq0$, the occurrence of $\delta_{0}>1$ thus contributes to accelerate the truncation process.

}
\subsubsection{Truncation versus decoherence: a general phenomenon}
\label{section.5.1.2}

\ZeText{

 It is often said ~\cite{Schlosshauer,zurek,Guilini,Blanchard,walls,walls_book,Braun} that ``von Neumann's reduction is a decoherence effect''. 
 (The traditional word ``reduction'' covers in the literature both concepts of ``truncation'' and ``reduction'' as defined in \S~1.3.2.) 
   As is well known, decoherence is the rapid destruction of coherent superpositions of distinct pure states induced by a random environment, 
 such as a thermal bath. In the latter seminal case, the characteristic decoherence time has the form of $\hbar/T$ divided by some power of the number 
 of degrees of freedom  of the system and by a dimensionless coupling constant between the system and the bath (see also our discussion of the decoherence 
 approach in section 2).  Here, things are different. As we have just seen and as will be studied below in detail, the initial truncation process involves only the magnet. 
 Although the bath is part of the apparatus, it has no effect here and the characteristic truncation time $\tau_{\rm trunc}$ does not depend on the 
 bath temperature. Indeed the dimensional factor of (5.6)  is $\hbar/g$, and not $\hbar/T$. The thermal fluctuations are replaced by the fluctuation $\Delta m$ of the
pointer variable, which does not depend on $T_0$ for $q=4$ and  which decreases with $T_{0}$ as (\ref{delta0=}) for $q=2$.

The fact that the truncation is controlled only by the coupling of the pointer variable $\hat m$ with ${\rm S}$\ is exhibited by the occurrence, in (\ref{taured}),
of its number $N$ of degrees of freedom of M. Registration of $s_{z}$ requires this variable to be \textit{collective}, so that $N\gg1$. However, long before
registration begins to take place in ${\rm A}$ through the influence on ${\hat {\cal D}}$ of $r_{\uparrow\uparrow}\left(  0\right)  $ and
$r_{\downarrow\downarrow}\left(  0\right)  $, the large size of the detector entails the loss of $r_{\uparrow\downarrow}\left(  0\right)  $ and
$r_{\downarrow\uparrow}\left(  0\right)  $.

Moreover, the \textit{basis} in which the truncation takes place is \textit{selected} by the very design of the apparatus. It depends on the
observable which is being measured. Had we proceeded to measure $\hat{s}_{x}$ instead of $\hat{s}_{z}$, we would have changed the orientation of the
magnetic dot; the part of the initial state $\hat{r}\left(  0\right)  $ of ${\rm S}$\ that gets lost would have been different, being related to the off-diagonal elements 
in the $x$-basis. Contrary to standard decoherence, truncation is here a controlled effect.

Altogether, it is only the pointer degrees of freedom \textit{directly coupled to} ${\rm S}$ that are responsible for the rapid truncation. 
As such, it is a {\it dephasing}. The effects of the bath are important (sections 6.2 and 7), but do not infer on the initial truncation process, 
on the time scale $\tau_{{\rm trunc}}$. We consider it therefore confusing to use the term \textquotedblleft decoherence\textquotedblright\ for
the decay of the off-diagonal blocks in a quantum measurement, since its mechanism can be fundamentally different from a standard
environment-induced decoherence. Here the truncation is a consequence of {\rm dephasing} between oscillatory terms which should be summed to generate the 
physical quantities\footnote{ In section 6.2 we shall discuss the effects of {\it decoherence} by the bath, which does take place, but long after the truncation time scale}.

\vspace{3mm}

The above considerations hold for the \textit{class of models} of quantum measurements for which {\it the pointer has many degrees
of freedom} directly coupled to S \cite{ABNqm2003,SpehnerHaake1,SpehnerHaake2} (see also \cite{Braun} in this context). 
We have already found for the truncation time a behavior analogous to (\ref{taured}) in a model where the
detector is a Bose gas \cite{ABNqm2001}, with a scaling in $N^{-1/4}$ instead of $N^{-1/2}$. More generally, 
suppose we wish to measure an arbitrary observable $\hat{s}$ of a microscopic system ${\rm S}$, with discrete
eigenvalues $s_{i}$ and corresponding projections $\hat{\Pi}_{i}$. The result should be registered by some pointer variable $\hat{m}$
of an apparatus ${\rm A}$\ coupled to $\hat{s}$. The full Hamiltonian has still the form (\ref{ham}), and it is natural to
assume that the system--apparatus coupling has the same form

\begin{equation}
\mytext{\textcurrency genint\textcurrency \qquad}
\hat{H}_{{\rm SA}}=-Ng\hat{s}\hat{m}, \qquad \textrm{(general operators $\hat s$, $\hat m$) }
 \label{genint}
 \end{equation}
as (\ref{HSA}). The coupling constant $g$ refers to each one of the $N$
elements of the collective pointer, so that a factor $N$ appears in
(\ref{genint}) as in (\ref{HSA}), if $\hat{m}$\ is dimensionless and
normalized in such a way that the range of its relevant eigenvalues is finite
when $N$ becomes large. The \truncated \ density matrix $\hat{r}\left(  t\right)
$\ is made of blocks $\left\langle i\alpha\right\vert \hat{r}\left\vert
j\beta\right\rangle $ where $\alpha$ takes as many values as the dimension of
$\hat{\Pi}_{i}$. It can be obtained as
\begin{equation}
\left\langle i\alpha\right\vert \hat{r}\left(  t\right)  \left\vert
j\beta\right\rangle =\sum_{m}\left\langle i\alpha\right\vert {\cal P}
\left(  m,t\right)  \left\vert j\beta\right\rangle { ,}
\end{equation}
where $\left\langle i\alpha\right\vert {\cal P}\left(  m,t\right)
\left\vert j\beta\right\rangle $, which generalizes $P_{\uparrow\downarrow
}\left(  m,t\right)  $, is defined by
\begin{equation}
\mytext{\textcurrency genP\textcurrency \qquad}
\left\langle i\alpha\right\vert
{\cal P}\left(  m,t\right)  \left\vert j\beta\right\rangle =\left\langle
i\alpha\right\vert {\rm tr}_{{\rm A}}\left(  \delta_{\hat{m}
,m}{\hat {\cal D}}\right)  \left\vert j\beta\right\rangle { .}
\label{genP}
\end{equation}
We have denoted by $m$\ the eigenvalues of $\hat{m}$, and by $\delta_{\hat
{m},m}$ the projection operator on $m$\ in the Hilbert space of ${\rm A}$.
The quantity (\ref{genP}) satisfies an equation of motion dominated by
(\ref{genint}):
\begin{equation}
\left[  i\hbar\frac{{\rm d}}{{\rm d}t}+Ng(  s_{i}-s_{j})m\right]  \left\langle i\alpha\right\vert {\cal P}\left(  m,t\right)
\left\vert j\beta\right\rangle \simeq0{ .}
\end{equation}
In fact, the terms arising from $\hat{H}_{{\rm S}}$ (which need no longer vanish but only commute with $\hat{s}$) and from $\hat{H}_{{\rm A}}$\ (which
commutes with the initial density operator ${\hat {\cal R}}\left(  0\right)$) are small during the initial instants compared to the term arising from the
coupling $\hat{H}_{{\rm SA}}$. We therefore find for short times

\begin{equation}
\mytext{\textcurrency interf\textcurrency \qquad}
\left\langle i\alpha\right\vert
\hat{r}\left(  t\right)  \left\vert j\beta\right\rangle =\left\langle
i\alpha\right\vert \hat{r}\left(  0\right)  \left\vert j\beta\right\rangle
{\rm tr}_{{\rm A}}{\hat {\cal R}}\left(  0\right)  e^{iNg(s_{i}-s_{j})  \hat{m}t/\hbar}{ .} \label{interf}
\end{equation}
The rapidly oscillating terms in the right-hand side interfere destructively as in (\ref{rofft}) on a short time, if $\hat{m}$ has a dense spectrum and an
initial distribution involving many eigenvalues. Each contribution is merely oscillating, but the summation over eigenvalues produces an extinction.
(We come back to this point in subsection 5.2 and in \S~\ref{fin12.2.3}.) This decrease takes place on
a time scale of order $\hbar/Ng\delta s\Delta m$, where $\delta s$ is the
level spacing of the measured observable $\hat{s}$\ and $\Delta m$ is the
width of the distribution of eigenvalues of $\hat{m}$\ in the initial state of
the apparatus. Leaving aside the later stages of the measurement process, we
thus acknowledge the generality of the present truncation mechanism, and that
of the expression (\ref{taured}) for the truncation time in the spin $\half$ situation where $\delta s=2$.

}

\subsubsection{Establishment and disappearance of correlations}
\label{section.5.1.3}

\hfill{\it The most rigid structures, the most impervious to change, }

\hfill{\it will collapse first}

\hfill{Eckhart Tolle}

\ZeText{

Let us now examine how the apparatus evolves during this first stage of the measurement process, described by Eqs.
(\ref{solPdiag}) and (\ref{solPoff}). The first equation implies that the marginal density operator 
$\hat {R}_{\rm M}\left(t\right)  =\hat{R}_{\uparrow\uparrow}\left(  t\right) +\hat{R}_{\downarrow\downarrow}\left(  t\right)  $ of
${\rm M}$\ remains unchanged. This property agrees with the idea that ${\rm M}$, {\it a large object}, has a {\it strong influence on}
${\rm S}$, a small object, but that conversely a {\it long time} is required {\it before ${\rm M}$\ is affected} by its interaction with
${\rm S}$. Eqs. (\ref{solPdiag}) also imply that no correlation is created between $\hat{s}_{z}$\ and ${\rm M}$.

However, although $\hat{R}_{\rm M}\left(  t\right)  =\hat{R}_{
}\left(  0\right)  $, correlations are created between ${\rm M}$\ and the
transverse component $\hat{s}_{x}$ (or $\hat{s}_{y}$) of ${\rm S}$. These
correlations are described by the quantities $C_{x}=P_{\uparrow\downarrow
}+P_{\downarrow\uparrow}$\ and $C_{y}=i\left(  P_{\uparrow\downarrow
}-P_{\downarrow\uparrow}\right)  $\ introduced in (\ref{P->C}). Since $\hat
{R}_{\uparrow\downarrow}$\ is a function of $\hat{m}$\ only, the components
$\hat{\sigma}_{x}^{\left(  n\right)  }$ and $\hat{\sigma}_{y}^{\left(
n\right)  }$\ of the spins of ${\rm M}$\ remain statistically independent,
with $\langle \hat{\sigma}_{x}^{\left(  n\right)  }\rangle=\langle \hat{\sigma}_{y}^{\left(  n\right)  }\rangle =0$ 
and with the quantum fluctuations $\langle \hat{\sigma}_{x}^{\left(  n\right)2}\rangle =
\langle \hat{\sigma}_{y}^{\left(  n\right)2}\rangle =1$. 
The correlations between ${\rm M}$\ and ${\rm S} $\ involve only the $z$-component of the spins 
$\mathbf{\hat{\sigma}}^{\left(n\right)  }$\ of the magnet and the $x$- or $y$-component of the tested spin
$\mathbf{s}$. We can derive them as functions of time from the generating function

\begin{equation}
\mytext{\textcurrency genf\textcurrency \qquad}
\Psi_{\uparrow\downarrow}(\lambda,t )  \equiv\sum_{k=0}^{\infty}\frac{i^k\lambda^{k}}{k!}
\langle \hat{s}_{-}\hat{m}^{k}(t) \rangle =\sum_{m}P^{\rm dis}_{\uparrow\downarrow}(m,t )  e^{i \lambda  m}
=r_{\uparrow\downarrow}(0 )  \sum_{m}P^{\rm dis}_{\rm M}(  m,0 )  e^{2iNgmt/\hbar+i \lambda  m}{ ,}
\label{genf}
\end{equation}
where $\hat{s}_{-}=\frac{1}{2}(\hat{s}_{x}-i\hat{s}_{y} )  $. In fact, whereas
$\Psi_{\uparrow\downarrow}\left(  \lambda,t\right)  $ generates
the expectation values $\langle\hat{s}_{-}\hat{m}^{k} \rangle $, the correlations
$\langle \hat{s}_{-}\hat{m}^{k} \rangle _{{\rm c}}$
are defined by the {\it cumulant expansion}

\begin{equation}
\mytext{\textcurrency genfc\textcurrency \qquad}
\Psi_{\uparrow\downarrow} ( \lambda,t )
=\sum_{k=0}^{\infty}\frac{i^k\lambda^{k}}{k!} \langle
\hat{s}_{-}\hat{m}^{k} \rangle _{{\rm c}}\left(
\sum_{k^{\prime }=0}^{\infty}
\frac{i^{k^\prime}\lambda^{k^{\prime}}}{k^{\prime}!} \langle \hat
{m}^{k^{\prime}} \rangle\right)
=\sum_{k=0}^{\infty}\frac{i^k\lambda^{k}}{k!} \langle
\hat{s}_{-}\hat{m}^{k} \rangle _{{\rm c}}\exp\left(
\sum_{k^{\prime }=1}^{\infty}
\frac{i^{k^\prime}\lambda^{k^{\prime}}}{k^{\prime}!} \langle \hat
{m}^{k^{\prime}} \rangle _{{\rm c}}\right)  { ,} \label{genfc}
\end{equation}
which factors out the correlations $\langle \hat{m}^{k^{\prime}}\rangle_{\rm c}$ within ${\rm M}$. The latter
correlations are the same as at the initial time, so that we shall derive the correlations between ${\rm S}$ and ${\rm M}$ from

\begin{equation}
\mytext{\textcurrency genfc2\textcurrency \qquad}
\sum_{k=0}^{\infty}
\frac{i^k\lambda^{k}}{k!} \langle \hat{s}_{-}\hat{m}^{k} (  t )
 \rangle _{{\rm c}}=r_{\uparrow\downarrow} (  0 )  \frac
{\Psi_{\uparrow\downarrow} (  \lambda,t )  }{\Psi_{\uparrow
\downarrow} (  \lambda,0 )  }{ .} \label{genfc2}
\end{equation}

For correlations involving not too many spins (we will discuss
this point in \S~\ref{section.5.3.2}), we can again replace the summation over
$m$ in (\ref{genf}) by an integral. Since $P_{\rm M}(m,0)$ is a Gaussian, the sole non-trivial cumulant
$ \langle \hat{m}^{k}\rangle _{{\rm c}}$ is
$ \langle \hat{m}^{2}\rangle =\Delta m^{2}$, given by
(\ref{Pm0}), (\ref{DELTAm}), and we get from (\ref{genf}) and
(\ref{genfc2})
\begin{equation}
\mytext{\textcurrency genfc3\textcurrency \qquad}
\sum_{k=0}^{\infty}\frac{i^k\lambda^{k}}{k!}\langle \hat{s}_{-}\hat{m}^{k} (t)\rangle _{{\rm c}}
=r_{\uparrow\downarrow} (0)\exp\left( -\frac{t^2}{\tau_{{\rm trunc}}^2}-\sqrt{2}\frac{t}{\tau_{{\rm trunc}}}\lambda\Delta m\right)
=r_{\uparrow\downarrow} (t)\exp\left(  -\sqrt{2}\frac{t}{\tau_{{\rm trunc}}}\lambda\Delta m\right){ .} \label{genfc3}
\end{equation}

At first order in $\lambda$, the correlations between ${\rm S}$ and any
single spin of ${\rm M}$\ are thus expressed by
\begin{eqnarray}
\mytext{\textcurrency corr1\textcurrency \qquad}
\langle \hat{s}_{x}\hat{\sigma}_{z}^{\left(  n\right)  }\left(  t\right)\rangle  &
=& \left\langle \hat{s}_{x}\hat{m}\left(  t\right)  \right\rangle _{{\rm c}
}=\sum_{m}C^{\rm dis}_{x}\left(  m,t\right)  m=\sqrt{2}\frac{t}{\tau_{{\rm trunc}}
}\langle \hat{s}_{y}\left(  t\right) \rangle \Delta m=\sqrt
{2}\frac{t}{\tau_{{\rm trunc}}}\langle \hat{s}_{y}\left(  0\right)\rangle e^{-\left(  t/\tau_{{\rm trunc}}\right)  ^{2}}\Delta m{
,}\nonumber\\
\langle \hat{s}_{y}\hat{\sigma}_{z}^{\left(  n\right)  }\left(  t\right)\rangle  & 
 =& \langle \hat{s}_{y}\hat{m}\left(  t\right)\rangle _{{\rm c}}=
 \sum_{m}C^{\rm dis}_{y}\left(  m,t\right)  m=-\sqrt{2}
\frac{t}{\tau_{{\rm trunc}}}\left\langle \hat{s}_{x}\left(  t\right)
\right\rangle \Delta m{ ,} \label{corr1}
\end{eqnarray}
 where we used (5.4). These correlations first increase, reach a maximum for $t=\tau_{{\rm trunc}
}/\sqrt{2}$, then decrease along with $\left\langle \hat{s}_{x}\left(
t\right)  \right\rangle $ and $\langle \hat{s}_{y}\left(  t\right)\rangle $ (Fig. 5.1). At this maximum, their values satisfy

\begin{equation}
\mytext{\textcurrency max corr1\textcurrency \qquad}
\frac{\left\langle \hat
{s}_{x}\hat{m}\left(  t\right)  \right\rangle }{\Delta m}=\langle \hat{s}_{y}\left(  t\right)\rangle =\frac{\langle \hat s_{y}\left(
0\right)  \rangle }{\sqrt{e}}{,\qquad\qquad}\frac{\langle
\hat{s}_{y}\hat{m}\left(  t\right)  \rangle }{\Delta m}=-\frac
{\left\langle \hat s_{x}\left(  0\right)  \right\rangle }{\sqrt{e}}{ .}
\label{maxcorr1}
\end{equation}
They do not lie far below the bound yielded by Heisenberg's inequality
\begin{equation}
\mytext{\textcurrency Heist\textcurrency \qquad}
\left\vert \left\langle \hat
{s}_{x}\hat{m}\right\rangle \right\vert ^{2}=\left\vert  \frac
{1}{2i}\langle [  \hat{s}_{y}-\langle \hat{s}_{y}\rangle ,\hat{s}
_{z}\hat{m}]  \rangle \right\vert ^{2}\leq\left(  1-\langle
\hat s_{y}\rangle ^{2}\right)  \Delta m^{2}{ ,} \label{Heis}
\end{equation}
which implies at all times
\begin{equation}
\mytext{\textcurrency Heist2\textcurrency \qquad}
\left(  \frac{2t^{2}}
{\tau_{{\rm trunc}}^{2}}+1\right) \langle \hat{s}_{y}(t) \rangle ^{2}\leq1{ ,} \label{Heis2}
\end{equation}
since the left-hand side of (\ref{Heis2}) is $2/e$ at the maximum of (\ref{corr1}). 

The next order correlations are obtained from (\ref{genfc3}) as $\left(a=x,y\right)  $

\begin{eqnarray}
\mytext{\textcurrency corr2\textcurrency \qquad}
\langle \hat{s}_{a}\hat {m}^{2}\left(  t\right)  \rangle _{{\rm c}}    \equiv\langle \hat{s}_{a}\hat{m}^{2}\left(  t\right)  \rangle -\langle \hat
{s}_{a}\left(  t\right)  \rangle \langle \hat{m}^{2}\rangle =-\frac{2t^{2}}{\tau_{{\rm trunc}}^{2}}\langle \hat{s}_{a}\left(t\right)  \rangle \Delta m^{2}{ .} 
\label{corr2}
\end{eqnarray}
These correlations again increase, but more slowly than (\ref{corr1}), reach 
(in absolute value) a maximum later, at $t=\tau_{{\rm trunc}}$, equal to $\left(  -2/e\right)$$\langle \hat{s}_{a}\left(  0\right)  \rangle \Delta m^{2}$, 
then decrease together with $\langle \hat{s}_{a}\left(  t\right)
\rangle $. Accordingly, the correlations between $\hat{s}_{x}$ and two
spins of ${\rm M}$, evaluated as in (\ref{sigmanp}), are given by

\begin{equation}
\mytext{\textcurrency corr2sigma\textcurrency \qquad}
\langle \hat{s}
_{x}\hat{\sigma}_{a}^{\left(  n\right)  }\hat{\sigma}_{b}^{\left(  p\right)
}\left(  t\right)  \rangle _{{\rm c}}=\langle \hat{s}_{x}\left(
t\right)  \rangle \frac{\delta_{a,z}\delta_{b,z}}{N-1}\left(
-\frac{2t^{2}}{\tau_{{\rm trunc}}^{2}}N\Delta m^{2}-1\right)  { ,}
\label{corr2sigma}
\end{equation}
which for large $N$ behaves as (\ref{corr2}).

Likewise, (\ref{genfc3}) together with (\ref{rofft}) provides the hierarchy of 
correlations through the real and imaginary parts of

\begin{equation}
\mytext{\textcurrency corr\textcurrency \qquad}
\langle (  \hat{s}_{x}-i\hat{s}_{y})  \hat{m}^{k}\left(  t\right)  \rangle
_{{\rm c}}=\langle(  \hat{s}_{x}-i\hat{s}_{y})  \left(0\right)  \rangle 
\left(  i\sqrt{2}\frac{t}{\tau_{{\rm trunc}}}\Delta
m\right)  ^{k}e^{-(t/\tau_{\rm trunc})^2} { ,} \label{corr}\label{corrk}
\end{equation}
 with $\Delta m$ from (\ref{delta0=}).
This expression also holds for more detailed correlations such as $\langle
\hat{s}_{a}\hat{\sigma}_{z}^{\left(  1\right)  }\hat{\sigma}_{z}^{\left(
2\right)  }{ }\cdots{ }\hat{\sigma}_{z}^{\left(  k\right)  }\left(
t\right)  \rangle _{{\rm c}}$ within corrections of order $1/N$ as in
eq. (\ref{corr2sigma}), provided $k/N$ is small.

Altogether (Fig. 5.1) the correlations (\ref{corr}) scale as 
$\Delta m^{k}=\left(  \delta_{0}/\sqrt{N}\right)  ^{k}$. If the rank $k$ is odd, $\langle \hat{s}
_{x}\hat{m}^{k}\left(  t\right)  \rangle _{{\rm c}}$ is proportional
to $\langle \hat{s}_{y}\left(  0\right)  \rangle $, if $k$ is even,
it is proportional to $\langle \hat{s}_{x}\left(  0\right)  \rangle
$, with alternating signs. The correlations of rank $k$ depend on time as
$\left(  t/\tau_{{\rm trunc}}\right)  ^{k}\exp[{-\left(  t/\tau_{{\rm trunc}
}\right)  ^{2}}]$. Hence, correlations of higher and higher rank begin to grow
later and later, in agreement with the factor $t^{k}$, and they reach a maximum later and later, 
at the time $t=\tau_{{\rm trunc}}\sqrt{k/2}$. For even $k$, the maximum of 
$\left\vert \langle \hat{s}_{x}\hat{m}^{k}\left(  t\right) \rangle _{{\rm c}}\right\vert $ is given by
\begin{equation}
\mytext{\textcurrency maxcorx\textcurrency \qquad}
\max\left\vert \frac
{\langle \hat{s}_{x}\hat{m}^{k}\left(  t\right)\rangle
_{{\rm c}}}{\langle \hat{s}_{x}\left(  0\right)  \rangle
\Delta m^{k} }\right\vert =\frac{1}{k!}\left(
\frac{2k}{e}\right)  ^{k/2}\left(  \frac{k}{2}\right)  !\simeq\frac{1}{\sqrt{2}}{ ,} \label{maxcorx}
\end{equation}
which is nearly independent of $k$.

\myskipfigText{
\begin{figure} \label{figABN5}
\centerline{\includegraphics[width=8cm]{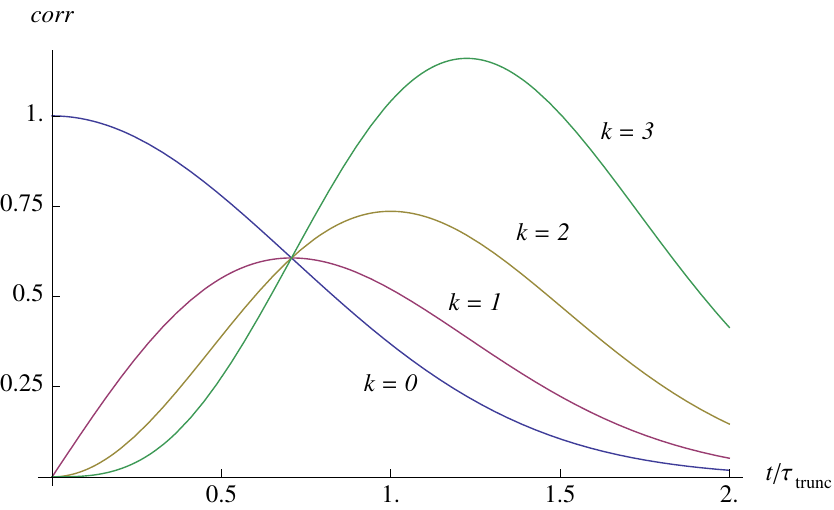}}
\caption{The relative correlations $corr =\langle(\hat s_x-i\hat s_y)\hat m^k(t)\rangle_{\rm c}
/\langle(\hat s_x-i\hat s_y)(0)\rangle(i\sqrt{2}\Delta m)^k$ from Eq. (\ref{corrk}), 
as function of $t/\tau_{\rm trunc}$.
For $k=0$,  $\langle\hat s_x(t)\rangle$ decreases as a Gaussian.
The curves for $k=1, 2$ and $3$ show that the correlations develop, 
reach a maximum, then disappear later and later.}
\end{figure}
}

}
\subsubsection{The truncation, a cascade process}
\label{section.5.1.4}

\hfill{\it Het viel in gruzelementen\footnote{ It fell and broke into tiny pieces}}

\hfill{Dutch saying}
\vspace{3mm}
\ZeText{

The mechanism of truncation in the present model is therefore
comparable to a current mechanism of irreversibility in
statistical mechanics (\S~\ref{section.1.2.2}). In a classical Boltzmann gas,
initially off-equilibrium with a non-uniform density, the
relaxation toward uniform density takes place through the
establishment of correlations between a larger and larger number
of particles, under the effect of successive collisions \cite{mayer_mayer,krylov,Callen}. Here, similar features occur although
quantum dynamics is essential. The relaxation (\ref{stransv}) of
the off-diagonal elements $r_{\uparrow\downarrow}=r_{\downarrow\uparrow}^{\ast}$ of the marginal state $\hat{r} $
of ${\rm S}$\ is accompanied by the generation, owing to the
coupling $\hat {H}_{{\rm SA}}$, of correlations between
${\rm S}$ and ${\rm M}$.

Such correlations, absent at the initial time, are {\it built up and fade out in a cascade}, as shown by eq. (\ref{corr}) and Fig. 5.1. 
Let us characterize the state $\hat{R}$\ of ${\rm S}+{\rm M}$\ by the
expectation values and correlations of the operators $\hat{s}_{a}$ and
$\hat{\sigma}_{a}^{\left(  n\right)  }$. The order of
${\rm S}$, initially embedded in the expectation values of the transverse
components $r_{\uparrow\downarrow}\left(  0\right)  $ of the spin
$\mathbf{\hat{s}}$, is progressively transferred to correlations (\ref{corr1})
between these components and one spin of ${\rm M}$, then in turn to
correlations (\ref{corr2sigma}) with two spins, with three spins, and so on.
The larger the rank $k$ of the correlations,  the smaller they are, as $\Delta
m^{k}\sim 1/N^{k/2}$ (Eq. (\ref{corrk})); but the larger their number is, as
$N!/k!\left(  N-k\right)!\approx N^k/k!  $. Their time-dependence, in $t^{k}\exp{[-\left(
t/\tau_{{\rm trunc}}\right)  ^{2}]}$, shows how they blow up and blow out successively.

As a specific feature of our model of quantum measurement, the interaction
process does not affect the marginal statistical state of ${\rm M}$. All
the multiple correlations produced by the coupling $\hat{H}_{{\rm SA}}$ lie
astride ${\rm S}$ and ${\rm M}$.

Truncation, defined as the disappearance of the off-diagonal blocks ${\hat {\cal R}}_{\uparrow\downarrow}={\hat {\cal R}}
_{\downarrow\uparrow}$\ of the full density matrix ${\hat {\cal D}}$ of ${\rm S}+{\rm A}$, or equivalently
 of the expectation values of all operators involving $\hat{s}_{x}$ or $\hat{s}_{y}$, results from the proportionality
of ${\hat {\cal R}}_{\uparrow\downarrow}\left( t\right)$ to $r_{\uparrow\downarrow}\left(  t\right)  $, within a polynomial
coefficient in $t$ associated with the factor $t^{k}$ in the $k$-th rank correlations. Initially, only few among the
$2^{N}\times2^{N}$ elements of the matrix $\hat{R}_{\uparrow\downarrow}\left(  0\right)  $ do not vanish,
those which correspond to $2r_{\uparrow\downarrow}\left(  0\right)
=\langle \hat{s}_{x}\left(  0\right) \rangle-i\langle \hat{s}_{y}\left( 0\right) \rangle $ and to
$P_{\rm M}\left(  m,0\right)  $ given by (\ref{Pm0}). The very many
elements of $\hat{R}_{\uparrow\downarrow}\left(  0\right)  $ which
describe correlations between $\hat{s}_{x}$ or $\hat{s}_{y}$ and
the spins of ${\rm M}$, absent at the initial time, grow, while
an overall factor $\exp[{-(  t/\tau_{{\rm trunc}})^{2}}]$ damps $\hat {R}_{\uparrow\downarrow}\left(  t\right)  $. At
times $\tau_{{\rm trunc}}\ll t\ll\hbar/g$, all elements of
$\hat{R}_{\uparrow\downarrow}\left(  t\right) $\ and hence of ${\hat {\cal R}}_{\uparrow\downarrow}\left(  t\right)$ have become negligibly 
small\footnote{The latter implication follows because the bath contributions cannot raise the S + A correlations}.
In principle, no information is lost since the equations of motion are reversible; in particular, the commutation of $\hat H$ with 
the projections $\hat\Pi_\up$ and $\hat\Pi_\down$, together with the equation of motion (4.1), implies that 
$i \hbar \d (\scriptR_{\up\down} \scriptR_{\down\up}) /\d t = [\hat H, \scriptR_{\up\down}\scriptR_{\down\up}]$, and hence that
tr$_{\rm A} \scriptR_{\up\down}\scriptR_{\up\down}^\dagger$ is constant in time and remains equal to 
$|r_{\up\down}(0)|^2{\rm tr}_{\rm A} [\scriptR(0)]^2$. However, the initial datum $r_{\uparrow \downarrow}\left(  0\right)  $\ gets spread
among very many matrix elements of ${\hat {\cal R}}_{\uparrow\downarrow}$ which nearly vanish, exactly as in the irreversibility paradox (\S~\ref{section.1.2.2}).

If $N$ could be made infinite, the progressive creation of correlations would
provide a rigorous mathematical characterization of the irreversibility of the
truncation process, as for relaxation processes in statistical mechanics.
Consider, for some fixed value of $K$, the set of correlations (\ref{corr})
of ranks $k$ such that $0\leq k\leq K$, including $\langle \hat{s}_{x}\rangle $ and $\langle \hat{s}_{y}\rangle $\ for $k=0$.
All correlations of this set vanish in the limit $N\rightarrow\infty$ for
fixed $t$, since $\tau_{{\rm trunc}}$\ then tends to $0$. (The coupling
constant $g$ may depend on $N$, in which case it should satisfy $Ng^{2}
\rightarrow\infty$.) This property holds even for infinite $K$, {\it provided}
$K\rightarrow\infty$ \textit{after} $N\rightarrow\infty$, a limit which
characterizes the irreversibility. However, such a limit is not uniform: the
reversibility of the underlying dynamics manifests itself through the
finiteness of high-order correlations for sufficiently large $t$\ (\S~\ref{section.5.3.2}).

Anyhow $N$ is not allowed in physics to go to infinity, since the time
$\tau_{{\rm trunc}}$\ would unrealistically vanish. For large but finite $N$,
there is no rigorous qualitative characterization of irreversibility, neither
in this model of measurement nor in statistical mechanics, but the above
discussion remains relevant. In fact, physically, it is legitimate to regard
as equal to zero a quantity which is less than some small bound, and to regard
as unobservable and irrelevant all correlations which involve a number $k$ of
spins exceeding some bound $K$ much smaller than $N$. We shall return to this issue in ~\S~12.2.3.

\clearpage
}

\subsection{Randomness of the initial state of the magnet}
\label{section.5.2}


\hfill{\it  Success isn't how far you got,}

\hfill{\it  but the distance you traveled from where you started}

\hfill{Greek proverb}

\vspace{3mm}

\ZeText{

Initial states ${\hat {\cal R}}\left(  0\right)  $ that can
actually be prepared at least in a thought experiment, such as
the paramagnetic canonical equilibrium distribution of \S~\ref{section.3.3.3},
involve large randomness. In particular, if the initialization
temperature $T_{0}$ is sufficiently large, the state (\ref{purePM}), i. e., $\hat{R}_{\rm M}\left(  0\right)  =2^{-N}\prod_n\hat\sigma_0^{(n)}$,
 is the most disordered statistical state of ${\rm M}$; in such a case,
$P_{\rm M}\left( m,0\right)  $\ is given by (\ref{Pus0}). We explore in
this subsection how the truncation process is modified for other,
less random, initial states of ${\rm M}$.

}
\subsubsection{Arbitrary initial states}
\label{section.5.2.1}
\ZeText{

The derivations of the equations of motion in subsections 4.1 and
4.2 were general, irrespective of the \ initial state. However, in
subsections 4.3 and 5.1 we have relied on the fact that
$\hat{R}_{\rm M}\left(  0\right) $\ depends only on
$\hat{m}$. In order to deal with an arbitrary initial state
$\hat{R}_{\rm M}\left(  0\right)  $, we return to eq.
(\ref{dRij}), where we can as above neglect for very short times
the coupling with the bath. The operators $\hat{R}_{ij}\left(
t\right)  $ and $\hat{H}_{i}$\ in the Hilbert space of
${\rm M}$\ no longer commute because $\hat{R}_{ij}$\ now
involves spin operators other than $\hat{m}$. However, the
probabilities and correlations $P_{ij}\left(  m,t\right)  $\
defined by (\ref{R->P}) still satisfy Eqs. (\ref{BE}) of Appendix B without
right-hand side. Hence the expressions (\ref{solPdiag}) and
(\ref{solPoff}) for $P_{ij}\left(  m\right)  $ at short times hold
\textit{for any initial state} $\hat{R}_{\rm M}\left(
0\right)  $, with $P_{\rm M}\left(  m,0\right)  $ given by$\ {\rm tr}
_{{\rm M}}\hat{R}_{\rm M}\left(  0\right)  \delta_{\hat{m},m}$.

The various expressions (\ref{stransv}), (\ref{slong}), (\ref{taured}),
(\ref{corr1}), (\ref{corr2}), (\ref{corrk}) relied only on the Gaussian shape
of the probability distribution $P_{\rm M}\left(  0,m\right)  $\ associated with the
initial state. They will therefore remain valid for any initial state $\hat
{R}_{\rm M}\left(  0\right)  $\ that provides a narrow distribution
$P_{\rm M}\left(  m,0\right)  $, centered at $m=0$ and having a width $\Delta
m$\ small ($\Delta m\ll1$) though large compared to the
level spacing, viz.  $\Delta m\gg2/N$. Indeed, within corrections of relative order $1/N$, such
distributions are equivalent to a Gaussian. The second condition ($\Delta
m\gg2/N$)\ ensures that $\tau_{{\rm trunc}}$\ is much shorter than $\hbar/g$,
another characteristic time that we shall introduce in \S~\ref{section.5.3.1}.

In fact, the behavior in $1/\sqrt{N}$\ for $\Delta m$\ is generic, so that the
truncation time has in general the same expression (\ref{taured}) as for a
paramagnetic canonical equilibrium state, with $\delta_{0}$ defined by
$\delta_{0}^{2}=N{\rm tr}_{\rm M}\hat{R}_{\rm M}\left(
0\right)  \hat{m}^{2}$. The dynamics of the truncation process described above
holds for most possible initial states of the apparatus: decay of
$\langle \hat{s}_{x}\left(  t\right) \rangle $\ and $\langle\hat{s}_{y}(t)\rangle $; 
generation of a cascade of correlations $\langle \hat{s}_{a}\hat{m}^{k}\left(  t\right)
\rangle $\ of order $\Delta m^{k}$\ between the transverse components of
the spin ${\rm S}$\ and the pointer variable $\hat{m}$; increase, then
decay of the very many matrix elements of $\hat{R}_{\uparrow\downarrow}\left(
t\right)  $, which are small as $\left( \sqrt{2}\, \Delta m\,t/\tau_{{\rm trunc}}\right)
^{k}\exp[{-\left(  t/\tau_{{\rm trunc}}\right)  ^{2}}]$ for $ t\ll\hbar/g$.

In case the initial density operator $\hat{R}_{\rm M}\left(  0\right)  $
is a symmetric function of the $N$ spins, the correlations between $\hat
{s}_{x}$\ or $\hat{s}_{y}$\ and the $z$-components of the individual spins of
${\rm M}$\ are still given by expressions such as (\ref{corr2sigma}).
However, in general, $\hat{R}_{\rm M}\left(  0\right)  $\ no longer
depends on the operator $\hat{m}$\ only; it involves transverse components
$\hat{\sigma}_{x}^{\left(  n\right)  }$\ or $\hat{\sigma}_{y}^{\left(
n\right)  }$, and so does $\hat{R}_{\uparrow\downarrow}\left(  t\right)  $,
which now includes correlations of $\hat{s}_{x}$\ or $\hat{s}_{y} $\ with $x$-
or $y$-components of the spins $\mathbf{\hat{\sigma}}^{\left(  n\right)  }$.
The knowledge of $P_{\uparrow\downarrow}\left(  m,t\right)  $ is in this case
not sufficient to fully determine $\hat{R}_{\uparrow\downarrow}\left(
t\right)  $, since (\ref{R->P}) holds but not (\ref{P->R}).

The proportionality of the truncation time $\tau_{{\rm trunc}}=\hbar/ \sqrt{2}\, N g \Delta m$ to the inverse of the fluctuation $\Delta m$ shows
that the truncation is a disorder effect, since $\Delta m$\ measures the randomness of the pointer variable in the initial state. This is
easy to understand: ${\rm S}$\ sees an effective magnetic field $Ngm$\ which is random through $m$, and it is this very randomness
which causes the relaxation. The existence of such a randomness in the initial state, even though it is small as $1/\sqrt{N}$, is
necessary to ensure the transfer of the initial order embodied in $r_{\uparrow\downarrow }\left(  0\right)  $\ into the cascade of
correlations between ${\rm S}$ and ${\rm M}$\ and to entail a brief truncation time $\tau_{{\rm trunc}}$. Boltzmann's
elucidation of the irreversibility paradox  also relied on statistical considerations about the initial state of a classical gas which will relax to equilibrium.

}
\subsubsection{Pure versus mixed initial state}
\label{section.5.2.2}

\ZeText{

It is therefore natural to wonder whether the truncation of the state would
still take place for pure initial states of ${\rm M}$, which are the least
random ones in quantum physics, in contrast to the paramagnetic state (\ref{purePM}) or  (\ref{Pus0}) which is the most
random one. To answer this question, we first consider the pure state with
density operator
\begin{equation}
\mytext{\textcurrency initial\textcurrency \qquad}
\hat{R}_{\rm M}\left(
0\right)  =\prod\limits_{n=1}^{N}\frac{1}{2}\left(  1+\hat{\sigma}
_{x}^{\left(  n\right)  }\right)  { ,} \label{initialx}
\end{equation}
in which all spins $\mathbf{\hat{\sigma}}^{\left(  n\right)  }$\ point in the
$x$-direction. This initialization may be achieved by submitting ${\rm M}
$\ to a strong field in the $x$-direction and letting it thermalize with a
cold bath ${\rm B}$\ for a long duration before the beginning of the measurement.
The fluctuation of $\hat{m}$\ in the state
(\ref{initialx}) is $1/\sqrt{N}$. Hence, for this pure initial state of
${\rm M}$, the truncation takes place exactly as for the fully disordered
initial paramagnetic state, since both yield the same probability distribution
(\ref{Pus0}) for $m$.

A similar conclusion holds for the most general factorized pure state, with
density operator
\begin{equation}
\mytext{\textcurrency initialu\textcurrency \qquad}
\hat{R}_{\rm M}\left(
0\right)  =\prod\limits_{n=1}^{N}\frac{1}{2}\left(  1+\mathbf{u}^{\left(n\right)  }\cdot
\textrm{ {\boldmath{$\hat\sigma$}}}^{(n)} \right)  { ,} 
\label{initialu}
\end{equation}
where the $\mathbf{u}^{\left(  n\right)  }$ are arbitrary unit vectors pointing in different directions\footnote{The consideration of such a state is academic 
since it would be impossible, even in a thought experiment, to set ${\rm M}$\ in it}. The fluctuation $\Delta m$, then given by

\begin{equation}
\delta_{0}^{2}=N\Delta m^{2}=\frac{1}{N}\sum_{n=1}^{N}\left[  1-\left(
u_{z}^{\left(  n\right)  }\right)  ^{2}\right]  { ,}
\end{equation}
is in general sufficiently large to ensure again the properties of subsection 5.1, which depend on $\hat{R}_{\rm M}\left(  0\right)  $\ only through $\Delta m$.

Incoherent or coherent superpositions of such pure states will yield the same effects. We will return to this point in \S~\ref{fin12.1.4},  noting 
conversely that an irreversibility which occurs for a mixed state is also statistically present in most of the pure states that underlie it.

Quantum mechanics brings in another feature: a given mixed state can be regarded as a superposition of pure states in many different ways. For instance, 
the completely disordered paramagnetic state (\ref{purePM}),  $\hat{R}_{\rm M}\left(  0\right)  =2^{-N}\prod_n\hat\sigma_0^{(n)}$, 
can be described by saying that each spin points at random in the $+z$ or in the $-z$-direction; it can also be described as an incoherent superposition of the
pure states (\ref{initialu}) with randomly oriented vectors $\mathbf{u}^{\left(  n\right)  }$. This ambiguity makes the analysis into pure 
components of a quantum mixed state unphysical (\S~\ref{fin10.2.3}).

Let us stress that the \textit{statistical or quantum nature of the
fluctuations} $\Delta m$ of the pointer variable in the initial state
\textit{is irrelevant} as regards the truncation process. In the most random state (\ref{purePM})
 this fluctuation $1/\sqrt{N}$\ appears as purely statistical; it would be just the same for
\textquotedblleft classical spins\textquotedblright\ having only a
$z$-component with random values $\pm1$. In the pure state (\ref{initialx}),
it is merely quantal; indeed, its value $1/\sqrt{N}$\ is the lower bound
provided by Heisenberg's inequality
\begin{equation}
\Delta m_{y}^{2}\Delta m_{z}^{2}\geq\frac{1}{4}\left\vert \left\langle \left[
\hat{m}_{y},\hat{m}_{z}\right]  \right\rangle \right\vert ^{2}=\frac{1}{N^{2}
}\left\langle \hat{m}_{x}\right\rangle ^{2}
\end{equation}
for the operators $\hat{m}_{a}=N^{-1}\sum_{n}\hat{\sigma}_{a}^{\left(n\right)  }\ $($a=x$, $y$, $z$), with here 
$\Delta m_{y}=\Delta m_{z}=1/\sqrt{N}$, $\left\langle \hat{m}_{x}\right\rangle =1$. Differences between these two
situations arise only at later times, through the coupling $\hat {H}_{{\rm MB}}$ with the bath.

}
\subsubsection{Squeezed initial states}
\label{section.5.2.3}

\hfill{\it He who is desperate will squeeze oil }

\hfill{\it out of a grain of sand}

\hfill{Japanese proverb}

\vspace{3mm}
\ZeText{

There exist states $\hat{R}_{\rm M}\left(  0\right)$, which we will term as ``{\it squeezed}'',  for which the fluctuation $\Delta m$\ is 
of smaller order than $1/\sqrt{N}$. An extreme case in which $\Delta m=0$ is, for even $N$, a pure state in which $N/2$\ spins point in the $+z$-direction, 
$N/2$\ in the $-z$-direction; then $P_{\rm M}\left(  m,0\right) =\delta_{m,0}$. Coherent or incoherent superpositions of such states yield the
same distribution $P_{\rm M}\left(  m,0\right)  =\delta_{m,0}$, in
particular the microcanonical paramagnetic state $\hat{R}_{\rm M}\left(  0\right)= \delta_{\hat m,0} [(N/2)!]^2/ N!$. 
 In all such cases, $m$ and $\Delta m$ exactly vanish so that the Hamiltonian and the initial state of S + M satisfy 
 ($\hat H_{\rm SA} + \hat H_{\rm M}) \hat D(0)=0$, $\hat D(0) (\hat H_{\rm SA} + \hat H_{\rm M} ) =0$. 
 According to Eq. (\ref{dRij}), nothing will happen, both in the diagonal and off-diagonal sectors, until the bath begins to act through 
 the weak terms of the right-hand side. The above mechanism of truncation based on the coupling between S and M thus fails 
 for the states $\hat D(0)$ such that $P_{\rm M}(m,0)=\delta_{m,0}$, whether these states are pure or not.

The situation is similar for slightly less squeezed states in which the fluctuation $\Delta m$ is of the order of the level spacing $\delta m=2/N$, with about
half of the spins nearly oriented in the $+z$-direction and half in the $-z$-direction. 
When the bath ${\rm B}$\ is disregarded, the off-diagonal block $\hat{R}_{\uparrow\downarrow}\left(  t\right)  $\ then evolves, 
as shown by (\ref{solPoff}), on a time scale of order $\hbar/2g$  instead of the much smaller truncation time (\ref{taured}), of order $1/\sqrt{N}$.

In such cases the truncation will appear (contrary to our discussion of \S~\ref{section.5.1.2}) as a phenomenon of the decoherence type, governed indirectly by
${\rm B}$\ through $\hat{H}_{{\rm SA}}$\ and $\hat{H}_{{\rm MB}}$,\ and taking place on a time scale much longer than $\tau_{{\rm trunc}}$.
This circumstance occurs in many models of measurement, see section~\ref{section.2}, in particular those for which S is not coupled with many degrees
of freedom of the pointer. It is clearly the large size of M which is responsible here for the fast  truncation.  We return to this point in \S~\ref{section.8.1.4}.

Here again, we recover ideas that were introduced to elucidate the irreversibility paradox. In a Boltzmann gas, one can theoretically imagine
initial states with a uniform density which would give rise after some time to a macroscopic inhomogeneity \cite{krylov,Callen}. But such states are 
extremely scarce and involve subtle specific correlations. Producing one of them would involve the impossible task of handling the particles one by one.
However, for the present truncation mechanism, the initial states of M such that the off-diagonal blocks of the density matrix fail to decay irreversibly 
are much less exceptional. While the simplest types of preparation of the apparatus, such as setting M in a canonical paramagnetic state through interaction 
with a warm bath (\S~3.3.3), yield a fluctuation $\Delta m$ of order $1/\sqrt {N}$, we can imagine producing squeezed states even through macroscopic means. 
For instance, a microcanonical paramagnetic type of initial state of M could be obtained by separating the sample of $N$ spins into two equal pieces, 
by setting them (using a cold bath and opposite magnetic fields) into ferromagnetic states with opposite magnetizations, and by mixing them again. 
Some spin-conserving interaction can then randomize the orientations before the initial time of the measurement process. 
We can also imagine, as in modern experiments on optical lattices, switching on and off a strong antiferromagnetic interaction to equalize 
the numbers of spins pointing up and down.

}
\subsection{Consequences of discreteness}
\label{section.5.3}

\hfill{{\it Hij keek of hij water zag branden}\footnote{He looked as if he saw water burn, i.e., he was very surprised}}

\hfill{Dutch proverb}

\vspace{3mm}
\ZeText{

 Somewhat surprisingly since $N$ is large, it appears that the discreteness of the pointer variable $m$ has specific implications in the decay of the off-diagonal blocks 
 of the density matrix. We shall later see that such effects do not occur in the diagonal sectors related to registration.

}
\subsubsection{The recurrence time}
\label{section.5.3.1}

\hfill{\it  It's no use going back to yesterday, }

\hfill{\it because I was a different person then}

\hfill{Lewis Carroll, Alice in Wonderland}

\vspace{3mm}
\ZeText{

Although we have displayed the truncation of the state as an irreversible process on the time scale $\tau_{{\rm trunc}}$, the 
dynamics of our model without the bath is so simple that we expect the reversibility of the equations of motion to manifest itself
for finite $N$. As a matter of fact, the irreversibility arises as usual (\S~\ref{section.1.2.2}) from an approximate treatment, justified only
under the conditions considered above: large $N$, short time, correlations of finite order. This approximation, which underlined
the results (\ref{stransv}) and (\ref{genfc3}) of subsections 5.1and 5.2, consisted in treating $m$ as a continuous variable. We
now go beyond it by returning to the expression (\ref{genf}), which is exact if the bath is inactive ($\gamma=0$), and by taking
into account the discreteness of the spectrum of $\hat{m}$.

For $N\gg1$, we can still use for $P_{\rm M}\left(  m,0\right)  $ the Gaussian form
(\ref{Pm0}) based on (\ref{deg}). The generating function (5.13) then reads
\begin{equation}
\mytext{\textcurrency gf=sum\textcurrency \qquad}
\Psi_{\uparrow\downarrow
}\left(  \lambda,t\right)  =r_{\uparrow\downarrow}\left(  0\right)
\sqrt{\frac{2}{\pi}}\frac{1}{N\Delta m}\sum_{m}\exp\left[  -\frac{m^{2}
}{2\Delta m^{2}}+i\pi Nm\frac{t}{\tau_{{\rm recur}}}+i \lambda  m\right]
{ ,} \label{gf=sum}
\end{equation}
where we have introduced the recurrence time
\begin{equation}
\mytext{\textcurrency taurec\textcurrency \qquad}
\tau_{{\rm recur}}
\equiv\frac{\pi\hbar}{2g}=\pi\sqrt{2}\frac{\Delta m}{\delta m}\tau
_{{\rm trunc}}{ .} \label{taurec}
\end{equation}
The values (\ref{eig}) of $m$  that contribute to the sum (5.28) are equally spaced, at distances $\delta m=2/N$.
The replacement of this sum by an integral, which was performed in \S~\ref{section.5.1.3}, is legitimate only if $t\ll \tau_{\rm recur}$ and $|\lambda|\ll N$. 
When the time $t$ increases and begins to approach $\tau_{\rm recur}$ within a delay of order $\tau_{\rm trunc}$, 
the correlations undergo an {\it  inverse cascade}: Simpler and simpler correlations are gradually generated from correlations 
involving a huge number of spins of M. This process is the time-reversed of the one described in \S~5.1.3. 
When $t$ reaches $\tau_{{\rm recur}}$, or a multiple of it, the various terms of (\ref{gf=sum}) add up, instead of
interfering destructively as when $t$ is of order of $\tau_{{\rm trunc}}$. In
fact, the generating function (\ref{gf=sum}) satisfies
\begin{equation}
\mytext{\textcurrency psiper\textcurrency \qquad}
\Psi_{\uparrow\downarrow
}\left(  \lambda,t+\tau_{{\rm recur}}\right)  =\left(  -1\right)  ^{N}
\Psi_{\uparrow\downarrow}\left(  \lambda,t\right)  { ,} \label{psiper}
\end{equation}
so that without the bath the state $\hat{D}\left(  t\right)  $ of
${\rm S}+{\rm M}$\ \textit{evolves periodically}, returning to its
initial expression $\hat{r}\left(  0\right)  \otimes\hat{R}_{
}\left(  0\right)  $ at equally spaced times: the Schr\"odinger cat terms revive.

This recurrence is a quantum phenomenon \cite{krylov,Callen}. It arises from
the discreteness and regularity of the spectrum of the pointer variable
operator $\hat{m}$, and from the oversimplified nature of the model
solved in the present section, which includes only the part
(\ref{HSA}) of the Hamiltonian. We will exhibit in
section~\ref{section.6} two mechanisms which, in less crude models,
modify the dynamics on time scales larger than $\tau_{{\rm trunc}}$\ and
prevent recurrences from occurring. 

The recurrence time (\ref{taurec}) is much longer than the truncation time, since $\Delta m/\delta m=\frac{1}{2}\delta_{0}\sqrt{N}$. Thus, long after the
initial order carried by the transverse components $\left\langle \hat{s}_{x}\right\rangle $\ and $\langle \hat{s}_{y}\rangle $\
of the spin ${\rm S}$\ has dissolved into numerous and weak correlations, {\it this order revives} through an inverse cascade. At
the time $\tau_{{\rm recur}} $, ${\rm S}$\ gets decorrelated from ${\rm M}$, with $r_{\uparrow \downarrow}\left(
\tau_{{\rm recur}}\right)  =\left(  -1\right)^{N}r_{\uparrow\downarrow}\left(  0\right)  $. The memory of the
off-diagonal elements, which was hidden in correlations,  was  only dephased, it was not lost for good, and it emerges back. 
Such a behavior of the transverse components of the spin ${\rm S}$\ is reminiscent of
the behavior of the transverse magnetization in spin echo experiments \cite{spin_echo,spin_echo1,spin_echo2a,spin_echo2b,spin_echo3,spin_echo4}.
By itself it is a dephasing which can cohere again, and will do so unless other mechanisms (see section 6) prevent this.

}
\subsubsection{High-order correlations}
\label{section.5.3.2}

\hfill{\it Vingt fois sur le m\'etier remettez votre ouvrage\footnote{Twenty times on the loom reset your handiwork}}

\hfill{Nicolas Boileau, L'Art po\'etique}

\vspace{3mm}

\ZeText{

We can write $\Psi_{\uparrow\downarrow}\left(  \lambda,t\right)  $ given by
(\ref{gf=sum}) more explicitly, for large $N$, by formally extending the summation over
$m$ beyond $-1$ and $+1$, which is innocuous, and by using Poisson's summation formula, which
reads
\begin{equation}
\mytext{\textcurrency Poisson\textcurrency \qquad}
\sum_{m}f\left(  m\right)
=\frac{N}{2}\sum_{p=-\infty}^{+\infty}\left(  -1\right)  ^{pN}\int
{\rm d}m\,e^{-i\pi Nmp}f\left(  m\right)  { .} \label{Poisson}
\end{equation}
As a result, we get

\begin{equation}
\mytext{\textcurrency psitotal\textcurrency \qquad}
\Psi_{\uparrow\downarrow}\left(  \lambda,t\right)  =r_{\uparrow\downarrow}\left(  0\right) \sum_{p=-\infty}^{+\infty}\left(  -1\right)^{pN}
 \exp\left(  \frac{i \lambda \Delta m}{\sqrt{2}}+i\frac{t-p\tau_{{\rm recur}}}{\tau_{{\rm trunc}}}\right)^{2}{ ,} \label{psitotal}
\end{equation}
which is nothing but a sum of contributions deduced from (\ref{genfc2}),
(\ref{genfc3}) and (\ref{rofft}) by repeated shifts of $t$ (with alternating
signs for odd $N$). This obviously periodic expression exhibits the
recurrences and the corrections to the results of subsections 5.1 and 5.2 due
to the discreteness of $m$.

In fact, $\Psi_{\uparrow\downarrow}\left(  \lambda,t\right)  $ is related to
the elliptic function $\theta_{3}$ \cite{AbramowitzStegun} through

\begin{eqnarray}
\mytext{\textcurrency theta3\textcurrency \qquad}
\frac{\Psi_{\uparrow\downarrow}\left(  \lambda,t\right)  }{r_{\uparrow\downarrow}\left(  0\right)}   &
=&\exp\left(  \frac{i \lambda \Delta m}{\sqrt{2}}+\frac{it}{\tau_{{\rm trunc}}
}\right)  ^{2}\theta_{3}\left[  \frac{1}{2}\left(  i \lambda \delta_{0}^{2}
+\eta+i\pi N^{2}\Delta m^{2}\frac{t}{\tau_{{\rm recur}}}\right)
,\frac{N^{2}\Delta m^{2}}{2}\right] \nonumber\\
&  =&\sqrt{\frac{2}{\pi}}\frac{1}{N\Delta m}\exp\left[  -\eta\left(  \frac{i\pi
t}{\tau_{{\rm recur}}}+\frac{i \lambda }{N}+\frac{1}{2N\delta_{0}^{2}}\right)\right] 
 \theta_{3}\left[  \frac{t}{\tau_{{\rm recur}}}-\frac{i}{N\pi}\left(
i \lambda +\frac{\eta}{\delta_{0}^{2}}\right)  ,\frac{2}{\pi^{2}N^{2}\Delta
m^{2}}\right]  { ,} \label{theta3}
\end{eqnarray}
with $\eta=0$ for even $N$, $\eta=1$ for odd $N$. It satisfies two periodicity properties, (\ref{psiper}) and

\begin{equation}
\mytext{\textcurrency psiper2\textcurrency \qquad}
\Psi_{\uparrow\downarrow
}\left(  \lambda-\frac{2i}{\delta_{0}^{2}},t \right)  =\exp\left({\frac{2\pi it}{\tau_{{\rm recur}}
}+\frac{2i \lambda }{N}+\frac{2}{N\delta_{0}^{2}}}\right)\,\Psi_{\uparrow\downarrow}\left(  \lambda
,t\right)  { .} \label{psiper2}
\end{equation}

According to (\ref{genfc2}) and (\ref{psitotal}) the dominant corrections to
the results of \S~\ref{section.5.1.3} are given for $t\ll\tau_{{\rm recur}}$ by the
terms $p=\pm1$ in $\Psi_{\uparrow\downarrow}\left(  \lambda,t\right)  $\ and
$\Psi_{\uparrow\downarrow}\left(  \lambda,0\right)  $, that is,
\begin{eqnarray}
\mytext{\textcurrency corrmod\textcurrency \qquad}
\langle \hat{s}_{-}\hat{m}^{k}\rangle _{{\rm c}}  &  =&r_{\uparrow\downarrow}\left(
0\right)  \exp\left(-\frac{  t^2}{\tau_{\rm trunc} ^2}\right)\left(  i\sqrt
{2}\Delta m\right)  ^{k}\left[ \left(  \frac{t}{\tau_{{\rm trunc}}}\right)^{k}+(-1)^NA_{k}(t)
\exp\left(-\frac{\tau_{\rm recur}^2}{\tau_{\rm trunc}^2}
\right)\right] , \nn\\
A_{k}(t)& \equiv& 
\left(\frac{t - \tau_{\rm recur}}{\tau_{\rm trunc}}\right)^k \exp \left(\frac{2t \tau_{\rm recur}}{\tau_{\rm trunc}^2}\right) 
+\left(\frac{t + \tau_{\rm recur}}{\tau_{\rm trunc}}\right)^k \exp \left(\frac{-2t \tau_{\rm recur}}{\tau_{\rm trunc}^2}\right) 
+\left[(-1)^{k+1}-1\right] \left(\frac{\tau_{\rm recur}}{\tau_{\rm trunc}}\right)^k. 
\label{corrmod}
\end{eqnarray}
For $t\rightarrow0$, the correction behaves as $t^2$ or $t$ depending on whether $k$ is even or odd,
whereas the main contribution behaves as $t^{k}$. However the coefficient is so small that this
correction is negligible as soon as $t>\tau_{{\rm trunc}}\exp({-\pi^{2}
N\delta_{0}^{2}/2k})$, an extremely short time for $k\ll N$.

We expected the expression (\ref{corrk}) for the correlations to become
invalid for large $k$. In fact, the values of interest for $t$ are of order
$\tau_{{\rm trunc}}$, or of $\tau_{{\rm trunc}}\sqrt{k}$ for large $k$,
since the correlations reach their maximum at $t=\tau_{{\rm trunc}}\sqrt
{k/2}$. In this range, the correction in (\ref{corrmod}) is dominated by the
first term of $A_k(t)$, which is negligibly small provided
\begin{equation}
\left(  \frac{t}{\tau_{{\rm trunc}}}\right)  ^{k}\gg\left(  \frac{\tau_{{\rm recur}}}{\tau_{{\rm trunc}}}\right)  ^{k}
\exp\left[-\frac{\tau_{\rm recur}(\tau_{\rm recur}-2t)}{\tau_{\rm trunc}^2}\right].
\end{equation}
Hence, in the relevant range $t\sim \tau_{{\rm trunc}}\sqrt{k}$, the
expression (\ref{corrk}) for the correlations of rank $k$ is valid provided
\begin{equation}
\mytext{\textcurrency corrvalid\textcurrency \qquad}
k\ll\frac{\pi^{2}N\delta
_{0}^{2}}{2\ln\left(  \tau_{{\rm recur}}/t\right)  }{ ,}
\label{corrvalid}
\end{equation}
but the simple shape (\ref{corrk}) does not hold for correlations between a number $k$ of particles violating (\ref{corrvalid}). 

In fact, when $t$ becomes sizeable compared to $\tau_{\rm recur}$, the generating function (\ref{psitotal}) 
is dominated by the terms $p=0$ and $p=1$. The correlations take, for arbitrary $k$, the form

  \BEQ \label{smkp01}
           \langle\hat s_- \hat m^k\rangle_{\rm c} = r_{\up\down}(0) 
          \left (i \pi \delta_0^2\right)^k \left\{ \left(\frac{t}{\tau_{\rm recur}}\right)^k \exp\left (-\frac{ t^2}{\tau_{\rm trunc}^2}\right) + 
           \left(\frac{\tau_{\rm recur}- t}{\tau_{\rm recur}}\right)^k \exp \left[ - \frac{(\tau_{\rm recur} - t)^2}{\tau_{\rm trunc}^2}\right]\right\}.  
\EEQ
They are all exponentially small for $N\gg1$ since $\tau_{\rm recur}^2/\tau_{\rm trunc}^2$ is large as $N$. The large rank correlations dominate. 
If for instance $t$ is half the recurrence time, both terms of (\ref{smkp01}) have the same size, and apart from the overall exponential 
$\exp(- N \pi^2 \delta_0^2/8)$   the correlations increase with $k$ by the factor $(\pi \delta_0^2/2)^k$, where $\delta_0\ge1$.

}
\renewcommand{\thesection}{\arabic{section}}
\section{Irreversibility of the truncation}
\setcounter{equation}{0}\setcounter{figure}{0}\renewcommand{\thesection}{\arabic{section}.}
\label{section.6}

\hfill{{\it Quare fremuerunt gentes, }

\hfill{\it  et populi meditati sunt inania?}\footnote{Why do the heathen rage, and the people imagine a vain thing?}}

\hfill{Psalm 2}

\vspace{0.3cm}
\ZeText{

The sole consideration of the  interaction between the tested spin S and the pointer M has been sufficient to explain and analyze the truncation of the
state, which takes place on the time scale $\tau_{{\rm trunc}}$, at the very early stage of the measurement process. However this Hamiltonian (Eq. (\ref{HSA}))
 is so simple that if it were alone it would give rise to recurrences around the times $\tau_{{\rm recur}}$, $2\tau_{{\rm recur}}$, ... . In fact the evolution
is modified by other processes, which as we shall see hinder the possibility of recurrence and render the truncation irreversible on any reachable time scale.

}
\subsection{Destructive interferences}
\label{section.6.1}

\hfill{{\it Bis repetita (non) placent}\footnote{Repetitions are (not) appreciated}}

\hfill{ diverted from Horace}
\vspace{0.3cm}

\ZeText{

We still neglect in this subsection the effects of the phonon bath
(keeping $\gamma =0$), but will show that the recurrent behavior exhibited
in \S~\ref{section.5.3.1} is suppressed by a small change in the model, which
makes it a little less idealized.

}
\subsubsection{Spread of the coupling constants}
\label{section.6.1.1}

\ZeText{

When we introduced the interaction (\ref{HSA}) between ${\rm S}$ and ${\rm A}$, we assumed that the coupling constants between the tested spin $\mathbf{\hat{s}}$ and each 
of the spins $\mathbf{\hat{\sigma}}^{\left(  n\right)  }$\ of the apparatus were all the same. However, even though the range of the forces is long compared to the
size of the magnetic dot, these forces can be different, at least slightly.  This is similar to the inhomogeneous broadening effect well known in NMR physics 
\cite{spin_echo,spin_echo1,spin_echo2a,spin_echo2b,spin_echo3,spin_echo4}. We thus replace here $\hat{H}_{{\rm SA}}$\ by the more general interaction
\begin{equation}
\mytext{\textcurrency HSAgn \textcurrency \qquad}
\hat{H}_{{\rm SA}}^{\prime
}=-\hat{s}_{z}\sum_{n=1}^{N}\left(  g+\delta g_{n}\right)  \hat{\sigma}
_{z}^{\left(  n\right)  }{ ,} \label{HSAgn}
\end{equation}
where the couplings $g+\delta g_{n}$\ are constant in time and have the small dispersion
\begin{equation}
\mytext{\textcurrency deltag\textcurrency \qquad}
\delta g^{2}=\frac{1}{N}
\sum_{n=1}^{N}\delta g_{n}^{2}{,\qquad\qquad}\sum_{n=1}^{N}\delta
g_{n}=0{ .} \label{deltag}
\end{equation}

The equations of motion (\ref{dRij}) for $\hat{D}$, the right-hand
side of which we disregard, remain valid, with the effective
Hamiltonian
\begin{equation}
\hat{H}_{i}=-s_{i}\sum_{n}\left(  g+\delta g_{n}\right)  \hat{\sigma}
_{z}^{\left(  n\right)  }-\sum_q\frac{NJ_q}{q}\hat{m}^{q}
\end{equation}
instead of (\ref{Hi}). This Hamiltonian, as well as the initial conditions
$\hat{R}_{ij}\left(  0\right)  =r_{ij}\left(  0\right)  \hat{R}_{\rm M}\left(  0\right)$, 
depends only on the commuting observables $\hat{\sigma}_{z}^{\left(  n\right)  }$. 
Hence the latter property is also satisfied by the operators $\hat{R}_{ij}\left(  t\right)  $\ at all times. 
Accordingly, $\hat{R}_{\uparrow\uparrow}\left(  t\right)  $ and $\hat{R}_{\downarrow
\downarrow}\left(  t\right)  $\ remain constant, and the part $\hat
{H}_{{\rm M}}$\ of $\hat{H}_{i}$ does not contribute to the equation for
$\hat{R}_{\uparrow\downarrow}\left(  t\right)  $, which is readily solved as
\begin{equation}
\mytext{\textcurrency Rrand\textcurrency \qquad}
\hat{R}_{\uparrow\downarrow
}\left(  t\right)  =r_{\uparrow\downarrow}\left(  0\right)  \hat
{R}_{\rm M}\left(  0\right)  \exp\frac{2i}{\hbar}\left(  Ng\hat{m}t+\sum
_{n=1}^{N}\delta g_{n}\hat{\sigma}_{z}^{\left(  n\right)  }t\right)
{ ,} \label{Rrand}
\end{equation}
with $\hat{R}_{\rm M}\left(  0\right)  $\ given in terms of $\hat{m}$\ by (\ref{HF0}).
Notice that here the operator $\hat R_{\uparrow\downarrow}$ does not depend only on $\hat m$.

If $\hat{R}_{\rm M}\left(  0\right)  $ is the most random paramagnetic
state (\ref{purePM}), produced for $q=2$ by initializing the apparatus with $T_{0}\gg J$ or with a strong RF field, 
or for $q=4$ with any temperature higher than the transition, (\ref{Rrand}) takes the form

\begin{equation}
\mytext{\textcurrency Rrand2\textcurrency \qquad}
\hat{R}_{\uparrow\downarrow
}\left(  t\right)  =r_{\uparrow\downarrow}\left(  0\right)  \prod_{n=1}
^{N}\frac{1}{2}\left[  \hat\sigma_0^{(n)}\cos\frac{2\left(  g+\delta g_{n}\right)  t}{\hbar}
+i\hat{\sigma}_{z}^{\left(  n\right)  }\sin\frac{2\left(  g+\delta g_{n}\right)t}{\hbar}\right]  { .} \label{Rrand2}
\end{equation}
The off-diagonal elements of the state of ${\rm S}$\ thus evolve according
to
\begin{equation}
\mytext{\textcurrency r2\textcurrency \qquad}
r_{\uparrow\downarrow}\left(
t\right)  =r_{\uparrow\downarrow}\left(  0\right)  \prod_{n=1}^{N}\cos\frac{2\left(g+\delta g_{n}\right)  t}{\hbar}{ .} 
\label{r2}
\end{equation}
The right-hand side behaves as (\ref{stransv}) for $\delta g\ll g$ as long as
$t$ is of order $\tau_{{\rm trunc}}$. However, it is expected to remain extremely small at
later times since the factors of (\ref{r2}) interfere destructively unless $t$
is close to a multiple of $\pi\hbar/2\left(  g+\delta g_{n}\right)  $ for most
$n$. In particular, the successive recurrences which occurred in \S~\ref{section.4.3.1} at
the times $\tau_{{\rm recur}}$, $2\tau_{{\rm recur}}$, ...\ for $\delta
g=0$ and $\gamma=0$ are now absent provided the deviations $\delta g_{n}$ are
sufficiently large. We thus obtain a permanent truncation if we have at the
time $t=\tau_{{\rm recur}} $

\begin{equation}
1\gg\prod_{n=1}^{N}\cos\frac{ \pi\delta g_{n}}{g}  \approx\prod_{n=1}^{N} e^{-\pi^{2}\delta g_{n}^{2}/2g^{2}}
=e^{-\pi^{2}\sum_n\delta g_{n}^{2}/2g^{2}} =e^{-N\pi^{2}\delta g^{2}/2g^{2}}{,}
\end{equation}
that is,
\begin{equation}
\mytext{\textcurrency conddeltag\textcurrency \qquad}
\frac{\delta g}{g}\gg
\frac{1}{\pi}\sqrt{\frac{2}{N}}{ .} \label{conddeltag}
\end{equation}
Provided this condition is satisfied, all results of subsections 5.1 and 5.2
hold, even for large times. The whole set of correlations $\langle\hat{s}_{-}\hat{m}^{k}\rangle _{{\rm c}}$, first created by the
coupling (\ref{HSAgn}), disappear for not too large $k$ after a time of order
$\tau_{{\rm trunc}}\sqrt{k}$, and \textit{do not revive} as $t$ becomes
larger. As in usual irreversible processes of statistical mechanics, it is
mathematically not excluded that (\ref{r2}) takes significant values around
some values of $t$, if $N$ is not too large and if many deviations $\delta
g_{n}$ are arithmetically related to one another; but this can occur only for
extremely large times, physically out of reach, as shown in \S~6.1.2.

These conclusions hold for an arbitrary initial state (\ref{HF0}). The
expression (\ref{Rrand}) is the product of $\hat{R}_{\uparrow\downarrow
}\left(  t\right) $, as evaluated in section~\ref{section.5} for $\delta g=0$,  by the phase
factor
\begin{equation}
\mytext{\textcurrency factor\textcurrency \qquad}
\prod_{n=1}^{N}\exp\left(\frac{
2i\delta g_{n}\sigma_{z}^{\left(  n\right)  }t}{\hbar}\right)  { .}
\label{factor}
\end{equation}
A generic set of coupling constants satisfying (\ref{deltag}) provides the
same results as if they were chosen at random, with a narrow gaussian
distribution of width $\delta g$. Replacing then (\ref{factor}) by its
expectation value, we find that the whole statistics of ${\rm S}
+{\rm M}$\ (without the bath) is governed by the product of the generating
function (\ref{psitotal}) by \cite{spin_echo}
\begin{equation}
\mytext{\textcurrency damp\textcurrency \qquad}
\prod_{n=1}^{N}\overline
{\exp\left(  2i\delta g_{n}\sigma_{z}^{\left(  n\right)  }t/\hbar\right)
}=e^{-\left(  t/\tau_{{\rm irrev}}^{{\rm M}}\right)  ^{2}}{ ,}
\label{damp}
\end{equation}
which introduces a characteristic decay time
\begin{equation}
\mytext{\textcurrency tauirrev\textcurrency \qquad}
\tau_{{\rm irrev}
}^{{\rm M}}=\frac{\hbar}{\sqrt{2N}\delta g}{ .} \label{tauirrev}
\end{equation}
This damping factor suppresses all the recurrent terms with $p\neq0$ in (\ref{psitotal}) if $\delta g$ satisfies the condition (\ref{conddeltag}).
Since the exponent of (\ref{damp}) is $\left(  \delta g/g\delta_{0}\right)^{2}(t/\tau_{{\rm trunc}})^{2}$, the first correlations
$\langle \hat{s}_{-}\hat{m}^{k}\left(  t\right)\rangle_{{\rm c}}$\ are left unchanged if $\delta g\ll g$, while those of higher
order are overdamped as $\exp(- k \delta g/2g \delta_0)$ for large $k$ since $(  t/\tau_{{\rm trunc}})  ^{2}=k/2 $ at their maximum.

\vspace{3mm}

Thus, the truncation of the state produced on the time scale $\tau
_{{\rm trunc}}$ by the coupling $\hat{H}_{{\rm SA}}^{\prime}$ of eq.
(\ref{HSAgn}), characterized by the decay (\ref{stransv}) of $\left\langle
\hat{s}_{x}\left(  t\right)  \right\rangle $ and $\langle \hat{s}
_{y}\left(  t\right)\rangle $\ and by the time dependence
(\ref{corrk}) of $\langle \hat{s}_{-}\hat{m}^{k}\left(  t\right)
\rangle _{{\rm c}}$, is fully irreversible. The time $\tau
_{{\rm irrev}}^{{\rm M}}$ characterizes this \textit{irreversibility
induced by the magnet} ${\rm M}$ alone, caused by the dispersion
of the constants $g+\delta g_{n}$ which couple $\mathbf{\hat{s}}$ with the
elements $\hat{\sigma}_{z}^{\left(  n\right)  }$ of the pointer variable. If
$\tau_{{\rm irrev}}^{{\rm M}}$ is such that $\tau_{{\rm trunc}}\ll
\tau_{{\rm irrev}}^{{\rm M}}\ll\tau_{{\rm recur}}$, that is, when
(\ref{conddeltag}) is satisfied, the off-diagonal blocks $\hat{R}
_{\uparrow\downarrow}(t)$ of $\hat{D}(t)$ remain negligible on time scales of
order $\tau_{{\rm recur}}$. We will show in \S~\ref{section.6.1.2} that
recurrences might still occur, but at inaccessibly large times.

}
\subsubsection{Generality of the direct damping mechanism}
\label{section.6.1.2}

\hfill{}\footnote{On a rainy day, many people will say: Ask for my water to bathe your chickens}

\vspace{-6.7mm}

\myskipfigText{
\begin{figure}[h!h!h!]
\label{ArmProv}
\hfill{\includegraphics[width=10.5cm]{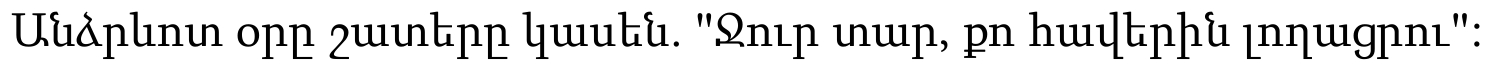}\hspace{2mm}}
\end{figure}}

\vspace{-0.5cm}

\hfill{Armenian proverb}

\vspace{3mm}

\ZeText{

We have just seen that a modification of the direct coupling between the tested spin ${\rm S}$\ and the magnet ${\rm M}$, without any
intervention of the bath, is sufficient to prevent the existence of recurrences after the initial damping of the off-diagonal blocks of $\hat{D}$. In fact,
recurrences took place in \S~\ref{section.5.3.1} only because our original model was peculiar, involving a complete symmetry between the $N$\ spins which
constitute the pointer. We will now show that the mechanism of irreversibility of \S~\ref{section.6.1.1}, based merely on the direct coupling between the
tested system and the pointer of the apparatus, is quite general: it occurs as soon as the pointer presents no regularity.

Let us therefore return to the wide class of models introduced in \S~\ref{section.5.1.2}, characterized by a coupling
\begin{equation}
\mytext{\textcurrency gen2\textcurrency \qquad}
\hat{H}_{{\rm SA}}=-Ng\hat{s}\hat{m} , \qquad \textrm{ (general operators $\hat s$, $\hat m$)}
\label{gen2}
\end{equation}
between the measured observable $\hat{s}$\ of the system ${\rm S}$\ and the pointer observable $\hat{m}$\ of the apparatus ${\rm A}$. We assume that
the pointer, which has $N$\ degrees of freedom, has no symmetry feature, so that the spectrum of $\hat{m}$\ displays neither systematic degeneracies nor
arithmetic properties. We disregard the other degrees of freedom of ${\rm A}$, in particular the indirect coupling with the bath. The model
considered above in \S~\ref{section.6.1.1} enters this general frame, since its Hamiltonian (\ref{HSAgn}) takes the form (\ref{gen2})
if we identify our $\hat s_z$ with the general $\hat s$ and if we redefine $\hat m$ as

\begin{equation}
\mytext{\textcurrency mgen\textcurrency \qquad}
\hat{m}=\frac{1}{N}\sum_{n=1}
^{N}\left(  1+\frac{\delta g_{n}}{g}\right)  \hat{\sigma}_{z}^{\left(
n\right)  }{ .} \label{mgen}
\end{equation}
Indeed, provided the condition (\ref{conddeltag}) is satisfied, the $2^{N}
$\ eigenvalues of (\ref{mgen}) are randomly distributed over the interval
$(-1,1)$\ instead of occurring at the values (\ref{eig}) with the huge
multiplicities (\ref{deg}).

In all such models governed by the Hamiltonian (\ref{gen2}), the off-diagonal
elements of $\hat{r}$\ behave as (\ref{interf}) so that their time-dependence,
and more generally that of the off-diagonal blocks of $\hat{R}$, has the form
\begin{equation}
\label{Ft=}
F\left(  t\right)  =\frac{1}{Q}\sum_{q=1}^{Q}e^{i\omega_{q}t}{ .}
\end{equation}
Indeed, the matrix element (5.11) is a sum of exponentials involving the eigenfrequencies 

\BEQ\label{omq=}
\omega_q \equiv \frac{Ng(s_i-s_j)m_q}{\hbar},
\EEQ
 where $m_q$ are the eigenvalues of $\hat  m$. The number $Q$ of these eigenfrequencies is large as an
exponential of the number $N$ of microscopic degrees of freedom of the
pointer, for instance $Q=2^{N}$ for (\ref{mgen}). To study a generic
situation, we can regard the eigenvalues $m_{q}$\ or the set $\omega_{q}$ as
independent random variables. Their distribution is governed by the density of
eigenvalues of $\hat{m}$\ and by the initial density operator $\hat{R}\left(
0\right)  $\ of the apparatus which enters (\ref{interf}) and which describes
a metastable equilibrium. For sufficiently large $N$, we can take for each
dimensionless $m_{q}$\ a narrow symmetric gaussian distribution, with width of
relative order 1/$\sqrt{N}$. The statistics of $F\left(  t\right)  $\ that we
will study then follows from the probability distribution for the frequencies
$\omega_{q}$,
\begin{equation}
\label{pomq}
p\left(  \omega_{q}\right)  =\frac{1}{\sqrt{2\pi}\Delta\omega}\exp\left(
-\frac{\omega_{q}^{2}}{2\Delta\omega^{2}}\right)  { ,}
\end{equation}
where $\Delta\omega$\ is of order $\sqrt{N}$ due to the factor $N$\ entering the definition (\ref{omq=}) of $\omega_{q}$. 
This problem has been tackled long ago by Kac~\cite{Kac}.

We first note that the expectation value of $F\left(  t\right)  $\ for this
random distribution of frequencies,
\begin{equation}
\mytext{\textcurrency expF\textcurrency \qquad}
\overline{F\left(  t\right)
}=e^{-\Delta\omega^{2}t^{2}/2}{ ,} \label{expF}
\end{equation}
decays exactly, for all times, as the Gaussian (\ref{rofft}) with a truncation
time $\tau_{{\rm trunc}}=\sqrt{2}/\Delta\omega$, encompassing the expression
(\ref{taured}) that we found for short times in our original model. This
result holds for most sets $\omega_{q}$, since the statistical fluctuations
and correlations of $F\left(  t\right)  $,\ given by
\begin{eqnarray}
\overline{F\left(  t\right)  F\left(  t^{\prime}\right)  }-\overline{F\left(
t\right)  }\ \overline{F\left(  t^{\prime}\right)  }  &  =\frac{1}{Q}\left(
e^{-\Delta\omega^{2}\left(  t+t^{\prime}\right)  ^{2}/2}-e^{-\Delta\omega
^{2}\left(  t^{2}+t^{\prime2}\right)  /2}\right), \\
\mytext{\textcurrency corrF\textcurrency \qquad}
\overline{F\left(  t\right)
F^{\ast}\left(  t^{\prime}\right)  }-\overline{F\left(  t\right)  }
\ \overline{F^{\ast}\left(  t^{\prime}\right)  }  &  =\frac{1}{Q}\left(
e^{-\Delta\omega^{2}\left(  t-t^{\prime}\right)  ^{2}/2}-e^{-\Delta\omega
^{2}\left(  t^{2}+t^{\prime2}\right)  /2}\right) , \label{corrF}
\end{eqnarray}
are small for large $Q$.

Nevertheless, for any specific choice of the set $\omega_{q}$, nothing prevents the real part of $F\left(  t\right)$ from reaching significant values at
some times $t$\ large as $t\gg\Delta\omega$, due to the tail of its probability distribution. Given some positive number $f$\ (less than $1$), say $f=0.2$, 
we define the \textit{recurrence time} $\tau_{{\rm recur}}$ as the typical delay we have to wait on average before $\Re F\left(  t\right) $\ rises back up to $f$.
We evaluate this time in Appendix~\ref{AppendixC}. For $f$ sufficiently small so that $\ln[ I_{0}\left(2f\right)]\simeq f^{2}$, a property which holds for $f=0.2$, we find

\begin{equation}  \label{taurecur=}
\tau_{{\rm recur}}=\frac{2\pi}{\Delta\omega}\,\exp\left({Qf^{2}}\right)=\pi\sqrt{2}
\tau_{{\rm trunc}\,}\exp\left({Qf^{2}}\right){ .}
\end{equation}
    
As $Q$\ behaves as an exponential of $N$, this generic recurrence time is \textit{inaccessibly large}. Even for a pointer involving only $N=10$\ spins,
in which case $Q=2^{N}=2^{10}$, and for $f=0.2$, we have $\tau_{{\rm recur}}/\tau_{{\rm trunc}}=2.7\cdot10^{18}$. The destructive interferences taking place
between the various terms of (\ref{interf}) explain not only the truncation of the state (\S~\ref{section.5.1.2})  but also, owing to the randomness of the coupling, 
the irretrievable nature of this decay process over any reasonable time lapse, in spite of the unitarity of the evolution.

Although we expect the eigenfrequencies $\omega_q$ associated with a large pointer to be distributed irregularly, the distribution (\ref{pomq}) chosen above,
 for which they are completely random and uncorrelated, is not generic. Indeed, according to  (\ref{omq=}), these eigenfrequencies are quantum objects,
 directly related to the eigenvalues $m_q$ of the operator $\hat m$. A more realistic model should therefore rely on the idea that $\hat m$ is a complicated operator, 
 which is reasonably represented by a random matrix. As well known, the eigenvalues of a random matrix are correlated: they repell according to Wigner's law. 
 The above study should therefore be extended to {\it random matrices} $\hat m$ instead of random uncorrelated frequencies $\omega_q$, 
 using the techniques of the random matrix theory~\cite{Mehta}.  We expect the recurrence time thus obtained to be shorter than above, 
 due to the correlations among the set $\omega_q$, but still to remain considerably longer than with the regular spectrum of \S~5.3.1. 

}
\subsection{Effect of the bath on the initial truncation}
\label{section.6.2}

\hfill{{\it You can't fight City Hall}}

\hfill{American saying}
\vspace{0.3cm}
\ZeText{

Returning to our original model of subsection~\ref{section.3.2} with a uniform coupling $g$ between ${\rm S}$ and the spins of ${\rm M}$, \ we now take into account
the effect, on the off-diagonal blocks of ${\cal D}$, of the coupling $\gamma$ between ${\rm M}$ and ${\rm B}$. We thus start from eq.
(\ref{dPoff2}), to be solved for times of the order of the recurrence time. We will show that the damping\ due to the bath can prevent 
$P_{\up\down}$ and hence $\hat R_{\up\down}$  from becoming significant at all times $t$ larger than $\tau_{{\rm trunc}}$, in spite of the regularity of the spectrum 
of $\hat{m}$ which leads to the anomalously short recurrence time $\pi\hbar/2g$ of (\ref{taurec})\footnote{For the related, effective decay 
of ${\cal R}_{\uparrow\downarrow}\left(t\right)$ and ${\cal R}_{\downarrow\uparrow}\left(t\right)$, see \S~12.2.3}.

Readers interested mainly in the physics of the truncation may jump to \S~9.6.1, where the mathematics is simplified using 
insights gained about the behavior of the equation of motion for $t \gg \hbar/T$ through the rigorous approach of \S~6.2.1 and of appendix D.

}
\subsubsection{Determination of $P_{\uparrow\downarrow}(t)$}
\label{section.6.2.1}

\ZeText{

We have found recurrences in $P_{\uparrow\downarrow}(m,t)$ by solving
(\ref{dPoff}) without right-hand side and by taking into account the
discreteness of $m$ (\S~\ref{section.5.3.1}). The terms arising from the bath will modify
for each $m$ the modulus and the phase of $P_{\uparrow\downarrow}^{\rm dis}(m,t)=(2/N)P_{\uparrow\downarrow}(m,t)$.

In order to study these changes, we rely on the equation of motion (\ref{dPoff}), the right-hand side of which has been obtained in the large 
$N$ limit while keeping however the values of $m$ discrete as in \S~5.3.1.
 Note first that the functions $\tilde{K}_{t>}(\omega)$ and $\tilde{K}_{t<}(\omega)$ defined by Eqs. (\ref{K>}) and (\ref{K<}), respectively, 
 are complex conjugate for the same  value of $\omega$. 
 It then results from Eq. (\ref{dPoff}) together with its initial condition that 
\footnote{Changing $g$\ into $-g$\ would also change $P_{\uparrow\downarrow}\left(  m,t\right)
$\ into $P_{\uparrow\downarrow}^{\ast}\left(  m,t\right)  $, but we shall stick to the ferromagnetic interaction $g>0$}

\begin{equation}
\mytext{\textcurrency m,i\textcurrency \qquad}
P_{\uparrow\downarrow}(-m,t)=P_{\uparrow\downarrow}^{\ast}(m,t)=P_{\downarrow\uparrow}(m,t){ .}
\label{m,i}
\end{equation}

  For $\gamma=0$, the solution of (\ref{dPoff}) with the initial condition (\ref{purePM}) is given by (5.2). Starting from this expression, we parametrize 
  $P_{\up\down}(m,t)$ as
  
  \BEQ  \label{6.21}
   P_{\up\down}(m,t) =r_{\up\down}(0) \sqrt{\frac{N}{2\pi \delta_0^2}} \, \exp\left [ - \frac{Nm^2}{2\delta_0^2}+ \frac{2i Ngmt}{\hbar} - N A(m,t)\right],  
   \EEQ
in terms of the function $A(m,t)$, to be determined at first order in $\gamma$ from Eq. (\ref{dPoff2}) with the initial condition $A(m,0)=0$. 
For large $N$, $A(m,t)$ contains contributions of orders $1$ and $1/N$. Its complete expression is exhibited in Appendix D in terms of the 
autocorrelation function $K(t)$ of the bath (Eq. (\ref{D.c})).

 The distribution $P_{\up\down}(m,t)$ takes significant values only within a sharp peak centered at $m=0$ with a width of order $1/\sqrt{N}$. 
 We can therefore consistently expand $A(m,t)$ in powers of $m$ up to second order, according to
 
 \BEQ
  A(m,t) \approx B(t) - i\Theta(t) m + \half D(t) m^2,      \label{6.22}
  \EEQ
so that we can write from (\ref{6.21}) and (\ref{6.22}) the expression for $P_{\up\down}^{\rm dis}=(2/N)P_{\up\down}$ in the form

\BEQ
P_{\up\down}^{\rm dis}(m,t) =r_{\up\down}(0) \sqrt{\frac{2}{\pi N \delta_0^2}} 
\exp \left\{- NB(t) + i N\left [\frac{2gt}{\hbar} + \Theta(t)\right]m - N \left[\frac{1}{\delta_0^2}+D(t)\right]\frac{m^2}{2} \right\}.  
\label{6.23}
\EEQ
The functions $B(t)$, $\Theta(t)$ and $D(t)$, proportional to $\gamma$, describe the effect of the bath on the off-diagonal blocks of the density 
matrix of S + M. They are real on account of (\ref{m,i}). The overall factor $\exp[-NB(t)]$ governs the amplitude of  $P_{\up\down}^{\rm dis}$. 
The term $\Theta(t)$ modifies the oscillations which arose from the coupling between S and M.
The term $D(t)$ modifies the width of the peak of $P_{\up\down}^{\rm dis}$. The explicit expressions of these functions, given by  (\ref{D.13a})
for $B(t)$,  (\ref{ThetaRes}) for $\Theta(t)$ and (\ref{DRes}) for $D(t)$, are derived in appendix D from the equation of motion (D.3) for $A(m, t)$, 
which itself results directly from Eq. (\ref{dPoff2}) for $P^{\rm dis}_{\up\down}$. We analyze them below.

\myskipfigText{
\begin{figure} \label{figB(t)}
\centerline{\includegraphics[width=8cm]{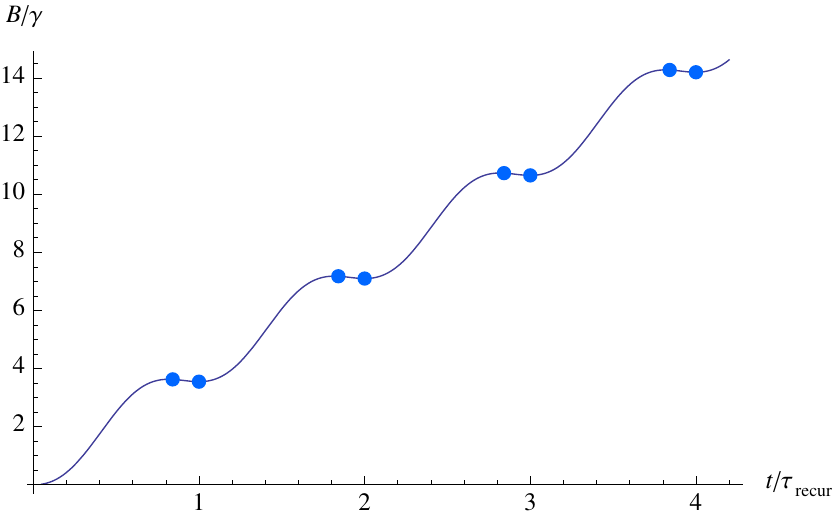}}
\caption{The damping function $B(t)$ issued from the interaction of the magnet with the bath. This function is measured in units of the dimensionless magnet-bath 
coupling constant $\gamma$, and the time is measured in units of the recurrence time $\tau_{\rm recur}=\pi \hbar/2g$. The parameters are 
$T = 0.2 J$ and $g = 0.045 J$ and $\hbar \Gamma=50\sqrt{\pi/2}\,J$. After an initial $t^4$ growth, the curve is quasi linear with periodic oscillations. 
``Anti-damping'' with $\d B/\d t<0$  occurs during the delay $\alpha \tau_{\rm recur}$ before each recurrence (Eq.(6.33)). 
The condition $N B(\tau_{\rm recur})\gg1$ entails the irreversible suppression of all the recurrences.
Bullets denote the local maxima (see (\ref{Bmaxima})) and the local minima at integer values of $t/\tau_{\rm recur}$. 
}
\end{figure}
}

}
\subsubsection{The damping function}

\ZeText{

The main effect of the bath is the introduction in (6.23) of the overall factor $\exp[- N B(t)]$, which produces a damping of the off-diagonal blocks 
$\hat R_{\up\down}$ and $\hat R_{\down\up}$ of the density matrix $\hat D$ of S + M. The expression for $B(t)$ derives from Eq. (\ref{D.f}) and is given explicitly by 

\BEA
B(t) =\gamma \int_{0}^\infty\frac{ \d \omega\, \omega}{\pi} \coth \frac{\hbar\omega}{2T}\,  \exp\left({-\frac{\omega}{\Gamma}}\right)
\left\{ \frac{\sin^2\Omega t }{ 2(\omega^2 - \Omega^2)} +\frac{ \Omega^2 (1 - \cos \omega t \, \cos\Omega t) -\omega\Omega \sin\omega t\, \sin\Omega t }
{ (\omega^2- \Omega^2)^2  } \right\},
 \label{D.8}
 \EEA
 with $\Omega=2g/\hbar$.
The $\omega$-integral can be carried out analytically if one replaces in the spectrum of phonon modes (\ref{Ktilde}) the Debye cutoff by a quasi Lorentzian one, 
see Eq. (\ref{newcutoff}) and the connection (\ref{GamGamtilde}) between the cutoff parameters; the result for $B$ is given in (\ref{D.13a}). 
The function $B(t)$ of Eq.  (\ref{D.13a}), or, nearly equivalently, Eq. (\ref{D.8}), 
 is  illustrated by fig. 6.1. We discuss here its main features in the limiting cases of interest. 
 
Consider first the short times $t\ll 1/\Gamma$. This range covers the delay $\tau_{\rm trunc}$ during which the truncation takes place, but it is much shorter than the 
recurrence time. We show in Appendix D that $B(t)$ behaves for $t\ll1/\Gamma$ as

\BEQ
\label{Bsimt4}
B(t) \sim \frac{\gamma \Gamma^2 g^2}{2 \pi \hbar^2}\,t^4,    
\EEQ              
increasing slowly as shown by fig. 6.1. If $N B(t)$ remains sufficiently small during the whole truncation process so that $\exp[- N B(t)]$ remains close to 1, 
the bath is ineffective over the delay $\tau_{\rm trunc}$. This takes place under the condition

\BEQ \label{6.26cond}
N B(\tau_{\rm trunc}) = N \frac{\gamma \Gamma^2 g^2}{ 2 \pi \hbar^2}\tau_{\rm trunc}^4 = \frac{\gamma \hbar^2 \Gamma^2}{ 8 \pi N\delta_0^4 g^2}  \ll 1, \qquad 
\EEQ
which is easily satisfied in spite of the large value of $\hbar \Gamma/g$, since $\gamma\ll 1$ and $N\gg1$. 
Then the coupling with the bath does not interfere with the truncation by the magnet studied in section 5. 
Otherwise, if $N B(\tau_{\rm trunc})$ is finite, the damping by $B$, which behaves as an exponential of  $- t^4$, enhances the truncation 
effect in $\exp[- (t/\tau_{\rm trunc})^2]$ of M, and reduces the tails of the curves of fig. 5.1.

  Consider now the times t larger than $\hbar/2 \pi T$, which is the memory time of the kernel $K(t)$. We are then in the Markovian regime. 
  This range of times encompasses the recurrences which in the absence of the bath occur periodically at the times $t=p \tau_{\rm recur}$, 
  with $\tau_{\rm recur}=\pi \hbar/2g$. Under the condition $t\gg\hbar/2 \pi T$, we show in Appendix D Eq. (\ref{D.19})), that $B(t)$ has the form

  \BEQ  
  \label{6.26}
  B(t) = \frac{\gamma\pi}{4} \coth\frac{g}{T}\left(\frac{ t}{\tau_{\rm recur}} -\frac{1}{2\pi}  \sin \frac{2 \pi t }{ \tau_{\rm recur}}\right) + 
  \frac{\gamma}{4 \pi}\ln \frac{\hbar \Gamma }{2\pi T} \left( 1- \cos \frac{2 \pi t }{ \tau_{\rm recur}} \right).   
 \EEQ
On average, $B(t)$ thus increases linearly along with the first term of (\ref{6.26}), as exhibited by fig. 6.1. Hence, the bath generates in this region
$t\gg\hbar/2 \pi T$ {\it the exponential damping}

\BEQ                \exp[- N   B(t) ]\sim  \exp\left({- \frac{t}{\tau^{\rm B}_{\rm irrev}}}\right), 
\label{Bexpdec}
\EEQ
where the decay is characterized by the bath-induced irreversibility time

\BEQ                   
 \tau^{\rm B}_{\rm irrev} =\frac{2 \hbar \tanh g/T }{  N \gamma g}.  
\EEQ
The recurrences, at $t=p \tau_{\rm recur}$, are therefore attenuated by the factor

\BEQ           
\exp \left( - \frac{p \tau_{\rm recur}}{\tau^{\rm B}_{\rm irrev}}\right)=\exp \left( -\frac{ p \pi N \gamma}{4 \tanh g/T}\right ).   \label{xxfuture6.31}
\EEQ
Thus, {\it all recurrences are irreversibly suppressed}, so that the initial truncation becomes definitive, provided the coupling between M and B is sufficiently strong so as to satisfy
 $N B(\tau_{\rm recur}) \gg 1$, or equivalently $\tau^{\rm B}_{\rm irrev}\ll \tau_{\rm recur}$, that is:
 
\BEQ \label{gammamin}
\gamma \gg \frac{4 \tanh g/T}{ \pi N} . 
\EEQ
In case $T\gg g$, the irreversibility time

\BEQ \tau^{\rm B}_{\rm irrev} \sim \frac{2\hbar}{ N \gamma T}      \label{6.30}
\EEQ
depends only on the temperature of the bath, on the number of spins of the magnet, and on the magnet-bath coupling, irrespective of the system-magnet coupling.

 In spite of the smallness of $\gamma$, the large value of $N$ makes the condition (\ref{gammamin}) easy to satisfy. 
 In fact, if the hardly more stringent condition $NB(\hbar/2 \pi T)\gg 1$, that is, $N \gamma\gg4 \pi$, is satisfied, we have $NB(t)\gg1$ in the region
 $ t\gg \hbar/2 \pi T$ where the approximation (\ref{6.26}) holds. Thus, although $B(t)$ is quasi linear in this region, the exponential shape of the decay (\ref{Bexpdec}), 
 with its characteristic time $\tau^{\rm B}_{\rm irrev}$, loses physical relevance since $\exp[- NB(t)]$ is there practically zero.

In this same region $t\gg\hbar/2 \pi T$, the expression (\ref{6.26}) of $B(t)$ involves oscillatory contributions superimposed to the linear increase considered above 
(fig. 6.1). In fact, the time derivative

\BEQ                       
\frac{\tau_{\rm recur}}{\gamma}\,\frac{\d B}{\d t} = \left(
\frac{\pi}{2} \coth \frac{g}{T}\sin \frac{\pi t}{\tau_{\rm recur}} 
+  \ln \frac{\hbar \Gamma }{ 2\pi T} \cos \frac{ \pi t }{ \tau_{\rm recur}} \right) \, \sin\frac{\pi t}{\tau_{\rm recur}}.  
\EEQ
of $B(t)$ is periodic, with period $\tau_{\rm recur}$, and it vanishes at the times $t$ such that

\BEQ          
\label{Bminima}
      \sin\frac{\pi t }{ \tau_{\rm recur}} =0  \quad \textrm{or}\quad    \tan\frac{\pi t }{\tau_{\rm recur}} = - \frac{2}{\pi} \ln \frac{\hbar \Gamma}{2\pi T}\, \tanh \frac{g}{T}. 
\EEQ
The first set of zeros occur at the recurrence times $p \tau_{\rm recur}$, which are local minima of $B(t)$. The second set provide local maxima, 
which occur somewhat earlier than the recurrences (fig. 6.1), at the times

\BEQ \label{Bmaxima}
  t = (p - \alpha)\tau_{\rm recur} , \qquad     \alpha=\frac{1}{\pi} \arctan \left( \frac{2}{\pi} \ln \frac{\hbar \Gamma }{ 2\pi T}\, \tanh \frac{g}{T}\right).    
  \EEQ
An unexpected quantum effect thus takes place in the off-diagonal blocks of the density matrix of S + M. Usually, a bath produces a monotonous relaxation. 
Here, the damping factor $\exp[-N B(t)]$, which results from the coupling of M with the bath, {\it increases} between the times $(p - \alpha)\tau_{\rm recur}$ and $p \tau_{\rm recur}$. 
During these periods, the system S + M undergoes an ``{\it anti-damping}''. This has no incidence on our measurement process, since the recurrences are anyhow 
killed under the condition (6.29) and since their duration, $\tau_{\rm trunc}$, is short compared to the delay $\alpha \tau_{\rm recur}$. 
One may imagine, however, other processes that would exhibit a similar effect.
 
}
\subsubsection{Time-dependence of physical quantities}
\label{section.6.2.2}

\ZeText{

 All the off-diagonal physical quantities, to wit, the expectation values $\langle\hat s_x(t)\rangle$,  $\langle\hat s_y(t)\rangle$, and the correlations 
 between $\hat s_x$ or  $\hat s_y$ and any number of spins of the apparatus are embedded in the generating function $\Psi_{\up\down}(\lambda, t)$ 
 defined as in (\ref{genf}).  As we recalled in \S~6.2.1, we must sum over the {\it discrete values} (\ref{eig}) of $m$, rather than integrate over $m$; 
 the distinction between summation and integration becomes crucial when the time $t$ reaches $\tau_{\rm recur}$ , since then the period in $m$ of the 
 oscillations of $P_{\up\down}^{\rm dis}(m, t)$ becomes as small as the level spacing. From (6.23), we see that the characteristic function, 
 modified by the bath terms, has the same form as in 
 \S~5.3.2 within multiplication by $\exp[- NB(t)]$ and within modification of the phase and of the width of $P_{\up\down}^{\rm dis}(m,t)$.
 
 Let us first consider the effect of $\Theta(t)$. Its introduction changes the phase of $P_{\up\down}$ according to
 
 \BEQ                   \frac{2 i N g m t }{\hbar}\mapsto \frac{ 2 i N g m t }{\hbar} + i N \Theta(t) m.     
 \EEQ
Hence, the occurrence of the term $\Theta(t)$ might shift the recurrences, which take place when

\BEQ                    \frac{2 g t }{\hbar} + \Theta(t) = p \pi.   
\EEQ
However, the expression of $\Theta(t)$ derived in the appendix D, Eq. (\ref{ThetaRes}),

\BEQ \Theta(t) \sim  - \frac{\gamma}{8g}\left [ \left(\frac{2 }{ \delta_0^2}  - 1\right) T+ J_2\right ] \left[ 1 -  \cos\frac{2 \pi t }{ \tau_{\rm recur}}\right]. 
\EEQ
vanishes for $t=p \tau_{\rm recur} =p \pi \hbar / 2 g$, so that the replacement (6.34) does not affect the values of the recurrence times. Between these recurrence times,
 the truncation makes all correlations of finite rank negligible even in the absence of the bath, as if $P_{\up\down}^{\rm dis}$ did vanish; then, the phase of 
 $P_{\up\down}^{\rm dis}$ is irrelevant. Altogether, $\Theta (t)$ is completely ineffective.
 
                Likewise, the term $D(t)$ is relevant only at the recurrence times. We evaluate it in Eq. (\ref{DRes}) as

\BEQ                D(p\tau_{\rm recur})  \simeq p\eta,\qquad \eta= \frac{\pi \gamma }{2} \frac{J_2}{g}\left(\frac{J_2}{3T} - 1\right). 
 \EEQ
This term changes the width of the distribution $P_{\up\down}^{\rm dis} (m, t)$ by a small relative amount of order $\gamma\ll1$, according to

\BEQ                
\Delta m = \frac{\delta_0}{ \sqrt{ N}}   \mapsto   \Delta m_p  = \frac{\delta_0}{\sqrt{N(1 +p\eta \delta_0 ^2)} } = \Delta m (1 - \half p\eta \delta_0^2).  
\EEQ
The width therefore increases if $J_2 < 3T$, or decreases if $J_2 > 3T$, but this effect is significant only if 
 the recurrences are still visible, that is, if the condition (\ref{gammamin}) is not satisfied.
 
 The expression (5.33) of the generating function is thus modified into
 
 \BEQ 
 \Psi_{\up\down}(\lambda, t) = r_{\up\down}(0) e^{-NB(t)} 
 \sum_{p=-\infty}^\infty  (-1)^{pN} \exp\left(\frac{i \lambda  \Delta m_p}{\sqrt{2}} +i \frac{t- p \tau_{\rm recur} }{\tau_{\rm trunc}}\right)^2.   
 \label{6.39} 
\EEQ
The crucial change is the presence of the damping factor $\exp[-NB(t)]$, which invalidates the periodicity (5.30) of $\Psi_{\up\down}(\lambda, t)$ and which inhibits the
 recurrences. Moreover, for any $t >0$, the terms $p<0$ in (\ref{6.39}) are negligible, since they involve (for $t=0$) the factor $\exp[- (p \tau_{\rm recur} / \tau_{\rm trunc})^2]$. 
Thus, under the conditions (\ref{6.26cond}) and (\ref{gammamin}), the sum (\ref{6.39}) reduces at all times to its term $p=0$. Accordingly, it is legitimate to express for arbitrary times 
$P_{\up\down}$ as

\BEQ
P_{\up\down}(m, t) = P_{\up\down}(m, 0) \exp \left[\frac{2i Ngmt }{\hbar} - N B(t)\right] ,    
\EEQ                
and to treat $m$ as a continuous variable. As a consequence, the full density matrix of S + M, which results from (\ref{P->R}), has off-diagonal blocks given by
  
  \BEQ              \hat R_{\up\down}(t) = r_{\up\down}(0) \hat R_{\rm M} (0)   \exp \left[\frac{2i Ngmt }{\hbar} - N B(t)\right] , 
  \EEQ
where we recall the expressions (\ref{D.13a}), (\ref{Bsimt4})  and (\ref{6.26}) for $B(t)$. 

 Altogether, as regards the evolution of the physical quantities $\langle\hat s_a \hat m^k (t)\rangle$ ($a=x$ or $y$), nothing is changed in the results of 
 \S~5.1.3 on the scale $t\ll\tau_{\rm recur}$ ; these results are summarized by Eq. (5.22) and illustrated by fig. 5.1.   For $t\gg\tau_{\rm irrev}^{\rm B}$,  
 the factor $\exp[-NB(t)]$ makes all these off-diagonal quantities {\it vanish irremediably}, including the high-rank correlations of \S~5.3.2.         
 
 \vspace{3mm}
 
 In spite of the simplicity of this result, our derivation was heavy because we wanted to produce a rigorous proof. It turned out that the interaction between the spins 
 of M, which occurs both through $\delta_0$ in the initial state of M and through $J_2$ in the dynamics generated by the bath,  has a negligible effect. Taking this 
 property for granted, treating M as a set of independent spins and admitting that for $t\gg\hbar/2 \pi T$ the autocorrelation function of the bath enters 
 the dynamical equation through (\ref{D.22}), we present in \S~9.6.1 a simpler derivation, which may be used for tutorial purposes and which 
 has an intuitive interpretation:  Both the precession of $\hat s$ and the damping of $\hat R_{\up\down}(t)$ by the bath arise from a dynamical process in which 
 each spin of M is independently driven by its interaction with S and independently relaxes under the effect of the bath B.
 

}

\subsubsection{The off-diagonal bath effect,  an ongoing decoherence process regulated by the tested observable}

\label{section.6.2.4}


\hfill{\it  \c Ca s'en va et \c ca revient\footnote{It goes away and back}}

\hfill{Song written by Claude Fran\c cois}

\vspace{3mm}
\ZeText{

The damping described above has two unusual features: on the one hand (fig. 6.1), its coefficient does not monotonically decrease; 
on the other hand, it is governed by a resonance effect. However, it has also clearly
the features of a standard decoherence ~\cite{Schlosshauer,zurek,Guilini,Blanchard,walls,walls_book,Braun}. It takes place 
in the compound system S + M under the influence of B which plays the role of an environment. The decay (\ref{Bexpdec}) is quasi-exponential, apart from non-essential oscillations.
The expression (\ref{6.30}) of the irreversibility time $\tau^{\rm B}_{\rm irrev}=2\hbar/N \gamma T$ (for $T\gg g$) is typical of a bath-induced decoherence: 
It is inversely proportional to the {\it temperature} $T$ of B, to the number $N$ which characterizes the {\it size} of the system S + M, and to the {\it coupling} 
$\gamma$ of this system with its environment, which is here the bath.

Nevertheless, we have stressed (\S~5.1.2) that the fundamental mechanism of the initial truncation of the state $\hat D(t)$ of S + M has not such a status of decoherence. 
It takes place in the brief delay  $\tau_{\rm trunc} = \hbar/\sqrt {N} \delta_0 g$, during which the bath does not yet have any effect.
Contrary to {\it decoherence}, this {\it dephasing}  process is internal to the system S + M, and does not involve its environment B. 
It is governed by the direct coupling $g$ between S and the pointer M, 
as shown by the expression of the truncation time. It is during delays of order $\tau_{\rm trunc}$ that the phenomena described in section 5 occur -- decay of the average 
transverse components of the spin S, creation then disappearance of correlations with higher and higher rank (\S~5.1.3 and fig. 5.1).
 The bath has no effect on this truncation proper.

When the bath begins to act,  that is, when $NB(t)$ becomes significant, 
the truncation can be considered as {\it practically achieved}  since Eq (6.27) is easily satisfied. 
The only tracks that remain from the original blocks $\hat R_{\up\down}(0)$ and $\hat R_{\down\up}(0)$ of $\hat D(0)$ 
are correlations of very high rank (\S~5.3.2), so that the state $\hat D(t)$ cannot be distinguished at such times from a state without off-diagonal blocks. However, if the Hamiltonian 
did reduce solely  to $\hat H_{\rm SA}$ (Eq. (\ref{HSA})), the simplicity of the dynamics would produce, from these hidden correlations, a revival of the initial state $\hat D(0)$, 
taking place just before $\tau_{\rm recur}$ , during a delay of order $\tau_{\rm trunc}$. The weak interaction $\gamma$ with the bath {\it wipes out the high rank correlations}, 
at times $t$ such that $\tau_{\rm trunc}\ll t\ll \tau_{\rm recur}$ for which they are the only remainder of $r_{\up\down}(0)$. 
Their destruction prevents the inverse cascade from taking place and thus  suppresses all recurrences.

The interaction between S and M does not only produce the initial truncation of $\hat D$ described in section 5. 
It is also an essential ingredient in the very mechanism of decoherence by the bath B. 
Indeed, the interaction (\ref{ham4}) between M and B is isotropic, so that it is the coupling between S and M which should govern the selection of the basis in which the suppression 
of recurrences will occur after the initial truncation. To understand how this ongoing {\it preferred basis problem} is solved, let us return to the derivation of the expression (\ref{D.8}) 
for the damping term $B(t)$, valid in the time range of the bath-induced irreversibility. This expression arose from the integral (\ref{D.f}), to wit,

   \BEQ
\frac{ \d B}{\d t} =  \frac{4\gamma 
 \sin \Omega t}{ \pi\hbar^2} \int_{-\infty}^\infty { \d \omega}\,\tilde K(\omega)
\,\frac{  \Omega( \cos \Omega t - \cos \omega t)  }{\omega^2 - \Omega^2}
\label{6.n}
   \EEQ  
which analyzes the influence, on the damping, of the various frequencies $\omega$ of the autocorrelation function $\tilde K(\omega)$ of the phonon bath. 
The effect of the system-magnet interaction $g$ is embedded in the frequency $\Omega=2g/\hbar=\pi/\tau_{\rm recur}$, directly related to the period of the recurrences. 
In appendix D, we show that the quasi-linear behaviour of $B(t)$ results from the approximation (D.20) for the last factor of (\ref{6.n}): 
This factor is peaked around $\omega=\pm\Omega$ for $t\gg \hbar/2 \pi T$. The integral (\ref{6.n}) then reduces to

\BEQ                   \frac{ \d B(t)}{\d t}=\frac{\gamma}{\hbar^2} \left[ \tilde K(\Omega)+\tilde K(- \Omega)\right] (1 - \cos 2 \Omega t) ,
\EEQ
the constant part of which produces the dominant, linear term $B\propto t$ of (\ref{6.26}). In the autocorrelation function $\tilde K(\omega)$
 which controls the damping by B in the equation of motion of S + M, 
$\hbar \omega$ is the energy of the phonon that is created or annihilated by  interaction with a spin of the magnet (\S~3.2.2). 
Thus, through a resonance effect  arising from the peak of the integrand in (6.44), the frequency $\omega$ of the phonons that contribute to the damping adjusts itself 
onto the frequency $\Omega=2g/\hbar$ associated with the precession of the spins of the magnet under the influence of the tested spin. Owing to this {\it resonance effect},
 the bath acts mainly through the frequency of the recurrences. Accordingly, phonons with energy $\hbar \omega$ close to the energy $\hbar\Omega=2g$ of a spin flip in M 
 (see Eq. (\ref{HSA})) are continuously absorbed and emitted, and this produces the shrinking of the off-diagonal blocks $\hat R_{\up\down}$ and 
 $\hat R_{\down\up}$. The effect is cumulative, since $B \propto t$.
   {\it The decoherence by the bath is thus continuously piloted by the coupling of the magnet with} S.

\vspace{3mm}
 
 In conclusion, the initial truncation and its further consolidation are in the present model the results of an interplay between the three interacting objects, S, M and B. 
 The {\it main effect}, on the time scale $\tau_{\rm trunc}$, arises from the coupling between S and the many degrees of freedom of M, and it should not be 
 regarded as decoherence. Rather, it is a {\it dephasing} effect as known in nuclear magnetic resonance. 
 Viewing the magnet M as ``some kind of bath or of environment'', as is often done, disregards the essential role of M: to act as the pointer that indicates the outcome 
 of the quantum measurement. Such an idea also confers too much extension to the concept of bath or environment. 
 Decoherence usually requires some randomness of the environment, and we have seen (\S~5.2.2) that truncation may occur even if the initial state of M is pure.
 
 The mechanisms that warrant, on a longer time scale $\tau^{\rm M}_{\rm irrev}$  or $\tau^{\rm B}_{\rm irrev}$, the permanence of the truncation can be regarded 
 as {\it adjuvants of the main initial truncation process}, since they become active after all accessible off-diagonal expectation values and correlations have (provisionally) 
 disappeared. We saw in subsection 6.1 that the intervention of B is not necessary to entail this irreversibility, which can result from a dispersion of the coupling constants 
 $g_n$. For the more efficient mechanism of suppression of recurrences of subsection 6.2, we have just stressed that it is a decoherence process arising 
 from the phonon thermal bath but {\it steered by the spin-magnet coupling}.
 
 In section 7, we turn to the most essential role of the bath B in the measurement, to allow the registration of the outcome by the pointer.

}

\renewcommand{\thesection}{\arabic{section}}
\section{Registration: creation of system-pointer correlations}
\setcounter{equation}{0}\setcounter{figure}{0}\renewcommand{\thesection}{\arabic{section}.}

\label{section.7}

\hfill{{\it Wie schrijft, die blijft}\footnote{ Who writes, stays}}

\noindent\noindent
\hfill{{\it Les paroles s'envolent, les \'ecrits restent}\footnote{Words fly away, writings stay}}

\hfill{Dutch and French proverbs}

\vspace{0.3cm}
\ZeText{

The main issue in a measurement process is the establishment of correlations between S and A, which will allow us to gain
information on S through observation of A \cite{blokhintsev1,blokhintsev2,deMuynck,BallentineBook,landau}. As shown in \S~\ref{section.5.1.3},
the process creates correlations in the off-diagonal blocks $\hat{{\cal R}}_{\uparrow\downarrow}(t)$ and
$\hat{{\cal R}}_{\downarrow\uparrow}(t)$ of the density matrix $\hat{{\cal D}}(t)$ of S + A, but those which survive after the
brief truncation time $\tau_{{\rm trunc}}$ involve a large number of spins $\mathbf{\hat{\sigma}}^{\left(  n\right)  }$ of M and are
inobservable. The considered quantum measurement thus cannot provide information on the off-diagonal elements
$r_{\uparrow\downarrow}(0)$ of the density matrix $\hat{r}(0)$ of S. We now show, by studying the dynamics of the diagonal blocks of $\hat{{\cal D}}(t)$, 
how M can register the statistical information embedded in $r_{\uparrow\uparrow}(0)$ and $r_{\downarrow \downarrow}(0)$ through creation of
system-apparatus correlations. This ``registration'' concerns a {\it large set of runs} of the measurement and has a statistical nature.
In order to retrieve the information thus transferred {\it  from} S {\it to the pointer} so as to read, print or process it, we need the indication of the pointer to be
 {\it well-defined for each run} (in spite of the quantum nature of A). We discuss this question in section 11.

If we can select the outcome, a question discussed in section 11, 
the process can be used as a preparation of S in the pure state $|\hspace{-1mm} \uparrow\rangle$ or $|\hspace{-1mm} \downarrow\rangle$.

The registration process presents two qualitatively different behaviors, depending on the nature of the phase transition 
of the magnet, of second order if the parameters of its Hamiltonian (\ref{HM=}) satisfy $J_2 > 3J_4$, of first order if they satisfy 
$3J_4 > J_2$. Recalling our discussion in \S~3.3.2, we will exemplify these two situations with the two pure cases $q=2$ and $q=4$. 
In the former case, for  $J_2\equiv J$ and $J_4=0$, the Hamiltonian is expressed by (\ref{HM2=}); in the latter case, for $J_4\equiv J$ and $J_2=0$,
 it is expressed by (\ref{HM4=}). We summarize these two cases by $H_{\rm M}=-(NJ/q)\hat m^q$ with $q=2$ and $4$, respectively.

}
\subsection{Properties of the dynamical equations}
\label{section.7.1}

\vspace{3mm}

\ZeText{

The dynamics of the diagonal blocks
$\hat{{\cal R}}_{\uparrow\uparrow}(t)$ of
$\hat{{\cal D}}(t)$ results for large $N$ from the equation
(\ref{dPdiag2}) for the scalar function $P_{\uparrow\uparrow}(t)$,
with initial condition
$P_{\uparrow\uparrow}(0)=r_{\uparrow\uparrow}(0)P_{\rm M}(m,0)$. The
initial distribution $P_{\rm M}(m,0)$ for the magnetization of M, given by
(\ref{Pm0}), is a Gaussian, peaked around $m=0$ with the small
width $\delta_{0}/\sqrt{N}$. We have noted (subsection 4.4) the analogy of the
equation of motion (\ref{dPdiag2}) with a Fokker-Planck equation \cite{risken}
for the random variable $m$
submitted to the effects of the thermal bath B. In this equation, which reads
\begin{equation}
\mytext{\textcurrency dPdt\textcurrency \qquad}
\frac{\partial P_{\uparrow
\uparrow}}{\partial t}=\frac{\partial}{\partial m}\left(  -vP_{\uparrow\uparrow}\right)  +\frac{1}{N}\frac{\partial^2}{\partial
m^2}\left(  w P_{\uparrow\uparrow}\right)  { ,} \label{dPdt}
\end{equation}
the first term describes a \textit{drift}, the second one a \textit{diffusion}  \cite{risken}. 
The drift velocity $v(m,t)$ is a function of $m$ and $t$
defined by (\ref{vupup}), whereas the diffusion coefficient $w_{\uparrow
\uparrow}(m,t)$ is defined by (\ref{wupup}). The normalization of
$P_{\uparrow\uparrow}$ remains unchanged in time:

\begin{equation}
\mytext{\textcurrency nomP\textcurrency \qquad}
\int\d{m}P_{\uparrow\uparrow
}(m,t)=\int\d{m}P_{\uparrow\uparrow}(m,0)=r_{\uparrow\uparrow}(0){ ,}
\label{nomP}
\end{equation}
so that the ratio $P(m,t)=P_{\uparrow\uparrow}(m,t)/r_{\uparrow\uparrow}(0)$ can be
interpreted as a conditional probability of $m$ if $s_{z}=1$.

}
\subsubsection{Initial and Markovian regimes}
\label{section.7.1.1}

\ZeText{

For very \textit{short times} such that $t\ll1/\Gamma$, we have

\begin{equation}
\tilde{K}_{t}(\omega)\sim2tK(0)=\frac{\hbar^{2}}{4\pi}\Gamma^{2}t{ ,}
\end{equation}
and hence

\begin{equation}
\mytext{\textcurrency vonMarkov\textcurrency \qquad}
v \sim-\frac{\gamma}{\pi}\Gamma^{2}mt{,\qquad}w\sim
\frac{\gamma}{\pi}\Gamma^{2}t\ {.} \label{vonMarkov}
\end{equation}
The solution of (\ref{dPdt}) then provides a Gaussian which remains centered
at $m=0$. Its width $\sqrt{D/N}$ decays for $q=2$, $\delta_{0}>1$ as
\begin{equation}
\mytext{\textcurrency narrowing\textcurrency \qquad}
D(t)=\delta_{0}^{2}
-(\delta_{0}^{2}-1)\left[1-\exp\left(-\frac{\gamma}{\pi}\Gamma^{2}t^{2}\right)\right]{ ,}
\label{narrowing}
\end{equation}
and is constant ($D=\delta_{0}^{2}=1$) for $q=4$. Anyhow, on the
considered time scale, the change in $P_{\uparrow\uparrow}(m,t)$ is not
perceptible since $\gamma\ll1$. The registration may begin to take place only
for larger times.

The weakness of the magnet-bath coupling $\gamma$ implies that the time scale of the registration is larger than the memory time $\hbar/2\pi T$ of $K(t)$.
Then $\tilde K_{t}(\omega)$ defined by (\ref{Ktomega}) reduces to $\tilde K(\omega)$, that is, to (\ref{Ktilde}). The equation of motion (\ref{dPdt})
for $P_{\uparrow\uparrow}$ becomes Markovian \cite{Weiss,Gardiner,petr}, with $v$ and $w$ depending only on $m$ and not  on $t$.  As soon as $t\gg\hbar/2\pi T$,
$P_{\uparrow\uparrow}$ thus evolves in a short-memory regime. Its equation of motion is invariant under time translation.

The explicit expressions (\ref{vupup}) and (\ref{wupup}) of $v_{\uparrow \uparrow}$ and $w$ become in this regime

\begin{eqnarray}
\mytext{\textcurrency Vsm\textcurrency \qquad}
v(m) &
=\gamma\omega_{\uparrow}  (1-m\coth\beta\hbar\omega_{\uparrow}){ ,}\label{Vsm}\\
\mytext{\textcurrency Wsm\textcurrency \qquad}
w(m) &
=\gamma\omega_{\uparrow}(\coth\beta\hbar\omega_{\uparrow}-m){
,}\label{Wsm}
\end{eqnarray}
where $\hbar\omega_{\uparrow}=g+J_2m+J_4m^3$ (including both $q=2$ and $q=4$) from the definition (\ref{omi=}).
These functions contain in fact an extra factor $\exp(-2|\omega_{\uparrow}|/\Gamma)$, which we disregard since the Debye cutoff is large:
\begin{equation}
\mytext{\textcurrency condgamma\textcurrency \qquad}
\hbar\Gamma\gg
g{,\qquad}\hbar\Gamma\gg J{ .}\label{condgamma2}
\end{equation}
While the diffusion coefficient $w(m)$ is everywhere
positive, the drift velocity $v(m)$ changes sign at the
values $m=m_{i}$ that are solutions of (\ref{MF}). We illustrate the behavior
of $v(m)$ in Figs. 7.1 for $q=2$ and 7.2 for $q=4$.

\myskipfigText{
\begin{figure}\label{figABN6}
\centerline{\includegraphics[width=8cm]{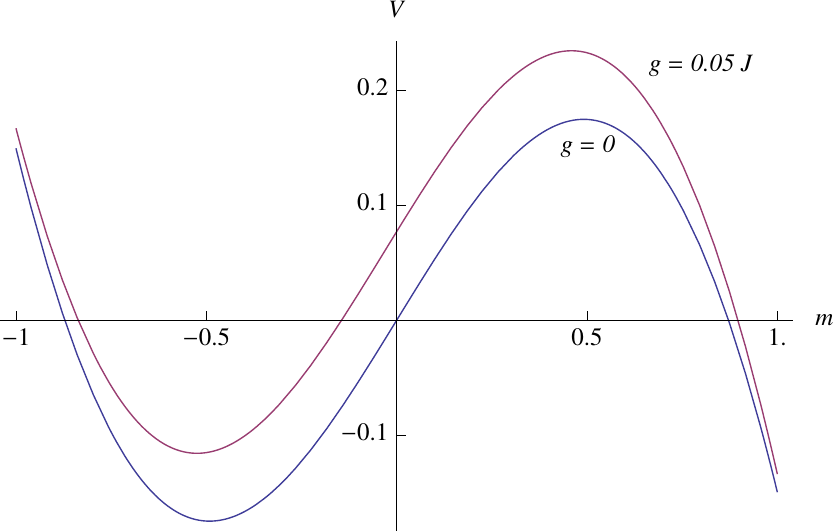}}
\caption{The drift velocity field $V(m)=\hbar v(m)/\gamma T=\beta(Jm+g)[1-m\,{\rm coth}\,\beta(Jm+g)]$ 
for second-order transitions ($q=2$, i. e., $J_2=J$, $J_4=0$), at the temperature $T=0.65J$. The fixed points, the zeroes
 of  $V(m)$, are the extrema of the free energy $F(m)$.  For g=0, the attractive fixed points lie at 
$\pm m_{\rm F}= \pm0.87$. For g=0.05J, the two attractive fixed points lie at $m_\Uparrow=0.90$ 
 and $m_\Downarrow=-0.84$, and the repulsive bifurcation lies at $m= - m_{\rm B}= - 0.14$.
For $g=0$ the attractive fixed points lie at $\pm m_{\rm F}=\pm 0.91$.
}
\end{figure}
}

\myskipfigText{
\begin{figure}\label{figABN7}
\centerline{\includegraphics[width=8cm]{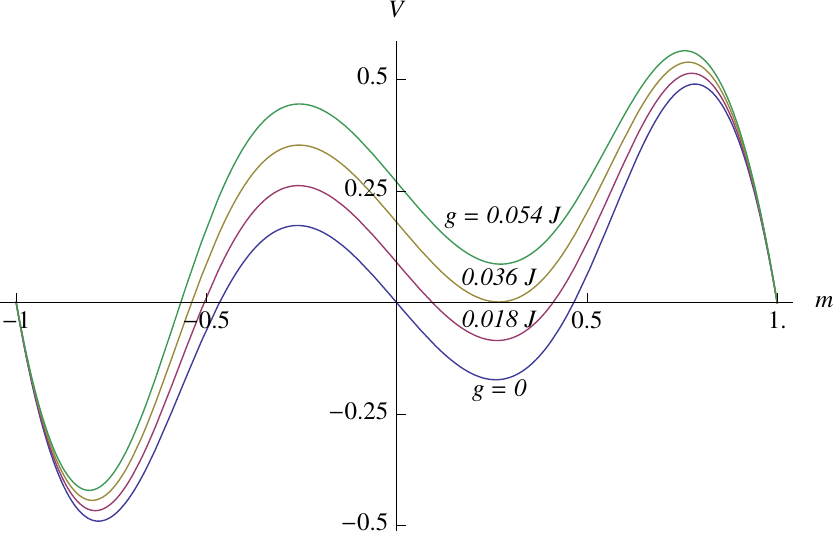}}

\caption{
The drift velocity field $V(m)=\hbar v(m)/\gamma T=\beta(Jm^3+g)[1-m\,{\rm coth}\,\beta(Jm^3+g)]$ for first-order transitions
($q=4$, i. e., $J_2=0$, $J_4=J$) at $T=0.2J$ and for various couplings $g$. The zeroes of $V(m)$ are the extrema of the free energy $F(m)$ 
(see Figs. 3.3 and 3.4). For $g=0$ there are three attractive fixed points, $m_{\rm P}=0$ and $\pm m_{\rm F}$ with $m_{\rm F}=1-9.1\cdot10^{-5}$
 and two repulsive fixed points, at $\pm 0.465$, close to $\pm\sqrt{T/J}=\pm0.447$. 
For increasing $g$, $m_{\rm P}$ increases up to $m_{\rm c}=0.268$ until  $g$ reaches $h_{\rm c}=0.0357\,J$.
 For larger $g$, the paramagnetic fixed point $m_{\rm P}$ disappears together with the positive repulsive point, and, since
 $V$ is positive for all $m>0$, the distribution of $m$ can  easily move from values near $0$ to values near $m_{\rm F}$,
 ``rolling down the hill'' of $F(m)$.
If $g$ is too small, $V(m)$ vanishes with a negative slope at the attractive paramagnetic fixed point $m_{\rm P}$ near the origin;
 the distrbution of $m$ then ends up around $m_{\rm P}$ and the apparatus returns to its paramagnetic state when 
 $g$ is switched off so that the registration fails.
}
\end{figure}
}

}
\subsubsection{Classical features}
\label{section.7.1.2}

\ZeText{

We have stressed (subsection 4.4) that the drift term in (7.1) is ``classical'', in the sense that it comes out for large $N$ by taking the continuous 
limit of the spectrum of $\hat m$, and that the diffusion term, although relevant in this large $N$ limit, results from the discreteness of the spectrum of 
$\hat m$ and has therefore a quantum origin. We can, however, forget this origin and regard this diffusion term as a ``classical'' stochastic effect. 
As a preliminary exercise, we show below that an empirical classical approach of the registration provides us at least with a drift, similar to the one occurring in eq. (7.1).
 
For times $t\gg\tau_{{\rm trunc}}$ it is legitimate to disregard the off-diagonal blocks $\hat{{\cal R}}_{\uparrow\downarrow}$
and $\hat{{\cal R}}_{\downarrow\uparrow}$ of $\hat{{\cal D}}$, and the process that takes place later on
involves only $P_{\uparrow\uparrow}$ and $P_{\downarrow\downarrow}$.  (In our present model the blocks evolve independently anyhow).
This process looks like the measurement of a \textquotedblleft classical discrete spin\textquotedblright\ which would take only two values $+1$ and
$-1$ with respective probabilities $r_{\uparrow \uparrow}(0)$ and $r_{\downarrow\downarrow}(0)$; the $x$- and $y$-components play no role. 
The magnet M also behaves, in the present diagonal sectors, as a collection of $N$ classical spins $\sigma_{z}^{(n)}$, the $x$- and $y$-components of which can be disregarded. 
The dynamics of M is governed by its coupling with the thermal bath B. If this coupling is treated classically, we recover a standard problem in classical statistical 
mechanics \cite{spin_echo,SuzukiKubo, BoltonLeng,GoldsteinScully,Wang}. Indeed, the dynamics of

\BEQ 
P(m,t)=\frac{P_{\uparrow\uparrow}(m,t)}{r_{\uparrow\uparrow}(0)} 
\EEQ
is the same as the relaxation of the random order parameter $m$ of an
Ising magnet, submitted to a magnetic field $h=g$ and weakly
coupled to the bath B at a temperature lower than the transition
temperature. Likewise, $P_{\downarrow
\downarrow}(m,t)/r_{\downarrow\downarrow}(0)$ behaves as the
time-dependent probability distribution for $m$ in a magnetic
field $h=-g$.

Such dynamics have been considered long since, see e.g. \cite{spin_echo,spin_echo1,spin_echo2a,spin_echo2b,spin_echo3,spin_echo4,SuzukiKubo, BoltonLeng,GoldsteinScully,Wang}. 
The variables $\sigma_{z}^{(n)}$ are regarded as $c$-numbers, which can take the two values $\pm1$. Due to the presence of transverse spin components at the quantum level
they may flip with a transition rate imposed by the bath. Since $N$ is large, it seems natural to assume that the variance of $m$ remains weak at all times, as
$D(t)/N$. (In fact, this property fails in circumstances that we shall discuss in subsection~\ref{section.7.3}.) The probability distribution $P$ is then equivalent to a Gaussian,

\begin{equation}
\mytext{\textcurrency Gauss\textcurrency \qquad}
P(m,t)= \sqrt{\frac{N}{2\pi D(t)}}\exp\left\{  -\frac{N[m-\mu(t)]^{2}}{2D(t)}\right\}  { ,}
\label{Gauss}
\end{equation}
In the present classical approximation we neglect $D$, assuming that $m$ is nearly equal to the expectation value $\mu(t)$. 
This quantity is expected to evolve according to an equation of the form

\begin{equation}
\mytext{\textcurrency dmu\textcurrency \qquad}
\frac{{\rm d}\mu(t)}
{{\rm d}t}=v(\mu(t)){ .} \label{dmu}
\end{equation}
 In our case $v$ is given by Eq.  (\ref{Vsm}). This type of evolution has been considered many times in the literature. 
In order to establish this law and to determine the form of the function $v$,
most authors start from a balance equation governing the probability that each
spin $\sigma_{z}^{(n)}$ takes the values $\sigma_{i}=\pm1$ (with $i=\uparrow$ or $\downarrow$). 
The bath induces a transition probability $W_{i}(m)$ per unit time, which governs the possible flip of each spin from $\sigma_{i}$ to
$-\sigma_{i}$, in a configuration where the total spin is $\sum_{i}\sigma_{i}=Nm$. A detailed balance property must be satisfied, 
relating two inverse processes, that is, relating $W_i$ and $W_{-i}$; it ensures that the Boltzmann-Gibbs distribution for the magnet at the
temperature of the bath is stationary, to wit,

\begin{equation}
\frac{W_{-i}[m-(2/N)\sigma_{i}]}{W_{i}(m)}=\exp[{-\beta\Delta E_i(m)}]{ ,}\qquad
\end{equation}
where $\Delta E_i\left(  m\right)  $ is the energy brought in by one spin flip
from $\sigma_{i}$ to $-\sigma_{i}$. For large $N$, we have $\Delta E_i=2\sigma_{i}(h+Jm^{q-1})$ 
(which reads for general couplings $\Delta E_i=2\sigma_{i}(h+J_2m+J_4m^3)$), so that $W_{i}(m)$ depends on $\sigma_{i}$ as
\begin{equation}
\mytext{\textcurrency Wim\textcurrency \qquad}
W_{i}(m)=\frac{1}{2\theta
(m)}[1+\tanh\beta\sigma_{i}(h+Jm^{q-1})]{ ,} \label{Wim}
\end{equation}
including a transition time $\theta(m)$ which may depend on $m$ and on the temperature $T=\beta^{-1}$ of ${\rm B}$. 
(Indeed,  $W_{-i}(m)$, obtained from $W_i$ by changing $\sigma_i$ into $\sigma_{-i}=-\sigma_i$, satisfies (7.12).)
As explained in \S~4.4.3, a balance provides the variation during the time $\d t$ of the probabilities $P^{\rm dis}(m,t)$ as function of the flipping probability 
$W_i(m)\d t$ of each spin. The continuous limit then generates, as in the derivation of Eq. (4.30), the drift coefficient  

\begin{equation}
\mytext{\textcurrency vm\textcurrency \qquad}
v(m)=\frac{1}{\theta(m)}[\tanh
\beta(h+Jm^{q-1})-m]{ .} \label{vm}
\end{equation}

Various forms for $\theta(m)$ can be found in the works devoted to this subject; they are based either on phenomenology or on an approximate solution
of models \cite{SuzukiKubo,BoltonLeng,GoldsteinScully,Wang}. In all cases the stable fixed points of the motion (\ref{dmu}), at
which $v(m)$ vanishes, are the values $m_{i}$ given for large $N$ by (\ref{MF}), where the free energy (\ref{F=}) is minimal. However, the
time-dependence of $\mu(t)=\langle m\rangle$ as well as the behavior of higher order cumulants of $m$ depend on the coefficient $\theta(m)$. For instance,
while $\theta$ is a constant in \cite{SuzukiKubo}, it is proportional to $\tanh\beta(h+Jm^{q-1})$ in \cite{GoldsteinScully} and \cite{Wang}; it still
has another form if $v(m)$ is taken to be proportional to $-{\rm d} F/{\rm d}m$. 

In the present, fully quantum approach, which relies on the
Hamiltonian introduced in subsection~\ref{section.3.2}, the drift velocity $v(m)$ has been
found to take the specific form (\ref{Vsm}) in the Markovian regime $t\gg
\hbar/2\pi T$. We can then identify the coefficient $\theta(m)$ of (\ref{vm})
with
\begin{equation}
\mytext{\textcurrency thetam\textcurrency \qquad}
\theta(m)=\frac{\hbar\tanh
\beta(h+Jm^{q-1})}{\gamma(h+Jm^{q-1})}{ .} \label{thetam}
\end{equation}
With this form of $\theta(m)$, which arises from a quantum microscopic theory,
the dynamical equation (\ref{dmu}) keeps a satisfactory behavior when $h$ or
$m$ becomes negative, contrary to the ad hoc choice $\theta(m)\propto
\tanh\beta(h+Jm^{q-1})$. It provides, for $q=2$, as shown in \S~\ref{section.7.3.2}, a
long lifetime for the paramagnetic state, and better low temperature features
than for $\theta(m)=$constant.

Altogether, our final equations for the evolution of the diagonal
blocks of $\hat{{\cal D}}$ are, at least in the Markovian
regime, similar to equations readily found from a classical
phenomenology. However, the quantum starting point and the rather
realistic features of our model provide us unambiguously with the
form (\ref{Vsm}) for the drift velocity, which meets several
natural requirements in limiting cases. The occurrence of Planck's
constant in (\ref{thetam}) reveals the quantum origin of our
classical-like equation. 
Moreover, quantum mechanics is also at the origin of the diffusion term and it provides the explicit form (7.7) for $w$. 
Finally, by varying the parameters of the model, we can discuss the validity of this equation and explore other regimes.

}
\subsubsection{$H$-theorem and dissipation}
\label{section.7.1.3}

\ZeText{

In order to exhibit the dissipative nature of our quantum equations of motion for $P_{\uparrow\uparrow}$ and $P_{\downarrow\downarrow}$ in the Markovian regime, 
we establish here an associated $H$-theorem \cite{risken}. This theorem holds for any Markovian dynamics, with or without detailed balance.
We start from the general, discrete equation (\ref{dPdiag}), valid even for small $N$, where
$\tilde{K}_{t}(\omega)$ is replaced by $\tilde{K}(\omega)$. We consider the probability $P^{\rm dis}(m,t)=(2/N)P(m,t)$,  normalized under summation,
which encompasses $P^{\rm dis}_{\uparrow\uparrow}(m,t)/r_{\uparrow\uparrow}(0)$  for $h=g>0$ and $P^{\rm dis}_{\downarrow\downarrow}
(m,t)/r_{\downarrow\downarrow}(0)$ for $h=-g<0$, and denote as $E({m})=-hN{m}-JNq^{-1}{m}^{q}$ 
the Hamiltonian (\ref{Hi}) with $h=\pm g$.  We associate with $P^{\rm dis}(m,t)$ the time-dependent entropy

\begin{equation}
\mytext{\textcurrency entM\textcurrency \qquad}
S(t)=-\sum_{m}P^{\rm dis}(m,t)\ln\frac{P^{\rm dis}(m,t)}{G(m)}{ ,} \label{entM}
\end{equation}
where the denominator $G(m)$ accounts for the multiplicity (\ref{deg}) of $m$, and the average energy
\begin{equation}
\mytext{\textcurrency enM\textcurrency \qquad}
U(t)=\sum_{m}P^{\rm dis}(m,t)E(m){ .}
\label{enM}
\end{equation}

The time-dependence of the {\it dynamical free energy} $F_{\rm dyn}(t)=U(t)-TS(t)$ is found by inserting
the equations of motion (\ref{dPdiag}) for the set $P(m,t)$ into
\begin{equation}
\frac{{\rm d}F_{\rm dyn}}{{\rm d}t}=\sum_{m}\frac{{\rm d}P^{\rm dis}(m,t)}{{\rm d}
t}\left[  E(m)+T\ln\frac{P^{\rm dis}(m,t)}{G(m)}\right]  { .}
\end{equation}
The resulting expression is simplified through summation by parts, using
\begin{equation}
\sum_{m}\left[  \Delta_{+}f_{1}(m)\right]  f_{2}(m)=\sum_{m}f_{1}
(m)[\Delta_{-}f_{2}(m)]=-\sum_{m}f_{1}(m_{+})[\Delta_{+}f_{2}(m)]{ ,}
\end{equation}
with the notations (\ref{delta+-}). (No boundary term arises here.)  This yields

\begin{eqnarray}
\frac{{\rm d}F_{\rm dyn}(t)}{{\rm d}t}  =-\frac{N\gamma }{\beta\hbar^{2}}\sum_{m}\left[  (1+m_{+})e^{\beta\Delta_{+}E(m)}P^{\rm dis}(m_{+},t)-(1-m)P^{\rm dis}(m,t)\right]
\tilde{K}[\hbar^{-1}\Delta_{+}E(m)]\,\Delta_{+}\left[  \ln
\frac{P^{\rm dis}(m,t)e^{\beta E(m)}}{G(m)}\right]  { ,}
\end{eqnarray}
where we used $\tilde{K}(-\omega)=\tilde{K}(\omega)\exp{\beta\hbar\omega}$.
Noting that $(1-m)G(m)=(1+m_{+})G(m_{+})$, we find
\begin{eqnarray}
\mytext{\textcurrency dis\textcurrency \qquad}
\frac{{\rm d}F_{\rm dyn}(t)}
{{\rm d}t}   =-\frac{\gamma N}{4\beta\hbar}\sum_{m}(1-m)G(m)\frac
{\Delta_{+}E(m)}{\Delta_{+}\exp\beta E(m)}e^{-|\Delta_{+}E(m)|/\hbar\Gamma}
\Delta_{+}\left[  \frac{P^{\rm dis}(m,t)e^{\beta E(m)}}{G(m)}\right]
\Delta_{+}\left[ \ln \frac{P^{\rm dis}(m,t)e^{\beta E(m)}}{G(m)}\right]  { .}
\label{dis}
\end{eqnarray}

The last two factors in (\ref{dis}) have the same sign, while the previous ones are positive, so that each term in the sum is negative. Thus the dynamical free
energy is a decreasing function of time. The quantity $-\beta{\rm d} F_{\rm dyn}/{\rm d}t$ can be interpreted as the \textit{dissipation rate} (or the entropy production) 
of the compound system M+B, that is, the increase per unit time of the entropy (\ref{entM}) of the magnet plus the increase $-\beta \d U/{\rm d}t$ 
of the entropy of the bath. In fact the entropy of M is lower in the final state than in the initial state, but the
increase of entropy of B associated with the energy dumping dominates the balance. The negativity of (\ref{dis}) characterizes the irreversibility of the registration.

The right-hand side of (\ref{dis}) vanishes only if all its terms vanish, that is, if $P^{\rm dis}(m,t)\exp[{\beta E(m)}]/G(m)$ does not depend on
$m$. This takes place for large times, when the {\it dynamical} free energy $F(t)$ has decreased down to the minimum allowed by the definitions
(\ref{entM}), (\ref{enM}). We then reach the limit $P^{\rm dis}(m)\propto G(m)\exp[{-\beta E(m)}]$, which is the distribution associated with
the canonical equilibrium of M for the Hamiltonian $E(\hat{m})$, that is, with the {\it static free energy}\footnote{The notions of dynamical (moderate time) and static
(infinite time) free energy are well known in the theory of glasses and spin glasses,  see e.g. \cite{Palmer,MPV,LeuzziNieuwenhuizen}.
In corresponding mean field models, they differ strongly; here, however, the dynamical free energy 
simply refers to processes close to equilibrium and decreases down to the static equilibrium free energy in agreement with the macroscopic 
Clausius--Duhem inequality~\cite{Callen,BalianBook}}.
We have thus proven for our model the following property, often
encountered in statistical physics \cite{petr,risken}. The same probability distribution for $m$ arises in two different circumstances. ($i$)
In {\it equilibrium statistical mechanics}, (\S~\ref{section.3.3.4}), $P^{\rm dis}\left( m\right)  $ follows from the {\it Boltzmann-Gibbs distribution} $\hat
{R}_{\rm M}\propto \exp[{-\beta\hat{H}_{{\rm M}}}]$ for the magnet alone. ($ii$) In {\it non-equilibrium statistical mechanics}, it
comes out as the {\it asymptotic distribution} reached in the long time limit when M is weakly coupled to the bath.

It is only in the Markovian regime that the dynamical free energy is ensured to decrease. Consider in particular, for the quadratic coupling $q=2$, the
evolution of $P^{\rm dis}(m,t)$ on very short times, which involves the narrowing (\ref{narrowing}) of the initial peak. The free energy associated with a
Gaussian distribution centered at $m=0$, with a time-dependent variance $D(t)/N$, is

\begin{eqnarray}
\label{Fdynq=2}
F_{\rm dyn}(t)  =\sum_{m}P^{\rm dis}(m,t)\left[  -gNm-\frac{1}{2}JNm^{2}+T\ln\frac{P^{\rm dis}(m,t)}{G(m)}\right]=-\frac{1}{2}(JD+T-TD+T\ln D){ .}
\end{eqnarray}
The time-dependence of $D$ is expressed for short times $t\ll\Gamma^{-1}$ by (\ref{narrowing}). The initial value $\delta_{0}^{2}$ of $D(t)$ being given by
(\ref{delta0=}), we find 

\begin{equation}
\frac{{\rm d}F_{\rm dyn}}{{\rm d}t}=\frac{\gamma\Gamma^{2}t}{\pi}\,\frac
{J^{2}(T_{0}-T)}{T_{0}(T_{0}-J)}{ .}
\end{equation}
Thus at the very beginning of the evolution, $F_{\rm dyn}$ slightly increases, whereas
for $t\gg\hbar/2\pi T$ it steadily decreases according to (\ref{dis}). In
fact, the negative sign of $v$ in the initial non-Markovian
regime (\ref{vonMarkov})\ indicates that, for very short times, the fixed
point near $m=0$ is stable although the bath temperature is lower than $J$.


}

\subsubsection{Approach to quasi-equilibrium}
\label{section.7.1.4}

\ZeText{

The above proof that the system eventually reaches the canonical equilibrium state $\hat{R}_{\rm M}\propto\exp({-\beta\hat{H}_{\rm M}})$ is mathematically correct
for finite $N$ and $t\rightarrow\infty$. However, this result is not completely relevant physically in the large $N$ limit. Indeed, the times that we
consider should be attainable in practice, and \textquotedblleft large times\textquotedblright\ does not mean \textquotedblleft infinite
times\textquotedblright\ in the mathematical sense \cite{krylov,Callen}.

In order to analyze this situation, we note that the summand of (\ref{dis}) contains a factor $P^{\rm dis}(m,t)$; thus the ranges of $m$ over which $P^{\rm dis}(m,t)$ 
is not sizeable should be disregarded. When the time has become sufficiently large so that the rate of decrease of $F(t)$ has slowed down, a regime is reached where
$P^{\rm dis}(m,t)\exp[{\beta E(m)}]/G(m)$ is nearly time-independent and nearly constant (as function of $m$) in any interval where $P^{\rm dis}(m,t)$ is not small. Within a
multiplicative factor, $P^{\rm dis}(m,t)$ is then locally close to $\exp[{-\beta F(m)}]$ where $F(m)=U(m)-T\ln G(m)$ is given by (\ref{F=}). It is thus concentrated in
peaks, narrow as $1/\sqrt{{N}}$ and located in the vicinity of points $m_{i}$ where $F(m)$ has a local minimum. Above the transition temperature, or when
the field $h=\pm g$ is sufficiently large, there is only one such peak, and the asymptotic form of $P^{\rm dis}(m,t)$ is unique. However, below the critical
temperature, two separate peaks may occur for $q=2$, and two or three peaks for $q=4$, depending on the size of $h$.

In such a case, $P^{\rm dis}(m,t)$ can be split into a sum of non-overlapping contributions $P^{\rm dis}_{{\rm M}i}(m,t)$, located respectively near $m_{i}$
and expected to evolve towards the equilibrium distributions $P_{{\rm M}i}^{\rm dis}(m)$ expressed by (\ref{Pim}).  
Since for sufficiently long times  $P^{\rm dis}(m,t)$ is concentrated around its maxima  $m_{i}$ with a shape approaching the Gaussian   (\ref{Pim}),
its equation of motion (\ref{dPdt}) does not allow for transfers from one peak to another
over any reasonable delay. (Delays exponentially large with $N$ are physically inaccessible.) Once such a regime has been
attained, each term $P^{\rm dis}_{{\rm M}i}(m,t)$ \textit{evolves independently} according to (\ref{dPdt}). Its normalization remains constant, and
its shape tends asymptotically to (\ref{Pim}). Hence, below the transition temperature, \textit{ergodicity is broken}
in the physical sense. (A breaking of ergodicity may occur in a mathematically rigorous sense only for infinite N or zero noise.)
If the system starts from a configuration close to some $m_{i}$, it explores, during a physically large time, only the configurations
for which $m$ lies around $m_{i}$. Configurations with the same energy but with values of $m$ around other minima of $F(m)$ remain
out of reach. This phenomenon is essential if we want to use M as the pointer of a measurement apparatus. If the spin S lies
upwards, its interaction with A should lead to values of $m$ that fluctuate weakly around $+m_{{\rm F}}$, not around
$-m_{{\rm F}}$. Ergodicity would imply that A spends the same average time in all configurations having the same energy,
whatever the sign of $m$ \cite{krylov,Callen}, once the interaction $\hat{H}_{{\rm SA}}$ is turned off. The breaking of invariance
is thus implemented through the dynamics: unphysical times, exponentially large with $N$, would be needed to reach the symmetric state $\exp({-\beta\hat {H}_{{\rm M}}})$.

In analogy with what happens in glasses and spin glasses \cite{Palmer,MPV,LeuzziNieuwenhuizen},
for physical large times $t$, the asymptotic value of $F_{\rm dyn}(t)$ is not necessarily the absolute minimum of $F(m)$. It is a weighted average of the
free energies of the stable and metastable states, with magnetizations $m_{i}$. The weights, that is, the normalizations of the contributions $P^{\rm dis}_{{\rm M}i}(m,t)$
to $P^{\rm dis}(m,t)$ are determined by the initial distribution $P^{\rm dis}(m,0)$, and they depend on $N$ and on the couplings $g$ and $J$ which enter the equations of
motion. For an ideal measurement, we require the process to end up at a single peak, $+m_{{\rm F}}$ for $P^{\rm dis}_{\uparrow\uparrow}$, $-m_{{\rm F}}$ for
$P^{\rm dis}_{\downarrow\downarrow}$ (subsection 7.2). Otherwise, if M may reach either one of the two ferromagnetic states $\pm m_{{\rm F}}$,
 the measurement is not faithful; we will determine in \S~\ref{section.7.3.3} its probability of failure.

In the present regime where the variations with $m$ of $Pe^{\beta E}/G$ are slow, we can safely write the continuous limit of the $H$-theorem (\ref{dis})
by expressing the discrete variations $\Delta_{+}$ over the interval $\delta m=2/N$ as derivatives. We then find the dissipation rate as
(we switch to the function $P(m)=(N/2)P^{\rm dis}(m)$ and to an integral over $m$)

\begin{eqnarray}
\mytext{\textcurrency dis2\textcurrency \qquad}
-\frac{1}{T}\,\frac{{\rm d} F_{\rm dyn}}{{\rm d}t}  &  =&\frac{\gamma NT}{\hbar}\int\d m \,P(m,t)\phi
(m)[\coth\phi(m)-m]\nonumber\\
&  \times&\left[  \frac{1}{NP}\frac{\partial P}{\partial m}-\frac{\tanh
\phi(m)-m}{1-m\tanh\phi(m)}\right]  \left[  \frac{1}{NP}\frac{\partial
P}{\partial m}-\phi(m)+\frac{1}{2}\ln\frac{1+m}{1-m}\right]  { ,}
\label{dis2} 
\end{eqnarray}

\noindent 
where we use the notation 

\begin{equation}
\mytext{\textcurrency Phim\textcurrency \qquad}
\phi(m)=\beta(h+Jm^{q-1}
){,\qquad}h=\pm g.
 \label{Phim}
\end{equation}
For large $N$, the term $(1/NP)\d P/\d m$ is not negligible in case $\ln P$ is proportional to $N$,  that is, in the vicinity of a narrow peak with width $1/\sqrt{N}$. 
The expression (\ref{dis2}) is not obviously positive. However, once $P(m,t)=\sum_{i=\pm1} P_{{\rm M}i}(m,t)$ has evolved into a sum 
of separate terms represented by peaks around the values $m_{i}$, we can write the dissipation as a sum of contributions, each of which we expand around $m_{i}$. 
The last two brackets of (\ref{dis2}) then differ only at order $(m-m_{i})^{3}$, and we get the obviously positive integrand

\begin{eqnarray}
&&  -\frac{1}{T}\,\frac
{{\rm d}F_{\rm dyn}}{{\rm d}t}=\frac{\gamma NT}{\hbar}\sum_{i=\pm1}\int\d m\,
P_{{\rm M}i}(m,t)\phi(m)[\coth\phi(m)-m] \\
&  \times&\left\{  \frac{1}{NP_{{\rm M}i}}\frac{\partial P_{{\rm M}i}}{\partial m}+\left[
\frac{1}{1-m_{i}^{2}}-(q-1)\beta Jm_{i}^{q-2}\right]  (m-m_{i})+\left[
\frac{m_{i}}{(1-m_{i}^{2})^{2}}-\frac{(q-1)(q-2)}{2}\beta Jm_{i}^{q-3}\right]
(m-m_{i})^{2}\right\}  ^{2}{ .} \nn \label{dis3}
\end{eqnarray}

We thus check that $F_{\rm dyn}(t)$ decreases, down to the weighted sum of free energies
associated with the stable or metastable equilibrium distributions
(\ref{Pim}). In fact, among the stationary solutions of (\ref{dPdt}), those
which satisfy
\begin{equation}
vP-\frac
{1}{N}\frac{{\rm d}(wP)}{{\rm d}m}=0{ ,}
\label{station}
\end{equation}
with $v$ and $w$ given by (\ref{Vsm})
and (\ref{Wsm}), coincide with (\ref{Pim}) around the values of $m_{i}$ given
by (\ref{MF}), not only in the mean-field approximation but also including the
corrections that we retained in those formulae.

}
\subsection{Registration times}
\label{section.7.2}

\hfill{\it Quid est ergo tempus? Si nemo ex me quaerat, scio;}

\hfill{\it si quaerenti explicare velim, nescio
\footnote{What then is time? If no one asks me, I know what it is; if I wish to explain it to him who asks, I do not know}}


\hfill{Saint Augustine}

\vspace{3mm}

\ZeText{
In the present subsection, we study the evolution of the distribution $P_{\uparrow\uparrow}(m,t)$,
 which is such that $(1/N) \ln P_{\uparrow\uparrow}$ is finite for large $N$. This property holds at $t=0$ and hence at all times. 
 As a consequence, $P_{\uparrow\uparrow}$ presents a narrow peak with width of order $1/\sqrt{N}$, and it is equivalent to a Gaussian. 
 We first note that the evolution (7.1) conserves its normalization $r_{\uparrow\uparrow}(0)$. 
 The ratio (7.9) can then be parametrized as in (7.10) by the position $\mu(t )$ of the peak and by its width parameter $D(t)$, 
 which are both finite for large $N$. 

}

\subsubsection{Motion of a single narrow peak}
\label{section.7.2.1}

\ZeText{

The equations of motion for $\mu(t)$ and $D(t)$ are derived 
 by taking the first moments of the equation (7.1) for $P_{\uparrow\uparrow}(m,t)$. Integration over $m$ of
(\ref{dPdt}) first entails the conservation in time of the normalization
$r_{\uparrow\uparrow}(0)$ of $\int\d{m}P_{\uparrow\uparrow}(m,t)$. We then
integrate (\ref{dPdt}) over $m$ after multiplication, first by $m-\mu(t)$, second by $N[m-\mu(t)]^{2}-D(t)$,
using on the right-hand side an integration by parts and the steepest descents method. To wit, 
expanding $v(m,t)$ and $w(m,t)$ in powers of $m-\mu(t)$, we rely on the vanishing of the integrals of $m-\mu(t)$ and of
$N[m-\mu(t)]^{2}-D(t)$ when weighted by $P_{\uparrow\downarrow}(m,t)$, and we
neglect for $k>1$ the integrals of $[m-\mu(t)]^{2k}$, which are small as
$N^{-k}$. This yields for sufficiently large $N$
\begin{eqnarray}
\mytext{\textcurrency dmu3\textcurrency \qquad}
\frac{{\rm d}\mu(t)}{{\rm d}t} &  =&v[\mu(t),t]{ ,}\label{dmu3}\\
\mytext{\textcurrency dD3\textcurrency \qquad}
\frac{1}{2}\frac{{\rm d}
D(t)}{{\rm d}t} &  =&\frac{\partial v[\mu(t),t]}{\partial\mu}D(t)+w[\mu
(t),t]{ .}\label{dD3}
\end{eqnarray}

At the very beginning of the evolution, when $t$ is not yet large compared to
$\hbar/2\pi T$, Eqs. (\ref{dmu3}) and (\ref{dD3}) should be solved self-consistently, using the
expressions (\ref{vupup}) for $v$ and (\ref{wupup}) for $w$. However, if the
coupling $\gamma$ is weak, the Markovian regime is reached before the shape of
$P_{\uparrow\uparrow}$ is significantly changed. We can thus solve
(\ref{dmu3}) and (\ref{dD3}) with the time-independent forms (\ref{Vsm}) and
(\ref{Wsm}) for $v$ and $w$, the initial conditions being $\mu(0)=0$ ,
$D(0)=\delta_{0}^{2}$.

The solution of (\ref{dmu3}) is then, for $t\gg\hbar/2\pi T$,
\begin{equation}
\mytext{\textcurrency t(mu)\textcurrency \qquad}
t=\int_{0}^{\mu}\frac{{\rm d}\mu^{\prime}}{v(\mu^{\prime})}
=\frac{\hbar}{\gamma T}\int_{0}^{\mu}\frac{{\rm d}\mu^{\prime}}{\phi(\mu^{\prime})[1-\mu^\prime\coth
\phi(\mu^{\prime})]}{ ,} \label{t(mu)}
\end{equation}
where the function $\phi$ is defined by (\ref{Phim}) with $h=+g$. Inversion of
(\ref{t(mu)}) provides the motion $\mu(t)$ of the peak of $P_{\uparrow
\uparrow}(m,t)$. For $P_{\downarrow\downarrow}$, we have to change $g$ into $-g$
in (\ref{Phim}), and $\mu\left(  t\right)  $ expressed by (\ref{t(mu)}) is
then negative.

If $N$ is very large, the probabilistic nature of the registration process fades out and the magnetization is located at $\mu(t)$ with near certainty. 
The evaluation of the time dependence of $\mu(t)$ may be proposed to students as an exercise (\S~9.6.2). 
Results for quadratic coupling ($q=2$) and for quartic coupling ($q=4$), which exemplify second and first-order transitions, respectively,
 are illustrated by Fig. 7.3 and by Fig. 7.4, respectively. 
The evolution from the initial paramagnetic state to the final ferromagnetic state exhibits several stages, which will be studied in 
\S~7.2.3 for $q=2$ and in \S~7.2.4 for $q=4$.

 \myskipfigText{
\begin{figure}\label{figABN8}
\centerline{\includegraphics[width=8cm]{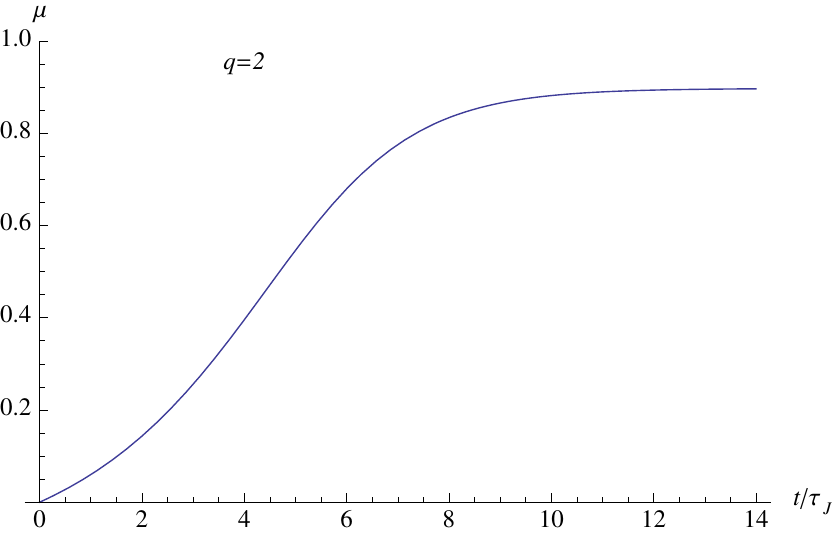}}
\caption{
The average magnetization $\mu(t)$ for a quadratic interaction ($q=2$)  goes from zero to $m_\Uparrow$. The time dependence, given by 
(7.30), results from the local velocity of Fig. 7.1. The parameters are $T=0.65J$ and $g=0.05J$, while the time scale is 
$\tau_J=\hbar/\gamma J$. One can distinguish the three stages of \S 7.2.3, characterized by the first registration time 
$\tau_{\rm reg}=[J/(J-T)]\, \tau_J=2.86\, \tau_J$ (eq. (7.44)) and the second registration time  $\tau_{\rm reg}'=8.4 \tau_J$ (eq. (7.48)): 
(i) Increase, first linearly as $(g/J)(t/\tau_J)=0.05 t/\tau_J$, then exponentially according to (7.42), with a coefficient 
$m_{\rm B}=g/(J - T)=0.143$ and a time scale $\tau_{\rm reg}$.  After a delay of a few $\tau_{\rm reg}$, the coupling may be 
switched off without spoiling the registration. 
(ii) Rise, according to (7.47), up to $m_{\rm F}- \frac{1}{2} m_{\rm B}=0.80$ 
reached at the second registration time $\tau'_{\rm reg}$. 
(iii) Exponential relaxation towards $m_\Uparrow=0.90$ (or $m_{\rm F}=0.87$ if $g$ is switched off) 
according to (7.49) with the time scale $1.6 \tau_J$.
}
\end{figure}
}

\myskipfigText{
\begin{figure}\label{figABN9}
\centerline{\includegraphics[width=8cm]{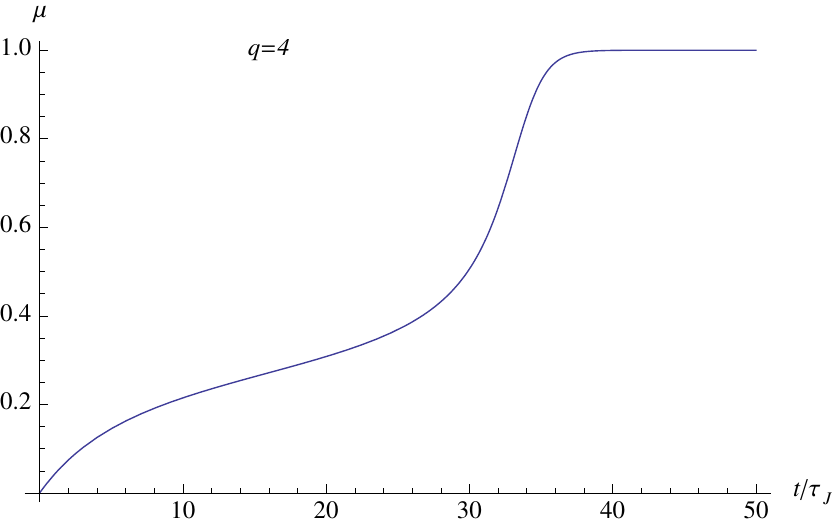}}
\caption{ 
The average magnetization $\mu(t)$ for a quartic interaction ($q=4$) goes from zero to $m_\Uparrow\simeq1$. The time dependence, given by (7.30), 
results from the local velocity of Fig. 7.2. The parameters are $T=0.2J$ and $g=0.045J$ while the time scale is 
$\tau_J=\hbar/\gamma J$. The characteristic registration time $\tau_{\rm reg}=38 \tau_J$ is now given by (7.52). 
(Note that it is much larger than for a quadratic interaction.) The initial increase of $\mu(t)$ takes place, first linearly as 
$(g/J)\,t/\tau_J=0.045 t/\tau_J$, then slows down according to (7.51), with a coefficient $g/J=0.045$ and a time scale $\tau_1=(g/J)\, \tau_J$. 
The region of $m_{\rm c}=0.268$, where the drift velocity is small, is a bottleneck: around this point, reached at the time
$ t=\frac{1}{2}\tau_{\rm reg}$, the average magnetization $\mu(t)$ lingers according to (7.53) where $\delta m_{\rm c}=0.11$. 
It then increases rapidly so as to reach at the time $\tau_{\rm reg}$ a value close to 
$m_{\rm F}\simeq 1$, and finally reaches $m_{\rm F}$ exponentially on the time scale $\tau_J$.     }    
\end{figure}
}

The width of the peak is obtained by regarding $D$ as a function of $\mu(t)$
and by solving the equation for ${\rm d}D/{\rm d}\mu$ that results from (\ref{dmu3}) and (\ref{dD3}).  This yields 

\begin{eqnarray}
D(\mu) =v^{2}(\mu)\left[\frac{\delta_{0}^{2}}{v^{2}(0)}+\int_{0}^{\mu}\frac{{\rm d}\mu^{\prime}\,2w(\mu^{\prime})}{v^{3}(\mu^{\prime})}\right] 
=\phi^{2}(\mu)[1-\mu\,\coth\phi(\mu)]^{2}\left\{  \frac{\delta_{0}
^{2}}{\beta^{2}g^{2}}+\int_{0}^{\mu}\frac{{\rm d}\mu^{\prime
}\,2[\coth\phi(\mu^{\prime})-\mu^{\prime}]}{\phi^{2}(\mu^{\prime
})[1-\mu^{\prime}\coth\phi(\mu^{\prime})]^{3}}\right\}  { .}
\label{D(mu)}
\end{eqnarray}
To analyze this evolution of $D(t)$, we first drop the term in $w$ from the
equation of motion (\ref{dPdt}) of $P_{\rm M}(m,t)$. This simplified equation
describes a deterministic flow in the space of $m$, with a local drift
velocity $v(m)$. For any initial condition, its solution is the mapping
\begin{equation}
\mytext{\textcurrency Pw=0\textcurrency \qquad}
P_{\rm M}(m,t)=\frac{1}{v(m)}
\int{\rm d}m^{\prime}P_{\rm M}(m^{\prime},0)\,\delta\left(  t-\int_{m^{\prime}}
^{m}\frac{{\rm d}m^{\prime\prime}}{v(m^{\prime\prime})}\right)  { ,}
\label{Pw=0}
\end{equation}
where $m^{\prime}$ is the initial point of the trajectory that reaches $m$ at
the time $t$. For a distribution (\ref{Gauss}) peaked at all times, we recover
from (\ref{Pw=0}) the motion (\ref{t(mu)}) of the maximum of $P_{\uparrow
\uparrow}(m,t)$ and the first term of the variance (\ref{D(mu)}). If only the
drift term were present, the width of the peak would vary as $v(\mu)$:
Indeed, in a range of $m$ where the drift velocity increases with $m$, the
front of the peak progresses more rapidly than its tail so that the width
increases, and conversely.

The second term of (\ref{D(mu)}) arises from the term in $w$. Since $w(m)$ is
positive, it describes a diffusion which widens the distribution. This effect
of $w$ is enhanced when $v$ is small. In particular, by the end of the
evolution when $\mu(t)$ tends to a zero $m_{i}$ of $v(m)$ with $\partial
v/\partial m<0$, the competition between the narrowing through $v$ and the
widening through $w$ leads to the equilibrium variance $D=-({\rm d}
v/{\rm d}\mu)^{-1}w$, irrespective of the initial width. This value is
given by $D^{-1}=(1-m_{i}^{2})^{-1}-(q-1)\beta Jm_{i}^{q-2}$, in agreement
with (\ref{Pim}) and with (\ref{station}).

We have noted that the drift velocity $v(m)$ has at each point the same sign
as $-{\rm d}F/{\rm d}m$, where $F$ is the free energy (\ref{F=}), and
that the zeroes $m_{i}$ of $v(m)$, which are the fixed points of the drift
motion, coincide with the extrema of $F$. At such an extremum, given by
(\ref{MF}), we have
\begin{equation}
-\frac{{\rm d}v}{{\rm d}m}=\frac{\gamma}{N\hbar}\frac{2\phi(m_{i}
)}{\sinh2\phi(m_{i})}\frac{{\rm d}^{2}F}{{\rm d}m^{2}}{ .}
\end{equation}
The minima of $F$ correspond to \textit{attractive} fixed points, with
negative slope of $v(m)$, its maxima to repulsive points, that is,
\textit{bifurcations}. In the present case of a narrow distribution, $\mu(t)$
thus increases from $\mu(0)=0$ to the smallest positive minimum $m_{i}$\ of
$F(m)$, which is reached asymptotically for large times. However, the present hypothesis of a single 
narrow peak is valid only if $P(m,t)$ lies entirely and at all times in a region of $m$ free of  bifurcations.
We will discuss in subsection~\ref{section.7.3} the situation where $P$ lies 
astride a bifurcation, either at the initial time or a little later on, if a tail due
to diffusion crosses the bifurcation.

}
\subsubsection{Threshold for the system-apparatus coupling; possibilities of failure}
\label{section.7.2.2}


\hfill{\it If you are not big enough to lose,}

\hfill{ \it you are not big enough to win}

\hfill{ Walter Reuther}

\vspace{3mm}

\ZeText{

The measurement is successful only if $P\left(  m,t\right)  \equiv P_{\uparrow\uparrow}(m,t)/r_{\uparrow\uparrow}(0)$, which is interpreted as
the conditional probability distribution for $m$ if $s_{z}=+1$, approaches for large times the narrow normalized peak (\ref{Pim}) located at the positive
ferromagnetic solution $m_{\Uparrow}$ of (\ref{MF}) with $h=+g$, close to $m_{{\rm F}}$ for $g\ll T$\footnote{
{We recall Eq. (\ref{mUpDown}) where  $m_\Uparrow\simeq+1$ and $m_\Downarrow\simeq-1$ are defined as the fixed points at finite $g$,
and $m_{\rm F}$ and $- m_{\rm F}$ as their $g\to0$ limits, respectively}}.
This goal can be achieved only if \textit{(i)} the center $\mu(t)$ of 
the peak approaches $m_{\Uparrow}$; \textit{(ii)} its width remains small at all times so that the above derivation is valid.

(\textit{i})  The first condition is relevant only for a first-order transition ($q=4$), since $m_\Uparrow$ is anyhow the only attractive fixed point in the region $m>0$ 
for a second-order transition ($q=2$). For quartic interactions, the first minimum of $F(m)$ that occurs for increasing $m$ is not necessarily $m_{\Uparrow}$  (Fig. 7.2).
Indeed, we have seen (end of \S~\ref{section.3.3.4} and  Fig. 3.4) that for a field lower than

\begin{equation}
\mytext{\textcurrency hc\textcurrency \qquad}
h_{{\rm c}}=T{\rm arctanh}\,
m_{{\rm c}}-Jm_{{\rm c}}^{3}\approx\frac{2}{3}Tm_{{\rm c}
}{,\qquad}m_{{\rm c}}^{2}=\frac{1}{2}-\frac{1}{2}\sqrt{1-\frac{4T}
{3J}}\approx\frac{T}{3J}+\frac{T^{2}}{9J^{2}}{ ,} \label{hc}
\end{equation}
 the free energy $F(m)$ has not only a ferromagnetic minimum at $m_{\Uparrow}$, but also a local paramagnetic minimum $m_{{\rm P}}$ at a smaller value of $m$. Hence, if the
spin-apparatus coupling $g$ is smaller than $h_{{\rm c}}$, $\mu(t)$ reaches for large times the locally stable point $m_{{\rm P}}$ in the sector
$\uparrow\uparrow$. It reaches $-m_{{\rm P}}$ in the sector $\downarrow\downarrow$, so that the apparatus seems to distinguish the values $s_{z}
=\pm1$ of S. However, if the coupling is switched off at the end of the process, the magnetization $m$ of M returns to $0$ in both
cases. The result of the measurement thus cannot be registered robustly for $g<h_{{\rm c}}$. 

The center $\mu(t)$ of the peak may escape the region of the origin only if $g>h_{{\rm c}}$  (Fig. 7.2).
Relying on the smallness of $T/3J$ (equal to 0.121 at the transition temperature), we can simplify the expression of $h_{{\rm c}}$ as in (\ref{hc}), so that this
threshold for $g$ is ($q=4$):

\begin{equation}
\mytext{\textcurrency thrg\textcurrency \qquad}
g>h_{{\rm c}}\simeq\frac{2T}{3}\sqrt{\frac{T}{3J}}{ .} \label{thrg}
\end{equation}
Under this condition, the peak $\mu(t)$ of $P_{\uparrow\uparrow}(m,t)$ reaches for large times $m_{\Uparrow}$, close to the magnetization $m_{{\rm F}}$
of the ferromagnetic state. If the coupling $g$ is removed  sufficiently after $\mu(t)$ has passed the maximum of $F(m)$, the peak is expected to end up 
at $m_{{\rm F}}$. Likewise, the peak of $P_{\downarrow\downarrow}(m,t)$ reaches $-m_{{\rm F}}$ at the end of the same process. The apparatus is non-ergodic 
and the memory of its triggering by S may be kept forever under the necessary (but not sufficient) condition (\ref{thrg}).

(\textit{ii}) The second requirement involves the width of the distribution $P_{\uparrow\uparrow}(m,t)$ and the location
$-m_{{\rm B}}<0$ of the repulsive fixed point, at which $F(m)$ is maximum. Consider first the pure drift flow (\ref{Pw=0})
without diffusion, for which $-m_{{\rm B}}$ is a bifurcation. The part $m>-m_{{\rm B}}$ of $P_{\uparrow\uparrow}(m,0)$  is
properly shifted upwards so as to reach eventually the vicinity of the positive ferromagnetic value $+m_{{\rm F}}$; however its
tail $m<-m_{{\rm B}}$ is pushed towards the negative magnetization $-m_{{\rm F}}$. If the relative weight of this
tail is not negligible, \textit{false measurements}, for which the value $-m_{{\rm F}}$ is registered by A although $s_{z}$ equals
$+1$, can occur with a sizeable probability. Such a failure is excluded for $q=4$, because $m_{{\rm B}}$ is then much
larger than the width $1/\sqrt{N}$ of $P_{\uparrow\uparrow}(m,0)$; for instance, in the case $q=4$ we have $m_{{\rm B}}=0.544$ for
the parameters $T=0.2J$ and $g=0.045J$ (which satisfy (\ref{thrg})).  However, in the case $q=2$ and $g\ll J-T$, the point $-m_{\rm B}$ with

\begin{equation}
\mytext{\textcurrency mB\textcurrency \qquad}
m_{{\rm B}}\simeq \frac{g}{J-T}\label{mB},
\end{equation}
lies close to the origin  (Fig. 7.1),  and a risk exists that the initial Gaussian
distribution in $\exp({-Nm^{2}/2\delta_{0}^{2}})$ extends below $-m_{{\rm B}
}$ if $g$ is too small. The probability of getting a wrong result
is significant if the condition $\delta_{0}\ll m_{{\rm B}}\sqrt{N}$ is not fulfilled.
We return to this point in \S~\ref{section.7.3.3}.

Moreover, in this case $q=2$, the lower bound thus guessed for the coupling,
\begin{equation}
\mytext{\textcurrency gmin0\textcurrency \qquad}
g=(J-T)m_{\rm B}\gg\frac{(J-T)\delta_{0}}
{\sqrt{N}}{ ,}\label{gmin0}
\end{equation}
is not sufficient to ensure a faithful registration. The diffusive
process, which tends to increase $D(t)$ and thus to thicken the
dangerous tail $m<-m_{{\rm B}}$ of the probability distribution
$P_{\uparrow\uparrow }(m,t)$, raises the probability of a false
registration towards $-m_{{\rm F}}$ instead of
$+m_{{\rm F}}$. In order to trust the Ansatz (\ref{Gauss}) and
the ensuing solution (\ref{t(mu)}), (\ref{D(mu)}) for
$P_{\uparrow\uparrow}(m,t)$, we need $D(t)$ to remain at all times
sufficiently small so that $P_{\uparrow\uparrow}(m,t)$ is
negligible for $m<-m_{{\rm B}}$. This is expressed, when taking
$\mu(t)$ as a variable
instead of $t$, as
\begin{equation}
\mytext{\textcurrency Dmax\textcurrency \qquad}
\frac{D(\mu)}{N(m_{{\rm B}
}+\mu)^{2}}\ll1\label{Dmax}
\end{equation}
for any $\mu$ between $0$ and $m_{{\rm F}}$: The width $\sqrt{D/N}$ of the
peak of $P_{\uparrow\uparrow}(m,t)$ should not increase much faster than its
position $\mu$. For sufficiently small $g$, we have $m_{{\rm B}}\ll
m_{{\rm F}}$, and we only need to impose (\ref{Dmax}) for times such that
$\mu\left(  t\right)  $\ lies in an interval $0<\mu\left(  t\right)
<\mu_{{\rm max}}$ such that $m_{{\rm B}}\ll\mu_{{\rm max}}\ll T/J$.
In this range we can evaluate $D(\mu)$ from (\ref{D(mu)}) by simplifying
$\tanh\phi(\mu)$ into $\phi(\mu)$, which yields
\begin{equation}
\mytext{\textcurrency D1\textcurrency \qquad}
\frac{D(\mu)}{(m_{{\rm B}}
+\mu)^{2}}=\frac{\delta_{0}^{2}}{m_{{\rm B}}^{2}}+\frac{T}{J-T}\left[
\frac{1}{m_{{\rm B}}^{2}}-\frac{1}{(m_{{\rm B}}+\mu)^{2}}\right]  {
.}\label{D1}
\end{equation}
This ratio increases in time from $\delta_{0}^{2}/m_{{\rm B}}^{2}$ to
$\delta_{1}^{2}/m_{{\rm B}}^{2}$, where
\begin{equation}
\mytext{\textcurrency delta1\textcurrency \qquad}
\delta_{1}^{2}=\delta_{0}
^{2}+\frac{T}{J-T}=\frac{T_{0}}{T_{0}-J}+\frac{T}{J-T}{ ,}\label{delta1}
\end{equation}
so that the left-hand side of (\ref{Dmax}) remains at all times smaller than
$\delta_{1}^{2}/Nm_{{\rm B}}^{2}$. The lower bound on $g$ required to
exclude false registrations is therefore ($q=2$)
\begin{equation}
\mytext{\textcurrency gmin\textcurrency \qquad}
g\gg\frac{(J-T)\delta_{1}}
{\sqrt{N}}{ ,}\label{gmin}
\end{equation}
a condition more stringent than (\ref{gmin0}) if $J-T\ll J$. Altogether, for
$q=2$ the system-apparatus coupling may for large $N$ be small, for instance
as $N^{1/3}$, provided it satisfies (\ref{gmin}). 

For $q=4$,  and more generally for a first-order transition ($3J_4 > J_2$), the
lower bound found as (\ref{thrg})\ remains finite for large $N$: A free energy
barrier of order $N$ has to be overpassed.
Moreover, the diffusion hinders the trend of $m$ to increase and may push part of the
distribution $P_{\uparrow\uparrow}(m,t)$ leftwards, especially its left tail,
while its peak moves rightwards. The widening of $P_{\uparrow\uparrow}(m,t)$ when the
barrier is being reached should not be too large, and this effect raises further the
threshold for $g$. We shall show in \S~\ref{section.7.2.4}  that the condition (\ref{thrg}) should thus be strengthened
into (\ref{gmin4}).

Another difference between first- and second order transitions lies in the
possible values of the temperature. For $q=2$, if $T$ lies near the critical
temperature $J$, the minima $m_{i}$ of $F(m)$ are very sensitive to $g$ and
the ferromagnetic value $m_{{\rm F}}$ in the absence of a field is small as
$\sqrt{3(J-T)/J}$. Using M as the pointer of a measurement apparatus requires
the temperature to lie sufficiently below $J$. For $q=4$, registration is
still possible if $T$ lies near the transition temperature, and even above,
although in this case the ferromagnetic states are not the most stable ones
for $h=0$. However, the coupling $g$ should then be sufficiently strong.

}
\subsubsection{The registration process for a second-order transition}
\label{section.7.2.3}

\ZeText{

Assuming $g$ to satisfy (\ref{gmin}) and $m_{{\rm F}}$ to be significantly
large, we resume the dynamics of $P_{\uparrow\uparrow}(m,t)$ for $q=2$
so as to exhibit its characteristic times. After
a short delay of order $\hbar/T$, most of the process takes place in the
Markovian regime, and the Gaussian Ansatz (\ref{Gauss}) is justified. We can
distinguish three stages in the evolution of $P_{\uparrow\uparrow}(m,t)$, which are exhibited on the example of Figs. 7.3 and 7.5.

(\textit{i}) During the first stage, as long as $\mu(t)\ll m_{{\rm F}}$, we can replace $\phi(m)\coth\,\phi(m)$ by $1$ in $v$ and $w$, so that
the drift velocity $v$ behaves  (Fig. 7.1) as 

\begin{equation}
\mytext{\textcurrency vappr\textcurrency \qquad}
v(m)\approx\frac{\gamma T}
{\hbar}\left[  \frac{g+Jm}{T}-m\right]  =\frac{\gamma(J-T)(m_{{\rm B}}
+m)}{\hbar}{ ,} \label{vappr}
\end{equation}
and the diffusion coefficient as $w\approx\gamma T/\hbar$. Integration of
(\ref{t(mu)}) then yields the motion
\begin{equation}
\mytext{\textcurrency mu1\textcurrency \qquad}
\mu(t)\sim m_{{\rm B}
}(e^{t/\tau_{{\rm reg}}}-1)=\frac{g}{J-T}(e^{t/\tau_{{\rm reg}}}-1)
\label{mu1}
\end{equation}
for the center of the peak, with the characteristic time
\begin{equation}
\mytext{\textcurrency taureg2\textcurrency \qquad}
\tau_{{\rm reg}}
=\frac{\hbar}{\gamma(J-T)}{ .} \label{taureg2}
\end{equation}
After beginning to move as $\mu\sim\gamma gt/\hbar$, the distribution shifts away from the origin faster and faster. Once $\mu$ has reached values of the
order of several times $m_{{\rm B}}$, $\left(  J-T\right)  \mu$ becomes larger than $g$, so that $v(\mu)$ does not depend much on $g$. It little
matters for the subsequent evolution whether the coupling $g$ is present or not. Thus, after $t/\tau_{{\rm reg}}$ reaches $2$\ or $3$, 
{\it the spin-apparatus coupling may be switched off} and the increase of $\mu$ goes on nearly unchanged. In fact, the distribution moves towards $m_{{\rm F}}$
rather than $m_{\Uparrow}$, but $m_{{\rm F}}-m_{\Uparrow}$ is small, less than $g/J$. We shall call $\tau_{{\rm reg}}$ the \textit{first registration time}. 
After it, M will necessarily reach the ferromagnetic state $+m_{{\rm F}}$, independent of ${\rm S}$, although the evolution is not achieved yet.

We have seen that during this first stage the width (\ref{D1}) is governed both by the drift which yields the factor $(m_{{\rm B}}+\mu)^{2}$,
increasing as $e^{2t/\tau_{{\rm reg}}}$, and by the diffusion which raises $\delta_{0}$ up to $\delta_{1}$.

(\textit{ii}) During the second stage $\mu(t)$ {\it rises rapidly} from $m_{{\rm B}}$ to $m_{{\rm F}}$, since the drift velocity $v(\mu)$ is no longer
small. The distribution has become wide, and its width is now governed mainly by the drift term. Matching $D(\mu)$ with (\ref{D1}) for $\mu$ larger than
$m_{{\rm B}}$ yields the width

\begin{equation}
\mytext{\textcurrency Dofv\textcurrency \qquad}
\sqrt{\frac{D(\mu)}{N}}\sim
\frac{\tau_{{\rm reg}}\delta_{1}}{m_{{\rm B}}\sqrt{N}}v(\mu)=\frac
{\hbar\delta_{1}}{\gamma g\sqrt{N}}v(\mu){ ,} \label{Dofv}
\end{equation}
which varies proportionally to $v\left(  \mu\right)  $.The drift velocity $v(m)$ first increases and then decreases as function of $m$  (Fig. 7.1), 
down to $0$ for $m=m_{\Uparrow}\simeq m_{{\rm F}}$.  Accordingly, the width $D(t)$ increases as function of time, then decreases (Fig. 7.5).
The time dependence (\ref{t(mu)}) of $\mu(t)$ and hence of $D(t)$ is evaluated explicitly in the Appendix E.1,
where $\mu$ is related to $t$ through Eq. (\ref{Atofmu}), that is, 

\begin{equation}
\mytext{\textcurrency tofmu\textcurrency \qquad}
\frac{t}{\tau_{{\rm reg}}
}=\ln\frac{m_{{\rm B}}+\mu}{m_{{\rm B}}}+a\ln\frac{m_{{\rm F}}^{2}
}{m_{{\rm F}}^{2}-\mu^{2}}{ ,} 
\label{tofmu}
\end{equation}
where the coefficient $a$, given by

\begin{equation}
\mytext{\textcurrency a=\textcurrency \qquad}
a=\frac{T(J-T)}{J[T-J(1-m_{{\rm F}}^{2})]  }{ ,} \label{a=}
\end{equation}
lies between $\frac{1}{2}$ and $1$.

We define the {\it second registration time} $\tau_{\rm reg}'$ as the delay taken by the average magnetization $\mu(t)$ to go from the paramagnetic value 
$\mu=0$ to the value $m_{{\rm F}}-\frac{1}{2} m_{{\rm B}}$ close to $m_{\rm F}$. From the equation (\ref{tofmu}) that relates $\mu$ to $t$,
we find this second registration time, the duration of the second stage, much longer than the first, as 

\begin{equation}
\mytext{\textcurrency tauprime\textcurrency \qquad}
\tau_{{\rm reg}}^{\prime}=\tau_{{\rm reg}}(1+a)\ln\frac{m_{{\rm F}}}{m_{{\rm B}}},  \label{tauprime}
\end{equation}

(\textit{iii}) The third stage of the registration, the {\it establishment of thermal equilibrium}, 
has been studied in \S~\ref{section.7.1.3} and \S~\ref{section.7.1.4}. 
While $\mu(t)$ tends exponentially to $m_{\Uparrow}$ (or to $m_{{\rm F}}$ if the
coupling $g$ has been switched off), we saw that the equilibrium width of $P_{\uparrow\uparrow}(m,t)$ 
is reached as a result of competition between the drift, which according to
(\ref{Dofv}) narrows the distribution, and the diffusion which becomes again
relevant and tends to widen it. It is shown in the Appendix E.1 that the final relaxation takes place, 
for times $t-\tau'_{\rm reg}\sim\tau_{\rm reg}$, according to

\BEQ
\mu(t)=m_{\rm F}\left[1-\half\left(\frac{m_{\rm F}}{m_{\rm B}}\right)^{1/a}\,\exp\left(-\frac{t}{a\tau_{\rm reg}}\right)\right].
\EEQ
At low temperatures, $T\ll J$, we have $m_{\rm F}\sim1$, $m_{\rm B}\sim g/J$, $a\sim1$.
If $T$\ lies close to the transition temperature, $J-T\ll J$, we have $m^2_{{\rm F}}
\sim{3(J-T)/J}$, $m_{\rm B}=g/(J-T)$ and $a\sim\frac{1}{2}$. 

The above scenario for the registration process is illustrated by Fig. 7.5 which represents a numerical solution of the equation for $P\left(  m,t\right)
=P_{\uparrow\uparrow}(m,t)/r_{\uparrow\uparrow}\left(  0\right)  $. The curves exhibit the motion from $0$ to $m_{{\rm F}}$ of the center $\mu(t)$ of the
peak (also shown by Fig. 7.3), its large initial widening, the intermediate regime where the width
$\sqrt{D(t)/N}$ is proportional to $\mu(t)$, and the final adjustment of $\mu$
and $D$ to their equilibrium values in the ferromagnetic state. Except near
the initial and final state, the width is not small although we have taken a
fairly large value $N=1000$, but one can see that the Gaussian approximation
used for $P_{\uparrow\uparrow}(m,t)$ is sufficient and that the resulting
formulae given above for $\mu(t)$ and $D(t)$ fit the curves. But while a
mean-field theory neglecting the fluctuations is satisfactory at equilibrium,
the dynamics entails large fluctuations of $m$ at intermediate times.

\myskipfigText{
\begin{figure}[h!]
\label{figABN10}
\centerline{\includegraphics[width=8cm]{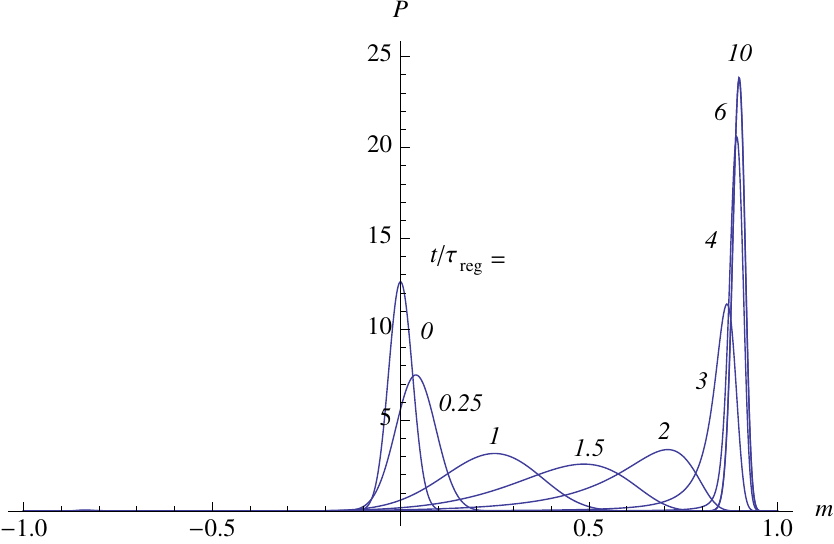}}
\caption{
The registration process for a quadratic interaction ($q=2$). 
The probability density $P(m,t)=P_{\uparrow\uparrow}(m,t)/r_{\uparrow\uparrow}(0)$ 
for the magnetization $m$ of M is represented at different times. The parameters were chosen as $N=1000$, $T=0.65J$ and $g=0.05J$ as in  Figs. 7.1 and 7.3.
The time scale is here the registration time $\tau_{\rm reg}=\hbar/\gamma(J-T)= 2.86\, \tau_J$. After a few times  $\tau_{\rm reg}$ the evolution
is no longer sensitive to the system-apparatus coupling $g$. In the initial fully disordered paramagnetic state ($T_0=\infty$), $P(m,0)$ is a Gaussian centered at 
$m=0$ with width $1/\sqrt{N}$. In the course of time, the peak of $P$ considerably widens, then narrows and reaches eventually the
equilibrium ferromagnetic distribution with positive magnetization $m_\Uparrow=0.90$,  which is given by (\ref{Pim}).
The repulsive fixed point lies at $-m_{\rm B}$ with $m_{\rm B}=0.14$ and no weight is found below this. 
The second registration time, at which $\mu(t)$ reaches $0.80$, is $\tau_{\rm reg}'=3\tau_{\rm reg}$.
It is seen that beyond this, the peak at $m_\Uparrow$ quickly builds up.
}
\end{figure}
}

}

\newpage

\subsubsection{The registration process for a first-order transition}
\label{section.7.2.4}

 \hfill{\it There is a star above us which unites souls of the first order,}

\hfill{\it  though worlds and ages separate them}
 
  \hfill{Christina, Queen of Sweden}

\vspace{3mm}

\ZeText{

The process is different when the interaction is quartic ($q=4$), a case that we chose to exemplify the first-order transitions which occur when $3J_4 > J_2$. 
 The spin-apparatus coupling $g$ must then at least be larger than the threshold (\ref{thrg}) to ensure that $v(m)$ remains positive up to $m_{\Uparrow}$,
which now lies near $m_{{\rm F}}\simeq1$  (Figs. 3.3 and 7.2). At the beginning of the evolution, we find from $v(m)\approx
(\gamma/\hbar)(g-Tm)$, using $g\ll T$, the motion

\BEQ
\mytext{muttau1}\qquad 
\label{muttau1}
\mu(t)\approx\frac{g}
{T}(1-e^{-t/\tau_{1}}){ ,\qquad}\tau_{1}=\frac{\hbar}{\gamma T}{ .} 
\label{begin} \EEQ
Like for $q=2$, the peak shifts first as $\mu\sim\gamma gt/\hbar$, but here
its motion slows down as $t$ increases, instead of escaping more
and more rapidly off the paramagnetic region, as exhibited on the example of Figs. 7.4 and 7.6. 
Extrapolation of (\ref{muttau1}) towards times larger than $\tau_1$
is not possible, since $\mu$ would then not go beyond $g/T$, and could not reach $m_{\rm F}$. 
In fact, $v(m)$ does not vanish at $m=g/T$ as implied by the above approximation but
only decreases down to a positive minimum near  $m_{{\rm c}}\simeq3h_{\rm c}/2T$ according to (\ref{hc}). The vicinity
of $m_{{\rm c}}$ is thus a \textit{bottleneck} for the motion from $\mu=0$
to $\mu=1$ of the peak of $P_{\uparrow\uparrow}(m,t)$: This motion is the
slowest around $m_{{\rm c}}$. The determination of the evolution of $P_{\uparrow\uparrow}(m,t)$,
embedded in $\mu(t)$ and $D(t)$, and the evaluation of the registration time thus require a 
control of the shape of $v(m)$, not only near its zeroes, but also near its minimum  (Fig. 7.2).

Let us recall the parameters which characterize $v(m)$. For $g=0$, it has 5 zeroes. Three of them
correspond to the attractive fixed points $\pm m_{\rm F}\simeq\pm1$ and 0 associated with
the ferromagnetic and paramagnetic states. The other two are repulsive, producing a bifurcation
in the flow of $P(m,t)$; they are located at $m\simeq\pm\sqrt{T/J}$, that is, at 
$m\simeq\pm m_{\rm c}\sqrt{3}$ according to (\ref{hc}).
When $g$ increases and becomes larger than $h_{\rm c}$, there remain the two ferromagnetic points,
while the repulsive point $-m_{\rm c}\sqrt{3}$ is shifted towards $-m_{\rm B}\simeq-2m_{\rm c}$. The paramagnetic
point and the repulsive point $m_{\rm c}\sqrt{3}$ converge towards each other, giving rise to the
minimum of $v(m)$ near $m=m_{\rm c}$. The value of $v(m)$ at this minimum is expressed by

\BEQ
\label{zeminimum}
\mytext{zeminimum}\qquad
\frac{\hbar}{\gamma T}v(m_{\rm c})\simeq\frac{\delta m_{\rm c}^2}{m_{\rm c}},\qquad
\delta m_{\rm c}\simeq\sqrt{\frac{(g-h_{\rm c})m_{\rm c}}{T}},\qquad h_{\rm c}\simeq\frac{2}{3}Tm_{\rm c}, 
\qquad m_{\rm c}=\sqrt{\frac{T}{3J}}.
\EEQ

We construct in Appendix E.2, for $\delta m_{\rm c}\ll m_{\rm c}$ and $m_{\rm c}$ small, a parametrization
of $v(m)$ which reproduces all these features, so as to derive an algebraic approximation (\ref{Aevol4})
which expresses the time dependence of $\mu(t)$ over all times. After the initial evolution (\ref{muttau1})  of $\mu(t)$
for $t\ll\tau_1=\hbar/\gamma T$, the motion of the peak $P_{\uparrow\uparrow}(m,t)$ is characterized by a much 
larger time scale. We define the {\it registration time} as

\BEQ
\mytext{tauregMT;taureg4}\qquad
\label{tauregMT}\label{taureg4}
 \tau_{\rm reg}=\frac{\pi\hbar}{\gamma T}\sqrt{\frac{m_{\rm c}T}{g-h_{\rm c}}}.
\EEQ
The bottleneck stage takes place around $\half\tau_{\rm reg}$. Between the times $t=\frac{1}{4}\tau_{\rm reg}$
and $t=\frac{3}{4}\tau_{\rm reg}$, the average magnetization $\mu(t)$ lingers in the narrow range 
$m_{\rm c}\pm\delta m_{\rm c}$, according to (Fig. 7.4)

\BEQ \label{mutcotan}
\mu(t)=m_{\rm c}-\delta m_{\rm c}{\rm cotan}\frac{\pi t}{\tau_{\rm reg}}.
\EEQ
It is shown in Appendix E.2 that, under the considered conditions on the parameters, 
$\mu(t)$ rises thereafter rapidly according to (\ref{fullt=tau}), and that the full time taken by the peak $\mu(t)$
of $P_{\uparrow\uparrow}(m,t)$ to go from 0 to the close vicinity of 1 is $\tau_{\rm reg}$ (Eq. (\ref{tauregMT})). 
It is also shown in Appendix E.2 that the final relaxation takes place on the short time scale $\hbar/\gamma J$.

We have focused on the location of the peak of $P_{\uparrow\uparrow}(m,t)$.  The consideration of its width $D(t)$
is essential to determine when S and A may be decoupled. During the bottleneck stage, the sole drift effect would 
produce a narrowing of $D(t)$ around $t=\half\tau_{\rm reg}$ expressed by the first term of  (\ref{D(mu)}),  
but the smallness of $v(m)$ enhances the second term, so that the diffusion acts during a long time and 
produces a large widening of $D(t)$. By using the parabolic approximation for $v(m)$,
which is represented by the first term of  (\ref{ML4}), and by replacing $w(m)$ by $\gamma T/\hbar$, 
we obtain, with $\mu(t)$ expressed by (\ref{mutcotan}),

\begin{equation}
\mytext{Dmu4}\qquad
\label{Dmu4}
D(\mu)\sim2m_{{\rm c}}\left[  (\mu-m_{{\rm c}})^{2}+\delta
m_{{\rm c}}^{2}\right]  \int_{0}^{\mu}\frac{{\rm d}\mu^{\prime}}
{[(\mu^{\prime}-m_{{\rm c}})^{2}+\delta m_{{\rm c}}^{2}]^{3}}.
\end{equation}
After the bottleneck has been passed, the diffusion may again be neglected. 
From (\ref{D(mu)}) and (\ref{Dmu4}),
 we find for all values of $\mu(t)$ such that $\mu-m_{{\rm c}}\gg\delta m_{{\rm c}}$
 
\begin{equation}
D(\mu)\sim
\frac{3\pi\hbar^{2}m_{{\rm c}}^{3}}{4\gamma^{2}T^{2}\delta m_{{\rm c}}^{5}}v^{2}(\mu)
=\frac{ 3\pi\sqrt{ T m_{\rm c}} }{4(g - h_{\rm c})^{5/2}}  (J\mu^3 + g)^2  [1 - \mu \coth\,\beta (J\mu^3 + g)]^2,
\end{equation}
where we used (\ref{Vsm}), (\ref{zeminimum}) and (\ref{omi=}).
Without any diffusion, the coefficient of $v^{2}(\mu)$ would have been $1/v^2(0)=9\hbar
^{2}/4\gamma^{2}T^{2}m_{{\rm c}}^{2}$;  both factors $v(\mu)$ are 
multiplied by the large factor $\sqrt{\pi/3} (m_{{\rm c}}/\delta m_{{\rm c}})^{5/2}$ due to diffusion. 

The distribution $P_{\uparrow\uparrow}(m,t)$  thus extends, at times larger than $\frac{3}{4}\tau_{\rm reg}$,
over the region $\mu(t)\pm \sqrt{D(t)/N}$. The {\it first registration time} has been defined  in \S~7.2.3 as the time after which
S {\it and} A {\it can be decoupled} without affecting the process.
When $g$ is switched off ($g\to0$), a repulsive fixed point appears at the zero $m=m_{\rm c}\sqrt{3}$ of $v(m)$.
In order to ensure a proper registration we need this decoupling to take place after the whole distribution
$P_{\uparrow\uparrow}(m,t)$ has passed this bifurcation, that is, at a time $t_{\rm off}$ such that

\BEQ  \mytext{(mutoff)}\qquad
 \label{mutoff} 
\mu(t_{\rm off})-\sqrt{D(t_{\rm off})/N}> m_{\rm c}\sqrt{3}.
\EEQ
The time dependence (\ref{fullt=tau}) of $\mu$ shows that the lower bound of $t_{\rm off}$ is equal to $\tau_{\rm reg}$
within a correction of order $\tau_1\ll\tau_{\rm reg}$. Moreover, we need the distribution to be sufficiently
narrow so that (\ref{mutoff}) is satisfied after $g$ is switched off.
Taking for instance $\mu(t_{\rm off})=2m_{{\rm c}}$, which according to (E.15)  is reached at the time 
$t_{\rm off} =\tau_{\rm reg} (1 - 0.25\, \delta m_{\rm c}/m_{\rm c})$,
we thus find, by inserting (7.55) with $\mu=2m_{\rm c}$ and $g\simeq h_{\rm c}$ into (7.56), by using (7.51) and evaluating
 the last bracket of (7.55) for $m_{\rm c}=0.268$, a \textit{further lower bound} for the coupling $g$ in our first order case $q=4$:

\begin{equation}
\mytext{\textcurrency gmin4\textcurrency \qquad}
\frac{g-h_{{\rm c}}
}{h_{{\rm c}}}\gg 8\left(  \frac{J}{NT}\right)  ^{2/5} 
{ .}
\label{gmin4}
\end{equation}

The first registration time, which governs the possibility of decoupling, and the second one, which is the
delay after which the pointer variable approaches the equilibrium value, are therefore nearly the same,
namely $\tau_{\rm reg}$, contrary to the case $q=2$ of a second order transition (\S~\ref{section.7.2.3}).

The registration process for $q=4$ is illustrated by  Figs. 7.4 and 7.6,
obtained through numerical integration. The time dependence of $\mu(t)$ as well as the widening of the distribution are
influenced by the existence of the minimum for the drift velocity.
Although in this example $g$ lies above the threshold $h_{\rm c}$, $N$ is not sufficiently large to fulfil the condition (7.57). 
The widening is so large that a significant part of the weight $P(m,t)$ remains for a long time below the bifurcation 
$m_{\rm c}\sqrt{3}$ which appears when $g$ is switched off. 
The bound (7.57) was evaluated by requiring that such a switching off takes place after the average magnetization $\mu$ passes 
$2m_{\rm c}=0.54$. Here however, for $N=1000$, $T=0.2J$ and $g=0.045J$, the bound is very stringent, 
since we cannot switch off $g$ before $\mu$ has reached (at the time $1.09 \tau_{\rm reg}$ found from (E.15)) 
the value $1 - 13 \cdot 10^{-5}$, close to the equilibrium value $m_{\rm F}=1- 9 \cdot 10^{-5}$.   

\vspace{3mm}

 Altogether, for $q=2$ as well as for $q=4$, we can check that the approximate algebraic treatment of \S\S~7.2.3 and 7.2.4 
 fits the numerical solution of Eq. (7.1) exemplified by the figures 7.3 to 7.6. In both cases, the registration times
(\ref{taureg2}) and (\ref{tauprime}) for $q=2$ or (\ref{taureg4}) for $q=4$, 
which characterize the evolution of the diagonal blocks of the
density matrix of the total system $\hat{{\cal D}}$, are much longer than
the truncation time (\ref{taured}) over which the off-diagonal blocks decay.
Two reasons conspire to ensure this large ratio: the weakness of the coupling
$\gamma$ between magnet and bath, which makes $\tau_{{\rm reg}}$ large; and
the large value of $N$, which makes $\tau_{{\rm trunc}}$ small.

\myskipfigText{
\begin{figure}[h!]
\label{figABN11}
\includegraphics[width=8cm]{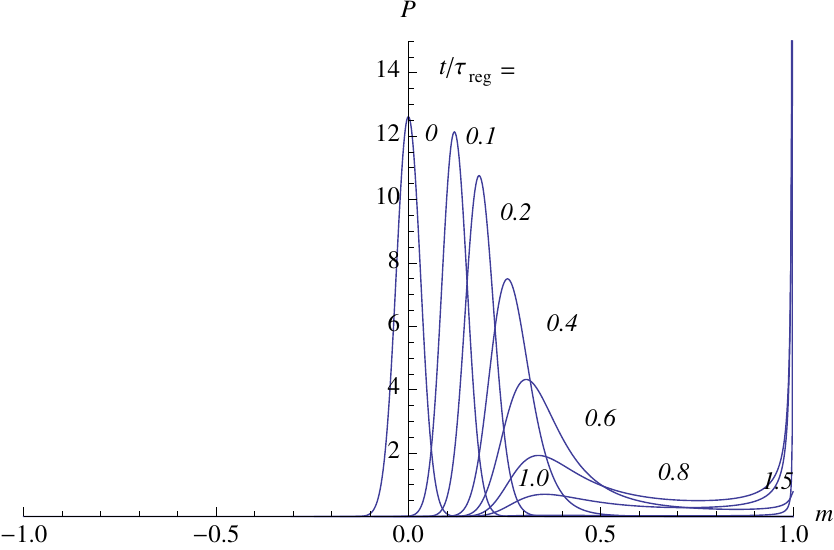}
\caption{
The registration process for quartic interactions $(q=4)$.
The probability density $P(m,t)=P_{\uparrow\uparrow}(m,t)/r_{\uparrow\uparrow}(0)$ 
as function of $m$ is represented at different times up to $t=1.5\,\tau_{\rm reg}$.
The parameters are chosen as $N=1000$,  $T=0.2J$ and $g=0.045J$ as in  Fig 7.4.
The time scale is here the registration time $\tau_{\rm reg}=38 \tau_J=38 \hbar/\gamma J$, which is large due to the existence of 
a bottleneck around $m_{\rm c}=0.268$. The coupling $g$ exceeds the critical value $h_{\rm c}=0.0357J$ needed for proper registration, 
but since $(g-h_{\rm c})/h_{\rm c}$ is small, the drift velocity has a low positive minimum at $0.270$ near $m_{\rm c}$ (Fig. 7.2). 
Around this minimum, reached at the time $\frac{1}{2}\tau_{\rm reg}$, the peak shifts slowly and widens much. 
Then, the motion fastens and the peak narrows rapidly, coming close to ferromagnetism around the time 
$\tau_{\rm reg}$, after which equilibrium is exponentially reached.}
\end{figure}
}

}

\subsection{Giant fluctuations of the magnetization}
\label{section.7.3}

\ZeText{

We have studied in subsection~\ref{section.7.2} the evolution of the probability distribution 
$P(m,t)=P_{\uparrow\uparrow}(m,t)/r_{\uparrow\uparrow}(0)$  of the magnetization of M in case this distribution
presents a single peak (\ref{Gauss}) at all times. This occurs when $P(m,t)$
always remains entirely located, except for negligible tails, on a single side
of the bifurcation $-m_{{\rm B}}$ of the drift flow $v(m)$. We will now
consider the case of an \textit{active bifurcation} \cite{suzuki,suzuki1,suzuki2,suzuki3,suzuki4}: The initial distribution
is split during the evolution into two parts evolving towards $+m_{{\rm F}
}$ and $-m_{{\rm F}}$. This situation is relevant to our
measurement process for $q=2$ in regard to two questions: (\textit{i}) How fast should
one perform the cooling of the bath before the initial time, and the switching
on of the system-apparatus interaction around the initial time? (\textit{ii})
What is the percentage of errors of registration if the coupling $g$ is so
small that it violates the condition (\ref{gmin})?

}

\newpage

\subsubsection{Dynamics of the invariance breaking}
\label{section.7.3.1}

\hfill{\it Be the change}

\hfill{\it  that you want to see in the world}

\hfill{Mohandas Gandhi }

\vspace{3mm}

\ZeText{

In order to answer the above two questions, we first determine the Green's
function for the equation of motion (\ref{dPdt}) which governs $P(m,t)$ for
$q=2$ in the Markovian regime. This will allow us to deal with an arbitrary
initial condition. The \textit{Green's function} $G(m,m^{\prime},t-t^{\prime
})$ is characterized by the equation

\begin{equation}
\mytext{\textcurrency dGfor\textcurrency \qquad}
\frac{\partial}{\partial t}G(m,m^{\prime},t-t^{\prime})+\frac{\partial}{\partial m}[v(m)G(m,m^{\prime},t-t^{\prime})]
-\frac{1}{N}\frac{\partial^2}{\partial m^2}[w(m)G(m,m^{\prime},t-t^{\prime})]=\delta(m-m^{\prime})\delta(t-t^{\prime}){ ,}
\label{dGfor}
\end{equation}
with $G(m,m^{\prime},t-t^{\prime})=0$ for $t<t^{\prime}$. We have replaced the initial time $0$ by a running time $t^{\prime}$ in
order to take advantage of the convolution property of $G$. The functions $v(m)$ and $w(m)$ defined by (\ref{Vsm}) and (\ref{Wsm})
involve a field $h$ which stands either for an applied external
field if ${\rm A}={\rm M}+{\rm B}$ evolves alone (a case that could appear but which we do not consider),  or for
$\pm g$ if we consider $P_{\uparrow\uparrow}$ or $P_{\downarrow\downarrow }$ if ${\rm A}$ is coupled to ${\rm S}$ during the measurement. 
We wish to face the situation in which $P(m,t)$ lies, at least after some time, astride the bifurcation point
$-m_{{\rm B}}=-h/(J-T)$. Such a situation has extensively been studied \cite{suzuki,suzuki1,suzuki2,suzuki3,suzuki4}, and we adapt the existing methods
to the present problem which is similar to Suzuki's model.

We first note that the initial distribution $P(m,t^{\prime}=0)$ is
concentrated near the origin, a property thus satisfied by the variable
$m^{\prime}$ in $G(m,m^{\prime},t)$. In this region, it is legitimate to
simplify $v(m^{\prime})$ and $w(m^{\prime})$ into

\begin{equation}
\mytext{\textcurrency VWsimp\textcurrency \qquad}
v(m^{\prime})\approx
\frac{\gamma}{\hbar}[h+(J-T)m^{\prime}]{,\qquad}w(m^{\prime})\approx
\frac{\gamma T}{\hbar}{ ,}\label{VWsimp}
\end{equation}
where we also used $h\ll T$. In order to implement this simplification which holds only
for $m'\ll1$, we replace the forward equation (\ref{dGfor}) in terms of $t$ which characterizes
$G(m,m^{\prime},t-t^{\prime})$ by the equivalent \textit{backward equation}, for 
$\partial G(m,m',t-t')/\partial t'$, in terms of the initial time $t^{\prime}$ which runs down from $t$ to $0$.
This equation is written and solved in Appendix F. The distribution $P(m,t)$ is then given by 

\begin{equation}
\mytext{\textcurrency P=GP\textcurrency \qquad}
P(m,t)=\int{\rm d}m^{\prime}G(m,m^{\prime},t)P(m^{\prime},0)
\label{P=GP}.
\end{equation}
We derive below several approximations for $P(m,t)$, which are valid in limiting cases. These various results are 
encompassed by the general expression (\ref{AbestP})--(\ref{Aksi0}) for $P(m,t)$, 
obtained through the less elementary approach of Appendix F.

As in \S~\ref{section.7.2.3}, the evolution takes place in three stages \cite{suzuki,suzuki1,suzuki2,suzuki3,suzuki4}: (i)
\textit{widening} of the initial distribution, which here takes place over the
bifurcation $-m_{{\rm B}}$; (ii) \textit{drift} on both sides of
$-m_{{\rm B}}$ towards $+m_{{\rm F}}$ and $-m_{{\rm F}}$; (iii)
narrowing around $+m_{{\rm F}}$ and $-m_{{\rm F}}$ of the two final
peaks, which evolve \textit{separately towards equilibrium}. We shall not need
to consider here the last stage, the approach to quasi-equilibrium, that we
studied in \S~\ref{section.7.1.4}. {\bf or ($i$) ($ii$) ($iii$)??}

The probability distribution $P(m,t)$ is thus expressed in terms of the initial
distribution $P(m,0)$ by (\ref{P=GP}), at all times, except during the final
equilibration. If $P(m,0)$ is a narrow Gaussian peak centered at $m=\mu_{0}$
with a width $\delta_{0}/\sqrt{N}$,  we can use the expression (\ref{AfinalG}) of $G$, which yields

\begin{equation}
\mytext{\textcurrency Psplit\textcurrency \qquad}
P(m,t)=\frac{v(\mu^{\prime})}{v(m)}\sqrt{\frac{N}{2\pi}}\frac{1}{\delta_{1}(t)}\exp\left[  -\frac{N}
{2}\frac{(\mu^{\prime}-\mu_{0})^{2}}{\delta_{1}^{2}\left(  t\right)  }\right]
{ .}\label{Psplit}
\end{equation}
The function $\mu^{\prime}(m,t)$ is defined for arbitrary values of $m$ by

\begin{equation}
\mytext{\textcurrency mumt\textcurrency \qquad}
t=\int_{\mu^{\prime}(m,t)}
^{m}\frac{{\rm d}m^{\prime\prime}}{v(m^{\prime\prime})}{ ,}
\label{mumt}
\end{equation}
while the variance that enters (\ref{Psplit}) is determined by

\begin{equation}
\mytext{\textcurrency delta1t\textcurrency \qquad}
\delta_{1}^{2}(t)\equiv \delta_{0}^{2}
+\frac{T}{J-T}(1-e^{-2t/\tau_{{\rm reg}}})\equiv\delta_{1}^{2}-\frac{T}{J-T}e^{-2t/\tau_{{\rm reg}}}, \qquad 
\delta_{1}^{2}\equiv\frac{T_{0}}{T_{0}-J}+\frac{T}{J-T}{ .}
\label{delta1t}
\end{equation}
With time, it increases from $\delta_{0}^{2}/N$\ to $\delta_{1}^{2}/N$.

}

\subsubsection{Spontaneous relaxation of the initial paramagnetic state}
\label{section.7.3.2}

\hfill{\it Co se doma uva\u{r}\'i, to se doma sn\'i \footnote{What is cooked home  is eaten home}}

\hfill{Czech proverb}

\vspace{3mm}

\ZeText{

The registration process that we studied in \S~\ref{section.7.2.3} is the same as the
relaxation, for $q=2$ and $T<J$, of the initial paramagnetic state (\ref{Pm0}) towards
the positive ferromagnetic state $+m_{{\rm F}}$ in the presence of a
sufficiently large positive external field $h$. We now consider the
situation in which ${\rm A}$ evolves \textit{in the absence of a field}.
The process will describe the \textit{dynamics of the spontaneous symmetry
breaking}, which leads from the unstable symmetric paramagnetic distribution
$P_{\rm M}(m,0)$ to the ferromagnetic distribution (\ref{Pim}) for $+m_{{\rm F}}$
and $-m_{{\rm F}}$, occurring with equal probabilities. We present below an
approximate analytic solution, and illustrate it by Fig. 7.7 which relies on a numerical solution.

\myskipfigText{
\begin{figure}\label{figABN12}
\centerline{\includegraphics[width=8cm]{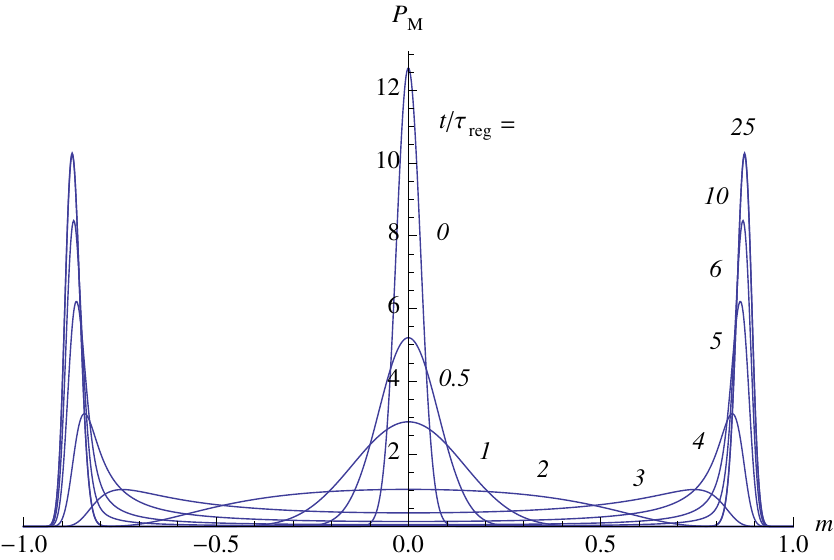}}
\caption{Relaxation of an unstable paramagnetic state $(q=2)$ in the absence of a field ($g=0$). 
The probability distribution $P_{\rm M}(m,t)$ is represented at several times.
As in  Figs. 7.3 and 7.5 the parameters are $N=1000$ and $T= 0.65J$.
First the Gaussian paramagnetic peak around $m=0$ with width ${1}/{\sqrt N}$ widens considerably. 
Around $t=\tau_{\rm flat}=2.2 \tau_{\rm reg}$, the distribution extends over most of
the interval $-m_{\rm F},+m_{\rm F}$  ($m_{\rm F}=0.872$) and is
nearly flat. Then, two peaks progressively build up, moving towards
$-m_{\rm F}$ and $+m_{\rm F}$. Finally each peak tends to the
Gaussian ferromagnetic equilibrium shape, 
the curves at $t=10\tau_{\rm reg}$ and $25\tau_{\rm reg}$ basically coincide.}
\end{figure}
}

Apart from the final stage, the result is given by (\ref{Psplit}) with
$\mu_{0}=0$, $\delta_{0}^{2}=T_{0}/(T_{0}-J)$, and $v(m)=(\gamma
/\hbar)Jm[1-m\coth(Jm/T)]$. During the first stage, we have
$v(m)\sim m/\tau_{{\rm reg}}$ and hence $\mu^{\prime}\sim me^{-t/\tau
_{{\rm reg}}}$, so that (\ref{Psplit}) reduces to
\begin{equation}
\mytext{\textcurrency Pfirst\textcurrency \qquad}
P_{\rm M}(m,t)=\sqrt{\frac{N}{{2\pi }}
}\frac{e^{-t/\tau_{{\rm reg}}}}{\delta_{1}\left(  t\right)  }\exp\left[
-\frac{Nm^{2}e^{-2t/\tau_{{\rm reg}}}}{2\delta_{1}^{2}(t)}\right]  {
.}\label{Pfirst}
\end{equation}
On the time scale $\tau_{{\rm reg}}=\hbar/\gamma(J-T)$, this distribution
widens exponentially, with the variance
\begin{equation}
\mytext{\textcurrency delta1=\textcurrency \qquad}
\frac{1}{N}\left[  \delta
_{1}^{2}e^{2t/\tau_{{\rm reg}}}-\frac{T}{J-T}\right]  {,\qquad}
\delta_{1}^{2}\equiv\frac{T_{0}}{T_{0}-J}+\frac{T}{J-T}{ .}
\label{delta1=}
\end{equation}
As in \S~\ref{section.7.2.3}, the widening is first induced by the diffusion term, which
is then relayed by the gradient of the drift velocity $v(m)$. However, the
effect is much stronger here because the distribution remains centered around
$m=0$.

In fact, at times of order $\tau_{{\rm reg}}\ln\sqrt{N}$, the width of the
peak of $P_{\rm M}(m,t)$ is no longer of order $1/\sqrt{N}$, but it is \textit{finite for large} $N$. 
If we define the lifetime $\tau_{\rm para}$ of the paramagnetic state as the delay 
during which this width is less that $\alpha$, say $\alpha=1/10$, it equals 
(for $\alpha\sqrt{N}\gg1$) 

\begin{equation}  \label{taupara} \tau_{\rm para}= \tau_{{\rm reg}}\ln\alpha\sqrt{N}=
\frac{\hbar}{\gamma(J-T)}\ln\alpha\sqrt{N}.
\end{equation} 

The second stage of the evolution is then reached (Fig. 7.7). An analytic
expression of $P_{\rm M}(m,t)$ can then be found by using the Mittag-Leffler approximation (\ref{ML})
for $v(m)$ (with $h=0$). The relation between $\mu^{\prime}$, $m$ and $t$ becomes

\begin{equation}
\frac{t}{\tau_{{\rm reg}}}=\ln\frac{m}{\mu^{\prime}}+a\ln\frac
{m_{{\rm F}}^{2}-\mu^{\prime}{}^{2}}{m_{{\rm F}}^{2}-m^{2}}{ ,}
\end{equation}
where the coefficient $a$, defined by (\ref{a=}), lies between $\frac{1}{2}$
for $J-T\ll J$ and $1$ for $T\ll J$. In this stage, when $t\gg\tau
_{{\rm reg}}$, the distribution
\begin{equation}
P_{\rm M}(m,t)=\frac{\mu^{\prime}(m_{{\rm F}}^{2}-\mu^{\prime}{}^{2})}
{m(m_{{\rm F}}^{2}-m^{2})}\frac{m_{{\rm F}}^{2}+(2a-1)m^{2}
}{m_{{\rm F}}^{2}+(2a-1)\mu^{\prime}{}^{2}}\sqrt{\frac{N}{2\pi \delta
_{1}^{2}}}\exp\left(  -\frac{N\mu^{\prime}{}^{2}}{2\delta_{1}^{2}}\right)
{ ,}
\end{equation}
depends on time only through $\mu^{\prime}$. It flattens while widening. In
particular, around the time
\begin{equation}
\mytext{\textcurrency tauflat\textcurrency \qquad}
\tau_{{\rm flat}}
=\tau_{{\rm reg}}\ln\left(  \frac{m_{{\rm F}}}{\delta_{1}}\sqrt{\frac
{N}{6a}}\right)  { ,} \label{tauflat}
\end{equation}
it behaves for small $m$ as
\begin{equation}
\mytext{\textcurrency Pflat\textcurrency \qquad}
P_{\rm M}(m,t)\approx\frac{1}{m_{{\rm F}}}\sqrt{\frac{3}{\pi}}e^{-(t-\tau_{{\rm flat}}
)/\tau_{{\rm reg}}}\left\{  1+\frac{3am^{2}}{m_{{\rm F}}^{2}}\left[
1-e^{-2(t-\tau_{{\rm flat}})/\tau_{{\rm reg}}}\right]  +{\cal O}
\left(  \frac{m^{4}}{m_{{\rm F}}^{4}}\right)  \right\}  { .}
\label{Pflat}
\end{equation}
When $t$ reaches $\tau_{{\rm flat}}$, the distribution $P_{\rm M}(m,\tau_{{\rm flat}})$ has widened so much 
that it has become \textit{nearly flat}: The probabilities of the possible values (\ref{eig}) of $m$ are nearly
the same on a range which extends over most of the interval $-m_{{\rm F}}$,
$+m_{{\rm F}}$. This property agrees with the value of $\frac{1}
{2}NP_{\rm M}(0,\tau_{{\rm flat}})=0.98/m_{{\rm F}}$; the coefficient of the
term in $(m/m_{{\rm F}})^{4}$, equal to $-a(8a-\frac{5}{2})$, yields a
correction $-(0.93\,m/m_{{\rm F}})^{4}$ for small $m_{{\rm F}}$,
$-(1.53\,m/m_{{\rm F}})^{4}$ for large $m_{{\rm F}}$.

When $t$ increases beyond $\tau_{{\rm flat}}$, the distribution begins to
deplete near $m=0$ and two originally not pronounced maxima appear there (Fig. 7.7),
which move apart as
\begin{equation}
m=\pm m_{{\rm F}}\sqrt{\frac{6(t-\tau_{{\rm flat}})}{(16a-5)\tau
_{{\rm reg}}}}{ .}
\end{equation}
They then become sharper and sharper as they move towards $\pm m_{{\rm F}}
$. When they get well separated, $P_{\rm M}(m,t)$ is concentrated in two symmetric
regions, below $m_{{\rm F}}$ and above $-m_{{\rm F}}$, and it reaches a
\textit{scaling regime} \cite{suzuki,suzuki1,suzuki2,suzuki3,suzuki4}
in which (for
$m>0$)
\begin{equation}
\mu^{\prime}(m,t)\sim m_{{\rm F}}e^{-t/\tau_{{\rm reg}}}\left[
\frac{m_{{\rm F}}}{2(m_{{\rm F}}-m)}\right]  ^{2}
\end{equation}
is small, of order $1/\sqrt{N}$. If we define, with $a$ $\,(\frac{1}{2}<a<1$) given by Eq. (\ref{a=}),
\begin{equation}
\xi(m,t)\equiv\sqrt{\frac{N}{2}}\frac{\mu^{\prime}(m,t)}{\delta_{1}}=\sqrt
{3a}\left[  \frac{m_{{\rm F}}}{2(m_{{\rm F}}-m)}\right]  ^{a}
e^{-(t-\tau_{{\rm flat}})/\tau_{{\rm reg}}}{ ,}
\end{equation}
$P_{\rm M}(m,t)$ takes in the region $m>0$, $\xi>0$, the form
\begin{equation}
\mytext{\textcurrency Pscal\textcurrency \qquad}
P_{\rm M}(m,t)\approx\frac{1}{\sqrt
{\pi}}\frac{\partial\xi}{\partial m}\,e^{-\xi^{2}}{ .} \label{Pscal}
\end{equation}
Its maximum lies at the point $m_{{\rm max}}$ given by
\begin{equation}
\xi(m,t)=\sqrt{\frac{a+1}{2a}}{,\qquad}\frac{m_{{\rm F}}
-m_{{\rm max}}}{m_{{\rm F}}}=\frac{1}{2}\left(  \frac{6a^{2}}
{a+1}\right)  ^{1/(2a)}e^{-(t-\tau_{{\rm flat}})/a\tau_{{\rm reg}}
}{ ,}
\end{equation}
which approaches $m_{{\rm F}}$ exponentially, and its shape is strongly
asymmetric. In particular, its tail above $m_{{\rm max}}$ is short, whereas
its tail below $m_{{\rm max}}$ extends far as $1/(m_{{\rm max}}
-m)^{a+1}$; only moments $\langle(m_{{\rm F}}-m)^{k}\rangle$ with $k<a$ exist.

After a delay of order $a\tau_{{\rm reg}}\ln\sqrt{N}$, the width of the
peaks of $P_{\rm M}(m,t)$ and their distance to $\pm m_{{\rm F}}$ reach an order of
magnitude $1/\sqrt{N}$. The diffusion term becomes active, and each peak tends
to the Gaussian shape (\ref{Pim}) as in \S~\ref{section.7.1.4}. This crossover could be
expressed explicitly by writing the Green's function for $m$ and $m^{\prime}$
near $m_{{\rm F}}$ (as we did near $0$ in \S~\ref{section.7.3.1}) and by taking
(\ref{Pscal}) as initial condition. All the above features fit the numerical solution shown by Fig. 7.7.

In our measurement problem, $q=2$, the above evolution begins to take place at the
time $-\tau_{{\rm init}}$ at which the apparatus is initialized
(\S~\ref{section.3.3.3}). Before $t=-\tau_{{\rm init}}$, paramagnetic equilibrium has
been reached at the temperature $T_{0}>J$, and the initial distribution of $m$
is given by (\ref{Pm0}), (\ref{DELTAm}) (\ref{delta0=}). The sudden cooling of
the bath down to the temperature $T<J$ lets the evolution (\ref{Pfirst}) start
at the time $-\tau_{{\rm init}}$. We wish that, at the time $t=0$ when the
coupling $g$ is switched on and the measurement begins, the distribution
$P_{\rm M}(m,0)$ is still narrow, close to (\ref{Pm0}). We thus need $\delta_{1}
(\tau_{{\rm init}})$ to be of the order of $\delta_{0}$, that is,
\begin{equation}
\mytext{\textcurrency condtauinit\textcurrency \qquad}
\frac{2\tau_{{\rm init}
}}{\tau_{{\rm reg}}}<\delta_{0}^{2}\frac{J-T}{T}=\frac{T_{0}}{T_{0}-J}
\frac{J-T}{T}{ .}\label{condtauinit}
\end{equation}
The bath should be cooled down and the system-apparatus interaction $\hat {H}_{{\rm SA}}$ should be switched on \textit{over a delay }$\tau
_{{\rm init}}$ \textit{not larger than the registration time} $\tau_{{\rm reg}}=\hbar/\gamma(J-T).$

 The situation is more favorable in case the initial depolarized state of the spins of M is generated by a radiofrequency field rather than through 
 equilibration with the phonon bath at a high but finite temperature $T_0$. In this case, a sudden cooling of the bath at the time  $- \tau_{\rm init}$ is not needed. 
 The bath can beforehand be cooled at the required temperature $T$ lower than $T_c=J$. At the time $- \tau_{\rm init}$, the spins are suddenly set by the field 
 into their most disordered state, a process which hardly affects the bath since $\gamma\ll1$. The above discussion then holds as if $T_0$ were infinite.

If a weak field $h_{0}$ is accidentally present during the preparation by thermalization  of the initial paramagnetic state, it should not produce a bias 
in the measurement. This field shifts the initial expectation value $\langle m\rangle$ of $m$ from $0$
to $\mu_{0}=h_{0}/(T_{0}-J)$, which enters (\ref{Psplit}). At the time $0$, $\langle m\rangle$ has become $\mu_{0}\exp({\tau_{{\rm init}}/\tau_{{\rm reg}}})$, 
so that the residual field $h_{0}$ is ineffective provided $\mu_{0}<\delta_{0}$, that is for

\begin{equation}
\mytext{\textcurrency condh0\textcurrency \qquad}
h_{0}<\sqrt{\frac{T_{0}
(T_{0}-J)}{N}}{ .} \label{condh0}
\end{equation}
The success of the measurement process thus requires the conditions
(\ref{condtauinit}) and (\ref{condh0}) on the parameters $\tau_{{\rm init}}$, $T_{0}$, $h_{0}$
that characterize the preparation of the initial state of the apparatus.

 For a quartic interaction ($q=4$),  the initial paramagnetic state is metastable rather than unstable. Its spontaneous decay in
  the absence of a field requires $m$ to cross the potential barrier of the free energy which ensures metastability, as shown by  Fig. 3.3. 
 At temperatures $T$ below the transition point but not too low, the dynamics is governed by an activation process, 
 with a characteristic duration of order ($\hbar/\gamma J) \exp(\Delta F/T$), where $\Delta F$ is the height of the barrier, 
 for instance $\Delta F= 0.054NT$ for $T=0.2J$. The lifetime of the paramagnetic state is thus exponentially larger than the 
 registration time for large $N$, so that there is no hurry in performing the measurement after preparation of the initial state.

\vspace{3mm}

The above derivation holds for a large statistical ensemble $\scriptE$ of systems in both classical and quantum statistical mechanics. In the former case, the doubly peaked 
probability $P(m)$ reached at the final time can be interpreted in terms of the individual systems of $\scriptE$: the magnetization of half of these systems is expected to reach
$m_{\rm F}$, the other half $-m_{\rm F}$. However, this seemingly natural assertion requires a proof in quantum physics, due to the ambiguity of the decomposition of the ensemble 
$\scriptE$ into subensembles (\S~10.2.3). Such a proof is displayed in the last part of \S~\ref{fin11.2.3}; it relies on a relaxation process generated by specific interactions 
within the magnetic dot.

}

\subsubsection{Probability of wrong registrations for second order phase transitions of the magnet}
\label{section.7.3.3}


\hfill{\it Je suis malade, }

\hfill{\it compl\`etement malade\footnote{I am sick, completely sick}}

\hfill{Written by Serge Lama, sung by Dalida} 
\vspace{3mm}

\ZeText{

We have seen (\S~\ref{section.7.2.3}) how the magnet M, under the conjugate effect of B
and S, reaches quasi certainly the final magnetization $+m_{{\rm F}}$ in
the sector $\uparrow\uparrow$ where $s_{z}=+1$, provided $g$ is not too small.
We expect that if the condition (\ref{gmin}) on $g$ is violated, the apparatus
will indicate, with some probability ${\cal P}_{-}$, the wrong
magnetization $-m_{{\rm F}}$, although $s_{z}=+1$. The evolution of
$P_{\uparrow\uparrow}(m,t)$ in such a situation is illustrated by Fig. 7.8. 
A similar failure may occur if the average magnetization $\mu_{0}$ in the
initial state is not $0$ but takes a negative value due to a {\it biased preparation}.

\myskipfigText{
\begin{figure}[h!]
\label{figABN13}
\centerline{\includegraphics[width=8cm]{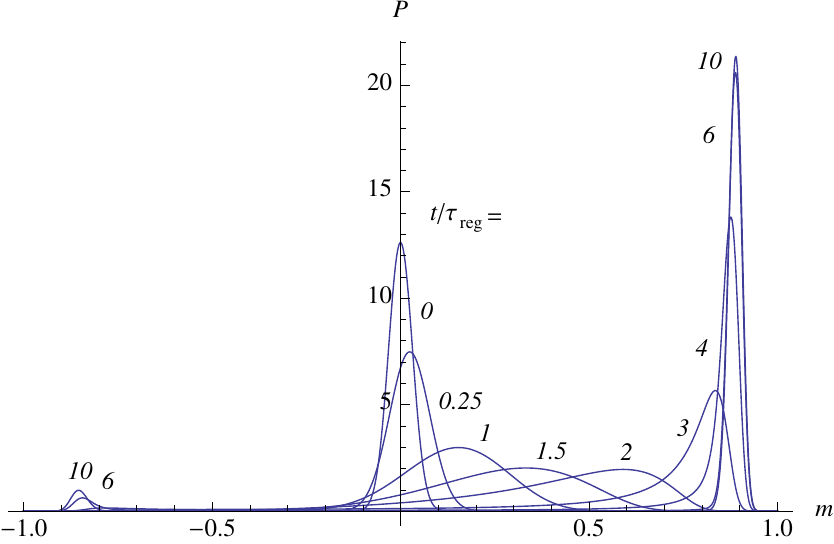}}
\caption{Wrong registration for quadratic interactions ($q=2$). 
The probability distribution $P(m,t)$ is represented at different times for the same parameters $N=1000$ and $T=0.65$ as in Fig. 7.5, 
but the coupling $g=0.03J$ is now sufficiently weak so that the apparatus registers the magnetization $-m_{\rm F}$ with a significant probability
${\cal P}_-$, although the system has a spin $s_z=+1$. Like in  Fig. 7.7,
the probability distribution flattens before the two ferromagnetic peaks emerge (with weights ${\cal P}_+$ and ${\cal P}_-$).}
\end{figure}
}

The \textit{probability} ${\cal P}_{-}$ \textit{of a wrong registration} $-m_{{\rm F}}$ for $s_{z}=+1$ arises from values $m<m_{\rm B}<0$ and reads
\begin{equation}
\mytext{\textcurrency Pminus\textcurrency \qquad}
{\cal P}_{-}=\int_{-1}^{m_{{\rm B}}}\d m\, \frac{P_{\uparrow\uparrow}(m,t)}{r_{\uparrow\uparrow}\left(
0\right)}  \equiv\int_{-1}^{m_{{\rm B}}}\d m\, P\left(  m,t\right)  { ,}
\label{Pminus}
\end{equation}
where the time $t$ is in principle such that
$P_{\uparrow\uparrow}(m,t)$ has reached its equilibrium shape,
with two peaks around $+m_{{\rm F}}$ and$\ -m_{{\rm F}}$. In
fact, we do not need the final equilibrium to have been reached
since (\ref{Pminus}) remains constant after $P_{\uparrow\uparrow
}$ has split into two separate parts. And even the latter
condition is not necessary: After the time $\tau_{{\rm reg}}$
the diffusion term becomes inactive and the evolution of
$P_{\uparrow\uparrow}(m,t)$ is governed by the pure drift Green's
function (\ref{AGw=0}); then there is no longer any transfer of
weight across the bifurcation $-m_{{\rm B}}=-g/(J-T)$. We can
therefore evaluate (\ref{Pminus}) at the rather early stage when
the distribution has not yet spread out beyond the small $m$
region where (\ref{VWsimp}) holds, provided we take
$t\gg\tau_{{\rm reg}}$.

We thus use the expression (\ref{Psplit}) of $P_{\uparrow\uparrow}(m,t)$ valid during the first stage of the process, 
which reads

\begin{equation}
\mytext{\textcurrency Psmallg\textcurrency \qquad}
P(m,t)=e^{-t/\tau_{{\rm reg}}}\sqrt{\frac{N}{2\pi }}\frac{1}{\delta_{1}(t)}\exp\left\{
-\frac{N}{2\delta_{1}^{2}(t)}\left[  (m+m_{{\rm B}})e^{-t/\tau
_{{\rm reg}}}-m_{{\rm B}}-\mu_{0}\right]  ^{2}\right\}  {
.}\label{Psmallg}
\end{equation}
By taking $(m+m_{{\rm B}})e^{-t/\tau_{{\rm reg}}}$ as variable we check that the integral  (\ref{Pminus}) depends on time only through the exponential in
(\ref{delta1t}), so that it remains constant as soon as $t\gg\tau_{{\rm reg}}$, when the second stage of the evolution is reached. We eventually find:

\begin{equation}
\mytext{\textcurrency Pminus=\textcurrency \qquad}
{\cal P}_{-}=\frac{1}
{2}{\rm erfc}\,\lambda{,\qquad}\lambda\equiv\sqrt{\frac{N}{2}}\frac
{1}{\delta_{1}}(m_{{\rm B}}+\mu_{0}){ ,}\label{Pminus=}
\end{equation}
where the error function, defined by
\begin{equation}
\mytext{\textcurrency erfc\textcurrency \qquad}
{\rm erfc}\,\lambda=\frac
{2}{\sqrt{\pi}}\int_{\lambda}^{\infty}{\rm d}\xi e^{-\xi^{2}}{
,}\label{erfc}
\end{equation}
behaves for $\lambda\gg1$ as
\begin{equation}
{\rm erfc}\,\lambda\sim\frac{1}{\sqrt{\pi}\lambda}e^{-\lambda^{2}}{ .}
\end{equation}
The diffusion which takes place during the first stage of the evolution has
changed in (\ref{Pminus=}) the initial width $\delta_{0}$ into $\delta_{1}$,
given by (\ref{delta1}).

For $\mu_{0}=0$, the probability of error becomes sizeable when $\sqrt{N}g/J$ is not sufficiently large.
For example, for $T=0.65J$ and $g=0.03J$, we find numerically 
${\cal P}_-= 21\%, \,13\%,\, 5.4\%, \,1.15\%$ and $0.065\%$ for $N = 250,\, 500,\, 1000,\, 2000$ and $4000$, respectively. 
These data are reasonably fitted by the approximation ${\cal P}_-(N) =1.2\, N^{-1/4} \exp(-0.0014 N)$ for (7.80). 
The result for $N = 1000$ is illustrated by the weight of the peak near $- m_{\rm F}$  in Fig. 7.8. 
False registrations were also present with the data of Fig. 7.5 ($N=1000$, $T=0.65 J$, $g=0.05J$), with a probability 
${\cal P}_- =0.36\%$, but the effect is too small to be visible on the scale of the figure.

The occurrence of a negative $\mu_{0}$ increases
${\cal P}_{-}$, an effect which, with the above data, becomes
sizeable for $\left\vert \mu _{0}\right\vert \sim0.05$. For
$P_{\downarrow\downarrow}$ the percentage of errors is given by
(\ref{Pminus=}) with $\mu_{0}$ changed into $-\mu_{0}$ in
$\lambda$.

We write for completeness in Appendix F the evolution of the shape of $P(m,t)$. This is not crucial for the measurement problem (for which
$P_{\uparrow\uparrow}\left(  m,t\right)  =r_{\uparrow\uparrow}\left(  0\right)  P\left(m,t\right)  $), 
but it is relevant for the dynamics of the phase transition, depending on the initial conditions and on the presence of a parasite field.
Here again, Suzuki's regime \cite{suzuki,suzuki1,suzuki2,suzuki3,suzuki4}, where the distribution is no longer peaked, is reached for
$t\gg\tau_{\rm reg}$. Now $P(m,t)$ is asymmetric, but it still has a quasi linear behavior in a wide range 
around $m=0$ when $\tau\simeq \tau_{\rm flat}$ (see Eqs. (\ref{tauflat}), (7.70)).

}

\subsubsection{Possible failure of registration for first order transitions}
\label{section.7.3.4}

\hfill{\it Quem n\~ao tem c\~ao, }

\hfill{\it ca\c ca como gato
\footnote{Who has no dog, hunts as a cat} }

\hfill{Portuguese proverb}

\vspace{3mm}

\ZeText{

The situation is quite different for first-order transitions ($q = 4$) as regards the possibility of wrong registrations. 
Note first that $F(m)$ has a high maximum for negative $m$ between $0$ and $m_\Downarrow<0$ (Figs. 3.3 and 3.4),
 which constitutes a practically impassable barrier that diffusion is not sufficient to overcome. 
 Accordingly, the zero of $v(m)$ at $m= - m_{\rm B}\simeq -2m_{\rm c}$ is a repulsive fixed point 
 (Fig.7.2 and \S~7.2.4), which prevents the distribution from developing a tail below it. 
 We shall therefore never find any registration with negative ferromagnetic magnetization in the sector $s_z=+1$.

\myskipfigText{
\begin{figure}[h!]
\label{figABN14}
\centerline{\includegraphics[width=8cm]{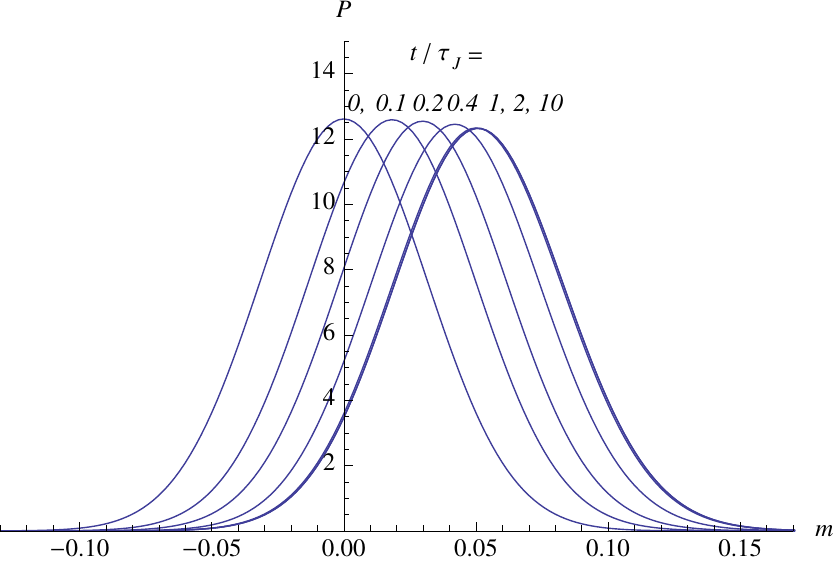}}
\caption{ 
 Failure of measurement for quartic interactions ($q=4$). The probability distribution $P(m,t)$ is represented at times
 up to $10\,\tau_J$, where $\tau_J=\hbar/\gamma J$.  The parameters are $N=1000$ and $T=0.2$ as in Figs. 7.4 and 7.6,
 but here $g=0.01J$ lies below the threshold $h_{\rm c}$. The peak evolves towards metastable paramagnetic equilibrium in the presence 
 of the field $g$, but $g$ is too small to allow crossing the barrier and reaching the more stable ferromagnetic equilibrium around 
 $m_{\rm F} \simeq 1$.  Switching off the coupling $g$ brings back the distribution to its original place around $0$, so that no proper
 registration is achieved.
} 
\end{figure}
}

Nevertheless, we have seen (\S~7.2.2) that registration is possible only if the coupling $g$ exceeds $h_{\rm c}$. 
For $g<h_{\rm c}$, the peak of $P(m)$ initially at $m=0$ moves upwards in the sector $\uparrow\uparrow$ associated with 
$s_z=+1$ (Fig. 7.9), and ends up by stabilizing at the first attractive point encountered, at $m=m_{\rm P}$ (Fig. 7.2). 
Symmetrically, the distribution of $P_{\downarrow\downarrow}(m)$ ends up at $- m_{\rm P}$. 
However this difference between the two values of $s_z$ cannot be regarded as a registration since switching off the coupling 
$g$ between S and A brings back both distributions $P_{\uparrow\uparrow}$ and $P_{\downarrow\downarrow}$ 
to the initial Gaussian shape around $m=0$. The apparatus A then always relaxes back to the locally stable paramagnetic state.

Finally, if the coupling $g$, although larger than the threshold $h_{\rm c}$, is close to it, the registration takes place correctly 
provided this coupling remains active until the distribution $P_{\uparrow\uparrow}(m)$ has completely passed the bifurcation 
$m_{\rm c} \sqrt{3}$ occurring for $g=0$ (Eq. (7.56)). 
The lower bound $t_{\rm off}$ of the time when $g$ can thus be safely switched off is close to 
$\tau_{\rm reg}$ (which is also close to the time needed to reach ferromagnetic equilibrium). 

In case S and A are decoupled too early, so that the condition (7.56) is violated, the tail of $P_{\uparrow\uparrow}(m,t)$ 
lying below the bifurcation $m = m_{\rm c}\sqrt{3}$ is pushed back towards the paramagnetic region 
$m\approx 0$.  If the decoupling $g\to0$ is made suddenly at the time $t_{\rm off}$, the probability ${\cal P}_0$ of such 
events can be evaluated as in \S~7.3.3 in terms of the error function 
by integration of $P_{\uparrow\uparrow}(m,t_{\rm off})$  from $m=-1$ up to $m_{\rm c}$. 
It represents the {\it probability of aborted measurement processes}, for which the apparatus returns to its neutral paramagnetic 
state without giving any indication, while S is left in the state $|\hspace{-1mm}\uparrow\rangle$. 
In a set of repeated measurements, a proportion ${\cal P}_0$ of runs are not registered at all, the other ones being registered correctly.

}

\subsubsection{Erasure of the pointer indication}

\hfill{\it De dag van morgen deelt met zijn eigen zorgen\footnote{The day of tomorrow addresses its own worries}}

\hfill{Dutch proverb}

\vspace{3mm}

\ZeText{
 
As shown in \S\S~7.2.3 and 7.2.4, the registration is achieved at a time $t_{\rm f}$ sufficiently larger than the delay $\tau_{\rm reg}$
after which S and M have been decoupled. The state $\hat{\cal D}(t_{\rm f})$ of S + A is then given by the expected expression (\ref{Dtf=}). 
Within the considered approximations, the distributions $P_{\uparrow\uparrow}(m, t)$ and $P_{\downarrow\downarrow}(m, t)$ 
no longer evolve for $t>t_{\rm f}$, and remain fully concentrated near $m_{\rm F}$ and $- m_{\rm F}$, respectively, 
so that the results can be read out or processed at any observation time $t_{\rm obs} > t_{\rm f}$. 
However, the breaking of invariance, on which we rely to assert that the two ferromagnetic states of the pointer are stationary, 
is rigorous only in the large $N$ limit. Strictly speaking, for finite $N$, the states $\hat R_{{\rm M}\Uparrow}$ and 
$\hat R_{{\rm M}\Downarrow}$ reached by M at this stage in each sector are not in equilibrium (though they may have a long lifetime). 
Indeed, in the Markovian regime, we have shown in \S~\ref{section.7.1.3} that the evolution of M under the influence of the thermal bath 
cannot stop until $P_{\rm M}(m,t)$ becomes proportional to $G(m) \exp[-\beta E(m)]$, with $E(m)=-JNq^{-1} m^q$. 
Otherwise, the time-derivative (7.21) of the free energy $F(m)$ of the state $\hat R_{{\rm M}}(t)$ cannot vanish. 
The limit reached by $\hat R_{\uparrow\uparrow}(t)/r_{\uparrow\uparrow}(0)$ (and of 
$\hat R_{\downarrow\downarrow}(t)/r_{\downarrow\downarrow}(0))$ 
is then  $\half(\hat R_{{\rm M}\Uparrow}+\hat R_{{\rm M}\Downarrow})$. 
 Hence, when the latter true equilibrium state for finite $N$ is attained, the indication of the pointer is completely random. 
 We have lost all information about the initial state of S, and the spin S has been completely depolarized whatever its initial state:  the result of the measurement 
 has been washed out. We denote as $\tau_{\rm eras}$ the characteristic time which governs this erasure of the indication of the pointer.

It is therefore essential to read or process the registered data before such a loss of memory begins to occur\footnote{Photographs on film or paper fade out after some time}. 
The observation must take place at a time $t_{\rm obs}$ much shorter than the erasure time:

\BEQ                         \tau_{\rm reg}   <   t_{\rm f}  <  t_{\rm obs}  \ll  \tau_{\rm eras} .  
\EEQ

The dynamics of the erasure, a process leading M from $\hat R_{{\rm M}\Uparrow}$ or $\hat R_{{\rm M}\Downarrow}$
to the state $\half(\hat R_{{\rm M}\Uparrow}+\hat R_{{\rm M}\Downarrow})$ of complete equilibrium, is governed by the Eq. (\ref{dPdiag})
for $P_{\rm M}(m,t)$  (with $\tilde K_t(\omega)$ replaced by $\tilde K(\omega)$ and $g=0$), which retains the quantum character 
of the apparatus. We will rely on this equation in subsection 8.1 where studying the Curie--Weiss model in the extreme case of $N=2$. 
For the larger values of $N$ and the temperatures considered here, we can use its continuous semi-classical limit (7.1), to be solved for 
an initial condition expressed by (\ref{F=}) with $m_i=m_{\rm  F}$ or $-m_{\rm F}$. Here we have to deal with the progressive, 
very slow leakage of the distribution $P_{\rm M}(m,t)$ from one of the ferromagnetic states to the other through the free energy 
barrier that separates them. This mechanism, disregarded in \S\S~7.2.3, 7.2.4, 7.3.3 and 7.3.4, is controlled by the weak tail of 
the distribution $P_{\rm M}(m,t)$ which extends into the regions of $m$ where $F(m)$ is largest. The drift term of (7.1) alone 
would repel the distribution $P_{\uparrow\uparrow}(m, t)$ and keep it concentrated near $m_{\rm F}$. 
An essential role is now played by the diffusion term, which tends to flatten this distribution over the whole range of $m$, 
and thus allows the leak towards $-m_{\rm F}$. Rather than solving this equation, it will be sufficient for our purpose to rely 
on a semi-phenomenological argument: Under the considered conditions, the full equilibration is an activation process 
governed by the height of the free energy barrier. Denoting as $\Delta F$ the difference between the maximum of 
$F(m)$ and its minimum, $F_{\rm ferro}$, we thus estimate the time scale of erasure as:

\BEQ
\tau_{\rm eras} \sim \frac{\hbar}{\gamma J} \exp \frac{\Delta F}{T}, 
\EEQ
which is large as an exponential of $N$. In order to use the process as a measurement, we need this time to be much larger than the registration time 
so that we are able to satisfy (7.83), which yields

\BEQ                    \frac{J}{J-T}\ll  \exp \frac{\Delta F}{T},\quad  (q=2) ;  \qquad  \frac{J}{T} \sqrt{\frac{ m_{\rm c}T}{ g-h_{\rm c}}}
\ll  \exp \frac{\Delta F}{T}
,\quad  (q=4).
\EEQ

 From (\ref{F=}) (taken for $h=0$), we find the numerical value of $\Delta F/T$ for the examples of figs. 7.5 and 7.6, namely 
 $0.130 N$ for $q=2$, $T=0.065J$, and $0.607 N$ for $q=4$, $T=0.2J$ (see fig. 3.3). The condition (7.85) sets again a 
 lower bound on $N$ to allow successful measurements, $N \gg 25$ for the example with quadratic interactions, 
 $N \gg 7$ for the example with quartic interactions. Such a condition is violated for a non-macroscopic apparatus,
  in particular in the model with $N=2$ treated below in subsection 8.1 which will require special care to ensure registration.

}

\subsubsection{\textquotedblleft Buridan's ass\textquotedblright effect: hesitation}

\hfill{\it  Do not hesitate,}

\hfill{\it  or you will be left in between doing something,}

\hfill{\it  having something and being nothing}

\hfill{Ethiopian proverb}

\vspace{3mm}

\label{section.7.3.5}

\ZeText{

In the case of a second-order transition ($q=2$), the subsections 7.2 and 7.3, illustrated by Figs. 7.5, 7.7 and 7.8, 
show off the occurrence, for the evolution of the probability distribution $P(m,t)$, of two contrasted regimes, depending
whether the bifurcation $-m_{{\rm B}}$ is active or not. The mathematical problem is the same as for many problems of
statistical mechanics involving dynamics of instabilities, such as directed Brownian motion near an unstable fixed point, and it has
been extensively studied \cite{suzuki,suzuki1,suzuki2,suzuki3,suzuki4}. The most remarkable feature is the behavior exemplified by Fig. 7.7:
For a long duration, the random magnetization $m$ hesitates so much between the two stable values
$+m_{{\rm F}}$ and $-m_{{\rm F}}$ that a wide range of values of $m$ in the interval $-m_{{\rm F}}$, $+m_{{\rm F}}$
have nearly equal probabilities. We have proposed to term this anomalous situation {\it Buridan's ass effect} ~\cite{ABNconf4}, 
referring to the celebrated argument attributed to Buridan, a dialectician of the first half of the XIVth century: An ass placed just half way between
two identical bales of hay would theoretically stay there indefinitely and starve to death, because the absence of causal reason to choose one bale or
the other would let it hesitate for ever, at least according to Buridan\footnote{The effect was never observed, though, at the farm where the last author of the present work grew up}.

In fact, major qualitative differences distinguish the situation in which the final state $+m_{{\rm F}}$ is reached with probability ${\cal P}_{+}=1$
(subsection 7.2) from the situation in which significant probabilities ${\cal P}_{+}$ and ${\cal P}_{-}$ to reach either $+m_{{\rm F}}$ or
$-m_{{\rm F}}$ exist (subsection 7.3). In the first case, the peak of $P(m,t)$ moves simply from $0$ to $+m_{{\rm F}}$; the \textit{fluctuation
of} $m$ \textit{remains of order} ${1}/\sqrt{N}$ at all times, even when it is largest, at the time when the average drift velocity $v(\mu)$ is maximum (Eq.
(\ref{Dofv})). In the second case, the exponential rise of the fluctuations of $m$ leads, during a long period, to a broad and flat distribution $P(m,t)$,
with a \textit{shape independent of} $N$.

In both cases, we encountered (for $q=2$) the same \textit{time scale} $\tau_{{\rm reg}}=\hbar/\gamma(J-T)$, which characterizes the first stage
of the motion described by either (\ref{mu1}), (\ref{D1}) or (\ref{Psmallg}). However, in the first case, ${\cal P}_+\simeq1$,
 the duration (\ref{tauprime}) of the whole process is just the product of $\tau_{{\rm reg}}$ by a factor independent of $N$,
of order $2\ln[m_{{\rm F}}(J-T)/g]$, as also shown by (\ref{muallt}), whereas in the second case, ${\cal P}_+<1$, 
the dynamics becomes infinitely slow in the large $N$ limit. The characteristic time $\tau_{{\rm flat}}$ at which the
distribution is flat, given by (\ref{tauflat}), is of order $\tau_{{\rm reg}}\ln[\sqrt{N}m_{{\rm F}}(J-T)/J]$. Suzuki's scaling regime 
\cite{suzuki,suzuki1,suzuki2,suzuki3,suzuki4} is attained over times of order $\tau_{{\rm flat}}$. Then $P(m,t)$ does not
depend on $N$ for $N\rightarrow\infty$, but the duration of the relaxation process is \textit{large as} $\ln\sqrt{N}$. It is this long delay which allows
the initial distribution, narrow as ${1}/\sqrt{N}$, to broaden enormously instead of being shifted towards one side.

Buridan's argument has been regarded as a forerunner of the idea of probability. The infinite time during which the ass remains at
$m=0$ is recovered here for $N\rightarrow\infty$. An infinite duration of the process is also found in the absence of diffusion
in the limit of a narrow initial distribution ($\delta_{0}\rightarrow0$). The flatness of $P(m,t)$ at times of
order $\tau_{{\rm flat}}$ means that at such times \textit{we cannot predict} at all where the ass will be on the interval
$-m_{{\rm F}}$, $+m_{{\rm F}}$, an idea that Buridan could not emit before the elaboration of the concept of probability. The
counterpart of the field $h$, for Buridan's ass, would be a strong wind which pushes it; the counterpart of $\mu_{0}$ would be a
different distance from the two bales of hay; in both of these cases, the behavior of the ass becomes predictable within small fluctuations.

Since the slowing factor which distinguishes the time scales in the two regimes is \textit{logarithmic}, very large values of $N$ are required to
exhibit a large ratio for the relaxation times. In Figs. 7.5, 7.7 and 7.8 we have taken $N=1000$ so as to make the fluctuations in ${1}/\sqrt{N}$ visible. As a
consequence, the duration of the registration is hardly larger in Fig. 7.7 ($q=2$, bias) than in Fig. 7.5 ($q=2$, no bias).

Except during the final equilibration, the magnet keeps during its evolution some memory of its initial state through $\delta_{1}$ (Eq. (\ref{delta1})). If
the bifurcation is inactive (\S~\ref{section.7.2.3}), this quantity occurs through the variance (\ref{Dofv}) of the distribution. If it is active (\S~\ref{section.7.3.2}), 
it occurs through the time scale $\tau_{{\rm flat}}$, but not through the shape of $P(m,t)$.

Our model of the ferromagnet is well-known for being exactly solvable at equilibrium in the large $N$ limit by means of a static mean-field approach.
In the single peak regime, the dynamics expressed by (\ref{t(mu)}) is also the same as the outcome of a time-dependent mean-field approach. However, 
in the regime leading to two peaks at $+m_{{\rm F}}$ and $-m_{{\rm F}}$, \textit{no mean-field approximation can describe the dynamics} even for large
$N$, due to the giant fluctuations. The intuitive idea that the variable $m$, because it is macroscopic, should display fluctuations small as ${1}/\sqrt{N}$
is then wrong, except near the initial time or for each peak of $P(m,t)$ near the final equilibrium.

The giant fluctuations of $m$ which occur in Buridan's ass regime may be regarded as a dynamic counterpart of the fluctuations that occur at
equilibrium at the critical point $T=J$ \cite{ll_stat,lavis}. In both cases, the order parameter, although macroscopic, presents large
fluctuations in the large $N$ limit, so that its treatment requires statistical mechanics. Although no temperature can be associated with
$M$ during the relaxation process, the transition from $T_{0}>J$ to $T<J$ involves intermediate states which behave as in the critical
region. The well-known critical fluctuations and critical slowing down manifest themselves here by the large uncertainty on $m$ displayed
during a long delay by $P(m,t)$. 

Suzuki's slowing down and flattening \cite{suzuki,suzuki1,suzuki2,suzuki3,suzuki4} take place not only in the symmetric case (\S~\ref{section.7.3.2}),  
but also in the asymmetric case (\S~\ref{section.7.3.3}), provided ${\cal P}_{-}$ is sizeable. Thus the occurrence of Buridan's ass effect is governed by the
non vanishing of the probabilities ${\cal P}_{+}$ and ${\cal P}_{-}$ of $+m_{{\rm F}}$ and $-m_{{\rm F}}$ \textit{in the final state}.
Everything takes place as if the behavior were governed by final causes: The process is deterministic if the target is unique; it displays large
uncertainties and is slow if hesitation may lead to one target or to the other. These features reflect in a probabilistic language, first, the slowness of the pure drift 
motion near the bifurcation which implies a long random delay to set $m$ into motion, and, second, the importance of the diffusion term there. 

}


\renewcommand{\thesection}{\arabic{section}}
\section{Imperfect measurements, failures and multiple measurements}
\setcounter{equation}{0}\setcounter{figure}{0}
\renewcommand{\thesection}{\arabic{section}.}

\label{section.8}

\hfill{\it Niet al wat blinkt is goud \footnote{\label{Gold}All that glitters is not gold}}

\hfill{\it Tout ce qui brille n'est pas or $^{\ref{Gold}}$}

\hfill{Dutch and french proverbs}

\ZeText{

\vspace{3mm}

In sections 5 to 7, we have solved our model under conditions on the various parameters which ensure that the measurement is ideal.
We will resume these conditions in section~\ref{section.9.1.3}. We explore beforehand some situations in which they may be violated, so as to
set forth how each violation prevents the dynamical process from being usable as a quantum measurement. We have already seen that,
in case the spin-apparatus interaction presents no randomness, the {\it magnet-bath interaction} should not be too small; otherwise,
recurrence would occur in the off-diagonal blocks of the density operator, and would thus prevent their truncation (\S~\ref{section.5.1}). 
We have also shown how a spin-apparatus coupling that is too weak may prevent the registration to take place for $q=4$ (\S~\ref{section.7.2.2} 
and \S~\ref{section.7.2.4}), or may lead to wrong results for $q=2$ (\S~\ref{section.7.3.3}). We study below what happens if the number 
of {\it degrees of freedom of the pointer} is small, by letting $N=2$ (subsection \ref{section.8.1});  see \cite{cini,gaveau_schulman} for model 
studies along this line. We then examine the importance of the {\it commutation} \cite{vNeumann,Wigner,Yanase,WAY,ozawa_way,meister_way} 
of the measured observable with the Hamiltonian of the system (subsection \ref{section.8.2}). Finally we exhibit a process
which might allow imperfect {\it simultaneous measurements of non-commuting observables}  (subsection \ref{section.8.3})
\cite{arthurs_complo,martens_complo,appleby_complo,luis_complo,lorenzo_complo,busch_complo}.

The solution of these extensions of the Curie--Weiss model involves many technicalities that we could not skip. 
The reader interested only in the results will find them in subsection 9.5.

}

\subsection{Microscopic pointer}
\label{section.8.1}
\hfill{{\it Ce que je sais le mieux,}

\hfill{\it  c'est mon commencement}\footnote{ What I know the best I shall begin with}}

\hfill{ Jean Racine, Les Plaideurs} 
\vspace{3mm}

\ZeText{

In the above sections, we have relied on the large number $N$ of degrees of freedom of the magnet M. As the statistical
fluctuations of the magnetization $m$ are then weak, the magnet can behave as a macroscopic pointer with classical features.
Moreover the truncation time $\tau_{\rm trunc}$ is the shortest among all the characteristic times (section 5) because it behaves as
${1}/\sqrt{N}$. The large value of $N$ was also used (section 7) to describe the registration process  by means of a partial
differential equation. It is natural to wonder whether a small value of $N$ can preserve the characteristic properties of a
quantum measurement. Actually the irreversibility of any measurement process (subsection 6.2) requires the apparatus to be
large. In subsection~\ref{section.6.1}, we showed that the irreversibility of the truncation can be ensured by a large value of $N$ and a
randomness in the couplings $g_n$, $n=1,\cdots, N$ (subsection 6.2); but this irreversibility, as well as that of the registration
(section 7), can also be caused by the large size of the bath. For small $N$, the irreversibility of both the truncation and the
registration should be ensured by the bath. We now study the extreme situation in which $N=2$.

}

\subsubsection{Need for a low temperature}
\label{section.8.1.1}

\ZeText{

For $N=2$ the magnetization $\hat m$ has the eigenvalue $m=0$ with multiplicity 2, regarded as ``paramagnetic'', and two non-degenerate eigenvalues 
$m=+1$ and $m=-1$ regarded as ``ferromagnetic".  Since $\hat m^4=\hat m^2$, we may set $J_4=0$ and denote $J_2=J$.
The corresponding eigenenergies of $\hat H_M$ are 0 and $-J$, and those of the Hamiltonian $\hat H_i$ of Eq. (\ref{Hi}) are $-2gs_im-Jm^2$.

The equations of motion of \S~\ref{section.4.3.2} involve only the two frequencies $\omega_\pm$, defined by

\begin{equation} 
\hbar\omega_\pm\equiv J\pm 2g, \nonumber
\end{equation} 
and they have the detailed form (notice that $P_{ij}\equiv\half N P^{\rm dis}_{ij}=P^{\rm dis}_{ij}$ for $N=2$)

\BEA \label{0upup}
\frac{{\rm d}P_{\uparrow\uparrow}(0,t)}{{\rm
d}t}&=&\frac{2\gamma}{\hbar^2}
\left\{2P_{\uparrow\uparrow}(1,t)\tilde
K_t(\omega_+)+2P_{\uparrow\uparrow}(-1,t)\tilde K_t(\omega_-)
-P_{\uparrow\uparrow}(0,t)\left[\tilde K_t(-\omega_+)
+\tilde K_t(-\omega_-)\right]\right\}, \\
\label{1upup} 
\frac{{\rm d}P_{\uparrow\uparrow}(\pm1,t)}{{\rm d}t}&=&\frac{2\gamma}{\hbar^2}
\left[P_{\uparrow\uparrow}(0,t)\tilde K_t(-\omega_\pm)
-2P_{\uparrow\uparrow}(\pm1,t)\tilde K_t(\omega_\pm)\right], \\
\label{0updown} 
\frac{{\rm d}P_{\uparrow\downarrow}(0,t)}{{\rm d}t}&=&\frac{2\gamma}{\hbar^2}
\left\{ 2P_{\uparrow\downarrow}(1,t)\left[\tilde
K_{t>}(\omega_+)+\tilde K_{t<}(\omega_-)\right]
+2P_{\uparrow\downarrow}(-1,t)\left[\tilde K_{t>}(\omega_-)+\tilde
K_{t<}(\omega_+)\right] 
\right.\nn\\&&\qquad\left.
-P_{\uparrow\downarrow}(0,t)\left[\tilde K_{t}(-\omega_+)
+\tilde K_{t}(-\omega_-)\right]\right\}, \\
\label{1updown} 
\frac{{\rm d}P_{\uparrow\downarrow}(\pm1,t)} {{\rm
d}t}&\mp&\frac{4ig}{\hbar}P_{\uparrow\downarrow}(\pm1,t)
=\frac{2\gamma}{\hbar^2} \left\{
P_{\uparrow\downarrow}(0,t)\left[\tilde K_{t>}(-\omega_\pm)
+\tilde K_{t<}(-\omega_\mp)\right]
-2P_{\uparrow\downarrow}(\pm1,t)\left[\tilde K_{t>}(\omega_\pm)
+\tilde K_{t<}(\omega_\mp)\right]\right\}. \EEA 
As initial state for $M$ we take the ``paramagnetic'' one, $P_{\rm M}(0)=1$, $P_{\rm M}(\pm1)=0$,
prepared by letting $T_0\gg J$ or with a radiofrequency field as in \S~\ref{section.3.3.3}. 
(We recall that $P_{\rm M}=P_{\uparrow\uparrow}+P_{\downarrow\downarrow}$.)
The initial conditions are thus $P_{ij}(m,0)=r_{ij}(0)\delta_{m,0}$.

In order to identify the process with an ideal measurement, we need at least to find at sufficiently large times ($i$) the truncation, 
expressed by $P_{\uparrow\downarrow}(m,t)\to 0$, and ($ii$) the system-pointer  correlations  expressed by 
$P_{\uparrow\uparrow}(m,t)\to r_{\uparrow\uparrow}(0)\delta_{m,1}$
and $P_{\downarrow\downarrow}(m,t)\to r_{\downarrow\downarrow}(0)\delta_{m,-1}$. 
This requires, for the magnet in contact with the bath, a long lifetime for the
``ferromagnetic" states $m=+1$ and $m=-1$. However, the breaking
of invariance, which for large $N$ allows the ferromagnetic state
where $m$ is concentrated near $+m_{\rm F}$ to be stable, cannot occur
here: Nothing hinders here the coupling with the bath to induce
transitions from $m=+1$ to $m=-1$ through $m=0$, so that for large
times $P(+1,t)$ and $P(-1,t)$, where $P(m,t) \equiv P_{\uparrow\uparrow} (m,t)/r_{\uparrow\uparrow}(0)$, 
tend to a common value close to $\half$ for $T\ll J$.

This is made obvious by the expression of (\ref{dis}) of the $H$-theorem.
The dissipation in the Markovian regime \cite{petr,Weiss,Gardiner} reads here

\BEA
\frac{{\rm d}F(t)}{{\rm d}t}&=&-\frac{\gamma}{2\beta}\frac{\omega_+
e^{-|\omega_+|/\Gamma}}{e^{\beta\hbar\omega_+}-1}
\left[P(0,t)e^{\beta\hbar\omega_+}-2P(1,t)\right]\ln\frac{P(0,t)e^{\beta\hbar\omega_+}}{2P(1,t)}
+\left[\omega_+\mapsto\omega_-,\quad P(1,t)\mapsto P(-1,t)\right],
\EEA 
and the free energy decreases until the equilibrium
$2P_{\rm M}(\pm1)=P_{\rm M}(0)\exp{\beta\hbar\omega_\pm}$ is reached. The only
possibility to preserve a long lifetime for the state $m=+1$ is to
have a low transition rate from $m=+1$ to $m=0$, that is,
according to (\ref{0upup}), a small $\tilde K_t(\omega_+)$. This
quantity is dominated in the Markovian regime by a factor
$\exp({-\beta\hbar\omega_+})$. Hence, unless $T\ll J$, the apparatus
cannot keep the result of the measurement registered during a
significant time, after the interaction with S has been switched
off. If this condition is satisfied, we may expect to reach for
some lapse of time a state where
$P(1,t)=P_{\uparrow\uparrow}(1,t)/r_{\uparrow\uparrow}(0)$ remains
close to $1$ while $P(0,t)$ is small as $P(-1,t)$.

Moreover, a faithful registration requires that the coupling $g$ with S
is sufficiently large so that the final state, in the evolution of
$P_{\uparrow\uparrow}(m,t)$, has a very small probability to yield
$m=-1$. Since in the Markovian regime the transition probabilities in
(\ref{0upup}) and (\ref{1upup}) depend on $g$ through $\omega_\pm=J\pm
2g$ in $\tilde K(\omega_\pm)$ and $\tilde K(-\omega_\pm)$
\cite{petr,Weiss,Gardiner}, and since this dependence arises mainly from
$\exp{\beta\hbar\omega_\pm}$, we must have $\exp{4\beta g}\gg1$. The
coupling $g$ should moreover not modify much the spectrum, so that we
are led to impose the conditions

\BEA
\label{condN=2} 
T\ll4g\ll J.
\EEA

}

\subsubsection{Relaxation of the apparatus alone} 
\label{section.8.1.2}

\hfill{\it Laat hem maar met rust\footnote{Better leave him alone}}

\hfill{Dutch expression}

\vspace{3mm}

\ZeText{

As we did in \S~\ref{section.7.3.2} for large $N$, we focus here on the evolution of the probabilities 
$P(m,t)\equiv P_{\uparrow\uparrow}(m,t)/r_{\uparrow\uparrow}(0)$ for the apparatus alone. 
It is governed by equations (\ref{0upup}) and (\ref{1upup}) in which $\omega_+=\omega_-=J/\hbar$.
For a weak coupling $\gamma$ we expect that the Markovian regime, where $\tilde K_t(\omega)=\tilde K(\omega)$
will be reached before the probabilities have deviated much from their initial value.
The equations of motion then reduce to
\BEA
\label{0Mark} 
\tau\frac{{\rm d}P(0,t)}{{\rm d}t}=e^{-J/T}\left[P(1,t)+P(-1,t)\right]-P(0,t),
\qquad
\label{1Mark}
\tau\frac{{\rm d}P(\pm1,t)}{{\rm d}t}=\half P(0,t)-e^{-J/T}P(\pm1,t),
\EEA
where we made use of
\begin{equation} 
\tilde K\left(\frac{J}{\hbar}\right)=e^{-J/T}\tilde K\left(-\frac{J}{\hbar}\right)=\frac{\hbar J}{4} \frac{e^{-J/\hbar\Gamma}}{e^{J/T}-1},
\nn
\end{equation} 
as well as $J/T \gg1$ and $J/\hbar\Gamma\ll1$,
and where we defined a characteristic time related to the spin-spin coupling as

\begin{equation} 
\tau\equiv\tau_J=\frac{\hbar}{\gamma J}. 
\end{equation} 
The Markovian approximation is justified provided this characteristic time scale $\tau$ is longer
than the time $t$ after which $\tilde K_t(\omega)=\tilde K(\omega)$, that is, for
\begin{equation}  \gamma\ll\frac{T}{J}.\nn
\end{equation} 

The general solution of (\ref{0Mark}), 
 obtained by diagonalization, is expressed by
\begin{eqnarray}
&&P(0,t)+P(1,t)+P(-1,t)=1,\nn\\
&&P(0,t)-e^{-J/T}\left[P(1,t)+P(-1,t)\right]\propto \exp\left[-\frac{t}{\tau}(1+ e^{-J/T})\right]
\approx \exp\left(-\frac{t}{\tau}\right), \\
&&P(1,t)-P(-1,t)\propto \exp\left(-\frac{t}{\tau} e^{-J/T}\right). \nn
\end{eqnarray} 

Let us first consider the {\it relaxation of the initial paramagnetic state}, for which $P(0,0)=1$ and $P(\pm1,0)=0$.
We find from the above equations
\begin{equation} 
P(0,t)=\frac{e^{-t/\tau}+e^{-J/T}}{1+e^{-J/T}},\qquad
P(1,t)=P(-1,t)=\frac{1-e^{-t/\tau}}{2(1+e^{-J/T})}.
\nn
\end{equation} 
The lifetime of this initial unstable state is therefore $\tau=\hbar/\gamma J$.
In a measurement, the interaction $g$
between S and A must thus be switched on rapidly after the preparation (\S~\ref{section.3.3.3}),
in a delay $\tau_{\rm init}\ll\tau$
so that $P(0)$ is still close to $1$ when the measurement process begins.

We now evaluate the delay $\tau_{\rm obs}$ {\it during which the
pointer keeps its value and can be observed}, after the measurement is achieved and
after the coupling with S is switched off. If in the sector
$\uparrow\uparrow$ the value $m=1$ is reached at some time $t_1$ with a
near certainty, the probabilities evolve later on, according to the above equations, as 

\begin{equation} 
P(0,t_1+t)=\frac{(1-e^{-t/\tau})e^{-J/T}}{1+e^{-J/T}},\qquad
P(\pm1,t_1+t)=\half\left[\frac{1+e^{-t/\tau}e^{-J/T}}{1+e^{-J/T}}
\pm\exp\left(-\frac{t}{\tau}e^{-J/T}\right)\right].
\nn \end{equation}  As expected, the information is lost for $t\to\infty$,
or, more precisely, for $t\gg\tau \exp({J/T})$, since $P(1,t)$ and
$P(-1,t)$ then tend to $\frac{1}{2}$. However, during the time
lapse $\tau \ll t \ll \tau \exp({J/T})$, $P(1,t)$ retains a value
$1-\frac{1}{2}\exp({-J/T})$ close to $1$, so that the probability of a
false registration is then weak. Although microscopic, the pointer
is a rather robust and reliable device provided $T\ll J$, on the time
scale $t\ll\tau_{\rm obs}$ where the {\it observation time} is

\begin{equation} 
 \label{tauobs} 
 \tau_{\rm obs}=\tau e^{J/T}=\frac{\hbar}{\gamma J}e^{J/T}. 
 \end{equation} 

}

\subsubsection{Registration}
\label{section.8.1.3}

\ZeText{

We  now study the time-dependence of the registration process, and
determine the probability to reach a false result, that is, to
find $m=-1$ in the sector $\uparrow\uparrow$. In the Markovian
regime and under the conditions (\ref{condN=2}), the equations of
motion (\ref{0upup}), (\ref{1upup}) for the probabilities
$P(m,t)=P_{\uparrow\uparrow}(m,t)/r_{\uparrow\uparrow}(0)$ read

\BEA
\label{0reg} 
\tau\frac{{\rm d}P(0,t)}{{\rm d}t}&=&e^{-(J+2g)/T}P(1,t)+e^{-(J-2g)/T}P(-1,t)-P(0,t),
\\
\label{1reg} 
\tau\frac{{\rm d}P(\pm1,t)}{{\rm d}t}&=&\half P(0,t)-e^{-(J\pm 2g)/T}P(\pm1,t). 
\end{eqnarray} 
 We have disregarded in each term contributions of relative order
$\exp({-J/T})$ and $2g/J$. The general solution of Eqs. (\ref{0reg}),
(\ref{1reg}) is obtained by diagonalizing their $3\times 3$
matrix. Its three eigenvalues $-z$ are  the solutions of 

\begin{equation} 
z^3-z^2\left(1+2e^{-J/T}\cosh\frac{2g}{T}\right)+ze^{-J/T}
\left(\cosh\frac{2g}{T}+e^{-J/T}\right)=0,\nn
\end{equation}  
that is, apart from $z=0$, 

\begin{equation} 
z=\half+e^{-J/T}\cosh\frac{2g}{T}\pm
\half\sqrt{1+4e^{-2J/T}\sinh^2\frac{2g}{T}},\nn \end{equation} 
 which under the conditions (\ref{condN=2}) reduce to $z\simeq 1$ and $z\simeq
\exp({-J/T})\cosh{2g}/{T}\simeq\half \exp[{-(J-2g)/T}]$. The
corresponding characteristic times ${\tau}/{z}$ are therefore
$\tau=\hbar/\gamma J$ and 

\begin{equation}  \label{regN=2}  
\tau_{\rm reg}=2\tau e^{(J-2g)/T}=\frac{2\hbar}{\gamma
J}e^{(J-2g)/T}. \end{equation}  
The solutions of (\ref{0reg}) and (\ref{1reg}) are then given by

\BEA
\label{sola} 
&&P(0,t)+P(1,t)+P(-1,t)=1,\\
\label{solb} 
&&P(0,t)-e^{-(J+2g)/T}P(1,t)-e^{-(J-2g)/T}P(-1,t)\propto e^{-t/\tau},
\\
\label{solc}
&&P(1,t)-P(-1,t)-\tanh\frac{2g}{T}\propto e^{-t/\tau_{\rm reg}}.
\end{eqnarray} 
The decay time $\tau$ associated with the combination (\ref{solb})
is much shorter than the time
$\tau_{\rm reg}$ which occurs in (\ref{solc}).

With the initial condition $P(0,0)=1$ we obtain,
dropping contributions small as $\exp({-J/T})$,
\begin{eqnarray} 
\label{sol2a} 
P(0,t)&=&e^{-t/\tau} , \\
\label{sol2b}
P(1,t)&=&\half\left[
\left(1-e^{-t/\tau}\right)+\tanh\frac{2g}{T}
\left(1-e^{-t/\tau_{\rm reg}}\right)\right] , \\
\label{sol2c} 
\quad P(-1,t)&=&\half\left[
\left(1-e^{-t/\tau}\right)-\tanh\frac{2g}{T}
\left(1-e^{-t/\tau_{\rm reg}}\right)\right] . \end{eqnarray}  
The evolution takes place in two stages, first on the time scale
$\tau=\hbar/\gamma J$, then on the much larger time scale
$\tau_{\rm reg}=2\tau \exp[{(J-2g)/T}]$.

During the first stage, M relaxes from the paramagnetic initial
state $m=0$ {\it to both ``ferromagnetic" states} $m=+1$ and
$m=-1$, with equal probabilities, as in the spontaneous process
where $g=0$. At the end of this stage, at times $\tau\ll
t\ll\tau_{\rm reg}$ we reach a nearly stationary situation in
which $P(0,t)$ is small as $2\exp({-J/T})$, while $P(1,t)$ and
$P(-1,t)$ are close to $\half$. Unexpectedly, in spite of the
presence of the coupling $g$ which is large compared to $T$, the
magnet M remains for a long time in a state close to the
equilibrium state which would be associated to $g=0$, {\it without
any invariance breaking}. This behavior arises from the large
value of the transition probabilities from $m=0$ to $m=\pm 1$,
which are proportional to $\tilde K(-\omega_\pm)$. For $J\pm2g\gg
T$, the latter quantity reduces to $\hbar(J\pm2g)/4$, which is not
sensitive to $g$ for $2g\ll J$.

In contrast to the situation for large $N$, the magnet thus begins
to {\it lose memory} of its initial state. For $N\gg 1$, it was
the coupling $g$ which triggered the evolution of M, inducing the
motion of the peak of $P_{\uparrow\uparrow}(m,t)$, initially at
$m=0$, towards larger and larger values of $m$. Only an initial
state involving values $m<-m_B$ led to false results at the end of
the process. Here, rather surprisingly, the two possible results
$m=+1$ and $m=-1$ come out nearly symmetrically after the first
stage of the process, for $\tau\ll t\ll\tau_{\rm reg}$. In fact we
do not even need the initial state to be ``paramagnetic". On this
time scale, any initial state for which $P(1,0)=P(-1,0)$ leads to
$P(1,t)=P(-1,t)\simeq\half$. (An arbitrary initial condition would
lead to $P(\pm1,t)=P(\pm1,0)+\half P(0,0)$.)

Fortunately, when $t$ approaches $\tau_{\rm reg}$ the effect of
$g$ is felt. For $t\gg\tau_{\rm reg}$ the probabilities $P(m,t)$
reach the values

\begin{equation}  \label{finalprob} 
P(1,t)=\frac{1}{1+e^{-4g/T}}, \qquad
P(-1,t)=\frac{e^{-4g/T}}{1+e^{-4g/T}},\qquad P(0,t)=2e^{-J/T},
\end{equation}  which correspond to the thermal equilibrium of M in the field
$g$. Thus, the {\it probability of a false measurement} is here
\begin{equation}  {\cal P}_-=e^{-4g/T}, \nn \end{equation}  and it
 is small if the conditions (8.7) are satisfied. On the
other hand, the {\it registration time} is $\tau_{\rm reg}$, and
the registration can be achieved only if  the interaction $\hat
H_{\rm SA}$ remains switched on during a delay larger than
$\tau_{\rm reg}$. After this delay, if we switch off the coupling
$g$, the result remains registered for a time which allows
observation, since $\tau_{\rm obs}$, determined in \S~\ref{section.8.1.2}, is
much larger than $\tau_{\rm trunc}$.

Thus, not only the first stage of the registration process is odd,
but also the second one. The mechanism at play in section~\ref{section.7} was a
{\it dynamical breaking of invariance} whereas here we have to
rely on the {\it establishment of thermal equilibrium in the
presence of} $g$. The coupling should be kept active for a long
time until the values (\ref{finalprob}) are reached, whereas for
$N\gg 1$, only the beginning of the evolution of
$P_{\uparrow\uparrow}(m,t)$ required the presence of the coupling
$g$; afterwards $P_{\uparrow\uparrow}$ reached the ferromagnetic
peak at $m=m_{\rm F}$, and remained there stably.

For $N=2$ the possibility of registration on the time scale $\tau_{\rm reg}$ relies on the form of the transition probabilities from $m=\pm 1$ to $m=0$, 
which are proportional to $\tilde K(\omega_\pm)$. Although small as $\exp({-\beta\hbar\omega_\pm})\tilde K(-\omega_\pm)$, these transition
probabilities contain a factor $\exp({-\beta\hbar\omega_\pm})\propto \exp({\mp 2g/T})$ which, since $2g\gg T$, strongly distinguishes $+1$
from $-1$, whereas $\tilde K(-\omega_+)\simeq \tilde K(-\omega_-)$. Hence the transition rate from $m=-1$ to $m=0$, 
behaving as $\exp[{-(J-2g)/T}]$, allows $P(0)$ to slowly increase at the expense of $P(-1)$, then to rapidly decay symmetrically. 
Since the transition rate from $m=+1$ to $m=0$, behaving as $\exp[{-(J+2g)/T}]$, is much weaker, the resulting increase of $P(1)$ remains gained.
Altogether $P(1,t)$ rises in two steps, from $0$ to $\frac{1}{2}$ on the time scale $\tau$, then from $\frac{1}{2}$ to nearly $1$ on
the time scale $\tau_{\rm reg}$, as shown by (\ref{sol2b}).  Meanwhile, $P(-1,t)$ rises from $0$ to $\frac{1}{2}$, then decreases back to $0$, 
ensuring a correct registration only at the end of the process, while $P(0,t)$ remains nearly $0$ between $\tau$ and $\tau_{\rm reg}$.

}

\subsubsection{Truncation}
\label{section.8.1.4}


\hfill{\it Les optimistes \'ecrivent mal\footnote{Optimists do not write well}}

\hfill{Paul Val\'ery, Mauvaises pens\'ees et autres}

\vspace{3mm}

\ZeText{

It remains to study the evolution of the off-diagonal blocks of
the density operator $\hat D$, which are characterized by the
three functions of time $P_{\uparrow\downarrow}(m,t)$. Their
equations of motion (\ref{0updown}), (\ref{1updown}) involve
oscillations in $P_{\uparrow\downarrow}(\pm 1,t)$ with frequency
${2g}/{\pi\hbar}$ generated by the coupling $g$ with S and by a
relaxation process generated by the bath. Since the oscillations
are not necessarily rapid, and since $\gamma$ is small, the
damping effect of the bath is expected to occur over times large
compared to ${\hbar}/{T}$, so that we can again work in the
Markovian regime. Moreover, since $g\ll J$, we are led to replace
$\omega_+$ and $\omega_-$ in $\tilde K_{t>}$ and $\tilde K_{t<}$
by ${J}{\hbar}$. Hence, we can replace, for instance, $\tilde
K_{t>}(\omega_+)+\tilde K_{t<}(\omega_-)$ by $\tilde K(J/\hbar)$.

The equations of motion for the set $P_{\uparrow\downarrow}(m,t)$
are thus simplified into \BEA \label{0simp} 
\tau\frac{{\rm d}P_{\uparrow\downarrow}(0,t)}{{\rm d}t}&=&
\varepsilon\left[
P_{\uparrow\downarrow}(1,t)+P_{\uparrow\downarrow}(-1,t)\right]
-P_{\uparrow\downarrow}(0,t),\\
\label{1simp} 
\tau\frac{{\rm
d}P_{\uparrow\downarrow}(\pm1,t)}{{\rm d}t}&=& \pm i\lambda
P_{\uparrow\downarrow}(\pm1,t) +\half
P_{\uparrow\downarrow}(0,t)-\varepsilon
P_{\uparrow\downarrow}(\pm1,t), \EEA where $\varepsilon$ and
$\lambda$ are defined by \begin{equation}  \varepsilon=e^{-J/T},\qquad
\lambda=\frac{4g}{\gamma J},\nn \end{equation}  with $\gamma\ll1$, $g\ll T\ll
J$. The truncation process is governed by the interplay between the
oscillations in $P(\pm1,t)$, generated by the coupling $g$ between
M and S, and the damping due to the bath. The two dimensionless
parameters $\lambda$ and $\varepsilon$ characterize these effects.

The eigenvalues of the matrix relating $-\tau {\rm
d}P_{\uparrow\downarrow}(m,t)/{\rm d}t$ to
$P_{\uparrow\downarrow}(m,t)$ are the solutions of the equation

\begin{equation} 
(z-1)[(z-\varepsilon)^2+\lambda^2]-\varepsilon(z-\varepsilon)=0.
\nn \end{equation}  The largest eigenvalue behaves for $T\ll J$ as

\begin{equation}  z_0\approx 1+\frac{\varepsilon}{1+\lambda^2}
+\frac{\varepsilon^2\lambda^2(1-\lambda^2)}{(1+\lambda^2)^3} ,
\end{equation}  whereas the other two eigenvalues $z_1$ and $z_2$, obtained
from \begin{equation}  z^2-z\varepsilon\left(\frac{1+2\lambda^2}{1+\lambda^2}+
\frac{\varepsilon^2\lambda^2(\lambda^2-1)}{(1+\lambda^2)^3}
\right)+ \lambda^2\left(1-\frac{\varepsilon}{1+\lambda^2}
+\frac{\varepsilon^2(1+\lambda^4)}{(1+\lambda^2)^3}\right)=0, \nn
\end{equation}  have a real part small as $\varepsilon$. The solution of
(\ref{0simp}), (\ref{1simp}), with the initial condition
$P_{\uparrow\downarrow}(m,0)=r_{\uparrow\downarrow}(0)\delta_{m,0}$
is given by

\BEA &&\label{solupdown0} 
 P_{\uparrow\downarrow}(0,t)=r_{\uparrow\downarrow}(0)\left[ e^{-z_0t/\tau}
-\frac{(z_1-\varepsilon)^2+\lambda^2}{(z_0-z_1)(z_1-z_2)}(e^{-z_1t/\tau}-e^{-z_0t/\tau})
-\frac{(z_2-\varepsilon)^2+\lambda^2}{(z_0-z_2)(z_2-z_1)}(e^{-z_2t/\tau}-e^{-z_0t/\tau})
\right],
\quad 
\\&& \label{solupdown1} 
P_{\uparrow\downarrow}(\pm1,t)=r_{\uparrow\downarrow}(0)\left[
\frac{z_1-\varepsilon\mp i\lambda}{2(z_0-z_1)(z_1-z_2)}
\left(e^{-z_1t/\tau}-e^{-z_0t/\tau}\right)
+\frac{z_2-\varepsilon\mp i\lambda}{2(z_0-z_2)(z_2-z_1)}
\left(e^{-z_2t/\tau}-e^{-z_0t/\tau}\right) \right]. \EEA

According to (\ref{solupdown0}), the first term of $P_{\uparrow\downarrow}(0,t)$ is damped for $\varepsilon\ll1$ over
the time scale $\tau={\hbar}/{\gamma J}$, just as $P_{\uparrow\uparrow}(0,t)$ in the registration process. However,
here again, the other two quantities $|P_{\uparrow\downarrow}(\pm 1,t)|$ increase in the
meanwhile and the truncation of the state is far from being achieved after the time $\tau$. In fact, all three components
$P_{\uparrow\downarrow}(m,t)$ survive over a much longer delay, which depends on the ratio ${2\lambda}/{\varepsilon}$.

In the overdamped situation $2\lambda<\varepsilon$ or $8g<\gamma J \exp({-J/T})$, the eigenvalues

\begin{equation}  z_{1,2}=\half\varepsilon\pm \half\sqrt{\varepsilon^2-4\lambda^2} \nn 
\end{equation}  
are real, so that we
get, in addition to the relaxation time $\tau$, two much longer
off-diagonal relaxation times, $\tau_{1,2}= {\tau}/{z_{1,2}}$. The
long-time behavior of $P_{\uparrow\downarrow}(m,t)$, governed by
$z_2$, is

\begin{equation}  P_{\uparrow\downarrow}(0,t)\sim r_{\uparrow\downarrow}(0)
\frac{\varepsilon(\varepsilon+\sqrt{\varepsilon^2-4\lambda^2})}
{2\sqrt{\varepsilon^2-4\lambda^2}}e^{-t/\tau_{\rm trunc}},\qquad
P_{\uparrow\downarrow}(\pm1,t)\sim r_{\uparrow\downarrow}(0)
\frac{\varepsilon \pm 2i\lambda+\sqrt{\varepsilon^2-4\lambda^2}}
{4\sqrt{\varepsilon^2-4\lambda^2}}e^{-t/\tau_{\rm trunc}}. \nn \end{equation} 
The truncation time

\begin{equation}  \label{red2over} 
\tau_{\rm trunc}=\frac{\tau}{2\lambda^2}\left(\varepsilon+
\sqrt{\varepsilon^2-4\lambda^2}\right)=\frac{\hbar \gamma J}{32
g^2} \left( e^{-J/T}+\sqrt{e^{-2J/T}-\frac{64g^2}{\gamma^2J^2}}\,\right),
 \end{equation} 
which characterizes the decay of $\langle\hat s_x\rangle$, $\langle\hat s_x\rangle$, and of their correlations with $\hat m$,
is here much longer than the registration time (10), since $\tau_{\rm trunc}/\tau_{\rm reg}$ is of order $(\varepsilon/2\lambda)^2\exp({2g/T})$, 
and even larger than $\tau_{\rm obs}$. The quantities $P_{\uparrow\downarrow}(m,t)$ remain for a long time proportional to
$r_{\uparrow\downarrow}(0)$, with a coefficient of order $1$ for $P_{\uparrow\downarrow}(\pm1,t)$, of order $\varepsilon$ for
$P_{\uparrow\downarrow}(0,t)$. Truncation is thus here a much slower process than registration: equilibrium is reached much
faster for the diagonal elements (\ref{finalprob}) than for the off-diagonal ones which are long to disappear. Let us stress that
for the present case of a small apparatus, they disappear due to the bath (``environment-induced decoherence''  
\cite{Schlosshauer,zurek,Guilini,Blanchard,walls,walls_book,Braun}) rather than, as in our previous discussion of a large apparatus, due to fast
dephasing caused by the large size of M.

For $2\lambda>\varepsilon$, we are in an oscillatory situation, where the eigenvalues

\begin{equation}  z_{1,2}=\frac{\varepsilon}{2}\frac{1+2\lambda^2}{1+\lambda^2}
\pm i\sqrt{\lambda^2-\frac{\varepsilon^2}{4}-\frac{\varepsilon\lambda^2}{1+\lambda^2}}
\nn \end{equation}  are complex conjugate. (Nothing prevents $\lambda=4g/\gamma J$ from
being large.) The long-time behavior is given by

\BEA
\label{P0pm1expres}
&&P_{\uparrow\downarrow}(0,t)\sim\frac{\varepsilon
r_{\uparrow\downarrow}(0)}{(1+\lambda^2)^2} e^{-t/\tau_{\rm trunc}}
\left[(1-\lambda^2)\cos\frac{2\pi t}{\theta} +
\frac{2\lambda^2}{\sqrt{\lambda^2-\varepsilon^2/4}} \sin\frac{2\pi
t}{\theta}\right],\nn\\&&
P_{\uparrow\downarrow}(\pm1,t)\sim\frac{r_{\uparrow\downarrow}(0)}{2(1\pm
i\lambda)} e^{-t/\tau_{\rm trunc}} \left[\cos\frac{2\pi t}{\theta}
\pm  \frac{i\lambda}{\sqrt{\lambda^2-\varepsilon^2/4}}
\sin\frac{2\pi t}{\theta}\right],\end{eqnarray} 
 with a truncation time

\begin{equation}  \label{red2osc} 
\tau_{\rm trunc}=\frac{2(1+\lambda^2)}{\varepsilon(1+2\lambda^2)}\tau
=\frac{2\hbar e^{J/T}(1+\lambda^2)}{\gamma J
(1+2\lambda^2)}=\frac{1+\lambda^2}{1+2\lambda^2}\tau_{\rm
reg}e^{2g/T}, \end{equation}  again much larger than the registration time.
While being damped, these functions oscillate with a period

\begin{equation}  \theta=\frac{2\pi\tau}{\sqrt{\lambda^2-\varepsilon^2/4}}
\nn\end{equation}  shorter than $\tau_{\rm trunc}$ if
$2\lambda>\varepsilon\sqrt{4\pi^2+1}$. The {\it truncation time} (\ref{red2osc}) 
practically does not depend on $g$ (within a factor $2$ when
$2\lambda$ varies from $\varepsilon$ to $\infty$), in contrast to
both the truncation time of section~\ref{section.5} and the irreversibility time
of section~\ref{section.6}. The present truncation time is comparable to the
lifetime $\tau_{\rm obs}$ of an initial pure state $m=+1$ when it
spontaneously decays towards $m=\pm1$  with equal probabilities
(\S~\ref{section.8.1.2}). Hence in both cases the truncation takes place over
the delay during which the result of the measurement can be
observed.

For $\lambda\gg \varepsilon$ and $t\gg \tau$, the off-diagonal
contributions (\ref{P0pm1expres}) to $\hat D$ are governed by

\BEA P_{\uparrow\downarrow}(\pm1,t)\sim \frac{
r_{\uparrow\downarrow}(0)}{2(1\pm i\lambda)} e^{-t/\tau_{\rm
red}\pm i\lambda t/\tau}.  \EEA 
The effects on M of S and B are well separated: the oscillations are the same as for $\gamma=0$,
while the decay, with characteristic time
${\tau}/{\varepsilon}=(\hbar/\gamma J) \exp({J/T})$, is a pure effect of
the bath. The amplitude becomes small for $\lambda\gg1$, that is,
$g\gg \gamma J$.

}

\subsubsection{Is this process with bath-induced decoherence a measurement?}
\label{section.8.1.5}

\hfill{{}\footnote{Every crow promotes her baby bird}}

\vspace{-0.95cm}
\myskipfigText{
\begin{figure}[h!h!h!]
\label{ArmProv2}
\hfill{\includegraphics[width=6cm]{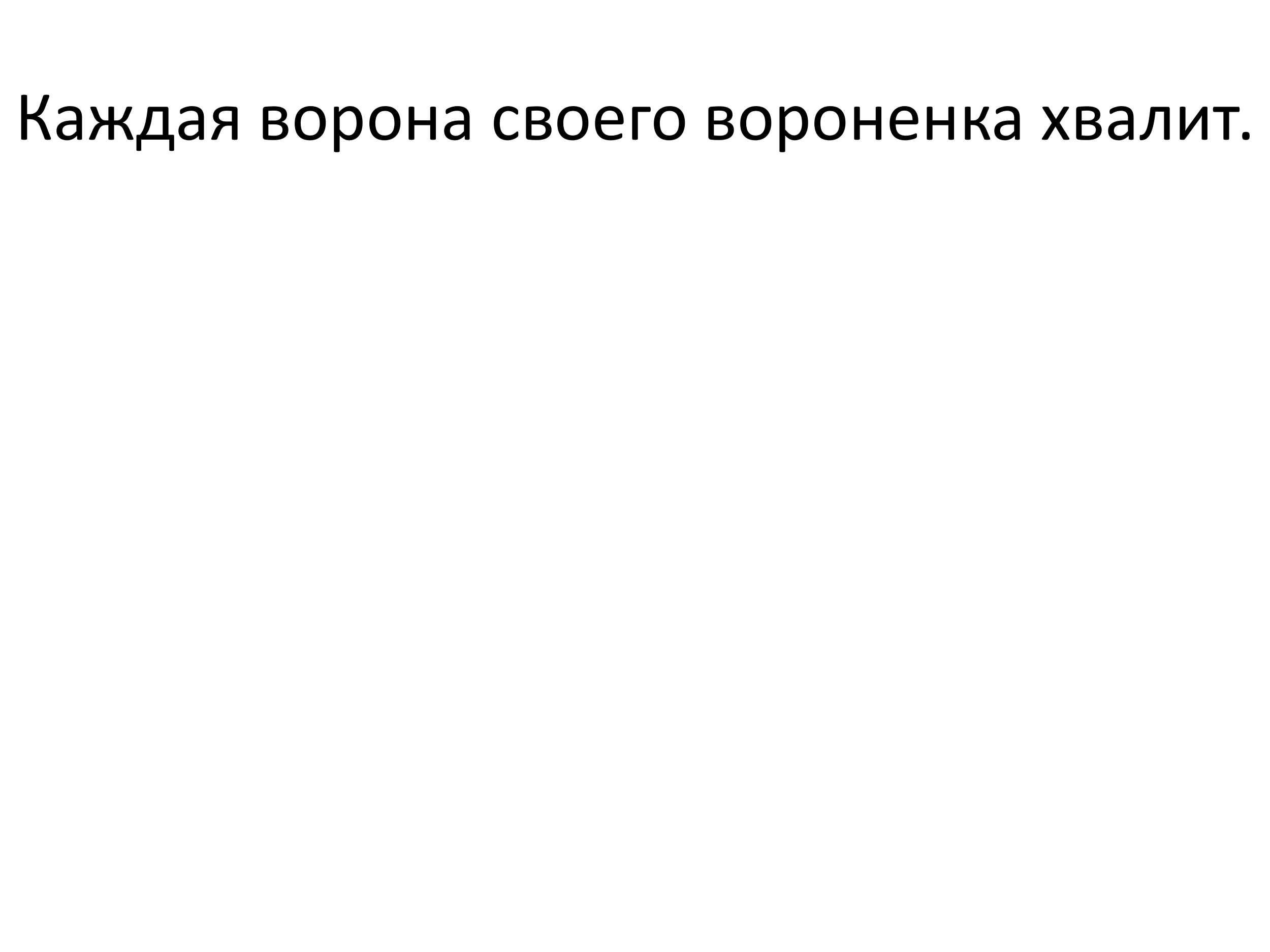}\hspace{2mm}}
\end{figure}}

\vspace{-4.25cm}

\hfill{Russian proverb}

\vspace{3mm}

\ZeText{

When the number $N$ of degrees of freedom of the pointer is small as here, the present model appears as a specific example among the general
class of models considered by Spehner and Haake ~\cite{SpehnerHaake1,SpehnerHaake2}.
As shown by these authors, the truncation is then governed by the large number of degrees of freedom {\it of the bath},
{\it not of the pointer}; the truncation is then not faster than the registration. Our detailed study allows us to compare the mechanisms of
two types of processes,  for large $N$ and for small $N$.

We have seen (\S~\ref{section.8.1.3}) that for $N=2$ as for $N\gg1$ both couplings $g$ and $\gamma$ between $S$, $M$ and $B$ establish the diagonal
correlations between $\hat s_z$ and $\hat m$ needed  to establish Born's rule. This result is embedded in the values reached by
$P_{\uparrow\uparrow}$ and $P_{\downarrow\downarrow}$ after the time $\tau_{\rm reg}=({2\hbar}/{\gamma J})\exp[{(J-2g)/T}]$, much
longer than the lifetime $\tau={\hbar}/{\gamma J}$ of the initial state in the absence of a field or a coupling. Although this
property is one important feature of a quantum measurement, its mechanism is here only a relaxation towards thermal equilibrium.
The registration is fragile and does not survive beyond a delay $\tau_{\rm obs}=({\hbar}/{\gamma J})\exp({J/T})$ once the coupling
with S is switched off. For larger $N$, the existence of a spontaneously broken invariance ensured the long lifetime of the
ferromagnetic states, and hence the robust registration of the measurement.

Another feature of a quantum measurement, the truncation of the state that represents a large set of runs, has also been recovered for $N=2$, but
with an unsatisfactorily long time scale. For large $N$, the truncation process took place rapidly and was achieved before the registration in the apparatus
 really began, but here,  whatever the parameters $\varepsilon$ and $\lambda$, the expectation values 
$\langle\hat s_x\rangle$, $\langle\hat s_y\rangle$ and the off-diagonal correlations embedded in $P_{\uparrow\downarrow}$ and
$P_{\downarrow\uparrow}$ fade out over a truncation time $\tau_{\rm trunc}$ given by (\ref{red2over}) or (\ref{red2osc}), which is
longer than the registration time and even than the observation time if $2\lambda\ll\varepsilon$. It
is difficult to regard such a slow decay as the ``collapse" of the state.

By studying the case $N=2$, we wished to test whether an {\it environment-induced decoherence} \cite{Schlosshauer,zurek,Guilini,Blanchard,walls,walls_book,Braun} 
might cause truncation. Here the ``environment" is the bath B, which is the source of irreversibility. It imposes thermal equilibrium
to S + M, hence suppressing gradually the off-diagonal elements of $\hat D$ which vanish at equilibrium, a suppression that we defined as ``truncation''. 
However, usually, decoherence time scales are the shortest of all; here, for $N=2$, contrary to what happened for $N\gg1$, the truncation time is not shorter
 than the registration time.

The effect of the bath is therefore quite different for large and for small $N$.  For $N\gg1$, we have seen in \S\S~5.1.2 and 6.2.4 that the rapid initial truncation 
was ensured by the large size of the pointer M, whereas bath-induced decoherence played only
 a minor role, being only one among the two possible mechanisms of suppression of recurrences. For $N=2$, the truncation itself is caused by the bath, but we cannot 
 really distinguish decoherence from thermal equilibration: Although the dynamics of the diagonal and off-diagonal blocks of $\hat D$ are decoupled, 
 there is no neat separation of time scales for the truncation and the registration.
 
 A last feature of measurements, the uniqueness of the outcome of individual runs, is essential as it conditions 
 both Born's rule and  von Neumann's reduction. We have stressed 
 (\S\S~1.1.2 and 1.3.2) that truncation, which concerns the large set of runs of the measurement, does not imply reduction, which concerns individual runs. 
 The latter property will be proven in section 11 for the Curie--Weiss model; its explanation will rely on a coupling between the large number of eigenstates of M 
 involved for $N\gg1$ in each ferromagnetic equilibrium state. Here, for $N=2$, the ``ferromagnetic'' state is non degenerate, and that mechanism cannot be invoked.

Anyhow, the process that we described cannot be regarded for $N=2$ as a full measurement. Being microscopic, the pair of spins M is not a ``pointer" that can 
be observed directly. In order to get a stable signal, which provides us with information and which we may use at a macroscopic level, we need to couple M to
 a genuine macroscopic apparatus. This should be done after the time
$\tau_{\rm reg}=({2\hbar}/{\gamma J})\exp[{(J-2g)/T}]$ when the correlations
$P_{\uparrow\uparrow}(m,t)=r_{\uparrow\uparrow}(0)\delta_{m,1}$ and
$P_{\downarrow\downarrow}(m,t)=r_{\downarrow\downarrow}(0)\delta_{m, -1}$
have been created between S and M. Then, S and M should be decoupled,
and the measurement of $m$ should be performed in the delay $\tau_{\rm obs}
=({\hbar}/{\gamma J})\exp({J/T})$.
In this hypothetical process, the decoupling of S and M will entail truncation,
the correlations which survive for the duration $\tau_\trunc =2\tau_{\rm obs}$ in
$P_{\uparrow\downarrow}$ and $P_{\downarrow\uparrow}$ being destroyed.

Altogether, it is not legitimate for small $N$ to regard $M+B$ as a ``measurement apparatus'', since nothing can be said about individual runs. Anyhow, 
registering robustly the outcomes of the process so as to read them during a long delay requires a {\it further apparatus} involving a {\it macroscopic pointer}. 
The system M, even accompanied with its bath, is not more than a quantum device coupled to S. However, its marginal state is
represented by a diagonal density matrix, in the basis which diagonalizes $\hat m$, so that the respective probabilities of
$m=0$, $m=+1$ and $m=-1$, from which we may infer $r_{\uparrow\uparrow}(0)$ and $r_{\downarrow\downarrow}(0)$, can
be determined by means of an apparatus with classical features.

}

\subsubsection{Can one simultaneously ``measure'' non-commuting variables?}
\label{section.8.1.6}


\hfill{{}\footnote{Two hopeful dreams cannot coexist}}

\vspace{-1.cm}
\myskipfigText{
\begin{figure}[h!h!h!]
\label{ArmProv2}
\hfill{\includegraphics[width=6cm]{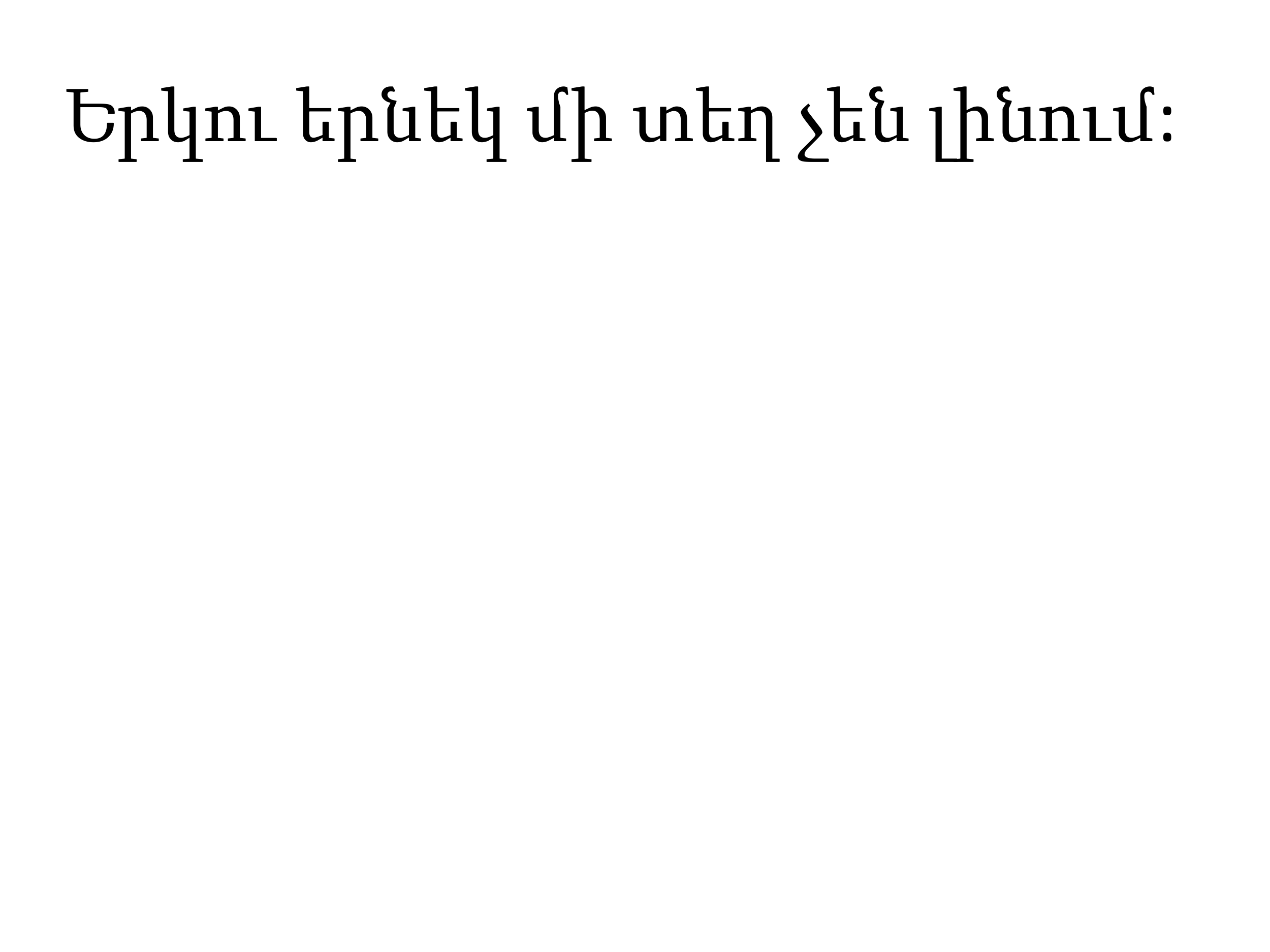}\hspace{2.6mm}}
\end{figure}}

\vspace{-4.2cm}

\hfill{Armenian proverb}

\vspace{3mm}

\ZeText{

Although the process described above cannot be regarded as an ideal measurement,  we have seen that it allows us to determine the diagonal elements
$r_{\uparrow\uparrow}(0)$ and $r_{\downarrow\downarrow}(0)$ of the density matrix of S at the initial time. Surprisingly, the same device may also give us 
access to the off-diagonal elements, owing to the pathologically slow truncation. Imagine S and M are decoupled at some time $\tau_{\rm dec}$ of order $\tau_\trunc $.
For $2\lambda\ll\varepsilon$, this time can be shorter than the observation time, so that a rapid measurement of $m$ will inform us statistically on
$r_{\uparrow\uparrow}(0)$ and $r_{\downarrow\downarrow}(0)$. However, the transverse components of the spin S have not disappeared on average,
and $r_{\uparrow\downarrow}(\tau_{\rm dec})$ is given at the decoupling time and later on by

\begin{equation}  
\label{rdec}
r_{\uparrow\downarrow}(\tau_{\rm dec})=\sum_m P_{\uparrow\downarrow}(m,\tau_{\rm dec})=
r_{\uparrow\downarrow}(0)\frac{\varepsilon+\sqrt{\varepsilon^2-4\lambda^2}} {2\sqrt{\varepsilon^2-4\lambda^2}} e^{-\tau_{\rm dec}/\tau_\trunc }. 
\end{equation} 
 A subsequent measurement on S in the $x$-direction at a time $t>\tau_{\rm dec}$ will then provide
$r_{\uparrow\downarrow}(t)+r_{\downarrow\uparrow}(t) =2\Re r_{\downarrow\uparrow}(\tau_{\rm dec}) $. If the various
parameters entering (\ref{rdec}) are well controlled, we can thus, through repeated measurements, determine indirectly
$r_{\uparrow\downarrow}(0)+r_{\downarrow\uparrow}(0)$, as well as $r_{\uparrow\uparrow}(0)$ and $r_{\downarrow\downarrow}(0)$.

Note, however, that such a procedure gives us access only to the {\it statistical} properties of the {\it initial state} of S, and that von Neumann's reduction is precluded.

Thus a unique experimental setting may be used to determine the statistics of the non-commuting observables $\hat s_x$ and $\hat s_z$.
This possibility is reminiscent of a general result \cite{ABNprl2004}; see also \cite{bahar,gerardo}. Suppose we wish to determine all the matrix elements $r_{ij}$ 
of the unknown $n\times n$ density matrix of a system S at the initial time. Coupling during some delay S with a similar auxiliary system S$^\prime$,
the initial state of which is known, leads to some density matrix for the compound system. The set $r_{ij}$ is thus mapped onto the $n^2$
diagonal elements of the latter. These diagonal elements may be measured simultaneously by means of a single apparatus, and inversion of the
mapping yields the whole set $r_{ij}$. Here the magnet M plays the role of the auxiliary system S$'$; we can thus understand the paradoxical
possibility of determining the statistics of both $\hat s_x$ and $\hat s_z$ with a single device. 

In this context we note that the simultaneous measurement of non-commuting observables is an important chapter of modern quantum
mechanics. Its recent developments are given in \cite{arthurs_complo,martens_complo,appleby_complo,luis_complo,lorenzo_complo}
(among other references) and reviewed in \cite{busch_complo}. 

With this setup we can also repeat measurements in the $z$-direction and see how much lapse should be in between to avoid non-idealities.

}

\subsection{Measuring a non-conserved quantity}
\label{section.8.2}

\hfill{{\it L'homme est plein d'imperfections, mais ce n'est pas} 

\hfill{\it \'etonnant si l'on songe \`a  l'\'epoque o\`u il a \'et\'e cr\'e\'e}\,\footnote{Man is full of imperfections, 
but this is not surprising if one considers when he was created}}

\hfill{ Alphonse Allais}

\vspace{0.3cm}
\ZeText{

It has been stressed by Wigner \cite{Wigner52} that an observable that does
not commute with some conserved quantity of the total system
(tested system S plus apparatus A) cannot be measured exactly, and
the probability of unsuccessful experiments has been estimated by Araki and Yanase
\cite{Yanase,Araki}. (Modern developments of this Wigner-Araki-Yanase limitation are given in \cite{WAY,ozawa_way,meister_way}.)
However, neither the irreversibility nor the dynamics of the measurement process were considered. We focus here
on the extreme case in which Wigner's conserved quantity is the energy itself. We have assumed till now that the measured observable
$\hat s_z$ commuted with the full Hamiltonian of S + A. This has allowed us to split the dynamical analysis into two separate
parts: The diagonal blocks $R_{\uparrow\uparrow}$, $R_{\downarrow\downarrow}$ of the full density matrix of S + A are
not coupled to the off-diagonal blocks $R_{\uparrow\downarrow}$, $R_{\downarrow\uparrow}$. This gives rise, for $N\gg1$, to a large
ratio between the time scales that characterize the truncation and the registration.

We will discuss, by solving a slightly modified version of our model, under which conditions one can still measure a quantity
which is not conserved. We allow therefore transitions between different eigenvalues of $\hat s_z$, by introducing a magnetic
field that acts on S. The part $\hat H_{\rm S}$ of the Hamiltonian, instead of vanishing as in (\ref{HS=0}), is taken as

\begin{equation} 
\label{HS} 
\hat H_{\rm S}=-b\hat s_y.
\end{equation} 
We take as measuring device a large, Ising magnet, with $q=2$ and $N\gg1$.

We wish to study how the additional field affects the dynamics of the measurement. We shall therefore work out the equations at lowest order in $b$, 
which however need not be finite as $N\to\infty$. In fact, a crucial parameter turns out to be the combination $b/g\sqrt{N}$.

}

\subsubsection{The changes in the dynamics}
\label{section.8.2.1}

\hfill{\it  Plus \c ca change, plus c'est la m\^eme chose\footnote{The more it changes, the more it remains the same}}

\hfill{French saying}

\hspace{3mm}

\ZeText{

The formalism of subsection~\ref{section.4.1} remains unchanged, the unperturbed Hamiltonian being now

\begin{equation}  \label{H0b} 
\hat H_0=\hat H_{\rm S}+\hat H_{\rm SA}+\hat H_{\rm M}= -b\hat s_y-Ng\hat m\hat s_z-\half JN\hat m^2.
\end{equation}  
The additional contribution (\ref{HS}) enters the basic equation (\ref{dD}) in two different ways.

($i$) On the left-hand side, the term $-\left[\hat H_{\rm S},\hat D\right]/i\hbar$ yields a contribution 

\begin{equation} 
\frac{b}{\hbar}
\left( \begin{array}{ccc@{\ }r} \hat R_{\uparrow\downarrow}+\hat R_{\downarrow\uparrow} &\hat R_{\downarrow\downarrow}-\hat R_{\uparrow\uparrow}   \\
\hat R_{\downarrow\downarrow}-\hat R_{\uparrow\uparrow}  &-\hat R_{\uparrow\downarrow}-\hat R_{\downarrow\uparrow}\end{array} \right)
\end{equation}  
to ${\rm d}\hat D/{\rm d} t$ which couples the diagonal and
off-diagonal sectors of (\ref{Rij}). Accordingly, we must add to
the right-hand side of the equation of motion (\ref{dPdiag}) for
${\rm d} P_{\uparrow\uparrow}/{\rm d} t$ the term
$\hbar^{-1}b(P_{\uparrow\downarrow}+P_{\downarrow\uparrow})$, and
subtract it from the equation for $\d
P_{\downarrow\downarrow}/{\rm d}t$; we should add to the equations
(\ref{dPoff}) for ${\rm d}P_{\uparrow\downarrow}/d t$ and ${\rm
d}P_{\downarrow\uparrow}/{\rm d} t$ the term
$\hbar^{-1}b(P_{\downarrow\downarrow}-P_{\uparrow\uparrow})$.

($ii$) The presence of $\hat H_{\rm S}$ in $\hat H_0$ has another,
indirect effect. The operators $\hat\sigma_a^{(n)}(u)$ defined by
(\ref{sigmau}), which enter the right-hand side of eq. (\ref{dD}),
no longer commute with $\hat s_z$. In fact, while
$\hat\sigma_z^{(n)}(u)$ still equals $\hat\sigma_z^{(n)}$, the
operators \begin{equation} 
\hat\sigma_+^{(n)}(u)=\left[\hat\sigma_-^{(n)}(u)\right]^\dagger=
\hat\sigma_+^{(n)}e^{-i\hat H_0(\hat m+\delta m)u/\hbar}e^{i\hat
H_0(\hat m)u/\hbar} =e^{-i\hat H_0(\hat m)u/\hbar}e^{i\hat
H_0(\hat m-\delta m)u/\hbar}\hat\sigma_+^{(n)} \nn \end{equation}  now
contain contributions in $\hat s_x$ and $\hat s_y$, which can be
found by using the expression (\ref{H0b}) of $\hat H_0$ and the
identity $\exp( i\,{\bf a}\cdot\hat{\bf s})=\cos (a)+i\sin( a)\,{\bf
a}\cdot\hat{\bf s}/a$. For $N\gg1$ and arbitrary $b$, we should
therefore modify the bath terms in ${\rm d}P_{ij}/{\rm d}t$ by
using the expression

\BEA \label{sigmab} 
\hat\sigma^{(n)}_+(u)=\hat\sigma^{(n)}\exp\left[\frac{2i\hat mu}{\hbar}\left(J+
Ng^2 \frac{ Ng\hat m\hat s_z+b\hat s_y}{N^2g^2\hat m^2+b^2}\right)\right],
\EEA 
instead of (\ref{sigmaplus}); we have dropped in the square bracket contributions that oscillate rapidly as
$\exp\left(2iu\sqrt{N^2g^2\hat m^2+b^2}/\hbar\right)$ with factors $\hat s_x$ and coefficients of order $1/N$.

Except in \S~\ref{section.8.2.5} we assume that S and A {\it remain coupled} at all times. Their joint distribution $\hat D(t)$ is then
expected to be driven by the bath B to an equilibrium $\hat D(t_{\rm f})\propto \exp(-\hat H_0/T)$ at large times. The
temperature $T$ is imposed by the factor $K(u)$ that enters the equation of motion (\ref{dD}), while $\hat H_0$ is imposed by the
form of $\hat\sigma_a^{(n)}(u)$. The additional terms in (\ref{sigmab}) are needed to ensure that S + M reaches the required
equilibrium state. As discussed in \S~\ref{section.7.1.4}, invariance is broken in the final state. Its density operator involves two
incoherent contributions, for which the magnetization of M lies either close to $+m_{\rm F}$ or close to $-m_{\rm F}$. In the
first one, the marginal state of S is $\hat r(t_{\rm f})\propto \exp\left[\left(b\hat s_y+Ngm_{\rm F}\hat s_z\right)/T\right]$. If
$b\ll Ng$, a condition that we will impose from now on, this state cannot be distinguished from the projection on $s_z=+1$. As when
$b=0$, the sign of the observed magnetization $\pm m_{\rm F}$ of M is fully correlated with that of the $z$-component of the spin S
in the final state, while $\langle \hat s_x(t_{\rm f})\rangle=\langle \hat s_y(t_{\rm f})\rangle=0.$  If dynamical stability of subensembles is ensured 
as in \S~\ref{fin11.2.3}, the process is {\it consistent with von Neumann's reduction}, and it can be used as a preparation.

Nevertheless, nothing warrants the weights of the two possible outcomes, $+m_{\rm F}, \,s_z=+1$ and $-m_{\rm F},\,s_z=-1$, to be
equal to the diagonal elements $r_{\uparrow\uparrow}(0)$ and $r_{\downarrow\downarrow}(0)$ of the {\it initial density matrix:
Born's rule may be violated.} A full study of the dynamics is required to evaluate these weights, so as to determine whether the
process is still a faithful measurement.

This study will be simplified by noting that the expression (\ref{sigmab}) depends on $b$ only through the ratio $b/Ng\hat m$.
Once the registration has been established, at times of order $\tau_{\rm reg}$, the relevant eigenvalues of $\hat m$, of order
$m_{\rm B}$, are finite for large $N$ and the field $b$ does not contribute to $\hat \sigma^{(n)}_a(u)$ since $b\ll Ng$. For short
times, during the measurement process, the distribution of $m$ is Gaussian, with a width of order $1/\sqrt{N}$, so that $b$ may
contribute significantly to $\hat \sigma^{(n)}_a(u)$ if $b$ is of order $g\sqrt{N}$. However, we have shown (section 6) that for the
off-diagonal blocks the bath terms in (\ref{dPoff2}) have the sole effect of inhibiting the recurrences in $P_{\uparrow\downarrow}(m,t)$. 
Anyhow, such recurrences are not seen when $m$ is treated as a continuous variable. We shall therefore rely on the simplified equations of motion

\begin{equation}  \label{offb} 
\frac{\partial P_{\uparrow\downarrow}}{\partial t}-\frac{2iNgm}{\hbar}
P_{\uparrow\downarrow}= \frac{\partial
P_{\downarrow\uparrow}}{\partial t}+\frac{2iNgm}{\hbar}
P_{\downarrow\uparrow}=\frac{b}{\hbar}\left(
P_{\downarrow\downarrow}-P_{\uparrow\uparrow}\right). 
\end{equation} 

As regards the diagonal blocks we shall disregard $b$ not only at times of order $\tau_{\rm reg}$, but even earlier. This is legitimate if $b\ll g\sqrt{N}$; 
if $b$ is of order $g\sqrt{N}$, such an approximation retains the main effects of the bath, driving the distributions $P_{\uparrow\uparrow}(m,t)$ and
$P_{\downarrow\downarrow}(m,t)$ apart from $-m_{\rm B}$ and $+m_{\rm B}$, respectively, and widening them. We write therefore:

\BEA \label{upb} 
\frac{\partial P_{\uparrow\uparrow}}{\partial t} +\frac{\partial}{\partial
m}\left( v_{\uparrow\uparrow} P_{\uparrow\uparrow}\right)
-\frac{1}{N}\frac{\partial^2}{\partial m^2}\left( w P_{\uparrow\uparrow}\right)=
\frac{b}{\hbar}\left(P_{\uparrow\downarrow}+P_{\downarrow\uparrow}\right), \\
\label{downb} 
\frac{\partial P_{\downarrow\downarrow}}{\partial t} +\frac{\partial}{\partial m}
\left( v_{\downarrow\downarrow} P_{\downarrow\downarrow}\right)
-\frac{1}{N}\frac{\partial^2}{\partial m^2}\left(w P_{\downarrow\downarrow}\right)=
-\frac{b}{\hbar}\left(P_{\uparrow\downarrow}+P_{\downarrow\uparrow}\right).
\EEA
(Here we should distinguish the drift velocities $v_{\uparrow\uparrow}$ and $v_{\downarrow\downarrow}$, but the diffusion coefficients are equal.)
 Since the outcome of the registration is governed by the
first stage studied in \S~\ref{section.7.2.3}($i$), and since the Markovian
regime (\S~\ref{section.7.1.1}) is reached nearly from the outset, we shall
use the simplified forms

\begin{equation}  v_{\up\up}=\frac{\gamma}{\hbar}[g+(J-T)m]
=\frac{1}{\tau_{\rm reg}}(m_{\rm B}+m), \qquad w=\frac{\gamma
T}{\hbar} ,\nn \end{equation} 
for the drift velocity and the diffusion coefficient;
 $v_{\down\down}$ follows from $v_{\up\up}$ by changing $g$ into $-g$.

We have to solve (\ref{offb}), (\ref{upb}) and (\ref{downb}) with initial conditions $P_{ij}(m,0)/r_{ij}(0)=P_{\rm M}(m,0)$ expressed by
(\ref{Pm0}). The drift and diffusion induced by the bath terms are slow since $\gamma\ll 1$, and the distribution
$P_{\rm M}(m,t)=P_{\uparrow\uparrow}(m,t)+P_{\downarrow\downarrow}(m,t)$ of the magnetization of M can be regarded as constant on the
time scales $\tau_\trunc $ and $\tau_{\rm Larmor}=\pi\hbar/b$, which is the period of the precession of the spin S when it does
not interact with A. Over a short lapse around any time $t$, the coupled equations for $C_x=P_{\uparrow\downarrow}+P_{\downarrow\uparrow}$,
$C_y=iP_{\uparrow\downarrow}-iP_{\downarrow\uparrow}$ and $C_z=P_{\uparrow\uparrow}-P_{\downarrow\downarrow}$ simply
describe, for each $m$, a Larmor precession of S \cite{spin_echo,spin_echo1,spin_echo2a,spin_echo2b,spin_echo3,spin_echo4} submitted to the
field $b$ along $\hat y$ and to the field $Ngm$ along $\hat z$, where $m$ is a classical random variable governed by the
probability distribution $P_{\rm M}(m,t)$. The slow evolution of $P_{\rm M}(m,t)$ is coupled to this rapid precession through (\ref{upb}) and (\ref{downb}).


}

\subsubsection{Ongoing truncation}
\label{section.8.2.2}

\hfill{\it Het kind met het badwater weggooien\footnote{\label{Badwater}To throw the baby out of the bath water}} 

\hfill{\it Jeter l'enfant avec l'eau du bain$^{\ref{Badwater}}$}

\hfill{Dutch and French expressions}

\vspace{3mm}

\ZeText{

We first eliminate the off-diagonal contributions by formally
solving (\ref{offb}) as 

\begin{equation}  
\label{elim}
P_{\uparrow\downarrow}(m,t)=P_{\downarrow\uparrow}^\ast=
r_{\uparrow\downarrow}(0)e^{2iNgmt/\hbar}P_{\rm M}(m,0)-\frac{b}{\hbar}
\int_0^t{\rm d}t'\,e^{2iNgm(t-t')/\hbar}
\left[P_{\uparrow\uparrow}(m,t')-P_{\downarrow\downarrow}(m,t')\right].
\end{equation}  
The physical quantities (except for correlations involving a large number of spins of M, see \S~\ref{section.5.1.3}) are obtained by
summing over $m$ with a weight smooth on the scale $1/\sqrt{N}$. The first term of (\ref{elim}), the same as in section~\ref{section.5} then
yields a factor decaying as $\exp[-(t/\tau_\trunc )^2]$, with $\tau_\trunc =\hbar/g\delta_0\sqrt{2N}$,  due to destructive interferences.

However, the second term survives much longer because the precession induced by the field $b$ along $\hat y$ couples
$2P_{\uparrow\downarrow}=C_x-iC_y$ to $C_z=P_{\uparrow\uparrow}-P_{\downarrow\downarrow}$ at all times $t$. truncation takes place
through the oscillatory factor in the integral, which hinders the effect of precession except at times $t'$ just before $t$.
Truncation is an {\it ongoing process}, which may take place (if $b$ is sufficiently large) for $t\gg\tau_\trunc $: The
non-conservation of the measured quantity $s_z$ tends to {\it feed up the off-diagonal components} $\hat R_{\uparrow\downarrow}$  and
$\hat R_{\downarrow\uparrow}$ of the density matrix $\hat D$ of S + M. In compensation,  $\hat R_{\uparrow\uparrow}$  and $\hat
R_{\downarrow\downarrow}$ may be progressively eroded through the right-hand side of (\ref{upb}) and (\ref{downb}).

At lowest order in $b$, we can rewrite explicitly the second term of (\ref{elim}) by replacing $P_{\uparrow\uparrow}$ by

\begin{eqnarray} 
\label{Pb=0} 
P^{(0)}_{\uparrow\uparrow}(m,t)&=&r_{\uparrow\uparrow}(0)
\sqrt{\frac{N}{2\pi D(t)}}\exp\left[-\frac{N}{2D(t)}
\left(m+m_{\rm B}-m_{\rm B}e^{t/\tau_{\rm reg}}\right)^2\right],\\
D(t)&=&\delta_0^2e^{2t/\tau_{\rm reg}}+\frac{T}{J-T}
\left(e^{2t/\tau_{\rm reg}}-1\right), \qquad \tau_{\rm
reg}=\frac{\hbar}{\gamma(J-T)}, \nn \end{eqnarray} 
that we evaluated for
$b=0$ in section~\ref{section.7}. We have simplified the general expression
(\ref{Psplit}) by noting that the final outcome will depend only
on the first stage of the registration, when $t$ is of order
$\tau_{\rm reg}$. For $P^{(0)}_{\downarrow\downarrow}$ we have to
change $r_{\uparrow\uparrow}(0)$ into
$r_{\downarrow\downarrow}(0)$ and $m_{\rm B}=g/(J-T)$ into
$-m_{\rm B}$.

}

\clearpage

\subsubsection{Leakage}
\label{section.8.2.3}


\hfill{\it Tout ce qui est excessif est insignifiant\footnote{Everything that is excessive is insignificant}}

\hfill{Talleyrand
}

\vspace{3mm}

\ZeText{

The expectation values of $\hat s_x$ or $\hat s_y$ and their
correlations with the pointer variable $\hat m$ are now found as
in \S~\ref{section.5.1.3} through summation over $m$ of
$P_{\uparrow\downarrow}(m,t)e^{i\lambda m}$. At times $t$ long
compared to $\tau_\trunc $ and short compared to $\tau_{\rm
reg}$, we find the characteristic function

\BEA
\label{Psib} 
\Psi_{\uparrow\downarrow}(\lambda,t)&\equiv&\langle \hat s_-e^{i\lambda \hat m}(t)\rangle=
\int\d m \,P_{\uparrow\downarrow}(m,t)e^{i\lambda m} \simeq
-\frac{b}{\hbar}\int{\rm d}m\int_0^t{\rm d}t' e^{2iNgm(t-t')/\hbar+i\lambda m}
\left[P_{\uparrow\uparrow}^{(0)}(m,t')-P_{\downarrow\downarrow}^{(0)}(m,t')\right] \\
&=&-\frac{b}{\hbar}\int_0^t{\rm d}t'r_{\uparrow\uparrow}(0)
\exp\left[-\left(\frac{t-t'}{\tau_\trunc }+\frac{\lambda\delta_0}{\sqrt{2N}}\right)^2
+\frac{2it'}{\tau_{\rm leak}}
\left(\frac{t-t'}{\tau_\trunc }+\frac{\lambda\delta_0}{\sqrt{2N}}\right)\right]
-\left\{r_{\uparrow\uparrow}\mapsto
r_{\downarrow\downarrow},\, \tau_{\rm leak}\mapsto -\tau_{\rm leak}\right\}. \nn 
\EEA
We have recombined the parameters so as to
express the exponent in terms of two characteristic times, the
truncation time $\tau_\trunc =\hbar/g\delta_0\sqrt{2N}$
introduced in (\ref{taured}) and the {\it leakage time}

\begin{equation}  \tau_{\rm leak}=\sqrt{\frac{2}{N}}\frac{\hbar\delta_0}{\gamma
g}= \sqrt{\frac{2}{N}}\frac{\tau_\trunc \delta_0}{m_{\rm B}}
=\frac{2\tau_\reg \, \delta_0^2}{\gamma}. \end{equation} 
Integration over $t'$ can be performed  in the limit $\tau_{\rm leak}\gg\tau_\trunc $, by noting that the dominant contribution
arises from the region $t-t'\ll t$, which yields in terms of the error function (\ref{erfc})

\BEA
\label{Psib2} 
\Psi_{\uparrow\downarrow}(\lambda,t)&=&
-\frac{b}{2g\delta_0}\sqrt{\frac{\pi}{2N}}e^{-(t/\tau_{\rm leak})^2}
\left[r_{\uparrow\uparrow}(0)
{\rm erfc}\left(-\frac{it}{\tau_{\rm leak}}+\frac{\lambda\delta_0}{\sqrt{2N}}\right)
 -r_{\downarrow\downarrow}(0)
{\rm erfc}\left(\frac{it}{\tau_{\rm
leak}}+\frac{\lambda\delta_0}{\sqrt{2N}}\right)\right]. \end{eqnarray} 
The leakage time characterizes {\it the dynamics of the transfer of
polarization} from the $z$-direction towards the $x$- and
$y$-directions. It is much shorter than the registration time,
since $N\gg1$ and $\gamma\ll1$. It also characterizes the delay
over which the distribution $P^{(0)}_{\uparrow\uparrow}(m,t)$
keeps a significant value at the origin: The peak of
$P^{(0)}_{\uparrow\uparrow}$ with width $\delta_0/\sqrt{N}$, moves
as $m_{\rm B}(e^{t/\tau_{\rm reg}}-1)\sim m_{\rm B}t/\tau_{\rm
reg}$, and at the time $t=\tau_{\rm leak}$ we have
$P^{(0)}_{\uparrow\uparrow}(0,\tau_{\rm
leak})/P^{(0)}_{\uparrow\uparrow}(0,0)=1/e$.

Using the properties of the error function we can derive 
from Eq. (\ref{Psib2}), which is valid at times $t\gg\tau_\trunc $ such that the memory
of $2r_{\uparrow\downarrow}(0)=\langle \hat s_x(0)\rangle-i\langle
\hat s_y(0)\rangle$ is lost, by expanding the first equality of  (\ref{Psib}) in powers of $\lambda$, 
the results

\BEA
\label{sxb}
\langle\hat s_x(t)\rangle&=&-\frac{b}{g\delta_0}\sqrt{\frac{\pi}{2 N}}
\langle \hat s_z(0)\rangle \exp\left[{-\left(\frac{t}{\tau_{\rm leak}}\right)^2}\right], \\
\label{syb1} 
\langle \hat s_y(t)\rangle&\approx&\frac{b}{g\delta_0}\sqrt{\frac{2}{N}}
\frac{t}{\tau_{\rm leak}}
\left[1-\frac{2}{3}\left(\frac{t}{\tau_{\rm leak}}\right)^2\right],
\qquad t\ll\tau_{\rm leak},\\
\label{syb2} 
\langle\hat s_y(t)\rangle&\sim&\frac{b}{g\delta_0}\frac{1}{\sqrt{2 N}}
\frac{\tau_{\rm leak}}{t},\qquad t\gg \tau_{\rm leak}.
\EEA
where we also used that $r_{\uparrow\uparrow}(0) -r_{\downarrow\downarrow}(0)=\langle\hat s_z(0)\rangle$ and
$r_{\uparrow\uparrow}(0) +r_{\downarrow\downarrow}(0)=1$. For $t$ of order $\tau_{\rm leak}$ these results are of order $b\Delta m/g\delta_0^2$, 
with $\Delta m=\delta_0/\sqrt{N}$ (see Eq. (\ref{DELTAm})).
Because $1-{\rm erfc}(z)={\rm erf}(z)$ is imaginary for imaginary values of  $z$, the correlations $\langle \hat s_x\hat m^k(t)\rangle$, $k\ge 1$, vanish
in this approximation, while $\langle \hat s_y\hat m^k(t)\rangle$
involves an extra factor $\Delta m^k$, for instance:

\BEA
\langle\hat s_y\hat m(t)\rangle&=&\frac{b}{gN}\langle\hat s_z(0)\rangle=\frac{b\,\Delta m^2}{g\delta_0^2}\langle\hat s_z(0)\rangle , \qquad
\langle\hat s_y\hat m^2(t)\rangle =\frac{b\delta_0\sqrt{2}}{gN^{3/2}}\frac{t}{\tau_{\rm leak}} =\frac{b\sqrt{2}\,\Delta m^3}{g\delta_0^2}\, \frac{t}{\tau_{\rm leak}}.
 \EEA 
 To understand these behaviors, we remember that the spin S is
submitted to the field $b$ in the $y$-direction and to the random
field $Ngm$ in the $z$-direction, where $m$ has a fluctuation
$\delta_0/\sqrt{N}$ and an expectation value which varies as $\pm
m_{\rm B}t/\tau_{\rm reg}=\pm\sqrt{2/N}\delta_0t/\tau_{\rm leak}$
if the spin S is polarized in the $\pm z$-direction.  The stationary value of $\langle\hat s_y\hat m(t)\rangle$ agrees with the value of the random field applied to S. 
The precession around $\hat y$ explains the factor $-b\langle \hat
s_z(0)\rangle$ in $\langle \hat s_x(t)\rangle$. The rotation
around $z$ hinders $\langle \hat s_x(t)\rangle$ through randomness
of $m$, its effects are characterized by the parameter $Ngm$, of
order $g\delta_0\sqrt{N}$. This explains the occurrence of this
parameter in the denominator. Moreover, this same rotation around
$\hat z$ feeds up $\langle \hat s_y(t)\rangle$ from $\langle \hat
s_x(t)\rangle$, and it takes place in a direction depending on the
sign of $m$; as soon as registration begins, this sign of $m$ is
on average positive for $s_z=+1$, negative for $s_z=-1$.Thus the
two rotations around $\hat y$ and $\hat z$ yield a polarization
along $\hat x$ with a sign opposite to that along $\hat z$,
whereas the polarization along $\hat y$ is positive whatever that
along $\hat z$. When $t\gg \tau_{\rm leak}$, the random values of
$m$ are all positive (for $P_{\uparrow\uparrow}$) or all negative
(for $P_{\downarrow\downarrow}$), with a modulus larger than
$1/\sqrt{N}$. Hence S precesses around an axis close to $+\hat z$
or $-\hat z$, even if $b$ is of order $g\sqrt{N}$, so that the
leakage from $C_z$ towards $C_x$ and $C_y$ is inhibited for such
times. Altogether, the duration of the effect is $\tau_{\rm
leak}$, and its size is characterized by the dimensionless parameter
$b/g\sqrt{N}$.

}

\subsubsection{Possibility of an ideal measurement}
\label{section.8.2.4}

\hfill{\it The loftier and more distant the ideal,}

\hfill{\it  the greater its power to lift up the soul}

\hfill{Hebrew proverb}

\vspace{3mm}

\ZeText{

We wish to find an upper bound on the field $b$ such that the
process can be used as a measurement. Obviously, if the Larmor
period $\tau_{\rm Larmor}=\pi\hbar/b$ is longer than the
registration time $\tau_{\rm reg}=\hbar/\gamma(J-T)$, we can
completely disregard the field. However, we shall see that this
condition, $b\ll\pi\gamma(J-T)$, is too stringent and that even
large violations of the conservation law of the measured quantity
$\hat s_z$ do not prevent an ideal measurement.

We therefore turn to the registration, still assuming that S and A
remain coupled till the end of the process. At lowest order in
$b$, the right-hand side of (\ref{upb}) and (\ref{downb}) is
expressed by (\ref{elim}) with (\ref{Pb=0}). The Green's functions
$G_\uparrow$ and $G_\downarrow$ of the left-hand sides are given
by (\ref{AfinalG}) with $h=+g$ and $h=-g$, respectively. We thus
find $P_{\uparrow\uparrow}(m,t)$ through convolution of
$G_\uparrow(m,m',t-t')$ with the initial condition
$\delta(t')P_{\uparrow\uparrow}^{(0)}(m,t')$, 
 which involves a Dirac $\delta(t')$, and the factor
 
\begin{equation} 
\frac{b}{\hbar}C_x^{(0)}(m',t')=\frac{b}{\hbar}\left [
P_{\uparrow\downarrow}^{(0)}(m',t')+P_{\downarrow\uparrow}^{(0)}(m',t')\right], \nn
\end{equation} 
and with

\begin{equation}  \frac{b}{\hbar}C_x^{(1)}(m',t')=-\frac{2b^2}{\hbar^2}
\int_0^{t'}{\rm d}t''\cos\left[2Ngm'(t'-t'')/\hbar\right]
\left[P_{\uparrow\uparrow}^{(0)}(m',t')-P_{\downarrow\downarrow}^{(0)}(m',t')\right].
\nn \end{equation} 
For $P_{\downarrow\downarrow}$ we change $G_\uparrow$
into $G_\downarrow$ and $C_x$ into $-C_x$. The zeroth-order
contribution, evaluated in section~\ref{section.7}, corresponds to an ideal
measurement. The first-order correction in $b$,
$P_{\uparrow\uparrow}^{(1)}$ issued from $C_x^{(0)}$, depends on
the transverse initial conditions $r_{\uparrow\downarrow}(0)$,
while the second-order correction, $P_{\uparrow\uparrow}^{(2)}$
issued from $C_x^{(1)}$, depends, as the main term, on
$r_{\uparrow\uparrow}(0)=1-r_{\downarrow\downarrow}(0)$.

Performing the Gaussian integrals on $m'$, we find:

\BEA
\label{P1upb}
&&P_{\uparrow\uparrow}^{(1)}(m,t)=\frac{2b}{\hbar}\Re\int{\rm d}m'{\rm d}t'
G_\uparrow(m,m',t-t')P_{\uparrow\downarrow}^{(0)}(m',t')
 \nn\\ &&
=\frac{2b}{\hbar}\Re\int_0^t{\rm d}t'r_{\uparrow\downarrow}(0)
\sqrt{\frac{N}{2\pi}}
\frac{e^{-(t-t')/\tau_{\rm reg}}}{\delta_1(t-t')}
 \exp\left\{-\frac{N}{2\delta_1^2(t-t')}
\left[ \mu'{}^2+4g^2\delta_0^2\delta_2^2\frac{t'{}^2}{\hbar^2}
-4ig\delta_0^2\mu'\frac{t'}{\hbar}
\right]\right\},  \\
\label{P2upb} 
&&P_{\uparrow\uparrow}^{(2)}(m,t)=-\frac{2b^2}{\hbar^2}
\Re\int_0^t{\rm d}t'\int_0^{t'}{\rm d}t'' r_{\uparrow\uparrow}(0)\sqrt{\frac{N}{2\pi}}
\frac{e^{-(t-t'+t'')/\tau_{\rm reg}}}{\delta_1(t-t'+t'')}
\nn\\&&\times 
\exp\left\{ -\frac{Ne^{-2t''/\tau_{\rm
reg}}}{2\delta_1^2(t-t'+t'')} \left [
(\mu'-\mu'')^2+4g^2e^{2t''/\tau_{\rm
reg}}\delta_1^2(t'')\delta_2^2 \frac{(t-t'')^2}{\hbar^2}
-4ig\left( e^{2t''/\tau_{\rm reg}}\delta_1^2(t'')\mu'
+\delta_2^2\mu''\right)\frac{t'-t''}{\hbar} \right]\right\}, \nn
\\&& -\left\{r_{\uparrow\uparrow}\mapsto r_{\downarrow\downarrow};
\quad \mu'' \mapsto -\mu''\right\}, \EEA
where $\delta_1(t)$ was defined by (\ref{delta1t}), where $\delta_2^2\equiv\delta_1^2(t-t')-\delta_0^2=T[1-e^{-2(t-t')/\tau_{\rm
reg}}]/(J-T)$, where $\mu'\equiv-m_{\rm B}+(m+m_{\rm B})\exp[{-(t-t')/\tau_{\rm reg}}]$ and where $\mu''\equiv m_{\rm
B}[\exp({t''/\tau_{\rm reg}})-1]$. These expressions hold for times $t$ of order $\tau_{\rm reg}$. For later times, the part of
$P_{\uparrow\uparrow}(m,t)$ for which $m$ is above (below) the bifurcation $-m_{\rm B}$  (with $m_{\rm B}=g/(J-T)$) develops a peak around 
$+m_{\rm F}$ $(-m_{\rm F})$. For $P_{\downarrow\downarrow}$, we have to change
the sign in $P_{\uparrow\uparrow}^{(1)}$ and $P_{\uparrow\uparrow}^{(2)}$ and to replace $m_{\rm B}$ by $-m_{\rm B}$ in $\mu'$; the bifurcation in $+m_{\rm B}$. 
The probability of finding $s_z=+1$ and $m\simeq m_{\rm F}$ at the end of the measurement is thus $ \int_{-m_{\rm B}}^1{\rm d}m\,P_{\uparrow\uparrow}(m,t)$, 
while $ \int_{-1}^{-m_{\rm B}}{\rm d}m P_{\uparrow\uparrow}(m,t)$ corresponds to $s_z=1$ and $m\simeq -m_{\rm F}$. Since $ \int_{-m_{\rm B}}^1{\rm
d}m\,P_{\uparrow\uparrow}^{(0)}(m,t)= r_{\uparrow\uparrow}(0)$ and $ \int_{-1}^{-m_{\rm B}}{\rm d}m\,P_{\uparrow\uparrow}^{(0)}(m,t)=0$, the contributions
$P_{\uparrow\uparrow}^{(0)}$ to $P_{\uparrow\uparrow}$ and $P_{\downarrow\downarrow}^{(0)}$ to $P_{\downarrow\downarrow}$
correspond to an ideal measurement. The  corrections of order $b$ and $b^2$ to $P_{\uparrow\uparrow}$ and $P_{\downarrow\downarrow}$
give thus rise to violations of Born's rule, governed at first order in $b$ by the off-diagonal elements $r_{\uparrow\downarrow}(0)$, $r_{\downarrow\uparrow}(0)$ of the
initial density matrix of S, and at second order by the diagonal elements $r_{\uparrow\uparrow}(0)$, $r_{\downarrow\downarrow}(0)$.
For instance, $ \int_{-m_{\rm B}}^1{\rm d}m\,P_{\uparrow\uparrow}^{(2)}(m,t)$ and $ \int_{-1}^{-m_{\rm
B}}{\rm d}m\,P_{\uparrow\uparrow}^{(2)}(m,t)$ are the contributions of these initial diagonal elements to the wrong
counts $+m_{\rm F}$ and $-m_{\rm F}$, respectively, associated with $s_z=+1$ in the final state of S.

In order to estimate these deviations due to non-conservation of
$\hat s_z$, we evaluate, as we did for the transverse quantities
(\ref{sxb}-\ref{syb2}), the expectation values $\langle \hat
s_z(t)\rangle$, $\langle \hat m(t)\rangle$, $\langle \hat s_z\hat
m(t)\rangle$ issued from (\ref{P1upb}) and (\ref{P2upb}). For
times $t\gg\tau_\trunc $ and $t$ not much longer than $\tau_{\rm
reg}$, we find

\BEA
\label{sztb} 
\langle \hat s_z(t)\rangle&=&\int{\rm d}m\left[
P_{\uparrow\uparrow}(m,t)-P_{\downarrow\downarrow}(m,t)\right]=
r_{\uparrow\uparrow}(0)-r_{\downarrow\downarrow}(0)
+\frac{4b}{\hbar}\Re\int_0^t{\rm d}t' r_{\uparrow\downarrow}(0)
\exp\left[-\left(\frac{t'}{\tau_\trunc }\right)^2\right]\nn\\
&&
-\frac{4b^2}{\hbar^2}\Re\int_0^t{\rm d}t'\int_0^{t'}{\rm d}t''
[r_{\uparrow\uparrow}(0)-r_{\downarrow\downarrow}(0)]
\exp\left[-\left(\frac{t'-t''}{\tau_\trunc }\right)^2+2i\frac{t''}{\tau_{\rm leak}}
\left(\frac{t'-t''}{\tau_\trunc }\right)\right] \nn\\
&=& \langle \hat s_z(0)\rangle
+\frac{b}{g\delta_0}\sqrt{\frac{\pi}{2N}}\langle \hat s_x(0)\rangle
-\frac{b^2}{2N\gamma g^2} \langle \hat s_z(0)\rangle
\left[1-{\rm erfc}\left(\frac{t}{\tau_\trunc }\right)\right];
\EEA
we noted that only short times $t'$, $t''$ and $t'-t''$ contribute.
A similar calculation provides

\begin{equation}  \label{mtb} 
\langle \hat m(t)\rangle=\int{\rm d}m\,m
\left[P_{\uparrow\uparrow}(m,t)+P_{\downarrow\downarrow}(m,t)\right]=
\langle \hat s_z(t)\rangle m_{\rm B}\left(e^{t/\tau_{\rm
reg}}-1\right). \end{equation}  For $t\gg \tau_{\rm leak}$, $\langle \hat
s_z(t)\rangle$ tends to a constant which differs from the value
$\langle \hat s_z(0)\rangle$ expected for an ideal measurement.
The ratio $\langle \hat m(t)\rangle/\langle \hat s_z(t)\rangle$
is, however, the same as in section~\ref{section.7} where $b=0$. Finally the
correlation is obtained as

\BEA
\label{szmtb} 
\langle \hat s_z\hat m(t)\rangle&=&
\int{\rm d}m\, m
\left[P_{\uparrow\uparrow}(m,t)-P_{\downarrow\downarrow}(m,t)\right]=
m_{\rm B}\left(e^{t/\tau_{\rm reg}}-1\right) 
+\frac{4b}{\hbar}\Re\int_0^t{\rm d}t' r_{\uparrow\downarrow}(0)
2ig\delta_0^2\frac{t'}{\hbar}e^{t/\tau_{\rm reg}}
\exp\left[-\left(\frac{t'}{\tau_\trunc }\right)^2\right]\nn\\
&&
-\frac{4b^2}{\hbar^2}\Re\int_0^t{\rm d}t'\int_0^{t'}{\rm d}t''
\left(\mu''+2ig\delta_0^2\frac{t'-t''}{\hbar}e^{t/\tau_{\rm reg}}\right)
\exp\left[-\left(\frac{t'-t''}{\tau_\trunc }\right)^2+2i\frac{t''}{\tau_{\rm leak}}
\left(\frac{t'-t''}{\tau_\trunc }\right)\right] \nn\\
&=& m_{\rm B}\left(e^{t/\tau_{\rm reg}}-1\right) +\frac{b}{Ng}
\langle \hat s_y(0)\rangle e^{t/\tau_\trunc}; \EEA the terms in
$b^2$ cancel out. As in (\ref{syb1}), (\ref{syb2}) the correlation
$\langle\hat s_z\hat m(t)\rangle$ is weaker by a factor $\sqrt{N}$
than the expectation value $\langle\hat s_z(t)\rangle$.

Altogether, the field $b$ enters all the results (\ref{sigmab}),
(\ref{sxb}-\ref{syb2}) and (\ref{sztb}-\ref{szmtb}) through the
combination $b/g\sqrt{N}$. However, the dominant deviation from
Born's rule, arising from the last term of (\ref{sztb}), also
involves the coupling $\gamma$ of M with B. The process can
therefore be regarded as an ideal measurement provided 

\begin{equation} 
\label{condb} 
b\ll g\sqrt{N\gamma}. \end{equation}  
Contrary to the probability of an unsuccessful measurement found in
\cite{Yanase}, which depended solely on the size of the apparatus,
the present condition involves $b$, which characterizes the
magnitude of the violation, as well as the couplings, $g$ between
S and M, and $\gamma$ between M and B, which characterize the
dynamics of the process. A large number $N$ of degrees of freedom
of the pointer and/or a large coupling $g$ inhibit the transitions
between $s_z=+1$ and $s_z=-1$ induced by $\hat H_{\rm S}$, making
the leakage time short and rendering the field $b$ ineffective. If
$g$ is small, approaching the lower bound (\ref{gmin}), the
constraint (\ref{condb}) becomes stringent, since $\gamma\ll1$.
Too weak a coupling $\gamma$ with the bath makes the registration
so slow that $b$ has time to spoil the measurement during the
leakage delay.

}

\subsubsection{Switching on and off the system-apparatus interaction}
\label{section.8.2.5}



\hfill{\it Haastige spoed is zelden goed $\hspace{-1mm}$\footnote{Being quick is hardly ever good}}

\hfill{Dutch proverb}

\vspace{0.3cm}
\ZeText{

The condition (\ref{condb}), which ensures that the process behaves as an ideal measurement although $\hat s_z$ is not
conserved, has been established by assuming that S and A interact from the time $t=0$ to the time $t=t_{\rm f}$ at which the pointer
has reached $\pm m_{\rm F}$. However, in a realistic ideal measurement, S and A should be decoupled both before $t=0$ and
after some time larger than $\tau_{\rm reg}$. At such times, the observable $\hat s_z$ to be tested suffers oscillations with
period $\tau_{\rm Larmor}=\pi b/\hbar$, which may be rapid. Two problems then arise.

($i$) The repeated process informs us through reading of M about the diagonal elements of the density matrix $\hat r$ of S, not at
any time, but {\it at the time when the coupling $g$ is switched on}, that we took as the origin of time $t=0$. Before this time,
the diagonal elements $r_{\uparrow\uparrow}(t)$ and $r_{\downarrow\downarrow}(t)$ oscillate freely with the period
$\tau_{\rm Larmor}$. If we wish the outcomes of M to be meaningful, we need to control, within a latitude small compared
to $\tau_{\rm Larmor}$, the time at which the interaction is turned on. Moreover, this coupling must occur suddenly: The time
during which $g$ rises from $0$ to its actual value should be short, much shorter than the leakage time.

($ii$) Suppose that the coupling $g$ is switched off at some time $\tau_{\rm dec}$ larger than $\tau_{\rm reg}$, the condition
(\ref{condb}) being satisfied. At this decoupling time $P_{\uparrow\uparrow}(m,\tau_{\rm dec})$ presents a peak for
$m>0$, with weight $\int\d m\,m \,P_{\uparrow\uparrow}(m,\tau_{\rm dec})=r_{\uparrow\uparrow}(0)$,
$P_{\downarrow\downarrow}(m,\tau_{\rm dec})$ a peak for $m<0$ with weight $r_{\downarrow\downarrow}(0)$, while
$P_{\uparrow\downarrow}(m,\tau_{\rm dec})$ vanishes. Afterwards the system and the apparatus evolve independently. The Larmor
precession of S \cite{spin_echo,spin_echo1,spin_echo2a,spin_echo2b,spin_echo3,spin_echo4} manifests itself through oscillations of 
$\int\d m\, \left[P_{\uparrow\uparrow}(m,t)-P_{\downarrow\downarrow}(m,t)\right]$ and of $\int\d m\,
\left[P_{\uparrow\downarrow}(m,t)+P_{\downarrow\uparrow}(m,t)\right]$, while M relaxes under the influence of the bath B. The two peaks
of the probability distribution $P_{\rm M}(m,t)=P_{\uparrow\uparrow}(m,t)+P_{\downarrow\downarrow}(m,t)$
move apart, towards $+m_{\rm F}$ and $-m_{\rm F}$, respectively. At the final time $t_{\rm f}$, once the apparatus has reached
equilibrium with broken invariance, we can observe on the pointer the outcomes $+m_{\rm F}$ with probability
$r_{\uparrow\uparrow}(0)$, or $-m_{\rm F}$ with probability $r_{\downarrow\downarrow}(0)$. Thus the counting rate agrees with
Born's rule. However the process is not an ideal measurement in von Neumann's sense: 
Even if the outcome of A is well-defined at each run (section 11), it is correlated not with the state of S at the final reading time, but only with its state
$\hat r(\tau_{\rm dec})$ at the decoupling time, a state which has been kept unchanged since the truncation owing to the interaction
of S with M. Selecting the events with $+m_{\rm F}$ at the time $t_{\rm f}$ cannot be used as a preparation of S in the state
$| \hspace{-1mm} \uparrow\rangle$, since $\hat r(t)$ has evolved after the decoupling.

}


\subsection{Attempt to simultaneously measure non-commutative variables}
\renewcommand{\thesection}{\arabic{section}.}
\label{section.8.3}

\hfill{{\it Je moet niet teveel hooi op je vork nemen}\footnote{ You should not put too much hay on your fork}}

\hfill{{\it   Qui trop embrasse mal  \'etreint }\footnote{He who embraces too much fails to catch}}

\hfill{Dutch and French proverbs}

\vspace{0.3cm}
\ZeText{

Books of quantum mechanics tell that a {\it precise} simultaneous
measurement of non-commuting variables is impossible
\cite{blokhintsev1,blokhintsev2,deMuynck,BallentineBook,landau}. It is, however,
physically sensible to imagine a setting with which we would try to
perform such a measurement approximately~\cite{arthurs_complo,martens_complo,appleby_complo,luis_complo,lorenzo_complo,busch_complo,stenholm2}.
It is interesting to analyze the corresponding dynamical process so as
to understand how it differs from a standard measurement. 

Consider first successive measurements. In a first stage the component $\hat s_z$ of the spin S is tested by coupling S to A
between the time $t=0$ and some time $\tau_{\rm dec}$ at which $\hat H_{\rm SA}$ is switched off. If $\tau_{\rm dec}$ is larger
than the registration time $\tau_{\rm reg}$, the apparatus A produces $m=m_{\rm F}$ with probability $r_{\uparrow\uparrow}(0)$
and $m=-m_{\rm F}$ with probability $r_{\downarrow\downarrow}(0)$. An interaction $\hat H_{\rm SA^\prime}$ is then switched on
between S and a second apparatus A$^\prime$, analogous to A but coupled to the component $\hat s_v$ of $\hat{\bf s}$ in some
$v$-direction. It is the new diagonal marginal state $\hat r(\tau_{\rm dec})$, equal to the diagonal part of $\hat r(0)$,
which is then tested by A$^\prime$. In this measurement of $\hat s_v$ the probability of reading $m^\prime=+m_{\rm F}$ on
A$^\prime$ and finding $\hat {\bf s}$ in the $v$-direction is $r_{\uparrow\uparrow}(0)\cos^2\frac{1}{2}\theta
+r_{\downarrow\downarrow}(0)\sin^2\frac{1}{2}\theta$, where $\theta$ and $\phi$ are the Euler angles of ${\bf v}$. The
measurement of $\hat s_v$ alone would have provided the additional contribution $\Re r_{\uparrow\downarrow}(0)\sin\theta e^{i\phi}$.
We therefore recover dynamically all the standard predictions of quantum mechanics.

Things will be different if the second apparatus is switched on too soon after the first one or at the same time. 

}

\subsubsection{A model with two apparatuses}

\hfill{\it Life is transparent,}

\hfill{\it  but we insist on making it opaque}

\hfill{Confucius}

\vspace{3mm}

\label{section.8.3.1}
\ZeText{

Let us imagine we attempt to measure simultaneously the
non-commuting components $\hat s_z$ and $\hat s_x$ of the spin
$\hat{\bf s}$. To this aim we extend our model by assuming that,
starting from the time $t=0$, S is coupled with two apparatuses A
and A$^\prime$ of the same type as above, A$^\prime$ being suited
to the measurement of $\hat s_x$. We denote by $\gamma'$, $g'$,
$N'$, $J'$, $T'$, \ldots, the parameters of the second apparatus.
The overall Hamiltonian $\hat H=\hat H_{\rm SA}+\hat H_{\rm
SA'}+\hat H_{\rm A}+\hat H_{\rm A'}$ thus involves, in addition to
the contributions defined in subsection~\ref{section.3.2}, the Hamiltonian $\hat
H_{\rm A'}$ of the second apparatus A$^\prime$, analogous to $\hat H_{\rm
A}=\hat H_{\rm M}+\hat H_{\rm B}+\hat H_{\rm MB}$, with
magnetization $m'=(1/N')\sum_{n=1}^{N'} \hat\sigma_x'{}^{(n)}$, and
the coupling term

 \begin{equation} 
 \label{HSAprime}
  \hat H_{\rm SA'}=-N'g'\hat s_x\hat m' \nn 
 \end{equation}  
 of A$^\prime$ and S. The solution of the Liouville--von Neumann equation for S + A+A$^\prime$ should determine how the indications of A and A$^\prime$ 
can inform us about the initial state $\hat r(0)$ of S, and how the final state of S is correlated with these indications.

We readily note that such a dynamical process can not behave as an ideal measurement, since we expect that, whatever the
initial state $\hat r(0)$ of S, its final state will be perturbed.

 The equations of motion are worked out as in section~\ref{section.4}.  After elimination of the baths B and B$^\prime$
at lowest order in $\gamma$ and $\gamma^\prime$, the density operator $\hat D$ of S + M +  M$'$ can be parametrized as in \S~\ref{section.3.3.1}
and \S~\ref{section.4.3.1} by four functions $P_{ij}(m,m',t)$, where $i,j=\uparrow,\downarrow$ refer to S, and where the magnetizations
$m$ and $m'$ behave as random variables. However, since the functions $P_{ij}$ are now coupled, it is more suitable to express the dynamics in 
terms of $P_{{\rm MM}'}(m,m',t)=P_{\uparrow\uparrow}+P_{\downarrow\downarrow}$, which describes the joint probability distribution of $m$ and $m'$, and
of the set $C_a(m,m',t)$ defined for $a=x,y$ and $z$ by (\ref{P->C}), which describe the correlations between $\hat s_a$ and the two magnets M and  M$'$. 
The density operator $\hat D(t)$ of S + M +  M$'$ generalizing (\ref{Rij}), with (\ref{Pdis2cont}), (\ref{RijP}) and  (\ref{P->C}), is

\begin{equation}  \label{dop} 
\hat D(t)=\frac{2}{NN'G(\hat m)G(\hat m')}\left[ P_{{\rm MM}'}(\hat m,\hat m',t)+{\bf C}(\hat m,\hat m',t)\cdot\hat{\bf s}\right]. \nn
\end{equation}  
(There is no ambiguity in this definition, since $\hat m$ and $\hat m'$ commute.) The full dynamics are thus governed by coupled equations for the
 functions $P_{{\rm MM}'}(m,m',t)$ and ${\bf C}(m,m',t)$ which parametrize $\hat D(t)$. The initial state $\hat D(0)$ is factorized as $\hat r(0)\otimes
\hat R_{\rm M}(0)\otimes \hat R_{\rm M^\prime}(0)$, where $\hat R_{\rm M}(0)$ and $\hat R_{\rm M^\prime}(0)$ describe the
metastable paramagnetic states (\ref{HF0}) of M and M$^\prime$, so that the initial conditions are

\begin{equation}  \label{initP} 
P_{{\rm MM}'}(m,m',0)=P_{\rm M}(m,0)P_{{\rm M}'}(m',0),\qquad {\bf C}(m,m',0)=P_{{\rm MM}'}(m,m',0)\langle\hat {\bf s}(0)\rangle, 
\end{equation}  
where $P_{\rm M}(m,0)$ and $P_{{\rm M}'}(m^\prime,0)$ have the Gaussian form (\ref{Pm0}) and where $\langle \hat{\bf s}(0)\rangle$ is the
initial polarization of S.

Two types of contributions enter $\partial P_{{\rm MM}'}/\partial t$ and $\partial{\bf C}/\partial t$, the first one active on the time scale $\tau_\trunc$, 
and the second one on the time scale $\tau_{\rm reg}$, but these time scales need not be very different here. 
On the one hand, for given $m$ and $m'$, the coupling $\hat H_{\rm SA}+\hat H_{\rm SA'}$ of S with the 
magnets M and M$^\prime$ behaves as a magnetic field ${\bf b}$ applied to S. This effective field is equal to

\begin{equation}   \label{efffield} 
{\bf b}(m,m')=\frac{2Ngm}{\hbar}\hat{\bf z}+\frac{2N'g'm'}{\hbar}\hat{\bf x}=b\hat {\bf u},  \qquad 
b(m,m')\equiv  |{\bf b}(m,m')|=\frac{2}{\hbar}\sqrt{N^2g^2m^2+N'{}^2g'{}^2m'{}^2},
 \end{equation}
 where $\hat{\bf z}$ and $\hat{\bf x}$ are the unit vectors in the $z$- and $x$-direction, respectively. This yields to $\partial{\bf C}/\partial t$ the contribution

\begin{equation}  
\left[\frac{\partial{\bf C}(m,m',t)}{\partial t}\right]_{{\rm MM}'}= -{\bf b}(m,m')\times{\bf
C}(m,m',t).
\label{RotC}
 \end{equation}  
Both the Larmor frequency $b$ and the precession axis, characterized by the unit vector $\hat {\bf u}={\bf b}/b$ in the $x-z$ plane, depend on $m$ and $m'$
(whereas the precession axis was fixed along $\hat{\bf z}$ or $\hat{\bf x}$ for a single apparatus). The distribution $P_{{\rm MM}'}(m,m',t)$ is insensitive to
the part $\hat H_{\rm SA}+\hat H_{\rm SA'}$ of the Hamiltonian, and therefore evolves slowly, only under the effect of the baths.

On the other hand, $\partial P_{{\rm MM}'}/\partial t$ and $\partial {\bf C}/\partial t$ involve contributions from the baths B and B$'$,
which can be derived from the right-hand sides of (\ref{dPdiag2}) and (\ref{dPoff2}). They couple all four functions $P_{{\rm MM}'}$ and ${\bf
C}$, they are characterized by the time scales $\tau_{\rm reg}$ and $\tau'_{reg}$, and they depend on all parameters of the model. In contrast
with what happened for a single apparatus, the effects of the precession (\ref{RotC}) and of the baths can no longer be separated.
 Indeed, the precession tends to eliminate the components
of ${\bf C}(m,m',t)$ that are perpendicular to ${\bf b}$, but the baths tend to continuously activate the creation of such components. 
The truncation, which for a single apparatus involved only the off-diagonal sectors and was achieved after a brief delay, is now replaced by an overall 
damping process taking place along with the registration, under the simultaneous contradictory effects of the couplings of M
and M$^\prime$ with S and with the baths.

Such an interplay, together with the coupling of four functions $P_{{\rm MM}'}$, ${\bf C}$ of three variables $m$, $m^\prime$, $t$, make the
equations of motion difficult to solve, whether analytically or numerically. A qualitative analysis will, however, suffice to provide us with
some interesting conclusions.

}

\subsubsection{Structure of the outcome}
\label{section.8.3.2}

\ZeText{

Note first that the positivity of the density operator
(\ref{dop}), maintained by the dynamics, is expressed by the
condition

\begin{equation}  \label{posit}
P_{{\rm MM}'}(m,m',t)\ge |{\bf C}(m,m',t)|,
\end{equation}  which holds at any time.

The outcome of the process is characterized by the limit, for $t$ larger than the registration time $\tau_{\rm reg}$, of the
distributions $P_{{\rm MM}'}$ and ${\bf C}$. In this last stage of the evolution, the interaction of M with the bath B is expected to
drive it towards either one of the two equilibrium states at temperature $T$, for which the normalized distribution
$P_{{\rm M}\Uparrow}(m)$ (or $P_{{\rm M}\Downarrow}(m)$) expressed by (\ref{Pim}) is concentrated near $m=+m_{\rm F}$ (or $m=-m_{\rm F}$). 
In order to avoid the possibility of a final relaxation of M towards its metastable paramagnetic state, which may produce failures as in \S~7.3.4, 
we consider here only a quadratic coupling $J_2$. Likewise, M$^\prime$ is stabilized into either one of the ferromagnetic states $P_{{{\rm M}'}\Uparrow}'(m')$ 
(or $P_{{{\rm M}'}\Downarrow}(m')$) with $m'\simeq+m'_{\rm F}$ (or $m'=-m'_{\rm F}$). Hence, $P_{{\rm MM}'}(m,m',t)$, which describes the statistics of the 
indications of the pointers, ends up as a sum of four narrow peaks which settle at $m=\varepsilon m_{\rm F}$, $m'=\varepsilon' m'_{\rm F}$, with
$\varepsilon=\pm1$, $\varepsilon'=\pm1$, to wit,

\begin{equation} \label{8.78}
P_{{\rm MM}'}(m,m',t)\mapsto\sum_{\varepsilon=\pm1}\sum_{\varepsilon'=\pm1}
{\cal P}_{\varepsilon\varepsilon'}P_{{\rm M}\varepsilon}(m)
P_{{{\rm M}'}\varepsilon'}(m').\end{equation}  The weights ${\cal P}_
{\varepsilon\varepsilon'}$ of these peaks characterize the
proportions of counts detected on M and M$^\prime$ in repeated
experiments; they are the only observed quantities.

The precession (\ref{RotC}) together with smoothing over $m$ and $m'$
eliminates the component $C_y$ of ${\bf C}$, so that the subsequent evolution keeps no memory of $C_y(m,m',0)$. Thus, among
the initial data (\ref{initP}) pertaining to S, only $\langle \hat s_x(0)\rangle$ and $\langle \hat s_z(0)\rangle$ are relevant to
the determination of the final state: the frequencies ${\cal P}_{\varepsilon\varepsilon'}$ of the
outcomes depend only on $\langle \hat s_x(0)\rangle$ and $\langle \hat s_z(0)\rangle$ (and on the parameters of the apparatuses).

If $\langle \hat s_x(0)\rangle=\langle \hat s_z(0)\rangle=0$ we have ${\cal P}_{\varepsilon\varepsilon'}=\frac{1}{4}$ due to the
symmetry $m\leftrightarrow -m$, $m'\leftrightarrow -m'$. Likewise, if $\langle \hat s_x(0)\rangle=0$, the symmetry $m'\leftrightarrow
-m'$ implies that ${\cal P}_{++}={\cal P}_{+-}$ and ${\cal P}_{-+}={\cal P}_{--}$. Since the equations of motion are linear,
${\cal P}_{++}-{\cal P}_{-+}$ is in this situation proportional to $\langle \hat s_z(0)\rangle$; we define the proportionality
coefficient $\lambda$ by ${\cal P}_{\varepsilon+}=\frac{1}{4}(1+\varepsilon\lambda\langle \hat
s_z(0)\rangle)$. In the situation $\langle \hat s_z(0)\rangle=0$ we have similarly   ${\cal P}_{+\varepsilon'}={\cal
P}_{-\varepsilon'}=\frac{1}{4}(1+\varepsilon'\lambda'\langle \hat s_x(0)\rangle)$. Relying on the linearity of the equations of
motion, we find altogether for an arbitrary initial state of S the general form for the probabilities ${\cal P}_{\varepsilon\varepsilon'}$:

\begin{equation}  
\label{Pepseps'} \label{8.79}
{\cal P}_{\varepsilon\varepsilon'}=\frac{1}{4}\left(1+\varepsilon\lambda\langle \hat s_z(0)\rangle+\varepsilon'\lambda'\langle \hat
s_x(0)\rangle\right) , \end{equation}  
where $\langle \hat s_z(0)\rangle=r_{\uparrow\uparrow}(0)- r_{\downarrow\downarrow}(0)$, $\langle \hat s_x(0)\rangle=r_{\uparrow\downarrow}(0)+
r_{\downarrow\uparrow}(0)$. We term $\lambda$ and $\lambda'$ the {\it efficiency factors}.

In the long time limit, the functions ${\bf C}(m.m',t)$ also tend to sums of four peaks located at $m=\pm m_{\rm F}$, $m'=\pm m'_{\rm F}$, as implied by (\ref{posit}). 
With each peak is associated a direction ${\bf u}_{\varepsilon\varepsilon'}$, given by (\ref{efffield}) where $m=\varepsilon m_{\rm F}$, 
$m'=\varepsilon' m'_{\rm F}$, around which the precession (\ref{RotC}) takes place. The truncation process eliminates the component of ${\bf C}$ perpendicular 
to ${\bf u}_{\varepsilon\varepsilon'}$, for each peak. Thus, if in their final state the apparatuses M and M$^\prime$ indicate $\varepsilon m_{\rm F}$, 
$\varepsilon' m'_{\rm F}$, the spin S is lead into a state partly polarized in the direction ${\bf u}_{\varepsilon\varepsilon'}$ of the effective
field ${\bf b}$ generated by the two ferromagnets.

}

\subsubsection{A fully informative statistical process}
\label{section.8.3.3}

\hfill{\it You may look up for inspiration or look down in desperation,}

\hfill{\it  but do not look sideways for information}

\hfill{Indian proverb}

\vspace{3mm}

\ZeText{

A well-defined indication for both pointers M and M$'$ can be obtained here in each individual run, because the argument of \S~\ref{fin11.2.3}
 holds separately for the apparatuses A and A$'$ at the end of the process. A mere counting of the pair of outcomes $\varepsilon$, 
 $\varepsilon'$ then provides experimentally the probability  (\ref{Pepseps'}).
	
However, the present process cannot be regarded as an ideal measurement. On the one hand, the above-mentioned correlations between the final state of S and 
the indications of the apparatus are not complete; they are limited by the inequality (\ref{posit}). In an ideal measurement the correlation must be {\it complete}:
if the apparatuses are such that they provide well-defined outcomes at each run (section 11), and if for a given run we read $+m_{\rm F}$ on the apparatus 
M measuring $\hat s_z$, the spin S must have been led by the ideal process into the pure state $|\hspace{-1mm}\uparrow\rangle$. 
Here we cannot make such assertions about an individual system, and we cannot use the process as a preparation.

On the other hand, in an ideal measurement, the outcome of the process is {\it unique} for both S and M in case S is initially in an eigenstate of the tested quantity. 
Suppose the spin S is initially oriented up in the $z$-direction, that is, $\hat r(0)=|\hspace{-1mm}\uparrow\rangle\langle\uparrow\hspace{-1mm}|$.  
The response of the apparatuses M and M$'$ is given by (\ref{Pepseps'}) as

\begin{equation}  {\cal P}_{++}={\cal P}_{+-}=\frac{1}{4}(1+\lambda),\qquad
{\cal P}_{-+}={\cal P}_{--}=\frac{1}{4}(1-\lambda), \nn 
\end{equation}  
so that there exists a probability $\frac{1}{2}(1-\lambda)$ to {\it read the wrong result} $-m_{\rm F}$ on M. Indeed, without even
solving the equations of motion to express the efficiency factors $\lambda$ and $\lambda'$ in terms of the various parameters of the model, we can
assert that $\lambda$ is  smaller than $1$: Because all ${\cal P}_{\varepsilon\varepsilon'}$ must be non-negative for any initial state of S, and because 
(\ref{Pepseps'})  has the form  $\frac{1}{4}(1+{\bf a}\cdot \langle\hat{\bf s}(0)\rangle)$,  we must have $|{\bf a}|<1$, so that $\lambda$ and $\lambda'$ should satisfy

\begin{equation}  \lambda^2+\lambda'{}^2\le 1, \nn
\end{equation}  
\ni and because not only $\hat s_z$ but also $\hat s_x$ are tested, $\lambda'$ should be non zero so that the probability of failure $\half(1-\lambda)$ is finite.
It is therefore clear  why the attempt to perform a simultaneous {\it ideal} measurement of $\hat s_x$ and $\hat s_z$ fails. Both Born's rule and von Neumann's 
truncation are violated, because A and A$^\prime$ influence each other though their couplings to S.

Nevertheless, consider a set of repeated experiments in which we read simultaneously the indications of the two apparatuses M and 
M$^\prime$. If the runs are sufficiently numerous, we can determine the probabilities ${\cal P}_{\varepsilon\varepsilon'}$ from the frequencies of occurrence 
of the four possible outcomes $\pm m_{\rm F}$, $\pm m_{{\rm F}'}$.  Let us assume that the coefficients $\lambda$, $\lambda'$, which depend on the parameters
of the model, take significant values. This requires an adequate choice of these parameters. In particular, the couplings $g$ and $g'$, needed to trigger the 
beginning of the registration, should however be small and should soon be switched off so as to 
reduce the blurring effect of the precession around ${\bf b}$. This smallness is consistent with the choice of a second order transition for M, already noted. 
Finally, the couplings $\gamma$, $\gamma'$ should ensure registration before disorder is settled. Under such conditions, inversion of eq. (\ref{Pepseps'}) yields

\BEA\label{Cl->Qu} 
 \langle \hat s_z(0)\rangle&=&r_{\uparrow\uparrow}(0)-r_{\downarrow\downarrow}(0)=\frac{1}{\lambda}( {\cal P}_{++}
+{\cal P}_{+-}-{\cal P}_{-+}-{\cal P}_{--}), \nn\\
\langle\hat s_x(0)\rangle&=&r_{\uparrow\downarrow}(0)+r_{\downarrow\uparrow}(0)=\frac{1}{\lambda'}({\cal P}_{++} -{\cal P}_{+-}+{\cal P}_{-+}-{\cal P}_{--}). 
\EEA 
Thus, a sequence of repeated experiments {\it reveals the initial expectation values of both} $\hat s_z$ {\it and} $\hat s_x$, although these observables do not commute.

Paradoxically, as regards the determination of an unknown initial density matrix, the present process is {\it more informative than
an ideal measurement} with a single apparatus~\cite{ABNprl2004}. Repeated measurements of $\hat s_z$ yield $r_{\uparrow\uparrow}(0)$ (and
$r_{\downarrow\downarrow}(0)$) through counting of the outcomes $\pm m_{\rm F}$ of M. Here we moreover find through repeated
experiments the real part of $r_{\uparrow\downarrow}(0)$. However, more numerous runs are needed to reach a given precision if
$\lambda$ and $\lambda'$ are small.  (If the parameters of the model are such that $\lambda$ and $\lambda'$ nearly vanish, the
relaxation of M and M$^\prime$ is not controlled by S, all ${\cal P}_{\varepsilon\varepsilon'}$ lie close to
$\frac{1}{4}$, and the observation of the outcomes is not informative since they are fully random.)

More generally, for a repeated process using three apparatuses M, M$^\prime$ and M$''$ coupled to $\hat s_z$, $\hat s_x$ and $\hat
s_y$, respectively, the statistics of readings allows us to determine simultaneously all matrix elements of the initial
density operator $\hat r(0)$. The considered {\it single compound apparatus} thus provides full statistical information about the state $\hat
r(0)$ of S. Our knowledge is gained indirectly, through an expression of the type (\ref{Cl->Qu}) which involves both {\it
statistics} and {\it calibration} so as to determine the parameters $\lambda$, $\lambda'$ and $\lambda''$. A process of the
present type, although it violates the standard rules of the ideal measurement, can be regarded as a {\it complete statistical
measurement} of the initial state of S. The knowledge of the efficiency factors allows us to determine simultaneously the
statistics of the observables currently regarded as incompatible. The price to pay is the loss of precision due to the fact that the
efficiency factors are less than 1, which requires a large number of runs.

The dynamics thus establish a one-to-one {\it correspondence between the initial density matrix of} S, which embeds the whole quantum
probabilistic information on S, and the {\it classical probabilities of the various indications} that may be registered by the apparatuses
at the final time. The possibility of such a mapping was considered in \cite{ABNprl2004}. The size of the domain in which the
counting rates may lie is limited; for instance, if S is initially polarized along $z$ in (\ref{Pepseps'}), no ${\cal P}_{\varepsilon\varepsilon'}$ can lie 
beyond the interval $[\frac{1}{4}(1-\lambda),\frac{1}{4}(1+\lambda)]$.  The limited size of the domain for the probabilities of the apparatus
indications is needed to reconcile the classical nature of these probabilities with the peculiarities of the quantum probabilities of
S that arise from non commutation.  It also sets limitations on the  precision of the measurement.

\vspace{3mm}

Motivated by the physics of spin-orbit interaction in solids,  Sokolovski and Sherman recently studied a model related to (\ref{HSAprime})~\cite{A7}.
Two components of the spin $\half$ couple not with collective magnetizations as in (\ref{HSAprime}), but with the
components of the momentum (the proper kinetic energy is neglected so that these are the only two terms in the Hamiltonian). The motivation
for studying this model is the same as above: to understand the physics of simultaneous measurement for two non-commuting observables \cite{A7}.
The authors show that, as a result of interaction, the average components of the momentum get correlated with the
time-averaged values of the spin  [instead of the initial values of the spin as in (\ref{8.78}), (\ref{8.79})]. This difference relates to the fact that the
model by Sokolovski and Sherman does not have macroscopic measuring apparatuses that would enforce relaxation in time.

}

\subsubsection{Testing Bell's inequality}
\label{section.8.3.4}

\hfill{\it Love levels all inequalities}

\hfill{Italian proverb}

\vspace{3mm}

\ZeText{

Bell's inequality for an EPR ~\cite{EPR} pair of spins is expressed in the 
CHSH form as \cite{CHSH}

\begin{equation}  \label{CHSH}
| \langle\hat s^{(1)}_a\hat s^{(2)}_{a'}\rangle
+\langle\hat s^{(1)}_b\hat s^{(2)}_{a'}\rangle
+\langle\hat s^{(1)}_a\hat s^{(2)}_{b'}\rangle
-\langle\hat s^{(1)}_b\hat s^{(2)}_{b'}\rangle|
\le 2,\end{equation} 

\ni which holds for classical random variables $s=\pm1$. If $\hat s^{(1)}_a$ and $\hat s^{(1)}_b$ are the components of a quantum spin
$\hat {\bf s}^{(1)}$ in the two fixed directions $a$ and $b$, and $\hat s^{(2)}_{a'}$ and $\hat s^{(2)}_{b'}$ the components the other spin
$\hat {\bf s}^{(2)}$ in directions $a'$ and $b'$,  the left-hand side of (\ref{CHSH}) can rise up to $2\sqrt{2}$
\footnote{For the establishment of Bell-type equalities for SQUIDs, see Jaeger et al. \cite{Jaeger}}.

Standard measurement devices allow us to test simultaneously a pair of commuting observables, for instance $\hat s^{(1)}_a$ and $\hat s^{(2)}_{a'}$.
At least theoretically, the counting rates in repeated runs {\it directly} provide their correlation, namely  $\langle\hat s^{(1)}_a\hat s^{(2)}_{a'}\rangle$.
However, since $\hat s^{(1)}_a$ and $\hat s^{(1)}_b,$ as well as $\hat s^{(2)}_{a'}$ and $\hat s^{(2)}_{b'}$ do not commute, we need four different settings 
to determine the four terms of (\ref{CHSH}). Checking the violation of Bellуs inequalities thus requires combining the outcomes of 
{\it four incompatible experimental contexts} \cite{Accardi1981,Accardi,Khrennikov}, in each of which the spin pair is being tested through repeated runs. 
This necessity may be regarded as a ``contextuality loophole''~\cite{TheoBell1,TheoBell2}.  Either hidden variables exist,
 and they cannot be governed by ordinary probabilities and ordinary logic, since there is no global 
 distribution function that would yield as marginals the partial results tested in the four different contexts. 
 Or we must admit that quantum mechanics forbids us to put together the results of these different measurements. 
 The latter alternative is favored by the solution of models, in which the values of physical quantities do not 
 pre-exist but are produced during a measurement process owing to the interaction between the system and 
 the apparatus. Since these values reflect the reality of the system only within its context, it appears inconsistent 
 to put them together~\cite{deMuynck,Accardi1981,Accardi,Khrennikov,TheoBell1,TheoBell2}.

In the present situation it is tempting to imagine using a combination of apparatuses of the previous type so as to {\it simultaneously test 
all four non-commuting observables} $\hat s^{(1)}_a$, $\hat s^{(1)}_b$, $\hat s^{(2)}_{a'}$, and $\hat s^{(2)}_{b'}$ through repeated runs. 
Such a unique experimental setting would bypass the  contextuality loophole. However, as shown in \S~\ref{section.8.3.3}, the counting rates of the
two apparatuses associated with the components $\hat s^{(1)}_a$ and $\hat s^{(1)}_b$ of the first spin are not directly related to the statistics of these
components, but only reflect them through an efficiency factor $\lambda$ at most equal to $1/\sqrt{2}$. For the pair of spins, one 
can generalize Eqs. (8.79) , (8.82), and  {\it deduce} a correlation 
such as $\langle \hat s^{(1)}_a\hat s^{(2)}_{a'}\rangle$ from the statistical indications of the corresponding apparatuses, but this {\it quantum
correlation} is at least equal to {\it twice} the associated {\it observed correlation} (since $1/\lambda^2>2$). 

Thus, with this experimental setting which circumvents the contextuallity loophole, the correlations directly exhibited by the counting rates 
{\it satisfy Bell's inequality};  this is natural since the outcomes of the macroscopic apparatus are measured simultaneously and therefore have a 
{\it classical} nature \cite{HessMichielsenDeRaedt}. However, from these very observations,
we can use standard quantum mechanics to analyse the results. We thus infer indirectly from the observations, by using a mapping of the type (8.82), 
the tested quantum correlations (8.83) between spins components. Within a {\it single set of repeated experiments} where the various data are simultaneously 
registered, we thus acknowledge the {\it violation} of Bell's inequality. Here this violation no longer appears as a consequence of merging incompatible sets 
of measurements, but as a consequence of a theoretical analysis of the ordinary correlations produced in the apparatus.

 }


 \renewcommand{\thesection}{\arabic{section}}
\section{Analysis of the results}
  \setcounter{equation}{0}\setcounter{figure}{0}\renewcommand{\thesection}{\arabic{section}.}

\label{section.9}
\label{fin9}

\hfill{{\it And the rain from heaven was restrained} }

\hfill{Genesis 8.2}

\vspace{0.3cm}

\ZeText{

 In section 3 we have introduced the Curie--Weiss model for the quantum measurement of a spin $\half$ and in sections 4--8 we have discussed
the dynamics of the density operator characterizing a large set of runs. For the readers who have not desired to go through all the details, and for those who did, 
we resume here the main points as a separate reading guide, and add pedagogical hints for making students  familiar with the matter and techniques.
We will discuss the solution of the quantum measurement problem for this model in section 11 by considering properties of individual runs. 

}

\subsection{Requirements for models of quantum measurements}
\label{section.9.1.1a}
\label{fin9.1}


\hfill{{\it J'ai perdu mon Eurydice\footnote{I lost my Euridice}}}

\hfill{{\it Che far\`o senza Euridice?}\footnote{What shall I do without Euridice?}}

  \hfill{
  Christoph Willibald Gluck,  Orph\'ee et Eurydice; Orfeo ed Euridice}

\vspace{3mm}
\ZeText{

A model for the apparatus A and its coupling with the tested system S that accounts for the various properties of ideal quantum measurements should
 in principle  satisfy the following requirements (``R''):  

\vspace{0.25cm}

\noindent\noindent
{\bf R1}: simulate as much as possible nearly ideal real experiments, and be sufficiently flexible to allow discussing imperfect processes;  

\vspace{0.25cm}

\noindent\noindent
{\bf R2}: ensure unbiased, robust and permanent registration by the pointer of A, which should therefore be macroscopic;

\vspace{0.25cm}

\noindent\noindent
{\bf R3}: involve an apparatus initially in a metastable state and evolving towards one or another stable state
under the influence of S, so as to amplify this signal; the transition of A, instead of occurring spontaneously, is triggered by S;

\vspace{0.25cm}

\noindent\noindent
{\bf R4}: include a bath where the free energy released because of the irreversibility of the process may be dumped;

\vspace{0.25cm}

\noindent\noindent
{\bf R5}: be solvable so as to provide a complete scenario of the joint evolution of S + A and to exhibit the characteristic times;

\vspace{0.25cm}

\noindent\noindent
{\bf R6}:  conserve the tested observable;

\vspace{0.25cm}

\noindent\noindent
{\bf R7}: lead to a final state devoid of ``Schr\"odinger cats"; for the whole set of runs (truncation, \S~1.3.2), and to a von Neumann 
reduced state for each individual run;

\vspace{0.25cm}

\noindent\noindent
{\bf R8}: satisfy Born's rule for the registered results;

\vspace{0.25cm}

\noindent\noindent
{\bf R9}:
produce, for ideal measurements or preparations, the required diagonal correlations between the tested system S 
and the indication of the pointer, as coded in the expression (9.1) for the final state of S + A;  

\vspace{0.25cm}

\noindent\noindent
{\bf R10}:  ensure that the pointer gives at each run a well-defined indication; this requires sufficiently complex interactions within the apparatus
 (dynamical stability and hierarchic structure of subensembles, see subsection \ref{fin11.2}).
\vspace{0.25cm}

\vspace{0.25cm}

\noindent\noindent

These features need not be fulfilled with mathematical rigor. A physical scope is sufficient, where violations
may occur over unreachable time scales or with a negligible probability.

}


\subsection{Features of the Curie--Weiss model}
\label{section.9.1.1}
\label{fin9.2}

\hfill{\it When you can measure  what you are speaking about,}

\hfill{\it and express it in numbers,  you know something about it}

\hfill{Lord Kelvin}

\vspace{3mm}

\ZeText{

The above Curie--Weiss model is satisfactory in this respect (except for the requirement {\bf  R10} which will be discussed in \S~\ref{fin11.2.1}). 
Its choice  (section 3) has relied on a compromise between two conflicting requirements. On the one hand, the apparatus A simulates a {\it real object},
a magnetic dot which behaves as a magnetic memory. On the other hand, the Hamiltonian
of S + A is sufficiently simple so as to afford an explicit and detailed {\it dynamical solution}.
The registration device is schematized as a set M of $N$ Ising spins (the magnet).
The size of the dot is supposed to be much smaller than the range of the interactions, both 
among the $N$ spins and between them and the tested spin S. We further simplify by taking into 
account only interactions between the $z$-components of the spins of M and S.
Finally, as in a real magnetic dot, phonons (with a quasi-ohmic behavior \cite{Weiss,Gardiner,ms,caldeira,petr}) 
behave as a thermal
bath B which ensures equilibrium in the final state (Fig. 3.1). 
In spite of the schematic nature of the model, its solution turns out to exhibit a rich structure and to display 
the various features listed in subsection \ref{section.9.1.1a}.

In particular, the choice for A = M + B of a system which can undergo a phase transition implies many
properties desirable for a measuring apparatus.  The weakness of the interaction $\gamma$ 
between each spin of the magnet M and the phonon bath B, maintained at a temperature $T$
lower than $T_{\rm c}$, ensures a long lifetime for the initial {\it metastable} paramagnetic state.
By itself, the system M+B would ultimately relax spontaneously towards a stable state, but here its transition is triggered by S. 
The symmetry breaking in the dynamics of the measurement produces either one of the two possible final 
{\it stable} ferromagnetic states, in {\it one-to-one correspondence}
with the eigenvalues of the tested observable $\hat s_z$ of the system S, so that the sign of
the final magnetization can behave as a pointer. It is this breaking of symmetry which underlies registration, entailing the 
{\it irreversibility} of the transition from the paramagnetic to either one of the ferromagnetic macroscopic states. Moreover, the built-in symmetry 
between the  two possible outcomes of A prevents the appearance of {\it bias}.

An essential property of a measurement, often overlooked, is the ability of the apparatus A
to {\it register} the indication of the pointer. Here this is ensured by the large value of the
number $N$ of spins of M, which entails a neat separation between the two ferromagnetic 
states of M and their extremely long lifetime. This stability warrants a {\it permanent} and {\it robust} registration.
The large value of $N$ is also an essential ingredient in the proof of the uniqueness of the indication fo the pointer in each run (\S~\ref{fin11.2.3}).
In both the paramagnetic state and the ferromagnetic states, the pointer variable $m$ presents
statistical fluctuations negligible as $1/\sqrt{N}$. Moreover, breaking of invariance makes quantum coherences ineffective (\S~\ref{fin11.2.3}).
The nature of the order parameter, a macroscopic magnetization, also makes the result
{\it accessible to reading, processing or printing}. These properties cannot be implemented in models
for which the pointer is a microscopic object.

The coupling between the tested spin S and  the apparatus A has been chosen in such a way
that the observable $\hat s_z$ is {\it conserved}, $[\hat s_z,\hat H]=0$, so as to remain
unperturbed during its measurement. This coupling {\it triggers} the beginning of the registration
process, which thereby ends up in a situation which {\it informs us} about the the physical state of S at the final moment,
so that the process might be used as a measurement.  This requires a sufficiently large value of the coupling
constant $g$ which characterizes the interaction of S and M.

Once the probability distribution of the magnetization $m$ has left the vicinity of $m=0$
to move towards either $+m_{\rm F}$ or $-m_{\rm F}$, the motion of this pointer is 
{\it driven by the bath} through the coupling $\gamma$ between M and B.
Somewhat later the interaction $g$ between S and A becomes ineffective and can as well be switched off.
It is the interplay between the metastability of the initial state of A, the initial triggering of M by S,
and the ensuing action of B on M which ensures an {\it amplification} of the initial perturbation.
This amplification is necessary since the indication of the pointer M, which is macroscopic,
should reflect an effect caused by the tested system S, which is microscopic ---  the very essence of a measurement.

Such a number of adequate properties makes this model attractive, but technical developments 
were needed to elaborate in sections 4 to 7 a rigorous proof that the final state of S + A has the form (\ref{Dtf=}), viz. 

  \begin{equation}
  \mytext{\textcurrency Dtf=\textcurrency \qquad}
  {\hat{{\cal D}}}\left(
  {t_{{\rm f}}}\right)  =\sum_{i}\left(  \hat{\Pi}_{i}\hat{r}\left(
  0\right)  \hat{\Pi}_{i}\right)  \otimes\hat{{\cal R}}_{i}
  =\sum_ip_i\hat r_i\otimes\hat{\cal R}_i{ ,}
  \label{Dtf9=}
  \end{equation}
 where $\hat r$ describes S and $\hat {\cal R}$ describes A. This form encompasses most among the required specific features of ideal quantum measurements, 
 in particular the absence of off-diagonal terms. These developments have allowed us to {\it discuss the conditions} under which the process might be used as 
a measurement, and also to explore what happens if one or another condition is violated.

Note, however, that the final form (9.1) of the density operator of S + A concerns the statistics of a large set of runs of the measurement. 
This form is {\it necessary, but not sufficient}, to ensure that the interaction process can be regarded as an ideal measurement.
 It remains to elaborate the physical interpretation of this result by turning to individual measurements. We postpone his task to section 11. 

}

\subsection{Scenario of  the Curie--Weiss  ideal measurement: the characteristic time scales}
\label{section.9.1.2}
\label{fin9.3}

\hfill{\it When God made time, he made enough of it}

\hfill{Irish proverb}

\vspace{3mm}

\ZeText{

The above study (sections \ref{section.4}--\ref{section.7}) of the dynamical process undergone by S + A 
has revealed several successive steps involving different time scales. These steps will be resumed in section 11 (table 1).

}

\subsubsection{Preparation}
\label{9.3.1}

\ZeText{

Before S and A are coupled, A should be prepared in a metastable state.
Indeed, in the old days of photography the unexposed film was metastable and could not be prevented
from evolving in the dark on a time scale of months. In our magnetic case, for quartic interactions within M,
the {\it lifetime of the paramagnetic initial state} is extremely large, exponentially large in $N$.
For quadratic interactions with coupling constant $J$, it was evaluated in section \S~\ref{section.7.3.2}
(eq. (\ref{taupara})) as

\begin{equation}  
\tau_{\rm para}=\frac{\hbar}{\gamma(J-T)}\ln\alpha \sqrt{N},
\end{equation} 
where $\alpha$ is typically of order $1/10$, and it is larger than all other characteristic times for
$\alpha\sqrt{N}\gg1$.
We can thus engage the measurement process by switching on the interaction between S and M
during the delay $\tau_{\rm para}$ after preparation of A,
before the paramagnetic state is spontaneously spoiled.

}

\subsubsection{Truncation}
\label{fin9.3.2}

\ZeText{

Let us recall (\S~3.3.2 and Fig. 3.2) our decomposition of the density matrix $\hat{{\cal D}}$ of the total system S + A
into blocks with definite value $s_z=\uparrow,\downarrow$ of the tested spin component $\hat s_z$:
\begin{equation}
\hat{\cal D}=
\left( \begin{array}{ccc@{\ }r}
{\hat{\cal R}}_{\uparrow\uparrow}& {\hat{\cal R}}_{\uparrow\downarrow}\\
{\hat{\cal R}}_{\downarrow\uparrow}& {\hat{\cal R}}_{\downarrow\downarrow}
\end{array} \right).
\label{CDelemts9}
\end{equation}
The first stage of the measurement process is the truncation, defined as the disappearance 
of the off-diagonal blocks $\hat{\cal R}_{\uparrow\downarrow}$ and $\hat{\cal R}_{\downarrow\uparrow}$
of the full density matrix (section ~\ref{section.5}). It takes place during the {\it truncation time}

\begin{equation}  \label{taured9}
\tau_\trunc=\frac{\hbar}{\sqrt{2N}\delta_0g}, 
\end{equation} 
which is governed by the coupling constant $g$ between S and M and the size $N$ of the pointer 
(the fluctuation of M in the paramagnetic state is $\delta_0/\sqrt{N}$). 
This characteristic time is the shortest of all; its briefness reflects an effect produced by a 
{\it macroscopic object}, the pointer M, on a {\it microscopic one}, the tested system S. 
During the delay $\tau_\trunc$, the off-diagonal components $a=x,y$ of the spin S decay on
average as $\langle \hat s_a(t)\rangle=\langle\hat s_a(0)\rangle \exp[-(t/\tau_\trunc)^2]$.

Over the time scale $\tau_\trunc$, only the off-diagonal blocks 
$\hat{\cal R}_{\uparrow\downarrow}=\hat{\cal R}_{\downarrow\uparrow}^\dagger$ of the overall density
matrix $\hat{\cal D}$ of S + A are affected by the evolution.
{\it Correlations} between S and M, involving larger and larger numbers $k=1,2,\cdots$ of spins of M,
such as $\langle \hat s_a \hat m^k(t)\rangle_{\rm c}\propto t^k\exp[-(t/\tau_\trunc)^2]$ ($a=x,y$) are successively 
created in a {\it cascade}: They develop later and later, each one reaches a small maximum for 
$t=\tau_\trunc \sqrt{k/2}$ and then tends to zero (\S~\ref{section.5.1.3} and Fig. 5.1).
The information originally carried by the off-diagonal elements of the initial density matrix
of S are thus transferred towards correlations which couple the system S with more and more spins of M and
eventually decline (\S~\ref{section.5.1.4}). When $t$ increases far beyond $\tau_\trunc$,  all the matrix elements of $\hat{\cal R}_{\uparrow\downarrow}$
that contribute to correlations of rank $k\ll N$ tend to zero. Correlations of higher rank $k$, for large but
finite $N$, are the residue of reversibility of the microscopic evolution generated by
$\hat H_{\rm SA}$ (\S~\ref{section.5.3.2}).

If the total Hamiltonian of S + A did reduce to the coupling $\hat H_{\rm SA}=-Ng\hat s_z\hat m$ which 
produces the above behavior, the truncation would be provisional,  since S + A would periodically
return to its initial state with the {\it recurrence time}

\begin{equation}  \tau_{\rm recur}=\frac{\pi\hbar}{2g},
\end{equation} 
much larger than $\tau_\trunc$ (\S~\ref{section.5.3.1}).  As in spin-echo experiments, the extremely small but extremely numerous correlations
created by the interaction between S and the many spins of M would conspire to progressively reconstruct
the off-diagonal blocks of the initial uncorrelated state of S + A: The reversibility and simplicity of the dynamics 
would ruin the initial truncation.

Two possible mechanisms can prevent such recurrences to occur. In subsection \ref{section.6.1}
we slightly modify the model, taking into account the (realistic) possibility of a spread $\delta g$ 
in the coupling constants $g_n$ between S and each spin of the magnet M. The Hamiltonian (\ref{HSAgn}) with the conditions
(\ref{deltag}) then produces the same initial truncation as with constant $g$, over the same
characteristic time $\tau_\trunc$, but recurrences are now ruled out owing to the dispersion of
the $g_n$, which produces an extra damping as $\exp[-(t/\tau_{\rm irrev}^{\rm M})^2]$.
The {\it irreversibility time induced by the spreading} $\delta g$ in the spin-magnet couplings, 

\begin{equation}  \tau_{\rm irrev}^{\rm M}=\frac{\hbar}{\sqrt{2N}\delta g}, 
\end{equation} 
is intermediate between $\tau_\trunc$ and $\tau_{\rm recur}$ provided $\delta g$ is sufficiently large, viz. $g/\hspace{-1mm}\sqrt{N}\ll\delta g \ll g$. 
As usual for a reversible linear evolution, a recurrence phenomenon still occurs here,
but the recurrence time is inaccessibly large as shown in \S~\ref{section.6.1.2} (see eq. (\ref{taurecur=})).
The numerous but weak correlations between S and M, issued from the off-diagonal blocks of the 
initial density matrix of S, are therefore completely ineffective over any reasonable time lapse.

An alternative mechanism can also rule out any recurrence, even if the couplings  between S and the spins are
all equal (subsection \ref{section.6.2}). In this case, the required irreversibility is induced by the bath,
which produces an extra decay, as $\exp[- N B(t)]$, of the off-diagonal blocks (the shape of $B(t)$ is shown in Fig 6.1). 
The initial truncation of section 5, for $t\ll1/\Gamma$, is not affected by the interaction with the bath if $N B(\tau_\trunc)\ll1$, that is, if 

\BEQ
                \frac{\gamma \hbar^2 \Gamma^2}{ 8 \pi N \delta_0^4 g^2}\ll1, 
                \label{9.6bis}
\EEQ
where $\Gamma$ is the Debye cutoff on the phonon frequencies. 
At times $t$ such that $t\gg\hbar/2 \pi T$, $B(t)$ is quasi linear and the bath produces an exponential decay, 
as $\exp(- t/\tau^{\rm B}_{\rm irrev})$, where the {\it bath-induced irreversibility time} is defined as

\begin{equation}  \label{tauirrB}
\tau_{\rm irrev}^{\rm B}= \frac{2 \hbar \tanh g/T }{  N \gamma g}\simeq  \frac{ 2\hbar}{ N \gamma T}.
\end{equation} 
This expression is a typical decoherence time, inversely proportional to the temperature $T$ of B,  to the bath-magnet coupling $\gamma$ and to the number 
$N$ of degrees of freedom of the system S + M. (Note that $\tau_{\rm irrev}^{\rm B} < \hbar/2 \pi T$.)  
The $p$-th recurrence is then damped by a factor $\exp (- p \tau_{\rm recur}/\tau^{\rm B}_{\rm irrev})$, 
so that the phonon bath eliminates all recurrences if $\tau^{\rm B}_{\rm irrev}\ll \tau_{\rm recur}$.

At this stage, {\it the truncation is achieved} in the sense that the off-diagonal blocks 
$\hat{\cal R}_{\uparrow\downarrow}(t)$ and $\hat{\cal R}_{\downarrow\uparrow}(t)$ of the density
operator (\ref{CDelemts9}) of S + A have practically disappeared in a definitive way. 
The off-diagonal correlations created during the truncation process have been irremediably destroyed at the end of this process,
 whereas the diagonal correlations needed to register in A the tested properties of S are not yet created. 
 See also \S~\ref{fin11.2.3} below.

}

\subsubsection{Registration by the pointer}
\label{fin9.3.3}

\hfill{\it Our fates are as registered in the scripts of heaven}

\hfill{Japanese proverb}

\vspace{3mm}

\ZeText{

Just after the above processes are achieved, the diagonal blocks $\hat{\cal R}_{\uparrow\uparrow}(t)$ and $\hat{\cal R}_{\downarrow\downarrow}(t)$ as well as the 
marginal density operator  $\hat {\cal R} (t)={\rm tr}_{\rm S}\hat{\cal D}(t)=\hat{\cal R}_{\uparrow\uparrow}(t)+\hat{\cal R}_{\downarrow\downarrow}(t)$ 
of A {\it remain nearly unaffected}. The process cannot yet be regarded as a measurement: The pointer gives no indication,
$m$ is still small, and no correlation exists between A and the initial state of S.
The registration then starts and proceeds on time scales much larger than the above ones. It is a slower process because it leads to a change of a
 {\it macroscopic object}, the apparatus, {\it triggered  by the microscopic} S.
 We term as ``registration'' a process which modifies the density operator of S + A associated with a {\it large set of measurements}. 
 To take advantage of the information stored thereby in the pointer of A, we need that for {\it each  individual measurement} the indication of this pointer be well-defined (see section 11).
 
   After a brief transient regime, the process becomes Markovian (\S~\ref{section.7.1.1}).
 The evolution of each of the two diagonal blocks $\hat{\cal R}_{\uparrow\uparrow}(t)$ or $\hat{\cal R}_{\downarrow\downarrow}(t)$
can be expressed in terms of that of the corresponding probability distribution $P_{\uparrow\uparrow}(m,t)$ or $P_{\downarrow\downarrow}(m,t)$ for the 
magnetization of M, which obeys an equation of the Fokker-Planck type \cite{risken}. This equation, presenting classical
features (\S~\ref{section.7.1.2}), is governed  for $P_{\up\up}(m,t)$ by a {\it drift velocity} $v(m)$ given by (\ref{Vsm}) and illustrated by 
Figs. 7.1 and 7.2, and by a {\it diffusion} coefficient given by (\ref{Wsm}). The irreversibility of the process is exhibited by an {\it H-theorem} (\S~\ref{section.7.1.3})
which implies the decrease of the free energy of M. Thus, the total entropy of M + B increases, and some energy is dumped from M into B, while the transition 
leads from the paramagnetic to either one of the ferromagnetic states. The existence of two possible final states is associated
with breaking of ergodicity, discussed for finite but large $N$ in \S~\ref{section.7.1.4} and subsection \ref{section.7.3}.

For purely quadratic interactions within M  (the coupling   (\ref{HM=}) having the form $J\hat m^2$),  the registration proceeds in three stages  (\S~\ref{section.7.2.3}), 
illustrated by Figs. 7.3 and 7.5. Firstly the distribution $P_{\uparrow\uparrow}(m,t)$, initially a paramagnetic symmetric peak around $m=0$, is shifted faster and 
faster towards the positive direction of $m$ and it widens, under the conjugate effects of both S and B. For suitably chosen parameters, after a delay given 
by Eq. (\ref{taureg2}), 

\begin{equation}  \label{tauregq=29}
\tau_{\rm reg}=\frac{\hbar}{\gamma(J-T)},
\end{equation} 
that we term the {\it first registration time}, $P_{\uparrow\uparrow}(m,t)$ is entirely located in the positive region of $m$, its tail in the region $m<0$
has then become negligible. Symmetrically,  $P_{\downarrow\downarrow}(m,t)$  lies entirely in the $m<0$ region for $t>\tau_{\rm reg}$. 
Thereafter the coupling between M and S becomes ineffective  and may be {\it switched off}, so that the registration is virtually, but not yet fully, 
achieved at this time  $\tau_{\rm reg}$.

The last two stages describe a standard relaxation process for which the tested system S is no 
longer relevant. The stochastic motion of $m$ is first governed mainly by the contribution of B to the 
drift of the magnetization $m$. The distribution $P_{\uparrow\uparrow}(m,t)$ moves rapidly towards
$+m_{\rm F}$, first widening, then narrowing. We term as {\it second registration time} 
$\tau_{\rm reg}'$ the delay needed for the average magnetization to go from 0 to the vicinity of $m_{\rm F}$.
It is expressed by Eq. (\ref{tauprime}), together with (\ref{a=}) and (\ref{mB}).
During the third stage of the registration, both the drift and the diffusion generated by B establish
thermal equilibrium of the pointer in an exponential process, and stabilize the distribution
$P_{\uparrow\uparrow}(m,t)$ around $+m_{\rm F}$. 
Thus, $\hat{\cal R}_{\uparrow\uparrow}(t)$ ends up as $r_{\uparrow\uparrow}(0)\hat{\cal R}_\Uparrow$,
where $\hat{\cal R}_\Uparrow$ denotes the ferromagnetic equilibrium state with positive magnetization,
and, likewise, $\hat{\cal R}_{\downarrow\downarrow}(t)$ ends up as $r_{\downarrow\downarrow}(0)\hat{\cal R}_\Downarrow$.

For purely quartic interactions within M (coupling as $J \hat m^4$), or for $3 J_4>J_2$, the transition is of first order. We can again distinguish in the 
registration the above three stages (\S~\ref{section.7.2.4}), illustrated by Figs 7.4 and 7.6.
Here the first  stage is slowed down by the need to pass through the bottleneck $m\simeq m_{\rm c}$ given by (\ref{hc}).
The widening of the distribution $P_{\uparrow\uparrow}(m,t)$ is much larger than for 
quadratic interactions, because diffusion is effective during the large duration of the bottleneck stage.
Both the first and the second registration times defined above are nearly equal here, and given by 
(\ref{zeminimum}), that is,

\begin{equation}  \label{tauregq=49}
\mytext{\textcurrency taureg9\textcurrency \qquad}
\tau_{{\rm reg}}=\frac{\pi\hbar}{\gamma T}\sqrt{\frac{m_{{\rm c}}T}{g-h_{{\rm c}}}}{ ,}
\qquad m_{\rm c}\simeq \sqrt{\frac{T}{3J}},\qquad h_{\rm c}\simeq \frac{2}{3}Tm_{\rm c}.
\end{equation}
The last stage is again an exponential relaxation towards the ferromagnetic state $+m_{\rm F}$ for 
  $P_{\uparrow\uparrow}(m,t)$.
  
  The ratio $\tau_{\rm reg}/\tau_\trunc$ between the registration and truncation times, proportional to  $\sqrt{N}/\gamma$,
  is large for two reasons, the weakness of $\gamma$ and the large value of $N$. As usual in statistical 
  mechanics, the coexistence of very {\it different time scales} is associated here with exact and approximate
 {\it  conservation laws}, expressed by $[\hat s_z,\hat H]=0$ and $[\hat m,\hat H]=[\hat m,\hat H_{\rm MB}]  \propto\sqrt{\gamma}$,
 which is small because $\gamma\ll1$.
  
 If $N$ is finite, the registration is not permanent. However, the characteristic time of erasure $\tau_{\rm eras}$
  is much larger than the registration time $\tau_{\rm reg}$ by a factor behaving as an exponential of $N$ (\S~7.3.5). 
 
 \vspace{3mm}
 
 The time scales involved in this Curie--Weiss measurement process present some analogy with the relaxation times in nuclear magnetic resonance 
  \cite{Abragam1961, Abragam1982}.
 The truncation, i. e., the disappearance of the transverse components $\langle\hat s_x\rangle$ and $\langle\hat s_y\rangle$ 
 and of their correlations with A, can be compared to the transverse relaxation in nuclear magnetic resonance (NMR). 
 The truncation time $\tau_\trunc$, as well as is the relaxation time 
 ${\cal T}^\ast_2$ associated in NMR with a dispersion in the precession frequencies of the spins of a sample due to a non-uniformity of the field along $z$, 
 are durations of dephasing processes in which complex exponentials interfere destructively. By themselves, these phenomena give rise to recurrences
(in our model of measurement) or to spin echoes (in NMR). The bath-induced irreversibility time $\tau^{\rm B}_{\rm irrev}$ is comparable to the relaxation time ${\cal T}_2$: 
 both characterize decoherence effects, namely the damping of recurrences in the measurement, and the complete transverse relaxation which damps the echoes in NMR. 
 Finally the registration time characterizes the equilibration of the diagonal blocks of the density matrix $\hat{\cal D}$, in the same way as the relaxation time 
 ${\cal T}_1$ characterizes the equilibration of the longitudinal polarization of the spins submitted to the field along $z$.

 \subsubsection{Reduction}
 
 The stages of the measurement process described in \S\S~9.3.1--9.3.3 are related to the evolution of the density operator $\scriptD(t)$ describing the statistics of the observables of 
 S + A {\it for the full ensemble} $\scriptE$ of runs. Consideration of {\it individual runs} requires a study of the dynamics {\it for arbitrary subensembles} $\scriptE_\sub$ of $\scriptE$. 
 This study will be achieved in section 11, where we will show that a last stage is required, near the end of the scenario (table 1). The model will then be supplemented with a weak
  interaction within the apparatus, which produce transitions conserving $m$ between the states of the pointer M. These interactions have a size $\Delta$, and the duration of the
   relaxation of the subensembles towards equilibrium is characterized by the very short time scale $\tau_\sub=\hbar/\Delta$ (Eq 11.17), much shorter than the registration time.

 The above summary exhibits the different roles played by the various coupling constants. On the one hand, truncation is ensured entirely by the coupling $g$ between S and M. 
 Moreover, the beginning of the registration is also governed by $g$, which selects one of the alternative ferromagnetic states and which should therefore be sufficiently large. 
 On the other hand, the coupling $\gamma$ between M and B governs the registration, since the relaxation of M towards ferromagnetic equilibrium requires a dumping of energy in 
 the bath. Finally, the weak interaction $\Delta$ within A governs the subensemble relaxation, which ensures the uniqueness of the outcome of each run and allows reduction.
  
}

\subsection{Conditions for ideality of the measurement}
\label{section.9.1.3}
\label{fin9.4}

\hfill{\it What you do not wish for yourself,}

\hfill{ \it do not do to others}

\hfill{Confucius}

\vspace{3mm}

\ZeText{

Strictly speaking, for finite values of the parameters of the model, the process that we have studied cannot be an ideal measurement in a mathematical sense. 
However, in a physical sense, the situation is comparable to the solution of the irreversibility paradox, which is found by disregarding correlations
between inaccessibly large numbers of particles and by focusing on time scales short compared to the inaccessible Poincar\'e recurrence time. 
Here (after having achieved the solution in section 11) we will likewise identify physically the process with an ideal measurement, 
within negligible deviations, provided the parameters of the model satisfy some conditions. 

The definition of the apparatus includes a {\it macroscopic pointer}, so that 

\begin{equation}  N\gg1.\end{equation} 

The temperature $T$ of the bath B should lie below the transition temperature of the magnet M,
which equals $J$ for quadratic interactions ($q=2$) and $0.363\,J$ for purely quartic interactions ($q=4$).

Our solution was found by retaining only the {\it lowest order} in the coupling between B and M.
Neglecting the higher order terms is justified provided

\begin{equation}  \gamma\ll\frac{T}{J}.
\end{equation} 
This condition ensures that the autocorrelation time of the bath, $\hbar/T$, is short compared to the 
registration time (\ref{tauregq=29}) or (\ref{tauregq=49}).
We have also assumed a large value for the {\it Debye cutoff}, a natural physical constraint expressed by

\begin{equation}  \hbar\Gamma\gg J.
\end{equation} 

The {\it irreversibility of the truncation}, if it is ensured by a dispersion $\delta g$ of the couplings between tested spin 
and apparatus spins, requires a neat separation of the time scales  $\tau_\trunc\ll\tau^{\rm M}_{\rm irrev}\ll \tau_{\rm recur}$, that is

\BEQ                      \label{9.14}
\delta_0 \gg \frac{\delta g}{g} \gg \frac{1}{\pi} \sqrt{\frac{2}{N}}.    
\EEQ
The coefficient $\delta_0$, the width of the initial paramagnetic distribution of $m \sqrt{ N}$, 
is somewhat larger than 1 for $q = 2$ (quadratic Ising interactions, Eq. (\ref{delta0=}) and equal to 1 for $q = 4$ (quartic interactions)
or when using a strong RF field to initialize the magnet, so that the condition (9.14) is readily satisfied.

If the irreversibility of the truncation is ensured by the bath, we should have $NB(\tau_{\rm recur})=\tau_{\rm recur}/\tau^{\rm B}_{\rm irrev} \gg1$, that is

\begin{equation}  \label{irrevB}
\gamma \gg \frac{4}{ \pi N} \tanh \frac{g}{T} . 
\end{equation} 
This condition provides a lower bound on the bath-magnet coupling.  An upper bound is also provided by (\ref{9.6bis})
 if we wish the initial truncation to be controlled by M only. Both bounds are easily satisfied for $N\gg1$.

The {\it coupling} $g$ between S and M has been assumed to be rather weak,

\begin{equation}  g< T.
 \end{equation} 
However, this coupling should be sufficiently strong to initiate the registration, and to ensure that the final
indication of the pointer  after decoupling will be $+m_{\rm F}$ if S lies initially in the state $|\hspace{-1mm}\uparrow\rangle$,
$-m_{\rm F}$ if it lies initially in the state $|\hspace{-1mm}\downarrow\rangle$. For $q=2$, this condition is not very 
stringent. We have seen in \S~\ref{section.7.2.2} that it is expressed by (\ref{gmin}), namely

\begin{equation}  \mytext{\textcurrency condg2\textcurrency} \qquad
\label{condg2}
g\gg\frac{(J-T)\delta_1}{\sqrt{N}},\qquad \delta_1^2=\delta_0^2+\frac{T}{J-T}=\frac{T_0}{T_0-J}+\frac{T}{J-T}.
\end{equation} 

 For purely quartic interactions $-\frac{1}{4}J\hat m^4$ (or for $3J_4>J_2$) 
the paramagnetic state is locally stable in the absence of interaction with S. 
The coupling $g$ should therefore be larger than some threshold, finite for large $N$,

\BEQ 
 \label{condg4}
g > h_{\rm c}\simeq   \sqrt{\frac{4T^3 }{27 J}}, 
 \EEQ
so as to trigger the phase transition from $m=0$ to $m=\pm\, m_{\rm F}$ during the delay (\ref{tauregq=49}).
Moreover, if we wish the decoupling between S and A to take place before the magnet has reached 
ferromagnetic equilibrium, $g$ must lie sufficiently above $h_{\rm c}$ (see Eq. (7.57)).

If all the above conditions are satisfied, the final state reached by S + A for the full set of runs of the measurement is physically indistinguishable from the surmise (\ref{Dtf9=}), 
which encompasses necessary properties of ideal measurements, to wit, truncation and unbiased registration, that is, full correlation between the indication of the apparatus 
and the final state of the tested system.  However, these properties are not sufficient to ensure the uniqueness of the outcome of individual runs (section 11).

}

\subsection{Processes differing from ideal measurements}
\label{section.9.1.4}
\label{fin9.5}

\hfill{\it In de beperking toont zich de meester\footnote{Conciseness exposes the master}}

\hfill{\it Le mieux est l'ennemi du bien\footnote{Best is the enemy of good}}

\hfill{Dutch and French sayings}
\vspace{3mm}

\ZeText{

Violations of some among the conditions of subsection~\ref{section.9.1.3} or modifications of the model allow us
to get a better insight on quantum measurements, by evaluating deviations from ideality and exploring
processes which fail to be measurements, but are still respectable evolutions of coupled quantum
mechanical systems.

In subsection \ref{section.5.2}, we modify the initial state of the apparatus, assuming that it is not
prepared in an equilibrium paramagnetic state. This discussion leads us to understand truncation as a
consequence of the {\it disordered nature of the initial state} of M, whether or not this state is pure (\S~\ref{section.5.2.2}). 
For ``squeezed'' initial states, the rapid truncation mechanism can even fail  (\S~\ref{section.5.2.3}).

Imperfect preparation may also produce another kind of failure. In \S~\ref{section.7.3.3} we consider a {\it bias} 
{\it in the initial state} due to the presence during the preparation stage of a parasite magnetic field which produces a 
paramagnetic state with non-zero average magnetization. Wrong registrations, for which M reaches for instance
a negative magnetization $-m_{\rm F}$ in the final state although it is coupled to a tested spin in the state
$s_z=+1$, may then occur with a probability expressed by (\ref{Psmallg}).

Section \ref{section.6} shows that {\it recurrences} are not washed out if the conditions Eq. (\ref{9.14}) or (\ref{irrevB}) are not fulfilled.
The probability for the $p$-th recurrence to occur is $\exp[-(p\tau_{\rm recur}/\tau_{\rm irrev}^{\rm M})^2]$
in the first case,  $\exp(-p\tau_{\rm recur}/\tau_{\rm irrev}^{\rm B})$ in the second case.
The process is not an ideal measurement if recurrences are still present when the outcome is read.  
 
The violation of the condition (\ref{condg2}) for $q=2$  or (\ref{condg4}) for $q=4$ prevents the registration
from taking place properly. For $q=2$, if the coupling $g$ is too weak to satisfy (\ref{condg2}), the apparatus
does relax towards either one of the ferromagnetic states $\pm m_{\rm F}$, but it may provide a 
{\it false indication}. The probability for getting wrongly $-m_{\rm F}$ for an initial state $|\hspace{-1mm}\uparrow\rangle$ of S,
evaluated in \S~\ref{section.7.3.3}, is given by (\ref{Psmallg}). For $q=4$, 
the registration is aborted if  (\ref{condg4}) is violated: the magnet M does not leave the paramagnetic region, 
and its magnetization returns to $0$ when the coupling is switched off.

The {\it large number} $N$ of elements of the pointer M is essential to ensure a faithful and long-lasting registration for each individual run. 
It also warrants a brief truncation time, and an efficient suppression of recurrences by the bath. 
We study in subsection \ref{section.8.1} the extreme situation with $N=2$, for which $\hat m$ has only two 
``paramagnetic'' eigenstates  with $m=0$ and two ``ferromagnetic'' eigenstates with $m=\pm1$. Although correlations 
can be established at the time (\ref{regN=2}) between the initial state of S and the magnet M in agreement with 
Born's rule, there is no true registration. The indication of M reached at that time is lost after a delay
$\tau_{\rm obs}$ expressed by (\ref{tauobs}); moreover, a macroscopic extra apparatus is needed to observe M itself during this delay. On the other hand, 
the truncation process, governed here by the bath, is more akin to equilibration than to decoherence; 
it has an anomalously long characteristic time, longer than the registration time. 
These non-idealities of the model with $N=2$ are discussed in \S~\ref{section.8.1.5}. 
However, such a device might be used (\S~\ref{section.8.1.6}) to implement the idea of determining
all four elements of the density matrix of S by means of repeated experiments using a single apparatus
\cite{ABNprl2004,bahar,gerardo}.

In subsection \ref{section.8.2} we tackle the situation in which the measured observable $\hat s_z$ 
is {\it not conserved} during the evolution. An ideal measurement is still feasible under the condition
(\ref{condb}), but it fails if S and A are not decoupled after some delay (\S~\ref{section.8.2.5}).

The model can also be extended (subsection \ref{section.8.3}) by simultaneously coupling S with
{\it two apparatuses} A and A$^\prime$ which, taken separately, would measure $\hat s_z$ and $\hat s_x$,
respectively. The simultaneous measurement of such non-commuting observables  is of course
impossible. However, here again, repeated runs can provide full 
information on the statistics of both $\hat s_z$ and $\hat s_x$ in the initial state $\hat r(0)$ (\S~\ref{section.8.3.3}). 
More generally, all the elements of the density matrix $\hat r(0)$ characterizing an ensemble of identically prepared 
spins S can be determined by repeated experiments involving a compound apparatus A+A$'$+A$''$,  where 
A, A$'$ and A$''$ are simultaneously coupled to the observables $\hat s_x$, $\hat s_y$ and $\hat s_z$, respectively.
Indirect tests of Bell's inequalities may rely on this idea (\S~\ref{section.8.3.4}).

}

\subsection{Pedagogical hints}

\label{fin9.6}

\hfill{\it The path is made by walking}

\hfill{\it Le mouvement se prouve en marchant}

\hfill{African and French proverbs}

\vspace{3mm}

\ZeText{

Models of quantum measurements give rise to many exercises of tutorial interest, which help the 
students to better grasp quantum (statistical) mechanics. We have encountered above several 
questions which may inspire teachers. The exercises that they suggest require the use of density operators. 
As quantum mechanics is often taught only in the language of pure states, we present in appendix G 
an introduction for students on this topic.
 
For instance, the {\it treatment of a thermal bath} at lowest order in its coupling with the rest of the system
(subsection ~\ref{section.4.1} and Appendix A), although standard, 
 deserves to be worked out by advanced students.

For a general class of models of measurement involving a pointer with many degrees of freedom, the {\it truncation mechanism}
exhibited in \S~\ref{section.5.1.2} shows how dephasing can eliminate the off-diagonal blocks of the density matrix of S + A 
over a short time through interferences.

The evaluation of the {\it recurrence time} for the pointer coupled with the tested system, or more generally for an arbitrary quantum system
(or for a linear dynamical system) having a random spectrum (\S~\ref{section.6.1.2} and Appendix C) is also of general interest.

We now give two further examples of exercises  for students which highlight the central steps of the 
quantum measurement.

}

\subsubsection{End of ``Schr\"odinger cats"}

\label{section.9.6.1}
\label{section.9.5.1}
\label{fin9.6.1}

\ZeText{

Focusing on the Curie-Weiss model, we present here a simpler derivation of the processes
which first lead to truncation and which prevent recurrences from occurring. We showed in section 6 and Appendix D that
the interactions $J_2$ and $J_4$ between the spins $\hat\sigma^{(n)}$ of M play little role here, so that we neglect them. 
We further assume that M lies initially in the most disordered state (\ref{purePM}), that we write out, using the notation (\ref{sigmaxyz0}),  as

\BEQ 
\hat{R}_{\rm M}(0)=\frac{1}{2^N}\hat\sigma_0^{(1)}\otimes\hat\sigma_0^{(2)}\otimes\cdots\otimes\hat\sigma_0^{(N)}.
\EEQ
 This occurs for $q=4$  and in the general case of $J_2>0$ provided the temperature of preparation
$T_{0}$ in (\ref{delta0=}) is much higher than $J_2$, so that $\delta_{0}=1$.
Then, since the Hamiltonian $\hat{H}_{{\rm SA}}+\hat{H}_{{\rm B}}
+\hat{H}_{{\rm MB}}$ is a sum of independent contributions associated with
each spin  {\boldmath{$\hat\sigma$}}$^{(n)}$, the spins of M behave independently at all
times, and the off-diagonal block $\hat{R}_{\uparrow\downarrow}(t)$ of
$\hat{D}(t)$ has the form
\begin{equation}
\mytext{\textcurrency Rfact\textcurrency \qquad}
\hat{R}_{\uparrow\downarrow}(t)=r_{\uparrow\downarrow}(0)\,\hat{\rho}^{(1)}(t)\otimes\hat{\rho}
^{(2)}(t)\otimes\cdots\otimes\hat{\rho}^{(N)}(t){ ,} \label{Rfact}
\end{equation}
where $\hat{\rho}^{(n)}(t)$ is a $2\times2$ matrix in the Hilbert space of the spin 
 {\boldmath{$\hat\sigma$}}$^{(n)}$.
 This matrix will depend on $\hat{\sigma
}_{z}^{(n)}$ but not on $\hat{\sigma}_{x}^{(n)}$ and $\hat{\sigma}_{y}^{(n)}$,
and it will neither be hermitean nor normalized.

The task starts with keeping the effect of the bath as in subsection  \ref{section.6.2}, but leaves
 open the possibility for the coupling $g_n$ to be random as in subsection \ref{section.6.1}, whence
the coupling between S and A reads  $\hat H_{\rm SA}=-\hat s_z\sum_{n=1}^Ng_n\hat \sigma_z^{(n)}$ instead of (\ref{HSA}).
  (As simpler preliminary exercises, one may keep the $g_n=g$ as constant, and/or disregard the bath.)
Each factor $\hat{\rho}^{(n)}(t)$, initially equal to $\frac{1}{2}\hat\sigma_0^{(n)}$, evolves according to the same equation as (\ref{dRij}) for
$\hat{R}_{\uparrow\downarrow}(t)$, rewritten with $N=1$.  (To convince oneself of the product structure (\ref{Rfact}),
 it is instructive to work out the cases $N=1$ and $N=2$ in Eq. (\ref{dRij}) or (\ref{dPoff}).) 
Admit, as was proven in subsection 6.2 and appendix D, that the effect of the bath is relevant only at times $t \gg\hbar/2 \pi T$, 
and that in this range $\hat \rho^{(n)}$ evolves according to
\begin{equation}
\label{rhoneq}
\frac{{\rm d}\hat{\rho}^{(n)}(t)}{{\rm d}t}-\frac{2ig_n}{\hbar}\hat{\rho}^{(n)}\hat{\sigma}_{z}^{(n)}
= - \frac{2 \gamma}{\hbar^2}\left[ \tildeK_- \left(\frac{2g_n}{\hbar}\right) + \tildeK_+ \left(- \frac{2g_n}{\hbar}\right)\right] 
\left[ \hat{\rho}^{(n)}-\frac{1}{2}\hat\sigma_0^{(n)}{\rm tr}\,\hat{\rho}^{(n)}\right]{ .}
\end{equation}
(Advanced students may derive this equation by noting that for $N=1$, $\hat\rho^{(n)}$ can be identified with  $P_{\up\down}(\hat m=\hat\sigma_z)$; 
starting then from Eq. (4.17) for $N = 1$, keeping in mind that $P_{\up\down}(\pm3) = 0$ and verifying that, in the non-vanishing terms, 
Eq. (4.13) implies that $\Omega^\pm_i = \mp 2g_n s_i /\hbar$, they should show that the factors $\tilde K_{t>}(\Omega^-_\up) + \tilde K_{t<}(\Omega^-_\down)$ 
and $\tilde K_{t>}(\Omega^+_\up) + \tilde K_{t<}(\Omega^+_\down)$ of (4.17) reduce for $t \gg\hbar/2 \pi T$
and for $J_2=0$ to the symmetric part of $\tilde K (2g_n/\hbar)$ according to (4.18) and (D.21).)

Next parameterize $\hat{\rho}^{(n)}$ as 

\begin{equation}\label{rhonpar}
\hat{\rho}^{(n)}(t)=\frac{1}{2}\exp\left[  -B_n(t)+i\Theta_n(t)\hat{\sigma}_{z}^{(n)}\right]  { ,}
\end{equation}
and derive from (\ref{rhoneq}) the equations of motion

\begin{eqnarray}
\frac{{\rm d}\Theta_n}{{\rm d}t}&=&
\frac{2g_n}{\hbar}-\frac{\gamma}{\hbar^{2}}\left[\tilde K\left (\frac{2g_n}{\hbar}\right) + \tilde K \left(- \frac{2g_n}{\hbar}\right)\right]\,\sin2\Theta_n
{ ,}\nonumber\\
         \frac{{\rm d}B_n}{{\rm d}t}&=&
         \frac{2\gamma }{\hbar^2}\left[\tilde K\left (\frac{2g_n}{\hbar}\right) + \tilde K \left(- \frac{2g_n}{\hbar}\right)\right]\, \sin^{2}\Theta_n ,
\label{Theta,Beq}
\end{eqnarray}
with initial conditions $\Theta_n(0)=0$, $B_n(0)=0$. 
 Keeping only the dominant contributions for $\gamma\ll 1$, use the expression (\ref{Ktilde}) for  $\tilde K$,  find the solution

 \BEQ   \label{Theta,B=}     
   \Theta_n(t)\simeq \frac{ 2g_n t}{\hbar},\qquad   B_n(t) \simeq  \frac{\gamma g_n }{2 \hbar}\coth \frac{g_n}{T}\left (t - \frac{\hbar}{4g_n}\sin \frac{4 g_n t}{\hbar}\right),
 \EEQ
and compare $B_n$ with (6.28) for $B$.

Eqs. (\ref{rhonpar}), (\ref{Theta,B=}) provide the evolution of the density matrix of the spin $n$ from the paramagnetic initial state
$\hat{\rho}^{(n)}(0)=\frac{1}{2}{\rm diag}(1,1)$ to

\begin{equation}
\hat{\rho}^{(n)}(t)=\frac{1}{2}{\rm diag}\left(e^{{2ig_nt}/{\hbar}},e^{{-2ig_nt}/{\hbar}}\right)
  \exp\left[ - \frac{\gamma g_n}{2 \hbar}\coth \frac{g_n}{T} \left(t - \frac{\hbar}{4g_n} \sin \frac{4 g_n t }{\hbar}\right)\right].
  \label{9.26}
\end{equation}
By inserting (\ref{9.26}) into (\ref{Rfact}) and tracing out the pointer variables, one finds the transverse polarization of S as
\BEA
&&\half\langle\hat s_x(t)- i \hat s_y(t)\rangle\equiv {\rm tr}_{\rm S,A}\hat{\cal D}(t)\half(\hat s_x- i \hat s_y)=
r_{\uparrow\downarrow}(t)\equiv r_{\uparrow\downarrow}(0)\,{\rm Evol}(t),\qquad \EEA
where the temporal evolution is coded in the function
\BEA\label{Evol}
{\rm Evol}(t)\equiv
\left({ \prod}_{n=1}^N\cos\frac{2g_nt}{\hbar}\right)\,
\exp\left[ - \sum_{n=1}^N\frac{\gamma g_n}{2 \hbar}\coth \frac{g_n}{T} \left(t - \frac{\hbar}{4g_n} \sin \frac{4 g_n t }{\hbar}\right)\right].
 \EEA

To see what this describes, the student can first take $g_n=g$, $\gamma=0$ and plot the factor $|{\rm Evol}(t)|$ 
from $t=0$ to $5\tau_{\rm recur}$, where $\tau_{\rm recur}= \pi\hbar/2g$ 
is the time after which $|r_{\uparrow\downarrow}(t)|$ has recurred to its initial value $|r_{\uparrow\downarrow}(0)|$. 
By increasing $N$, e.g.,  $N=1,2,10,100$,  he/she can convince him/herself that the decay near $t=0$ becomes 
close to a Gaussian decay, over the characteristic time $\tau_\trunc$ of Eq. (\ref{taured9}). The student may demonstrate this analytically
by setting $\cos2g_nt/\hbar\approx\exp(-2g_n^2t^2/\hbar^2)$  for small $t$.
This time characterizes decoherence, that is, disappearance of the off-diagonal blocks of the density matrix; 
we called it ``truncation time'' rather than ``decoherence time'' to distinguish it from usual decoherence, which is induced by a thermal environment
and coded in the second factor of Evol($t$).  

The exercise continues with the aim to show that $|$Evol$|\ll1$ at $t=\tau_{\rm recur}$ in order that the model describes
 a faithful quantum measurement. To this aim,  keeping $\gamma=0$, the student can in the first factor of Evol decompose $g_n=g+\delta g_n$,  
where $\delta g_n$ is a small Gaussian random variable with $\langle\delta g_n\rangle=0$ and
 $\langle\delta g_n^2\rangle\equiv\delta g^2\ll g^2$, and average over the $\delta g_n$. 
The Gaussian decay (\ref{damp}) will thereby be recovered, which already prevents recurrences.
The student may also take e.g. $N=10$ or $100$, and plot the function to show this decay and to estimate the size of Evol at later times.

Next by taking $\gamma>0$ the effect of the bath in (\ref{Evol}) can be analyzed. For values $\gamma$ such that $\gamma N\gg1$ the bath
will lead to a suppression.
 Several further  tasks can be given now: Take all $g_n$ equal and plot the function Evol($t$); take a small spread in them and compare the results;
make the small-$g_n$ approximation $g_n \coth g_n/T\approx T$, and compare again. 

At least one of the two effects (spread in the couplings or suppression by the bath) should be strong enough to prevent 
recurrences, that is, to make $|r_{\uparrow\downarrow}(t)|\ll |r_{\uparrow\downarrow}(0)|$
at any time $t\gg \tau_\trunc$, including the recurrence times. The student can recover the conditions  (\ref{9.14}) or (\ref{irrevB}) under which
 the two mechanisms achieve to do so.  The above study will show him/her that, in the
dynamical process for which each spin {\boldmath{$\hat\sigma$}}$^{(n)}$ of M independently
rotates and is damped by the bath, the truncation, which destroys the expectation values $\langle\hat s_a\rangle$ and all correlations 
$\langle\hat s_a\hat m^k(t)\rangle$ ($a=x$ or $y$, $k\ge 1$), arises from the precession of the tested spin $\hat s$
around the $z$-axis; this is caused by the conjugate effect of the many spins $\hat\sigma^{(n)}$ of M,  while the suppression of recurrences is either due
to dephasing if the $g_n$ are non-identical, or due to damping by the bath.

A less heavy exercise is to derive (5.27) from (5.26); hereto the student first calculates $\langle m\rangle$ and then $\langle m^2\rangle$.
Many other exercises may be inspired by sections 5 and 6, including the establishment and disappearance of the off-diagonal spin-magnet correlations (\S~5.1.3); 
the numerical or analytical derivation of the damping function $B(t)$ (Appendix D); its short-time behavior obtained either as for (D.9) or from the first two terms 
of the short-time expansion of $K(t)$; the analytical study of the autocorrelation functions $K(t)$, $K_{>t}$ and $K_{<t}$ of the bath for different time scales 
using the complex plane technique of Appendix D.

}

\subsubsection{Simplified description of the registration process}
\label{section.9.6.2}
\label{fin9.6.2}

\ZeText{

We have seen in \S~\ref{section.7.1.2} that the registration process looks, for the diagonal block
$R_{\uparrow\uparrow}(t)$, as a classical relaxation of the magnet M towards the stable state with
magnetization $+m_{\rm F}$ under the effect of the coupling $g$ which behaves in this sector as a positive field.
This idea can be used to describe the registration by means of the classical 
Fokker-Planck equation (\ref{dPdt}) which governs the evolution of the probability distribution 
$P(m,t)=P_{\uparrow\uparrow}(m,t)/r_{\uparrow\uparrow}(0)$.

By assuming explicit expressions for the drift and the diffusion coefficient which enter this equation of motion, 
one can recover some of the results of section \ref{section.7} in a form adapted to teaching.

In particular, if we keep aside the shape  and the width of the probability distribution, which has a narrow peak 
for large $N$ (\S~\ref{section.7.2.1}), the center $\mu(t)$ of this peak moves according to the mean-field equation

\BEQ 
\label{mudot=}
\frac{\d\mu(t)}{\d t}=v[\mu(t)],
\EEQ
where $v(m)$ is the local drift velocity of the flow of $m$.,
This equation can be solved once $v(m)$ is given, and its general properties do not depend on the
precise form of $v(m)$. The first choice is phenomenological:  we take $v(m)$
proportional to $- \d F/\d m$, where $F$ is the free energy (\ref{F=}), resulting in

\BEQ \label{v91}
v(m)=\frac{C(m)}{\hbar} \,\left(Jm^{q-1}+g-\frac{T}{2}\ln\frac{1+m}{1-m}\right),\EEQ
with a dimensionless, positive function $C(m)$ which may depend smoothly on $m$ in various ways 
(\S~\ref{section.7.1.2}),  or even be approximated as a constant.
An alternative phenomenological choice consists in deriving from detailed balance, as in \S~\ref{section.7.1.2}, 
the expression (7.14) for $v(m)$, that is, within a multiplicative factor $\theta(m)$,     

\begin{equation}
v(m)=\frac{1}{\theta(m)}\left(\tanh\frac{g+Jm^{q-1}}{T}-m\right){ .} \label{vmvm}%
\end{equation}
possibly approximating $\theta$ as a constant. 
A more precise way is to derive $v(m)$ from the autocorrelation function of the bath (Eq. (7.6)) as

\BEQ
\label{v92}
v(m)=\frac{\gamma}{ \hbar} (g+Jm^{q-1}) \left(1 -m\,\coth \frac{g+Jm^{q-1}}{ T}\right).
\EEQ
An introductory exercise is to show that the $C(m)$ (or the $\theta(m)$) 
obtained from equating  (\ref{v91}) (or (\ref{vmvm})) to  (\ref{v92}) is a smooth positive function, finite at the 
stable or unstable fixed points of  Eq. (\ref{mudot=}), given by the condition $v(m)=0$, which can in all three
cases be written as $m=\tanh[(g+Jm^{q-1})/{ T}]$.

If the coupling $g$ is large enough, the resulting dynamics will correctly describe the transition of the magnetization from 
the initial paramagnetic value $m=0$ to the final ferromagnetic value $m=m_{\rm F}$.
Comparison between quadratic interactions ($q=2$) and quartic interactions ($q=4$) is instructive. 
The student can determine in the latter case the minimum value of the coupling $g$ below 
which the registration cannot take place, and convince him/herself that it does not depend on the form of $C(m)$.
Approaching this threshold from above, one observes the slowing down 
of the process around the crossing of the bottleneck.     
This feature is made obvious by comparing the Figs 7.3 and 7.4 which illustrate the two situations $q=2$ and $q=4$, respectively, 
and which were evaluated by using the form (\ref{vmvm})) of $v(m)$.

The above exercise overlooks the broadening and subsequent narrowing of the profile at intermediate times, which is relevant for 
finite values of $N$. More advanced students may be proposed to numerically solve  the time evolution of $P(m,t)$, i. e.,
the whole registration process, at finite $N$, taking in the rate equations  Eq. (\ref{dPdiag}) 
e.g. $N=10,\,100$ and $1000$. For the times of interest, $t\gg \hbar/\Gamma$, one is allowed to employ the simplified form of the
 rates from (\ref{KOmipm=}) and (\ref{ompmi=}), and to set $\Gamma=\infty$.
The relevant rate coefficients are listed at the end of Appendix B.

}


  \renewcommand{\thesection}{\arabic{section}}
\section{Statistical interpretation of quantum mechanics}
  \setcounter{equation}{0}\setcounter{figure}{0}
  \renewcommand{\thesection}{\arabic{section}.}
\label{section.9.2}
\label{fin10}

\hfill{\it A man should first direct himself in the way he should go.}

\hfill{\it Only then should he instruct others}           

\hfill{                                  Buddha}

\ZeText{

\vspace{0.3cm}

Measurements constitute privileged tools for relating experimental reality and quantum theory. The solution of models of quantum measurements is therefore 
expected to enlighten the foundations of quantum mechanics, in the same way as the elucidation of the paradoxes of classical statistical mechanics has provided 
a deeper understanding of the Second Law of thermodynamics, either through an interpretation 
of entropy as missing information at the microscopic scale \cite{BalianPoincare03,BalianUtrecht,BrillouinBook,BalianBook,vedral,Jaynes,Jaynes_Book,Katz},
or through a microscopic interpretation of the work
and heat concepts \cite{lindblad_book, X5, X6, X7, X8,  X9, X10, X13, X14}. In fact,
the whole literature devoted to the quantum measurement problem has as a background the interpretation of quantum mechanics.
Conversely, some specific formulation of the principles and some interpretation are needed to understand the meaning of calculations about models. The use of quantum 
statistical mechanics (sections 2 and 9) provides us with a density operator of the form (9.1) at the final time; before drawing physical conclusions (section 11) we have to make 
clear what such a technical tool really means. We prefer, 
among the various interpretations of quantum mechanics \cite{deMuynck,Laloe,bellac,alter}, 
the statistical one which we estimate the most adequate. We review below the main features
of this statistical interpretation, as underlined by Park \cite{X4} and supported by other authors. It is akin to the one advocated
by Ballentine \cite{BallentineRMP,BallentineBook}, but it does not coincide with the latter in all aspects.
For a related historic perspective, see Plotnitsky~\cite{Plotnitsky}.

}

\subsection{Principles}

\label{fin10.1}

\ZeText{

In its statistical interpretation (also called ensemble interpretation), 
quantum mechanics presents some conceptual analogy with statistical mechanics. It has a dualistic nature, involving two types 
of mathematical objects, associated with a system and with possible predictions about it, respectively.
On the one hand, the ``observables'', non-commutative random operators, describe the physical quantities related to the studied system. 
On the other hand, a ``state'' of this system, represented by a density operator, gathers the whole probabilistic information available about it under given circumstances. 

}

\subsubsection{Physical quantities: observables}
\label{section.9.2.2}
\label{fin10.1.1}

\hfill{\it Hello, Dolly! }

\hfill{\it  It's so nice to have you back where you belong}

\hfill{Written by Jerry Herman, sung by Louis Armstrong}

\vspace{0.3cm}
\ZeText{

In classical physics, the physical quantities are represented by $c$-numbers, that is, scalar commuting variables, possibly random in
stochastic dynamics or in classical statistical mechanics. In quantum physics, the situation is different. The physical quantities cannot be directly observed or manipulated; 
hence we refrain from the idea that they might take well-defined scalar values. The microscopic description of a system requires counterintuitive concepts, 
which nevertheless have a precise mathematical representation, and which will eventually turn out to fit experiments.

The physical quantities that we are considering are, for instance, the position, the momentum, or the components of the spin of each particle constituting 
the considered system, or a field at any point. The mathematical tools accounting for such quantities in unspecified circumstances have a random nature. 
Termed as ``observables'', they are elements of some algebra which depends on the specific system. One should not be misled by the possibly subjective connotation 
of the term ``observable'': the ``observables'' of quantum mechanics pertain only to the system, and do not refer to any external observer or measuring device.
 Along the lines of Heisenberg's matrix theory  \cite{blokhintsev1,blokhintsev2,deMuynck,Laloe,bellac,BallentineBook,landau,alter}, they can be represented as linear 
 operators acting in a complex Hilbert space ${\cal H}$, or as matrices once a basis is chosen in this space,  which exhibits the algebraic structure. 

The present more abstract approach is also more general, as it encompasses other representations, termed as Liouville representations 
\cite{BalianAJPh1999,BalaszJennings,HilleryOConnellScullyWigner}
in which the product is implemented differently; an example of these, the Wigner representation, is useful in the semi-classical limit. 
The structure of the set of observables, a  $C^\ast$-algebra~\cite{EmchBook}, 
 involves addition, multiplication by complex $c$-numbers, hermitean conjugation, and non-commutative product\footnote{In mathematical terms, a $C^\ast$-algebra is defined 
 as a closed associative algebra, including an involution $x\leftrightarrow x^\ast$ (with $(xy)^\ast=y^\ast x^\ast$) and a norm (with $||x+y||\le ||x||+||y||$ and $||xy||\le ||x||\, ||y||$) 
 which satisfies the identity $||x^\ast x||=||x||^2=||x^\ast ||^2$. In quantum mechanics or quantum field theory, we deal with a $C^\ast$-algebra over complex numbers including 
 unity; an observable is a self-adjoint element of $C^\ast$ and a state is a positive linear functional on $C^\ast$}.
 The physical observables $\hat O$ are hermitean. They play in quantum mechanics the same
   r\^{o}le as {\it  random variables} in classical statistical mechanics, except for the essential
  fact that they belong to a {\it non-commutative algebra}, the structure of which
  fully characterizes the system \cite{EmchBook}.   Ordinary reasoning and macroscopic experience
  do not help us to develop intuition about such non-commuting physical
  quantities, and this is the main incentive for proposals of alternative
  interpretations of quantum mechanics \cite{Bassi_Ghirardi,BohmHolland,BohmCushing,adler,pearle_review,genovese}.

     In some circumstances, when the observables of interest constitute
a commutative subset, the peculiar aspects of quantum mechanics that
raise difficulties of interpretation do not appear \cite{EmchBook,X2,X3}. For
instance, the classical probability theory is sufficient for working out
the statistical mechanics of non-interacting Fermi or Bose gases at
thermal equilibrium. This simplification occurs because we deal there
only with commuting observables, the occupation number operators $\hat
n_k$ for the single particle states $|k\rangle$, which can be treated as
random $c$-numbers taking the discrete values $n_k=0$ or $1$ for
fermions, $n_k=0, \,1,\, 2,\cdots$ for bosons. However, even in this
simple case, it is the underlying non-commutative algebra of the
creation and annihilation operators $\hat a^\dagger_k$ and $\hat a_k$
which explains why the eigenvalues of $\hat n_k=\hat a^\dagger_k \hat a_k$ are those integers.  A similar situation occurs for macroscopic systems,
 for which classical behaviors emerge from the hidden microscopic fundamental quantum theory. 
 The variables controlled in practice then commute, at least approximately, so that classical concepts are sufficient. 
Macroscopic properties such as electronic conduction versus insulation, magnetism, heat capacities, superfluidity, or the very existence of crystals 
all have a quantum origin but obey equations of a ``classical'' type, in the sense that they involve only commutative variables.
Non commutation, the essence of quantum mechanics, may manifest itself only exceptionally in systems that are not microscopic, see
~\cite{Kofler} and references therein.

What one calls ``quantum'' and ``classical'' depends, though,  on which quantities are observed and how the difference with respect to their classical limit is quantified 
(if such a limit exists at all). We have  identified above a ``truly quantum'' behavior
 with non-commutativity, a deep but restrictive definition. Other viewpoints are currently expressed, such as dependence on $\hbar$.
 Quantum electrodynamics have two classical limits, wave-like when the non-commutation of the electric and magnetic fields is not effective, 
 and particle-like when the number of photons is well defined.
 Moreover, the quantal or classical nature of a given concept may depend on the specific situation. 
The center of mass of a small metallic grain can be described by its ``classical'' value, while the shape of its heat capacity requires a quantum description, 
such as the Debye model although the concept of specific heat, its measurement, its thermodynamic aspects, are all ``pre-quantal''. 
On the other hand, in atomic clocks one needs to control the quantum fluctuations of the position of the center of mass, which is therefore not so classical. 
An extreme case of quantal center of mass is a mechanical resonator in its ground state or excited by one phonon   \cite{Cleland}. 

}

  \subsubsection{Dynamics}
\label{fin10.1.2}

\ZeText{

  Dynamics is currently implemented in the quantum theory through the Schr\"odinger picture, where the observables remain constant while the states (pure or mixed) evolve 
  according to the Schr\"odinger or the Liouville-von Neumann equation. Following the tradition, we have relied on this procedure in sections 4 to 8, and will still use it in section 11. 
  The evolution then bears on the wave function or the density operator, objects which characterize our information on the system. 
  However, dynamics should be regarded as a property of the system itself, regardless of its observers. It is therefore conceptually enlightening to account for 
  the evolution of an {\it isolated system} in the Heisenberg picture, as a change in time of its observables which pertain to this system.

We should then implement the dynamics as a transformation of the set of observables, represented by a linear mapping that leaves invariant the algebraic 
relations between the whole set of observables \cite{blokhintsev1,blokhintsev2,deMuynck,Laloe,bellac,BallentineBook,landau}.
In the Hilbert space representation, this implies that the {\it transformation is unitary}. 
(In Liouville representations, where observables behave as vectors, their evolution is generated by the Liouvillian superoperator.) 
Denoting by $t_0$ the reference time at which the observables $\hat O$ 
are defined,  we can thus write the observables $\hat O(t, t_0)$  at the running time $t$ as $\hat O(t, t_0)=\hat U^\dagger(t, t_0) \hat O \hat U(t, t_0)$,  
where the unitary transformation $\hat U(t, t_0)$ carries  the set of observables from $t_0$ to $t$. 
(In the Schr\"odinger picture, it is the density operator which depends on time, according to $\hat U(t, t_0) \hat{\cal D} \hat U^\dagger(t, t_0)$.) 
The infinitesimal generator of this transformation being the Hamiltonian $\hat H$, the time-dependent 
observable $\hat O(t, t_0)$ is characterized either by the usual Heisenberg equation $i\hbar \partial\hat O(t, t_0)/\partial t = [\hat O(t, t_0), \hat H]$ 
with the boundary condition $\hat O(t_0, t_0)=\hat O$
or by the backward equation $i\hbar \partial\hat O(t, t_0)/\partial t_0 = [\hat H, \hat O(t, t_0)]$ with the boundary condition $\hat O(t, t)=\hat O$. 
The backward equation, more general as it also holds if $\hat H$ or $\hat O$  depend explicitly on time, is efficient for producing dynamical 
approximations, in particular for correlation functions ~\cite{balianveneroni1993}.
The interest of the backward viewpoint for the registration in a measurement  is exhibited in \S~7.3.1, Appendix F and \S~\ref{fin13.1.3}. 

Note that the observables and their evolution in the Heisenberg picture can be regarded 
as non commutative, one-time random objects that may be ascribed to {\it a single system}. 
 We do not speak yet of information available about these time-dependent observables in some specific circumstance. This will require the introduction 
of statistical ensembles of similarly prepared systems (\S~\ref{fin10.1.3}) and of ``states'' that encompass the information and from which probabilistic predictions 
about measurements can be derived (\S~\ref{fin10.1.4}).

The Heisenberg picture thus defines time-dependent algebraic structures that are dynamically invariant  \cite{EmchBook}.  
For instance, the $x-p$ commutation relation acquires a definite kinematical status, irrespective of the statistics of these physical quantities.
Whereas the Schr\"odinger picture tangles the deterministic and probabilistic aspects of quantum mechanics within the time-dependent states 
$|\psi(t)\rangle$ or $\scriptD(t)$, these two aspects are well separated in the Heisenberg picture, deterministic dynamics of the observables, 
probabilistic nature of the time-independent states. We will rely on this remark in subsection 13.1. The Heisenberg picture also allows to define 
correlations of observables taken at different times and pertaining to the same system  \cite{alter,balianveneroni1993}. 
  Such autocorrelations, as the Green's functions in field theory, contain detailed information about the dynamical probabilistic behavior 
  of the systems of the considered ensemble, but cannot be directly observed through ideal measurements.

}

\subsubsection{Interpretation of probabilities and statistical ensembles of systems}
\label{section.9.2.1}
\label{fin10.1.3}

\hfill{\it What is true is no more sure than the probable}

\hfill{Greek proverb}

\vspace{3mm}

\ZeText{


  While the observables and their evolution appear as properties of the objects under study, our knowledge about them is probabilistic. 
  The statistical interpretation highlights the fact that quantum mechanics provides us only with probabilities 
\cite{BallentineRMP,blokhintsev1,blokhintsev2,X4,HomeWhitaker,deMuynck,BalianAJPh1989,BalianUtrecht}.  
Although a probabilistic theory may produce some predictions with certainty, most quantities that
we deal with at the microscopic scale are subject to statistical fluctuations: expectation values, correlations at a given time, or
autocorrelations at different times when we observe for instance the successive transitions of a trapped ion \cite{X0,X1}. 
Exact properties of individual systems can be found only in special circumstances,  such as the ideal measurement of some observable (section 11).
Thus, explicitly or implicitly, our descriptions refer to statistical ensembles of systems and to repeated experiments \cite{BallentineRMP,X4,deMuynck}. 
Even when we describe {\it a single object} we should imagine that it belongs to {\it a thought ensemble} ${\cal E}$ \cite{alter},
all elements of which are considered to be prepared under similar conditions characterized by the same set of data\footnote{When
  accounting probabilistically for the cosmic microwave spectrum, one imagines the Universe to belong to an ensemble of possible universes. 
  With a single Universe at hand, this leads to the unsolvable {\it cosmic variance} problem. The same ideas hold whenever probabilities are applied to 
  a single system or event ~\cite{Cox},  and this is the subject of standard and thorough developments in books of probabilities, including already Laplace's}.
Notice the similarity with ensemble theory in classical statistical physics, which also allows probabilistic predictions on single systems \cite{krylov,Callen}. 
However, there is no quantum system devoid of any statistical fluctuations \cite{BallentineRMP,deMuynck}. Individual events
resulting from the same preparation are in general not identical but obey some probability law, even when the preparation is as complete as possible. 
  
The concept of probability, inherent to quantum mechanics, is subject to several interpretations, two of which are currently used in physics\footnote{Kolmogorov's axioms, 
the starting point of many mathematical treatises, do not prejudge how probabilities may be interpreted in applications}. On the one hand, 
in the ``frequentist'' interpretation, a probability is identified with the {\it relative frequency} of occurrence of a given event. This conception of probabilities, the current one in 
the XVII$^{\rm th}$ and XVIII$^{\rm th}$ centuries, has been given a mathematical foundation, on which we will return in \S~11.2.2, by Venn~\cite{venn} and von Mises~\cite{mises}. 
On the other hand, in the ``logical Bayesian'' approach, initiated by Bayes and Laplace, and later on formalized by Cox \cite{Cox} and advocated by Jaynes ~\cite{Jaynes_Book}, 
probabilities are defined as a mathematical {\it measure of likelihood} of events; they are not inherent to the considered object alone, but are tools for making reasonable
 predictions about this object through consistent inference\footnote{We keep aside the ``subjective Bayesian'' interpretation, developed by de Finetti  \cite{deFinetti}, 
 and suited more to ordinary life or economy  than to science. There, probabilities are associated with the state of mind of an agent, and help him to take rational decisions. 
 Prior probabilities reflect their subjectivity, whereas priors are provided by a physical invariance in quantum mechanics 
 (unitary invariance in Hilbert space) or in statistical mechanics (invariance under canonical transformations in phase space), so that the entropy is then defined uniquely}.
  Both interpretations are relevant to quantum theory, and their equivalence has been established \cite{BalianBalazs}, in the context of assigning a quantum probability 
 distribution to a system (\S~10.2.2).  In fact, understanding the conceptual quantum issues (including measurement) does not demand that one 
  adheres to one rather than to the other. Possible mistakes committed in discussing these issues should not be assigned to a specific (Bayesian or frequency) 
  interpretation of probability~\cite{Vervoort, KhrennikovFop02}. 
     
 Depending on the circumstances, one of these interpretations may look more natural than the other. In measurement theory, Born's probabilities $p_i$ can be regarded as 
 relative frequencies, since $p_i$ is identified, for a large set $\scriptE$ of runs, with the relative number of runs having produced the outcome $A_i$ of the pointer. 
 We will rely on the same idea in section 11, where we consider arbitrary subensembles of $\scriptE$: For a given subensemble, the weight $q_i$ associated with each outcome 
 $A_i$ will then be interpreted as a probability in the sense of a proportion of runs of each type. On the other hand, according to its definition in \S~10.1.4, the concept of 
 quantum state has a Bayesian aspect. In this approach, the prior needed for assigning a state to a system in given circumstances is provided by unitary invariance (\S~10.2.2). 
 A state does not pertain to a system \textit{in itself}, but characterizes \textit{our information} on it or on the ensemble to which it belongs.
   In fact, information has turned out to be a central concept in statistical physics 
  \cite{BalianPoincare03,BalianUtrecht,BrillouinBook,BalianBook,vedral,Jaynes,Jaynes_Book,Katz}.  This idea is exemplified by spin-echo 
  experiments \cite{spin_echo,spin_echo1,spin_echo2a,spin_echo2b,spin_echo3,spin_echo4,Abragam1961,Abragam1982}.
    After the initial relaxation, an observer not aware of the history of the system cannot describe  its spins better than
  by means of a completely random probability distribution. However, the experimentalist, who is able to manipulate the
  sample so as to let the original magnetization revive, includes in his probabilistic description the hidden correlations that keep
  track of the ordering of the initial state. Likewise,  we can assign different probabilities to the content of a coded message that we have
  intercepted, depending on our knowledge about the coding \cite{BrillouinBook}. Since quantum theory is irreducibly 
  probabilistic, it has thus a partly subjective nature --- or rather  \textquotedblleft inter-subjective\textquotedblright
  \ since under similar conditions all observers, using the same knowledge, will describe a quantum system
  in the same way and will make the same probabilistic predictions about it. The
  recent developments about the use of quantum systems as information processors \cite{galindo}
  enforce this information-based interpretation \cite{Jaeger2009Book,caves} (see the end of \S~\ref{fin12.4.2}).

\vspace{3mm}

It is important to note that, depending on the available information, a given system may be embedded in different statistical ensembles, and hence may be described by 
different probability distributions. This occurs both in classical probability theory and in quantum physics. Such a distinction between an ensemble and one of its subensembles, 
both containing the considered system, will turn out to be essential in measurements (\S~11.1.2). There, a single run may be regarded as an event chosen 
{\it among all possible runs} issued from the initial state $\scriptD(0)$ of S + A, but may also be regarded as {\it belonging to some subset of runs} -- in particular the subset that will be 
tagged after achievement of the process by some specific indication of the pointer (\S~\ref{finfin11.3.2}). The study of the {\it dynamics of subensembles} (subsection 11.2) will therefore be 
a crucial issue in the understanding of reduction in measurements. 
  
}

\subsubsection{States}
\label{section.9.2.3}
\label{section.10.1.3n}
\label{fin10.1.4}

\hfill{{\it L'Etat c'est moi}\footnote{The State, that's me}}

\hfill{Louis XIV}

\vspace{0.3cm}
\ZeText{

  In the present scope, the definition of a quantum state is conceptually the same as in statistical mechanics \cite{Thirring,BalianBook,X4}:  
  A state of the considered system (or more precisely a state of the real or virtual statistical ensemble  ${\cal E}$  (or subensemble) of systems to which it belongs) is 
  characterized by specifying the correspondence $\hat{O}\mapsto$ $\langle \hat{O}\rangle$
  between the elements $\hat O$ of the $C^\ast$-algebra of observables and $c$-numbers $\langle \hat{O}\rangle$. 
  This correspondence has the following properties \cite{BalianAJPh1989,BalianUtrecht}: it is linear, it associates
  a real number to hermitean operators, a non-negative number to the square of
  an observable, and the number $1$ to the unit operator.   Such properties entail in particular that  $\langle\hat O^2\rangle-\langle\hat O\rangle^2$ cannot be negative.
  
  The $c$-number $\langle\hat O\rangle$ associated through the above mathematical definition with the observable $\hat O$ will eventually be interpreted as the expectation
   value of the physical quantity represented by $\hat O$, and this interpretation will emerge from the ideal measurement process of $\hat O$ (\S~11.3.1). 
   Accordingly, $\langle\hat O^2\rangle-\langle\hat O\rangle^2$ appears as the variance of  $\hat O$; likewise, the probability of finding for $\hat O$ some eigenvalue $O_i$ is 
   the expectation value $\langle\hat\Pi_i\rangle$  of the projection operator $\hat\Pi_i$ over the corresponding eigenspace of $O_i$. 
   A {\it quantum state} has thus a probabilistic nature,  as it is identified with the {\it collection of expectation values of all the observables}. 
   However, if two observables $\hat O_1$ and $\hat O_2$ do not commute and thus cannot be 
   measured simultaneously, $\langle\hat O_1\rangle$ and $\langle\hat O_2\rangle$, taken together, should not be regarded as expectation values in the ordinary 
   sense of probability theory (\S~10.2.1).
		
For infinite systems or fields, this definition of a state as a mapping of the algebra of observables onto
commuting $c$-numbers has given rise to mathematical developments in the theory of $C^\ast$-algebras \cite{EmchBook}. 
Focusing on the vector space structure of the set of observables, one then considers the states as elements of the dual vector space. 
For finite systems the above properties are implemented in an elementary way once the observables are represented as operators in a Hilbert space. 
The mapping is represented by a {\it density operator} ${\hat{{\cal D}}}$ in this Hilbert space, which is hermitean, non-negative and
normalized, and which generates all the expectation values through \cite{BalianAJPh1989,BalianUtrecht}

  \begin{equation}
  \mytext{\textcurrency EV\textcurrency \qquad}
  \hat{O}\mapsto\langle \hat{O}\rangle ={\rm {\rm tr}}{\hat{{\cal D}}}\hat{O}{.} 
  \label{EV}
  \end{equation}
  In fact, according to Gleason's theorem \cite{gleason}, the linearity of this correspondence
  for any pair of commuting observables is sufficient to ensure the existence of ${\hat{{\cal D}}}$.
  (We use the notation  ${\hat{{\cal D}}}$ for the generic system considered here; no confusion should arise with the state of
   S + A in the above sections.)

A tutorial introduction to density operators is presented in Appendix G.
  
  A density operator which characterizes a state
  plays the r\^{o}le of a probability distribution for the non-commuting
  physical quantities $\hat O$ since it gathers through (\ref{EV}) our whole information
  about an ensemble of quantum systems \cite{X4,BalianAJPh1989,BalianUtrecht,Jaynes_Book}.  
  As in probability theory, the amount of {\it missing information} associated 
  with the state $\hat{\cal D}$ is measured by its von Neumann entropy
\cite{BalianAJPh1989,BalianUtrecht,Jaynes_Book}. 

\BEQ 
S(\hat{\cal D})=-{\rm tr}
 \hat{\cal D}\ln \hat{\cal D}. \EEQ  
  
For time-dependent predictions on an isolated system, Eq. (\ref{EV}) holds both in the Schr\"odinger picture, 
with fixed observables and the Liouville--von Neumann evolution for $\scriptD(t)$, and in the Heisenberg picture,  with fixed $\hat {\cal D}$ 
and observables evolving unitarily. However, two-time (and multi-time) autocorrelation functions cannot be defined within the Schr\"odinger picture. 
They are obtained as ${\rm tr}\,\hat{\cal D}\hat O_1(t_1,t_0)\hat O_2(t_2,t_0)$, where the observables in the Heisenberg picture refer to the physical quantities of interest
 and their dynamics, and where the state accounts for our knowledge about the system \cite{EmchBook}.  
 In particular, when defining in \S~3.3.2 the autocorrelation function $K(t-t')$ of the bath, it was necessary to express the time-dependent  bath operators in the 
 Heisenberg picture (although we eventually inserted $K(t-t')$ into the Liouville--von Neumann equations of motion of S + M in section 4 and appendix A).
 
   In this interpretation, what we call the state {\it ``of a system''}, whether it is pure or not, is not a property of \textit{the considered
system in itself}, but it characterizes the statistical properties of the real or virtual \textit{ensemble} (or subensemble) to which this system belongs
\cite{X4,BalianAJPh1989,BalianUtrecht,Jaynes_Book}.  The word \textquotedblleft {\it state}\textquotedblright\ itself is also
misleading, since we mean by it the summary of {\it our knowledge} about the ensemble, from which we wish to make probabilistic predictions.
The conventional expression \textquotedblleft the state of the system\textquotedblright\ is therefore doubly improper in quantum physics, especially within the
statistical interpretation \cite{X4,BalianAJPh1989,BalianUtrecht}, and we should not be misled by this wording --- although we cannot help to use it when teaching. 
  
 Density operators \textit{differ} from distributions of the probability theory taught in mathematical courses and from densities in phase space of classical
  statistical mechanics,  because the quantum  physical quantities have a \textit{non-commutative} nature 
  \cite{blokhintsev1,blokhintsev2,deMuynck,Laloe,bellac,BallentineBook,BalianAJPh1989,BalianUtrecht,landau,alter}.
This algebraic feature, compelled by experiments in microphysics,  lies at the origin of the odd properties which make quantum mechanics
  counterintuitive. It implies {\it quantization}. It also implies {\it the superposition principle}, which is embedded in the matrix nature of $\hat{\cal D}$.
It entails {\it Heisenberg's inequality} $\Delta\hat O \Delta\hat O'\ge \frac{1}{2}|\langle [\hat O,\hat O'] \rangle|$ 
  and hence Bohr's complementarity: since the product of the variances of two
  non-commuting observables has a lower bound, it is only in a fuzzy way that we
  can think simultaneously of quantities such as the position and the momentum
  (or the wavelength) of a particle, contrary to what would happen in classical statistical mechanics. 
Thus the non-commutation of observables implies the existence of
  intrinsic fluctuations, and the quantum theory is irreducibly probabilistic
  \cite{blokhintsev1,blokhintsev2,X4,deMuynck,Laloe,bellac,BallentineBook,landau,alter}.

One should note, however, that the non-commutation of two observables does not necessarily imply that they present quantum fluctuations.
For instance, if two operators do not commute, there may exist states (their common eigenstates) in which both have well-defined values. 
As an example, in states with orbital momentum zero, the components $\hat L_x$ and $\hat L_y$ vanish without any statistical fluctuation. 
(This does not contradict the Heisenberg inequality
 $\Delta \hat L_x  \Delta \hat L_y \ge \half \hbar |\langle \hat L_z\rangle|$, because both sides vanish in this case; 
 more general uncertainty relations for orbital momentum are given in \cite{RivasLuis2008}.)
Conversely, two commuting observables may fluctuate in some states, even pure ones.

  In the statistical interpretation, we should refrain from imagining that the
  observables might take well-defined but undetectable values in a given state, and that the
  uncertainties about them might be a mere result of incomplete knowledge. The very
  concept of physical quantities has to be dramatically changed.
  We should accept the idea that quantum probabilities, as represented by a density operator, do not 
  simply reflect as usual our ignorance  about supposedly preexisting values of physical quantities (such as the
  position and the momentum of a particle), but arise because {\it our very conception of physical quantities as scalar numbers}, 
 inherited from macroscopic experience, {\it is not in adequacy with microscopic reality} 
\cite{blokhintsev1,blokhintsev2,deMuynck,Laloe,bellac,BallentineBook,landau,alter}.
 Macroscopic physical quantities take scalar values that we can observe, in particular for a pointer, but the scalar values that we are led to 
 attribute to microscopic (non-commuting) observables are the outcome of inferences which are indirectly afforded by our measurement processes.
 
 \vspace{3mm}
 
 From an epistemological viewpoint, the statistical interpretation of quantum theory has a dualistic nature, both objective and subjective. 
 On the one hand, observables are associated with the physical properties of a real system. On the other hand, in a given circumstance,
  the reality of this system is ``veiled'' \cite{DEspagnat}, in the sense that our knowledge about these physical properties cannot be better than probabilistic, 
  and what we call ``state'' refers to the information available to observers.

  }

\subsection{Resulting properties}

\label{fin10.2}

\subsubsection{Contextuality}
\label{10.1.4}
\label{fin10.2.1}

\ZeText{

Information about quantum systems can be gained only through complex measurement processes, involving interaction with instruments and selection of the outcomes. 
What we observe when testing the ``state of the system'' is in fact a joint property of the system S and the apparatus A. Moreover, due to the non-commutation of the observables 
which implies their irreducibly probabilistic nature, we cannot assign well defined numerical values to them before achievement of the process. These values do not belong to S 
alone, but also to {\it its experimental context}. They have no existence before measurement, but emerge indirectly from interaction with a given instrument A and are defined only 
with reference to the setting which may determine them.

In a theoretical analysis of a measurement process, we have to study the density operator that describes a statistical ensemble $\scriptE$ of joint systems S+A. If we use 
another apparatus A$^\prime$, the ensemble described is changed into $\scriptE^\prime$. Putting together results pertaining to $\scriptE$ and $\scriptE'$
may produce paradoxical consequences although the tested system S is prepared in the same state. The statements of quantum mechanics  are meaningful and can 
be logically combined {\it only} if one can imagine a {\it unique experimental context}  in which the quantities involved might be simultaneously measured.

These considerations are illustrated by various odd phenomena that force us to overturn some of our ways of thinking. A celebrated example is the violation of Bell's 
inequalities, recalled in \S~2.2.1. Other quantum phenomena, involving properties satisfied exactly rather than statistically, may be regarded as failures of ordinary logic. 
They are exemplified by the GHZ paradox [34, 36, 298], recalled below\footnote{The GHZ setup is as follows: Consider six observables $\hat{B}_{i}$ and $\hat{C}_{i}$ ($i=1$,
  $2$, $3$) such that $\hat{B}_{i}^{2}=\hat{C}_{i}^{2}\equiv\hat{I}$, $\hat
  {C}_{1}\hat{C}_{2}\hat{C}_{3}\equiv\hat{I}$,  and with commutators $[  \hat{B}_{i},\hat{B}_{j}]  =[\hat{C}_{i},\hat{C}_{j}] =0$, 
  $[  \hat{B}_{i},\hat{C}_{i}]  =0$ and $\hat{B}_{i}\hat{C}_{j}=-\hat{C}_{j}
  \hat{B}_{i}$ for $i\neq j$.   A physical realization with 3 spins is provided by taking
 $\hat{B}_{1}=\hat{\sigma}_{x}^{(1)}$,  $\hat{C}_{1}=\hat{\sigma}_{z}^{(2)}\hat{\sigma}_{z}^{(3)}$ 
 (or, more precisely, $\hat{B}_{1}=\hat{\sigma}_{x}^{(1)}\hat{\sigma}_{0}^{(2)}\hat{\sigma}_{0}^{(3)}$,  $\hat{C}_{1}=\hat\sigma_0^{(1)}\hat{\sigma}_{z}^{(2)}\hat{\sigma}_{z}^{(3)}$),
 and likewise,  in a cyclic manner.
  In the pure state $\left\vert \varphi
  \right\rangle $ characterized by $\hat{B}_{i}\hat{C}_{i}\left\vert
  \varphi\right\rangle =\left\vert \varphi\right\rangle $, each one of the three
  statements \textquotedblleft$B_{i}$ takes the same value as $C_{i}
  $\textquotedblright, where $B_{i}=\pm1$ and $C_{i}=\pm1$\ are\ the values
  taken by the observables $\hat{B}_{i}$\ \^{a}nd $\hat{C}_{i}$,\ is
  \textit{separately} true, and can be experimentally checked. However, these
  three statements cannot be true \textit{together}, since the identity $\hat{C}_{1}\hat
  {C}_{2}\hat{C}_{3}\equiv\hat{I}$ seems to entail that $B_{1}B_{2}B_{3}=+1$ in
  the considered state, whereas the algebra implies ${B_{1}B_{2}B}_{3}{=-1}$, which is confirmed experimentally ~\cite{Pan2000}.
  Indeed we are not even allowed to think simultaneously about the values of $B_{1}$ and $C_{2}$, for instance, since these observables do not commute.
  It is not only impossible to measure them simultaneously but it is even ``forbidden'' (i. e., devoid of any physical meaning) to imagine in a given system the 
  simultaneous existence of numerical values for them, since these numerical values should be produced through interaction with different apparatuses}.

}

\subsubsection{Preparations and assignment of states}

\label{fin10.2.2}

\hfill{\it Que sera, sera\footnote{What will be, will be}}

 \hfill{ Jay Livingston and Ray Evans; sung by Doris Day in {\it The man who knew too much}}

\vspace{3mm}
\ZeText{

In order to analyze theoretically quantum phenomena, we need to associate with the considered situation the state that describes adequately the system 
(or rather the set of systems of the considered ensemble). In particular, to study a dynamical process in the Schr\"odinger picture,  we must 
specify the initial state. Such an assignment can be performed in various ways, depending on the type of preparation of the system \cite{X2,X3}.

Textbooks often stress {\it complete preparations}, in which a complete set of commuting observables is controlled; 
 see Refs. \cite{X2,X3} for a recent conceptual discussion that goes beyond the average text-book level. The state $\hat{\cal D}$
is then the projection on the common eigenvector of these observables determined by their given eigenvalues. 
(This unambiguous determination of $\hat{\cal D}$ should not hide its probabilistic nature.) The control of a single observable may in fact be 
sufficient to allow a complete preparation of a pure state, in case one is able to select a non-degenerate eigenvalue that characterizes this state. 
Atoms or molecules are currently prepared thereby in their non-degenerate ground state \cite{X1}.

As indicated in \S~1.1.4, the ideal measurement of an observable $\hat s$  (like the spin component $\hat s_z$ in the Curie--Weiss model considered 
in the bulk of the present work) of a system S, followed by the selection of the outcome $A_i$ of the pointer constitutes a {\it preparation through measurement}. 
If the density operator of S before the process is $\hat r(0)$, this selection produces the filtered state $\Pi_i \hat r(0) \Pi_i$, where $\Pi_i$ denotes the projection
operator onto the eigenspace associated with the eigenvalue $s_i$ of $\hat s$ (see \S~\ref{finfin11.3.2}).  
This theoretical scheme of preparing states via measurements 
was realized experimentally \cite{X1,X11}. 

There are however other, {\it macroscopic} methods of preparing quantum states that  are much more incomplete \cite{krylov,Callen}.
Usually they provide on the quantum system of interest a number of data much too small to characterize a single density operator. As in ordinary
probability theory, for describing a macroscopic preparation, one can rely on some criterion to select among the
allowed $\hat{\cal D}$'s the least biased one \cite{Jaynes_Book}.
A current criterion is Laplace's  ``principle of insufficient reason'': when nothing else is known than the set of possible events, we should assign to them equal probabilities. 
In fact, this assignment relies implicitly on the existence of some invariance group. For a discrete set of ordinary events, this is the group of their permutations, as they should 
be treated a priori on the same footing. In quantum theory, the required prior invariance group is afforded by physics, it is the unitary group in Hilbert space. 
When some data are known, namely the expectation values of some observables, Laplace's principle cannot be directly applied since these data constrain the density operator, 
but one can  show that it yields, as least biased density operator among all those compatible with the available data, the one that maximizes the
entropy (10.2) \cite{BalianBalazs,Uffink,partovi_entropy}.  In particular, the energy
of a small object can be controlled by macroscopic means, exchange of heat or of work; depending on the type of control, the maximum entropy
criterion leads us to assign a different distribution to this object \cite{BalianUtrecht,BalianBook,Jaynes}. This distribution should be verified experimentally. 
For instance, if one controls only the expectation value of its energy, which is free to fluctuate
owing to exchanges with a large bath, the least biased state is the canonical one.  Alternatively, for a non-extensive system such that the
logarithm of its level density is not concave, another type of thermal equilibrium (locally more stable) can be established \cite{Campa2009}
through a different preparation involving the confinement of the energy in a narrow range. Within this range, the maximum entropy criterion
leads us to attribute the same probability to all allowed levels and to adopt a microcanonical distribution. 
  
The fact that states of macroscopic systems cannot be characterized completely entails that in measurement models the apparatus should be supposed to
have initially been prepared in a mixed state. Thus, the discussion of the quantum measurement problem within the statistical interpretation
does need the existence of macroscopic preparations that are different from preparations via quantum measurements.

  }

\subsubsection{Mixed states and pure states}
\label{section.10.1.4}
\label{fin10.2.3}

\hfill{\it Something is rotten in the state of Denmark}

\hfill{Shakespeare, Hamlet}

\vspace{0.3cm}
\ZeText{

Most textbooks introduce the principles of quantum mechanics by relying on pure states $|\psi\rangle$, which evolve according to the
Schr\"odinger equation and from which the expectation value of any observable $\hat O$ can be evaluated as $\langle\psi|\hat O|\psi\rangle$
\cite{vNeumann,landau}. Mixed states, represented by density operators, are then constructed from pure states \cite{vNeumann,landau}. 
This form of the principles entail the above-mentioned laws, namely, the Liouville--von Neumann (or the Heisenberg) equation of motion and 
the properties of the mapping (10.1) (linearity, reality, positivity and normalization). 
 
Within the statistical interpretation, there is at first sight little conceptual difference between pure states and mixed states,
since in both cases the density operator behaves as a non-Abelian probability distribution that realizes the correspondence (\ref{EV})
\cite{BallentineRMP,blokhintsev1,blokhintsev2,deMuynck,BallentineBook,BalianAJPh1989,BalianUtrecht,BalianBook}.  
As a mathematical specificity, pure states are those for which all eigenvalues but one of the density operator $\hat{\cal D}$ vanish, or
equivalently those for which the von Neumann entropy $S(\hat{\cal D})$ vanishes.  They appear thus as extremal among the set of Hermitean
positive normalized operators, in the form $|\psi\rangle\langle\psi|$.  However, a major physical difference\footnote{Another essential 
difference between pure and mixed states is especially appealing to intuition \cite{partovi_uncertainty,luis_complo}.  Consider a system in a state
represented by a density operator $\hat{\cal D}$ whose eigenvalues are non-degenerate and differ from zero.  Consider next a set of observables
that have non-degenerate spectra. Then none of such observables can produce definite results when measured in the state $\hat{\cal D}$
\cite{luis_complo}. In other words, all such observables have non-zero dispersion in $\hat{\cal D}$. This statement has been suitably
generalized when either $\hat{\cal D}$ or the observables have degeneracies in their spectra; see Appendix C of Ref.
\cite{luis_complo}.  In contrast, for a pure density operator $|\psi\rangle\langle \psi|$ all observables that have $|\psi\rangle$ as
eigenvector are dispersionless.  Pure and mixed states also differ as regards their preparation and as regards their determination
via measurements (e.g., the number of observables to be measured for a complete state determination)~\cite{Rozonoer}}, stressed by 
Park \cite{X4}, exists, {\it the ambiguity in the decomposition of a mixed state into pure states}. 
This question will play an important r\^ole in section 11, and we discuss it below.

Let us first note that a mixed state $\scriptD$ can always be decomposed into a weighted sum of projections over pure states, according to

\begin{equation} 
\label{9*}
\hat{\cal D}=\sum_k |\phi_k\rangle \nu_k\langle \phi_k|.
\end{equation}
It is then tempting to interpret this decomposition as follows. Each of the pure states $|\phi_k\rangle$ would describe systems belonging to an ensemble $\scriptE_k$, 
and the ensemble $\scriptE$ described by $\scriptD$ would be built by extracting a proportion $\nu_k$ of systems from each ensemble $\scriptE_k$. 
Such an interpretation is consistent with the definition of quantum states as mappings (10.1) of the set of observables onto their expectation values, 
since (10.3) implies $\langle\hat O\rangle=\sum_k \nu_k \langle \phi_k|\hat O|\phi_k\rangle ={\rm tr}\, \hat{\cal D} \hat O$.  
 It is inspired by classical statistical mechanics, where a mixed state, represented by a density in phase space, can be regarded in a unique fashion 
 as a weighted sum over pure states localized at given points in phase space. 
However, in quantum mechanics, the state $\scriptD$ (unless it is itself pure) can be decomposed as (10.3) {\it in an infinity of different ways}.
For instance, the $2\times2$ density operator $\hat{\cal D}=\half\hat\sigma_0$ which represents an unpolarized spin $\half$ might be interpreted as
describing a spin polarized either along $+z$ with probability $\half$ or along $-z$ with probability $\half$; but these two possible
directions of polarization may also be taken as $+x$ and $-x$, or as $+y$ and $-y$; the same isotropic state $\hat{\cal D}=\half\hat\sigma_0$ can
also be interpreted by assuming that the direction of polarization is fully random \cite{deMuynck,BallentineBook}. 
Within the statistical interpretation of quantum mechanics, this ambiguity of the decompositions of $\scriptD$ prevents us from selecting a 
``fundamental'' decomposition and to give a sense to the pure states $|\phi_k\rangle$ and the weights $\nu_k$ entering (10.3).

More generally, we may decompose the given state $\scriptD$ into a weighted sum

 \begin{equation}
\label{9x} 
\hat{\cal D}=\sum_k \nu_k\hat{\cal D}_k
\end{equation} 
of density operators $\scriptD_k$ associated with subensembles $\scriptE_k$. But here again, such a decomposition can always be performed
{\it in an infinity of different ways, which appear as contradictory}. Due to this ambiguity, splitting the ensemble $\scriptE$ described by $\scriptD$ 
into subensembles $\scriptE_k$, described either by pure states as in (10.3) or by mixed states as in (10.4), is physically meaningless 
(though mathematically correct) if no other information than $\scriptD$ is available.
	
The above indetermination leads us to acknowledge an important difference between pure and mixed quantum states
\cite{BallentineRMP,X4,deMuynck,BallentineBook,landau,X4,bkj}. If a statistical ensemble ${\cal E}$ of systems is described by a pure state,
any one of its subensembles is also described by {\it the same pure state}, since in this case (\ref{9x}) can include only a single term. 
If for instance a set of spins have been prepared in the polarized state $|\hspace{-1mm}\up\rangle$, the statistical prediction about any subset are embedded in
$|\hspace{-1mm}\up\rangle$ as for the whole set.  In contrast, the existence of many decompositions (\ref{9*}) or
(\ref{9x}) of a mixed state $\hat{\cal D}$ describing an ensemble ${\cal E}$ implies that there exists many ways of splitting this ensemble into
subensembles ${\cal E}_k$ that would be described by different states $\hat{\cal D}_k$. In particular, pure states $|\phi_k\rangle$ that would underlie as
in (\ref{9*}) a mixed $\hat{\cal D}$ cannot {\it a posteriori} be identified unambiguously by means of experiments performed on the ensemble of systems. In the
statistical interpretation, such {\it underlying pure states have no physical meaning}.
More generally, decompositions of the type (10.4) can be given a meaning only if the knowledge of $\scriptD$ is completed with extra information, 
allowing one to identify, within the considered ensemble $\scriptE$ described by $\scriptD$, subensembles $\scriptE_k$ 
that do have a physical meaning  \cite{deMuynck,BallentineBook,bkj}.  

According to this remark, since the outcome of a large set of measurements is represented by a mixed state $\hat{\cal D}(t_{\rm f})$, 
this state can be decomposed in many different ways into a sum of the type  (\ref{9x}). The decomposition (9.1), each term of which is associated 
with an indication $A_i$ of the pointer, is not the only one. This ambiguity of $\hat{\cal D}(t_{\rm f})$, as regards the splitting of the ensemble 
${\cal E}$ that it describes into subensembles, will be discussed in \S~\ref{fin11.1.3},  and we will show subsections 11.2 and 13.1 how 
 {\it the dynamics of the process  removes this ambiguity} by privileging the decomposition (9.1) and yielding a physical meaning to each 
of its separate terms, thus allowing us to make statements about individual systems.

} 

\subsubsection{Ensembles versus aggregates}
\label{fin10.2.4}

\ZeText{

We have assumed above that the density operator $\scriptD$ and the corresponding ensemble $\scriptE$ were given {\it a priori}. 
In practice, the occurrence of a mixed state $\scriptD$ can have various origins. An incomplete preparation (\S~10.2.2) always yields a mixed state, 
for instance, the initial state $\scriptR(0)$ of the apparatus in a measurement model. The mixed nature of a state may be enhanced by dynamics, 
when some randomness occurs in the couplings or when approximations, justified for a large system, are introduced; this is illustrated by the final state 
$\scriptD(t_f)$ of a measurement process. 

Density operators have also been introduced by Landau in a different context \cite{deMuynck,BallentineBook,landau}. Consider a compound system S$_1$ + S$_2$.  
Its observables are the operators that act in the Hilbert space $ {\cal H}={\cal H}_1\otimes{\cal H}_2$, and its states $\hat{\cal D}$ are
characterized by the correspondence (10.1) in the space ${\cal H}$. If we are interested only in the subsystem S$_1$, disregarding the
properties of S$_2$ and the correlations between S$_1$ and S$_2$, the relevant observables constitute the subalgebra of operators acting in
${\cal H}_1$, and the correspondence (10.1) is implemented in the subspace ${\cal H}_1$ by means of the mixed density operator $\hat{\cal
D}_1={\rm tr}_2\hat{\cal D}$.  Suppose for instance that in an ensemble ${\cal E}$ of pairs S$_1$, S$_2$ of spins $\half$ prepared in the singlet pure state 
$2^{-1/2}(\hspace{0.7mm} |\hspace{-0.8mm}\up\rangle_1|\hspace{-0.8mm}\up\rangle_2-|\hspace{-0.8mm}\down\rangle_1|\hspace{-0.8mm}\down\rangle_2)$, 
we wish to describe only the spin S$_1$. Its marginal state in the considered ensemble ${\cal E}$ is again the unpolarized state,
represented by $\hat{\cal D} _1=\half\hat\sigma_0^{(1)}$. Isotropy is here built in, from this definition of the state of the spin S$_1$.

In all such cases, the state $\scriptD$ describes a statistical ensemble $\scriptE$, and the argument of \S~10.2.3 entails the impossibility of splitting unambiguous 
this ensemble into subensembles described by well defined pure states.

Another approach to density operators, initiated by von Neumann \cite{vNeumann}, consists in constructing them from pure states, by following a path converse 
to that of  \S~10.2.3. We start from a collection of statistical ensembles $\scriptE_k$ of systems prepared in pure states $|\phi_k\rangle$. We build a new set 
$\scriptE$ by {\it extracting randomly} each individual system of $\scriptE$ from one among the ensembles $\scriptE_k$, the probability of this extraction being $\nu_k$. 
If we have lost track of the original ensemble $\scriptE_k$ from which each drawing was performed, we have no means to acknowledge in which pure state
$|\phi_k\rangle$ a given system of $\scriptE$ was originally lying. The expectation value, for this system, of any quantity is then given by 
$\langle\hat O\rangle=\sum_k \nu_k \langle\phi_k|\hat O|\phi_k\rangle={\rm tr}\, \scriptD \hat O$, and we are led to assign to it the mixed state 
defined by (10.3). Here again, the ambiguity of \S~10.2.3 is present: If two different constructions have lead to the same state $\scriptD$, 
they cannot be distinguished from each other in any measurement.
 
 A further important point should be stressed. The above procedure of {\it randomly selecting} the elements extracted from the ensembles $\scriptE_k$ produces 
 a set $\scriptE$ of systems which is a {\it bona fide ensemble}. Indeed, a statistical ensemble must have an essential property, the {\it statistical independence} 
 of its elements, and this property is here ensured by the randomness of the drawings. Thus, our full information about the ensemble, and not only about each 
 of its individual systems, is embedded in the density operator $\scriptD$. In the ensemble $\scriptE$ obtained {\it after mixing}, the pure states $|\phi_k\rangle$ 
 have been {\it completely lost},  although they were originally meaningful. In other words, no observation of an ensemble $\scriptE$ obtained by merging 
  subensembles $\scriptE_k$ can reveal the history of its elaboration.

Another, slightly different construction, also inspired by von Neumann's idea, is preferred by some authors, see, e.g., \cite{DEspagnat}.
In this alternative procedure, a (non random) number $n_k$ of systems is extracted from each ensemble $\scriptE_k$ so as to constitute a set $\scriptA$ 
having $n=\sum_k n_k$ elements, which we term as an {\it aggregate}. Losing again track of the origin of each system of $\scriptA$, we have to assign to any 
individual system of $\scriptA$ the density operator (10.3), with $\nu_k=n_k/n$. However, in spite of this analogy with the ensemble $\scriptE$ constructed above, 
we will acknowledge an important difference between the two situations, due to the nature of the numbers $\nu_k$, which are {\it probabilities} for $\scriptE$, 
{\it proportions} for $\scriptA$.

As an illustration, let us consider the aggregate $\scriptA_z$ built by gathering $n_1=\half n$ spins prepared in the pure state $|\hspace{-1.6mm}\up\rangle$ 
($s_z=+1$) and $n_2= \half n$ spins prepared in the pure state $|\hspace{-1.6mm}\down\rangle$ ($s_z=-1$), and by forgetting the original state of each spin. 
Each individual spin of the aggregate $\scriptA_z$ in then described by the unpolarized density operator $\scriptD= \half\hat\sigma_0$, exactly as each spin 
of the ensemble $\scriptE_z$, obtained by picking up states $|\hspace{-1.4mm}\up\rangle$  or $|\hspace{-1.4mm}\down\rangle$  randomly with equal probabilities. 
Nevertheless, the joint statistics of two systems belonging to the aggregate $\scriptA_z$ differs from that of two spins belonging to the ensemble $\scriptE_z$ 
(which are statistically independent). Indeed, {\it the systems of an aggregate are correlated}, due to the construction procedure. In our spin example, this is flagrant 
for $n_1=n_2=1$: if we measure the first spin down, we know for sure that the second is up.
More generally, if $\hat\sigma_z$ is simultaneously measured on all $n$ spins of the aggregate $\scriptA_z$, the correlations will be expressed 
by the equality of the number of outcomes $\up$ and $\down$. If the ideal measurement bears on $n-1$ spins, we can predict for the last spin the sign of $\sigma_z$ with 
100\% confidence. For an ensemble $\scriptE_z$ containing $n$ spins, we cannot infer anything about the $n$'th spin from the outcomes of previous 
measurements on the $n-1$ other ones.

\vspace{3mm}

Altogether, {\it an aggregate is not a statistical ensemble}, because its elements are {\it correlated with one another}. 
A random selection is needed in von Neumann's procedure of defining mixed states, so as to ensure the statistical independence required for ensembles.
	
The above point was purely classical (since we dealt with the $z$-component only), but it can have quantum implications. Prepare another aggregate 
$\scriptA_x$ with $n_1$ spins oriented in the  $+x$-direction and $n_2$ spins oriented in the $-x$-direction. Consider likewise the ensemble $\scriptE_x$ 
built by randomly selecting spins in the $+x$- and $-x$-directions, with equal probabilities. Any single system belonging to either $\scriptA_z$ or $\scriptE_z$ 
or $\scriptA_x$ or $\scriptE_x$ is described by the same unpolarized density operator  $\half\hat\sigma_0$. However, differences occur when correlations 
between systems are accounted for. We first remember that the ensembles $\scriptE_z$ and $\scriptE_x$ are undistinguishable. In contrast, the two aggregates 
$\scriptA_z$ and $\scriptA_x$ have different properties. Measuring for $\scriptA_x$, as above for $\scriptA_z$, the components $\hat\sigma_z$ of all the $n$ 
spins of $\scriptA_x$ does not show up the correlations that were exhibited for $\scriptA_z$: Instead of finding exactly $\half n$ spins up and $\half n$ spins down, 
we find outcomes that are statistically independent, and characterized by a same binomial law as in the case of the ensembles $\scriptE_z$ or $\scriptE_x$. 
Within $\scriptA_x$ the correlations occur between $x$-components. 

Hence, failing to distinguish aggregates from ensembles leads to the inevitable conclusion that ``two ensembles having the same density matrix can be 
distinguished from each other''  \cite{DEspagnat}.  This statement has influenced similar conclusions by other authors, see e.g.  \cite{o_cohen}.
The persistent occurrence of such an idea in the literature (see \cite{fratini}) demonstrates that the difference between ensembles and aggregates is 
indeed far from being trivial. In the light of the above discussion, and in agreement with  \cite{terno_2000} and \cite{bodor},  
we consider such statements as incorrect. Indeed, {\it two aggregates having for a single system the same density matrix can be distinguished from 
each other} via two-system (or many-system) measurements, but {\it two statistical ensembles cannot}.

 }

\renewcommand{\thesection}{\arabic{section}}
\section{Solving the quantum measurement problem within the statistical interpretation}
\setcounter{equation}{0}
\setcounter{figure}{0}
\setcounter{table}{0}
\renewcommand{\thesection}{\arabic{section}.}

\label{section.9.3.5}

 \hfill {{\it All's well that ends well} }
 
 \hfill{ Shakespeare}

\vspace{0.3cm}
\ZeText{

In section 9 we have resumed the detailed solution of the dynamical equation for the Curie--Weiss model. As other models of measurement treated in the framework of 
quantum  statistical dynamics (section 2), it yields, for the compound system S + A at the end of the process, a density operator $\scriptD(t_{\rm f})$
 which satisfies the properties required for ideal measurements. However, we  have already stressed that such a result, although necessary, is not sufficient to 
 afford a complete understanding of quantum measurements. Indeed, the statistical interpretation of quantum mechanics emphasizes the idea that this theory, 
 whether it deals with pure or mixed states, {\it does not govern individual systems but statistical ensembles} of systems (\S~\ref{fin10.1.3}). 
 Within this statistical interpretation, the density operator $\scriptD(t_{\rm f})$ accounts in a probabilistic way for a large set $\scriptE$ 
 of similar runs of the experiment, whereas a measurement involves properties of individual runs.  Can we then make assertions about
 the individual runs?  This question is the core of the present section.
 
 The remaining challenge is to elucidate the {\it quantum measurement problem}, that is, to explain why {\it each individual run provides a definite outcome}, 
 for both the apparatus and the tested system. As we will discuss, this property is not granted by the knowledge of $\scriptD(t_{\rm f})$, 
 an object associated with the full set $\scriptE$. Since we deal only with ensembles, the individual systems that we wish to consider within the statistical interpretation 
 should be embedded in some subensembles  of $\scriptE$, which should eventually be characterised by a specific outcome. Our strategy will rely on a study of the 
 dynamics of such subensembles under the effect of interactions within the apparatus.
 It will be essential in this respect to note that, within its statistical interpretation, {\it standard quantum mechanics} applies not only to the full ensemble $\scriptE$ 
 of runs, but also to {\it any one of its subensembles} -- even though {\it we are unable to identify a priori which state corresponds to a given subset of physical runs}.

}

\subsection{Formulating the problem: Seeking a physical way out of a mathematical embarrassment}

\hfill{\it ``There must be some way out of here'', said the joker to the thief}

\hfill{ from Bob Dylan's song All Along the Watchtower, re-recorded by Jimi Hendrix}

\label{fin11.1}

\vspace{0.3cm}

\ZeText{

The present subsection aims at introducing in a tutorial scope the specific difficulties encountered when facing the quantum measurement problem in the framework of the 
statistical interpretation. It also presents some ideas that look natural but lead to failures. It mainly addresses students; the readers aware of such questions may jump 
to subsection 11.2.

}

\subsubsection{A physical, but simplistic and circular argument}

\label{fin11.1.1}
\hfill{\it  Une id\'ee simple mais fausse s'impose toujours face \`a une id\'ee juste mais compliqu\'ee\footnote{
A simple but wrong idea always prevails over a right but complex idea}}

\hfill{ Alexis de Tocqueville}

\vspace{3mm}

\ZeText{

As shown by the review of section 2 and by the Curie--Weiss example of section 3, many models of ideal quantum measurements rely on the following ideas.
The apparatus A is a macroscopic system which has several possible stable states $\hat{\cal R}_i$ characterized by the value $A_i$ of the (macroscopic) 
pointer variable. If A is initially set into a metastable state $\hat{\cal R}(0)$, it may spontaneously switch towards one or another state $\hat{\cal R}_i$
after a long time.  In a measurement, this transition is triggered by the coupling with the tested object S, it happens faster,
and it creates correlations such that, if the apparatus reaches the state $\hat{\cal R}_i$, the tested observable $\hat s$ 
takes the value $s_i$. The neat separation between the states $\hat{\cal R}_i$ and their long lifetime, together with the lack of survival
  of ``Schr\"odinger cats'', suggest that each individual process has a unique outcome, characterized by the indication $A_i$ of the pointer and by the value 
  $s_i$ for the observable $\hat s$ of the system S. 

This intuitive argument, based on current macroscopic experimental observation and on standard classical theories of phase transitions, is nevertheless delusive. 
Although its outcome will eventually turn out to be basically correct, it postulates the very conclusion we wish to justify, namely that the apparatus 
reaches in each run one or another among the states $\hat{\cal R}_i$.  This idea is based on a classical type of reasoning applied blindly to subtle properties of quantum ensembles, 
which is known to produce severe mistakes  (prescribed ensemble fallacy)  \cite{molmer,erwi,RudolphSanders2001,Kok2000}.  In order to explain why the indication of the 
apparatus is unique in a single experiment, we ought to analyze quantum measurements by means of rigorous quantum theoretical arguments.

}

\subsubsection{Where does the difficulty lie?}
             \label{section.12.1.1}
\label{fin11.1.2}

\hfill{
{\it After all is said and done,  more is said than done}}

\hfill{Aesop}

      
\vspace{0.3cm} 

\ZeText{

 The most detailed statistical mechanical treatments of ideal measurement models provide the evolution of the density operator
 $\scriptD(t)$ of the compound system S + A, from the initial state

\begin{equation}  
\label{12.1}
\hat{\cal D}(0)=\hat r(0) \otimes\hat{\cal R}(0),\qquad
 \end{equation} 
to the final state

\begin{equation}   \label{12.2} 
\hat{\cal D}(t_{\rm f})=\sum_i p_i\hat{\cal D}_i,
\qquad \hat{\cal D}_i=\hat r_i\otimes\hat{\cal R}_i,
 \qquad p_i={\rm tr}_S\hat r(0)\hat\Pi_i,
 \qquad p_i \hat r_i = \hat\Pi_i  \hat r(0) \hat \Pi_i.
 \end{equation} 
  In the Curie--Weiss model its explicit form is the expression (\ref{CDelemtsfin}), that is,

\BEQ \label{12.3}
\hat{\cal  D}(t_{\rm f})=p_\uparrow \,  |\hspace{-1mm}\uparrow\rangle\langle \uparrow\hspace{-1mm}| \otimes   \hat { \cal R}_{\Uparrow}
+ p_\downarrow\, |\hspace{-1mm}\downarrow\rangle\langle \downarrow\hspace{-1mm}| \otimes {\hat {\cal R}}_{\Downarrow}.
\EEQ
As we wish to interpret this result physically, we recall its nature. The state $\scriptD(t)$ provides a faithful probabilistic account for 
the dynamics of the expectation values of all observables of S + A, for  {\it a large set ${\cal E}$ of runs} of similarly prepared experiments, but {\it nothing more}.
We need, however, to focus on individual runs so as to explain in particular why, at the end of each run, the pointer yields a well-defined indication $A_i$. 
This property agrees with our macroscopic experience and seems trivial, but it is not granted in the quantum framework.
Quantum mechanics is our most fundamental theory, but even a complete solution of its dynamical equations refers only to the statistics of an ensemble ${\cal E}$. 
The description of {\it individual processes} is {\it excluded}  (\S~\ref{fin10.1.3}): As any quantum state, (\ref{12.2}) is irreducibly probabilistic.
In fact, probabilities occur for many other reasons (\S~12.1.2), which have not necessarily a quantum origin.

 The specific form of the expression (\ref{12.2}) for the final state of S + A properly accounts for all the features of ideal measurements that are related 
 the large set $\scriptE$ of runs. Von Neumann's reduction implies that each individual run should end up one of the states $\scriptD_i$, which  exhibits
  in a factorized form the expected complete correlation between the final state $\hat r_i$ of S and that  $\scriptR_i$ of A characterized by the indication $A_i$ of the pointer.
The ensemble $\scriptE$ obtained by putting together these runs should thus be represented by a sum of these blocks  $\scriptD_i$,
weighted by Born's probabilities, in agreement with (11.2). The truncation of the off-diagonal blocks was also needed; as shown in \S~11.2.1, 
the presence of sizeable elements in them would forbid the pointer to give well-defined indications.
 
 Nevertheless, in spite of its suggestive form, the expression (11.2) does not imply all properties of ideal measurements, which require the consideration 
 of individual runs, or at least of subensembles of $\scriptE$. 
  The correlation existing in (\ref{12.2}) means that, {\it if $A_i$ is observed}, S will be described by $\hat r_i$. However, nothing in $\scriptD(t_{\rm f})$ 
  warrants that one can observe some well-defined value of the pointer in an individual run 
  ~\cite{ haake1,haake2,venugopalan,ABNqm2001,SpehnerHaake1,SpehnerHaake2,privman}, so that the standard classical interpretation cannot be
   given to this quantum correlation. Likewise, Born's rule means that a proportion $p_i$ of individual runs end up in the state $\scriptD_i$. 
   The validity of this rule requires $\scriptD(t_{\rm f})$ to have the form (\ref{12.2}); but conversely, as will be discussed in \S~\ref{fin11.1.3}, 
   the sole result (\ref{12.2}) is not sufficient to explain Born's rule which requires the counting of the individual runs tagged by the outcome $A_i$. 
 And of course von Neumann's reduction requires a selection of the runs having produced a given outcome.

 If quantum mechanics were based on the same kind of probabilities as classical physics, it would be obvious to infer statistically the properties of individual 
 systems from the probability distribution governing the statistical ensemble to which they belong.  At first sight, the description of a quantum ensemble by 
 a density operator seems analogous to the description of an ensemble of classical statistical mechanics by a probability density in phase space -- or to 
 the description of some ensemble of events by ordinary probabilities. We must acknowledge, however, a major difference. In ordinary probability theory, 
 one can distinguish exclusive properties, one of which unambiguously occurs for each individual event. When we toss a coin, we get either heads or tails. 
 In contrast, a quantum state is plagued by the impossibility of analyzing it in terms of an exclusive alternative, as demonstrated by the example of an 
 unpolarized spin $\half$ (\S~\ref{fin10.2.3}). We are not allowed to think, in this case, that the spin may lie either in the $+z$ (or the $-z$)  direction, 
 since we might as well have thought that it lay either in the $+x$ (or the $-x$) direction.
 
 This ambiguity of a mixed quantum state may also be illustrated, in the Curie--Weiss model, by considering the final state of the magnet M alone. 
For the ensemble $\scriptE$, it is described
by the density operator $\hat R_{\rm M}(t_{\rm f})=P^{\rm dis}_{\rm M} (\hat m, t_{\rm f})/G(\hat m)$, where the probability distribution 
$P^{\rm dis}_{\rm M} (m, t_{\rm f})$ is strongly peaked around the two values $m = m_{\rm F}$ and $m = -m_{\rm F}$ of the pointer variable $m$, 
with the weights $p_\up$ and $p_\down$. In standard probability theory this would imply that for a single system $m$ takes either the value
 $\mF$ or the value $-\mF$. However, in quantum mechanics, an individual system should be regarded as belonging to some subensemble 
 $\scriptE'_k$ of $\scriptE$. We may imagine, for instance, that this subensemble is described by a pure state $|\psi\rangle$
such that $|\langle m, \eta |\psi\rangle|^2$ 
presents the same two peaks as $P_{\rm M}(m, t_{\rm f})$, where we noted as $|m, \eta\rangle$  the eigenstates of $\hat m$ 
 (the other quantum number $\eta$ takes a number $G(m)$ of values for each $m$). This state lies astride the two ferromagnetic configurations, 
 with coherences, so that the magnetization of the considered individual system cannot have a definite sign.
  From the sole knowledge of $\hat R_{\rm M}(t_{\rm f})$, we cannot infer  the uniqueness of the macroscopic magnetization. 
 
 Thus, albeit both quantum mechanics and classical statistical mechanics can be formulated as theories dealing with statistical ensembles, going to individual 
 systems is automatic in the latter case, but problematic in the former case since it is impossible to characterize unambiguously the 
 subensembles of $\scriptE$. 
   
  }

  \subsubsection{A crucial task: theoretical identification of the subensembles of real runs}
  \label{fin11.1.3}
  
             \hfill{{\it Horresco referens}\footnote{ I shiver while I am telling it}}

      \hfill{Virgil, Aeneid}   
      
\vspace{0.3cm} 

  \ZeText{

Remember first that, when quantum mechanics is used to describe an individual system, the density operator that characterizes its state refers either to a real 
or to a thought ensemble (\S~\ref{fin10.1.3}). If we consider a real set $\scriptE$ of measurement processes, each individual outcome should be embedded 
in a {\it real subset} of $\scriptE$. We are thus led to study the various possible splittings of $\scriptE$ into subensembles.

A superficial examination of the final state (\ref{12.2}) suggests the following argument. In the same way as we may obtain an unpolarized spin state by merging 
two populations of spins separately prepared in the states $|\hspace{-1.2mm}\up\rangle$ and $|\hspace{-1.2mm}\down\rangle$, let us imagine that we have prepared 
many compound systems S + A in the equilibrium states $\scriptD_i$. We build ensembles $\scriptE_i$, each of which contains a proportion $p_i$ of systems 
in the state $\scriptD_i$, merge them into a single one $\scriptE$ and lose track of this construction. The resulting state for the ensemble $\scriptE$ is identified 
with (\ref{12.2}) and all predictions made thereafter about $\scriptE$ will be the same as for the state $\scriptD(t_{\rm f})$ issued from the dynamics of the 
measurement process.  It is  tempting to admit conversely that the set $\scriptE$ of runs of the measurement may be split into subsets $\scriptE_i$, 
each of which being characterized by the state $\scriptD_i$. 
This would be true in ordinary probability theory. If the reasoning were also correct in quantum mechanics, we would have proven
 that each run belongs to one of the subsets $\scriptE_i$, so that it leads S + A
 to one or another among the states $\scriptD_i$ at the time $t_{\rm f}$, and that its outcome is well-defined.

Here as in \S~\ref{fin11.1.1} the above argument is fallacious. Indeed, as stressed in \S~\ref{fin10.2.3}, 
and contrarily to a state in classical statistical mechanics, a mixed state $\scriptD$ can be split in {\it many different incompatible ways} into a weighted sum of 
 density operators which are more informative than $\scriptD$. Here, knowing the sole final state $\scriptD(t_{\rm f})$ for the set $\scriptE$ of runs,  
 we can decompose it not only according to (\ref{12.2}), but  alternatively into one out of many different forms
 
 \BEQ	
 \scriptD(t_{\rm f})=\sum_k \nu_k \scriptD'_k(t_{\rm f}) ,  \qquad (\nu_k\ge0;\quad \sum_k\nu_k=1),
 \label{12.4}
\EEQ
where the set of states $\scriptD'_k(t_{\rm f})$, possibly pure,  differ from the set $\scriptD_i$: The very concept of decomposition is {\it ambiguous}. 
 If we surmise, as we did above above when we regarded $\scriptE$ as the union of (thought) subensembles $\scriptE_i$ described by $\scriptD_i$, 
 that the density operator $\scriptD'_k(t_{\rm f})$ is associated with a (thought) subset $\scriptE'_k$ of $\scriptE$ containing a
fraction  $\nu_k$ of runs of the measurement, we stumble upon a physical contradiction:  The full set $\scriptE$ of runs could be partitioned in different ways, 
so that a given run would belong both to a subset $\scriptE_i$ and to a subset $\scriptE'_k$, but then we could not decide whether its final state is 
$\scriptD_i$ or $\scriptD'_k(t_{\rm f})$, which provide different expectation values. 

While $\scriptD(t_{\rm f})$ is physically meaningful, its various decompositions (\ref{12.2}) and (\ref{12.4}) are purely {\it mathematical properties}
without physical relevance. Unless we succeed to 
identify some physical process that selects one of them, the very fact that they formally exist precludes the task considered here, namely to explain
 the uniqueness of individual measurements, the quantum measurement problem. In the present context, {\it only  the decompositions involving the particular density 
 operators $\scriptD_i$ may be physically meaningful},
i. e., may correspond to the splitting of the real set $\scriptE$ of runs described by $\scriptD(t_{\rm f})$ into actually existing subsets. 
If we wish to remain within standard quantum mechanics we can only identify such a physical decomposition by studying dynamics of subensembles. 
Nothing a priori warrants that the set of states $\scriptD_i$  will then play a privileged role, 
and this specific ambiguity is the form taken here by {\it the quantum measurement problem}.

The above ambiguity is well known in the literature ~\cite{PeresBook,RudolphSanders2001,molmer,Kok2000}. 
Not paying attention to its existence, and then imposing by hand the desired separation into subensembles, was called the ``{\it prescribed ensemble fallacy}'' 
\cite{Kok2000}. The question does not seem to have yet been resolved in the context of proper measurement models, but we attempt to answer it below.

To illustrate the harmfulness of this ambiguity, consider the simple case of a Curie--Weiss model with $N=2$ (subsection 8.1). 
Although it cannot be regarded as an ideal measurement, it will clearly exhibit the present difficulty.
After elimination of the bath, after reduction and under the conditions (8.7), the state of S + M at a time $t_{\rm f}$ 
such that $\tau_{\rm reg}\ll t_{\rm f}\ll \tau_{\rm obs}$ has the form

\BEQ	
\label{12.5}	
\hat D(t_{\rm f})=p_\up \hat D_\up + p_\down \hat D_\down,     
\EEQ
where $\hat D_\up$ is the projection onto the pure state $|\hspace{-1mm}\up,\Uparrow\rangle$  
characterized by the quantum numbers $s_z=1$, $m=1$,  and likewise $\hat D_\down$ the projection onto $|\hspace{-1mm}\down, \Downarrow\rangle$ 
with $s_z=m=-1$. This form suggests that individual runs of the measurement should lead as expected either to the state $|\hspace{-1mm}\up,\Uparrow\rangle$ 
or to the state  $|\hspace{-1mm}\down, \Downarrow\rangle$ with probabilities $p_\up$ and $p_\down$ given by the Born rule. 
However, this conclusion is not granted since we can also decompose $\hat D(t_{\rm f})$ according, for instance, to
 
 \BEQ
 \label{12.6}		
 \hat D(t_{\rm f})=\nu_1 \hat D'_1 + \nu_2 \hat D'_2, 
 \EEQ
 where $\hat D'_1$ is the projection onto 
 $\sqrt{p_\up/\nu_1} \,\cos \alpha \  |\hspace{-1mm}\up,\Uparrow\rangle+ \sqrt{p_\down/\nu_1} \sin \alpha \  |\hspace{-1mm}\down, \Downarrow\rangle$
  and $\hat D'_2$ the projection onto 
   $\sqrt{p_\up/\nu_2} \,\sin \alpha \times \\  |\hspace{-1mm}\up,\Uparrow\rangle-\sqrt{p_\down/\nu_2} \cos \alpha \ |\hspace{-1mm}\down, \Downarrow\rangle$,
  with $\alpha$ arbitrary and $\nu_1=p_\up \cos^2\alpha+p_\down \sin^2\alpha =1 - \nu_2$. Nothing would then prevent the real runs of the measurement to 
  constitute two subensembles described at the final time by $\hat D'_1$ and $\hat D'_2$, respectively; in such a case, neither $m$ nor $s_z$ could take a 
  well-defined value in each run. In spite of the suggestive form of (\ref{12.5}), 
we cannot give any physical interpretation to its separate terms, on account of the existence of an infinity of
alternative formally similar decompositions (\ref{12.6}) with arbitrary angle $\alpha$.

	In order to interpret the results drawn from the solution of models, it is therefore essential to determine not only the state $\scriptD(t)$
 for the full ensemble $\scriptE$ of runs of the measurement, but also the final state of S + A for any real subensemble of runs. 
 Only then may one be able to assign to an individual system, after the end of the process, a density operator more informative than 
 $\scriptD(t_{\rm f})$ and to derive from it the required properties of an ideal measurement. 
 
To this end, one might {\it postulate that a measuring apparatus is a macroscopic device which produces at each run a well-defined value for the pointer 
 variable}, a specific property which allows registration.  (This idea is somewhat reminiscent of Bohr's view that the apparatus is classical.)
 Thus, the apparatus would first be treated as a quantum object so as to determine the solution 
 $\scriptD(t)$ of the Liouville--von Neumann equation for the full ensemble $\scriptE$, and would then be postulated to behave classically so as to determine 
 the states of the subensembles to which the individual runs belong. No contradiction would arise, owing to the reduced form found for $\scriptD(t_{\rm f})$. 
 (This viewpoint differs from that of the quantum--classical models of section 2.2.)
  
Although expedient, such a way of eliminating the ambiguity of the decomposition of $\scriptD(t_{\rm f})$ is unsatisfactory.
It is obviously unjustified in the above $N=2$ case. To really solve the measurement problem, we need to {\it explain the behavior of the 
  apparatus in individual runs} by relying on the sole principles of quantum mechanics, instead of supplementing them with a doubtful postulate. 
  We now show that the task of understanding from quantum dynamics the uniqueness of measurement outcomes is feasible,  at least for sufficiently 
  elaborate models of quantum measurements. In fact, we will prove in the forthcoming subsections that the quantum Curie--Weiss model for a magnetic 
  dot M + B can be modified so as to explain the {\it classical behavior of its ferromagnetic phases}, and hence the full properties of the measurement. 

}


\subsection{The states describing subensembles at the final time}

\label{fin11.2}

\label{section.12.1.2}

\hfill{\it De hond bijt de kat niet\footnote{\label{DogsCats}Dogs do not beget cats}}

\hfill{{\it  Les chiens ne font pas des chats}$^{\ref{DogsCats}}$}

\hfill{Dutch and French sayings}

\vspace{3mm}

\ZeText{

Quantum mechanics in its statistical interpretation does not allow us to deal directly with individual runs of the measurement. However, at least it accounts not only 
for the full ensemble, but also for {\it arbitrary subensembles} of runs. We first exhibit necessary properties that such subensembles should fulfill at the final time (\S~11.2.1), 
 then relate these properties to the ``collectives''  introduced in the frequency interpretation of probabilities (\S~11.2.2). We plan to establish that they are ensured by 
 a quantum relaxation process, relying for illustration on the Curie--Weiss model. We first present a seemingly natural but unsuccessful attempt (\S~11.2.3), 
 in order to show that the required process cannot be implemented before registration is achieved. We then present two alternative solutions. The first one (\S~\ref{fin11.2.3}) is 
 efficient but requires somewhat artificial interactions within the pointer. The second one (\S~\ref{fin11.2.4-5}) is more general and more realistic but less elementary.

}

\subsubsection{Hierarchic structure of physical subensembles}
\label{fin11.2.1}

\hfill{\it Un po\`eme n'est jamais fini, seulement abandonn\'e\footnote{A poem is never finished, just abandoned}}

\hfill{Paul Val\'ery}

\vspace{3mm}

\ZeText{

\myskip{We first note that the question of splitting the state $\scriptD(t)$ of S + A into a weighted sum of states $\scriptD'_k(t)$, each one associated with a 
subensemble $\scriptE'_k$, can be raised only after the truncation has taken place, at a time $\tau_{\split}$  such that $\tau_{\split}\gg\tau_\trunc$ and under
 conditions that exclude recurrences.  This is not a strong constraint, since we anyhow intend to wait until $t_{\rm f}$, the end  of the registration.
Among the various possible states $\scriptD'_k$ that one can imagine to associate with subensembles of $\scriptE$, 
 one should never find at least those which possess at the final time non-zero elements in off-diagonal blocks.
 (These elements only have to cancel out in the sum over $k$ of (\ref{12.4}).) 
 Such a situation is exemplified by (\ref{12.6}) for $\cos2\alpha\neq0$ {\bf I find:  sin 2alpha $\neq$0}, in which case $\hat{\cal D}'_1$ possesses off-diagonal terms 
$|\hspace{-1mm}\uparrow,\Uparrow\rangle\langle\downarrow,\Downarrow\hspace{-1mm}|$ and 
$|\hspace{-1mm}\downarrow,\Downarrow\rangle\langle\uparrow,\Uparrow\hspace{-1mm}|$, and hence contains 
(due to the positivity of $\scriptD'_1$) both the diagonal terms 
$|\hspace{-1mm}\uparrow,\Uparrow\rangle\langle\uparrow,\Uparrow\hspace{-1mm}|$ and 
$|\hspace{-1mm}\downarrow,\Downarrow\rangle\langle\downarrow,\Downarrow\hspace{-1mm}|$.
An individual run of this type could not yield a well-defined macroscopic value for the pointer.    
}

 A model suitable to fully explain an ideal measurement must yield for S + A, at the end of each run, one or another among the states 
 $\scriptD_i$ defined by (\ref{12.2}). We do not have direct access to individual runs, but should regard them as embedded in subensembles. 
 Consider then an arbitrary subensemble of real runs drawn from the full ensemble $\scriptE$ and containing a proportion $q_i$ of individual runs of the type $i$. 
 We expect this subensemble $\scriptE_\sub$ to be described at the end of the measurement process by a density operator of the form

\BEQ\label{12hex}\label{12.7}
 \scriptD_{\rm sub}(t_{\rm f})= \sum_i q_i  \scriptD_i.
\EEQ
The coefficients $q_i$ are non-negative and sum up to 1, but are otherwise arbitrary, depending on the subensemble\footnote{In particular, the state $\scriptD(t_{\rm f})$ describing
 the full ensemble $\scriptE$ has the form (11.7) where the coefficients $q_i$ are replaced by the probabilities $p_i$ of Born's rule. If this ensemble is split into some set 
 of disjoint subensembles, each $p_i$ of $\scriptE$ is a weighted sum of the corresponding coefficients $q_i$ for these subensembles}. 
 They satisfy the following {\it additivity property}. If two disjoint subensembles $\scriptE_\sub^{(1)}$ and $\scriptE_\sub^{(2)}$ containing $N_1$ and $N_2$ elements, 
 respectively, merge so as to produce the subensemble $\scriptE_\sub$ containing $N=N_1+N_2$ elements, the additivity of the corresponding coefficients is expressed by 
 $Nq_i=N_1q_i^{(1)} + N_2 q_i^{(2)}$, with weights proportional to the sizes of the subensembles.

We will refer to the essential property  (\ref{12hex}) as the {\it hierarchic structure of subensembles}. 
It involves two essential features, the occurrence of {\it the same building blocks} $\scriptD_i$ for all subensembles, and {\it the additivity of the coefficients} $q_i$. 
The existence of this common form for all subensembles  is a {\it consistency property}. It is trivially satisfied in ordinary probability theory within the frequency
 interpretation (\S~11.2.2), since there all subensembles are constructed from the same building blocks, 
but it is not granted in quantum mechanics due to the infinity of different ways of splitting the state of $\scriptE$ into elementary components as in 
(\ref{9*}), or into subensembles as in (\ref{9x}). The existence of {\it the hierarchic structure removes this ambiguity} stressed in \S~\ref{fin11.1.3}.
In fact, for an arbitrary decomposition (11.4) of $\scriptD(t_{\rm f})$, the state $\scriptD'_k(t_{\rm f})$ that describes at the final time some subset $\scriptE'_k$ 
of runs has no reason to take the form (11.7). We must therefore prove that the final state of {\it any subensemble of $\scriptE$ has the form} (\ref{12hex}). 

Since we will rely on the analysis of $\scriptE$ into subensembles in order to extrapolate quantum mechanics towards some properties of individual systems, 
we stress here that these subensembles must be {\it completely arbitrary}. Had we extracted from $\scriptE$ only large subensembles with elements selected randomly, 
their coefficients $q_i$ would most often have taken values close to the Born coefficients $p_i$. We want, however, to consider also more exceptional subensembles that involve 
arbitrary coefficients $q_i$, so as to encompass the limiting cases for which one $q_i$ reaches the value 1, a substitute to single systems which are not dealt with directly 
in the statistical interpretation. To take an image, consider a game in which we would be allowed only to toss many coins at a time. Most draws would provide nearly 
as many heads as tails; if however we wish to infer from these experiments that tossing a single coin would yield either heads or tails, we have to acknowledge the
 occurrence of exceptional draws where all coins fall on the same side. Admittedly, this example is improper as it disregards the quantum ambiguity of subensembles, 
 but it may give an idea of the reasoning that we have in mind. 

Truncation is the disappearance of off-diagonal blocks (\S~1.3.2).
Note that the allowed states (11.7) of subensembles are all truncated. Although the state $\scriptD(t_{\rm f})$ of the full ensemble has a truncated form, nothing 
prevents its decompositions (11.4) to involve non-zero elements in off-diagonal blocks. (These elements only have to cancel out in the sum over $k$ of (11.4).) 
 Such a situation is exemplified by (\ref{12.6}) for $\sin2\alpha\neq0$, in which case $\hat{\cal D}'_1$ possesses pairs of off-diagonal terms of the form
$|\hspace{-1mm}\uparrow,\Uparrow\rangle\langle\downarrow,\Downarrow\hspace{-1mm}|$ and 
$|\hspace{-1mm}\downarrow,\Downarrow\rangle\langle\uparrow,\Uparrow\hspace{-1mm}|$.
Due to the positivity of $\scriptD'_1$, the presence of these terms implies the simultaneous occurrence of the two corresponding diagonal terms 
$|\hspace{-1mm}\uparrow,\Uparrow\rangle\langle\uparrow,\Uparrow\hspace{-1mm}|$ and 
$|\hspace{-1mm}\downarrow,\Downarrow\rangle\langle\downarrow,\Downarrow\hspace{-1mm}|$, and hence of both indications of the pointer.
A well-defined indication of the pointer would therefore be unexplainable in such a situation.

\myskip{
 Such a situation is exemplified by (\ref{12.6}) for $\cos2\alpha\neq0$ {\bf I find:  sin 2alpha $\neq$0}, in which case $\hat{\cal D}'_1$ possesses off-diagonal terms 
$|\hspace{-1mm}\uparrow,\Uparrow\rangle\langle\downarrow,\Downarrow\hspace{-1mm}|$ and 
$|\hspace{-1mm}\downarrow,\Downarrow\rangle\langle\uparrow,\Uparrow\hspace{-1mm}|$, and hence contains 
(due to the positivity of $\scriptD'_1$) both the diagonal terms 
$|\hspace{-1mm}\uparrow,\Uparrow\rangle\langle\uparrow,\Uparrow\hspace{-1mm}|$ and 
$|\hspace{-1mm}\downarrow,\Downarrow\rangle\langle\downarrow,\Downarrow\hspace{-1mm}|$.
An individual run of this type could not yield a well-defined macroscopic value for the pointer.    }

Our strategy will again rely on a {\it dynamic analysis}, now not for the whole ensemble as before, but for an arbitrary subensemble.
Consider, at the time $t_\split$, some splitting of $\scriptE$ into subensembles $\scriptE'_k$. We select one of these, denoted as 
$\scriptE_\sub$  and described for $t>t_\split$ by the state $\scriptD_\sub (t)$. Since $\scriptD_\sub (t_\split)$ is issued from a decomposition of 
$\scriptD(t_\split)$ of the type (\ref{12.4}),  it presents some arbitrariness, but is constrained  by the positivity of  $\scriptD(t_\split) -\nu_k \scriptD_\sub(t_\split)$ 
for a sizable value of $\nu_k$. We will then study, at least in the Curie--Weiss model, the Liouville--von Neumann evolution of the state $\scriptD_\sub (t)$, 
starting from the time  $t=t_\split$  at which it was selected, and will prove that it relaxes towards the form (\ref{12hex}) at the final time $t_{\rm f}$.

Actually, we need the hierarchic structure (11.7) to hold for the subensembles of {\it real runs}. We have no means of identifying the decompositions of $\scriptE$ into subsets of real 
runs. However, by considering {\it all possible mathematical splittings} of $\scriptD (t_{\rm split})$, we can ascertain that the entire set of states that we are considering contains the 
states which describe real processes. Thus we do not have to care whether the subensemble $\scriptE_{\rm sub}$ described by $\scriptD_{\rm sub}(t)$ is virtual or real.
We could not decide beforehand whether $\scriptE_\sub$ was real or virtual, but all real subensembles will anyhow be accounted for by this treatment, 
which will therefore yield the desired conclusion.

}

\subsubsection{ Hierarchic structure from the viewpoint of  the frequency interpretation of the probability}


\hfill{\it C'est dans les vieux pots qu'on fait la meilleure soupe\footnote{The best soup is made in the old pots}}  

\hfill{French proverb}

\vspace{3mm}


\ZeText{

The notion of hierarchic structure for subsensembles can be enlightened by comparison with the
basic concepts of the frequency interpretation of probability,
as developed by Venn and von Mises \cite{venn,mises}. This
interpretation appeals to the physicist's intuition \cite{khinchin}, but
its direct usage in physics problems is not frequent (in 1929 when the
review paper \cite{khinchin} was written it was hoped to find wide
applications in physics). Only recently scholars started to use this interpretation 
for elucidating difficult questions of quantum mechanics
\cite{Vervoort,KhrennikovFop02}.

The major point of the frequency interpretation is that the usual notion
of an ensemble ${\cal E}$ is supplemented by two additional requirements,
and then the ensemble becomes a {\it collective} as defined in
\cite{mises}.  

({\it i})  The ensemble (of events characterized by some set of numerical values) allows choosing
specific subensembles, all elements of which have the same numerical value.
Provided that for each such value $x$ one chooses the maximally large
subensemble ${\cal E}$, the probability of $x$ is defined via ${\rm
lim}_{N\to\infty}N_x/N$, where $N_x$ and $N$ are, respectively, the
number of elements in ${\cal E}_x$ and ${\cal E}$.  
The limit is demanded to be unique. 

({\it ii}) Assuming that the elements $\sigma_k$ of ${\cal E}$ are
indexed, $k=1,2...N$, consider a set of integers $\phi(k)$, where the
function $\phi(k)$ is strictly increasing,i.e., $\phi(k_1)<\phi(k_2)$
for $k_1<k_2$. We stress that $\phi$ does not depend on the value of
$\sigma$, but it only depends on its index $k$. Select the elements
${\cal E}_{\phi(k)}$ so as to build a subensemble ${\cal E}[\phi]$ of
${\cal E}$. If for or instance, $\phi(k)=2k-1$, we select the elements
with odd indices.  For $N\to \infty$, one now demands that for all such
$\phi(...)$, ${\cal E}[\phi]$ produces the same probabilities as ${\cal
E}$. 

The first condition is needed to define probabilities, the second one
excludes any internal order in the ensemble so as to make it statistical
(or random).  This condition led to an extended criticism of the
frequency approach \cite{khinchin}, but it does capture the basic points
of defining the randomness in practice, e.g. judging on the quality of a
random number generator \cite{compagner}. It is clear that some
condition like ({\it ii}) is needed for any ensemble (not only a
collective) to have a physical meaning. For instance, keeping this
condition in mind, we see again why the aggregates are not proper
statistical ensembles; as instead of ({\it ii}) their construction
introduces correlations between their elements (see \S~10.2.4). 

The fact that within the frequency interpretation, the probability is
always defined with respect to a definite collective allows to avoid
many sophisms of the classical probability theory \cite{mises}. Likewise, it
was recently argued that the message of the violation of the Bell
inequalities in quantum mechanics is related to inapplicability of the
Kolmogorov's model of probability, but can be peacefully accommodated
into the frequency interpretation \cite{KhrennikovFop02}. 

Returning to our immediate purposes, we note that the hierarchic
structure of the subensembles that we wish to establish is a direct
consequence of the first condition on collectives recalled above.
Indeed, the additivity of the coefficients $q_i$, in the sense defined
after (11.7), is the same as the additivity of frequencies ${\rm
lim}_{N\to\infty}N_x/N$. If the frequencies would be non-additive,
one can separate ${\cal E}$ into two
subensembles such that the unique limit ${\rm lim}_{N\to\infty}N_x/N$ on ${\cal
E}$ does not exist. 

Thus the hierarchical structure of ensembles reconciles the logical Bayesian
approach to probabilities with the frequency interpretation. The former,
which underlies the definition of a state as a collection of expectation
values, allows us to speak of probabilities before constructing the full
theory of quantum measurement, while the frequency interpretation will
support the solution of the measurement problem (see section 11.3.1). A
similar bridge between the two interpretations is found in the purely
classical set-up of selecting the non-informative prior probability
distribution, the most controversial aspect of the Bayesian statistics
\cite{jaynes_prior,coolen,skilling} \footnote{The choice of the non-informative prior is straightforward for
a finite event space, where it amounts to the homogeneous probability
(all events have equal probability).  Otherwise, its choice is not
unique and can be controversial if approached formally
\cite{jaynes_prior,coolen,skilling}.}.

}

\subsubsection{Attempt of early truncation}
\label{fin11.2.2}

\hfill{\it No diguis blat fins que no estigui al sac i ben lligat\footnote{\rm Do not say it is wheat until it is in the bag and securely tied}}

\hfill{Catalan proverb}

\vspace{3mm}

\ZeText{

If we take the splitting time after achievement of the truncation ($t_{\rm split}\gg \tau_{\rm trunc}$), under conditions that exclude recurrences, we are at least ensured that 
all elements in off-diagonal blocks are eliminated from the density matrix $\scriptD(t_{\rm split})$ for the full ensemble $\scriptE$. In order to extend this property to the 
subensembles of $\scriptE$, it is natural to try to approach the problem as in section 5. We thus take a splitting time $t_\split$, satisfying $\tau_\trunc\ll t_\split\ll \tau_\reg$, 
sufficiently short so that $\scriptD(t_{\rm split})$ has the form $\sum_i \hat\Pi_i \hat r(0) \hat\Pi_i\otimes  \scriptR(0)$
issued from $\hat {\cal D}(0)$ by projecting out its off-diagonal blocks; the state $\hat {\cal R}(0)$ of the apparatus has not yet been significantly affected. 
The initial condition $\scriptD_\sub(t_{\rm split})$ for $\scriptD_\sub(t)$ is found from some decomposition of the simple truncated state $\scriptD(t_{\rm split})$.
To follow the fate of the subensemble $\scriptE_\sub$, we have to solve the equations of motion of section 4.
The situation is the same as in section 5, except for the replacement of the initial condition $\scriptD(0)$ by $\scriptD_\sub (t_\split)$. 
In case the truncation mechanisms of sections 5 and 6 are effective, the elements present in the off-diagonal blocks of 
$\scriptD_\sub(t)$ disappear over the short time scale $\tau_\trunc$, as they did for $\scriptD(t)$. The state $\scriptD_\sub(t)$ is thus dynamically 
unstable against truncation. Later on, the diagonal blocks that remain after this relaxation will evolve as in section 7,
and give rise to ferromagnetic states for M,  so that  $\scriptD_\sub(t_{\rm f})$ will eventually reach the form (\ref{12hex}).

Unfortunately, the truncation mechanism based on the coupling between S and M is not efficient for all possible initial states $\scriptD_\sub(t_{\split})$. 
We have seen in section 5 that truncation requires a sufficient width in the initial paramagnetic probability 
distribution $P_{\rm M}(m, 0)$ of the pointer variable, and that it may fail for ``squeezed'' initial states of M (\S~5.2.3). While the full state $\scriptD(t_{\rm split})$ 
involves a width of order $1/\sqrt{N}$ for $P_{\rm M}(m, 0)$, this property is not necessarily satisfied by  $\scriptD_{\rm sub}(t_{\rm split})$, 
which is constrained only by the positivity of $\scriptD(t_{\rm split})-\nu_k\scriptD_{\rm sub}(t_{\rm split})$ for a sizeable $\nu_k$ (\S 11.2.1). 
We thus fail to prove in the present approach that $\scriptD(t)$ finally reaches the form  (\ref{12hex}) for an arbitrary subensemble $\scriptE_{\rm sub}$.

One reason for this failure lies in the weakness on the constraint set upon  $\scriptD_{\rm sub}(t_{\rm split})$ by $\scriptD(t_{\rm split})$ at the time $t_{\rm split}$. 
Taking below a later value for $t_{\rm split}$ will entail more severe constraints on $\scriptD_{\rm sub}(t_{\rm split})$ so that the required relaxation will always take place. 
Moreover, the Curie--Weiss model as it stands was too crude for our present purpose since  only a single variable, the combination $\hat m=(1/N)\sum_n\hat\sigma^{(n)}_z$ 
of the pointer observables, enters its dynamics. The irreversible process that ensures the hierarchic structure 
of the subensembles is more elaborate than the truncation process of $\scriptD(t)$ and requires dynamics involving many variables.
 
}

\subsubsection{Subensemble relaxation of the pointer alone}
\label{fin11.2.3}

\hfill{\it Laat me alleen, alleen met al mijn verdriet\footnote{Leave me alone, alone with all my sorrows}}

\hfill{Lyrics by Gerrit den Braber, music by Giovanni Ullu, sung by Rita Hovink}

\vspace{3mm}

\ZeText{

As we just saw, the relaxation towards  (\ref{12hex}) of the states describing arbitrary subensembles cannot be achieved by the interaction $\hat H_{\rm SA}$, so that we 
have to rely on the Hamiltonian of the {\it apparatus itself}. 
We also noted that the dynamical mechanism responsible for this relaxation cannot work at an early stage.
The form of  (\ref{12hex})  suggests to distinguish the subensembles at a late time $t_{\rm split}$ such that M, after having been triggered by S, has reached for the full
 ensemble $\scriptE$ a {\it  mixture of the two ferromagnetic states}. The time $t_{\rm split}$ at which we imagine splitting $\scriptE$ into subensembles $\scriptE'_k$
  is thus taken {\it at the end of the registration}, just before the time $t_{\rm f}$, so that the new initial state $\scriptD_\sub(t_{\rm split})$ of the considered dynamical 
  process for an arbitrary subensemble $\scriptE_{\rm sub}$ is one element of some decomposition (\ref{12.4}) of $\scriptD(t_\split)\simeq \scriptD(t_{\rm f})$. 
  Note that the irreversibility of the evolution that has led to $\scriptD(t_{\rm split})$ prevents us from identifying the state $\scriptD_\sub(t)$ at earlier stages of 
  the process,  when $m$ has not yet reached $m_{\rm F}$ or $-m_{\rm F}$. For $t>t_\split$, $\scriptD_\sub(t)$ will be found by solving the Liouville--von Neumann equation 
  with the initial condition at $t=t_\split$. As the registration is achieved at the time $t_{\rm split}$, the interaction $\hat H_{\rm SA}$ 
 is then supposed to have been {\it switched off},  so that the apparatus will relax by itself, though its correlations already established with S will be preserved.

The decompositions  (\ref{12.4}) of $\scriptD(t_{\rm f})$ are made simpler if we replace, in the expression (\ref{12.2}),  each {\it canonical} ferromagnetic equilibrium state 
 $\scriptR_i$ by a {\it microcanonical} state\footnote{The proof below is readily extended to our original situation, where (11.8) and (11.9) involve canonical 
 equilibrium states $\hat R_\Uparrow$ and $\hat R_\Downarrow$ of the pointer instead of microcanonical ones. We merely have to imagine that the eigenstates
 $ |m_{\rm F}, \eta\rangle$ of M which occur in (11.10) denote the eigenvectors of $\hat m$ associated with the eigenvalues of $\hat m$ lying in a small interval of width 
 $1/\sqrt{N}$ around $m_{\rm F}$. The index $\eta$ then denotes these various eigenstates. Eq. (11.8) is replaced by a weighted sum over them, and $G$ 
 again denotes their number, now larger than $G(m_{\rm F})$.  However, as $G(m)$ behaves as an exponential of $N$, the two weights have the same order of magnitude 
 for large $N$.  The subsequent developments remain valid \label{prevfootn}};
  this is justified for large $N$. Tracing out the bath, which reduces  $\hat{\cal D}$ to $\hat D$,
 we will therefore consider arbitrary decompositions of the analogue for S+M of the state (\ref{12.3}), that is, of

\BEQ \label{A0}
\hat D(t_{\rm f})=p_\up \,  \hat r_\up 
\otimes  {\hat  R}_{\Uparrow}^{\mu}
+ p_\down \, \hat r_\down 
\otimes {\hat R}_{\Downarrow}^{\mu},
\EEQ
where $\hat r_\up = |\hspace{-1mm}\up\rangle\langle \up\hspace{-1mm}|$ and 
$\hat r_\down = |\hspace{-1mm}\down\rangle\langle \down\hspace{-1mm}|$. The two occurring microcanonical states of M are expressed as 
(with the index $\mu$ for microcanonical)

\BEQ \label{A}
{\hat R}_{\Uparrow}^{\mu}= \frac{1}{G}\sum_\eta |m_{\rm F},\eta\rangle\langle m_{\rm F},\eta |,\qquad 
{\hat R}_{\Downarrow}^{\mu}= \frac{1}{G}\sum_\eta  |\hspace{-0.7mm}-\hspace{-0.7mm}m_{\rm F},\eta\rangle\langle -m_{\rm F},\eta |,
\EEQ
where $|m,\eta\rangle$ denote the eigenstates  $|\sigma^{(1)}_z,\cdots,\sigma^{(n)}_z,\cdots,\sigma^{(N)}_z\rangle$ of $\hat H_{\rm M}$, with 
$m=(1/N)\sum_n \sigma^{(n)}_z$; 
the index $\eta$ takes a number $G(m)$ of values, and the degeneracy $G(m)$ of the levels, expressed by (\ref{deg}), is large as 
an exponential of $N$; for shorthand we have denoted $G(m_{\rm F})=G(-m_{\rm F})$ as $G$.

The density matrix $\hat D(t_\split) \simeq \hat D(\tf)$ associated with $\scriptE$ has no element outside the large eigenspace $m\neq m_{\rm F}$, 
$m\neq -m_{\rm F}$ associated with its vanishing eigenvalue. The same property holds for the density operator $\hat D_\sub(t_\split)$ associated with any 
subensemble $\scriptE_\sub$. More precisely, as (\ref{A0}) is an operator in the $2G$-dimensional space spanned by the 
basis $ |\hspace{-1mm}\up\rangle \otimes |m_{\rm F},\eta\rangle$,  $|\hspace{-1mm}\down\rangle \otimes |\hspace{-0.7mm}-\hspace{-0.7mm}m_{\rm F},\eta\rangle$, 
any density operator $\hat D_\sub(t_{\rm split})$ issued from the decomposition of $\hat D(t_{\rm split})=\hat D(t_{\rm f})$ is a linear combination of projections 
over pure states $|\Psi(t_{\rm split})\rangle$ of the form   \cite{Jaynes,erwi,wotti} 

\BEQ
 |\Psi(t_{\rm split})\rangle =\sum_\eta U_{\up\eta} \, |\hspace{-1mm}\up\rangle \otimes | m_{\rm F},\eta\rangle + \sum_\eta U_{\down\eta}
  \, |\hspace{-1mm}\down\rangle\otimes 
 |\hspace{-0.7mm}-\hspace{-0.7mm}m_{\rm F},\eta\rangle,
 \label{B}
 \EEQ
with arbitrary coefficients $U_{\up\eta}$, $U_{\down\eta}$ normalized as $\sum_\eta(\, |U_{\up\eta}|^2+ |U_{\down\eta}|^2)=1$. 
Having split the ensemble $\scriptE$ into subensembles after achievement of the registration has introduced a strong constraint on the states 
$\hat D _{\rm sub}$, since only the components for which $m=m_{\rm F}$ or $m=-m_{\rm F}$ occur. The 
``cat terms''  $ |\hspace{-1mm}\up\rangle\langle\down\hspace{-1mm}| \otimes  |m_{\rm F},\eta\rangle\langle -m_{\rm F},\eta '|$ in $|\Psi(\tf)\rangle\langle\Psi(\tf)|$, 
 and their hermitean conjugates, describe coherences of S + M, while the diagonal terms include correlations.
	
 Since any $\hat D _{\rm sub}$ is a linear combination of terms $|\Psi(t)\rangle\langle\Psi(t)|$,  our problem amounts to show that 
 $|\Psi(t)\rangle\langle\Psi(t)|$ decays on a time scale $\tau_{\rm sub}$ short compared to $t_{\rm f}$ towards an {\it  incoherent sum of microcanonical distributions},
 according to

\BEQ\label{B1}
  |\Psi(t)\rangle\langle\Psi(t)| \to   q_\up \hat r_\up\otimes  \hat R^\mu_\Uparrow +q_\down \hat r_\down\otimes \hat R^\mu_\Downarrow,\qquad
q_\up=  \sum_\eta |U_{\up\eta}|^2 ,\quad q_\down= \sum_\eta |U_{\down\eta}|^2 .
 \EEQ
 We will term this decay the {\it subensemble relaxation}. It is a generalization to a {\it pair of macroscopic equilibrium states} of the microcanonical relaxation process,
 which was discussed in literature several times and under various assumptions; see Ref. \cite{A8} for an early review and Refs. \cite{A9,A10,A11,A12} for further results. 
 Note that, in each subensemble $\scriptE_\sub$,  the expectation values of the observables of S + M may evolve 
according to (11.11) on the time lapse $\tau_{\rm sub}$; however, they remain constant for the full ensemble since $\hat D (t)$ has already reached its stationary value: 
When the subensembles $\scriptEу_k$ of some decomposition (11.4) of $\scriptE$ are put back together, the time dependences issued from (11.11) compensate one another.

Obviously, our simple Curie--Weiss model as defined in section 3 is inappropriate to produce this desired relaxation. Indeed, all the states 
$|\hspace{-1mm}\up\rangle\otimes |m_{\rm F},\eta\rangle$ 
and  $|\hspace{-1mm}\down\rangle\otimes |-m_{\rm F}, \eta\rangle$ are eigenstates with the same eigenvalue of both the coupling $\hat H_{\rm SM}$ and
 the Ising Hamiltonian $\hat H_{\rm M}$,  so that $\hat H_{\rm SA} + \hat H_{\rm M}$ has no effect on $|\Psi(t)\rangle\langle\Psi(t)|$.
Whether S is still coupled to M or not at the time $t_{\rm split}$ thus makes 
no difference. Moreover, the coupling $\hat H_{\rm MB}$ with the bath was adequate to allow dumping of energy from M to B during the registration, whereas we 
need here transitions between states $|m_{\rm F}, \eta\rangle$ and $|m_{\rm F}, \eta'\rangle$ with equal energies
(or nearly equal energies,  within a margin of order $1/\sqrt{N}$, for canonical equilibrium $^{\ref{prevfootn}}$). We must therefore extend the model, 
by supplementing the original Hamiltonian of subsection 3.2 with weak interactions $\hat V_{\rm M}$ which 
may induce the required transitions among the spins of M without affecting the previous results. 
As these transitions should not modify $m$, the perturbation $\hat V_{\rm M}$ has the form 
$\hat V_{\rm M}=\hat V_\Uparrow+\hat V_\Downarrow$, where $\hat V_\Uparrow$ and $\hat V_\Downarrow$
 act in the subspaces $|m_{\rm F},\eta\rangle$ and $|\hspace{-0.7mm}-\hspace{-0.7mm}m_{\rm F},\eta\rangle$, respectively, so that 
 $\hat V_\Uparrow|-m_{\rm F},\eta\rangle=\hat V_\Downarrow|m_{\rm F},\eta\rangle=0$. 

In order to find explicitly the time dependence of $|\Psi(t)\rangle\langle\Psi(t)|$, we have to specify $\hat V_{\rm M}$.  A simple possibility consists in taking 
$\hat V_\Uparrow$ and $\hat V_\Downarrow$ as {\it random matrices} \cite{Mehta}. 
This procedure does not describe a stochasticity that would be generated by some environment, but is simply founded as usual on Wigner's idea that complicated interactions 
will generate similar properties;  averaging thus appears as a means for deriving such generic results through feasible calculations.
 We shall regard $\hat V_{\rm M}$ as the sum of two independent random Hermitean matrices $\hat V_\Uparrow$ and $\hat V_\Downarrow$ of size $G$, with a weight 
proportional to\footnote{The only constraint on $\hat V_{\rm M}$ being hermiticity, the maximum entropy criterion yields, as least biased choice of probability distribution 
\cite{BalianNuovCim1968}, 
the Gaussian unitary ensemble (11.12), invariant under unitary transformations. Had we constrained $\hat V_{\rm M}$ to be invariant under time reversal, that is, 
to be represented by real symmetric matrices, we would have dealt with the Gaussian orthogonal ensemble, with a probability distribution invariant under 
orthogonal transformations;  the results would have been the same. We will rely in \S \ref{fin11.2.4-5} and in appendix H on another type of random matrices, 
which yields a more standard time dependence for the subensemble relaxation}

\BEQ \label{E}
\exp\left[-\frac{2G}{\Delta^2}\,  \left ({\rm tr} \,\hat V_\Uparrow^2+{\rm tr}\, \hat V_\Downarrow^2\right )\right],
\EEQ
and average  $|\Psi(t)\rangle\langle\Psi(t)|$ with the weight (11.12) over the evolutions generated by the various realizations of $\hat V_{\rm M}$. 
The matrix elements of $\hat V_{\rm M}$ have a very small typical size $\Delta/\sqrt{G}$, where we remind that $G$ is large as an exponential of $N$.
The $G$ energy levels of $\hat H_{\rm M} +\hat V_{\rm M}$ in the subspace $|m_{\rm F},\eta\rangle$ are now no longer degenerate, and, 
taking as origin for the energy $E$ the unperturbed value issued from $\hat H_{\rm M}$,  their density obeys Wigner's semi-circle law 
$(2/\pi\Delta^2) \sqrt{\Delta^2-E^2}$ since $G\gg1$. We do not wish the perturbation $\hat V_{\rm M}$ to spoil the above analysis of the original Curie--Weiss model
which led to $\hat D(t_{\rm f})$;
its effect, measured for large $G$  by the parameter $\Delta$, should therefore be sufficiently weak so as to produce a widening $\Delta$ small compared to the fluctuation 
of the energy in the canonical distribution. Since the fluctuation of $\hat m$ in the latter distribution is of order $1/\sqrt{N}$, we should take, 
according to (\ref{HM=}),

\BEQ \label{F}
 \Delta \ll \sqrt{N}(J_2+J_4).
\EEQ

\newcommand{\tp}{{t'}}

Returning to $|\Psi(t)\rangle\langle\Psi(t)|$, where we set $t=t_\split+\tp$ so as to take $t_\split$ as an origin of the time $\tp$, we notice that the system 
S behaves as a spectator, so that we need only to study, in the space of M, the time dependence of the operators 

 \BEA  
 \hat X_\Uparrow^ {\,\eta \eta'}(\tp)&\equiv& 
 \overline{\exp(-i\hat V_\Uparrow \tp  /\hbar) |m_{\rm F},\eta\rangle\langle m_{\rm F},\eta'|\exp(i\hat V_\Uparrow \tp /\hbar)},
    \label{C} \\
   \hat Y^{\,\eta\eta'}(\tp ) &\equiv& 
 \overline{\exp(-i\hat V_\Uparrow \tp  /\hbar) |m_{\rm F},\eta\rangle\langle -m_{\rm F},\eta'|\exp(i\hat V_\Downarrow \tp /\hbar)},
  \label{D}
\EEA
and of the operators $\hat X_\Downarrow^{\,\eta\eta'}(\tp )$ and $\hat Y^{\,\eta'\eta}(\tp )^\dagger$ obtained by interchanging $\Uparrow$ and $\Downarrow$. 
Because $\hat V_\Uparrow$ and $\hat V_\Downarrow$ are statistically independent, 
the evaluation of $\hat Y^{\eta\eta'}(\tp )$ simply involves the separate averages of $\exp(- i\hat V_\Uparrow \tp /\hbar)$ and of $\exp( i\hat V_\Downarrow \tp /\hbar)$,
 which for symmetry reasons are proportional to the unit operator. We can therefore evaluate the time dependence of $\hat Y^{\eta \eta'}(\tp )$ through the trace 

\BEQ \label{G}
\phi(\tp ) \equiv
\frac{1}{G} {\rm tr} \, \overline{ \exp(-i\hat V_\Uparrow \tp /\hbar)}=\frac{2}{\pi \Delta^2}
\int_{-\Delta}^\Delta  \d E \,\sqrt{\Delta^2-E^2} \exp(-iE\tp/\hbar)=\frac{2\tau_\sub}{\tp} J_1\left(\frac{\tp }{\tau_\sub}\right),
\EEQ
where we made use of the semi-circle law for the density of eigenvalues recalled above.

This expression exhibits the {\it characteristic time $\tau_\sub$ associated with the relaxation of the subensembles}:  

\BEQ \label{H}
\tau_\sub=\frac{\hbar}{\Delta}.
\EEQ
 Notice that $\tau_\sub$ does not depend on the huge size $G$ of our Hilbert space.
We wish $\tau_\sub$ to be short compared to the registration time $\tau_\reg$ given by (9.9) or (9.10). As $N\gg1$ and $\gamma\ll1$, the condition 
(\ref{F}) permits easily a value of $\Delta$ such that $\tau_\sub\ll \tau_\reg$, i.e., $\sqrt{N}\gg \Delta/J\gg \gamma$. 
From (\ref{D}) and (\ref{G}), we find that $\hat Y^{\,\eta \eta'}(\tp )$ behaves as $\hat Y^{\,\eta \eta'}(\tp )=f_Y(\tp )\hat Y^{\,\eta \eta'}(0)$, where

\BEA
f_Y(\tp )=\phi^2(\tp )= \left(\frac{2\tau_\sub}{\tp }\right)^2J_1^2\left(\frac{\tp }{\tau_\sub}\right)   &\approx &
 \left[1 -\left(\frac{\tp }{2\tau_\sub}\right)^2 \right]  , \hspace{2.2cm} (\tp \ll {\tau_\sub}), \nn\\ &\sim&
 \frac{8}{\pi}  \left(\frac{\tau_\sub}{\tp }\right)^3\sin^2\left( \frac{\tp }{\tau_\sub}-\frac{\pi}{4}\right) , \qquad (\tp\gg {\tau_\sub}).  
 \label{J}
\EEA
Accordingly, the off-diagonal blocks of  $|\Psi(t_{\rm split}+\tp )\rangle\langle\Psi(t_{\rm split}+\tp )|$, which involve both ferromagnetic states $m_{\rm F}$ and $-m_{\rm F}$,
decay for $\tp \gg \tau_\sub$ as Eq. (\ref{J}). It is thus seen that {\it the coherent contributions $\hat Y^{\,\eta \eta'}$ fade out over the short time} $\tau_\sub$.

The time dependence of $f_Y(\tp )$ includes a slow decrease as $1/\tp ^{3}$ and oscillations, unusual features for a physical decay. 
These peculiarities result from the sharp behavior of the level density at $E= \pm\Delta$. We will show in \S~\ref{fin11.2.4-5} how 
how a more familiar exponential decay comes out from more realistic models.

To evaluate $\hat X^ {\eta \eta'}_\Uparrow(\tp )$, we imagine that the two exponentials of (\ref{C}) are expanded in powers of $\hat V_\Uparrow$ 
and that Wick's theorem is used to express the Gaussian average over (\ref{E}) in terms of the averages
$\overline{\hat V_{\eta \eta'} \hat V_{\eta'\!\eta}} = \Delta^2/4G$.  We thus find a diagrammatic expansion \cite{tHooftPlanar,A9}
 for the matrix elements of $\hat X^{\eta\eta'}_\Uparrow(\tp )$  in the basis $|m_{\rm F}, \eta\rangle$.  
Apart from the factor $(-i)^n i^{n'}/n! n'! $ arising from the expansion of the exponentials, each line of a diagram carries a contraction

 \BEQ\label{11.19}
\frac{\tp ^2}{\hbar^2} \hspace{1mm}
\overline{\hat V_{\eta \eta'} \hat V_{\eta' \eta} }
=\frac{\Delta^2 \tp ^2}{4\hbar^2G}=\frac{1}{4G} \left(\frac{\tp }{\tau_\sub}\right)^2,    
\EEQ
and each summation over an internal index $\eta$ brings in a factor $G$. The structure of the contractions (\ref{11.19}) implies that each index must come in a 
right-left pair. Hence, for $\eta\neq\eta'$, the sole non-vanishing matrix element of $\hat X^{\eta\eta'}_{\Uparrow}(\tp )$ is 
$\langle m_{\rm F}, \eta | \hat X^{\eta\eta'}_\Uparrow(\tp )|m_{\rm F}, \eta'\rangle$. Among the contributions to this matrix element, the only diagrams that survive in
 the large-$G$ limit are those which involve as many summations over indices $\eta$ as contractions. This excludes in particular all diagrams containing contractions 
 astride the left and right exponentials of (\ref{C}). The evaluation of $\langle m_{\rm F}, \eta|\hat X^{\eta\eta'}_\Uparrow(\tp )|m_{\rm F}, \eta'\rangle$ 
 thus involves the same factorization as in $\langle m_{\rm F}, \eta | \hat Y^{\eta\eta'}(\tp )|-m_{\rm F}, \eta'\rangle$, and this simply produces the factor $[\phi(\tp )]^2$. 
 We therefore find, for $\eta\neq\eta'$, that $\hat X^{\eta\eta'}_\Uparrow(\tp )=\phi^2(\tp ) \hat X^{\eta\eta'}_\Uparrow(0)$ tends to $0$ just as (\ref{J}).

For $\hat X^{\eta \eta}_\Uparrow(\tp) $, the pairing of indices shows that the sole non-vanishing elements are
$\langle m_{\rm F}, \eta |\hat X^{\eta \eta}_\Uparrow (\tp) |m_{\rm F}, \eta\rangle$, the outcome of which does not depend on $\eta$, and, for $\eta\neq\eta'$,  
$\langle m_{\rm F}, \eta' |\hat X^{\eta \eta}_\Uparrow (\tp) |m_{\rm F}, \eta'\rangle$, 
which depends neither on $\eta$ nor on $\eta'$.  Moreover, according to the definitions (\ref{A}) and (\ref{C}), we note that 
tr $\hat X^{\eta\eta}_\Uparrow(\tp)  = 1$, so that $\hat X^{\eta\eta}_\Uparrow(\tp) $ must have the general form

\BEQ \label{K}
\hat X^ {\eta\eta}_\Uparrow(\tp) = f_X(\tp) \,\hat X^ {\eta\eta}_\Uparrow(0)   + \left[1- f_X(\tp) \right]\, \hat R^\mu_\Uparrow. 
\EEQ
In the large-$G$ limit, the same analysis as for 
$\langle m_{\rm F}, \eta |\hat X^{\eta \eta'}_\Uparrow (\tp) |m_{\rm F}, \eta'\rangle$ holds for 
$\langle m_{\rm F}, \eta |\hat X^{\eta \eta}_\Uparrow (\tp) |m_{\rm F}, \eta\rangle$,
and we find likewise $f_X(\tp) =\phi^2(\tp) $, so that the first term of (\ref{K}) again decays as  (\ref{J}).  (A direct evaluation of 
$\langle m_{\rm F}, \eta' |\hat X^{\eta \eta}_\Uparrow (\tp) |m_{\rm F}, \eta'\rangle$ for $\eta\neq\eta'$,
which contributes to the second term of  (\ref{K}), would be tedious since this quantity, small as $1/G$, involves correlations between the two 
exponentials of (\ref{C}).) Thus, on the time scale $\tau_\sub$, the operators $\hat X^{\eta\eta'}_\Uparrow(\tp) $ fade out for $\eta\neq\eta'$ and 
{\it tend to the microcanonical distribution} for  $\eta=\eta'$.

\vspace{3mm}

Let us resume the above results. Starting from the ensemble $\scriptE$ described by the state (\ref{A0}), we consider
at a time $t_\split$ slightly earlier than $t_{\rm f}$ and such that $t_{\rm f} - t_\split \gg  \tau_\sub$, any (real or virtual) subensemble 
$\scriptE_\sub$ described by a state $\hat D_\sub(t_\split)$ issued from a decomposition of $\hat D(t_\split)\simeq \hat D(t_{\rm f})$. In the present model $\hat D_\sub$ 
evolves according to 

\BEQ\label{11.21}
\hat D_\sub(t_\split+\tp)=\phi^2(\tp)  \hat D_\sub(t_\split) + [1 - \phi^2(\tp) ] \hat D_\split(t_{\rm f}),   
\EEQ
where

\BEQ\label{11.22}
\hat D_\split(t_{\rm f}) = q_\up \hat r_\up \otimes \hat R^\mu_\Uparrow + q_\down \hat r_\down\otimes  \hat R^\mu_\Downarrow,\qquad
q_{\up, \down}={\rm tr}\hspace{1mm} \hat D_\split(t_{\rm split}) \hat r_{\up, \down},    
\EEQ
and where $\phi^2(\tp) $, expressed by (\ref{J}), (\ref{H}), decreases over the very short time scale $\tau_{\rm sub}\ll t_{\rm f}-t_{\rm split}$. 
The state of any subensemble therefore relaxes rapidly to the expected asymptotic form (\ref{11.22}), fulfilling the hierarchic structure at the final 
time $t_{\rm f}$. {\it Truncation and equilibration take place simultaneously}. 

The final result (11.22) shows that S and A remain fully correlated while A evolves. We have thus proven in the present model
 that the surmise (\ref{12hex}) is justified for any subensemble, 
and that the set of subensembles possess at the final time $t_{\rm f}$ the hierarchic structure which removes the quantum ambiguity associated with the splitting 
of the full ensemble of runs. The solution of the measurement problem thus relies on specific properties of the apparatus, especially of its pointer M. We had already 
dwelt on the large number of degrees of freedom of M, needed to let it reach several possible equilibrium states. Now, we wish coherent states astride these 
equilibrium states to decay rapidly, so that the pointer can yield well-separated indications; the present model shows that this is achieved owing 
to the {\it macroscopic size} of M and to a {\it sufficient complexity of the internal interactions} $\hat V_{\rm M}$. 
Moreover these interactions equalize the populations of all levels within each microcanonical equilibrium state.

\vspace{3mm}

Note that the above relaxation is a property  {\it of the magnet alone}, if we deal with {\it broken invariance in the quantum framework}.
So we momentarily disregard the system S and measurements on it.
 Consider the perfectly symmetric process of \S~7.3.2 (fig 7.7) which brings a statistical ensemble $\scriptE$ of magnets M from the paramagnetic state 
 to the quantum mixture $\hat R_{\rm M}(t_{\rm f})=P^{\rm dis}_{\rm M}(\hat m, t_{\rm f})/G(\hat m)$ of both ferromagnetic states. To simplify the discussion we replace 
 the canonical distribution $P^{\rm dis}_{\rm M}(m, t_{\rm f})$ by a microcanonical distribution located at $m_{\rm F}$ and $-m_{\rm F}$. 
 The mixed state $\hat R_{\rm M}(t_{\rm f})$ can be decomposed, as indicated at the end of \S~\ref{fin11.1.2}, into a weighted sum of projections 
 $|\psi\rangle\langle\psi |$ onto pure states (notice that $\Psi$ in (\ref{B}) refers not only to M but also to S, that is absent here)
 
\BEQ \label{L}
 |\psi\rangle = \sum_\eta U_{\up\eta}' |m_{\rm F}, \eta\rangle + \sum_\eta U_{\down\eta}' |-m_{\rm F}, \eta\rangle,    
 \EEQ
each of which describes a subensemble of $\scriptE$ and contains coherent contributions astride $m=m_{\rm F}$ and $m=-m_{\rm F}$. 
 Let us imagine that such a pure state has been prepared at some initial time. Then, in the present model including random interactions, it is {\it dynamically 
 unstable} and decays into $\sum_\eta |U_{\up\eta}' |^2 \hat R^\mu_\Uparrow+\sum_\eta |U_{\down\eta}'|^2 \hat R^\mu_\Downarrow$ on the time scale $\tau_{\rm sub}$. 
 Starting from  $|\psi\rangle\langle\psi|$ we are left after a while with an incoherent superposition of microcanonical equilibrium states of M. 
 Contrary to the initial state, this final situation {\it can be interpreted classically} as describing individual events, in each of which $m$ {\it takes a well-defined value}, 
 either $m_{\rm F}$ or $-m_{\rm F}$. Quantum dynamics thus allows us, at least in the present model, to by-pass the postulate about the apparatus suggested 
 the end of \S~\ref{fin11.1.3}. Quantum magnets (and, more generally, macroscopic quantum systems having several equilibrium states)
 can just relax rapidly into well-defined unique macroscopic states,  and in that sense behave as classical magnets (systems), as one would expect.


}

\subsubsection{Subensemble relaxation in more realistic models}
\label{fin11.2.4-5}

\hfill{\it People who understand physics do not write many formulas}

\hfill{Nikolay Timofeev-Ressovsky quoting Niels Bohr}

\vspace{3mm}

\ZeText{

\newcommand{\tp}{{t'}}

As usual in the applications of the random matrix theory, the use of a random interaction $\hat V_{\rm M}$ 
was justified by the expected similitude of the dynamical effects of most interactions, which allows us to average $|\Psi(t)\rangle\langle\Psi(t) |$ 
over $\hat V_{\rm M}$. Nevertheless, although our choice of a Gaussian randomness (\ref{E}) was mathematically sensible and provided the desired result,
 this choice was artificial. We have noted above that it yields a non-exponential decay (\ref{J}) of $f_Y(\tp )= f_X(\tp )$, which is not satisfactory. 
 In fact, by assuming that all the matrix elements of $\hat V_\Uparrow$ have comparable sizes, we have put all the states $|m_{\rm F}, \eta\rangle$
on an equal footing and disregarded their structure in terms of the spins $\sigma^{(n)}_z$. Such a  $\hat V_\Uparrow$ is rather unphysical, as it produces
transitions from $|m_{\rm F}, \eta\rangle$ to $|m_{\rm F}, \eta'\rangle$ that involve 
flip-flops of many spin pairs, with the same amplitude as transitions that involve a single flip-flop. (The total spin remains unchanged in these dynamics.)

A more realistic model should involve, for instance, as interaction $\hat V_\Uparrow$ 
a sum of terms   $\hat \sigma^{(n)}_-  \hat\sigma^{(n')}_+$, which keep $m$ fixed and produce single flip-flops within the set 
 $|m_{\rm F},\eta\rangle = |\sigma^{(1)}_z,\cdots,\sigma^{(n)}_z,\cdots,\sigma^{(N)}_z\rangle$. The number of significant elements of the $G\times G$ matrix 
 $\hat V_\Uparrow$ is then of order $G$ rather than $G^2$ as for Gaussian ensembles. This idea can be implemented in a workable model by taking for 
$\hat V_\Uparrow$ and $\hat V_\Downarrow$ other types of random matrices. If, for instance, the level density associated with $\hat V_\Uparrow$ 
is Gaussian instead of satisfying the semi-circle law, the relaxation will be exponential.
One possible realization of this exponential relaxation scenario is achieved via a class of random matrices, where the distribution of eigenvalues is factorized 
from that of the eigenvectors. This case corresponds to the homogeneous (Haar's) distribution. The above Gaussian ensemble, with the distribution of
the eigenvalues satisfying the semi-circle law, belongs to this class \cite{risken}.  
Appendix H justifies that if the distribution of the eigenvalues is taken to be Gaussian (independent from the Haar distribution of the eigenvectors), 
the relaxation is indeed exponential.

One can justify the use of random matrices from a different, open-system perspective. 
We have assumed till now that the decay (\ref{11.21}) was due to 
interactions within the spins of M. Alternatively, a concrete physical mechanism involving the bath B can efficiently produce the same decay. Instead of being 
governed by $\hat V_{\rm M}$ as above, the evolution of $|\Psi(t)\rangle$ is now governed by an interaction $\hat V_{\rm MB}$ with the bath. 
In contrast to the spin-boson interaction $\hat H_{\rm MB}$ defined by (\ref{ham4}) which flips the spins of M one by one and which produces the registration, 
this interaction $\hat V_{\rm MB}$ does not affect the energy of M, and thus consists of flip-flops of spin pairs. It gives rise to transitions within the subspaces  
$|m_{\rm F}, \eta\rangle$ or $|\hspace{-0.7mm}-\hspace{-0.7mm}m_{\rm F},\eta\rangle$,  which can be described as a {\it quantum collisional process}. 
Successive brief processes take place within M + B. Each such ``collision'' may be produced by one among the various elements $k$ of the bath, which act 
independently. Its effect on M is thus described in the subspaces $|m_{\rm F}, \eta\rangle$  and $|\hspace{-0.7mm}-\hspace{-0.7mm}m_{\rm F},\eta\rangle$ by 
either one of the unitary transformations $\hat U_\Uparrow^k$ and $\hat U_\Downarrow^k$ associated with the element $k$ of B. It is then fully legitimate to treat 
the effective Hamiltonians for M entering each $\hat U_\Uparrow^k$ and $\hat U_\Downarrow^k$  as random matrices. Their randomness arises here from
 tracing out the bath.

This collisional approach is worked out in Appendix I. It is shown to produce the required decay (\ref{B1}) of  $|\Psi(t)\rangle\langle\Psi(t) |$ through the two effects 
already described in \S~\ref{fin11.2.3}: the {\it disappearance of the coherent contributions} of the marginal density matrix of M, and the {\it microcanonical relaxation}. 
The process is rapid, because the collisions produce transitions between kets having the same energy, and the decay is {\it exponential} as expected on physical grounds. 

\vspace{3mm}

 Altogether (table 1), simple models such as the Curie--Weiss model of section 3 can provide, for the full set $\scriptE$ of runs of a measurement issued from the initial 
state (\ref{12.1}), the final state (\ref{12.2}) issued from two relaxation processes, the {\it truncation}, and the {\it registration} which fully correlates the system and the 
pointer.  However, ideal measurements require a property, less easy to ensure, the {\it hierarchic structure} of the subensembles of $\scriptE$,  expressed by the special
 form (\ref{12hex}) of their states, which are constructed from the same building blocks $\scriptD_i$ as the state (\ref{12.2}) for the full ensemble. We have just seen 
 that more elaborate dynamical models involving suitable interactions within the apparatus must be introduced to establish this property
 for any subensemble, real or not,  issued from a mathematically allowed decomposition of $\scriptE$, and hence for any subensemble of real runs. 
 
 We have no means to identify, among all possible mathematical decompositions, which subensembles are the physical ones and which are their states at the time 
 $t_\split$, but their hierarchic structure is warranted by the above dynamical process. Then, if a physical state having a form different from  (\ref{12hex}) happens to 
 occur for some subensemble at the time $t_\split$ just before the end of the process, it is dynamically unstable and undergoes a {\it new type of rapid relaxation} 
  towards a form  (\ref{12hex}). This  mechanism {\it removes the quantum ambiguity} in the possible decompositions  (\ref{12.4}) of the state $\scriptD(t_{\rm f})$ describing
the full ensemble $\scriptE$: all subensembles of real runs will at the final time $t_{\rm f}$ be described by states of the form (\ref{12hex}), 
{\it the only physical ones at the end of the process}.

  \renewcommand{\thetable}{\arabic{table}}

\begin{table}[h]
\begin{center}     
\noindent     
\begin{tabular}{||l|l|l|l|l||}
\hline 
\hline 
 Step  & Result  & Time scale & Parameter & Mechanism(s)  $ \begin{array}{ccc@{\ }l} \textrm{} \\ \textrm{}\end{array}$   \\ \hline  \hline 
Preparation & Metastable apparatus&   $\tau_{\rm para}$    &  $\gamma$, $T$  & 
$ \begin{array}{ccc@{\ }l} \textrm{Cooling of bath} \\ \textrm{ or RF on magnet}\end{array}$
\\ \hline  
Initial truncation &  $ \begin{array}{ccc@{\ }l} \textrm{Decay of} \\ \textrm{off-diagonal blocks}\end{array}$ & $\tau_{\rm trunc}$   & $g$ & Dephasing\\\hline  
{Irreversibility of truncation} &  No recurrence &  $ \begin{array}{ccc@{\ }l} \tau^{\rm M}_{\rm irrev} \\  \, \\\tau^{\rm B}_{\rm irrev} \end{array}$
& $ \begin{array}{ccc@{\ }r} \delta g\\ \, \\ \gamma\cdot T\end{array}$ &$ \begin{array}{ccc@{\ }l} \textrm{Random S$-$M coupling } \\ \, \\
\hspace{-4mm}\textrm{Decoherence                            }\end{array}$ \\ \hline  
Registration & 
$ \begin{array}{ccc@{\ }l} \textrm{S$-$M correlation} \\ \textrm{ in diagonal blocks}\end{array}$
& $\tau_{\rm reg}$  & $\gamma$, $T$, $J$& 
$ \begin{array}{ccc@{\ }l} \textrm{Energy dumping} \\ \textrm{ into bath}\end{array}$ \\\hline  
Subensemble relaxation & Hierarchic structure & $\tau_{\rm sub}$    & $\Delta$ & 
$ \begin{array}{ccc@{\ }l} \textrm{Truncation} \\ \textrm{Equilibration}\end{array}$
 \\\hline  \hline
Reduction&
Gain of information& 
$ \begin{array}{ccc@{\ }l} \textrm{} \\ \textrm{}\end{array}$$<\tau_{\rm ergodic}$
   &  & Selection of outcome \\\hline  
\hline 
\end{tabular}
\end{center}     
\caption{The steps of an ideal measurement in the Curie--Weiss model. The preparation (\S 3.3.3 and \S 7.3.2) brings the magnetic dot into its metastable paramagnetic state. 
 The truncation eliminates the off-diagonal blocks of the density matrix of S + A describing the full set of measurements. It is governed initially (section 5) by the coupling $g$ 
 between S and M, and it becomes permanent later on owing to two alternative mechanisms, either randomness of the S--M coupling (subsection 6.1) or bath-induced 
 decoherence (subsection 6.2).  The registration (section 7), defined as the establishment of correlations in the diagonal blocks  between the system and the pointer for
  the full set of runs, accompanies the transition of the dot into one of its stable ferromagnetic states, depending 
 on the diagonal sector of the density matrix. The time scales (subsection 9.3) satisfy $\tau_{\rm trunc}^{\rm M;B} \ll\tau_{\rm irrev} \ll \tau_{\rm reg}$. 
 In contrast with the previous steps which refer to statistical properties of the full ensemble of runs of the measurement, the establishment of the 
 hierarchic structure refers to the dynamics of states associated with arbitrary subensembles of runs (\S~\ref{fin11.2.3} and \S~\ref{fin11.2.4-5}). It is governed by a specific type of 
 relaxation, and its time scale is very short, $\tau_{\rm sub}\ll \tau_{\rm reg}$. The resulting hierarchic structure entails the production of a well 
 defined outcome for each individual run.
 The last step, the reduction of the state of S+A (\S~11.3.1 and  \S~11.3.2), does not involve dynamics but consists in the selection of the indication of the pointer. It allows
 reading, printing or processing the result. (This should be done not too late, before the stable indication of the pointer is 
 finally erased due to thermal fluctuations, see \S~7.3.5.)}
\end{table} 


}

\subsection{Emergence of uniqueness and of classical features in measurements}
\label{fin11.3}

 \hfill{{\it Luctor et emergo}\footnote{I fight and emerge}}

\hfill{{\it  Fluctuat nec mergitur}\footnote{She is agitated by the stream, but does not sink}}

\hfill{Devices of the often flooded Dutch province Zeeland and of the city of Paris}

\vspace{3mm}

\ZeText{

\myskip{We are now in position to tackle the quantum measurement problem within the statistical interpretation of quantum mechanics. 
The dynamical establishment of the hierarchic structure (\ref{12hex}) for any state  $\scriptD_\sub(t_{\rm f})$ 
that may arise from any decomposition of $\scriptD(t_{\rm f})$, whether virtual as in (\ref{12.4}) or real, allows us to focus upon the 
{\it real subsets of runs} of the measurement. 
As regards these subsets, the large set $\scriptE$ has exactly the same structure as an ordinary, classical statistical ensemble. The quantum ambiguity of 
\S~11.1.3 has disappeared; each term of the decomposition (11.2) has become meaningful since a state $\scriptD_\sub(t_\split)$
 issued from an arbitrary decomposition (11.4) would anyhow relax to a state of the form (12.7). {\bf  (\ref{12hex})} The explanation of all properties of ideal quantum 
 measurements will rely on this {\it hierarchic structure} which results from the {\it elimination of the quantum ambiguity}
 and which gives an answer to  the quantum measurement problem.
}

We are now in position to tackle the quantum measurement problem within the statistical interpretation of quantum mechanics. Through a dynamical analysis 
based on the Liouville--von Neumann equation, we  have proved that, for suitable interactions ensuring the subensemble relaxation, the states 
 $\scriptD_\sub$, which describe S + A for all the {\it real sets of runs, reach the hierarchical structure}  (\ref{12hex})  at the time $t_{\rm f}$. 
 We will rely on this essential feature to explain the various properties of ideal measurement, including the uniqueness of outcomes.

}

\subsubsection{Individual processes,  ordinary probabilities and Born rule}
\label{section12.1.3}
\label{fin11.3.1}

\hfill{\it Dubito, ergo cogito. Cogito, ergo sum\footnote{I doubt, therefore I think. I think, therefore I exist. Ren\'e Descartes}}

\hfill{R. Cartesius}

\vspace{3mm}

\ZeText{

The principles of quantum mechanics, as recalled in section 10, apply not only to a large ensemble $\scriptE$ of systems, but {\it also to any subensemble} $\scriptE_\sub$. 
This remark has allowed us, through standard dynamical analyses, not only to find the final state $\scriptD(\tf)=\sum_i p_i \scriptD_i$ of S + A issued from 
$\scriptD(0)$, for the large set $\scriptE$ of runs of the measurement, but also to establish the general form $\scriptD_\sub(\tf) =\sum_i q_i \scriptD_i$ of the final 
states associated with arbitrary subsets $\scriptE_\sub$, with coefficients $q_i$ depending on $\scriptE_\sub$. Although the complete description of an individual run 
lies beyond the scope of the statistical interpretation, we have gathered the largest possible information about the outcome of this individual run, through the 
states $\scriptD_\sub(\tf)$ that describe the {\it statistics of all the possible subensembles $\scriptE_\sub$ in which it may be embedded}. Owing to the macroscopic 
size of the pointer, the rapid subensemble relaxation {\it has thus eliminated the quantum ambiguity of the decompositions} (11.4) of $\scriptD(\tf)$\footnote{Because the
states $\scriptD_i$ are generally not pure, the quantum ambiguity of their decompositions is not removed by the dynamics, 
 especially if this ambiguity occurs at a microscopic level, for instance in case $\hat r_i$ is a mixed state. This remaining ambiguity has no incidence on the solution 
 of the measurement problem, since we only need to find for each $i$ a well-defined value for the indication of the pointer}. 

We then note that all the states $\scriptD_\sub(\tf)$ contain the same building blocks $\scriptD_i$, and that when two disjoint subensembles merge, their coefficients 
$q_i$ are additive in the sense given after Eq. (11.7). This {\it additivity property is the same as for probabilities} in their frequency interpretation (\S~11.2.1 and \S~11.2.2). 
In order to infer from this analogy conclusions about individual systems, as can be done in ordinary probability theory, we supplement the statistical interpretation of 
quantum mechanics (section 10) with the following natural additional principle: If we can ascertain that {\it all possible splittings of a large ensemble $\scriptE$ 
into subensembles give rise to a hierarchic structure, we may regard $\scriptE$ as a ``collective''} in the sense of von Mises (\S~11.2.2). 
In other words, we assume that, if $\scriptE$ is large and is endowed with a hierarchical structure, it possesses physical subensembles that involve arbitrary 
values\footnote{Such an arbitrariness of the coefficients $q_i$ is obvious for the set of {\it mathematically allowed} decompositions of a mixed state $\scriptD$. 
It is exhibited, for instance, by Eq. (11.11), since any state of the form (11.10) can enter (for $p_\up \neq 0$, $p_\down \neq 0$) a decomposition of the state (11.8) 
which describes the full ensemble $\scriptE$. However, nothing warrants that a state such as (11.10) describes a {\it physically meaningful subensemble} of $\scriptE$}
 for the coefficients $q_i$, with $q_i \ge 0$ and $\sum_i q_i=1$. In particular, in agreement with the condition (i) of \S~11.2.2, there exist subensembles 
of $\scriptE$ such that $q_i=1$ for a given $i$ (and $q_{i'}=0$ for $i'\neq i$), and among them a maximal subensemble $\scriptE_i$.
          
  This identification of $\scriptE$ with a von Mises collective entitles us to {\it interpret the subensemble $\scriptE_i$ as the set of individual runs described by the state 
  $\scriptD_i$, and the coefficient $p_i$ in $\scriptD(\tf)=\sum_i p_i \scriptD_i$ as the relative frequency} of occurrence of such runs in the large ensemble $\scriptE$. 
  Likewise, the coefficient $q_i$ appears as {\it the proportion of runs with outcome $\scriptD_i$ in the subset} $\scriptE_\sub$, as in an ordinary probabilistic process
   (\S~11.1.2 and \S~11.2.2). The concept of quantum state, defined as a correspondence between the observables and their expectation values (\S~10.1.4), presents a 
  similitude with the concept of probability distribution, but this similitude is only formal since quantum expectation values cannot be given the same interpretation as 
  in classical probability theory. However, the full description, within the purely quantum framework, of ideal measurement processes does produce 
{\it   ordinary probabilities in the frequency interpretation}.

 A natural argument has thus allowed us to infer, from the hierarchical structure of the final states for {\it arbitrary subensembles} of $\scriptE$, that {\it each individual 
 run} has a {\it well-defined outcome}. We have therefore explained, at least in the present model and using the above principle, the phenomenon of {\it reduction}, 
 that is, the production, in each individual run, of {\it one} among the diagonal blocks $\scriptD_i$ of the truncated final density matrix $\scriptD(\tf)=\sum_ip_i\scriptD_i$ of S+A, 
 which describes the whole set $\scriptE$ and arises from $\scriptD(0)$. The possibility of making such a 
statement about individual processes in spite of the irreducibly probabilistic nature of quantum mechanics (in its statistical interpretation) is founded on the special 
dynamics of the apparatus, as shown in \S~\ref{fin11.2.3}.

Solving models provides the values of the probabilities $p_i$ as $p_i={\rm tr}_{\rm S} \hat r(0) \hat\Pi_i$. For a large number of runs, the census of the proportion 
of runs for which the apparatus has provided the outcome $A_i$ thus provides partial information about the initial state $\hat r(0)$ of S.
(In the Curie--Weiss model it yields its diagonal elements.) This fully justifies {\it Born's rule}.

\subsubsection{Reduction and preparations through measurement}
\label{finfin11.3.2}

The solution of the Curie--Weiss model has not only justified the hierarchic structure of the subensembles, but it has also provided the expression of each building block 
$\scriptD_i$: The density operator $\scriptD_i$ that describes the outcome of the runs belonging to the set $\scriptE_i$ is an equilibrium state of S + A, which has the 
{\it factorized form} $\scriptD_i = \hat r_i \otimes \scriptR_i$. The state $\hat r_i$ of S is associated with the eigenvalue $s_i$ of the tested observable $\hat s$, 
while the state $\scriptR_i$ of A is characterized by the value $A_i$ of the order parameter, taken as a pointer variable.

The information needed to partition $\scriptE$ into its subsets $\scriptE_i$ 
is embedded in the indication $A_i$ of the pointer. The uniqueness of this indication has been explained by the subensemble relaxation\footnote {\label{notevenwrong}The 
physical argument of \S~\ref{fin11.1.1} turns out to be ``not even wrong''. It also turns out that we do not need the additional postulate alluded to at of the end of 
\S~\ref{fin11.1.3}, owing to realistic interactions which act within the apparatus at the end of the proces, and which need not play a major role in the truncation and registration}.
The macroscopic size of the pointer then allows observing, storing or processing its outcome. The complete correlations established between S and A by the registration, 
exhibited in $\scriptD(\tf)=\sum p_i\hat r_i\otimes\hat D_i$, entail {\it uniqueness of the outcome} $s_i$ {\it for the tested observable} $\hat s$ of S in each run\footnote {But of course 
there are no well-defined results for observables that do not commute with $\hat s$}.
A {\it filtering of the runs of an ideal measurement}, which are tagged by the indication $A_i$ of the pointer, therefore constitutes a {\it preparation} of the system 
S, performed through the reduction of S + A (as was anticipated in \S~1.1.4) \cite{deMuynck}. Lying initially (for the ensemble ${\cal E}$) in the state $\hat  r(0)$, 
   this system lies, after measurement and selection of the subset ${\cal E}_i$, in the {\it final prepared} state $\hat r_i$.
   In the Curie--Weiss model, this final filtered state   is pure, $\hat r_\up=|\hspace{-0.7mm}\uparrow\rangle\langle\uparrow\hspace{-0.7mm}|$ or  
  $\hat r_\down=|\hspace{-0.7mm}\downarrow\rangle\langle\downarrow\hspace{-0.7mm}|$
  or, shortly, $|\hspace{-1mm}\uparrow\rangle$ or $ |\hspace{-1mm}\downarrow\rangle$. 
   
  In this circumstance, quantum mechanics, although irreducibly probabilistic and dealing with ensembles, {\it can provide certainty about $\hat s$ 
  for an individual system} S after measurement and selection of the indication of the pointer. While answering, within the statistical interpretation, 
  Bohr's modest query ``{\it What can we say  about...}?'' \cite{bohr_dialectica}, an ideal measurement gives a partial answer to Einstein's 
  query ``{\it What is....}?'' \cite{einstein_dialectica}. The solution of models involving not only the interaction of the microscopic object with a macroscopic apparatus 
   but also appropriate interactions within this apparatus thus explains the emergence of 
  a well-defined answer for the system S in a single measurement, a property interpreted as the emergence of a ``physical reality''.
 However, although the outcome of each individual process is unique, it could not have been predicted.
The current statement ``the measurement is responsible for the appearance of the uniqueness of physical reality'' holds only for the considered 
single system and for the tested observable, and {\it only after} measurement with selection of the result. 

\vspace{3mm}

Let us stress that, in agreement with the statistical interpretation of quantum ``states'' (\S 10.1.4), the state assigned to S + A or to S {\it at the end of a single run} 
depends on the ensemble in which this run is embedded, and which is {\it itself conditioned by our information}. Before acknowledging the outcome of 
the process,  we have to regard it as an element of the full set $\scriptE$ of runs issued from the initial state  $\scriptD(0)$ of Eq (\ref{12.1}), and we thus assign
to S + A the state  $\scriptD(t_{\rm f})$ of Eq. (\ref{12.2}) (which involves correlations between S and A). After having read the specific outcome $A_i$, 
we have learnt that the considered single run belongs to the subset $\scriptE_i$ which has emerged from the dynamics, so that we assign 
to S + A the more informative state $\scriptD_i$ (which has the uncorrelated form $\hat r_i\otimes \scriptR_i$). Predictions about experiments performed 
on S after the considered run should therefore be made from the weakly truncated state $\sum_j \hat\Pi_j \hat r(0) \hat\Pi_j$ 
if the result is not read, and from the reduced state $\hat r_i=\hat\Pi_i \hat r(0) \hat\Pi_i/p_i$ if the result $A_i$ has been read off and selected.

Whereas the transformation of the state $\scriptD(0)$ into $\scriptD(t_{\rm f})$ is a {\it real physical process}, 
{\it reduction} from the state $\scriptD(t_{\rm f})$  to the state $\scriptD_i$ has {\it no dynamical meaning}. 
It is simply an {\it updating of our probabilistic description},  allowed by the acquisition of the information $A_i$ which characterizes the new narrower ensemble $\scriptE_i$. 
The state $\scriptD_i$ retains  through $\hat r_i$ some features inherited from the initial state $\scriptD(0)$, but not all due to irreversibility 
of truncation and registration, and it accounts in addition for the knowledge of $A_i$.  Measurement can thus indeed be regarded as {\it information processing}; 
the amounts of information acquired and lost are characterized by the entropy balance of  \S~1.2.4.

}

\subsubsection{Repeatability of ideal measurements}
\label{fin11.3.2}

\hfill{\it  It is a bad plowman that quarrels with his ox}

\hfill{Korean proverb}

\vspace{3mm}

\ZeText{

A property that allows us to approach physical reality within the statistical interpretation is the repeatability of ideal measurements\footnote{It can be shown 
that the sole property of repeatability implies reduction in the weak sense, that is, reduction of the marginal state of S \cite{BalianAJPh1989} }.
Suppose two successive ideal measurements are performed on the same system S, first with an apparatus A, then, 
independently, with a similar apparatus A$'$. The second process does not affect A, and generates for S and A$'$ the same effect 
as the first one, as exhibited by Eq (\ref{12.2}). Hence, the initial state

\BEQ                            
\hat {\cal D}(0)=\hat r(0) \otimes  \hat {\cal R}(0) \otimes  \hat {\cal R}'(0)                   
\EEQ
of S + A + A$'$ becomes at the time $t_{\rm f}$ between the two measurements

\BEQ                            
\hat {\cal D}(t_{\rm f})=\sum_i p_i  \hat r_i \otimes  \hat {\cal R}_i \otimes  \hat {\cal R}'(0), 
\EEQ
and

\BEQ                            
\hat {\cal D}(t'_{\rm f})=\sum_i p_i  \hat r_i \otimes  \hat {\cal R}_i \otimes  \hat {\cal R}_{i}' 
\EEQ
at the final time $t'_{\rm f}$ following the second process. For the whole statistical ensemble ${\cal E}$, a complete correlation is therefore
exhibited {\it between the two pointers}. In an individual process, the second measurement does not affect S. We can even retrodict, from the 
observation of the value $A_i'$ for the pointer of the apparatus A$'$, that S lies in the state $\hat r_i$ not only at the final time 
$t'_{\rm f}$, but already at the time $t_{\rm f}$,  the end of the first measurement.

}

\myskip{
\subsubsection{Classical probabilities}
\label{section.12.1.4}
\label{fin11.3.3}

{\bf Armen's expression}

\hfill{\it De oudjes doen het nog goed $\hspace{-0mm}$\footnote{The oldies are doing well still}}

\hfill{Dutch expression}

\vspace{3mm}
\ZeText{

We have inferred from the hierarchic structure that, for any subensemble $\scriptE_\sub$, the coefficients $q_i$ entering (\ref{12.7}) can be interpreted 
as the relative numbers of runs with outcome $i$ (\S~\ref{fin11.3.1}). For the full ensemble $\scriptE$ of runs, described at the final time by 
$\scriptD(t_{\rm f})$, the weight $p_i={\rm tr}_{\rm S} \hat r(0) \hat \Pi_i$ appears as the proportion of individual runs described by the reduced 
state $\scriptD_i$. This {\it fully justifies Born's rule}.  The {\it census of the indications of the pointer} 
in a large number of runs affords {\it a partial determination of the initial density matrix} $\hat r(0)$ of S (its diagonal elements for the Curie--Weiss model).

The quantity $p_i$ can also be interpreted as the ordinary probability of finding the outcome $A_i$, $s_i$ in each run. From the density operator $\hat r(0)$,
a quantum probabilistic object which has no classical interpretation, the measurement process has thus extracted {\it classical probabilities}. 
Because of the one-by-one selection procedure  of the runs, these probabilities are defined within the {\it frequency interpretation} \cite{X4}. 

}
} 

\subsection{The ingredients of the solution of the measurement problem}
\label{fin11.4}

\hfill{\it Bring vor, was wahr ist;} 

\hfill{\it schreib'  so, da\ss \ es klar ist}

\hfill{\it und verficht's, bis es mit dir gar ist\footnote{Put forward what is true, write it such that it is clear,
and fight for it till it is finished with you}}

\hfill{Ludwig Boltzmann}

\vspace{3mm}

\ZeText{

Altogether, as in statistical mechanics \cite{krylov,Callen,BalianBook}, {\it qualitatively new features} 
emerge in an ideal  measurement process, with a {\it near certainty}.
The explanation of the appearance, within the quantum theory, of properties seemingly in contradiction with
this very theory relies on several ingredients, exhibited by the detailed solution of the Curie--Weiss model.
($i$) The {\it macroscopic size} of the apparatus
allows the pointer to relax towards one or another among some possible values, within weak statistical or quantum fluctuations; 
these outcomes remain unchanged for a long time and can be {\it read} or {\it processed}; 
the choice of the a priori equivalent alternatives is triggered by the tested system. 
($ii$) {\it Statistical considerations} help us to disregard unlikely events.
($iii$) The {\it special dynamics} of the process must produce several effects (Table 1). The {\it truncation}, initiated by the 
interaction between the tested system and the pointer, eliminates the off-diagonal blocks of $\scriptD$ which would prevent any classical interpretation. 
The {\it registration}, too often overlooked in theoretical considerations, which requires a triggering by the system and a dumping of energy towards the bath, 
creates the needed correlations between the system and the pointer.
The registration also lets the apparatus reach, in the state $\scriptD(t)$,  at large $t$, a mixture of the possible final states; this paves the way to the 
process of \S~\ref{fin11.2.3}, where more elaborate but possibly very small interactions within the apparatus ensure that all subensembles reach at the
 final time the {\it hierarchic structure} required for {\it reduction}. This last step,  together with the principle of \S~11.3.1, explains how statements  about individual 
systems and how classical features may emerge from measurement processes in spite of the quantum oddities (\S\S~\ref{fin10.2.1} and \ref{fin10.2.3}) 
associated with the irreducibly probabilistic nature of the theory~\cite{blokhintsev1,blokhintsev2,deMuynck,BallentineBook,BalianAJPh1989,BalianUtrecht}.

As the symmetry breaking for phase transitions, a breaking of unitarity takes place, entailing an apparent violation of the superposition principle 
for S + A\footnote{\label{foot89} As the tested system interacts with the apparatus, it is not an isolated system, so that the breaking of unitarity in its evolution is trivial}.
Here also, there cannot exist any breaking in the strict mathematical sense for a finite apparatus and for finite parameters. 
Nevertheless, this acknowledgement has no physical relevance: the approximations that underlie the effective breaking of unitarity are justified
for the evaluation of physically sensible quantities.

However, the type of emergence that we acknowledge here is more subtle than in statistical mechanics, although both arise from a change of scale. 
In the latter case, emergence bore on {\it phenomena} that have no microscopic equivalent, such as irreversibility,  phase transitions or viscosity. 
In quantum measurements, it bears on {\it concepts}.  Quantum theory, which is fundamentally probabilistic, deals with ensembles, but measurements reveal properties of
{\it individual} systems, a fact that we understand within this very theory. The tested physical quantity, random at the microscopic level, comes out with a well-defined value.
Ordinary probabilities, ordinary correlations, emerge from a non-commutative physics, and thus afford a {\it classical interpretation} for the outcome of the measurement. 
Thus, ideal measurements establish a bridge between the macroscopic scale, with its every day's life features, and the microscopic scale, giving us access to 
microscopic quantities presenting unusual quantum features and impossible to grasp directly\footnote{A more artificial link between microscopic and macroscopic 
scales was established by Bohr \cite{bohr_dialectica} -- see also \cite{landau,petersen1,petersen2} -- by postulating the classical behavior of the measuring apparatus.
Though we consider that the apparatus must be treated as a quantum object, we have noticed (\S~\ref{fin11.2.3}) that quantum dynamics lets 
 the pointer variable reach some classical features }.
In the measurement device we {\it lose track of the non-commutative nature of observables}, which constitutes the deep originality of quantum mechanics and 
which gives rise to its peculiar types of correlations and of probabilities, and we thus recover familiar macroscopic concepts.
(The disappearance of non-commuting observables will be seen to arise directly from the Heisenberg dynamics in \S~\ref{fin13.1.4}.)

  \renewcommand{\thetable}{\arabic{table}}

}

  \renewcommand{\thesection}{\arabic{section}}
\section{Lessons from measurement models}
  \setcounter{equation}{0}\setcounter{figure}{0}\renewcommand{\thesection}{\arabic{section}.}

\label{fin11}
\label{section.9.3}

\hfill{{\it Cette le\c{c}on vaut bien un fromage, sans doute}\footnote{Surely, this lesson is worth a cheese}}

\hfill{Jean de La Fontaine, Le Corbeau et le Renard}

\vspace {0.3cm}
\ZeText{

A microscopic interpretation of the entropy concept has been provided through the elucidation of the irreversibility paradox
\cite{mayer_mayer,krylov,Callen,lindblad_book}.
Likewise, most authors who solve models of quantum measurements (section 2) aim at elucidating the measurement problem so as to get insight on
the interpretation of quantum mechanics. We gather below several ideas put forward in this search, using as an illustration the detailed solution of the
Curie--Weiss model presented above, and we try to draw consequences on the interpretation of quantum physics. These ideas deserve to be taken
into account in future works on measurement models. 

\clearpage

}

\subsection{About the nature of the solutions}
\label{section.12.1}
\label{fin12.1}

\hfill{\it La Nature est un temple o\`u de vivants piliers}

\hfill{\it Laissent parfois sortir de confuses paroles;}

\hfill{\it L'homme y passe \`a travers des for\^ets de symboles}

\hfill{\it Qui l'observent avec des regards familiers\,\footnote{Nature is a temple where living pillars / 
Let sometimes emerge confused words; \\ \indent\indent \hspace{-1.2mm} 
Man passes there through forests of symbols / Which watch him with familiar glances}}

\hfill{	Charles Baudelaire, Les fleurs du mal, Correspondances}
\vspace{3mm}

\ZeText{

The most important conclusion that can be drawn from the solution of models is that one can reach a {\it full understanding of ideal measurements through 
standard quantum statistical mechanics}. Within a {\it minimalist} interpretation of quantum mechanics, the sole use of {\it Hamiltonian dynamics 
is sufficient to explain all the features of ideal measurements}. In particular, {\it uniqueness} of the outcome of each run and {\it reduction} can be derived
from the Hamiltonian dynamics of the macroscopic pointer alone. {\it Unconventional interpretations are not needed}.

}
 
\subsubsection{Approximations are needed}
\label{section.12.1.1}
\label{fin12.1.1}



\hfill{\it Fire could leave ashes behind}

\hfill{Arab proverb}

\vspace{3mm}

\ZeText{

As stressed in \S~1.2.1, a measurement is an {\it irreversible} process, though governed by the {\it reversible} von Neumann equation of motion for the
coupled system S + A. This apparent contradiction cannot be solved with mathematical rigor if the compound system S + A is finite and all its
observables are under explicit control.  As in the solution of the irreversibility paradox (\S~1.2.2), some approximations, justified on
physical grounds, should be introduced \cite{mayer_mayer,krylov,Callen,Weiss,petr}. 
We must accept the approximate nature of theoretical analyses of quantum measurements \cite{Laughlin}. 
 
For instance, when solving the Curie--Weiss model, we were led to neglect some contributions, which strictly speaking do not vanish for a
finite apparatus A = M + B, but which are very small under the conditions of subsection \ref{fin9.4}. For the diagonal blocks 
$\hat{\cal R}_{\uparrow\uparrow}$ and $\hat{\cal R}_{\downarrow\downarrow}$, the situation is the same as for ordinary thermal relaxation processes \cite{Weiss,Gardiner,petr}: 
the invariance under time reversal is broken through the {\it elimination of the bath}  B, performed by keeping only the lowest order terms in $\gamma$ and
by treating the spectrum of B as continuous (section 4).  Correlations within B and between B and M+S are thus disregarded, and an irreversible
nearly exact Fokker--Planck equation \cite{risken} for the marginal operators $\hat{R}_{\uparrow\uparrow}$ and $\hat{R}_{\downarrow\downarrow}$ thus
arises from the exact reversible dynamics. For the off-diagonal blocks $\hat{R}_{\uparrow\downarrow}$ and $\hat{R}_{\uparrow\downarrow}$, {\it correlations} between S 
and a  large number, of order $N$, of spins of M are also {\it discarded} (section 5).  Such correlations are ineffective, except for recurrences;  but these recurrences 
are damped either by a randomness in the coupling between S and M (subsection 6.1) or by the bath (subsection 6.2), at least on accessible time scales.
We will return to this point in \S~\ref{fin12.2.3}. We also showed that, strictly speaking, {\it false or aborted registrations} may occur but that they are  very rare (\S~7.3.4 and \S~7.3.5). 
 
Mathematically rigorous theorems can be proved in statistical mechanics by going to the thermodynamic limit of infinite systems
\cite{EmchBook}. In the Curie--Weiss model, the disappearance of $\hat{R}_{\uparrow\downarrow}$ and $\hat{R}_{\downarrow\uparrow}$ would
become exact in the limit where $ N \to\infty$ first, and then $t\to\infty$.  However, in this limit, we lose track of the time scale $\tau_{\rm trunc}$, which tends to 0. 
Likewise, the weak coupling condition $\gamma\to 0$, needed to justify the elimination of the bath, implies that $\tau_{\rm reg}$ tends to $\infty$.  
Physically sensible time scales are obtained only {\it without limiting process} and at the price of approximations. 
 
}

\subsubsection{Probabilities are omnipresent}
\label{section.11.1.2}

\hfill{\it O Fortuna, imperatrix mundi\footnote{Oh Fortune, empress of the world}}

\hfill{Carmina Burana}

\vspace{0.3cm}
\ZeText{

Although the dynamics of S + A is deterministic, randomness occurs in the solution of measurement models for several reasons.  On the one
hand, quantum physical quantities are blurred due to the {\it non-commutation} of the observables which represent them, so that quantum mechanics is
irreducibly probabilistic (section \ref{fin10} and  \cite{blokhintsev1,blokhintsev2,deMuynck,BallentineBook,BalianAJPh1989,BalianUtrecht}). 
On the other hand, the {\it large size} of the apparatus, needed to ensure registration, does not allow us to describe it at the microscopic scale; 
for instance it lies after registration in a thermal equilibrium (or quasi-equilibrium) state. Thus, both conceptually and technically, we are compelled to
analyse a quantum measurement by relying on the formalism of {\it quantum statistical mechanics}. 

Moreover, as shown in subsection 5.2, some randomness is needed in the {\it initial state} of the apparatus.  Indeed, for some specific initial pure states, 
 the truncation process may fail, in the same way, for instance, as some exceptional initial configurations of a
classical Boltzmann gas with uniform density may produce after some time a configuration with non uniform density. For realistic models of
quantum measurements, which are of rising interest for $q$-bit processing in quantum information theory, {\it experimental noise} and 
{\it random errors} should also be accounted for \cite{leonhardt}. 

Recognizing thus that a quantum measurement is a process of quantum statistical mechanics has led us to privilege the statistical
interpretation of quantum mechanics, in which an assertion is {\it ``certain" if its probability is close to one. }
For instance, the probability of a false registration does not vanish but is small for large $N$ (\S~7.3.3). 
Still, the statistical solution of the quantum measurement problem does not exclude the existence of a hidden variable theory that would describe
individual measurements, the statistics of which would be given by the probabilistic theory, that is, the standard quantum mechanics; see 
\cite{genovese} for a recent review of hidden variable theories.

}

\subsubsection{Time scales}
\label{section.11.1.3}

\hfill{\it De tijd zal het leren\footnote{Time will tell}}

\hfill{Dutch proverb}

\vspace{3mm}

\ZeText{

Understanding a quantum measurement requires mastering the {\it dynamics of the process} during its entire duration \cite{Bell}. This is also
important for experimental purposes, especially in the control of quantum information. Even when the number of parameters is small, a
measurement is a complex process which takes place over several time scales, as exhibited by the solution of the Curie--Weiss model
(subsection \ref{fin9.3}). There, the truncation time turns out to be much shorter than the registration time.  This feature arises from the large number
of degrees of freedom of the pointer M (directly coupled to S) and from the weakness of the interaction between M and B. The large ratio that we
find for $\tau_{\rm reg}/\tau_{\rm trunc}$ allows us to distinguish in the process a rapid disappearance of the off-diagonal blocks of the density matrix of S + A. 
After that,  the registration takes place as if the density matrix of S were diagonal.  The registration times are also not the same for quartic or quadratic
interactions within M. The final subsensemble relaxation of M, that allows reduction (\S~\ref{fin11.2.3}), is also rapid owing to the large size of the pointer.

In the variant of the Curie--Weiss model with $N=2$ (subsection 8.1), the orders of magnitude of the truncation and registration times
are reversed.  A large variety of results have been found in other models for which the dynamics was studied (section 2).
This should encourage one to explore the dynamics of other, more and more realistic models. 
However, it is essential that such models ensure a crucial property, the dynamical establishment of the hierarchic structure of subensembles (\S~\ref{fin11.2.1}).

 }

\subsubsection{May one think in terms of underlying pure states?}
\label{section.11.1.4}
\label{fin12.1.4}

\hfill{{\it Als de geest uit de fles is,}

\hfill{\it  krijg je hem er niet makkelijk weer in}\footnote{ When the genie is out of the bottle, it is not easy to get it in again}}

\hfill{Dutch proverb}

\vspace{0.3cm}

\ZeText{

The solutions of the Curie--Weiss model and of many other models have relied on the use of density operators.  We have argued
(\S~\ref{fin10.2.3}) that, at least in the statistical interpretation, the non uniqueness of the representations (10.3) of mixed states as superpositions of pure states 
makes the existence of such underlying pure states unlikely.  Here again, Ockam's razor works against such representations, which are not unique and are more 
complicated than the framework of quantum statistical mechanics, and which in general would not permit
explicit calculations. Moreover, it is experimentally completely unrealistic to assume that the apparatus has been initially prepared in a pure state.
Nevertheless, although pure states are probabilistic entities, it is not a priori wrong to rely on other interpretations in which they are regarded as more fundamental
 than density operators \cite{vKampen}, and to afford the latter a mere status of technical tools, used to describe both the initial state and the evolution.

We can compare this situation with that of the irreversibility paradox for a gas (\S~1.2.2). In that case, although it is technically simpler
to tackle the problem in the formalism of statistical mechanics, one may equivalently explain the emergence of irreversibility by regarding
 the time-dependent density in phase space as a mathematical object that synthesizes the trajectories and the random initial conditions \cite{krylov,Callen}. 
 The dynamics is then accounted for by Hamilton's equations instead of the Liouville equation, whereas the statistics bears on the initial 
 conditions. (We stressed, however, in \S~\ref{fin10.2.3}, that although density operators and densities in phase space have a similar status, 
 quantum pure states differ conceptually from points in phase space due to their probabilistic nature; see also \cite{krylov,Callen} in this context.)

Likewise, in the Curie--Weiss measurement model, one may theoretically imagine to take as initial state of A a pure state, S being also in a pure state. 
Then at all subsequent times S + A lies in pure states unitarily related to one another.  However it is impossible in any experiment to prepare A = M + B in a pure state. 
What can be done is to prepare M and B in thermal equilibrium states, at a temperature higher than the Curie temperature for M, lower for B. Even if one wishes 
to stick to pure states, one has to explain {\it generic experiments}. As in the classical irreversibility problem, this can be done by weighing the
possible initial pure states of A = M + B as in $\hat{\cal R}(0)$, assuming that M is a typical paramagnetic sample and B a typical sample
of the phonons at temperature $T$.  This statistical description in terms of weighted pure states governed by the Schr\"odinger equation is
technically the same as the above one based on the density operator $\hat{\cal D}(t)$, governed by the Liouville--von Neumann equation, so
that the results obtained above for the full ensemble $\scriptE$ of runs are recovered in a statistical sense for most relevant pure states. 
As regards the expectation values in the ensemble $\scriptE$  of physical quantities (excluding correlations between too many particles), 
the typical final pure states are equivalent  to $\hat{\cal D}(t_{\rm f})$. Very unlikely events will never be observed over reasonable times for most 
of these pure states (contrary to what happens for the squeezed initial states of M considered in \S~5.2.3).

However, it does not seem feasible to transpose to the mere framework of pure states the explanation of reduction given in section 11, 
based on the unambiguous splitting of the mixed state $\scriptD(t_{\rm f})$.
This splitting is needed to identify the real subsets of runs of the measurement, and it has no equivalent in the context of pure states.
 
}

\subsection{About truncation and reduction}
\label{section.11.2}

\hfill{{\it   Le diable est dans les d\'etails\footnote{\label{Duivel}The devil is in the details}}

\hfill{\it De duivel steekt in het detail}$^{\ref{Duivel}}$}

\hfill {French and Dutch proverb} 
\vspace{3mm}

\ZeText{

We have encountered two types of disappearance of off-diagonal blocks of the density matrix of S + A, which should carefully be distinguished. 
On the one hand, the truncation (sections 5 and 6) is the decay of the off-diagonal blocks of the density matrix $\scriptD(t)$, which is issued from the initial 
state $\scriptD(0)$, and which characterizes the statistics of the full set $\scriptE$ of runs of the measurement. On the other hand, the reduction (section 11) 
requires the establishment of the hierarchic structure for all the subsets of runs. There we deal with a decay of the off-diagonal blocks of the density matrix
$\scriptD_{\rm sub}(t)$ associated with {\it  every possible subensemble} $\scriptE_{\rm sub}$ of runs; this second type of decay may be effective only 
{\it at the end of the measurement process}.

}

\subsubsection{The truncation must take place for the compound system S + A}
\label{section.11.2.1}

\hfill{\it Het klopt als een bus\footnote{It really fits}}

\hfill{Dutch expression}

\vspace{3mm}

\ZeText{

In many approaches, starting from von Neumann \cite{vNeumann,Bassi_Ghirardi,daneri} the word ``collapse'' or ``reduction'' is taken in a weak acception, 
{\it referring to} S {\it alone}. Such theoretical analyses involve only a proof that, in a basis that diagonalizes the tested observable, the
off-diagonal blocks of the marginal density matrix $\hat r(t)$ of S fade out, but not necessarily those of the full density matrix $\hat{\cal
D}(t)$ of S + A. In the Curie--Weiss model, this would mean that $\hat r_{\uparrow\downarrow}(t)$ and $\hat{r}_{\downarrow\uparrow}(t)$, or the
expectation values of the $x$- and $y$-components of the spin S, fade out, but that $\hat{R}_{\uparrow\downarrow}(t)$ and
$\hat{R}_{\downarrow\uparrow}(t)$, which characterize the correlations between the pointer M and these components, do not necessarily disappear.

Let us show that the presence of non negligible elements in the off-diagonal blocks of the final state $\scriptD(t_{\rm f})$ of S + A is prohibited for ideal 
measurements. Remember first the distinction between truncation and reduction (\S~1.1.2 and \S~1.3.2). Both terms refer to the compound system S + A, 
but while the truncation is the disappearance of the off-diagonal blocks in the matrix $\scriptD(t_{\rm f})$ that describes the full ensemble $\scriptE$ 
of runs of the measurement, the reduction is the assignment of the final state $\scriptD_i$ to a subset $\scriptE_i$ of $\scriptE$. 
More precisely, once the uniqueness of the outcome of each run is ensured (subsections \ref{fin11.2} and \ref{fin11.3}), one can sort out the runs that have 
produced the specific indication $A_i$ of the pointer. In each such run, the system S lies in the state $\hat r_i$ and the apparatus A in the state $\scriptR_i$, 
hence S + A lies in the state $\scriptD_i$. Born's rule implies that the proportion of runs in $\scriptE_i$ is $p_i$. Collecting back the subsets 
$\scriptE_i$ into $\scriptE$, we find that this full set must be described by the state $\sum_i p_i\scriptD_i$, which is a truncated one. It is therefore essential, 
when solving a model of ideal measurement, to prove the strong truncation property, for S + A, as we did in sections 5 and 6, a prerequisite to the proof of reduction.  
A much more stringent result must thereafter be proven (\S~\ref{fin11.2.1}), the ``hierarchic property'' (\ref{12.7}), 
according to which the state $\scriptD_{\rm sub}$ of S+A must have the form $\sum_i q_i \scriptD_i$ 
for arbitrary, real or virtual, subensembles $\scriptE_{\rm sub}$ of $\scriptE$.

The weak type of truncation is the mere result of disregarding the off-diagonal correlations that exist between S and A.
This procedure of tracing out the apparatus has often been considered as a means of circumventing the existence of ``Schr\"odinger cats''
issued from the superposition principle \cite{Schlosshauer,zurek,Blanchard,walls,walls_book,Braun}.
However, this tracing procedure as such does not have a direct physical meaning  \cite{vKampen,ABNqm2003}.
While satisfactory for the statistical predictions about the final marginal state of S, which has the required form $\sum_i p_i \hat r_i$, the lack of a complete 
truncation for S + A keeps the quantum measurement problem open since the apparatus is left aside. Indeed, the proof of uniqueness of \S~\ref{fin11.2.3} 
takes as a starting point the state $\scriptD(t_{\rm f})$ for $\scriptE$ where truncation and registration have already taken place, and moreover this proof 
involves only the apparatus. Anyhow, tracing out the apparatus eliminates the correlations between the system and the indications of the pointer, which are 
the very essence of a measurement (subsection \ref{fin11.3}). Without them we could not get any information about S. 
This is why the elimination of the apparatus in a model is generally considered as a severe weakness of 
such a model \cite{Bassi_Ghirardi}, that even led to the commandment ``{\it Thou shalt not trace}'' ~\cite{zurek}.

So indeed, theory and practice are fundamentally related. The elimination of the apparatus in the theory of measurements is no less serious 
than its elimination in the experiment!

}


\subsubsection{The truncation is a material phenomenon; the reduction involves both dynamics and ``observers''}

\label{section.11.2.2}
\label{fin12.2.2}

\vspace{0.3cm}





\hfill{\it Weh! Ich ertrag' dich nicht\footnote{Beware, I can't stand you}}

\hfill{Johann Wolfgang von Goethe, Faust, part one}

\vspace{0.3cm}
\ZeText{

The truncation of the density matrix of S + A appears in measurement models as an irreversible change, occurring with a nearly unit probability 
during the dynamical process. It has a material effect on this compound system, modifying its properties as can be checked by subsequent 
measurements. In the Curie--Weiss model, this effect is the disappearance of correlations between the pointer and the components $\hat s_x$ and 
$\hat s_y$ of S. Though described statistically for an ensemble, the joint truncation of S + A thus appears as a purely dynamical, real phenomenon.

The reduction has a more subtle status. It also relies on a {\it dynamical process} governed by the Liouville--von Neumann equation (\S~\ref{fin11.2.3}), 
the {\it subensemble relaxation},
which takes place by the end of the measurement for any subensemble $\scriptE_{\rm sub}$ of $\scriptE$ (whereas the truncation took place earlier and for 
the full ensemble $\scriptE$). Moreover, reduction requires the {\it selection} of the subset $\scriptE_i$ of runs characterized by the value $A_i$ of the pointer variable
(\S~11.3.2). This selection, {\it based on a gain of information about} A, allows the {\it updating of the state} $\scriptD(t_{\rm f})$, which plays the role of a probability 
distribution for the compound system S + A embedded in the ensemble $\scriptE$, into the state $\scriptD_i$ which refers to the subensemble about which we have 
collected information. Subsequent experiments performed on this subensemble will be described by $\scriptD_i$ 
(whereas we should keep $\scriptD(t_{\rm f})$ for experiments performed on the full set $\scriptE$ without sorting).

The idea of an ``observer'', who selects the subset $\scriptE_i$ of systems so as to assign to them the density operator $\scriptD_i$, therefore underlies 
the reduction, as it underlies any assignment of probability. However, ``observation'' is meant here as ``identification and sorting of runs'' through discrimination 
of the outcomes of the pointer. Such a ``reading'' does not require any ``conscious observer''. The ``observer'' who selects the runs $A_i$ will in fact, 
in many experiments, be a macroscopic automatic device triggered by the pointer. An outstanding example is the sophisticated automatic treatment of the 
information gathered by detectors in particle physics, achieved in order to select the extremely rare events of interest.

}

\subsubsection{Physical extinction versus mathematical survival of the off-diagonal sectors}
\label{fin12.2.3}

\hfill{\it I have not failed.}

\hfill{\it I've just found 10,000 ways that won't work}

\hfill{Thomas A. Edison}

\vspace{3mm}

\ZeText{

 Many works on quantum measurement theory stumble over the following paradox. The evolution of the density matrix $\hat{\cal D}(t)$ of the isolated system S + A is unitary. 
 Hence, if $\hat{\cal D}$  is written in a representation where the full Hamiltonian $\hat H$ is diagonal, each of its matrix elements is proportional to a complex exponential 
 $\exp(i \omega t)$ (where $\hbar \omega$ is a difference of eigenvalues of $\hat H$), so that its modulus remains constant in time. 
 In the ideal case where the tested observable $\hat s$ commutes with $\hat H$, we can imagine writing $\hat{\cal D}(t)$ in a common eigenbasis of $\hat s$ and $\hat H$; 
 the moduli of the matrix elements of its off-diagonal block $\hat{\cal R}_{\up\down}(t)$ are therefore independent of time. 
 Such a basis was used in sections 5 and 6.1 where the bath played no r\^ole; in section 6.2, the term $\hat H_{\rm MB}$ does not commute with $\hat s$, 
 and likewise in most other models the full Hamiltonian is not diagonalizable in practice. In such a general case, the moduli of the matrix elements 
 of $\hat{\cal R}_{\up\down}(t)$, in a basis where only $\hat s$ is diagonal, may vary, but we can ascertain that the  {\it norm} 
 Tr $\hat{\cal R}_{\up\down}(t) \hat{\cal R}^\dagger_{\up\down}(t)$ {\it remains invariant}. 
 This {\it mathematically rigorous} property seems in glaring contradiction with the {\it physical}  phenomenon  of truncation, 
 but both are valid statements, the former being undetectable, the latter being important in practice for measurements.
 
In which sense are we then allowed to say that the off-diagonal block $\hat{\cal R}_{\up\down}(t)$ decays? The clue was discussed in \S~6.1.2: 
The physical quantities of interest are weighted sums of matrix elements of $\hat{\cal D}$, or here of its block $\hat{\cal R}_{\up\down}$. 
For instance, the off-diagonal correlations between $\hat s_x$ or $\hat s_y$ and the pointer variable $\hat m$ are embedded in the characteristic function (5.14), 
which reads

\BEQ                  
  \Psi_{\up\down}(\lambda, t)\equiv  \langle\hat s_-  e^{i \lambda \hat m}\rangle  ={\rm Tr}_{\rm A} \hat{\cal R}_{\up\down}(t)  e^{i\lambda \hat m},  \EEQ
where the trace is taken over A = B + M. Likewise, the elimination of the bath B, which is sensible since we cannot control B and have no access to its correlations with M and S, 
produces $\hat R_{\up\down}={\rm Tr}_{\rm B} \hat{\cal R}_{\up\down}$, which contains our whole off-diagonal information, and which is a sum of matrix elements 
of the full density matrix $\hat{\cal D}$. We are therefore interested only in weighted sums of complex exponentials, that is, in {\it almost periodic functions }
(in the sense of Harald Bohr\footnote{The mathematician and olympic champion Harald Bohr, younger brother of Niels Bohr,
 founded the field of almost periodic functions. For a recent discussion of his contributions, see the expository talk 
 ``The football player and the infinite series'' of H. Boas~\cite{HaraldBohr}}).
 For a large apparatus, these sums involve {\it a large number of terms}, which will usually have incommensurable frequencies. 
Depending on the model, their large number reflects the large size of the pointer or that of some environment.
The situation is the same as for a large set of coupled harmonic oscillators ~\cite{ms,caldeira,NplusA,petr,Weiss,Gardiner},
which in practice present damping although some exceptional quantities involving a single mode or a few modes oscillate. In \S~6.1.2 
we have studied a generic situation where the frequencies of the modes are random. 
The random almost periodic function $F(t)$ defined by (6.14) then exhibits a decay over a time scale proportional to $1/\sqrt{N}$;
Poincar\'e {\it recurrences} are not excluded, but occur only after {\it enormous times} ---  not so enormous as for chaotic evolutions but still large as $\exp(\exp N)$.
 
  The above contradiction is therefore apparent. The off-diagonal blocks cannot vanish in a mathematical sense since their norm is constant. 
  However, all quantities of physical interest in the measurement process combine many complex exponentials which interfere destructively, 
  so that everything takes place as if $\hat{\cal R}_{\up\down}$ did vanish at the end of the process. 
  The exact final state of S + A and its reduced final state are thus equivalent with respect to all physically reachable quantities in the sense of Jauch \cite{Jauch}. 
  Admittedly, one may imagine some artificial quantities involving few exponentials; or one may imagine processes with huge durations. 
  But such irrealistic circumstances are not likely to be encountered by experimentalists in a near future, nor even to be recognized if they would occur.
  
Note that the matrix elements of the marginal state $\hat r(t)$ of S, obtained by tracing out the apparatus, are again obtained by summing a very large number of 
matrix elements of $\hat{\cal D}(t)$. We can thus understand that the decay of the off-diagonal elements of $\hat r$  is easier to prove than the truncation of the full state 
of S + A.
 
  }

\subsubsection{The preferred basis issue}

\label{section.11.2.3}

\hfill{\it Lieverkoekjes worden niet gebakken\footnote{``I-prefer-this'' cookies are not baked, i.e., you won't get what you want}}

\hfill{Dutch saying}

\vspace{3mm}

\ZeText{

Realistic models must explain why the truncation does not take place in an arbitrary basis but in the specific basis in which the tested 
observable of S is diagonal. This leads to the question of determining which mechanism selects this basis; intuitively, 
it is the very apparatus that the experimentalist has chosen and, in the Hamiltonian, the form of the interaction 
between S and A. One has, however, to understand precisely for each specific apparatus how the dynamics achieve this property. 
For the Curie--Weiss model and for similar ones, the tested observable is directly coupled through (\ref{HSA}) with the pointer observable $\hat m$, 
and the preferred basis problem is readily solved because the initial truncation is a mere result of the form of this coupling and of the large 
number of degrees of freedom of the pointer M. The finite expectation values 
$\langle\hat s_x\rangle$ and $\langle\hat s_y\rangle$ 
in the initial state of S are thereby transformed into correlations with many spins of the pointer, which eventually vanish (sections 5 and 6). 
Pointer-induced reduction thus takes place, as it should, in the eigenbasis of the tested observable.

We have also shown (\S~6.2.4) that in this model the suppression of the recurrences by the bath, although a {\it decoherence} phenomenon, 
is {\it piloted by the spin-magnet interaction} which selects the decoherence basis.
When it is extended to a microscopic pointer, the Curie--Weiss model itself exhibits the preferred basis difficulty (\S~8.1.5).
In the large $N$ model, the final subensemble relaxation process (subsection 11.2) ensures that the reduction takes place in the same basis as the truncation. 
This basis should therefore have been determined by the dynamics at an earlier stage.
 
In other models, a decoherence generated  by a random environment would have no reason to select this basis
 \cite{Schlosshauer,zurek,Guilini,Blanchard,walls,walls_book,Braun}.  It is therefore essential, in models where truncation and registration are caused 
by some bath or some environment, to show how the interaction $\hat H_{\rm SA}$ determines the basis where these phenomena take place.
}

\subsubsection{Dephasing or bath-induced decoherence?}
\label{section.11.2.4}
\label{fin12.2.4-5}

\ZeText{

We reserve here the word ``decoherence'' to a truncation process generated by a random environment, such as a thermal bath. We have just recalled that, in the 
Curie-Weiss model with large $N$,  the initial truncation is ensured mainly by a dephasing effect, produced by the interaction between the system and the pointer;
 the bath only provides one of the two mechanisms that prevent recurrences from occurring after reduction (subsection 6.2).  
 We have contrasted this direct mechanism with bath-induced decoherence (\S~5.1.2). In particular, our truncation time 
 $\tau_{\rm trunc}$ does not depend on the temperature as does usually a decoherence time, and it is so short that the bath B is not yet effective.
 Later on, the prohibition of recurrences by the bath in this model is a subtle decoherence process, which involves resonance and which implies 
  all three objects, the tested spin,  the magnet and the bath (\S~6.2.4)

  We have shown (\S~5.1.2 and \S~6.1.2) that more general models with macroscopic pointers can also give rise to direct truncation by the pointer. 
  However, in models involving a microscopic pointer (see subsections 2.1, 2.4.1, 2.5 and 8.1), the truncation mechanism can only be a 
  bath-induced decoherence \cite{Schlosshauer,zurek,Guilini,Blanchard,walls,walls_book,Braun}, 
  and the occurrence of a preferred truncation basis is less easy to control.

As regards the subensemble relaxation mechanism, which ensures the hierarchical structure and thus allows reduction (section 11), it may either arise from 
interactions within the pointer itself (\S~\ref{fin11.2.3}), or be induced by the bath (\S~\ref{fin11.2.4-5} and appendix I). Although the latter process includes a kind 
of decoherence or self-decoherence, it presents very specific features associated with the breaking of invariance of the pointer. It involves two sets of levels 
associated with the two possible indications of the pointer, all at nearly the same energy. The coherences astride the two sets of levels rapidly disappear, 
but during the same time lapse, each set also reaches microcanonical ferromagnetic equilibrium.  

}

\subsection{About registration}
\label{section.11.3}

\hfill{\it J'\'evite d'\^etre long, et je deviens obscur\footnote{Avoiding lengths, I become obscure}}

\hfill{Nicolas Boileau, L'Art po\'etique}

\vspace{3mm}

\ZeText{

In order to regard a dynamical process as an ideal measurement, we need it to account for registration, a point too often overlooked. Indeed, we have seen 
(section 11)  that not only truncation but also registration are prerequisites for the establishment of the uniqueness of the outcome in each individual run. 
The mechanism that ensures this property relies on the dynamics of the {\it sole macroscopic apparatus} and on its {\it bistability}; 
it may therefore be effective only after registration. Of course, registration is also our sole access to the microscopic tested system.

}

\subsubsection{The pointer must be macroscopic}
\label{section.11.3.1}

\hfill{\it Iedere keer dat hij het verhaal vertelde, werd de vis groter\footnote{Every time he told the story, the caught fish became bigger} }

\hfill{Dutch expression}

\hspace{3mm}

\ZeText{

  Like  the truncation, the registration is a material process, which affects the apparatus and creates correlations between it and the 
  tested system. This change of A must be {\it detectable}:  We should be able to {\it read, print or process} the results registered by the pointer,
  so that they can be analyzed by ``automatic observers''. 
  In the Curie-Weiss model, the apparatus simulates a magnetic memory, and, under the conditions of subsection 9.4, 
  it  satisfies these properties required for registrations (section 7).  The apparatus is {\it faithful}, since the probability of a wrong registration, 
  in which the distribution $P_{\uparrow\uparrow}(m, \tau_{\rm reg})$ would be sizeable for negative values of $m$, 
  is negligible, though it does not vanish in a mathematical sense. The registration is {\it robust} since both ferromagnetic states 
  represented by density operators yielding magnetizations located around 
  $+m_{\rm F}$ and $-m_{\rm F}$  are stable against weak perturbations, such as the ones needed to read or to process the result. 
 
  The registration is also {\it permanent}. This is an essential feature, not only for experimental purposes but also because the solution of 
  the quantum measurement problem (section 11) requires the state $\scriptD(t_{\rm f})$ to have reached the form (11.2)
  and all the states $\scriptD_{\rm sub}(t_{\rm f})$ the similar form  (\ref{12hex})  for any subensemble. 
   However, this permanence, or rather quasi-permanence, may again be achieved only in a physical sense (\S~11.1.1),
   just as the broken invariance associated with phase transitions  is only displayed at physical times and not at ``truly infinite times''  for finite materials.
   Indeed, in the Curie-Weiss model, thermal fluctuations have some probability to induce in the magnetic dot transitions from one 
   ferromagnetic state to the other. More generally, information may spontaneously be erased after some delay in any finite 
   registration device, but this delay can be extremely long, sufficiently long for our purposes. 
   For our magnetic dot, it behaves as an exponential of $N$ owing to invariance breaking, see Eq. (7.84).
   
      The enhancement of the effect of S on A is ensured by the metastability of the initial state of A, and by the irreversibility of the process, 
      which leads to a stable final state.
         
All these properties require a {\it macroscopic} pointer (\S~1.2.1), and not only a macroscopic apparatus. In principle,
the models involving a large bath but a small pointer are therefore unsatisfactory for the aim of describing ideal measurements. 
In many models of quantum measurement (section 2),
including the Curie-Weiss model for $N = 2$ (subsection 8.1), the number of degrees of freedom of the pointer is not large. We
have discussed this situation, in which an ideal measurement can be achieved, but only if the small pointer is coupled at the
end of the process to a {\it further, macroscopic apparatus} ensuring amplification and true registration of the signal.

 Altogether the macroscopic pointer behaves in its final state as a {\it classical object} which may lie in either one or the other of the
states characterized within negligible fluctuation by the value $A_i$ of the pointer observable $\hat A$. 
(In the Curie-Weiss model, $A_i\simeq\pm \,m_{\rm F}$ is semiclassical, while $s_i=\pm 1$ is quantal).
This crucial point has been established in section 11. Theoretically, nothing prevents us from imagining that the pointer M lies in a quantum state 
including coherences across $m=m_{\rm F}$ and $-m_{\rm F}$. For the full ensemble $\scriptE$ (section 7), such a situation does not occur during 
the slow registration process due to the spin-apparatus interaction which creates complete correlations.  For any subensemble $\scriptE_{\rm sub}$, 
coherences might exist near the end of the process, but according to section 11, they would rapidly disappear, owing to the large size of M and to suitable weak 
interactions within the apparatus (\S~\ref{fin11.2.3} and \S~\ref{fin11.2.4-5}). The correlations between the signs of $s_z$ and of $m$ produced during the registration, and the uniqueness 
property (\S\S~\ref{fin11.3.1} and 11.3.2), then separate the two sectors. The large size of the pointer is therefore essential for a complete solution of the ideal measurement problem.

}

\subsubsection{Does the registration involve observers?}
\label{11.3.2}


 \hfill{\it Hij stond erbij en keek er naar\footnote{He stood there and watched, i.e., he did not attempt to assist}}

\hfill{Dutch saying}

\vspace{3mm}
\ZeText{

We have seen that truncation does not involve observers. Likewise, conscious observers are irrelevant for the registration, which is 
a physical process, governed by a Hamiltonian.  Once the registration of the outcome has taken place, the correlated values of 
$A_i\simeq\pm m_{\rm F}$ and $s_i=\pm1$ take an {\it objective} character, since any observer will read the same well-defined 
indication $A_i$ at each run.  ``Forgetting'' to read off the registered result will not modify it in any way. Anyhow, nothing prevents
 the {\it automatic processing} of the registered data, in view of further experiments on the tested system (\S~\ref{fin12.2.2}). 

We thus cannot agree with the idealist statement that
``the state is a construct of the observer''. Although we interpret the concept of probabilities as a means for making predictions from 
available data (\S~\ref{fin10.1.4}), a state reflects real properties of the physical system acquired through its preparation, 
within some undetermined effects due to the non-commutative nature of the observables.

  }

\subsubsection{What does ``measuring an eigenvalue'' mean?}  
\label{11.3.3}
\ZeText{
 
A measurement process is an experiment which creates in the apparatus an image of some property of the tested system. From a merely experimental viewpoint 
alone, one cannot know the observable of S that is actually tested, but experience as well as theoretical arguments based on the form of the interaction
 Hamiltonian may help to determine which one. From the observed value $A_i$ of the pointer variable, one can then infer the corresponding eigenvalue $s_i$ 
 of the measured operator (that appears in the interaction Hamiltonian), provided the correlation between $A_i$ and $s_i$ is complete (an example 
 of failure is given in \S~7.3.3).  In the Curie-Weiss model the observed quantity is the magnetization of M;  we infer from it the 
 eigenvalue of $\hat s_z$.  The statement  of some textbooks ``only eigenvalues of an operator can be measured''
 refers actually to the pointer values, which are in one-to-one correspondence with the eigenvalues of the tested observable 
 provided the process is an ideal measurement. The eigenvalues of an observable as well as the quantum state of S are 
 abstract mathematical objects associated with a microscopic probabilistic description, whereas the physical measurement 
 that reveals them indirectly relates to the macroscopic pointer variable.

 }

\subsubsection{Did the registered results preexist in the system?}
\label{11.3.4}




\ZeText{

   After the measurement process has taken place and after the outcome of the apparatus has been read, we can assert that the 
   apparatus lies in the state  $\hat{\cal R}_{i}$ characterized by the value $A_i$ of the pointer while the system lies in the final projected state
   $\hat r_i$  (Eq. (\ref{12.2})). We can also determine the weights $p_i$ from the statistics of the various outcomes $A_i$. 
   However a quantum measurement involves not only a change in A that reflects a property of S, but also a change in S (\S~1.1.2). 
   In an ideal measurement the latter change is minimal, but we have to know precisely which parts of the initial state $\hat r(0)$ 
   are conserved during the process so as to extract information about it from the registered data.
   
  Consider first the whole ensemble of runs of the experiment. Together with the theoretical analysis it provides the set of final states 
   $\hat r_i$ and their weights $p_i$. The corresponding marginal density operator $\sum_i p_i \hat r_i$ of S is obtained from $\hat r(0)$ 
   by keeping only the diagonal blocks, the off-diagonal ones being replaced by $0$. We thus find a partial statistical information 
   about the initial state: all probabilistic properties of the tested observable $\hat s$ remain unaffected, as well as those pertaining 
   to observables that commute with $\hat s$. (The amount of information retained is minimal, see \S~1.2.4.) 
   Some retrodiction is thus possible, but it is merely statistical and partial.
   
    Consider now a single run of the measurement, which has provided the result $A_i$. The fact that S is thereafter in the state $\hat r_i$
    with certainty does not mean that it was initially in the same state. In fact no information about the initial state $\hat{\cal D}(0)$ is provided by reading 
    the result $A_i$, except for the fact that the expectation value in $\hat{\cal D}(0)$ of the projection on the corresponding eigenspace of $\hat A$ does not vanish.
    For a spin $\half$, if we have selected at the end of a single run the value $s_z=1$, we can only ascertain that the 
    system was not in the pure 
    state $|\hspace{-1.2mm}\down\rangle$ at the initial time; otherwise its polarization could have been arbitrary. In contrast, a classical measurement 
    may leave the system invariant, in which case we can retrodict from the observation of $A_i$ that the measured quantity took initially the value $s_i$. {\it  For an 
    individual quantum measurement, retrodiction is impossible}, and devoid of physical meaning, due to the probabilistic nature of observables and to the irreversibility of the process. 
    The property ``$\hat s$ takes the value $s_i$" did not preexist the process.
    It is only in case all runs provide the outcome $A_i$  that we can tell that S was originally in the state $\hat r_i$.  One should therefore beware of some 
    realist interpretations in which the value $s_i$ is supposed to preexist the individual measurement\footnote{\label{BellFootnote} {In a hidden variable 
    description that enters discussions of Bell inequalities in the BCHSH setup, one should thus describe the measured variable not as a ``predetermined'' 
    value set only by the pair of particles (Bell's original setup) but as depending on the hidden variables of both the pair and the detector 
    (Bell's extended setup). See Ref. \cite{TheoBell2}  for a discussion of an assumption needed in that setup}}: 
    they do not take properly into account the perturbation brought in by the measurement \cite{deMuynck}.

}

 \subsection{Ideal measurements and interpretation of quantum mechanics}

\label{section.10.2}

\hfill{\it An expert is a man who has made all the mistakes which can be made,}

\hfill{\it  in a narrow field}

\hfill{ Niels Bohr}

\vspace{3mm}

\ZeText{

Quantum measurements throw bridges between the microscopic reality, that we grasp through quantum theory, and the macroscopic reality, 
easier to apprehend directly. The images of the microscopic world that we thus get appear more ``natural'' (i.e., more customary) than the counter-intuitive 
quantum laws, although they emerge from the underlying quantum concepts (subsection \ref{fin11.3}). However, the interpretation of the latter concepts 
is subject to ongoing debate. In particular, as a measurement is a means for gaining information about a physical quantity pertaining to some state of 
a system, the meaning of  ``physical quantity'' and of ``state'' should be made clear.

}

\subsubsection{The statistical interpretation is sufficient to fully explain measurements} 
\label{section.10.2.1}

\hfill{\footnote{ Better to be an ant's head than a lion's tail}}

\vspace{-7mm}

\myskipfigText{
\begin{figure}[h!h!h!]
\label{ArmProv}
\hfill{\includegraphics[width=7cm]{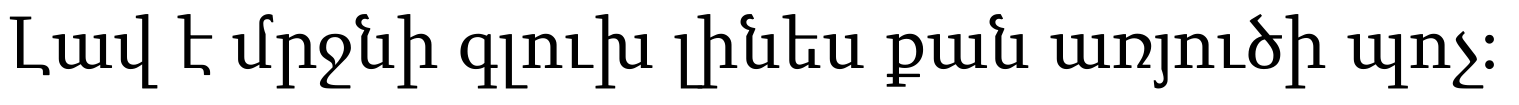}\hspace{2mm}}
\end{figure}
}

\vspace{-5.7mm}

\hfill{Armenian proverb}

\vspace{3mm}

\ZeText{

Many authors treat quantum measurements as irreversible processes of quantum statistical mechanics involving interaction between the tested system 
and a macroscopic apparatus or a macroscopic environment (section 2). The natural tool in such approaches is the density operator of the system S + A, 
which can be regarded as representing a state in the statistical interpretation of quantum mechanics (\S~\ref{fin10.1.4}).
 Implicitly or explicitly, we have relied throughout the present work on this interpretation, resumed in section 10.

A {\it classical} measurement can be regarded as a means to exhibit, through an apparatus A, some {\it pre-existing} property
of an {\it individual} system S. In the statistical interpretation of a quantum measurement, we deal
with the joint evolution of an {\it ensemble} of systems S + A, the outcome
of which indirectly reveals only {\it some probabilistic properties} of the initial state of S
\cite{blokhintsev1,blokhintsev2,deMuynck,BallentineBook,BalianAJPh1989,BalianUtrecht}.  
The ensemble $\scriptE$ considered in sections 4--9 encompasses the set of all possible processes issued from the original preparation; in section 11, 
we considered arbitrary subensembles $\scriptE_{\rm sub}$ of $\scriptE$, just before the final time.{\it  Neither these subensembles nor the  value} $s_i$ 
of the tested observable $\hat s$ inferred from the observation of the indication 
 $A_i$ of the pointer {\it did preexist} the process,  even though we can assert that it is taken by S after an ideal
  measurement where $A_i$ has been registered and selected.

A preliminary step in a measurement model is the assignment to the apparatus at the initial time of a density operator $\scriptR(0)$, namely, in the 
Curie--Weiss model (\S~3.3.2 and \S~3.3.3), a paramagnetic state for M and a thermal equilibrium state for B. The preparation of this initial state is of the 
macroscopic type, involving a control of only few variables such as energy. The {\it assignment of a density operator} is based, according to the statistical 
interpretation,  on probabilistic arguments (\S~10.2.2), in particular on the maximum entropy criterion which underlies the choice of canonical distributions. 
(A preparation of the apparatus through a measurement is excluded, not only because it is macroscopic, but also logically, since the measurement that we wish 
to explain by a model should not depend on a preceding measurement.)

The next stages of the solution, truncation and registration (sections 4 to 7), are mere relaxation processes of quantum statistical mechanics, governed by 
the Liouville--von Neumann equation, which lead the state of S + A from $\scriptD(0)$ to $\scriptD_{\rm exact}(t_{\rm f})$ for the large ensemble 
$\scriptE$ of runs. Approximations justified under the conditions of subsection  9.4 let us replace $\scriptD_{\rm exact}(t_{\rm f})$ by $\scriptD(t_{\rm f})$. 
The breaking of unitarity entailed by this replacement can be understood, in the interpretation of a state as a mapping (10.1) of the observables onto their 
expectation values, as a restriction of this mapping to the ``{\it relevant observables}'' \cite{BalianUtrecht}. Indeed, if we         
  disregard the ``irrelevant'' observables associated with correlations between an inaccessibly large number of particles, 
  which are completely ineffective if no recurrences occur, both states $\hat{\cal D}(t_{\rm f})$ and
  $\hat{\cal D}_{\rm exact}(t_{\rm f})$ realize the same correspondence (10.1) for all other, accessible observables $\hat{O}$.
  The entropy $S[\hat D(t_{\rm f})]$, larger than $S[\hat D_{\rm exact}(t_{\rm f})]=S[\hat D(0)]$, 
  (\S~1.2.4 and \cite{lindblad_book}),
  enters the framework of the general concept 
  of {\it relevant entropies} associated with a reduced description from which irrelevant variables have been eliminated~\cite{BalianUtrecht}.

Within the informational definition (10.1) of states in the statistical interpretation, we may acknowledge a restriction of information to relevant observables when eliminating 
either the environment in models for which this environment induces a decoherence, or the bath B in the Curie--Weiss model (subsection 4.1). In the latter case, 
the states $\scriptD(t)$ and $\hat D(t)\otimes \hat R_{\rm B}(t)$ (Fig. 3.2) should be regarded as equivalent if we disregard the inaccessible observables 
that correlate B with M and S.

Still another equivalence of ``states'' in the sense of (10.1) will be encountered in \S~\ref{fin13.1.5}, where the Curie--Weiss model is reconsidered in the 
Heisenberg picture. There the evolution of most off-diagonal observables lets them vanish at the end of the process, so that they become irrelevant. 
The (time-independent) density matrix and the resulting truncated one are therefore equivalent after the time $t_{\rm f}$, since they carry the same information 
about the only remaining diagonal observables. Note also that, in the statistical interpretation, it is natural to attribute the quantum specificities (\S~\ref{fin10.2.1}) 
to the non commutation of the observables; in the Heisenberg picture, the effective commutation at the time $t_{\rm f}$ of those which govern the measurement
 sheds another light on the emergence of classicality (\S~\ref{fin13.1.4}).
 
 We have stressed that, in the statistical interpretation, a quantum state does not describe an individual system, but an ensemble (\S~\ref{fin10.1.3}). 
 The solution $\scriptD(t)$ of the Liouville von-Neumann equation for S + A describes fully, but in a probabilistic way, a large set $\scriptE$ of runs originated 
 from the initial state $\scriptD(0)$: quantum mechanics treats statistics of processes, not single processes. However, the solution of the quantum measurement 
 problem requires to distinguish, at the end of the process, single runs or at least {\it subensembles} $\scriptE_i$ of $\scriptE$ having yielded the outcome $A_i$ for the 
 pointer.  A measurement is achieved only after reading, collecting, processing or selecting the result of each individual process, so as to interpret its results in
 every day's language \cite{bohr_dialectica}. It is essential to understand how ordinary logic, ordinary probabilities, ordinary correlations, as well as exact statements 
 about individual systems may emerge at our scale from quantum mechanics in measurement processes, even within the statistical interpretation which is foreign 
 to such concepts.  Although $\scriptD(t)$ appears as an adequate tool to account for truncation and registration, it refers to the full set $\scriptE$, and its mere 
 determination is not sufficient to provide information about subsets. The difficulty lies in the {\it quantum ambiguity} of the decomposition of the mixed state 
 $\scriptD(t_{\rm f})$ into states describing subensembles (\S~\ref{fin10.2.3} and \S~\ref{fin11.1.3}). We have achieved the task of understanding ideal measurements 
 in section 11 by relying on a dynamical relaxation mechanism of subensembles, according to which the macroscopic apparatus retains quantum features only over 
 a brief delay. This provides the unambiguous splitting of $\scriptE$ into the required subsets $\scriptE_i$.

}

\subsubsection{Measurement models in other interpretations}
\label{section.10.2.3}
\label{fin12.4.2}

 \hfill{\it Het kan natuurlijk ook anders\footnote{It can of course also be done differently}}

 \hfill{ Dutch expression} \vspace{3mm}

\ZeText{

As shown above, standard quantum mechanics within the statistical interpretation provides a satisfactory explanation 
of all the properties, including odd ones, of quantum measurement processes. Any other interpretation is of course admissible 
insofar as it yields the same probabilistic predictions.
However, the {\it statistical interpretation}, in the present form or in other forms, as well as alternative equivalent interpretations, {\it is minimalistic}.
Since it has been sufficient to explain the crucial problem of measurement, we are led   
to leave aside at least those interpretations which require additional postulates, while keeping the same probabilistic status.

In particular, we can eliminate the variants of the ``orthodox'' {\it Copenhagen interpretation} in which it is postulated that two different
types of evolution may exist, depending on the circumstances, a Hamiltonian evolution if the system is isolated, and a sudden change
producing von Neumann's reduction and Born's rule if the system S undergoes an ideal measurement \cite{vNeumann,krips}. We can rule out the second 
type of evolution, since we have seen in detail (section 11) that the standard Liouville--von Neumann evolution alone, when applied to arbitrary 
subensembles, 
is sufficient to explain the reduction. The apparent violation of the superposition principle is understood as the result 
of suitable interactions within the macroscopic apparatus, together with standard treatments of quantum statistical mechanics.
It is therefore legitimate to abandon the ``postulate of reduction'', in the same way as
the old ``quantum jumps'' have been replaced by transitions governed by quantum electrodynamics. 
It is also superfluous to postulate the uniqueness of the outcome of individual runs (\S~\ref{fin11.1.3}).

Interpretations based on {\it decoherence} by some environment underlie many models (subsection 2.7). The detailed study of section 11 shows, however, that 
a proper explanation of reduction requires a special type of decoherence, which accounts for the bistability (or multistability) of the apparatus (\S~11.2.4).
 Decoherence models in which a special mode of the environment is considered as ``pointer mode''  \cite{zurek,Schlosshauer}
are unrealistic, since, by definition, the environment cannot be manipulated or read off. See also the discussion of this issue in \cite{requardt}. 

Many interpretations are motivated by a wish to describe individual systems, and to get rid of statistical ensembles. The consideration of {\it conscious observers} 
was introduced in this prospect. However, the numerous models based on the S + A dynamics show that a measurement is a {\it real dynamical process},
in which the system undergoes a physical interaction with the apparatus, which modifies both the system and the apparatus, as
 can be shown by performing subsequent experiments. The sole role of the observer (who may be replaced by an automatic device) is to select the outcome 
of the pointer after this process is achieved.

Reduction in an individual measurement process has often been regarded as a kind of bifurcation which may lead the single initial state $\scriptD(0)$ towards 
several possible outcomes $\scriptD_i$, a property seemingly at variance with the linearity of quantum mechanics. In the interpretation of Bohm and de Broglie
 \cite{Bohm,deBroglie}, such a bifurcation occurs naturally. Owing to the introduction of {\it trajectories} piloted by the wave function, a one-to-one 
 correspondence exists between the initial and the final point of each possible trajectory; the initial point is governed by a classical probability law determined by
  the initial quantum wave function, while the set of trajectories end up as separate bunches, each of which is associated with an outcome $i$. Thus, the final subsets
  $\scriptE_i$ reflect pre-existing subsets of $\scriptE$ that already existed at the initial time. In spite of this qualitative explanation of reduction, the trajectories, 
  which refer to the coupled system S + A, are so complicated that models relying on them seem out of reach.
  
  At the other extreme, the reality of collapse is denied in Everett's many-worlds   interpretation 
  ~\cite{Everett,ManyWorlds}.  A measurement is supposed to create several branches in the ``relative state'', one of which only being observed, but no 
  dynamical mechanism has been proposed to explain this branching. 
  
  In our approach the density operator (or the wave function) does not represent a real systems, but our knowledge thereof. Branching does occur, but only at the classical level, 
  by separating a statistical ensemble into subensembles labelled by the outcome of the pointer, as happens when repeatedly throwing a dice. 
  We may call a ``branch'', among the 6 possible ones, the selection of the rolls in which the number 5, for instance, has come up . 
  
The same concern, describing individual quantum processes, has led to a search for рsub-quantum mechanicsс 
\cite{BeyondTheQuantum,deMuynck,CettodelaPenaBook}.  Although new viewpoints on measurements might thus emerge, such drastic changes 
do not seem needed in this context. Justifications should probably be looked for at scales where quantum mechanics would fail, hopefully at length 
scales larger than the Planck scale so as to allow experimental tests.

Of particular interest in the context of measurements are the information-based interpretations \cite{ 
BalianAJPh1989,BalianUtrecht,BrillouinBook,Brillouin,vedral,caves},
 which are related to the statistical interpretation (\S~\ref{fin10.1.3} and \S~\ref{fin10.1.4}). 
Indeed, an apparatus can be regarded as a device which {\it processes information} about the system S, or rather
about the ensemble ${\cal E}$ to which S belongs.  The initial density operator $\hat r(0)$, if given, gathers our information about some
preliminary preparation of S. During the process, which leads ${\cal E}$ to the final truncated state $\hat r(t_{\rm f})=\sum_i p_i \hat r_i$ (Eq. (\ref{rfin=})),
all the off-diagonal information are lost. However, the correlations created between S and A then allow us to gain indirectly information on
S by reading the outcome of the pointer, to select the corresponding subensemble $\scriptE_i$, and to update our information about 
$\scriptE_i$ as $\hat r_i$. The amounts of information involved in each step are measured by the entropy balance of \S~1.2.4.

\clearpage

}

 \subsubsection{Empiricism versus ontology: within quantum mechanics or beyond?}
\label{section.10.2.2}


 \hfill{{\it Einstein, stop telling God what to do}}

 \hfill{ Niels Bohr}

\vspace{0.3 cm}

\ZeText{

There is no general agreement about the purpose of science \cite{DEspagnat}\footnote{The present authors do not regard science as having a unique purpose}.
 Is our task only to explain and predict phenomena? Does theoretical physics provide only an imperfect mathematical image of reality? Or is it possible to uncover the
very nature of things? This old debate, more epistemologic than purely scientific, cannot be skipped since it may inflect our research. The
question has become more acute with the advent of quantum physics, which deals with a ``veiled reality'' \cite{DEspagnat}.  Physicists, including
the authors of the present article, balance between two extreme attitudes, illustrated by Bohr's pragmatic question \cite{bohr_dialectica}: {\it ``What can we
say about...?''} facing Einstein's ontological question \cite{einstein_dialectica}: {\it ``What is...?''} The latter position leads one to ask questions about
{\it individual systems} and not only about general properties, to regard quantum mechanics as an {\it incomplete} theory and to look for hidden ``elements of reality''.

 This opposition may be illustrated by current discussions about the status of pure states. In the statistical interpretation, there is no conceptual difference 
 between pure and mixed states (\S~\ref{fin10.1.4}); both behave as probability distributions and involve the observer. In order to reject the latter, many authors 
 with ontological aspirations afford pure states a more fundamental status, even though they acknowledge their probabilistic character,
  a point also criticized by van Kampen~\cite{vKampen}. 
  Following von Neumann's construction of density operators (in analogy to densities in phase space of classical statistical mechanics), they regard pure states 
  as building blocks rather than special cases of mixed states. In a decomposition (10.3) of a density operator $\hat {\cal D}$ associated with an ensemble 
  $\scriptE$,  they consider that each individual system of $\scriptE$ has its own ket.  In this {\it realist interpretation} \cite{deMuynck}, 
two types of probabilities are distinguished \cite{vKampen}: ``merely quantal'' probabilities are interpreted as properties of the
individual objects through $|\phi_k\rangle$, while the weights $\nu_k$ are interpreted as ordinary probabilities associated with
our ignorance of the structure of the statistical ensemble $\scriptE$. Such an interpretation might be sensible if the decomposition (10.3) were unique. 
We have stressed, however, its ambiguity (\S~\ref{fin10.2.3} and \S~\ref{fin11.1.3}); as a consequence, the very collection of pure states $|\phi_k\rangle$ among 
which each individual system is supposed to lie cannot even be imagined. It seems therefore difficult to imagine the existence of  ``underlying pure states'' 
which would carry more ``physical reality'' than $\hat {\cal D}$  \cite{bkj,molmer}. 
The distinction between the two types of probabilities on which decompositions (10.3) rely is artificial and meaningless \cite{blokhintsev1,blokhintsev2}. 

}

\ZeText{

Landau's approach to mixed states may inspire another attempt to regard a pure state as an intrinsic description of an individual system \cite{vKampen,landau}.
When two systems initially in pure states interact, correlations are in general established between them and the marginal state of each one becomes mixed. 
To identify a pure state, one is led to embed any system, that has interacted in the past with other ones, within larger and larger systems. 
Thus, conceptually, the only individual system lying in a pure state would be the whole Universe \cite{smolin,Tegmark}, a hazardous extrapolation  \cite{blokhintsev1,blokhintsev2}.   
Not to mention the introduction in quantum mechanics of  a hypothetic multiverse~\cite{Everett,Susskind}.

Such considerations illustrate the kind of difficulties to be faced in a search for realist interpretations, a search which, however, is legitimate since purely operational 
interpretations present only a blurred image of the microscopic reality and since one may long for a description that would uncover hidden faces of 
Nature  \cite{DEspagnat}. Among the proposed realist interpretations, one should distinguish those which provide exactly the same outcomes as the conventional 
quantum mechanics, and that can therefore neither be verified nor falsified. They have been extensively reviewed 
 \cite{Bassi_Ghirardi,BohmHolland,deMuynck,bellac,BohmCushing,adler,pearle_review,genovese} (see also references in \S~1.1.1), and we discussed 
 above some of them in connection with models of measurements. Many involve hidden variables of various kinds (such as Bohm and de Broglie's bunches of
  trajectories or such as stochastic backgrounds) or hidden structures (such as consistent histories, see subsection 2.9).
  
Other approaches attempt to go ``beyond the quantum''. They resort, for instance, to stochastic electrodynamics  \cite{CettodelaPenaBook,cetto,TheoSED1,TheoSED2},
to quantum Langevin equations \cite{deMuynck},  to nonlinear corrections to quantum mechanics such as in the GRW approach  \cite{Bassi_Ghirardi,pearle,GRW}, 
or to speculations about quantum gravitation \cite{PenroseEmperor}. The sole issue issue to close the Einstein--Bohr debate in such fields 
is a search for testable specific predictions \cite{einstein_dialectica,bohr_dialectica}.

For the time being, empirical approaches appear satisfactory ``for all practical purposes'' \cite{BellBook}. The statistical interpretation, either in the form put forward 
by Blokhintsev \cite{blokhintsev1,blokhintsev2} and Ballentine \cite{BallentineRMP,BallentineBook} or in the form presented above, is empirical and minimalist: 
It regards quantum mechanics only as a means for deriving predictions from available data. It is related to partly subjective interpretations that focus on information 
 \cite{caves},  since information is akin to probability. We have seen (section 11) that, although the statistical interpretation is irreducibly probabilistic, involving both 
 the system  (as regards the observables and their evolution) and the observers (as regards the state), 
  although it only deals with statistical ensembles, it suffices in conjunction with dynamics to account for individual behaviours in ideal 
 measurements. The same epistemological attitude is shared by phenomenological-minded people, and is advocated, for instance, by 
 Park \cite{X4}, van Kampen \cite{vKampen} and de Muynck \cite{deMuynck}. It can be viewed as a common ground for all physicists, as stressed 
 by Lalo\"e \cite{Laloe},  whose ``correlation interpretation'' emphasizes predictions as correlations between successive experiments. A more extreme philosophical 
 position,  the rejection of any interpretation, is even defended by Fuchs and Peres in \cite{FuchsPeres}. According to such positions, quantum theory has the 
 modest task of accounting for the statistics of results of experiments or of predicting them. It deals with what we know about reality, and does not claim 
 to unveil an underlying reality per se\footnote{This point may be illustrated on the double slit experiment. While  the particle-wave duality allows to imagine that electrons 
 or photons ``go through both slits simultaneously'', some authors find it hard to accept this for large objects such as bucky balls
 \cite{buckyballs} or viruses \cite{viruses}}.
 Quantum theory does not make any statement about going through both slits or not; As such it can be considered as incomplete. 
 Bohr himself shared \cite{bohr_dialectica} this conception when he said (see \cite{petersen1,petersen2} 
for a list of Bohr's quotations):  {\it ``There is no quantum world. There is only an abstract quantum physical description. It is wrong
to think that the task of physics is to find out how nature is. Physics concerns what we can say about nature.''}

   }


\renewcommand{\thesection}{\arabic{section}}
\section{What next?} 
\setcounter{equation}{0}
\setcounter{figure}{0}
\renewcommand{\thesection}{\arabic{section}.}
\label{fin13}

\hfill{\footnote{Preparing this porridge still requires much water}}

\vspace{-7.5mm}

\myskipfigText{
\begin{figure}[h!h!h!]
\label{ArmProv}
\hfill{\includegraphics[width=5.7cm]{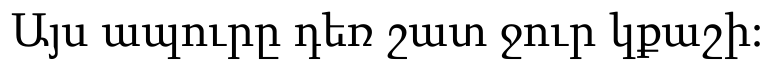}{\hspace{2mm}}}
\end{figure}
}

\vspace{-0.53cm}


\hfill{\it Il va couler encore beaucoup d'eau sous les ponts\footnote{Much water will still flow under the bridges}}

\hfill{\it Er zal nog heel wat water door de Rijn moeten\footnote{
Quite some water will still have to flow through the Rhine river}}

\hfill{Armenian, French and Dutch proverbs}

\vspace{3mm}

\ZeText{

Much can still be learnt from models, even about the ideal quantum measurements on which we have focused. Various features of measurements 
and their incidence on interpretations of quantum mechanics have been explained by the many models reviewed in section 2. However, the treatments 
based on quantum statistical mechanics provide, as final state describing the outcome of a large set of runs of the measurement, a mixed state. 
Such a state cannot be decomposed unambiguously into components that would describe subsets of runs (\S~11.1.3), so that a further study 
was required to explain the uniqueness of the outcome of each run. A dynamical mechanism that achieves this task has been proposed (\S\S~\ref{fin11.2.3} 
and \ref{fin11.2.4-5} and appendices H and I). Adapting it to further models should demonstrate the generality of such a solution of the measurement problem. 
 
Alternative approaches should also be enlightening. We suggest some paths below.

}

\subsection{Understanding ideal measurements in the Heisenberg picture}
\label{fin13.1}

\hfill{\it Nou begrijp ik er helemaal niets meer van\footnote{Now I don't understand anything of it anymore}}

\hfill{ Dutch expression}

\vspace{3mm}

\ZeText{

Some insight can be gained by implementing the dynamics of the measurement process in the Heisenberg
picture (\S~10.1.2) rather than in the more familiar Schr\"odinger picture.  Both pictures are technically equivalent 
but the Heisenberg picture will provide additional understanding. 
 It is then the observables $\hat O (t,t_0)$ which evolve, in terms of either the running time $t$ or of the reference 
 time $t_0$.  By taking $t_0$ as the initial time $t_0=0$, an observable $\hat O (t,t_0)$ is governed for an isolated system by the Heisenberg 
 equation
 
\BEQ 	i\hbar \frac{\d\hat O(t,0)}{\d t} = [\hat O (t,0), \hat H]    
\label{12.10}
\EEQ
with the initial condition $\hat O(0,0)=\hat O$, while the states assigned at the reference time $t_0=0$ remain constant. This formulation presents a conceptual advantage; 
it clearly dissociates {\it two features} of quantum mechanics, which in the Schr\"odinger picture are merged within the time-dependent density operator. 
Here, the {\it deterministic evolution} is carried by the observables, which represent random physical quantities; on the other hand, our whole 
{\it probabilistic information} about these quantities is embedded in the time-independent density operator\footnote{We 
use the term ``observables'' in the sense of ``operator-valued random physical quantities'' (\S~\ref{fin10.1.1}), 
 not of ``outcomes of observations''. The latter quantities (frequencies of occurrence, expectation values, variances) 
 are joint properties (10.1) of ``states'' (i. e., density operators playing the role of quantum probabilities) and observables}. 
 
We can thus account for the dynamics of a system in a general way, without having to specify its probabilistic description in the particular situation we wish to describe.
The use of the Heisenberg picture has therefore an incidence on the interpretation of quantum mechanics. Whereas the Schr\"odinger picture only allows 
us to describe dynamics of the statistical ensemble represented by the density operator, we can regard the equation of motion (13.1) as pertaining to an {\it individual system}\footnote{As understood, in the statistical interpretation, to belong to an ensemble of identically prepared members }.
 It is only when evaluating expectation values as tr$[\scriptD \hat O(t, 0)]$ that we have to embed the studied system in a statistical ensemble.

Moreover, when a measurement is described in the Schr\"odinger picture, the density operator of S + A undergoes {\it two types of changes}, 
the time dependence from $\scriptD(0)$ to $\scriptD(t_{\rm f})$, and the restriction to $\scriptD_i$ if the 
outcome $A_i$ is selected. The temptation of attributing the latter change to some kind of dynamics will be eluded 
in the Heisenberg picture, where only the observables vary in time.

Let us sketch how the Curie--Weiss model might be tackled in the Heisenberg picture. 

}

	\subsubsection{Dynamical equations}
\label{fin13.1.1}

\hfill{\it Rock around the clock tonight}

\hfill{Written by Max C. Freedman and James E. Myers, performed by Bill Haley and His Comets}

\vspace{3mm}

\ZeText{
	
	 The equations of motion (\ref{12.10}) which couple the 
 observables to one another have the same form as the Liouville--von Neumann equation apart from a sign change and from the boundary conditions. 
 Thus, their analysis follows the same steps as in section 4. Elimination of the bath takes place by solving at order $\gamma$ the equations (\ref{12.10}) 
 for the bath observables $\hat B_a^{(n)}(t,0)$, inserting the result into the equations for the observables of S + M and averaging over the state $\hat R_{\rm B}$ of B; 
 this provides integro-differential equations that couple the observables $\hat s_a(t,0)$ of S and those $\hat \sigma_a(t,0)$ of M ($a=x$, $y$ or $z$). 
 The conservation of $\hat s_z$ implies, instead of the decoupling between the four blocks 
 $\uparrow\uparrow$, $\downarrow\downarrow$,  $\uparrow\downarrow$,  $\downarrow\uparrow$ of the
 Schr\"odinger density matrix, the decoupling between four sets of observables, the {\it  diagonal observables proportional to }  
 $\hat\Pi_\uparrow\equiv \frac{1}{2}(1+\hat s_z)$ and  $\hat \Pi_\downarrow\equiv\frac{1}{2}(1-\hat s_z)$, 
 and the {\it   the off-diagonal observables proportional to} $\hat s_-$ and $\hat s_+$, respectively. 
 Finally the symmetry between the various spins of M allows us again to deal only with $\hat m$, 
 so that the dynamics bears on the observables $\hat \Pi_\up f(\hat m)$, $\hat \Pi_\down f(\hat m)$, $\hat s_- f(\hat m)$ and 
 $\hat s_+ f(\hat m)$, coupled within each sector.

}

\subsubsection{Dynamics of the off-diagonal observables}
\label{fin13.1.2}

\hfill{\it En spreid en sluit\footnote{And open and close (the legs)}}

\hfill{Dutch instruction in swimming lessons}

\vspace{3mm}

\ZeText{

The evolution (\ref{12.10})  of the off-diagonal observables generated over very short times
 $ t\ll \tau_{\rm recur}=\pi \hbar/2g$ by $\hat H_{\rm SA}=-Ng\hat s_z\hat m$ (section 5) is expressed by

\BEQ \hat s_-(t) = \hat s_-  \exp\frac{2iNg \hat m t}{\hbar},  \qquad \hat m(t)=\hat m.   
\label{12.11}
\EEQ
Instead of the initial truncation exhibited in the Schr\"odinger picture, we find here a rapid oscillation, which will entail a damping 
after averaging over the canonical paramagnetic state of M.

The suppression of recurrences through the non-identical couplings of subsection 6.1 replaces 
$Ng\hat m$ by $\sum_n (g+\delta g_m) \hat\sigma^{(n)}_z$ in  (\ref{12.11}),
 a replacement which after averaging over most states will produce damping. The bath-induced mechanism of subsection 6.2 introduces, 
 both in $\hat m(t)$ and in the right side of (\ref{12.11}), observables pertaining to the bath which are regarded as unreachable.
  Tracing out B then produces the damping of recurrences for the off-diagonal observables.
   
We have shown (\S\S~\ref{fin11.2.3} and \ref{fin11.2.4-5}) that reduction can result from a decoherence produced by a random interaction within M or by a collisional process. 
In the Heisenberg picture, the result is again the decay towards 0 of the off-diagonal observables 
$|\hspace{-0.7mm}\up\rangle\langle\down\hspace{-0.7mm}| \otimes |m_{\rm F}, \eta\rangle\langle -m_{\rm F}, \eta'|$ and 
$|\hspace{-0.7mm}\down\rangle\langle\up\hspace{-0.7mm}| \otimes |\hspace{-0.6mm}-\hspace{-0.6mm}m_{\rm F}, \eta\rangle\langle m_{\rm F}, \eta'|$.

}

\subsubsection{ Establishment of system--apparatus correlations}

 \label{fin13.1.3}


 \hfill{\footnote{The mule can swim over seven rivers, but as soon as it sees the water it forgets everything}}
 
\vspace{-7mm}

\myskipfigText{
\begin{figure}[h!]
 \label{ArmProv}
\hfill{\includegraphics[width=11cm]{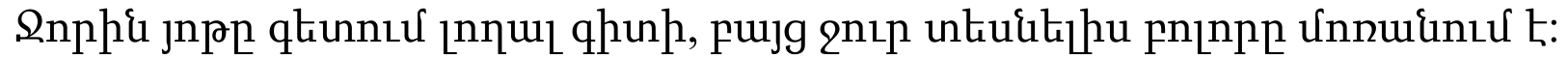}\hspace{3mm}}
\end{figure}
}

\vspace{-5mm}

 \hfill{Armenian proverb}

\vspace{3mm}

\ZeText{
 
  The evolution of the {\it diagonal observables} in the Heisenberg picture is analogous to the registration of section 7, but it is represented by more 
 general equations than in the Schr\"odinger picture.
 Indeed, denoting  by $\delta_{\hat m, m}$ the projection operator on the eigenspace associated with the eigenvalue $m$ of $\hat m$, 
 we now have in the sector $\up\up$ to look at the dynamics of 
 the time-dependent observables $\hat \Pi_\uparrow \delta_{\hat m, m} (t,0)$, instead of the dynamics of their expectation values 
 $P^{\rm dis}_{\uparrow\up} (m,t)$ in the specific state $\hat D(0)$ of S + M as in section 7. 
 The solution of the equations of motion has the form

\BEQ \label{12.15}
\hat \Pi_\up \delta_{\hat m, m} (t_{\rm f}, 0) = \sum_{m'} K_\up(m, m') \hat\Pi_\up \delta_{\hat m, m'}.  
\EEQ
The kernel $K_\up(m, m')$ represents the transition probability of the random order parameter $\hat m$ from its eigenvalue $m'$ at the time 
$0$ to its eigenvalue $m$ at the time $t_{\rm f}$,  under the effect of the bath and of a field $+g$. 
It is obtained by taking the long-time limit of the Green's function defined by Eq. (7.58), and we infer
its properties from the outcomes of section 7. As $m'$ is arbitrary, we must deal here with a 
{\it bifurcation} (as in subsection 7.3). For $m'$ larger than some negative threshold, $K_\up(m, m')$ is concentrated near $m \simeq +m_{\rm F}$;
 this will occur in particular if $m'$ is small, of order $1/\sqrt{N}$. However, if $m'$ is negative with sufficiently large $|m'|$,  it
 will be sent towards $m=-m_{\rm F}$. Likewise, $K_\down (m, m')$ is concentrated around $m \simeq -m_{\rm F}$ 
 if $|m'|$ is sufficiently small (or if $m'$ is negative), but around $m \simeq+m_{\rm F}$ if $m'$ is positive and sufficiently large. 
The complete correlations required for the process to be a measurement will be created only after averaging over a state of the pointer concentrated 
around $m'=0$.

At later times, around $t_{\rm f}$, the process of \S~\ref{fin11.2.3} produces the irreversible decay of the diagonal observables 
$|\hspace{-0.7mm}\up\rangle\langle\up\hspace{-0.7mm}| \otimes |m_{\rm F}, \eta\rangle\langle m_{\rm F}, \eta'|$ and 
$|\hspace{-0.7mm}\down\rangle\langle\down\hspace{-0.7mm}| \otimes |\hspace{-0.7mm}-\hspace{-0.7mm}m_{\rm F}, \eta\rangle\langle -m_{\rm F}, \eta'|$
towards $\delta_{\eta \eta'} \hat R^\mu_\Uparrow$ and $\delta_{\eta \eta'} \hat R^\mu_\Downarrow$, respectively.  
Notice that  while the initial observables involve here the full set $\hat \sigma^{(n)}_z$, their evolution narrows this set, leading it only towards 
the projection operators on $\hat m=m_{\rm F}$ and $\hat m=-m_{\rm F}$.
 
}

 \subsubsection{Fate of observables at the final time}


\hfill{{\it Carpe diem}\footnote{Seize the day}}

\hfill{Roman proverb}

\vspace{3mm}

 \label{fin13.1.4}
 
\ZeText{

 	Physical data come out in the form ${\rm tr}\, \scriptD^{\rm Heis} \hat O(t, 0)$ where 
 $\scriptD^{\rm Heis}= \scriptD^{\rm Schr} (0)=\hat r \otimes \scriptR$ is time-independent, namely just the initial state in the Schr\"odinger  picture.
 The success of an ideal measurement process now appears as the joint result of the algebraic properties that result in the expressions of the time-dependent 
 observables, and of some specific properties of the initial preparation of the apparatus embedded in $\scriptR(0)$. On the one hand, the width in $1/\sqrt{N}$
  of the initial paramagnetic distribution $P^{\rm dis}_{\rm M}(m, 0)$ is sufficiently large so that the oscillations (\ref{12.11}) of $\hat s_- (t)$ are numerous and 
  {\it interfere destructively} on the time scale $\tau_\trunc $. On the other hand, it is sufficiently narrow so as to {\it avoid wrong registrations}: The final probability 
  distribution $P^{\rm dis}_{\up\up}(m, t_{\rm f})$ for the pointer is the expectation value of (\ref{12.15}) over $\scriptD^{\rm Heis}$, and the concentration 
  near the origin of $\hat m=m'$ in $\scriptR(0)$ entails the concentration near $+m_{\rm F}$ of $P^{\rm dis}_{\up\up}(m, t_{\rm f})$.

	Some intuition about ideal measurements may be gained by acknowledging the {\it decay of the off-diagonal observables} during the process and their 
{\it effective  disappearance}\footnote{In fact, the disappearance of the off-diagonal observables
   is approximate for finite $N$ and is not complete: We disregard the inaccessible observables, whether they belong to the bath or they are associated with correlations
    of a macroscopic number of particles. The suppression of all the accessible off-diagonal observables relies on the mechanism of  \S~\ref{fin11.2.3}, 
    itself based on the concentration of $\hat m$ around $\pm m_{\rm F}$ } after the time  $t_{\rm f}$. 
 The evolution of the diagonal observables also implies that, under the considered circumstances, only the eigenspaces of 
  $\hat m$ associated with eigenvalues close to $m_{\rm F}$ and $-m_{\rm F}$ survive at $t_{\rm f}$.
  The only observables remaining at the end of the process, $\Pi_\up \delta_{\hat m, m} (t_{\rm f}, 0)$ with $m$ close to $m_{\rm F}$ 
  expressed by (Eq b) and $\hat \Pi_\down \delta_{\hat m, m} (t_{\rm f}, 0)$ with $m$ close to $-m_{\rm F}$, belong to an {\it abelian algebra}. 
   It is therefore natural to regard them as {\it ordinary random variables} governed by standard probabilities, and to use daily reasoning 
 which allows statements about individual events. The singular features of quantum mechanics which arose from non-commutativity (\S~10.2.1) 
 can be disregarded.  The {\it emergence of classicality} in measurement processes 
 now appears as a {\it property of the Heisenberg dynamics} of the observables.

}

\subsubsection{Truncation}
\label{fin13.1.5}

\hfill{\it En toen kwam een olifant met een hele grote snuit}

\hfill{\it En die blies het verhaaltje uit\footnote{And then came an elephant with a very big trunk, and it blew the story to an end}}

\hfill{The ending of Hen Straver's fairy tales}

\vspace{3mm}

\ZeText{

We now turn to the states describing the ensemble $\scriptE$ of runs and its subensembles. 
Remember that in the statistical interpretation and in the Heisenberg picture,  a ``state''
is a time-independent mathematical object that accounts for our information about the evolving observables (\S~\ref{fin10.1.4}). 
Equivalently, the density operator gathers the expectation values of all observables at any time.
The assignment of a density matrix $\scriptD^{\rm Heis}$ to the whole set $\scriptE$ of runs of the measurement relies on information acquired 
before the interaction process  (i.e., the measurement) and embedded in the states $\hat r$, $\hat R_{\rm M}$ and $\hat R_{\rm B}$ of S, M and B. 
These information allow us to describe the statistics of the whole process between the times 0 and $t_{\rm f}$ through 
the equations of motion (\ref{12.10}) and the density operator $\scriptD^{\rm Heis}=\hat r(0)\otimes\hat R_{\rm M}(0)\otimes\hat R_{\rm B}(0)$ 
describing the set $\scriptE$.

However, the vanishing at $t_{\rm f}$ of the off-diagonal observables (at least of all accessible ones) entails that their {\it expectation values vanish}, 
not only for the full set $\scriptE$ of runs of the measurement but also for {\it any subset}. The {\it information} about them, that was embedded at the beginning 
of the process in the off-diagonal blocks of $\scriptD^{\rm Heis}$, have been {\it irremediably lost} at the end, so that these off-diagonal blocks become
 irrelevant after measurement. For the whole ensemble $\scriptE$, and for any probabilistic prediction at times $ t>t_{\rm f}$, it makes no difference 
 to replace the state $\scriptD^{\rm Heis}$ by the sum of its diagonal blocks according to

\BEQ 
\label{12.17}
\scriptD^{\rm Heis}   \mapsto   \scriptD_\trunc ^{\rm Heis} = \sum_i p_i \scriptD_i^{\rm Heis},\qquad
 \scriptD_i^{\rm Heis} =\hat \Pi_i \hat r(0) \hat \Pi_i \otimes\hat R_{\rm M}(0)\otimes \hat R_{\rm B}(0),
 \qquad    (i=\up, \down).
  \EEQ
This reasoning sheds a new light on the interpretation of {\it truncation}, which in the Schr\"odinger picture appeared as the result of an irreversible evolution of the state. 
In the present Heisenberg picture, truncation comes out as the mere replacement (\ref{12.17}), which is nothing but an innocuous and convenient elimination of those 
{\it parts of the state} $\scriptD^{\rm Heis}$ {\it which have become irrelevant}, because the corresponding
{\it  observables have disappeared} during the measurement process. 

}

\subsubsection{Reduction}
\label{fin13.1.6}

\hfill{\it Joue de veau brais\'e▌e, couronn\'e▌e de foie gras po\^e░l\'e, r\'e▌duction de Pedro Xim\'e▌nez\footnote{Braised veal cheek, topping of foie gras, reduction of sweet sherry }}

\hfill{Recipe by the chef Alonso Ortiz}

\vspace{3mm}

\ZeText{

The argument given at the end of \S~13.1.2 then allows us to assign states to the {\it subensembles of} $\scriptE$ which can be distinguished at the time $t_{\rm f}$, 
after the observables have achieved their evolution and after decoupling of S and A. Here, however, the states, which do not depend 
on time, can be directly constructed from $\scriptD^{\rm Heis}$ or from the equivalent expression (\ref{12.17}). The vanishing of the off-diagonal observables themselves 
 simplifies the discussion since we can always eliminate the off-diagonal blocks of a state associated with any subensemble. The diagonal observables 
display the same correlations between the system and the pointer as in (\ref{12.17}), so that any subset of runs of the measurement can be represented by the state

\BEQ                        \scriptD^{\rm Heis}_{\rm sub} = \sum_i q_i \scriptD^{\rm Heis}_i .
\label{12.18} 
\EEQ
which is the basis for all predictions about the considered subensemble at times later than $t_{\rm f}$. (Note that the inclusion of elements in the 
off-diagonal blocks of (\ref{12.18}) would not change anything since there are no surviving observables in these blocks.)

The uniqueness of the outcome of each individual run comes out from (\ref{12.18}) through the same argument as in \S~11.3.1. A well-defined indication $i$ 
of the pointer is the {\it additional piece of information} that allows us to assign to the system S + A, which then belongs to the subensemble $\scriptE_i$, 
 the state $ \scriptD_i^{\rm Heis}$.
 Retaining only one diagonal block of $ \scriptD^{\rm Heis}$ in (\ref{12.17}) amounts to {\it upgrade our probabilistic description}
\footnote{In the Schr\"odinger picture, the expectation value of any (time-independent) observable for the subensemble $\scriptE_i$ was found from the state
 $\scriptD_i$ of Eq. (11.21). Here, it is obtained from the evolution (13.3) and the state $\scriptD_i^{\rm Heis}$
 of Eq (13.4). The state $\scriptD_i$  results from $\scriptD_i^{\rm Heis}$ by integrating the Liouville--von Neumann equation from $t=0$
  to $t_{\rm f}$}.

\vspace{3mm}

The specific features of the Heisenberg representation were already employed in literature for arguing that this representation (in
contrast to that by Schr\"odinger) has advantages in explaining the features of quantum measurements \cite{A3,A5,A6}. In particular, Rubin
argued that obstacles preventing a successful application of the Everett interpretation to quantum measurements are absent (or at least
weakened) in the Heisenberg representation \cite{A5,A6}. Certain aspects of the analysis by Rubin do not depend on the assumed Everett
interpretation and overlap with the presentation above (that does not assume this interpretation). Blanchard, Lugiewicz and Olkiewicz
employed the decoherence physics within the Heisenberg representation for showing that it accounts more naturally (as compared to the
Schr\"odinger representation) for the emergence of classical features in quantum measurements \cite{A3}. Their approach is phenomenological 
(and shares the criticisms we discussed in section 2.2), but the idea of an emergent Abelian (classical) algebra again overlaps with the
preliminary results reported above.  The emergent Abelian algebra is also the main subject of the works by Sewell 
\cite{sewell_2005,sewell_2007,sewell} and Requardt \cite{requardt}  that we already reviewed in section 2.4.3. 
In particular, Requardt explains that closely related ideas were already
expressed by von Neumann and van Kampen (see references in \cite{requardt}).

\vspace{3mm}

As shown by this reconsideration of the Curie--Weiss model, the Heisenberg picture enlightens the truncation, reduction and registration processes, 
by exhibiting them as a purely dynamical phenomena and by explaining their generality. 
Although mathematically equivalent to the Schr\"odinger picture, it suggests more transparent interpretations, 
owing to a separate description of the dynamics of quantum systems and of our probabilistic knowledge about them. 
A better insight on other models of measurement should therefore be afforded by their treatment in the Heisenberg picture.

}

\subsection{Other types of measurements}

\label{fin13.2}


\vspace{3mm}

\hfill{\it Corruptissima republica plurimae leges\footnote{The greater the degeneration of the republic, the more of its laws}}

\hfill{Tacitus}

\vspace{3mm}

\ZeText{

We have only dealt in this article with ideal quantum measurements, in which information about the initial state of the tested system S is
displayed by the apparatus at some later time, and in which the final state of the system S is obtained by projection. 
Other realistic setups, e.g. of particle detectors or of avalanche processes, deserve to be studied through models.
 Measurements of a more elaborate type, in which some quantum property of S is continuously followed in
time, are now being performed owing to experimental progress \cite{X0,X1,esteve_korotkov,haroche,X11}. 
For instance, non-destructive (thus non-ideal) repeated observations of photons allow the study of quantum jumps \cite{haroche}, 
and quantum-limited measurements, in which a
mesoscopic detector accumulates information progressively \cite{meso}, are of interest to optimize the efficiency of the processing of $q$-bits.
Quantum measurements are by now employed for designing feedback control processes \cite{X11,X12}, 
a task that in the classical domain is routinely done via classical measurements.

Such experiments seem to reveal properties of individual systems, in apparent contradiction with the statistical interpretation 
of quantum mechanics. However, as in ideal measurements, repeated observations of the above type on identically prepared 
systems give different  results, so that they do not give access to trajectories in the space of the tested variables, but only to 
{\it autocorrelation functions} presenting quantum fluctuations.
  It seems timely, not only for conceptual purposes but to help
the development of realistic experiments, to work out further models, in
particular for such quantum measurements in which the whole history of
the process is used to gather information. In this context we should
mention the so-called weak measurements \cite{AharonovAlbertVaidman} that (in a sense) minimize the
back-action of the measurement device on the measured system, and -- although they
have certain counterintuitive features -- can reveal the analogues of
classical concepts in quantum mechanics; e.g., state determination with
the minimal disturbance, classical causality \cite{holger,howard,lars,tollaksen,BauerBernard2011,BauerBenoistBernard2012}, 
and even mapping out of the complete wave function \cite{Lundeen} 
or of the average trajectories of single photons in a double-slit experiment~\cite{Kocsis2011}.

  Apart from such foreseeable research works, it seems desirable to make educational progress by taking into account the insights 
  provided by the solution of models of quantum measurement processes. The need of quantum statistical mechanics to explain 
  these processes, stressed all along this paper, and the central role that they play in the understanding of quantum phenomena, 
  invite us to a reformation of teaching at the introductory level. The statistical interpretation, as sketched in subsection 10.1, 
  is in keeping with the analysis of measurements. Why not introduce the concepts and bases of quantum mechanics within its framework.
  This ``minimal'' interpretation seems more easily assimilable by students than the traditional approaches. It thus appears desirable to 
  foster the elaboration of new courses and of new textbooks, which should hopefully preserve the forthcoming generations from 
  bewilderment when being first exposed to quantum physics...and even later!

}



 \addcontentsline{toc}{section}{Acknowledgments}

 \section*{Acknowledgments}

\ZeText{

We thank Bernard d'Espagnat, Franck Lalo\"e, Michel Le Bellac and Roland Omn\`es for having urged us to stress the explanation of uniqueness,
and especially F. Lalo\"e for many thorough and enlightening discussions and suggestions.  Fr\'ed\'eric Joucken and Marti Perarnau 
have kindly checked the calculations and Claudia Pombo has discussed various aspects.
 We thank Loic Bervas and Isabela Pombo Geertsma for taking part in the typesetting and 
 we are grateful for hospitality at CEA Saclay and the University of Amsterdam during the various stages of this work.
 The research of AEA was supported by the R\'egion des Pays de la Loire under the Grant 2010-11967.


}

 \addcontentsline{toc}{section}{Appendices}

 \section*{Appendices}

\appendix

 \renewcommand{\thesection}{\alph{section}}
  \renewcommand{\thesubsection}{\thesection\arabic{subsection}}
   \renewcommand{\theequation}{\thesection\arabic{equation}}

\setcounter{section}{0}

\hfill{\it Non scholae, sed vitae discimus\footnote{We learn not for school, but for life}}

\hfill{Seneca}

\renewcommand{\thesection}{\Alph{section}}
 \section{Elimination of the bath} 
 \label{AppendixA}

  \setcounter{equation}{0}
  \setcounter{figure}{0}
 \renewcommand{\thesection}{\Alph{section}.}

\hfill{\it Do not bathe if there is no water}

\hfill{Shan proverb}

\vspace{3mm}
\ZeText{

Taking $\hat H_0=\hat H_{\rm S}+\hat H_{\rm SA}+\hat H_{\rm M}$ and $\hat H_{\rm B}$
as the unperturbed Hamiltonians of S + M and of B, respectively, and denoting by
$\hat U_0$ and $\hat U_{\rm B}$ the corresponding evolution operators, we consider
the full evolution operator associated with $\hat H=\hat H_0+\hat H_{\rm B}+\hat H_{\rm MB}$ 
in the interaction representation. We can expand it as

\begin{equation}
\mytext{\textcurrency UUU\textcurrency \qquad}
\hat{U}_{0}^{\dagger}\left(t\right)  \hat{U}_{{\rm B}}^{\dagger}\left(  t\right)  
e^{-i\hat{H}t/\hbar}\approx\hat{I}-i\hbar^{-1}\int_{0}^{t}{\rm d}t^{\prime}\hat{H}
_{{\rm MB}}\left(  t^{\prime}\right)  +{\cal O}\left(  \gamma\right)
{ ,} \label{UUU}
\end{equation}
where the coupling in the interaction picture is
\begin{equation}
\mytext{\textcurrency HMBt\textcurrency \qquad}
\hat{H}_{{\rm MB}}\left(
t\right)  =\sqrt{\gamma}\sum_{n,a}\hat{U}_{0}^{\dagger}\left(  t\right)
\hat{\sigma}_{a}^{\left(  n\right)  }\hat{U}_{0}\left(  t\right)  \hat{B}
_{a}^{\left(  n\right)  }\left(  t\right)  { ,} \label{HMBt}
\end{equation}
with $\hat{B}_{a}^{\left(  n\right)  }\left(  t\right)  $ defined by (\ref{Bt}).

We wish to take the trace over ${\rm B}$ of the exact equation of motion eq. (\ref{vN}) for 
$\hat{\cal D}(t)$, so as to generate an equation of motion for the density operator $\hat{D}\left(
t\right)  $ of ${\rm S}+{\rm M}$. In the right-hand side the term
${\rm tr}_{{\rm B}}\left[  \hat{H}_{{\rm B}},{\hat {\cal D}
}\right]  $ vanishes and we are left with
\begin{equation}
\mytext{\textcurrency dD1\textcurrency \qquad}
i\hbar\frac{{\rm d}\hat{D}
}{{\rm d}t}=\left[  \hat{H}_{0},\hat{D}\right]  +{\rm tr}
_{{\rm B}}\left[  \hat{H}_{{\rm MB}},{\hat {\cal D}}\right]  { .}
\label{dD1}
\end{equation}

The last term involves the coupling $\hat{H}_{{\rm MB}}$ both directly and
through the correlations between ${\rm S}+{\rm M}$ and ${\rm B}$
which are created in ${\cal D}\left(  t\right)  $ from the time $0$ to the
time $t$. In order to write (\ref{dD1}) more explicitly, we first exhibit
these correlations. To this aim, we expand ${\cal D}\left(  t\right)  $ in
powers of $\sqrt{\gamma}$ by means of the expansion (\ref{UUU}) of its
evolution operator. This provides, using $\hat U_0(t)=\exp[-i\hat H_0t/\hbar]$, 
\begin{equation}
\mytext{\textcurrency UUDUU\textcurrency \qquad}
\hat{U}_{0}^{\dagger}\left(
t\right)  \hat{U}_{{\rm B}}^{\dagger}\left(  t\right)  {\hat {\cal D}
}\left(  t\right)  \hat{U}_{{\rm B}}\left(  t\right)  \hat{U}_{0}\left(
t\right)  \approx{\hat {\cal D}}\left(  0\right)  -i\hbar^{-1}\left[
\int_{0}^{t}{\rm d}t^{\prime}\hat{H}_{{\rm MB}}\left(  t^{\prime
}\right)  \hspace{-1mm},\hat{D}\left(  0\right)  \hat{R}_{{\rm B}}\left(  0\right)
\right]  +{\cal O}\left(  \gamma\right)  { .} \label{UUDUU}
\end{equation}

Insertion of the expansion (\ref{UUDUU}) into (\ref{dD1}) will allow us to
work out the trace over ${\rm B}$. Through the factor $\hat{R}_{{\rm B}
}\left(  0\right)  $, this trace has the form of an equilibrium
expectation value. As usual, the elimination of the bath variables
will produce memory effects as obvious from (\ref{UUDUU}). We wish
these memory effects to bear only on the bath, so as to have a
short characteristic time. However the initial state which enters
(\ref{UUDUU}) involves not only $\hat {R}_{{\rm B}}\left(
0\right)  $ but also $\hat{D}\left(  0\right)  $, so that a mere
insertion of (\ref{UUDUU}) into (\ref{dD1}) would let $\hat
{D}\left(  t\right)  $ keep an undesirable memory of
$\hat{D}\left(  0\right) $. We solve this difficulty by
re-expressing perturbatively $\hat{D}\left( 0\right)  $ in terms
of $\hat{D}\left(  t\right)  $. To this aim we note that
the trace of (\ref{UUDUU}) over ${\rm B}$ provides
\begin{equation}
\mytext{\textcurrency DtD0\textcurrency \qquad}
U_{0}^{\dagger}\left(  t\right)
\hat{D}\left(  t\right)  \hat{U}_{0}\left(  t\right)  =\hat{D}\left(
0\right)  +{\cal O}\left(  \gamma\right)  { .} \label{DtD0}
\end{equation}
We have used the facts that the expectation value over $\hat{R}_{{\rm B}
}\left(  0\right)  $ of an odd number of operators $\hat{B}_{a}^{\left(
n\right)  }$ vanishes, and that each $\hat{B}_{a}^{\left(  n\right)  }$ is
accompanied in $\hat{H}_{{\rm MA}}$ by a factor $\sqrt{\gamma}$. Hence the
right-hand side of (\ref{DtD0}) as well as that of (\ref{dD1}) are power
series in $\gamma$ rather than in $\sqrt{\gamma}$.

We can now rewrite the right-hand side of (\ref{UUDUU}) in terms of $\hat
{D}\left(  t\right)  $ instead of $\hat{D}\left(  0\right)  $ by means of
(\ref{DtD0}), then insert the resulting expansion of ${\hat {\cal D}}\left(
t\right)  $ in powers of $\sqrt{\gamma}$ into (\ref{dD1}). Noting that the
first term in (\ref{UUDUU}) does not contribute to the trace over ${\rm B}
$, we find
\begin{equation}
\mytext{\textcurrency dD2\textcurrency \qquad}
\frac{{\rm d}\hat{D}
}{{\rm d}t}-\frac{1}{i\hbar}\left[  \hat{H}_{0},\hat{D}\right]  =-\frac
{1}{\hbar^{2}}{\rm tr}_{{\rm B}}\int_{0}^{t}{\rm d}t^{\prime
}\left[  \hat{H}_{{\rm MB}},\hat{U}_{{\rm B}}\hat{U}_{0}\left[  \hat
{H}_{{\rm MB}}\left(  t^{\prime}\right)  ,\hat{U}_{0}^{\dagger}\hat{D}
\hat{U}_{0}\hat{R}_{{\rm B}}\left(  0\right)  \right]  \hat{U}_{0}
^{\dagger}\hat{U}_{{\rm B}}^{\dagger}\right]  +{\cal O}\left(
\gamma^{2}\right)  { ,} \label{dD2bis}
\end{equation}
where $\hat{D}$, $\hat{U}_{{\rm B}}$ and $\hat{U}_{0}$ stand for $\hat
{D}\left(  t\right)  $, $\hat{U}_{{\rm B}}\left(  t\right)  $ and $\hat
{U}_{0}\left(  t\right)  $. Although the effect of the bath is of order $\gamma$,
the derivation has required only the first-order term, in $\sqrt{\gamma}$, of
the expansion (\ref{UUDUU}) of ${\cal D}\left(  t\right)  $.

The bath operators $\hat{B}_{a}^{\left(  n\right)  }$ appear through $\hat
{H}_{{\rm MB}}$ and $\hat{H}_{{\rm MB}}\left(  t^{\prime}\right)  $, and
the evaluation of the trace thus involves only the equilibrium autocorrelation
function (\ref{Kt-s}). Using the expressions (\ref{ham4}) and (\ref{HMBt}) for
$\hat{H}_{{\rm MB}}$ and $\hat{H}_{{\rm MB}}\left(  t^{\prime}\right)
$, denoting the memory time $t-t^{\prime}$ as $u$, and introducing the
operators $\hat{\sigma}_{a}^{\left(n\right)  }\left(  u\right) $ 
defined by (\ref{sigmau}), we finally find the differential equation
(\ref{dD}) for $\hat D(t)$.

\clearpage

}

\renewcommand{\thesection}{\Alph{section}}
\section{Representation of the density operator of S + M by scalar functions}
\setcounter{equation}{0}
\setcounter{figure}{0}
\renewcommand{\thesection}{\Alph{section}.}
\label{AppendixB}

\hfill{\it Je moet je niet beter voordoen dan je bent\footnote{Don't pretend to be more than you are}}

\hfill{Dutch proverb}

\vspace{3mm}
\ZeText{

We first prove that, if the operators $\hat R_{ij}(t)$ in the Hilbert space of M depend only on $\hat m$,
the right hand side of (\ref{dRij}) has the same property.

The operators $\hat{\sigma}_{+}^{\left(  n\right)  }=\frac{1}{2}\left(
\hat{\sigma}_{x}^{(n)}+i\hat{\sigma}_{y}^{\left(  n\right)  }\right)  $ and
$\hat{\sigma}_{-}^{\left(  n\right)  }=\left(  \hat{\sigma}_{+}^{\left(
n\right)  }\right)  ^{\dagger}$ raise or lower the value of $m$ by $\delta
m=2/N$, a property expressed by
\begin{equation}
\mytext{\textcurrency msigma\textcurrency \qquad}
[\hat\sigma_+^{(n)},\hat\sigma_z^{(n)}]=-2\hat\sigma_+^{(n)},\qquad
\hat{\sigma}_{+}^{\left( n\right)  }\hat{m}=\left(  \hat{m}-\delta m\right)  \hat{\sigma}_{+}^{(n)}
{\rm .} \label{msigma}
\end{equation}
The last identity can be iterated to yield

\BEQ 
\hat{\sigma}_{+}^{\left( n\right)  }\hat{m}^k=\left(  \hat{m}-\delta m\right)  \hat{\sigma}_{+}^{(n)}\hat m^{k-1}
=\cdots=\left(  \hat{m}-\delta m\right) ^k \hat{\sigma}_{+}^{(n)},
\EEQ
so that for every function that can be expanded in powers of $\hat m$, but does not otherwise depend on the $\hat\sigma_a^{(k)}$, it holds that

\BEQ\label{frule}
\hat{\sigma}_{\pm}^{\left( n\right)  }f(\hat m) =f(  \hat{m}\mp\delta m)  \hat{\sigma}_{\pm}^{(n)}.
\EEQ
In order to write explicitly the time-dependent operators $\hat{\sigma}
_{a}^{\left(  n\right)  }\left(  u,i\right)  $ defined by (\ref{sigmaui}) with the definition
(\ref{Hi}), it is convenient to introduce the notations
\begin{eqnarray}\label{m+-}
\mytext{\textcurrency m+-\textcurrency \qquad}
m_{\pm}  &  =&m\pm\delta
m=m\pm\frac{2}{N}{\rm  ,}\label{m+-}\\
\mytext{\textcurrency delta+-\textcurrency \qquad}
\Delta_{\pm}f\left(  m\right)
&  =&f\left(  m_{\pm}\right)  -f\left(  m\right)  {\rm  .} \label{Adelta+-}
\end{eqnarray}
The time-dependent operators (\ref{sigmaui}) are then given by ($u=t-t'$ is the memory time; $i=\up,\down$)

\begin{equation}
\mytext{\textcurrency sigmaz\textcurrency \qquad}
\hat{\sigma}_{z}^{\left(
n\right)  }\left(  u,i\right)  =\hat{\sigma}_{z}^{\left(  n\right)  }{\rm  ,}
\label{sigmaz}
\end{equation}
\begin{eqnarray}
\mytext{\textcurrency sigmaplus\textcurrency \qquad}
\hat{\sigma}_{+}^{\left( n\right)  }\left(  u,i\right)   &  
=\frac{1}{2}\left[  \hat{\sigma}_{x}^{\left(  n\right)  }\left(  u,i\right)  +i\hat{\sigma}_{y}^{\left( n\right)  }\left(  u,i\right)  \right]  
=e^{-i\hat{H}_{i}u/\hbar}\hat{\sigma}_{+}^{\left(  n\right)  }e^{i\hat{H}_{i}u/\hbar}
=\hat{\sigma}_{+}^{\left(  n\right)  }e^{-i\hat{\Omega}_{i}^{+}u}
=e^{i\hat{\Omega}_{i}^{-}u}\,\hat{\sigma}_{+}^{\left(  n\right)  } 
=[\hat{\sigma}_{-}^{\left( n\right)  }\left(  u,i\right) ]^\dagger {\rm  ,} 
\label{sigmaplus}
\end{eqnarray}
where we used (\ref{frule}) and where the operators $\hat{\Omega}_{\uparrow
}^{+}$, $\hat{\Omega}_{\uparrow}^{-}$, $\hat{\Omega}_{\downarrow}^{+}$,
$\hat{\Omega}_{\downarrow}^{-}$ are functions of $\hat{m}$ defined by
$\hat{\Omega}_{i}^{\pm}=\Omega_{i}^{\pm}\left(  \hat{m}\right)  $ and by
\begin{equation}
\mytext{\textcurrency Aomega\textcurrency \qquad}
\hbar\Omega_{i}^{\pm}\left(
m\right)  =\Delta_{\pm}H_{i}\left(  m\right)  =H_{i}\left(  m\pm\delta
m\right)  -H_{i}\left(  m\right)  {\rm  .} \label{Aomega}
\end{equation}

If in the right-hand side of (\ref{dRij}) the operator $\hat{R}_{ij}$ depends
only on $\hat{m}$ at the considered time, the terms with $a=z$ cancel out on account of
(\ref{sigmaz}). The terms with $a=x$ and $a=y$, when expressed by means of
(\ref{sigmaplus}), generate only products of $\hat{\sigma}_{+}^{\left(
n\right)  }\hat{\sigma}_{-}^{\left(  n\right)  }$ or $\hat{\sigma}
_{-}^{\left(  n\right)  }\hat{\sigma}_{+}^{\left(  n\right)  }$ by functions
of $\hat{m}$. This can be seen by using (\ref{frule}) to bring $\hat{\sigma
}_{+}^{\left(  n\right)  }$ and $\hat{\sigma}_{-}^{\left(  n\right)  }$ next
to each other through commutation with $\hat{R}_{ij}$. Since $\hat{\sigma}
_{+}^{\left(  n\right)  }\hat{\sigma}_{-}^{\left(  n\right)  }=1-\hat{\sigma
}_{-}^{\left(  n\right)  }\hat{\sigma}_{+}^{\left(  n\right)  }=\frac{1}
{2}\left(  1+\hat{\sigma}_{z}^{\left(  n\right)  }\right)  $, we can then
perform the summation over $n$, which yields products of some functions of $\hat m$ by the factor
\begin{equation}
\mytext{\textcurrency sumsigma\textcurrency \qquad}
\sum_{n}\hat{\sigma}
_{+}^{\left(  n\right)  }\hat{\sigma}_{-}^{\left(  n\right)  }=N-\sum_{n}
\hat{\sigma}_{-}^{\left(  n\right)  }\hat{\sigma}_{+}^{\left(  n\right)
}=\frac{N}{2}\left(  1+\hat{m}\right),
  \label{sumsigma}
\end{equation}
itself depending only on $\hat m$. Hence, if $\hat{R}_{ij}$ is a function of the operator
$\hat{m}$ only, this property also holds for $\d \hat R_{ij}(t)/\d t$ given by (\ref{dRij}).
Since, except in section 5.2, it holds at the initial time, it holds at any time.

The equations of motion (\ref{dRij}) for $\hat R_{ij}(t)$ are therefore equivalent to the corresponding
equations for $P_{ij}(m,t)$ which we derive below. 
The matrices $\hat R_{ij}(t)$ which characterize the density operator of S + M are parametrized as $\hat R_{ij}(t)=R_{ij}(\hat m,t)=P_{ij}^{\rm dis}(\hat m,t)/G(\hat m)$; 
in the continum limit, we introduced $P_{ij}(m,t)=(N/2)P^{\rm dis}_{ij}(m,t)$. 
We first note that the autocorrelation function $K(t)$ enters (\ref{dRij}) through integrals of the form

\begin{eqnarray}
\mytext{AK>}
 &&  \tilde{K}_{t>}\left(  \omega\right)  =\int_{0}
^{t}{\rm d}ue^{-i\omega u}K\left(  u\right)  =\frac{1}{2\pi i}\int
_{-\infty}^{+\infty}{\rm d}\omega^{\prime}
\frac{e^{i(  \omega^{\prime}-\omega)  t} -1}{\omega^{\prime}-\omega}\tilde{K}\left(  \omega^{\prime
}\right)  {\rm  ,}\label{AK>}\nn \\
\mytext{ AK<}
&& {\rm  \tilde{K}}_{t<}\left(  \omega\right)= 
\int_{-t}^0{\rm d}ue^{-i\omega u}K\left(  u\right)
 =\int_{0}^{t}{\rm d}ue^{i\omega u}K\left(  -u\right)  
=\left[  \tilde{K}_{t>}\left(\omega\right)  \right]  ^{\ast}{\rm  .} \label{AK<}
\end{eqnarray}

As shown above, only the contributions to (\ref{dRij}) with $a=x$
or $a=y$ survive owing to (\ref{sigmaz}). The first term is transformed, by
relying successively on (\ref{sigmaplus}), (\ref{AK>}),  (\ref{frule}) and 
(\ref{sumsigma}), into
\begin{eqnarray}
\label{tech}
\int_{0}^{t}{\rm d}u\sum_{n}\sum_{a=x,y}K\left(  u\right)  \hat{\sigma}_{a}^{\left(  n\right)  }\left( u,i\right)  \hat{R}_{ij}\hat{\sigma}_{a}^{\left(  n\right)  }  
&  =&2\int_{0}^{t}{\rm d}u\sum_{n}K\left(  u\right)  \left[  e^{i\hat{\Omega}_{i}^{-}u}\hat{\sigma}_{+}^{\left(  n\right)  }R_{ij}\left(  \hat{m}\right)
\hat{\sigma}_{-}^{\left(  n\right)  }+e^{i\hat{\Omega}_{i}^{+}u}\hat{\sigma}_{-}^{\left(  n\right)  }R_{ij}\left(  \hat{m}\right)  \hat{\sigma}
_{+}^{\left(  n\right)  }\right]  \\
&  =&N\tilde{K}_{t>}\left(  -\hat{\Omega}_{i}^{-}\right) R_{ij}\left(  \hat{m}-\delta m\right)  \left(  1+\hat{m}\right) 
 +N\tilde{K}_{t>}\left(  -\hat{\Omega}_{i}^{+}\right) R_{ij}\left(  \hat{m}+\delta m\right)  \left(  1-\hat{m}\right)  {\rm  .}\nn
\end{eqnarray}

From  the relation $R_{ij}\left( m\right)=P^{\rm dis}_{ij}\left(m\right) /G(m)$ (see Eq. (\ref{RijP})), we get

\begin{equation}
 \label{Rm+-}
(1\mp m){R_{ij}\left( m_{\pm}\right)} =(1\mp m)\frac{P^{\rm dis}_{ij}\left(m_{\pm}\right) }{G(m_\pm)}
 =\frac{1\pm m_\pm}{G(m)}{P^{\rm dis}_{ij}\left(m_{\pm}\right) },
 \label{Rm+-}
\end{equation}
so that we can readily rewrite (\ref{tech}) in terms of $P_{ij}\left(  \hat m\right) =\half NP^{\rm dis}_{ij}\left(\hat  m\right)  $
instead of $\hat{R}_{ij}$. The same steps allow us to express the other three
terms of (\ref{dRij}) in a similar form. Using also $\Delta_+\Omega^-_i=\Delta_+[H_i(m-\delta m)-H_i(m)]=-\Omega^+_i$
and $\Delta_-\Omega^+_i=-\Omega^-_i$, where $\Delta_{+}$ and $\Delta_{-}$ were defined by (\ref{m+-}) and (\ref{Adelta+-}),
we find altogether, after multiplying by $G(m)$, 

\begin{eqnarray}
 \mytext{\textcurrency BE\textcurrency \qquad}
\frac{{\rm d}}{{\rm d} t}P_{ij}\left( m,t\right) -\frac{1}{i\hbar}\left[  H_{i}\left(  m\right)-H_{j}\left(  m\right)  \right]  P_{ij}\left(  m,t\right) 
&  =&\frac{\gamma N}{\hbar^{2}}\Delta_{+}\left\{  \left(  1+m\right)  \left[
\tilde{K}_{t>}\left(  \Omega_{i}^{-}\right)  +\tilde{K}_{t<}\left(  \Omega
_{j}^{-}\right)  \right]  P_{ij}\left(  m,t\right)  \right\} \nonumber\\
&+ & \frac{\gamma N}{\hbar^{2}}\Delta_{-}\left\{  \left(  1-m\right)  \left[
\tilde{K}_{t>}\left(  \Omega_{i}^{+}\right)  +\tilde{K}_{t<}\left(  \Omega
_{j}^{+}\right)  \right]  P_{ij}\left(  m,t\right)  \right\}  {\rm  ,}
\label{BE}
\end{eqnarray}
For $i=j$ this equation simplifies into Eq. (\ref{dPdiag}), due to both the cancellation in the left-hand side and the 
appearance of the combination (\ref{Ktomega}) in the right-hand side.

Since it is an instructive exercise for students to numerically solve the full quantum dynamics of the registration process 
at finite $N$,
we write out here the ingredients of the dynamical equation (\ref{BE}) for $ P_{\uparrow\uparrow}$ and $P_{\downarrow\downarrow}$. 
As we just indicated above, this equation simplifies for $i=j$ into (\ref{dPdiag}). Moreover, in the registration regime, we can replace 
$\tilde K_{t>}(\omega)+\tilde K_{t<}(\omega)=\tilde K_t(\omega)$ by $\tilde K(\omega)$, 
defined in (\ref{Ktilde}). The rates entering Eq.  (\ref{dPdiag}) or Eq. (\ref{BE}) for $i=j$ have therefore the form

\BEQ 
\label{Brate}
\frac{\gamma N}{\hbar^2}\tilde K(\omega)=
\frac{N\hbar\omega}{8J\,\tau_J}\left[\coth\left(\half\beta\hbar\omega \right)-1\right]\,
\exp\left(-\frac{|\omega|}{\Gamma}\right),
\EEQ
where the timescale $\tau_J=\hbar/\gamma J$ can be taken as a unit of time. The variable $\omega$ in $\tilde K(\omega)$ takes the values
 $\Omega^\pm _i$, with $i=j=\,\, \uparrow$ or $\downarrow$, which are explicitly given by (\ref{ompmi=}) in terms of the discrete variable $m$. 
 It can be verified that, for $\Gamma\gg J/\hbar$, the omission of the Debye cut-off in (\ref{Brate}) does not significantly affect the dynamics.

}

\renewcommand{\thesection}{\Alph{section}}
\section{Evaluation of the recurrence time for a general pointer}
\setcounter{equation}{0}
\setcounter{figure}{0}
\renewcommand{\thesection}{\Alph{section}.}
\label{AppendixC}

\hfill{\it For what cannot be cured, patience is best}

\hfill{Irish proverb}


\vspace{3mm}

\ZeText{

We consider here general models for which the tested observable $\hat s$ is coupled to a pointer through the Hamiltonian (\ref{gen2})
where the pointer observable $\hat m$ has $Q$ eigenvalues behaving as independent random variables. The probability
distribution $p(\omega_q)$ for the corresponding eigenfrequencies $\omega_q\equiv Ng(s_i-s_j)m_q/\hbar$ which enter the function 
$\Re F(t)=Q^{-1}\sum_q \cos\omega_q t$ is taken as (\ref{pomq}). 
For shorthand we denote from now on in the present appendix  by $F(t)$ the real part $\Re F$ 
of the function defined in \S~\ref{section.6.1.2} by (\ref{Ft=}).

We wish to evaluate the probability ${\cal P}\left(  f,t\right)  $\ for
$F\left(  t\right)  $\ to be larger than some number $f$\ at a given time $t\gg
\Delta\omega$. This probability is deduced from the characteristic function
for $F\left(  t\right)  $\ through

\begin{eqnarray}
\mytext{\textcurrency Pft1\textcurrency \qquad}
{\cal P}\left(  f,t\right)
=\overline{\theta\left[  F\left(  t\right)  -f\right]  }=\int_{-\infty
}^{+\infty}\frac{{\rm d}\lambda}{2\pi\left(  i\lambda+0\right)
}e^{-iQ\lambda f}\ \overline{e^{iQ\lambda F\left(  t\right)  }}
 =\int_{-\infty}^{+\infty}\frac{{\rm d}\lambda}{2\pi\left(
i\lambda+0\right)  }\left[  e^{-i\lambda f}\int{\rm d}\omega p\left(
\omega\right)  e^{i\lambda\cos\omega t}\right]  ^{Q}{ .} \label{Pft1}
\end{eqnarray}
Since $t\gg1/\Delta\omega$, the factor $p\left(  \omega\right)  $\ in the integrand varies slowly over the period $2\pi/t$\ of the exponential factor
$\exp{i \lambda \cos \omega t}$. This exponential  may therefore be replaced by its average on $\omega$\ over
one period, which is the Bessel function $J_{0}\left(  \lambda\right)  $.  The integral over $\omega$ then gives unity, and we end up with
\begin{equation}
\mytext{\textcurrency Pft2\textcurrency \qquad}
{\cal P}\left(  f,t\right)
=\int_{-\infty}^{+\infty}\frac{{\rm d}\lambda}{2\pi\left(  i\lambda
+0\right)  }\exp\{Q\left[  \ln J_{0}\left(  \lambda-i0\right)  -i\lambda
f\right]  \}{ .} \label{Pft2}
\end{equation}

For $Q\gg1$, the exponent has a saddle point $\lambda_{{\rm s}}$\ given as
function of $f$\ \ by
\begin{equation}
\mytext{\textcurrency saddle\textcurrency \qquad}
\lambda_{{\rm s}}
\equiv-iy{,\quad\quad}\frac{I_{1}\left(  y\right)  }{I_{0}\left(
y\right)  }=f{,\quad\quad}\frac{{\rm d}f}{{\rm d}y}=1-\frac{f}
{y}-f^{2}{ ,} \label{saddle}
\end{equation}
and we find
\begin{equation}
\mytext{\textcurrency Pft3\textcurrency \qquad}
{\cal P}\left(  f,t\right)
=\frac{1}{y}\left(  2\pi Q\frac{{\rm d}f}{{\rm d}y}\right)  ^{-1/2}
\exp\left\{  -Q\left[  yf-\ln I_{0}\left(  y\right)  \right]  \right\}  {
.} \label{Pft3}
\end{equation}

We now evaluate the average duration $\overline{\delta t}$\ of an excursion of
$F\left(  t\right)  $\ above the value $f$. To this aim, we determine the
average curvature of $F\left(  t\right)  $\ at a peak, reached for values of
the set $\omega_{q}$\ such that $F\left(  t\right)  >f$. The quantity
\begin{equation}
\mytext{\textcurrency d2f\textcurrency \qquad}
\overline{\theta\left[  F\left(
t\right)  -f\right]  \frac{{\rm d}^{2}F\left(  t\right)  }{{\rm d}t^{2}
}} \label{d2f}
\end{equation}
is obtained from (\ref{Pft1}) by introducing in the integrand a factor
\begin{equation}
\frac{-\int{\rm d}\omega p\left(  \omega\right)  \omega^{2}\cos\omega
t\,\,e^{-i\lambda\cos\omega t}}{\int{\rm d}\omega p\left(  \omega\right)
e^{i\lambda\cos\omega t}}=\frac{-i\Delta\omega^{2}J_{1}\left(  \lambda\right)
}{J_{0}\left(  \lambda\right)  }{ ,}
\end{equation}
where we used $t\Delta\omega\gg1$. The saddle-point method, using
(\ref{saddle}), then provides on average, under the constraint $F\left(
t\right)  >f$,
\begin{equation}
\frac{1}{F\left(  t\right)  }\frac{{\rm d}^{2}F\left(  t\right)
}{{\rm d}t^{2}}=-\Delta\omega^{2}{ .}
\end{equation}
A similar calculation shows that, around any peak of $F\left(  t\right)  $
emerging above $f$, the odd derivatives of $F\left(  t\right)  $\ vanish on
average while the even ones are consistent with the gaussian shape
(\ref{expF}), rewritten for $f^{-1}F\left(  t^{\prime}\right)  $ in terms of
$t^{\prime}-t<1/\Delta\omega$. This result shows that the shape of the
dominant term of (\ref{corrF}) is not modified by the constraint $F(t)>f$.
Hence, if $F\left(  t\right)  $\ reaches a maximum $f+\delta f$\ at some time,
the duration of its excursion above $f$\ is
\begin{equation} \label{deltat=}
\delta t=\frac{2}{\Delta\omega}\sqrt{\frac{2\delta f}{f}{ .}}
\end{equation}
From (\ref{Pft3}) we find the conditional probability density for $F\left(
t\right)  $ to reach $f+\delta f$\ if $F(t)>f$, as $Qye^{-Qy\delta f}$, and
hence
\begin{equation}
\overline{\delta t}=\frac{1}{\Delta\omega}\sqrt{\frac{2\pi}{Qyf}}{ .}
\end{equation}

Since the probability ${\cal P}\left(  f,t\right)  $\ for a recurrence to
occur at the time $t$\ does not depend on this time, and since the average
duration of the excursion is $\overline{\delta t}$, the average delay between
recurrences is here
\begin{equation}
\tau_{{\rm recur}}=\frac{\overline{\delta t}}{{\cal P}\left(
f,t\right)  }=\frac{2\pi}{\Delta\omega}\sqrt{\frac{y}{f}\frac{{\rm d}
f}{{\rm d}y}}e^{Q\left[  yf-\ln I_{0}\left(  y\right)  \right]},
\end{equation}
where $y$ is given by $I_1(y)=f\,I_0(y)$.

For $f$ sufficiently small so that $\ln\,I_0(f)\simeq f^2$ (for $f=0.2$ the relative error is $1\%$), we find from 
(\ref{saddle}) that $y\simeq 2f$, and this expression of the recurrence time reduces to (\ref{taurecur=}), that is exponentially large in Q. 

We notice that in this derivation the shape of the eigenvalue spectrum $p(\omega)$ hardly played any role,
we only used that it is smooth on the scale $2\pi/t$, where $t$ is the observation time.
So after times $t\gg 2 \pi / \Delta \omega$, where the individual 
levels are no longer resolved, there will be an exponentially long timescale for the pointer to recur.

}

\renewcommand{\thesection}{\Alph{section}}
\section{Effect of the bath on the off-diagonal sectors of the density matrix of S + M}
\setcounter{equation}{0}
\setcounter{figure}{0}
\renewcommand{\thesection}{\Alph{section}.}
\label{AppendixD}


 \hfill{{\it  Dop\'oty dzban wode\hspace{-1mm}\textrthook \ nosi, dop\'oki mu sie\hspace{-1mm}\textrthook \ ucho nie urwie} 
\footnote{A jug carries water only until its handle breaks off}}

\hfill{Polish proverb} 

\vspace{3mm}

\subsection{Full expression of $P_{\up\down}$ for large $N$}
  
\ZeText{

In Eq.  (\ref{6.21}) we have parametrized $P_{\uparrow\downarrow}(m,t)$ in terms of the function $A(m,t)$, which satisfies

\BEQ
\frac{\p A}{\p t}=\frac{2igm}{\hbar} - \frac{1}{NP_{\up\down}} \frac{\p P_{\up\down}}{\p t},       \label{D.a}
\EEQ       
with $A(m,0)=0$. In subsection 4.4, we have derived the equation (\ref{dPoff2}) for $P_{\up\down}$, from which $A(m,t)$ can be obtained for large 
$N$ at the two relevant orders (finite and in $1/N$). As we need $A(m,t)$ only at linear order in $\gamma$, we can replace 
in (\ref{dPoff2})  the quantity $X_{\up\down}(m,t)$ by its value for $\gamma=0$, 

\BEQ               X\equiv  X_{\up\down}(m,t)= \frac{2igt}{\hbar}- \frac{m}{\delta^2_0} ,      \label{D.b} 
\EEQ
which contains no $1/N$ term. We then insert (\ref{dPoff2}) in  (\ref{D.a}) to obtain

\BEA
\frac{\p A(m,t)}{\p t}\hspace{-1mm}=\hspace{-1mm}\frac{\gamma}{\hbar^2}\left\{\left(1-e^{2X}\right)(1+m)\tildeK_- +\left(1-e^{-2X}\right)(1-m)\tildeK_+ 
-\frac{2}{N}\left[   \frac{\p[ (1+m)\tildeK_-e^{X}]}{\p m}  e^X -\frac{\p [(1-m)\tildeK_+ e^{-X}]}{\p m}e^{-X} \right] \right\},
\label{D.c}
\EEA
where the combinations
$ \tildeK_\pm(m,t)=\tilde{K}_{t>}\left(  \Omega_{\uparrow}^{\pm}\right)  +\tilde{K}_{t<}\left(\Omega_{\downarrow}^{\pm}\right)$ were introduced in (\ref{Kpm}).
The functions $\tilde K _{t>}(\omega)$ and $\tilde K_{t<}(\omega)=\tilde K_{t>}^\ast(\omega)$ were defined by (\ref{TF}), (\ref{Ktilde}), 
(\ref{K>}) and (\ref{K<}), and the frequencies $\Omega^\pm_\up$ and $\Omega^\pm_\down$ by (\ref{ompmi=}).
 The initial condition is $A(m, 0) = 0$.

}

\subsection{Expansion for small $m$}
\ZeText{

The above result holds for arbitrary values of $m$ and $t$. However, since  in $P_{\up\down}(m,t)$ the values of $m$ remain small as $1/\sqrt{N}$,
only the first three terms in the expansion 
\BEQ 
A(m,t) \approx B(t) - i\Theta(t) m + \half D(t) m^2, \qquad         \label{D.d}
\EEQ
are relevant. 
 The time-dependence of these three functions, which vanish for $t=0$, will be elementary so that we will work out only their time derivatives, 
 which are simpler and which result from (\ref{D.c}).
 
 We note as $\Omega$ the frequency defined by
 \BEQ
	\Omega\equiv \frac{2g}{\hbar} \equiv\frac{\pi}{\tau_{\rm recur}}, \qquad          \label{Om=taurecur}
\EEQ
which is related to the period $\tau_{\rm recur}$  of the recurrences that arise from the leading oscillatory term 
$\exp(2iNg m t/\hbar)$  in  (\ref{6.21}) with $m$ taking the discrete values (\ref{eig}).
 We can then rewrite, up to the order $m^2$ and up to corrections in $1/N$,

\BEQ
\Omega_\up^\pm\approx \mp\Omega\mp \frac{2 J_2 m}{ \hbar} ,\qquad
\Omega_\down^\pm\approx \pm \Omega \mp \frac{2 J_2 m}{\hbar} ,\qquad  X= i\Omega t - \frac{m}{\delta_0^2}. 
  \label{D.7}
\EEQ
The expressions (\ref{K>}) and (\ref{K<}) for $\tilde K_{t>}(\omega)$ or $\tilde K_{t<}(\omega)$ then provide for their combinations (\ref{Kpm}) the expansion

\BEQ
 \tildeK_\pm(m,t)
  \approx {e^{\pm i \Omega t}} \int_{-\infty}^\infty \frac{\d \omega}{\pi} \, \tilde K\left(\omega\mp \frac{2 J_2 m}{\hbar}\right)\, 
 \frac{\omega \sin \omega t - \Omega \sin \Omega t 
 \mp  i \Omega(\cos \Omega t-\cos \omega t)}{\omega^2 -\Omega^2}+{\cal O}\left(\frac{1}{N}\right).          \label{dD.8} \EEQ
The required functions $B(t)$, $\Theta(t)$ and $D(t)$ are obtained by inserting (\ref{D.d}) and (\ref{dD.8}) into (\ref{D.c}). While the term of order $1/N$ in $B(t)$ 
provides a finite factor in $P_{\up\down}(m, t)$, the terms of order $1/N$ in $\Theta(t)$ and $D(t)$ provide negligible contributions. 
However that may be, it will be sufficient for our purpose to evaluate only the finite contribution to $B(t)$ and the large $t$ approximations for $\Theta(t)$ and $D(t)$.
 
 }
\subsection{The damping term $B(t)$}
\ZeText{

To find $B(t)$, we simply set $m=0$ in (\ref{D.c}) and (\ref{dD.8}). Next we employ the expression (\ref{Ktilde}) for $\tilde K(\omega)$ and
take advantage of the symmetry of the integrand with respect to $\omega$, which allows us to keep only the symmetric part 
of $\tilde K(\omega)$. This yields

 \BEQ
\frac{ \d B}{\d t} =  \frac{4\gamma 
\Omega \sin \Omega t}{ \hbar^2} \int_{-\infty}^\infty\frac{ \d \omega}{\pi}\,\tilde K(\omega)
\,\frac{   \cos \Omega t - \cos \omega t  }{\omega^2 - \Omega^2}
=  \frac{\gamma \Omega \sin \Omega t}{ 2  \pi } \int_{-\infty}^\infty \d \omega\, \omega \coth \frac{\hbar\omega}{2T}\,
 \exp\left({-\frac{|\omega|}{\Gamma}}\right)
\,\frac{   \cos \Omega t - \cos \omega t  }{\omega^2 - \Omega^2}. 
  \label{D.f}
\EEQ
where we  discarded corrections of order $1/N$. This entails the result for $B(t)$ presented in Eq. (\ref{D.8}) of the main text.
  For $t\ll1/\Gamma$, (\ref{D.f}) reduces to $\d B/\d t\sim ({\gamma \Gamma^2 \Omega^2}/{ 2 \pi})  t^3$ and hence
  
 \BEQ  \label{Bt4}
 B(t) \sim\frac{\gamma \Gamma^2 \Omega^2}{ 8 \pi} t^4 =\frac {\gamma \Gamma^2 g^2}{ 2 \pi \hbar^2} t^4.      \EEQ
 
 The $\omega$ integral in Eq. (\ref{D.8}) for $B(t)$ can be easily carried out numerically and the result is plotted in Fig 6.1 for typical values of the parameters. 
It is nevertheless instructive to carry out this integral explicitly. 
  This calculation is hindered by the non-analyticity of our Debye cutoff. However, since the result is 
 not expected to depend significantly on the shape of the cutoff ($\Gamma$ is the largest frequency of the model), we may 
replace the exponential cutoff in (\ref{Ktilde}) by a quasi Lorentzian cutoff, 
 
 \BEQ \label{newcutoff}
 \exp\left(-\frac{|\omega|}{\Gamma}\right)\mapsto \frac{4\tildeGamma^4}{4\tildeGamma^4+\omega^4};\qquad
\tilde{K}\left(\omega\right)  \mapsto \frac{\hbar^2\omega}{4(e^{\beta\hbar\omega}-1)} 
 \frac{4\tildeGamma^4}{4\tildeGamma^4+\omega^4}, 
 \EEQ
 where the factors 4 are introduced for later convenience. This expression ensures convergence while being analytic with simple poles. 
The cutoff  (\ref{newcutoff}) provides for $B(t)$ the same short time behavior as (\ref{Bt4}) if we make the connection

 \BEQ
  \tildeGamma=\sqrt{\frac{2}{\pi} }\,\Gamma.
  \label{GamGamtilde}
 \EEQ
 
  In order to integrate  the thus modified version of (\ref{D.f})
  over $\omega$, we first split $\cos \omega t$ into $\half\exp{i \omega t} +\half \exp{-i\omega t}$
 and then slightly rotate the integation contour  so that $\omega$ passes below $+\Omega$ and above $-\Omega$, instead of passing through these poles. 
 For each of the terms we can close the contour either in the upper or  lower half-plane, 
such that it decays for $|\omega|\to \infty$, and pick up the residues at the various poles. 
 The first set of poles, arising from the denominator of (\ref{D.f}), consist of $\pm\Omega$;
 since they lie on the real $\omega$-axis, they will produce a non-decaying long time behavior.
 The second set  of poles arise from the $\coth$, as exhibited by the expansion
 
\BEQ                  \coth \frac{\hbar \omega}{2T}=
 \sum_{ n=-\infty}^{\infty}  \frac{2T }{\hbar (\omega - i\Omega_n)},\qquad \Omega_n\equiv\frac{2\pi nT}{\hbar},
\EEQ
where the sum is meant as principal part for $n\to\pm \infty$; the frequencies $\Omega_n$ are known as Matsubara frequencies.
Thirdly, the cutoff (\ref{newcutoff}) provides the four poles $\pm\tildeGamma\pm i\tildeGamma$. 
We can also take advantage of the symmetry $\omega \to -\omega$, which associates pairwise complex conjugate residues. 
Altogether, we find

 \BEA  
 \frac{1}{\gamma\Omega} \frac{\d B}{\d t} &=&\coth \frac{\hbar\Omega}{2T}\,\frac{\tildeGamma^4}{4 \tildeGamma^4+\Omega^4} 
 (1-\cos 2 \Omega t)   +\frac{T}{\hbar} \sum_{n=1}^\infty \frac{\Omega_n}{\Omega_n^2+ \Omega^2}
 \frac{4\tildeGamma^4}{4 \tildeGamma^4+\Omega_n^4}  \left[\sin 2\Omega t  - 2 \exp({- \Omega_n  t}) \sin \Omega t \right]\nn\\ 
 &&+ \frac{1}{2}\Im \left\{
\coth \frac{(1+i)\hbar\tildeGamma}{ 2T}\, \frac{\tildeGamma^2}{2\tildeGamma^2+i\Omega^2} \, 
\left[\sin 2\Omega t - 2 \exp[{-(1-i)\tildeGamma t}]\,\sin \Omega t \right]\right\}\,
 + {\cal O}\left(\frac{1}{N}\right). 
\label{Bdotsum}
 \EEA
Now $B$ is easily obtained by integrating this from $0$ to $t$,

   \BEA  B(t)&=&\frac{\gamma}{2}\coth\frac{g}{T}\,
  \frac{\tildeGamma^4}{4\tildeGamma^4+\Omega^4}(2\Omega t-\sin 2\Omega t) \nn\\
 & &+ \sum_{n=1}^\infty \frac{4\gamma\tildeGamma^4\Omega_nT}{\hbar(4\tildeGamma^4+\Omega_n^4)} \,
  \left [\frac{\sin^2\Omega t}{\Omega^2+\Omega_n^2}+2\Omega\frac{(\Omega\cos\Omega t+\Omega_n\sin\Omega t)\exp({-\Omega_nt})-\Omega}{(\Omega^2+\Omega_n^2)^2}\right]  \\ && 
- \frac{\gamma\tildeGamma^2}{2}\Re\left\{ \coth \frac{(1+i)\hbar \tildeGamma}{2T} 
 \left[\frac{\sin^2\Omega t}{\Omega^2-2i\tildeGamma^2}+2\Omega
\frac{\left(\Omega\cos\Omega t+(1-i)\tildeGamma\sin\Omega t\right)\exp[{-(1-i)\tildeGamma t}]-\Omega}{(\Omega^2-2i\tildeGamma^2)^2}
\right]
\right\}, \nn
\label{D.13a}
\EEA
  where we made the residues at $(\pm1\pm i)\tildeGamma$  look as much as possible like the ones at $\Omega_n$.

  With these exact results in hand, let us discuss the relative sizes of the various terms.  
 The above complete formula exhibits some contributions that become exponentially small for sufficiently large $t$. 
 Such contributions are essential to ensure the behavior (\ref{Bt4}) of $B$ for $t\ll1/ \tildeGamma$, and also its behavior for 
 $t\ll\hbar / 2 \pi T$, but can be neglected otherwise. Moreover, we have $\hbar\tildeGamma\gg T$  and $\tildeGamma\gg\Omega$; 
 hence, within exponentially small corrections, the third term of (\ref{Bdotsum}) reduces, for $t\gg1/\tildeGamma$, to 
 $  -\Omega^2 \sin (2 \Omega t)/8 \tildeGamma^2$ and is therefore negligible compared to the first two terms. 
   In the first term of (\ref{Bdotsum}), the Debye cutoff is irrelevant, but it is needed in the second term to ensure convergence of the series.
 Restoring our exponential cutoff, we can write this series as
   
   \BEQ                       \frac{1}{2 \pi} \sum_{n=1}^\infty \frac{ n}{ n^2 + a^2} e^{-bn} ,\qquad   a\equiv \frac{\hbar \Omega}{2 \pi T} \ll 1,  
\qquad   b \equiv\frac{2 \pi T}{\hbar \Gamma}\ll 1,   
   \EEQ
which, within corrections of order $a^2$, is equal to

\BEQ                     
 \frac{1}{2 \pi} \sum_{n=1}^\infty \frac{ 1}{ n} e^{-bn}
= -  \frac{1}{2 \pi} \ln \left(1-e^{- b} \right) \sim \frac{1}{2 \pi} \ln \frac{\hbar \Gamma}{  2 \pi T }. 
    \EEQ
Altogether, returning to our original notations through use of (\ref{Om=taurecur}), we find from the first two terms of (\ref{Bdotsum}), for $t\gg\hbar/2 \pi T$:

\BEQ                       
\frac{\tau_{\rm recur}}{\gamma}\frac{\d B}{\d t} = \frac{\pi}{4} \coth \frac{g}{T}\left(1 - \cos \frac{2 \pi t}{\tau_{\rm recur}} \right)
+ \frac{1}{2} \ln \frac{\hbar \Gamma }{ 2\pi T} \sin \frac{2 \pi t }{ \tau_{\rm recur}} .  
\EEQ
Likewise, the function $B(t)$ itself behaves in this region as

  \BEQ  
  B(t) = \frac{\gamma\pi}{4} \coth\frac{g}{T}\left(\frac{ t}{\tau_{\rm recur}} -\frac{1}{2\pi}  \sin \frac{2 \pi t }{ \tau_{\rm recur}}\right) + 
  \frac{\gamma}{4 \pi}\ln \frac{\hbar \Gamma }{2\pi T} \left( 1- \cos \frac{2 \pi t }{ \tau_{\rm recur}} \right)
   -  \frac{\gamma \zeta(3) }{  \pi^3}\frac{ g^2}{T^2},
   \label{D.19}
  \EEQ
  where the last piece arises,  in the considered approximation, from the last term of the sum in (\ref{D.13a}).
  
}

\subsection{Approximations for $\Theta(t)$ and $D(t)$}

\ZeText{

We have just seen that the dominant contribution to $B(t)$ in the region $ t\gg\hbar/2 \pi T$ originates from the poles $\omega = \pm \Omega$
 of the integrand of (\ref{D.f}). Likewise, as we need only an estimate of $\Theta(t)$ and $D(t)$, we will evaluate approximately the integral in (\ref{dD.8})
 by picking up only the contributions of these poles. As we did for $B(t)$, we deform and close the integration contour in the upper or in the lower half-plane, 
 but we now disregard the singularities of $\tilde K(\omega\mp 2 J_2 m/\hbar)$. This approximation amounts to make the replacements

\BEA
&&\frac{\omega \sin \omega t - \Omega \sin \Omega t}{\omega^2-\Omega^2} \mapsto \frac{\pi}{2} \cos(\Omega t) [ \delta(\omega-\Omega) + \delta(\omega + \Omega) ],  \\  
&&\frac{\Omega(\cos \Omega t-\cos \omega t)}{\omega^2 -\Omega^2}\mapsto  \frac{\pi}{2} \sin(\Omega t) [ \delta(\omega-\Omega) + \delta( \omega+ \Omega) ] ,  
   \EEQ
which as we have seen are justified for $t\gg \hbar/2 \pi T$. As a result, we find the time-independent expressions for $\tildeK_\pm$,

\BEQ 
\tildeK_\pm \approx \half [ \tilde K(\Omega\mp  2J_2 m) + \tilde K( - \Omega \mp 2J_2 m) ].        
\label{D.22}
 \EEQ
 
We now return to our original notations by use of (\ref{Om=taurecur}) for $\Omega$ and (\ref{D.7}) for $X$, 
rewriting the dominant part  of (\ref{D.c}) as

\BEQ \frac{\tau_{\rm recur}}{\gamma} \frac{\d A}{\d t} = \frac{\pi }{2 \hbar g} \left[(1-e^{2X})(1+m) \tildeK_- + (1-e^{-2X})(1-m)\tildeK_+ \right].
\label{D.21}   
\EEQ
In order to generate $\Theta(t)$ and $D(t)$ through the expansion (\ref{D.d}) of $A(m,t)$ in powers of $m$, 
we insert into (\ref{D.21})  the expansions

\BEQ           \left [ 1 - e^{\pm 2X}\right] (1\pm m)\approx \left[1 - e^{\pm 2i\Omega t} \right] \pm  
\left[ 1 +  e^{\pm 2i\Omega t} \left(\frac{2}{ \delta_0^{2}}  - 1 \right) \right] m 
+ 2 e^{\pm  2i\Omega t}  \left(\frac{1}{\delta_0^{2}} -\frac{1}{ \delta_0^{4}}\right) m^2 ,               
\EEQ

\BEQ  
\frac{4}{\hbar}\tildeK_\pm  \approx  g \coth\frac{g}{T}\pm  J_2 m - \frac{J_2 ^2 }{ T^2 \sinh^2g/T} \left(\coth \frac{g}{T} - \frac{T}{g}\right) m^2 .    
\EEQ
Gathering, in the resulting expansion of $A(m,t)$, the terms in $m$, we find (for $g\ll T$)

\BEQ
 \frac{\tau_{\rm recur}}{\gamma} \frac{\d \Theta}{\d t} = - \frac{\pi}{4}
\left [ \left(\frac{2}{ \delta_0^2}  - 1\right) \coth \frac{g}{T} + \frac{J_2}{g} \right] \sin \frac{2 \pi t }{ \tau_{\rm recur}} \sim 
 - \frac{\pi}{4}\left[ \left(\frac{2}{ \delta_0^2}  - 1\right)\frac{ T}{g}+ \frac{J_2}{g}\right ] \sin \frac{2 \pi t }{ \tau_{\rm recur}} ,                        
 \EEQ
which is integrated as

\BEQ 
\label{ThetaRes}
\Theta(t) \sim  - \frac{\gamma}{8g}\left [ \left(\frac{2 }{ \delta_0^2}  - 1\right) T+ J_2\right ] \left[ 1 -  \cos\frac{2 \pi t }{ \tau_{\rm recur}}\right].    
\EEQ

Likewise, the terms in $m^2$ yield

\BEQ                 \frac{\tau_{\rm recur}}{\gamma}\,\frac{ \d D}{\d t} \sim \frac{\pi}{2}
\left[\frac{ J_2^2}{T^2\sinh^2 {g}/{T}}\left(\coth \frac{g}{T} -\frac{ T}{g}\right) - \frac{J_2}{g}\right]\left( 1 - \cos\frac{2 \pi t}{ \tau_{\rm recur}}\right) 
 + \frac{\pi}{2} \left[ 2 \coth \frac{g}{T}\left (\frac{1}{\delta_0^2} - \frac{1}{\delta_0^4}\right) 
 - \frac{2J_2 }{g \delta_0^2}\right ] \cos\frac{2 \pi t } {\tau_{\rm recur}} .  
\EEQ
The first bracket simplifies for $g\ll T$ into

\BEQ         \frac{ J_2^2}{T^2\sinh^2 {g}/{T}}\left(\coth \frac{g}{T} -\frac{ T}{g}\right) - \frac{J_2}{g}
 \sim \frac{J_2}{g}\left(\frac{J_2}{3T} - 1\right)    .   
\EEQ
We shall only need the values of $D(t)$ at the recurrence times $p \tau_{\rm recur}$. Integration of the factors $\cos2 \pi t / \tau_{\rm recur}$ 
generates $\sin2 \pi t / \tau_{\rm recur}$, which vanishes at these times. We have therefore the compact result

\BEQ        \label{DRes}        
D(p\tau_{\rm recur})  \simeq p\times D(\tau_{\rm recur})  = p\,\frac{\pi \gamma }{2} \frac{J_2}{g}\left(\frac{J_2}{3T} - 1\right).    
 \EEQ

}

\renewcommand{\thesection}{\Alph{section}}
\section{Time dependence of the registration process}
\setcounter{equation}{0}
\setcounter{figure}{0}
\renewcommand{\thesection}{\Alph{section}.}
\label{AppendixE}


\hfill{\it Time heals all wounds}

\hfill{Proverb}

\vspace{3mm}

\ZeText{

The location $\mu(t)$ of the peak of the distribution $P(m,t)$ increases in time according to 
(\ref{t(mu)}) where $\phi(m)$ is defined by (\ref{Phim}). We wish in \S~\ref{section.7.2.3} and 
\S~\ref{section.7.2.4} to obtain an algebraic approximation for $\mu(t)$ at all times.
To this aim, we will represent
$1/v(\mu)$  by its Mittag-Leffler expansion
\begin{equation}
\mytext{\textcurrency ML\textcurrency \qquad}
\frac{\gamma T}{\hbar v(m)}
\equiv \frac{1}{\phi(m)[1-m\coth\,\phi(m)]}=\sum_{i}\frac{m_{i}
}{[(1-m_{i}^{2})({\rm d}\phi/{\rm d}m_{i})-1]\phi(m_{i})}\,\frac
{1}{m-m_{i}}{ ,} \label{ML}
\end{equation}
which sums over all real or complex values $m=m_i$ where $v(m)=0$.

}

\subsection{Registration for second-order transition of {\rm M}}


\ZeText{

For $q=2$, it is sufficient for our purpose to keep in the expansion (\ref{ML}) only the real poles $m_i$. 
This truncation does not affect the
vicinity of the (stable or unstable) fixed points where the motion of $\mu(t)$
is slowest, and provides elsewhere a good interpolation provided $T/J$ is not
too small. Three values $m_{i}$ occur here, namely $-m_{{\rm B}}$,
$m_{\Uparrow}\simeq m_{{\rm F}}$ and $m_{\Downarrow}\simeq-m_{{\rm F}}$,
with $m_{{\rm B}}\ll m_{{\rm F}}$, so that we find over the whole range
$0<\mu<m_{{\rm F}}$, through explicit integration of (\ref{t(mu)}),

\begin{equation}
\mytext{\textcurrency Atofmu\textcurrency \qquad}
\frac{t}{\tau_{{\rm reg}}
}=\ln\frac{m_{{\rm B}}+\mu}{m_{{\rm B}}}+a\ln\frac{m_{{\rm F}}^{2}
}{m_{{\rm F}}^{2}-\mu^{2}}{ ,} \label{Atofmu}
\end{equation}
where the coefficient $a$, given by
\begin{equation}
\mytext{\textcurrency Aa=\textcurrency \qquad}
a=\frac{T(J-T)}{J[T-J(1-m_{{\rm F}}^{2})]  }{ ,} \label{Aa=}
\end{equation}
decreases with temperature from $a=1$ at $T=0$ to $a=\half$ for $T=J$. 
For short times, such that $\mu\ll m_{\rm F}$, we recover 
from the first term of (\ref{Atofmu}) the evolution (\ref{mu1}) of $\mu(t)$.
When $\mu$ approaches $m_{\rm F}$, the second term dominates, but as long as $m_{\rm F}-m$ is of order
$m_{\rm B}$ the time needed for $\mu$ to reach $m$ is of order $\tau_{\rm reg}\ln(m_{\rm F}/m_{\rm B})$.
We define the cross-over by writing that the two logarithms of (\ref{Atofmu}) are equal, which yields 
$\mu=m_{\rm F}-\half m_{\rm B}$.
The time $\tau_{\rm reg}'$ during which $\mu(t)$ goes from 0 to $m_{\rm F}-\half m_{\rm B}$,
termed the second characteristic registration time, is then given by (\ref{tauprime}), that is,

\begin{equation}
\mytext{\textcurrency tauprime\textcurrency \qquad}
\tau_{{\rm reg}}^{\prime
}=\tau_{{\rm reg}}(1+a)\ln\frac{m_{{\rm F}}
}{m_{{\rm B}}}{ .} \label{Atauprime}
\end{equation}

When $\mu$ approaches $m_{\rm F}$ in the regime $m_{\rm F}-\mu\ll m_{\rm B}$, we can invert (\ref{Atofmu}) as

\BEQ
\mu(t)=m_{\rm F}\left[1-\half\left(\frac{m_{\rm F}}{m_{\rm B}}\right)^{1/a}\,\exp\left(-\frac{t}{a\tau_{\rm reg}}\right)\right],
\EEQ
which exhibits the final exponential relaxation. We can also invert this relation in the limiting cases 
$T\to J$ and $T\to 0$. If $T$\ lies close to the transition temperature, we have $m_{{\rm F}}
\sim\sqrt{3(J-T)/J}$ and $a=\frac{1}{2}$. Provided the coupling is weak so
that $m_{{\rm B}}=g/(J-T)\ll m_{{\rm F}}$, we find

\begin{equation}
\mytext{\textcurrency muallt\textcurrency \qquad}
\mu(t)=\frac{m_{{\rm B}
}m_{{\rm F}}}{m_{{\rm B}}^{2}+m_{{\rm F}}^{2}e^{-2t/\tau
_{{\rm reg}}}}\left[  \sqrt{m_{{\rm B}}^{2}+(m_{{\rm F}}
^{2}-m_{{\rm B}}^{2})e^{-2t/\tau_{{\rm reg}}}}-m_{{\rm B}
}m_{{\rm F}}e^{-2t/\tau_{{\rm reg}}}\right]  { .} \label{muallt}
\end{equation}
This expression encompasses all three regimes of \S~\ref{section.7.2.3}, namely, $\mu\sim
m_{{\rm B}}t/\tau_{{\rm reg}}$ for $t\ll\tau_{{\rm reg}}$, $\mu$
running from $m_{{\rm B}}$ to $m_{{\rm F}}$ for $t$ between
$\tau_{{\rm reg}}$ and $\tau_{{\rm reg}}^{\prime}$, and

\begin{equation}
\mu(t)\approx m_{{\rm F}}\left(  1-\frac{m_{{\rm F}}^{2}}{2m_{{\rm B}
}^{2}}e^{-2t/\tau_{{\rm reg}}}\right)
\end{equation}
for $t-\tau_{{\rm reg}}^{\prime}\gg\tau_{{\rm reg}}$. In the low
temperature regime ($T\ll J$, with $m_{{\rm B}}\sim g/J$\ and $a\sim1$), we
can again invert (\ref{Atofmu}) as

\begin{equation}
\mu\left(  t\right)  =\frac{1}{2m_{{\rm B}}}\left[  \sqrt{4m_{{\rm B}
}^{2}\left(  m_{{\rm F}}^{2}-m_{{\rm B}}^{2}\right)  +\left(
2m_{{\rm B}}^{2}-m_{{\rm F}}^{2}e^{-t/\tau_{{\rm reg}}}\right)  ^{2}
}-m_{{\rm F}}^{2}e^{-t/\tau_{{\rm reg}}}\right]  { ,}
\end{equation}
encompassing the same three regimes; for $t-\tau_{{\rm reg}}^{\prime}
\gg\tau_{{\rm reg}}$, we now have
\begin{equation}
\mu(t)\approx m_{F}\left(  1-\frac{m_{{\rm F}}}{2m_{{\rm B}}}
e^{-t/\tau_{{\rm reg}}}\right)  { .}
\end{equation}

}

\subsection{Registration for first-order transition of {\rm M}}
\label{section.7.2.4A}

\ZeText{

For $J_4\neq0$, such as the $q=4$ case with $J_2=0$ and $J_4=J$, we need to account for the presence of the minimum of $v(m)$ at $m=m_{\rm c}$. 
To this aim, we still truncate the Mittag-Leffler expansion  (\ref{ML}) of $1/v(m)$. 
However, we now retain not only the real poles
but also the two complex poles near $m_{{\rm c}}$ which govern the
minimum of $v(m)$. These poles are located at
\begin{equation}
m_{{\rm c}}\pm i\delta m_{{\rm c}}{,\qquad}\delta m_{{\rm c}
}^{2}=\frac{m_{{\rm c}}(1-m_{{\rm c}}^{2})^{2}}{1+2m_{{\rm c}}^{2}
}\frac{g-h_{{\rm c}}}{T}\sim m_{{\rm c}}\left(  \frac{g}{T}
-\frac{2m_{{\rm c}}}{3}\right)  { .}
\end{equation}
The real pole associated with the repulsive fixed point lies at
$-m_{{\rm B}}\sim-2m_{{\rm c}}$, and the ferromagnetic poles lie close
to $\pm m_{{\rm F}}\sim\pm1$. We have thus, at lowest order in
$T/J\simeq3m_{{\rm c}}^{2}$ and in $g/T\sim2m_{{\rm c}}/3$, but with
$T/J$ sufficiently large so that we can drop the other complex poles,
\begin{equation}
\mytext{\textcurrency ML4\textcurrency \qquad}
\frac{\gamma T}{\hbar v(m)}
=\frac{m_{{\rm c}}-\frac{1}{2}(m-m_{{\rm c}})}{(m-m_{{\rm c}}
)^{2}+\delta m_{{\rm c}}^{2}}+\frac{1}{3(m+2m_{{\rm c}})}+\frac
{2Tm}{J(1-m^{2})}{ .} \label{ML4}
\end{equation}

Hence the time-dependence of the peak $\mu(t)$ of $P_{\uparrow\uparrow}(m,t)$
is given through integration of (\ref{t(mu)}) as

\begin{equation}
\mytext{\textcurrency evol4\textcurrency \qquad}
\frac{t}{\tau_{{\rm reg}}
}=\frac{1}{\pi}\left(  \frac{\pi}{2}+\arctan\frac{\mu-m_{{\rm c}}}{\delta
m_{{\rm c}}}\right)  +\frac{\delta m_{{\rm c}}}{\pi m_{{\rm c}}
}\left[  \frac{1}{4}\ln\frac{m_{{\rm c}}^{2}}{(\mu-m_{{\rm c}})^{2}+\delta
m_{{\rm c}}^{2}}+\frac{1}{3}\ln\frac{\mu+2m_{{\rm c}}}{2m_{{\rm c}}
}+\frac{T}{J}\ln\frac{1}{1-\mu^{2}}\right]  { ,} \label{Aevol4}
\end{equation}
where we introduced the registration time

\begin{equation}
\mytext{\textcurrency Ataureg4\textcurrency \qquad}
\tau_{{\rm reg}}\equiv
\frac{\pi\hbar m_{{\rm c}}}{\gamma T\delta m_{{\rm c}}}=\frac{\pi\hbar
}{\gamma T}\sqrt{\frac{m_{{\rm c}}T}{g-h_{{\rm c}}}}{ ,}
\label{Ataureg4}
\end{equation}
with $m_{{\rm c}}=\sqrt{T/3J}=3h_{{\rm c}}/2T$. 

The initial evolution (\ref{muttau1}) is recovered from (\ref{Aevol4}) for $\mu\ll m_{\rm c}$ and $t \ll \hbar/\gamma T$. 
It matches the bottleneck stage in which $\mu(t)$ varies slowly around the value $m_{{\rm c}}$ on the time
scale $\tau _{{\rm reg}}$. Then, the right-hand side of (\ref{Aevol4}) is dominated by its first term, so that the
magnetization increases from $m_{{\rm c}}-\delta m_{{\rm c}}$ to $m_{{\rm c}}+\delta m_{{\rm c}}$ 
between the times $t=\tau_{\rm reg}/4$ and $t=3\tau_{\rm reg}/4$,\ according to:

\begin{equation}
\mu(t)=m_{{\rm c}}-\delta m_{{\rm c}}{\rm cotan}\,\frac{\pi t}{\tau_{{\rm reg}}}.
\end{equation}

After $\mu$ passed the bottleneck, for $\mu-m_{\rm c}\gg\delta m_{\rm c}$, (\ref{Aevol4}) provides

\begin{equation}
\label{fullt=tau}
t=\tau_{\rm reg}+\tau_{1}
\left(-\frac{m_{{\rm c}}}{\mu-m_{{\rm c}}}+\frac{1}{2}\ln\frac{m_{{\rm c}}}{\mu-m_{{\rm c}}}
+\frac{1}{3}\ln\frac{\mu+2m_{{\rm c}}}{2m_{{\rm c}}}+\frac{T}{J}\ln
\frac{1}{1-\mu^{2}}\right){ ,}
\end{equation}
which is nearly equal to $\tau_{\rm reg}$ within corrections of order $\tau_1=\hbar/\gamma T$, 
as long as $\mu$ is not very close to 1. The final exponential relaxation takes place on the still shorter 
scale $\hbar/\gamma J$.

}

\renewcommand{\thesection}{\Alph{section}}
\section{Effects of bifurcations}
\setcounter{equation}{0}
\setcounter{figure}{0}
\renewcommand{\thesection}{\Alph{section}.}
\label{AppendixF}

\hfill{\it Of je door de hond of de kat gebeten wordt, het  blijft om het even\footnote{Whether bitten by the dog or the cat, the result is equal}}

\hfill{Dutch proverb}

\vspace{3mm}
\ZeText{

In subsection ~\ref{section.7.3} we consider situations in which Suzuki's slowing down is present,
namely the preparation of the initial metastable state for $q=2$ and the possibility of false
registrations. We gather here some derivations.

The Green's function $G(m,m^{\prime},t-t^{\prime})$ associated to the equation (\ref{dPdt})
for $P_{\rm M}(m,t)$ will be obtained from the backward equation 

\begin{eqnarray}
\frac{\partial}{\partial
t^{\prime}}G(m,m^{\prime},t-t^{\prime})+v(m^{\prime})\frac{\partial}{\partial
m^{\prime}}G(m,m^{\prime},t-t^{\prime})+\frac{1}{N}[w(m^{\prime})\frac{\partial^2}{\partial m^{\prime}{}^2}G(m,m^{\prime
},t-t^{\prime})]=-\delta(m-m^{\prime})\delta(t-t^{\prime}){,}\label{dGback}  
\end{eqnarray}
where $t'$ runs down from $t+0$ to $0$. Introducing the time scale $\tau_{{\rm reg}}$ defined by (\ref{taureg2}) and 
using the expression (\ref{vappr}) for $v(m')$ for small $m'$ together with the related  $w(m')\approx\gamma gt/\hbar $, 
we have to solve the equation
\begin{equation}
\mytext{\textcurrency dGsimp\textcurrency \qquad}
\left[  \tau_{{\rm reg}
}\frac{\partial}{\partial t^{\prime}}+(m_{{\rm B}}+m^{\prime}
)\frac{\partial}{\partial m^{\prime}}+\frac{1}{N}\frac{T}{J-T}\frac
{\partial^{2}}{\partial{m^{\prime}}^{2}}\right]  G(m,m^{\prime},t-t^{\prime
})=0{ ,}\label{dGsimp}
\end{equation}
with the boundary condition $G(m,m^{\prime},0)=\delta(m-m^{\prime})$. Its
solution in terms of $m^{\prime}$ has the Gaussian form
\begin{equation}
\mytext{\textcurrency G=\textcurrency \qquad}
G(m,m^{\prime},t)=A(m,t)\sqrt
{\frac{N}{2\pi D(m,t)}}\exp\left\{  -\frac{N[m^{\prime}-\mu^{\prime}
(m,t)]^{2}}{2D(m,t)}\right\}  { ,}\label{G=}
\end{equation}
where the coefficients $\mu^{\prime}$, $D$ and $A$ should be found by insertion into (\ref{dGsimp}).

As in \S~\ref{section.7.2.3}, the evolution of $P_{\rm M}(m,t)$ takes place in three stages: (i)
\textit{widening} of the initial distribution, which here takes place over the
bifurcation $-m_{{\rm B}}$; (ii) \textit{drift} on both sides of
$-m_{{\rm B}}$ towards $+m_{{\rm F}}$ and $-m_{{\rm F}}$; (iii)
narrowing around $+m_{{\rm F}}$ and $-m_{{\rm F}}$ of the two final
peaks, which evolve \textit{separately towards equilibrium}. We are interested here only in the first
two stages.  During the first stage, the relevant values of $m$ lie
in the region where the approximation (\ref{VWsimp}) holds. The functions of
$m$ and $t$: $\mu^{\prime}$, $D$ and $A$, satisfy according to (\ref{dGsimp})
the equations
\begin{equation}
\tau_{{\rm reg}}\frac{\partial\mu^{\prime}}{\partial t}=-m_{{\rm B}}
-\mu^{\prime}{,\qquad}\frac{1}{2}\tau_{{\rm reg}}\frac{\partial
D}{\partial t}=\frac{T}{J-T}-D{,\qquad}\tau_{{\rm reg}}\frac{\partial
A}{\partial t}=-A{ ,}
\end{equation}
and the boundary condition $G(m,m^{\prime},0)=\delta(m-m^{\prime})$ for
$t^{\prime}=t-0$ yields
\begin{equation}
\mytext{\textcurrency AmuD\textcurrency \qquad}
\mu^{\prime}=-m_{{\rm B}
}+(m+m_{{\rm B}})e^{-t/\tau_{{\rm reg}}}{,\qquad}D=\frac{T}
{J-T}(1-e^{-t/\tau_{{\rm reg}}}){,\qquad}A=e^{-t/\tau_{{\rm reg}}
}{ .}\label{AmuD}
\end{equation}
As function of $m$, the probability
\begin{equation}
\mytext{\textcurrency AP=GP\textcurrency \qquad}
P_{\rm M}(m,t)=\int{\rm d}m^{\prime
}G(m,m^{\prime},t)P_{\rm M}(m^{\prime},0)\label{AP=GP}
\end{equation}
given by (\ref{G=}), (\ref{AmuD}) involves fluctuations which increase exponentially as
$\exp({t/\tau_{{\rm reg}}})$.

In the second stage, the time is sufficiently large so that $P_{\rm M}(m,t)$ extends
over regions of $m$ where the linear approximation (\ref{VWsimp}) for $v(m)$
fails; we must account for the decrease of $|v(m)|$, which vanishes at $m=\pm
m_{F}$. We therefore cannot comply directly with the boundary condition for
$G(m,m^{\prime},t-t^{\prime})$ at $t^{\prime}=t$ since it requires $m^{\prime
}$ to be large as $m$. However, during this second stage $P_{\rm M}(m,t)$ is not
peaked, so that diffusion is negligible compared to drift. The corresponding
Green's function, with its two times $t$ and $t^{\prime}$ taken during this
stage, is given according to (\ref{Pw=0}) by

\begin{equation}
\mytext{\textcurrency AGw=0\textcurrency \qquad}
G(m,m^{\prime},t-t^{\prime})=\frac{1}{v(m)}\delta\left(  t-t^{\prime}-\int_{m^{\prime}}^{m}
\frac{{\rm d}m^{\prime\prime}}{v(m^{\prime\prime})}\right)  { .}
\label{AGw=0}
\end{equation}
We can now match the final time of (\ref{G=}), (\ref{AmuD}) with the initial
time of (\ref{AGw=0}), using the convolution law for Green's functions. 
This yields an approximation for $G(m,m^{\prime},t)$ valid up to the final
equilibration stage. We therefore define the function $\mu^{\prime}(m,t)$ by
the equation
\begin{equation}
\mytext{\textcurrency Amumt\textcurrency \qquad}
t=\int_{\mu^{\prime}(m,t)}
^{m}\frac{{\rm d}m^{\prime\prime}}{v(m^{\prime\prime})}{ ,}
\label{Amumt}
\end{equation}
of which (\ref{AmuD}) is the approximation for small $m$ and $\mu^{\prime}$.
For $m>-m_{{\rm B}}$, we have $m>\mu^{\prime}>-m_{{\rm B}}$ and
$v\left(  m^{\prime\prime}\right)  >0$; for $m<-m_{{\rm B}}$ we have
$m<\mu^{\prime}<-m_{{\rm B}}$ and $v\left(  m^{\prime\prime}\right)  <0$.
We also note that the convolution replaces $A=e^{-t/\tau_{{\rm reg}}}$ by
\begin{equation}
A(m,t)=\frac{v[\mu^{\prime}(m,t)]}{v(m)}=\frac{\partial\mu^{\prime}
(m,t)}{\partial m}{ .}
\end{equation}

Altogether the Green's function (\ref{G=}) reads
\begin{equation}
\mytext{\textcurrency AfinalG\textcurrency \qquad}
G(m,m^{\prime},t)=\frac
{v(\mu^{\prime})}{v(m)}\sqrt{\frac{N(J-T)}{2\pi T(1-e^{-2t/\tau_{{\rm reg}
}})}}\exp\left[  -\frac{N(J-T)(m^{\prime}-\mu^{\prime})^{2}}{2T(1-e^{-2t/\tau
_{{\rm reg}}})}\right]  { ,}\label{AfinalG}
\end{equation}
where $\mu^{\prime}=\mu^{\prime}(m,t)$ is found through (\ref{Amumt}). The resulting distribution function $P_{\rm M}(m,t)$,
obtained from (\ref{AP=GP}), (\ref{AfinalG}) and $P_{\rm M}(m,0)\propto 
\exp[-N(m-\mu_0)^2/2\delta_0^2]$, is expressed by (\ref{AfinalG}) or, in the main text, by (\ref{Psplit}) 
with (\ref{delta1t}) for $\delta_1(t)$. 
Notice that here we allowed for a finite value $\mu_0$ of the average magnetization in the initial state.

We have studied in \S~\ref{section.7.3.2} the evolution of $P_{\rm M}(m,t)$ for $g=0$ and for an
unbiased initial state. For $m_{\rm B}=g/(J-T)\neq 0$ and a non-vanishing expectation value of $\mu_0$ of $m$ 
in the initial state, the dynamics of $P_{\rm M}(m,t)$ is explicitly found from (\ref{AfinalG}) by noting that 
$m_{\rm B}\ll m_{\rm F}$; the expression (\ref{ML}) for $v(m)$ thus reduces to

\BEQ
\frac{1}{\tau_{\rm reg}v(m)}=\frac{1}{m+m_{\rm B}}+\frac{2am}{m_{\rm F}^2-m^2},
\EEQ
with $\tau_{\rm reg}=\hbar/\gamma(J-T)$ and $a$ defined by (\ref{Aa=}).
Hence, the relation (\ref{Amumt}) between $\mu'$, $m$ and $t$ reads

\BEQ
\frac{t}{\tau_{\rm reg}}=\ln\frac{m+m_{\rm B}}{\mu'+m_{\rm B}}+a\ln\frac{m_{\rm F}^2-\mu'{}^2}{m_{\rm F}^2-m^2}.
\EEQ
For large $N$, the quantities $\mu'$, $m_0$ and $m_{\rm B}$ are small as $1/\sqrt{N}$, except at the very
large times when $P_{\rm M}(m,t)$ is concentrated near $+m_{\rm F}$ and $-m_{\rm F}$. We can thus write (\ref{P=GP})
as 

\BEQ 
\label{AbestP}
P_{\rm M}(m,t)=\frac{1}{\sqrt{\pi}}\,\frac{\p\xi}{\p m}\,e^{-(\xi-\xi_0)^2},
\EEQ
where we introduced the functions

\BEQ
\label{Abestksi}
\xi(m,t)=\sqrt{3a}\frac{m+m_{\rm B}}{m_{\rm F}}\left(\frac{m_{\rm F}^2}{m_{\rm F}^2-m^2}\right)^a
\frac{\delta_1}{\delta_1(t)}e^{-(t-\tau_{\rm flat})/\tau_{\rm reg}},
\EEQ
\BEQ \label{Aksi0}
\xi_0(t)\equiv\sqrt{\frac{N}{2}}\frac{m_{\rm B}+\mu_0}{\delta_1(t)}.
\EEQ
The characteristic time $\tau_{\rm flat}$ is the same as (\ref{tauflat}), it is large as $\half\ln N$. 
The function $\delta_1(t)$ and the parameter $\delta_1$ are defined in (\ref{delta1t})

The expression (\ref{AbestP}) encompasses (\ref{Pfirst}), (\ref{Pflat}), (\ref{Pscal}) and (\ref{Psmallg}), 
which were established in the special case
where the distribution is symmetric ($m_{\rm B}=\mu_0=0$) and/or when $m$ is small as $1/\sqrt{N}$.
For $t\gg\tau_{\rm reg}$ we reach Suzuki's scaling regime characterized by the scaling parameter (\ref{Abestksi}),
in which $\delta_1(t)$ reduces to the constant $\delta_1$ and in which $m_{\rm B}$ can be disregarded.
The asymmetry of $P_{\rm M}(m,t)$ then arises only from the constant $\xi_0$.
Even in the presence of this assymetry, the time $t=\tau_{\rm flat}$ still corresponds to a flat $P_{\rm M}(m,t)$,
in the sense that the curvature of $P_{\rm M}(m,\tau_{\rm flat})$ at $m=0$ vanishes.

}

\renewcommand{\thesection}{\Alph{section}}
\section{Density operators for beginners}
\setcounter{equation}{0}
\setcounter{figure}{0}
\renewcommand{\thesection}{\Alph{section}.}
\label{AppendixG}

\hfill{\it Begin at the beginning}

\hfill{\it  and go on till you come to the end:}

\hfill{\it  then stop}

\hfill{Lewis Carroll, Alice's Adventures in Wonderland}


\vspace{3mm}
 \ZeText{

 In elementary courses of quantum mechanics, a state is usually represented by a vector  $|\psi\rangle$ in Hilbert space (or a ket,  or a wave function).  Such a definition is too
  restrictive. On the one hand, as was stressed by Landau \cite{landau,Landau1927}, if the considered system is not isolated and presents quantum correlations 
  with another system, its properties cannot be described by means of a state vector. On the other hand, as was stressed by von Neumann \cite{vNeumann}, 
   an incomplete preparation does 
  not allow us to assign a unique state vector to the system; various state vectors are possible, with some probabilities, and the formalism of quantum statistical mechanics is needed. 
  Both of these circumstances occur in a measurement process: The tested system is correlated to the apparatus, and the apparatus is macroscopic. 
   The opinion, too often put forward, that the (mixed) post-measurement state cannot be derived from the Schr\"odinger equation, originates from the will 
   to work in the restricted context of pure states. This is why we should consider, to understand quantum measurement processes, 
   the realistic case of a mixed initial state for the apparatus, and subsequently study the time-dependent mixed state for the tested system and the apparatus.

   The more general formulation of quantum mechanics that is needed requires the use of density operators, and is presented in section 10
    in the context of the statistical interpretation of quantum mechanics. We introduce here, for teaching purposes, an elementary introduction
     to \S~\ref{fin10.1.4}. In quantum (statistical) mechanics, a state is represented by a {\it density operator} $\hat{\cal D}$ or, 
in a basis $|i\rangle$  of the Hilbert space, by a {\it density matrix} with elements $\langle i|\hat{\cal D}|j\rangle$. 
The {\it  expectation value} in this state of an observable $\hat O$ (itself represented on the basis $|i\rangle$  by the matrix $\langle i|\hat O|j\rangle$) 
is equal to 

\BEQ  \langle \hat O\rangle ={\rm tr} \,\hat{\cal D} \hat O=\sum _{ij} \langle i|\hat{\cal D}|j\rangle  \langle j|\hat O|i\rangle .  
\EEQ
This concept encompasses as a special case that of state vector, as the expectation value of $\hat O$ in the state $|\psi \rangle$,

\BEQ 
 \langle \hat O\rangle = \langle \psi |\hat O|\psi \rangle =\sum_{ij} \langle \psi |j\rangle  \langle j|\hat O|i\rangle  \langle i|\psi \rangle,
 \EEQ
is implemented by associating with $|\psi \rangle$  the density operator $\hat{\cal D}= |\psi \rangle  \langle \psi |$ 
or the density matrix $\langle i|\hat{\cal D}|j\rangle =\langle i|\psi \rangle \langle \psi |j\rangle$, referred to as a ``pure state'' in this context.

 Density operators have several characteristic properties. (i) They are {\it Hermitean}, $\hat{\cal D}=\hat{\cal D}^\dagger$, 
 (i. e., $\langle j|\hat{\cal D}|i\rangle=\langle i|\hat{\cal D}|j\rangle^\ast$),  implying that the expectation
 value (G1) of a Hermitean observable is real. (ii) They are {\it normalized}, ${\rm tr}\,\,\hat{\cal D}=1$, meaning that the expectation value of the 
 unit operator is $1$. 
 (iii) They are {\it non-negative}, $\langle \phi|\hat{\cal D}|\phi\rangle\ge 0$ $\forall \,|\phi\rangle$, meaning that the variance 
 $\langle \hat O^2\rangle  - \langle \hat O\rangle^2$ of any Hermitean observable $\hat O$ is non-negative. 
 A density operator can be diagonalized; its eigenvalues are real, non negative, and sum up to 1. 
 For a pure state $\hat{\cal D}= |\psi \rangle  \langle \psi |$, all eigenvalues vanish but one, equal to 1.
 
 In the Schr\"odinger picture, the evolution of the time-dependent density operator $ \hat{\cal D}(t)$ is governed by the Hamiltonian
 $H$ of the system if it is isolated. The {\it Liouville--von Neumann equation of motion}, 
 
 \BEQ   i \hbar \frac{\d\hat{\cal D}(t)}{\d t} = [\hat H, \hat{\cal D}(t)],  
 \EEQ
generalizes the Schr\"odinger equation $i\hbar\d|\psi\rangle/\d t=\hat H|\psi\rangle$, 
or, in the position basis,  $i\hbar\d\psi(x)/\d t=\hat H\psi(x)$, 
which governs the motion of pure states. The evolution of $\hat{\cal D}(t)$ is unitary; 
it conserves its eigenvalues.
 
 In quantum statistical mechanics, the {\it von Neumann entropy}
 
 \BEQ                       S(\hat{\cal D})= - {\rm tr} \hat{\cal D} \ln \hat{\cal D} 
 \EEQ 
is associated with $\hat{\cal D}$. It characterizes the amount of information about the system that is missing when it is 
described by  $\hat{\cal D}$, the origin of values of $S$ being chosen as $S=0$ for pure states. If $S(\hat{\cal D})\neq 0$, 
$\hat{\cal D}$ can be decomposed in an infinite number of ways into a sum of projections onto pure states (\S~\ref{fin10.2.3}).

 The concept of density operator allows us to define the {\it state of a subsystem}, which is not feasible in the context of state 
 vectors or pure states. Consider a compound system S$_1$ + S$_2$, 
 described in the Hilbert space ${\cal H}_1\otimes {\cal H}_2$ by a density operator $\hat{\cal D}$. 
 This state is represented, in the basis $|i_1, i_2\rangle$  of ${\cal H}_1\otimes {\cal H}_2$, by the density matrix  
 $\langle i_1, i_2|\hat{\cal D}|j_1, j_2\rangle$. Suppose we wish to describe the subsystem S$_1$ alone, that is, 
 to evaluate the expectation values of the observables $O_1$ pertaining only to the Hilbert space ${\cal H}_1$ and thus 
 represented by matrices $\langle i_1|O_1|j_1\rangle$ in ${\cal H}_1$, or 
 $\langle i_1|O_1|j_1\rangle \delta_{i_2,j_2}$ in ${\cal H}_1\otimes{\cal H}_2$. These expectation values are given by

\BEQ 
\langle \hat O_1\rangle =tr_1 \hat{\cal D}_1 \hat O_1=\sum _{i_1,j_1} \langle i_1|\hat{\cal D}_1|j_1\rangle  \langle j_1|\hat O_1|i_1\rangle ,
\EEQ
where the matrix $\langle i_1|\hat{\cal D}_1|j_1\rangle$  in the Hilbert space ${\cal H}_1$ is defined by
 
 \BEQ                           \langle i_1|\hat{\cal D}_1|j_1\rangle =\sum_{i_2} \langle i_1, i_2|\hat{\cal D}|j_1, i_2\rangle .    
 \EEQ
The partial trace $\hat{\cal D}_1={\rm tr}_2 \hat{\cal D}$ on the space ${\cal H}_2$ is therefore, according to (G1), the density operator 
of the subsystem S$_1$. If the subsystems S$_1$ and S$_2$ interact, the evolution of $\hat{\cal D}_1$ should in principle be determined 
by solving (G3) for the the density operator $\hat{\cal D}$ of the compound system, then by taking the partial trace at the final time. 
The elimination of the bath (subsection 4.1) followed this procedure. The evolution of a subsystem is in general not unitary, 
because it is not an isolated system.

The formalism of density operators is more flexible than that of pure states: It affords the possibility not only of changing the basis in the
 Hilbert space, but also of performing linear transformations in the vector space of observables, which mix
 the left and right indices of observables $\langle i|\hat O|j\rangle$  
 and of density matrices $\langle i|\hat{\cal D}|j\rangle$. The resulting {\it Liouville representations of quantum mechanics }
 \cite{BalianAJPh1999,BalaszJennings,HilleryOConnellScullyWigner} are useful in many circumstances. 
 They include for instance the {\it Wigner representation}, suited to study the semi-classical limit,  and the {\it polarization representation} for a spin, currently used by experimentalists, 
 in which any operator is represented by its coordinates on the basis (\ref{sigmaxyz0}) of the space of operators;  in the present work, the parametrization of the
  state $\hat{D}$ of S + M by $P^{\rm dis}_{\rm M}(m)$ and $C^{\rm dis}_a(m)$ enters this framework (Eqs. (\ref{Rij}),  (\ref{P->R}), (\ref{RijP}), (\ref{P->C})). 
  
	
}



\renewcommand{\thesection}{\Alph{section}}
\section{Evolution generated by random matrices from the factorized ensemble}
\setcounter{equation}{0}
\setcounter{figure}{0}
\renewcommand{\thesection}{\Alph{section}.}
\label{AppendixH}


\hfill{\it For they have sown the wind, and they shall reap the whirlwind }

\hfill{   Hosea 8.7 }

\vspace{3mm}

\ZeText{

The purpose of this Appendix is to work out Eq.~(11.14) of the main
text, where the average is taken over an ensemble of random Hamiltonians
with the eigenvector distribution factorized from the eigenvalue
distribution. The eigenvectors are then distributed with the uniform
(Haar) measure, while we are free to choose the eigenvalue distribution 
(e.g. from some plausible physical arguments). The case where the random matrix elements are Gaussian
and distributed identically belongs to this class \cite{Mehta}. 
For simplicity we shall deal here with the microcanonical relaxation of one set of states.
The extension to two sets (the case discussed in the main text) is straightforward.

We thus need to determine the average evolution [inside this Appendix we take $\hbar=1$]
\BEA
\label{dsi1}
\overline{\hat U_{\Uparrow} \hat\rho\hat U^\dagger_{\Uparrow} }   =  
\overline{e^{-{it}\hat V_{\Uparrow}} \hat\rho\, e^{{it}\hat V_{\Uparrow}} }     ,
\EEA
where $\hat V_{\Uparrow}$ is a random matrix generated according to the above ensemble, and where $\hat\rho$ 
is an initial density matrix; see Eq.~(11.14) of the main text in this context. To calculate (\ref{dsi1}) we introduce
\BEA
\label{des}
\hat U_{\Uparrow} \hat\rho \hat U^{ \dagger}_{\Uparrow}
=\sum_{\alpha=1}^{G} \langle \psi_\alpha|\hat\rho|\psi_\alpha\rangle |\psi_\alpha\rangle\langle \psi_\alpha|
+\sum_{\alpha\not=\beta}^{G} \langle \psi_\alpha|\hat\rho|\psi_\beta\rangle\,\, |\psi_\alpha\rangle\langle \psi_\beta|
\,\,
e^{{it}(E_\beta-E_\alpha)},
\EEA
where
\BEA
\hat U_{\Uparrow}(t)=\sum_{\alpha=1}^{G} e^{{-it}E_\alpha} \, |\psi_\alpha\rangle\langle \psi_\alpha|
\EEA
is the eigenresolution of $\hat U_{\Uparrow}(t)$.

We now average (\ref{des}) over the states ${|\psi_\alpha\rangle}$ assuming that they are distributed uniformly (respecting the constraints 
of ortogonality and normalization). This averaging will be denoted by an overline,

\BEA
\overline{ \hat U_{\Uparrow} \hat\rho \hat U^{ \dagger}_\Uparrow } =G
\overline{   \langle \psi_1|\hat\rho|\psi_1\rangle \, |\psi_1\rangle\langle \psi_1| }
+ \overline{\langle \psi_1|\hat\rho|\psi_2\rangle\,\, |\psi_1\rangle\langle \psi_2| }
\sum_{\alpha\not=\beta}^{G} \,\,
e^{it(E_\beta-E_\alpha)}.
\label{gang}
\EEA
It suffices to calculate $\overline{ \langle \psi_1|\hat\rho|\psi_1\rangle \,\, |\psi_1\rangle\langle \psi_1|}$, 
since $\overline{\langle\psi_1|\hat\rho|\psi_2\rangle\,\, |\psi_1\rangle\langle \psi_2|}$ will  be deduced from putting $t=0$ in
(\ref{gang}). The calculation is straightforward:

\BEA
\label{des0}
\overline{   \langle \psi_1|\hat\rho|\psi_1\rangle \,\,  |\psi_1\rangle\langle \psi_1| }=(c_{40}-c_{22})\hat\rho+c_{22}\hat{1}
\EEA
where

\BEA
\label{des1}
c_{40}=\frac{\int_0^\infty \prod_{\alpha=1}^{G} \left(x_\alpha\d x_\alpha\right) x_1^4\,\delta\left[\sum_\alpha x^2_\alpha-1  \right]  }
{\int_0^\infty \prod_{\alpha=1}^{G} \left(x_\alpha\d x_\alpha\right)\,\delta\left[\sum_\alpha x^2_\alpha-1  \right] }, \qquad 
c_{22}=\frac{\int_0^\infty \prod_{\alpha=1}^{G} \left(x_\alpha\d x_\alpha\right) x_1^2\, x_2^2\,\delta\left[\sum_\alpha x^2_\alpha-1  \right]  }
{\int_0^\infty \prod_{\alpha=1}^{G} \left(x_\alpha\d x_\alpha\right)\,\delta\left[\sum_\alpha x^2_\alpha-1  \right] }.
\label{des2}
\EEA
The integration variables in (\ref{des1}) refer to random
components of a normalized vector. Expectedly, (\ref{des0}) is a linear
combination of $\hat\rho$ and the unit  matrix, because only this
matrix is invariant with respect to all unitary operators.

The calculation of (\ref{des1}) brings
\BEA
\label{des4}
c_{40}=2c_{22}, \qquad c_{22}=\frac{1}{G(G+1)}.
\EEA
Using (\ref{des4}, \ref{des0}) in (\ref{des}) we obtain:
\BEA
\overline{   \hat U_{\Uparrow} \hat\rho \hat U_{\Uparrow}^{ \dagger}}
=\frac{1}{(G+1)(G-1)}(G\hat 1-\hat\rho)
+ \frac{1}{(G+1)(G-1)} \left(\hat\rho-\frac{\hat 1}{G} \right)
\left|\sum_{\alpha=1}^{G} e^{it E_\alpha}\right|^2. 
\label{deso}
\EEA
For sufficiently large times the sum goes to zero. Neglecting terms of  ${\rm O}({\hat\rho}{G^{-2}})$ we obtain  from (\ref{deso}) that 

\BEA
\overline{   \hat U_{\Uparrow} \hat\rho \hat U_{\Uparrow}^{ \dagger}}
\to \frac{\hat1}{G}. 
\label{ds888}
\EEA
The considered arbitrary initial state $\hat\rho$ thus tends to the microcanonical distribution under the sole condition $G\gg 1$.

The relaxation in (\ref{ds888}) will be exponential, if we assume that the eigenvalues in (\ref{deso}) are Gaussian. Indeed, assuming that they are independently
distributed with zero average and dispersion $\Delta$ we get in the limit $G\gg 1$: $\sum_{\alpha=1}^{G} \exp({{it}
E_\alpha})\propto \exp({-{t^2\Delta^2}})$. Obviously, the same relaxation scenario (under the stated assumptions)
will hold for the off-diagonal components; see Eq.~(11.15) of the main text. 

The reason of the non-exponential relaxation for the Gaussian ensemble
is that all the non-diagonal elements of the random matrix are taken to
be identically distributed. This makes the distribution of the
eigenvalues bounded (the semi-circle law). If the elements closer to the
diagonal are weighted stronger, the distribution of the eigenvalues
will be closer to the Gaussian. The above factorized ensemble models
this situation. 

}

\renewcommand{\thesection}{\Alph{section}}
\section{Collisional relaxation of subensembles and random matrices } 
\setcounter{equation}{0}
\setcounter{figure}{0}
\renewcommand{\thesection}{\Alph{section}.}
\label{AppendixH}

\hfill{\it Collisions have a relaxing effect}

\hfill{Anonymous}

\vspace{3mm}

\ZeText
{

The purpose of this Appendix is to show that the evolution produced by a
random Hamiltonian|which is normally regarded as a description of a
closed, complex quantum system|may be generated within an open-system dynamics.
This enlarges the scope and applicability of the random matrix approach. 

\subsection{General discussion}

The ideas of collisional relaxation are well-known in the context of the classical Boltzmann equation. It is possible to extend the main ideas of
the linearized Boltzmann equation (independent collisions with a system in equilibrium) to the quantum domain \cite{partovi,mityugov,brailovskii}. We
shall first describe this scenario in general terms and then apply it to the specific situation described in \S~\ref{fin11.2.4-5}. 

Each collision is an interaction between the target quantum system $\mathbb{T}$ and a particle of the bath B. The interaction lasts a finite but short amount of time.
Then another collision comes, {etc}. The bath particles are assumed to be independent of one another and thermalized. Each collision is generated by the Hamiltonian

\BEA
\label{abner}
\hat H_{\rm \mathbb{T}+B} = \hat H_{\bf \mathbb{T}}+\hat H_{\rm B}+\hat H_{\rm I},
\EEA
where $\hat H_{\bf \mathbb{T}}$ and $\hat H_{\rm B}$ are the Hamiltonians of
$\mathbb{T}$ and B, respectively, and where $\hat H_{\rm I}$ is the
interaction Hamiltonian. Each collision is spontaneous and obeys the
strict energy conservation:
\BEA
\label{ds21}
[\hat H_{\rm I},\hat H_{\rm B}+\hat H_{\mathbb{T}}]=0.
\EEA
This condition guarantees that there are no energy costs for switching the collisional interaction 
$\hat H_{\rm I}$ on and off.

The initial density matrix of ${\rm B}$ is assumed to be Gibbsian (this assumption can be relaxed)
\BEA
\label{ds77}
\hat{\rho}_{\rm B}=\frac{1}{Z_{\rm B}}\exp[-\beta \hat H_{\rm B}] 
\EEA
with Hamiltonian $\hat H_{\rm B}$ and temperature $1/\beta=T>0$. The
target system starts in an arbitrary initial state $\rho_{\mathbb{T}}$ and
has Hamiltonian $\hat H_{\mathbb{T}}$. The initial state of ${\bf
\mathbb{T}}$ + B  is $\hat{\rho}_{\rm
\mathbb{T}+B}=\hat{\rho}_\mathbb{T}\otimes\hat{\rho}_{\rm B}$. The
interaction between them is realized via a unitary operator $\hat{{\cal
V}}$, so that the final state after the first collision is
\BEA
\hat\rho_{\rm \mathbb{T}+B}'=\hat{{\cal V}}\,\hat \rho_{\rm \mathbb{T}+B}\hat{{\cal V}}^\dagger, ~~~
\hat\rho_{\mathbb{T}}'={\rm tr}_{\rm B}\,\hat \rho_{\rm \mathbb{T}+B}'.
\label{ds444}
\EEA
For the second collision, the bath has lost memory of the first collision so that the new initial state of $\mathbb{T}$ + B is
$\hat \rho_{\mathbb{T}}'\otimes \hat \rho_{{\rm B}}$, and so on.

Let the energy levels of $\mathbb{T}$ involved in the interaction with ${\rm B}$
be degenerate: $\hat H_{\mathbb{T}}\propto \hat{1}$. 
Using (\ref{abner}--\ref{ds444}) and going to the eigenresolution of $\rho_{\rm B}$ we see 
that the evolution of $\mathbb{T}$ in this case can be described as a mixture of unitary processes
(note that in $\hat U^k$ below, $k$ is an index and not the power exponent)
\BEA
\label{ds91}
\hat \rho_\mathbb{T}'=\sum_k \lambda_k \hat U^k\, \hat \rho_\mathbb{T}\, \hat U^{k\,\dagger},\\
\hat U^k=\exp\left( {-i\delta\langle k |\hat H_{\rm I}|k\rangle}\right),\qquad  \hat U^k \hat U^{k\,\dagger}=\hat{1},
\EEA
where $\{\lambda_k\}$ and $\{|k\rangle\}$ are the eigenvalues and eigenvectors, respectively, 
of $\hat \rho_{\rm B}$ and $\delta$ is the interaction time.  Eq.~(\ref{ds91}) holds
for all subsequent collisions; now $k$
in (\ref{ds91}) is a composite index. Within this Appendix we put $\hbar=1$.

Note that the mixture of unitary processes increases the von Neumann entropy $S_{\rm vN}[\hat{\rho}_\mathbb{T}]=-{\rm
tr}[\hat{\rho}_\mathbb{T}\ln \hat{\rho}_\mathbb{T}]$ of $\mathbb{T}$; this is the concavity feature of $S_{\rm vN}$. 
Hence after sufficiently many collsiions $\mathbb{T}$ will relax to the microcanonical density matrix $\hat{\rho}_\mathbb{T}\propto \hat{1}$ that
has the largest entropy possible. 

The same process (\ref{ds91}) can be generated assuming the Hamiltonian $\langle k |\hat H_{\rm I}|k\rangle$ to be random, and then averaging
over it. This is closely related to \S~\ref{fin11.2.3} of the main text, where we postulated the random Hamiltonian $V_{\rm M}=\langle k |\hat
H_{\rm I}|k\rangle$ as a consequence of complex interactions. For the purpose of  \S~\ref{fin11.2.3}, $\mathbb{T}$ amounts to ${\rm S+M}$ (system +
magnet) and the complex interactions are supposed to take place in M.  In contrast, the averaging in (\ref{ds91}) arises due to tracing the bath out.
If the $\hat U^k$ mutually commute, (\ref{ds91}) means averaging over varying phases, i.e. it basically represents a (partial) dephasing in the common
eigenbasis of $\hat U^k$.

We shall apply the collisional relaxation to the target system $\mathbb{T} ={\rm S+M}$ after the measurement, so without the S-M
coupling {\bf $(g=0)$}.  We can directly apply mixtures of unitary processes for describing the relaxation; see (\ref{ds91}). Following to the discussion
in \S~\ref{fin11.2.3} of the main text [see the discussion before (11.12)], we assume that each unitary operator $\hat U^k$ in the mixture
(\ref{ds91}) will have the following block-diagonal form:

\BEA
\label{ds67}
\hat U^k=\hat{\Pi}_\Uparrow \hat U^{k}_{\Uparrow}\hat{\Pi}_\Uparrow +\hat{\Pi}_\Downarrow \hat U^{k}_{\Downarrow}\hat{\Pi}_\Downarrow ,
\EEA
where in view of (11.10) of the main text we defined the following projectors
\BEQ 
\hat{\Pi}_{\Uparrow} = \sum_\eta |m_F,\eta\rangle\langle m_F,\eta|,
\qquad
\hat{\Pi}_{\Downarrow} = \sum_\eta |\hspace{-0.7mm}-\hspace{-0.7mm} m_F,\eta\rangle\langle - m_F,\eta|.
\EEA

Eq.~(\ref{ds67}) is now to be applied to (11.9) of the main text, which yields

\BEA
\hat U^k |\Psi\rangle\langle\Psi|\hat U^{k\,\dagger} 
&=&\sum_{\eta\eta'}U_{\up\eta} U_{\up\eta'}^\ast  |\uparrow\rangle\langle\uparrow|\otimes 
\hat U^k_\Uparrow |m_F,\eta\rangle\langle m_F,\eta'| \hat U^{k\, \dagger}_\Uparrow
+\sum_{\eta\eta'} U_{\down\eta} U_{\down\eta'}^\ast
|\downarrow\rangle\langle\downarrow|\otimes 
\hat U^k_\Downarrow |\hspace{-0.7mm}-\hspace{-0.7mm}m_F,\eta\rangle\langle -m_F,\eta'| \hat U^{k\, \dagger}_\Downarrow\nonumber\\
&&+\left[\sum_{\eta\eta'}U_{\up\eta} U_{\down\eta'}^\ast |\uparrow\rangle\langle\downarrow|\otimes 
\hat U^k_\Uparrow |m_F,\eta\rangle\langle -m_F,\eta'| \hat U^{k\, \dagger}_\Downarrow +{\rm h.c.}\right],
\label{ds101}
\EEA
where ${\rm h.c.}$ means the hermitean conjugate of the last term. 

\subsection{Gaussian random matrix ensemble: characteristic time scale within the collisional relaxation scenario}

As we saw in the main text (\S~\ref{fin11.2.3}), the relaxation generated by the Gaussian
ensemble of random Hamiltonians (where the elements of the random matrix
Hamiltonian are identically distributed Gaussian random variables) is
not exponential. From the viewpoint of the collisional relaxation, the
averaging over a random matrix ensemble corresponds to a single
collision.  We now show that taking into account many short collisions
can produce exponential relaxation. 

Our technical task is to work out Eq.~(11.14) of the main text for multiple collisions. 
We introduce a shorthand $\hat\rho(0)=|m_F,\eta\rangle\langle m_F,\eta|$ and recall that within this Appendix $\hbar=1$.
Following the assumptions we made in \S~\ref{fin11.2.3} of the main text [see Eq.~(11.12)] we write 
$\hat U_{\Uparrow,\Downarrow}=e^{-it \hat{V}_{\Uparrow,\Downarrow}}$, where $\hat{V}_{\Uparrow}$ 
and $\hat{V}_{\Downarrow}$ are independent random matrices:
the elements $V_{\Uparrow,\,\, \eta\eta'}$ of $\hat V_{\Uparrow}$ in the basis $|m_F,\eta\rangle$ 
(and of $\hat V_{\Downarrow}$ in $|\hspace{-0.7mm}-\hspace{-0.7mm}m_F,\eta\rangle$)
are statistically independent, identically distributed random quantities with zero average and variance
\BEA
\label{ds72}
\overline{V_{\Uparrow,\, \eta_1\eta_2}V_{\Uparrow,\, \eta_3\eta_4}}=
\overline{V_{\Downarrow,\, \eta_1\eta_2}V_{\Downarrow,\, \eta_3\eta_4}}
=\frac{\Delta^2}{4G}
\delta_{\eta_1\eta_4}\delta_{\eta_2\eta_3}.
\EEA
Note that, for the Gaussian unitary ensemble characterized by the weight (11.12) for Hermitean matrices, the real and imaginary parts of the off-diagonal elements 
of $\hat V_\Uparrow$ ($\hat V_\Downarrow$) are statistically independent, and that (I.10) holds for both diagonal and off-diagonal elements.

We shall now assume that the duration $\delta$ of each collision is small and
work out the post-collision state
$\hat U_{\Uparrow}\hat\rho(t) U_{\Uparrow}^\dagger=e^{-i\delta \hat V_{\Uparrow}}\hat\rho(t) e^{i\delta \hat V_{\Uparrow}}$:
\BEA
e^{-i\delta \hat V_{\Uparrow}}\hat\rho(t) e^{i\delta \hat V_{\Uparrow}}=\hat\rho(t)-i\delta[\hat V_{\Uparrow}, \hat\rho]
-\frac{\delta^2}{2}\left\{\hat V_{\Uparrow}\hat\rho(t)+\hat\rho(t) \hat V_{\Uparrow}-2 \hat V_{\Uparrow}\hat\rho(t) 
\hat V_{\Uparrow}\right\}+{\cal O}(\delta^3).
\EEA
Averaging with help of (\ref{ds72}) produces
\BEA
&&\overline{\,\hat V_{\Uparrow}\hat\rho\,} =0, \qquad 
\overline{\,\hat V^2_{\Uparrow}\hat\rho\,} =\overline{\,\hat\rho \hat V^2_{\Uparrow}\,} 
=\frac{1}{4}\Delta^2\hat\rho,
\qquad \overline{\, \hat V_{\Uparrow}\hat\rho \hat V_{\Uparrow} \,}=\frac{\Delta^2}{4G}{\rm tr}(\hat\rho)\hat{1}.
\EEA
This brings
\BEA 
\hat\rho(t+\delta)=\hat\rho(t)- \frac{1}{4}\delta^2 \Delta^2\left[\hat\rho(t)-\frac{\hat{1}}{G}\right] +{\cal O}[\delta^4\Delta^4].
\label{ds90}
\EEA 
If the factor ${\cal O}[\delta^4\Delta^4]$ in (\ref{ds90}) is neglected, {\it i.e.} if
\BEA
\label{dskrys}
\delta^2\Delta^2\ll 1,
\EEA
(\ref{ds90}) can be extended to a recurrent relation for all subsequent collisions:
\BEA 
\label{dssol}
\hat\rho(n\delta)=\hat\rho((n-1)\delta)- \frac{1}{4}\delta^2 \Delta^2\left[\hat\rho((n-1)\delta)-\frac{\hat{1}}{G}\right],
\EEA 
where $n=1,2,\ldots$ is the number of collisions. Eq.~(\ref{dssol}) is solved as
\BEA 
\label{dsstop}
\hat\rho(n\delta)=(1-\frac{1}{4}\delta^2\Delta^2 )^n\hat\rho(0)+\frac{\hat{1}}{G}\left[1-(1-\frac{1}{4}\delta^2 \Delta^2)^n\right].
\EEA 
It is seen from (\ref{dsstop}) that the relaxation time of $\hat\rho(n\delta)\to \hat{1}/G$ is 
\BEA
-\frac{\delta}{\ln\left(1-\frac{1}{4}\delta^2 \Delta^2\right)}.
\label{dsrel}
\EEA

We now want to satisfy several conditions: ({\it i}) the magnitude $\sqrt{\,\,\overline{\,\hat V^2_{\Uparrow}\,}\,\,}=\Delta/2$ of the random
Hamiltonian has to be much smaller than $N$, because the random Hamiltonian has to be thermodynamically negligible. ({\it ii}) The relaxation
time (\ref{dsrel}) has to be very short for a large (but finite) $N$. ({\it iii}) Condition (\ref{dskrys}) has to hold. 

All these conditions can be easily satisfied simultaneously by taking,
e.g., $\Delta\propto N^{\gamma}$ and $\delta\propto N^{-\chi}$, where
\BEA
2\gamma>\chi>\gamma, \qquad \gamma<1. 
\EEA
Now the relaxation time will
be $\propto N^{\chi-2\gamma}\ll 1$, while (\ref{dskrys}) will
hold, because $N^{2(\gamma-\chi)}\ll 1$.

The same derivation applies to non-diagonal elements 
$\hat U_{\Uparrow}|m_F,\eta\rangle\langle -m_F,\eta'| \hat U_{\Downarrow}^\dagger
=e^{-i\delta \hat V_{\Uparrow}} |m_F,\eta\rangle\langle -m_F,\eta'| e^{i\delta \hat V_{\Downarrow}}$
in (\ref{ds101}). Instead of (\ref{dsstop}) we get
\BEA
\label{ds901}
\hat\rho(n\delta)=(1- \frac{1}{4}\delta^2 \Delta^2)\hat\rho((n-1)\delta), \qquad
\hat\rho(0)=|m_F,\eta\rangle\langle -m_F,\eta'|,
\EEA
with the same form of the characteristic time as for the exponential relaxation $\hat\rho(n\delta)\to 0$ for $n\to\infty$.

}

 \addcontentsline{toc}{section}{References}
 
 \section*{References}
 
\ZeText{

}


\end{document}